\documentclass[12pt, a4paper]{report}
\usepackage[utf8x]{inputenc}
\usepackage{amsmath,mathtools}
\usepackage{amsfonts}
\usepackage{amssymb}
\usepackage{graphicx}
\usepackage{appendix}
\usepackage{bm}
\usepackage{multicol}
\usepackage{geometry}
\usepackage{colortbl}
\usepackage{changepage}
\usepackage{color}
\usepackage{mathrsfs}
\usepackage{slashbox}
\usepackage{bigints}
\usepackage{pdflscape}
\usepackage{adjustbox}
\usepackage{ccaption}
\usepackage{tocloft}
\usepackage{lscape}
\usepackage[dvipsnames]{xcolor}
\usepackage[colorlinks=true,
            linkcolor=blue,
            urlcolor=blue,
            citecolor=blue]{hyperref}	

\newcommand{\ie}{{i.e.~}}

\newcommand{\eg}{\textsl{e.g.~}}

\newcommand{\order}[1]{\mathcal{O}\!\left(#1\right)}

\DeclareMathOperator{\erfc}{erfc}

\DeclarePairedDelimiterX\Basics[1](){ #1}

\newcommand{\uniform}{\mathcal{U}}
\newcommand{\normal}{\mathcal{N}}
\newcommand{\rdec}{r_{\rm dec}}

\DeclarePairedDelimiterX{\infdivx}[2]{(}{)}{%
  #1\;\delimsize\|\;#2%
}

\newcommand{\dkl}{D_{\rm KL}}
\newcommand{\djs}{D_{\rm JS}}

\newcommand{\dd}{\mathrm{d}}
\newcommand{\ee}{e}

\newcommand{\pisto}{\pi_\mathrm{sto}}
\newcommand{\pilog}{\pi_\mathrm{log}}
\newcommand{\Mmd}{M_{\mathrm{md}}}

\newcommand{\figpilogsto}{
\vspace{-0.1cm}
\hspace{3.8cm}
$\pilog$\hspace{6.2cm} 
$\pisto$
\vspace{-0.1cm}
}

\newcommand{\boldsigma}{\boldsymbol{\sigma}}
\newcommand{\boldz}{\boldsymbol{z}}

\newcommand{\sss}[1]{{\scriptscriptstyle{#1}}}

\newcommand{\uPl}{\mathrm{Pl}}

\newcommand{\umin}{\mathrm{min}}
\newcommand{\umax}{\mathrm{max}}
\newcommand{\uend}{\mathrm{end}}

\newcommand{\ueff}{\mathrm{eff}}

\newcommand{\ureh}{\mathrm{reh}}
\newcommand{\uereh}{\mathrm{ereh}}
\newcommand{\urad}{\mathrm{rad}}

\newcommand{\uS}{\mathrm{S}}

\newcommand{\usssS}{\sss{\uS}}

\newcommand{\usssPl}{\sss{\uPl}}

\newcommand{\nS}{n_\usssS}
\newcommand{\AmpS}{A_\usssS}

\newcommand{\alphaS}{\alpha_\usssS}

\newcommand{\uNL}{\mathrm{NL}}

\newcommand{\calP}{\mathcal{P}}

\newcommand{\calM}{\mathcal{M}}

\newcommand{\like}{{\cal L}}

\newcommand{\bmuf}{\boldsymbol{\mu}_{{}_{\rm F}}}

\newcommand{\muf}{\mu_{{}_{{\rm F}}}}

\newcommand{\MeV}{\mathrm{MeV}}
\newcommand{\GeV}{\mathrm{GeV}}

\newcommand{\boldx}{\boldsymbol{x}}
\newcommand{\boldy}{\boldsymbol{y}}

\newcommand{\Mp}{M_\usssPl}

\newcommand{\fnl}{f_\uNL}

\newcommand{\deci}{\mathscr{D}}

\newcommand{\efolds}{$e$-folds~}
\newcommand{\efold}{$e$-fold}

\newcommand{\beq}{\begin{equation}}
\newcommand{\eeq}{\end{equation}}
\newcommand{\bea}{\begin{eqnarray}}
\newcommand{\eea}{\end{eqnarray}}

\newlength{\wsingfig}
\newlength{\wdblefig}
\newlength{\wquadfig}
\newlength{\wtriplefig}
\newcommand{\barpi}{\bar{\pi}}
\newcommand{\barp}{\bar{p}}
\newcommand{\barlike}{\bar{{\cal L}}}
\newcommand{\hatpi}{\hat{\pi}}
\newcommand{\hatp}{\hat{p}}
\newcommand{\hatlike}{\hat{{\cal L}}}

\newcommand{\Eq}[1]{Eq.~(\ref{#1})}
\newcommand{\Eqs}[1]{Eqs.~(\ref{#1})}
\newcommand{\Fig}[1]{Fig.~{\ref{#1}}}
\newcommand{\Figs}[1]{Figs.~{\ref{#1}}}
\newcommand{\Ref}[1]{Ref.~{\cite{#1}}}
\newcommand{\Refs}[1]{Refs.~{\cite{#1}}}
\newcommand{\Sec}[1]{Sec.~\ref{#1}}
\newcommand{\Chap}[1]{Chapter~\ref{#1}}
\newcommand{\Secs}[1]{Secs.~\ref{#1}}
\newcommand{\App}[1]{Appendix~\ref{#1}}

\usepackage{tikz, lipsum}
\newcommand*{\chapnumfont}{\normalfont\sffamily\huge\bfseries}
\newcommand*{\printchapternum}{
  \begin{tikzpicture}
    \draw[fill,color=Plum] (0,0) rectangle (2cm,2cm);
    \draw[color=white] (1cm,1cm) node { \chapnumfont\thechapter };
  \end{tikzpicture}
}
\newcommand*{\chaptitlefont}{\normalfont\sffamily\Huge\bfseries}
\newcommand*{\printchaptertitle}[1]{\flushright\chaptitlefont#1}

\makeatletter
\def\@makechapterhead#1{%
  \vspace*{50\p@}%
  {\parindent \z@ \raggedleft
    \ifnum \c@secnumdepth >\m@ne
        \printchapternum
        \par\nobreak
        \vskip 20\p@
    \fi
    \interlinepenalty\@M
    \printchaptertitle{#1}\par\nobreak
    \vskip 40\p@
  }}
\def\@makeschapterhead#1{%
  \vspace*{50\p@}%
  {\parindent \z@ \raggedleft
    \interlinepenalty\@M
    \printchaptertitle{#1}\par\nobreak
    \vskip 40\p@
  }}
\makeatother

\newcommand{\niceline}{{\vskip-5ex} \rule[0.5ex]{\linewidth}{1pt} {\vskip+1ex}}

\begin{document}
\pagenumbering{alph}	
\thispagestyle{empty}	
\vspace*{5mm}	
\begin{center}
\linespread{1.3} 
\begin{figure}[t] 
\begin{center}
\includegraphics[width=5cm,trim=4 4 4 10,clip]{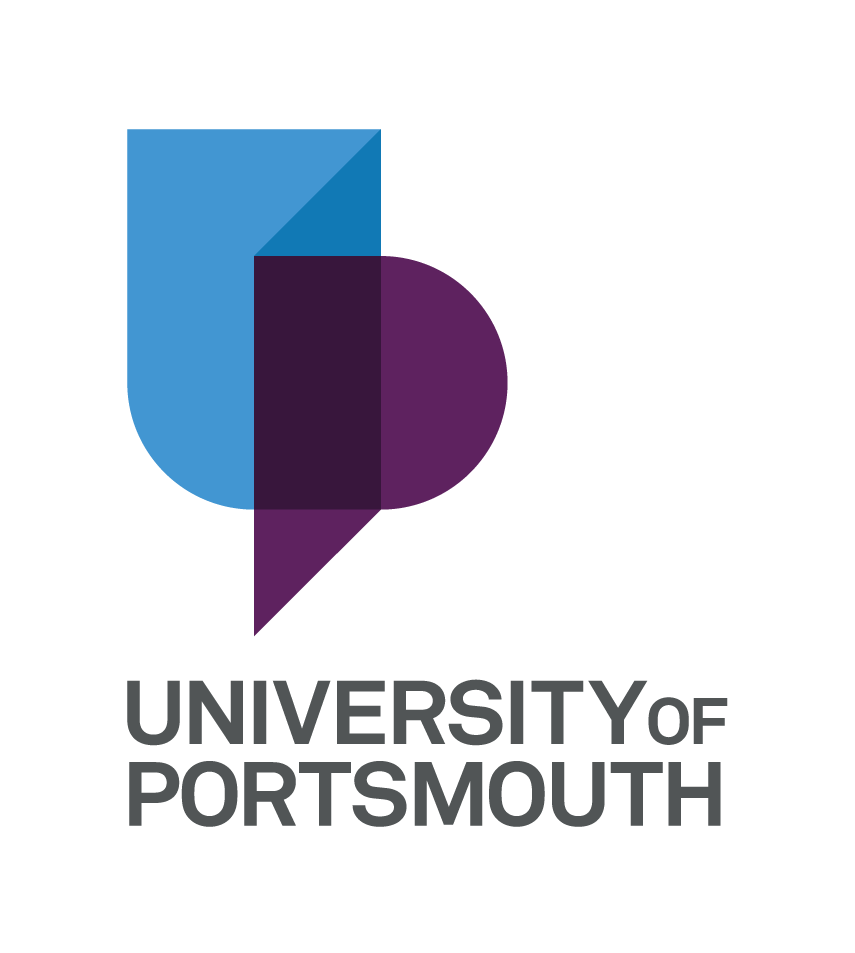} 
\end{center}
\end{figure}{\vskip-10ex}
\rule[0.5ex]{\linewidth}{1pt} \\
\vspace{2mm}
\huge\textbf{\textsf{Testing Inflationary Cosmology}}\\  
\vspace{2mm}
\rule[0.5ex]{\linewidth}{1pt} \\
\vspace{4mm}
\Large \textbf{\textsf{Robert J. Hardwick}}\\
\vspace{5mm}
\large Institute of Cosmology and Gravitation\\
\vspace{5mm}
\normalsize This thesis is submitted in partial fulfilment of\\
the requirements for the award of the degree of\\
Doctor of Philosophy of the University of Portsmouth.\\
\vspace{5mm}
\rule[0.5ex]{\linewidth}{1pt} \\
\begin{multicols}{2}
\begin{flushleft}
\textsc{Supervisors:} \\
Prof. David Wands \\
Dr. Vincent Vennin \\
Dr. Hooshyar Assadullahi
\end{flushleft}
\hfill
\begin{minipage}{.9\linewidth}
\vspace{5mm}
\begin{flushleft}
\includegraphics[width=6.0cm]{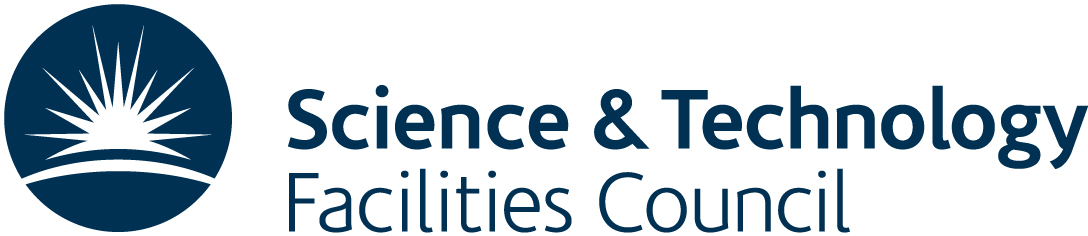}
\end{flushleft}
\end{minipage}
\end{multicols}
\rule[0.5ex]{\linewidth}{1pt} \\
\vspace{1mm}
\today	
\end{center}

\newpage
\pagenumbering{roman}	
\phantomsection	
\addcontentsline{toc}{chapter}{\textsf{Abstract}}	
\chapter*{\textsf{Abstract}} \niceline	


{\noindent Inflation is a period of accelerated expansion in the very early Universe that is typically invoked as a solution to the problem of originating the observed cosmic microwave background anisotropies, as well as those of the original hot big bang model. The particle content of the inflationary era is as-of-yet unknown, hence it is imperative that the best models of inflation are studied carefully for their potentially unique observational characteristics and then  compared to current observations in a statistically rigorous way.}

In this thesis we will primarily demonstrate how additional scalar degrees of freedom --- which are motivated from many high-energy embeddings --- open up new observational windows onto the physics of inflation. We construct a Bayesian framework to statistically compare models with additional fields given the current astronomical data. Putting inflation to the test, we perform our analysis on the quadratic curvaton accompanying a range of inflationary potentials, where we find that only one potential remains as a viable candidate. Furthermore, if the curvaton mechanism were to be confirmed by future non-Gaussianity measurements (from large scale structure surveys), the model could prove to be tremendously informative of the early inflationary history.

The initial conditions given to these scalar fields become apparent when considering their fundamentally quantum behaviour. Taking this physics into account leads us to develop detailed models for post-inflationary phenomenology (namely, the curvaton and freeze-in dark matter models) and to discover powerful new probes of inflation itself. We further demonstrate how this theoretical study complements our statistical approach by motivating the prior information in our Bayesian analyses.

The thesis finishes with a discussion of the future prospects for inflationary model selection. By hypothesising different toy survey configurations, we forecast different outcomes using information theory and our newly developed Bayesian experimental design formalism. In particular, we find that the most likely observable to optimise model selection between single-field inflationary models, through an order of magnitude precision improvement in the future, is the scalar spectral index. We conclude with a summary of the results obtained throughout.

\newpage
\renewcommand{\contentsname}{{\vskip-9ex}}
\chapter*{\textsf{Table of Contents}} \niceline
\pdfbookmark{\textsf{Table of Contents}}{Table of Contents}	
\tableofcontents 

\newpage

\renewcommand{\listtablename}{{\vskip-8ex}}
\chapter*{\textsf{List of Tables}} \niceline

\pdfbookmark{\textsf{List of Tables}}{List of Tables} 	
\listoftables 

\newpage

\renewcommand{\listfigurename}{{\vskip-8ex}}
\chapter*{\textsf{List of Figures}} \niceline

\pdfbookmark{\textsf{List of Figures}}{List of Figures} 	 
\listoffigures 


\newpage
\phantomsection
\addcontentsline{toc}{chapter}{\textsf{Declaration}}
\chapter*{\textsf{Declaration}} \niceline

{\noindent Whilst registered as a candidate for the above degree, I have not been registered for any other research award. The results and conclusions embodied in this thesis are the work of myself and have not been submitted for any other academic award.}

\Chap{sec:cosmological-intro} and \Chap{sec:statistical-intro} are introductory, written by myself and drawn from multiple references which are cited accordingly. 

\Chap{sec:curvaton-reheating} is primarily based on the work in: \href{http://iopscience.iop.org/article/10.1088/1475-7516/2016/08/042/pdf}{JCAP {\bf 1608} (2016), no.08, 042}. I am the primary author of this publication where the code, upon which the work relies, was partially written by and entirely run by myself.  

\Chap{sec:infra-red-divergences} is primarily based on the works in: \href{http://iopscience.iop.org/article/10.1088/1475-7516/2017/10/018}{JCAP {\bf 1710}, (2017)018}; and \href{http://iopscience.iop.org/article/10.1088/1475-7516/2018/05/054}{JCAP {\bf 1805}, no.05, 054(2018)}. I am the primary author and wrote the majority of text in both publications, where in the latter I was the sole author. The analytic and numerical calculations in both works were all performed by myself, where in some cases in the former work there were replications and additional checks on these by my co-authors. I was also the sole developer of the code in the latter publication. This chapter also contains some original calculations for non-minimally coupled spectator fields written by myself.

\Chap{sec:isocurvature-fields} is primarily based on the works in: \href{https://www.worldscientific.com/doi/abs/10.1142/S0218271817430258}{Int.\ J.\ Mod.\ Phys.\ D {\bf 26} (2017) no.12, 1743025}; \href{http://iopscience.iop.org/article/10.1088/1475-7516/2018/02/006}{JCAP {\bf 1802}, no.02, 006(2018)}; and \href{https://arxiv.org/abs/1712.05364}{arXiv:1712.05364}. I am the primary author of the first publication and a major co-author in the other two works. I wrote approximately half of the text in the first two works and a less, but still significant, component of the latter. All of the analytic and numerical calculations in these works were performed by myself, either in the first instance or as checks for my co-authors.

\Chap{sec:future-prospects} is primarily based on the work in: \href{http://iopscience.iop.org/article/10.1088/1475-7516/2018/05/070}{JCAP {\bf 1805}, no.05, 070(2018)}. The entire body of text was written and the code was developed by myself. 

\Chap{sec:conclusions-whole} is an original piece of writing by myself that is intended to summarise all previous sections. References are used where necessary. \\ 
\begin{center}
Word count: 49,095 words. \\
Ethical review code: 4C42-FF17-B7FF-2C76-32CC-FA5B-2ACD-593C
\end{center}

\newpage
\phantomsection
\addcontentsline{toc}{chapter}{\textsf{Acknowledgements}}
\chapter*{\textsf{Acknowledgements}} \niceline

I would like to sincerely thank all of my truly superb supervisors: Prof David Wands, Dr Vincent Vennin and Dr Hooshyar Assadullahi, for their expertise, advice and great humour throughout the last 3 years. In particular, I would like to thank: David for the huge amount of knowledge that you have imparted to me and the relaxed, encouraging way in which you imparted it; and Vincent, for both teaching me 
so much and for the extraordinary example you set for me in all aspects of research. 

To my examiners: Dr Roberto Trotta and Prof Robert Crittenden, I sincerely thank you both for your careful reading of the manuscript and insightful commments.  

I would also like to acknowledge and thank all of my collaborators, both past and present, from whom I have learned a huge amount: Christian Byrnes, Emanuela Dimastrogiovanni, Kari Enqvist, Matteo Fasiello, Kazuya Koyama, Tommi Markkanen, Sami Nurmi, Diederik Roest, Tommi Tenkanen and Jes\'{u}s Torrado.

To all of my fellow PhD colleagues: You are a fantastic bunch of people and I sincerely wish you to all achieve your dreams! Most notably to the pub crew --- Paul, Ben, Dan, Matt and Mike --- whose ridiculous conversations have always cheered me up (and inspired me). Thank you all so much for everything.

Lastly, and most importantly. To Camila and my family, who have always supported me through thick and thin: I love you all very dearly. I feel that this thesis sums up everything that you have all helped me to accomplish.

\newpage
\phantomsection
\addcontentsline{toc}{chapter}{\textsf{Dissemination}}
\chapter*{\textsf{Dissemination}} \niceline

\subsection*{\textsf{Publications}}

\textbf{R. J. Hardwick}, V. Vennin and D. Wands, 
``\emph{The decisive future of inflation},'' 
JCAP {\bf 1805}, no.05, 070(2018), \href{http://iopscience.iop.org/article/10.1088/1475-7516/2018/05/070}{doi:10.1088/1475-7516/2018/05/070}, [\href{https://arxiv.org/abs/1803.09491}{arXiv:1803.09491} [astro-ph.CO]]. \\
\textbf{R. J. Hardwick}, 
``\emph{Multiple spectator condensates from inflation},'' 
JCAP {\bf 1805}, no.05, 054(2018), \href{http://iopscience.iop.org/article/10.1088/1475-7516/2018/05/054}{doi:10.1088/1475-7516/2018/05/054}, [\href{https://arxiv.org/abs/1803.03521}{arXiv:1803.03521} [gr-qc]]. \\
J. Torrado, C. T. Byrnes, \textbf{R. J. Hardwick}, V. Vennin and D. Wands, 
``\emph{Measuring the duration of inflation with the curvaton},'' 
\href{https://arxiv.org/abs/1712.05364}{arXiv:1712.05364} [astro-ph.CO]. \\
K. Enqvist, \textbf{R. J. Hardwick}, T. Tenkanen, V. Vennin and D. Wands, 
``\emph{A novel way to determine the scale of inflation},'' 
JCAP {\bf 1802}, no.02, 006(2018), \href{http://iopscience.iop.org/article/10.1088/1475-7516/2018/02/006}{doi:10.1088/1475-7516/2018/02/006}, [\href{https://arxiv.org/abs/1711.07344}{arXiv:1711.07344} [astro-ph.CO]]. \\
\textbf{R. J. Hardwick}, V. Vennin and D. Wands,
``\emph{A Quantum Window Onto Early Inflation},'' 
Int.\ J.\ Mod.\ Phys.\ D {\bf 26} (2017) no.12, 1743025, \href{https://www.worldscientific.com/doi/abs/10.1142/S0218271817430258}{doi:10.1142/S0218271817430258}, [\href{https://arxiv.org/abs/1705.05746}{arXiv:1705.05746} [hep-th]]. \\
\textbf{R. J. Hardwick}, V. Vennin, C. T. Byrnes, J. Torrado and D. Wands, 
``\emph{The stochastic spectator},'' 
JCAP {\bf 1710}, (2017)018, \href{http://iopscience.iop.org/article/10.1088/1475-7516/2017/10/018}{doi:10.1088/1475-7516/2017/10/018}, [\href{https://arxiv.org/abs/1701.06473}{arXiv:1701.06473} [astro-ph.CO]]. \\
\textbf{R. J. Hardwick}, V. Vennin, K. Koyama and D. Wands,
``\emph{Constraining Curvatonic Reheating},'' 
JCAP {\bf 1608} (2016), no.08, 042, \href{http://iopscience.iop.org/article/10.1088/1475-7516/2016/08/042/pdf}{doi:10.1088/1475-7516/2016/08/042}, [\href{https://arxiv.org/abs/1606.01223}{arXiv:1606.01223} [astro-ph.CO]].

\newpage
\pagenumbering{arabic}	

\chapter{\textsf{Cosmological introduction}}
\label{sec:cosmological-intro} \niceline {\vskip+1ex} 


\begin{center}
\fbox{\parbox[c]{13cm}{\vspace{1mm}{\textsf{\textbf{Abstract.}}} In this chapter we will review the Friedmann-Lema\^{i}tre-Robertson-Walker (FLRW) Universe, cosmological inflation and reheating, emphasising the components in understanding that are necessary to read the main body of the thesis. More specifically, we shall focus on both the origin of divergences in inflationary correlation functions and the stochastic framework in which to calculate their observational effects, as well as the essential physics of perturbative reheating. For more pedagogical modern reviews, we suggest Refs.~\cite{Patrignani:2016xqp, Ade:2015lrj, Amin:2014eta, Drewes:2013iaa, Baumann:2009ds, Bassett:2005xm}.\vspace{1mm}}}
\end{center}

\section{\textsf{The FLRW Universe}}

\subsection{\textsf{Geometry}} \label{sec:geometry-flrw}

Through successive observations of the mass, distance and recessional velocity of astrophysical objects, we know that our Universe is expanding and cooling~\cite{peebles:1993,Liddle:452061,Mukhanov:991646,Peter:1208401}. It is also filled with a vast array of structures that are distributed on many length scales. Despite this complexity, on the largest (cosmological) length scales, the Universe appears to be statistically homogeneous and isotropic to all observers. Imposing these symmetries, one finds that the spacetime geometry of the Universe at these scales is well described by a Friedmann-Lema\^{i}tre-Robertson-Walker (FLRW) metric, with the following line element in spherical polar coordinates~\cite{Friedmann1924,doi:10.1093/mnras/91.5.483,1935ApJ....82..284R,doi:10.1112/plms/s2-42.1.90}
\begin{equation} \label{eq:FLRW-metric}
\dd s^2 = g_{\mu \nu}\dd x^\mu \dd x^\nu = -\dd t^2 + a^2(t) \left( \frac{\dd r^2}{1-Kr^2} + r^2\dd \Omega_2 \right) \,,
\end{equation}
where $r$ is the radial distance from a fundamental observer, $K$ is a constant scalar curvature,\footnote{This can be a positive number for spatially closed, 0 for spatially flat and negative for spatially open universes.} $\dd \Omega_2\equiv \dd \theta_1^2 + \sin^2\theta_1 \dd \theta_2^2$ is the 2-dimensional solid angle, $a(t)$ is the FLRW scale factor and $t$ is cosmic time: the proper time of the fundamental observer. \Eq{eq:FLRW-metric} is very simple due to the symmetries of homogeneity and isotropy. If the scale factor were to spatially vary $a(t)\rightarrow a(x^i,t)$ then homogeneity would be violated. 

A useful parameterisation $\dd \eta = \dd t / a(t)$ factors expansion out of the time elapsed for the observer, converting \Eq{eq:FLRW-metric} into
\begin{equation} \label{eq:FLRW-metric-conf}
\dd s^2 =  a^2(\eta ) \left( -\dd \eta^2 + \frac{\dd r^2}{1-Kr^2} + r^2\dd \Omega_2 \right) \,,
\end{equation}
where $\eta$ is known as `conformal time'. Note that throughout this thesis the convention of (Planck) natural units ($c=1$, $\hbar = 1$, $k_{\rm B}=1$) will be adopted.

\subsection{\textsf{Dynamics}}

In order to calculate how the Universe dynamically evolves, one must introduce a theory of gravitation. The Einstein-Hilbert action~\cite{Einstein:1915ca,Hilbert:1915tx} of General Relativity is given by
\begin{equation} \label{eq:einstein-hilbert}
\ S = \frac{\Mp^2}{2} \int \dd^4 x \sqrt{-g} R \,,
\end{equation}
in which $\Mp \simeq 2.435 \times 10^{18}{\rm GeV}$ (in natural units) is the reduced Planck mass, $R_{\mu \nu}$ is the Ricci tensor\footnote{In General Relativity the Ricci tensor is a contraction of the Riemann tensor 
\begin{equation*}
\ R_{\mu \nu}\equiv R^{\rho}{}_{\mu \rho \nu} = 2\Gamma^\rho{}_{\mu [\nu \rho]} + 2\Gamma^\rho{}_{\sigma [ \rho}\Gamma^\sigma{}_{\nu ]\mu} \,,
\end{equation*}
where we are using square brackets to denote antisymmetrising $A_{[ab]}=(A_{ab}-A_{ba})/2$ and the Christoffel symbols are 
\begin{equation*}
\Gamma^\rho{}_{\mu \nu} = \frac{1}{2}g^{\rho \sigma}\left( \partial_\mu g_{\sigma \nu} + \partial_\nu g_{\sigma \mu} - \partial_\sigma g_{\mu \nu} \right) \,.
\end{equation*}} and its contraction $R\equiv R^\mu{}_{\mu}$ is the Ricci scalar. Note that in \Eq{eq:einstein-hilbert} the integral comes equipped with a spacetime 4-volume element $\sqrt{-g}\equiv \sqrt{-\det (g_{\mu \nu})}$. If one varies \Eq{eq:einstein-hilbert} with respect to $g_{\mu \nu}$ 
\begin{equation}
\frac{\delta S}{\delta g_{\mu \nu}} = 0  \quad \Longleftrightarrow \quad R_{\mu \nu} = 0\,,
\end{equation}
which is the vacuum solution to the theory. We note here that the convention of Latin and Greek indices to represent 3 and 4-vectors, respectively, will be used throughout this section unless otherwise indicated.

Adding gravitating matter --- through its Lagrangian density ${\cal L}_{\rm m}$ --- and a cosmological constant $\Lambda$ to the Universe gives a new action
\begin{equation} \label{eq:efes-action}
\ S = \Mp^2 \int \dd^4 x \sqrt{-g}\left(  \frac{1}{2} R - \Lambda \right) + \int \dd^4 x \sqrt{-g} {\cal L}_{\rm m} \,,
\end{equation}
from which we can see that $\Mp$ plays the role of a coupling between matter and the gravitational field $g_{\mu \nu}$. From \Eq{eq:efes-action} the energy-momentum tensor $T^{\mu \nu}$ of the matter fields can be obtained 
\begin{equation}
\frac{1}{2}\sqrt{-g}T^{\mu \nu} = -\frac{\partial \left( \sqrt{-g} {\cal L}_{\rm m} \right)}{\partial g_{\mu \nu}} = \sqrt{-g}\frac{\partial {\cal L}_{\rm m}}{\partial g_{\mu \nu}} - \frac{1}{2}\sqrt{-g}g^{\mu \nu}{\cal L}_{\rm m} \,.
\end{equation}
Hence, by varying the overall action in \Eq{eq:efes-action} with respect to $g_{\mu \nu}$, one arrives at the Einstein field equations
\begin{equation} \label{eq:efes}
\frac{2}{\sqrt{-g}\Mp^2}\frac{\delta S}{\delta g_{\mu \nu}} = 0  \quad \Longleftrightarrow \quad R_{\mu \nu} -\frac{1}{2}Rg_{\mu \nu} + \Lambda g_{\mu \nu} = \frac{1}{\Mp^2}T_{\mu \nu}\,,
\end{equation}
which describe how the energy-momentum of matter sources the dynamics of $g_{\mu \nu}$.

For a fundamental observer moving with respect to the rest frame in a perfect fluid with density $\rho$, pressure ${\sf P}$ and four velocity vector $u^\mu$, one can identify $T_{\mu \nu} = (\rho + {\sf P})u_\mu u_\nu + {\sf P}g_{\mu \nu}$. When the observer is at rest, $u_0=1$ and $u_i=0$, so we find that $T_{\mu}{}^\nu = {\rm diag}(-\rho , {\sf P}, {\sf P}, {\sf P})$, which is consistent with the homogeneity and isotropy assumed by FLRW if $\rho$ and ${\sf P}$ do not spatially vary. One may then use \Eq{eq:FLRW-metric} and components of \Eq{eq:efes} to derive the following equations: The $00$-component
\begin{align}
\ H^2 &= \frac{\rho}{3\Mp^2} + \frac{\Lambda}{3} - \frac{K}{a^2}\,, \label{eq:friedmann} \\ \intertext{and the $ii$-component} 
\frac{1}{a}\frac{\dd^2 a}{\dd t^2} &= -\frac{\rho}{6\Mp^2}(1+3w) + \frac{\Lambda}{3} \label{eq:raychaud}\,,
\end{align}
where we have defined the Hubble parameter $H\equiv \dd{\ln}a/ \dd t$ and the equation of state parameter $w \equiv {\sf P}/\rho$. Eqs.~\eqref{eq:friedmann} and~\eqref{eq:raychaud} are also consistent with the continuity equation
\begin{equation} \label{eq:continuity}
\frac{\dd{\ln}\rho}{\dd t} = -3H (1+w) \,,
\end{equation}
which may also be derived from the 0-component of the conservation of energy-momentum,\footnote{In fact, \Eq{eq:efes} satisfies the more general energy-momentum conservation law $\nabla_\mu \left( R^{\mu \nu} - g^{\mu\nu}R/2 + \Lambda g^{\mu\nu}\right) = \nabla_\mu T^{\mu \nu}/\Mp^2 = 0$ where the geometric side of the relation follows from the Bianchi identities.} i.e., $\nabla_\mu T^{\mu 0}=0$, where $\nabla^\mu$ denotes a covariant derivative. The general solution of \Eq{eq:continuity} is straightforward
\begin{equation}
\rho (t) = \rho (t_0) \exp \left\{ \int^{t_0}_t 3H(\tilde{t})\left[ 1+w(\tilde{t})\right] \dd \tilde{t} \, \right\} \label{eq:gensoln-continuity} \,.
\end{equation}
Note that, in the limit where $w$ is constant: substituting \Eq{eq:friedmann} into \Eq{eq:gensoln-continuity} yields 
\begin{equation}
\frac{\rho}{\rho_0} = \left( \frac{a_0}{a}\right)^{3(1+w)} \label{eq:soln-continuity-constw} \,,
\end{equation}
where we have now implicitly dropped the time dependencies of each quantity $A$ such that $A\equiv A(t)$ and $A_0\equiv A(t_0)$. Familiar solutions to \Eq{eq:soln-continuity-constw} include: a vacuum energy ($w=-1$); a matter-like energy density ($w=0$); and a radiation-like (conformal) energy density ($w=1/3$). Note also that the cosmological constant coincides with a vacuum energy-like $w=-1$ and spatial curvature $K$ can be identified as a fluid with $w=-1/3$. 

\begin{figure}[h]
\begin{center}
\includegraphics[width=10cm]{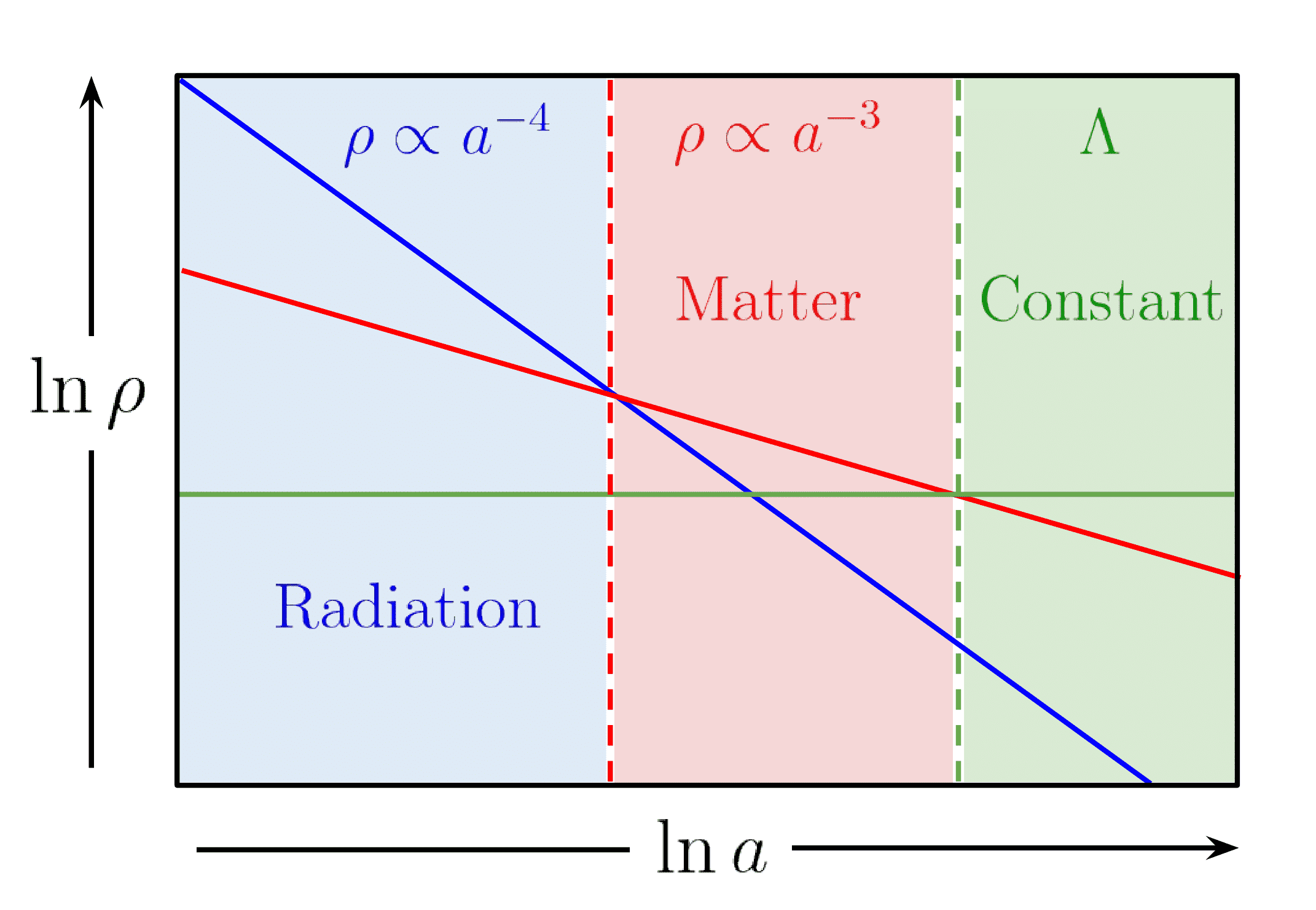}
\caption[Energy density scaling illustration]{Simple illustration of the `stitching together' of epochs with different scaling in energy density.}
\label{fig:endensscale}
\end{center}
\end{figure}

By replacing $\rho \rightarrow \sum_i\rho_i$ in \Eq{eq:friedmann} to account for all distinct constituents of gravitating matter in the Universe, one may account for a more complex cosmic history by `stitching together' separate epochs of $\rho_i$-dominated expansion. Each $\rho_i$-dominated epoch may dilute with energy density differently according to an equation of state $w_i$ and hence one may make a multiplicative chain to track the evolution of the total energy density using barotopic terms taking the form of \Eq{eq:soln-continuity-constw}. Such  calculation is illustrated in \Fig{fig:endensscale}, which corresponds to the true scaling in energy density that is expected in the cosmic past.

\subsection{\textsf{Past}} \label{sec:past}

\Fig{fig:endensscale} reveals an important characteristic of our expanding Universe: those components of matter which dilute more efficiently with expansion are, conversely, expected to dominate the total energy density in the distant past. Tracking the evolution backward in time, one can invert \Eq{eq:soln-continuity-constw} to find that the Universe must become both increasingly dense and thus, because it was radiation dominated, at a higher temperature. The oldest light detected from this era is known as the Cosmic Microwave Background (CMB) radiation. 

\begin{figure}[h]
\begin{center}
\includegraphics[width=10cm]{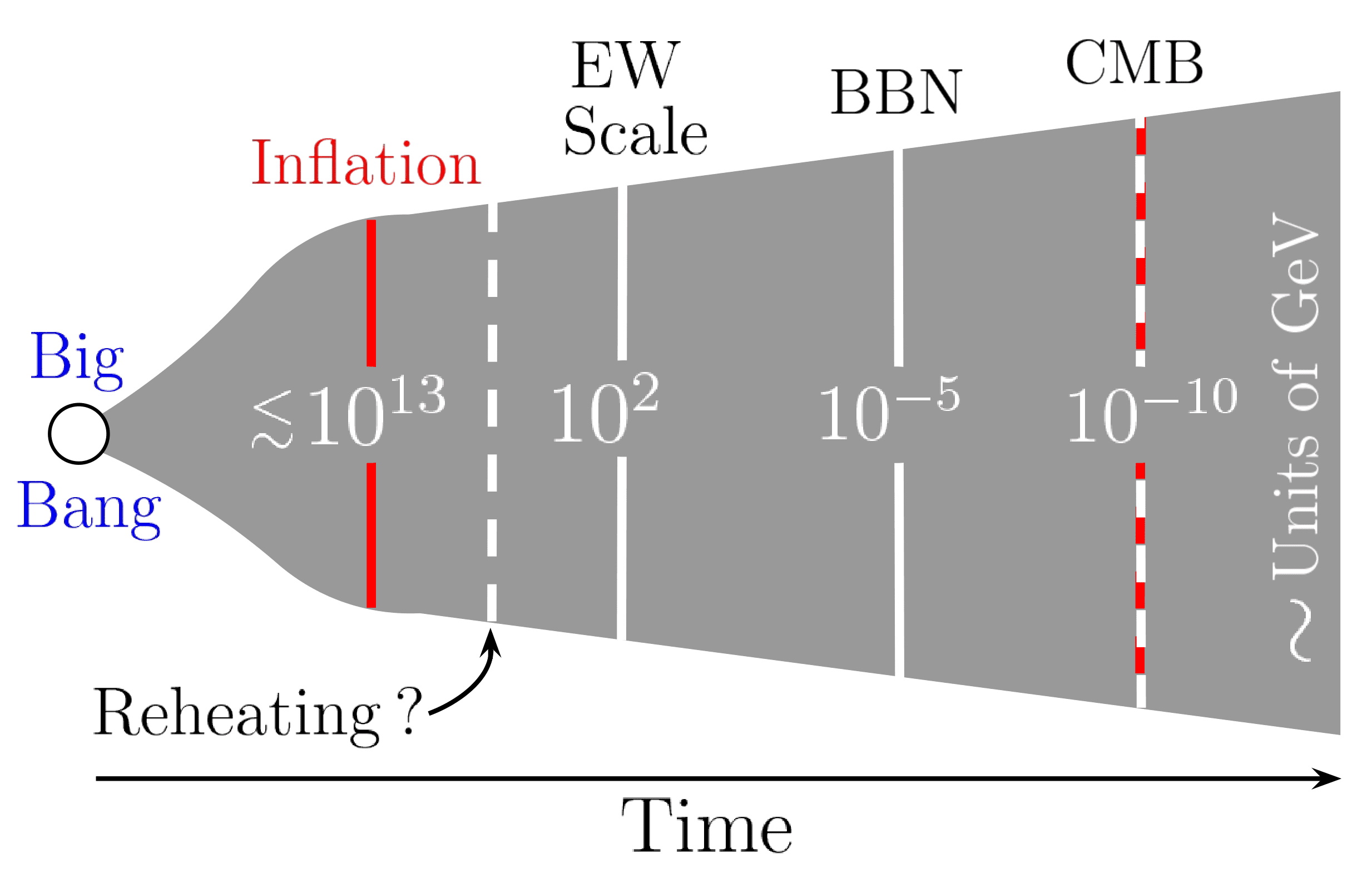}
\caption[Cosmic history illustration]{An illustrated timeline of some epochs in the cosmic thermal history that are key to understanding this thesis. The temperature of the Universe at each point in units of GeV is indicated in white. Inflation is typically the highest energy process, followed by reheating and the ElectroWeak (EW) phase transition=, Big Bang Nucleosynthesis (BBN) is the epoch when the Universe cools enough to form stable nuclei and the CMB subsequently forms. }
\label{fig:coshist-illus}
\end{center}
\end{figure}

The CMB is a near-perfect blackbody spectrum of radiation --- measured to have a temperature today of $T^{\rm CMB}_{0}\simeq 2.35\times 10^{-4}{\rm eV}$ --- which formed when the Universe cooled sufficiently such that free electrons and protons could bind to form neutral Hydrogen (a process known as recombination) during the matter era (labeled in \Fig{fig:endensscale}). We have indicated when the CMB forms relative to the earliest epochs in \Fig{fig:coshist-illus}. In order to better understand the key processes expected at earlier times, and how the CMB formed, one needs to understand the properties of a thermal bath of particles in a cosmological context. In light of this, we shall briefly review some of the required elements in statistical mechanics.

The central object in describing the state space of many-body systems is the distribution function $f({\rm state})$ --- from which observable quantities, such as pressure and temperature, may be calculated by integration with an appropriate function. The `state' of the system is generally a configuration in a time-dependent phase space, and hence we have $f=f(p_i,x_i,t)$, where we remind the reader that $p_i$ and $x_i$ denote the corresponding 3-vector components in momentum and space, respectively. Typically, if the interaction rate of a system is sufficiently high, it reaches thermodynamic equilibrium and hence equilibrium distribution functions $f_{\rm eq} = f_{\rm eq}(p_i,x_i)$ may be used. We note that $f_{\rm eq}(p_i,x_i)$ are often analytic functions where there is no longer any explicit temporal variation due to stationarity. For an adiabatically expanding Universe that retains thermal equilibrium, however, implicit time dependence is still present since the energy of the system will decrease with increasing volume of the thermal bath.\footnote{One can easily see this by considering the first law of thermodynamics $\dd E = T\dd {\cal S} - {\sf P}\dd V$, which is valid for a change in total energy $\dd E$ of a closed system in thermal equilibrium by either a total change in entropy $\dd {\cal S}$ or volume $\dd V$. }   

The equilibrium distribution function for a species of particle, with degrees of freedom $g$, that is relativistic (its rest mass $m\ll T$, where $T$ is the temperature of the thermal bath) is a stationary solution of the  relativistic Boltzmann equation. The relativistic Boltzmann equation takes the form
\begin{equation} \label{eq:boltz-rel}
\hat{L}[f] = \hat{C}[f] \,,
\end{equation} 
where the relativistic Liouville operator $\hat{L}$ is
\begin{equation}
\hat{L} = p^\mu\frac{\partial}{\partial x^\mu} - \Gamma^{\sigma}{}_{\mu \nu}p^\mu p^\nu \frac{\partial}{\partial p^\sigma} \label{eq:liouvilleop} \,,
\end{equation}
and $\hat{C}$ is the collision operator. In an FLRW Universe, and hence using \Eq{eq:FLRW-metric} and its Christoffel symbols, this operator reduces to  
\begin{equation} \label{eq:liouville-FLRW}
\hat{L} = - E\frac{\partial}{\partial t} + H\vert p_i \vert^2\frac{\partial}{\partial E} \,,
\end{equation} 
where we have assumed statistical homogeneity and $\vert p_i\vert^2 $ is the square magnitude of the 3-momentum vector.\footnote{Note that we have made use of the invariant $p^\mu p_\mu = -E^2+\vert p_i \vert^2$ and the fact that only $\Gamma^0_{\mu \nu}$ is non-vanishing.} In a semi-classical treatment $\hat{C}$ must contain the fact that the scattering species is either Fermionic or Bosonic, whose $2\rightarrow 2$ scattering collision operator will take the form~\cite{Escobedo2003aa}
\begin{align} \label{eq:collision-op-rel}
\hat{C}[f] &=  \int \int \int \dd^3 p' \dd^3 p'' \dd^3 p''' \, {\cal T}(f,f'\vert f'',f''') \, {\cal F}(f,f',f'',f''') \\
\ {\cal F}(f,f',f'',f''') &\equiv  f''f''' \left( 1 + \Upsilon f \right) \left( 1 + \Upsilon f' \right) - ff' \left( 1  + \Upsilon f'' \right) \left( 1 + \Upsilon f''' \right)  \label{eq:collision-op-rel-midterm} \,,
\end{align}
where ${\cal T}(f,f'\vert f'',f''')$ is the $p_i,p_i' \rightarrow p_i''p_i'''$ transition rate and in our notation $f^{(n)}=f(p^{(n)}_i,x^{(n)}_i,t)$ is the phase space distribution function over the $n$-th (denoting the number of primes $'$) particle. If the species is Fermionic, an initial or final scattering state cannot be occupied at the same time by both particles and hence $\Upsilon = -1$ should be chosen in \Eq{eq:collision-op-rel-midterm}. Equivalently, if the species were Bosonic, $\Upsilon = +1$ should be chosen due to the fact that a Boson can occupy any of the initial or any of the final states. Finally, to be governed by Maxwell-Boltzmann statistics, the value of $\Upsilon =0$ should be used.

Now consider the system in an FLRW background at equilibrium (stationary limit) so $\partial f_{\rm eq}/\partial t = 0$ such that the first term on the left hand side of \Eq{eq:liouville-FLRW} vanishes. The second term accounts for the expansion rate which limits the progress towards equilibrium by increasing the distance between scattering particles. However, in the high interaction rate limit this term is negligible to the collision term and so it too can be treated as vanishing. Hence, because now $\hat{L}[f_{\rm eq}]=0$, \Eq{eq:boltz-rel} leaves us with the requirement that 
\begin{gather} 
\ {\cal F}(f_{\rm eq},f'_{\rm eq},f''_{\rm eq},f'''_{\rm eq}) = 0 \nonumber \\
\Rightarrow \ln \left( \frac{f''_{\rm eq}}{1 +\Upsilon f''_{\rm eq}} \right) + \ln \left( \frac{f'''_{\rm eq}}{1+\Upsilon f'''_{\rm eq}} \right) = \ln \left( \frac{f_{\rm eq}}{1+\Upsilon f_{\rm eq}} \right) + \ln \left( \frac{f'_{\rm eq}}{1+\Upsilon f'_{\rm eq}} \right)  \,.
\label{eq:collision-condition-eq}
\end{gather}
\Eq{eq:collision-condition-eq} suggests that the quantity $\ln [f_{\rm eq}/(1+\Upsilon f_{\rm eq} )]$ is invariant under scattering, and hence is equal to a linear combination of other invariants
\begin{equation} \label{eq:deriv-fd-be}
\ln \left[ \frac{f_{\rm eq}(p_i)}{1+\Upsilon f_{\rm eq}(p_i)}\right] = C_1 + C_2 E \quad \Longrightarrow \quad f_{\rm eq}(p_i) = \frac{1}{\exp \left( C_1+C_2E \right) - \Upsilon}\,.
\end{equation}
The constants $C_1$ and $C_2$ in \Eq{eq:deriv-fd-be} can be determined such that we may identify the Fermi-Dirac/Bose-Einstein/Maxwell-Boltzmann distributions by integration over the total number of particles\footnote{A full derivation of \Eq{eq:rel-distf} requires a maximum probability analysis over the state space, such as the Darwin-Fowler method~\cite{doi:10.1080/14786440908565189}. Here we shall simply quote the result. } 
\begin{equation} \label{eq:rel-distf}
\tilde{f}_{\rm eq}(p_i) = \frac{\nu}{(2\pi )^3}\frac{1}{\exp \left( \frac{E-\mu}{T}\right) - \Upsilon} \,,
\end{equation}
where we note that $E=\sqrt{m^2+\vert p_i\vert^2}$ is the relativistic energy of the particle and $m$ is its rest mass. An additional factor of $\nu /(2\pi)^3$ is present in \Eq{eq:rel-distf} to account for the number of degenerate spin states per unit volume.

\Eq{eq:rel-distf} is a distribution from which one can extract number density ${\sf n}$, energy density $\rho$ and pressure ${\sf P}$ from the microphysics of the relevant species. For a Bosonic species 
\begin{align}
\ {\sf n}_{\rm B}  &= \int \dd^3 p  \left. \tilde{f}_{\rm eq} (p_i) \right\vert_{\Upsilon = +1} = \frac{\nu \zeta (3)}{\pi^2} T^3 \label{eq:num-dens-bos} \\
\rho_{\rm B} &= \int \dd^3 p \, E \left. \tilde{f}_{\rm eq} (p_i) \right\vert_{\Upsilon = +1} = \frac{\nu}{30}\pi^2 T^4 \label{eq:en-dens-bos} \\
\ {\sf P}_{\rm B} &= \int \dd^3 p \, \frac{\vert p_i\vert^2}{3E}  \left. \tilde{f}_{\rm eq} (p_i) \right\vert_{\Upsilon = +1} = \frac{\rho_{\rm B}}{3} \label{eq:pressure-bos} \,,
\end{align}
in the relativistic limit ($m,\mu \ll T$), where $\zeta (3) \simeq 1.202$ is a value of the Riemann zeta function. The corresponding number density, energy density and pressure for a Fermionic ($\Upsilon = -1$) species are ${\sf n}_{\rm F} = 3{\sf n}_{\rm B}/4$, $\rho_{\rm F} = 7\rho_{\rm B}/8$ and ${\sf P}_{\rm F}=\rho_{\rm F}/3$. Note that \Eq{eq:pressure-bos}\footnote{The factor of $\vert p_i\vert^2/(3E)$ is correct if one considers 3 spatial directions each with a magnitude in rate of change in momentum per unit area (or force per unit area) integrated over the spatial volume used by the motion of particles $\dd^3x_i \dd p_i / (\dd t \dd A) = \left[ \vert p_i \vert^2/(3E) \right] (\dd t \dd A)/(\dd t \dd A)$.} (and its Fermionic counterpart) correctly reproduce the equation of state for radiation ($w=1/3$) which is used in \Eq{eq:soln-continuity-constw}. Notice also that \Eq{eq:continuity} can now be confirmed by integrating \Eq{eq:boltz-rel} in the collisionless limit over $\dd^3p$ and combining with Eqs.~\eqref{eq:liouvilleop}, \eqref{eq:en-dens-bos} and \eqref{eq:pressure-bos}.  

For these relativistic species, as the temperature decreases with expansion, eventually they will fall out of thermal equilibrium. The quantities $({\sf n},\rho , {\sf P})$ will then become frozen in at their decoupling value, which is then diluted through the increase of volume during expansion. Note that because Eqs.~\eqref{eq:num-dens-bos}, \eqref{eq:en-dens-bos} and \eqref{eq:pressure-bos} all depend on temperature (and equivalently for the Fermions) this subsequent dilution can be accounted for by considering how the temperature reduces with expansion. Notice that this scaling can be easily connected to the equation of state of the thermal bath by comparing \Eq{eq:en-dens-bos} to \Eq{eq:soln-continuity-constw}. Hence, we find that $T\propto 1/a$.

\subsection{\textsf{Composition}}

\begin{figure}[h]
\begin{center}
\includegraphics[width=6cm]{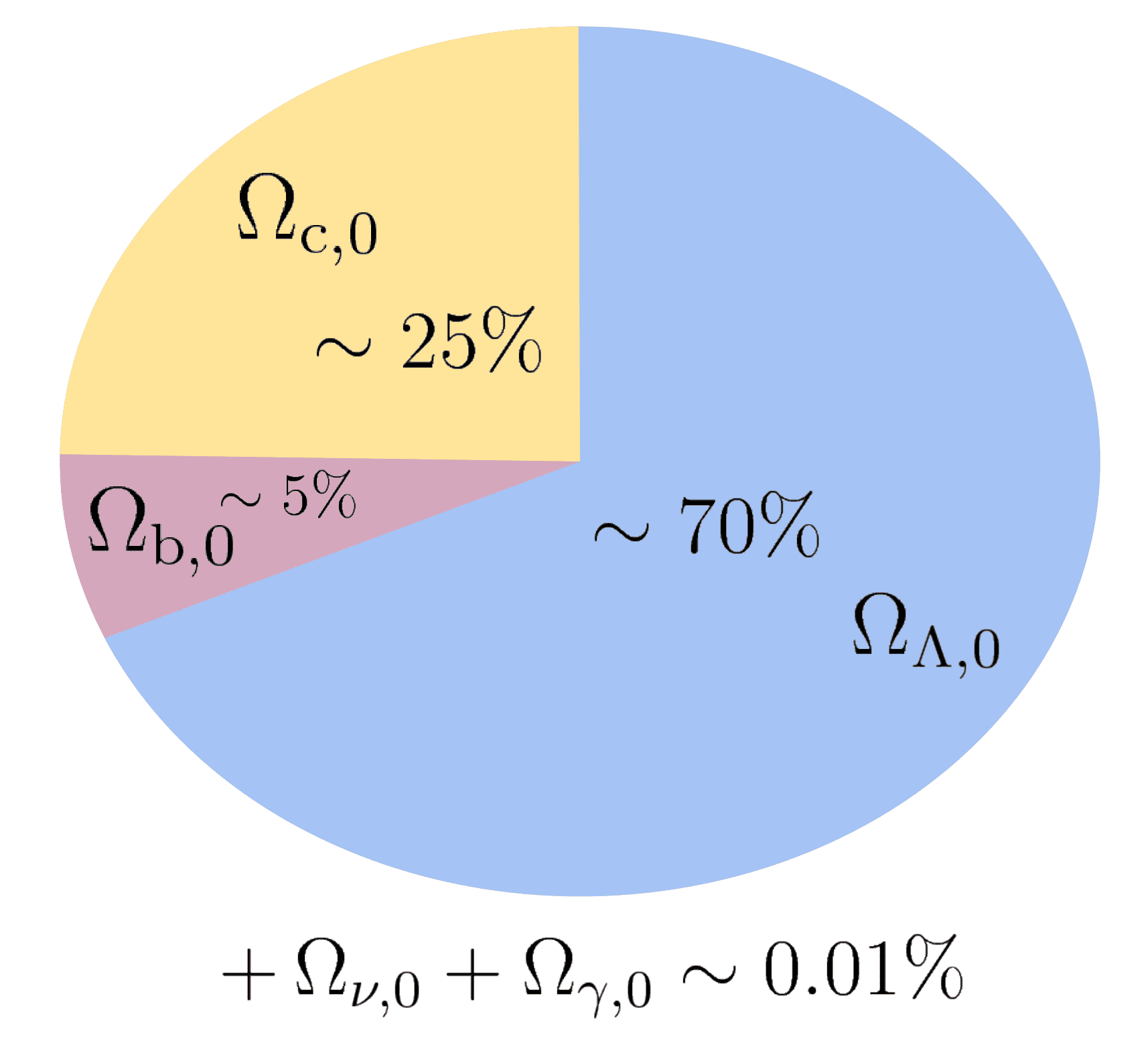}
\caption[Cosmic pie chart]{A `cosmic pie chart' indicating the approximate percentages of components that make up the Universe. Precise values with measurement uncertainties may be found in~\Ref{Aghanim:2018eyx}.}
\label{fig:cos-pie}
\end{center}
\end{figure}

Let us define $\rho_{\rm tot}$ as the total energy density of the Universe. Summing over: relativistic species in the Standard Model (SM), i.e., neutrinos $\rho_{\nu}$ and photons $\rho_{\gamma}$; vacuum energy density $\rho_\Lambda \equiv \Lambda \Mp^2$; Baryonic matter $\rho_{\rm b}$; and Dark matter $\rho_{\rm c}$ in \Eq{eq:friedmann} we find
\begin{align}
\ H^2 &= \frac{1}{3\Mp^2} \left( \rho_{\nu} + \rho_{\gamma} + \rho_{\rm b}+\rho_{\rm c}+\rho_\Lambda \right) - \frac{K}{a^2} \nonumber \\
\Rightarrow  1 - \Omega_{K} &= \Omega_{\nu} + \Omega_{\gamma} + \Omega_{\rm b}+\Omega_{\rm c}+\Omega_\Lambda  \nonumber \\
\Rightarrow  1 - \frac{\Omega_{K,0}}{a^2} &= \frac{\Omega^{\rm rel}_{\nu ,0}}{a^{4}} + \frac{\Omega^{\rm nrel}_{\nu ,0}}{a^{3}} + \frac{\Omega_{\gamma ,0}}{a^{4}} + \frac{\Omega_{{\rm b},0}}{a^{3}}+\frac{\Omega_{{\rm c},0}}{a^{3}}+\Omega_\Lambda  \label{eq:reduced-fried-evol}\,,
\end{align}
where $\Omega_X \equiv \rho_X/\rho_{\rm tot}$ and we have defined $\Omega_K \equiv -K/(a^2\rho_{\rm tot})$. The value of the reduced energy densities today are denoted with $\Omega_{X,0}$ and their approximate values are indicated in \Fig{fig:cos-pie}. The dilution factors, in powers of $a$, in \Eq{eq:reduced-fried-evol} are found using the known equations of state for each component of matter and \Eq{eq:soln-continuity-constw}. In obtaining \Eq{eq:reduced-fried-evol}, we have assumed that baryons are non-relativistic --- this is, of course, different depending on the temperature above which they are relativistic due to their interactions with the thermal bath (and hence $\rho_{\rm b}\propto a^{-4}$). In the case of neutrinos, we have included the possibility that some neutrinos could be either non-relativistic $\Omega^{\rm nrel}_{\nu ,0}$ or relativistic $\Omega^{\rm rel}_{\nu ,0}$ today.

In addition to its homogeneous matter constitution, today the Universe on large scales exhibits many inhomogeneities such as filaments, clusters and voids. Such structures must have been sourced by fluctuations in the total energy density of the Universe that subsequently collapsed under their own gravity. High-precision observations of the CMB radiation~\cite{Ade:2013sjv, Adam:2015rua, Array:2015xqh} have revealed temperature fluctuations $\delta T/T \sim {\cal O}(10^{-5})$ that seeded these collapsed structures, but the CMB itself must have been imprinted with perturbations in the primordial plasma energy density from a much earlier mechanism.

\section{\textsf{Inflation}}

Inflation~\cite{Starobinsky:1980te, Sato:1980yn, Guth:1980zm, Linde:1981mu, Albrecht:1982wi, Linde:1983gd} is the leading paradigm to describe the physical conditions that prevailed in the very early Universe. During this accelerated expansion epoch, cosmological perturbations are amplified from the vacuum quantum fluctuations of the gravitational and matter fields~\cite{Starobinsky:1979ty, Mukhanov:1981xt, Hawking:1982cz,  Starobinsky:1982ee, Guth:1982ec, Bardeen:1983qw} and, as implied in the previous section, measurements~\cite{Ade:2013sjv, Adam:2015rua, Array:2015xqh} of these inhomogeneities in the CMB have significantly improved our knowledge of inflation~\cite{Ade:2015lrj, Martin:2013tda, Martin:2013nzq, Vennin:2015eaa}. 

\subsection{\textsf{Classical inflationary dynamics}}

At its simplest (and perhaps most successful), inflation is driven by the slow roll of a quantum scalar field down its potential. To discuss the dynamics further, one must introduce a model by way of example. Let us consider all other matter fields (and $\Lambda$) to be negligible\footnote{This can also be made reasonable as an assumption in the language of effective field theory: all other fields may take masses which are too high to be excited at this energy scale, and thus may be `integrated out'.} and introduce the following canonical single scalar field $\varphi$ Langrangian density into the matter Lagrangian density ${\cal L}_{\rm m}$ of \Eq{eq:efes-action}
\begin{equation} \label{eq:varphi-matter-lagran}
\ {\cal L}_\varphi = -\frac{1}{2}g^{\mu \nu}\partial_\mu \varphi \partial_\nu \varphi - V(\varphi ) \,.
\end{equation}
The action for this canonical scalar field in a general cosmological background is thus
\begin{equation}\label{eq:efes-action-with-varphi}
\  S = \int \dd^4 x \sqrt{-g} \left[ \frac{\Mp^2}{2} R - \frac{1}{2}g^{\mu \nu}\partial_\mu \varphi \partial_\nu \varphi - V(\varphi ) \right] \,.
\end{equation}
The energy-momentum tensor of $\varphi$ is 
\begin{equation} \label{eq:enmom-lag}
\ T_{\mu \nu} = \frac{-2}{\sqrt{-g}}\frac{\partial }{\partial g^{\mu \nu}} \left( \sqrt{-g}{\cal L}_{\varphi}\right) = \partial_\mu \varphi \partial_\nu \varphi - g_{\mu \nu}\left[ \frac{1}{2}g^{\rho \lambda}\partial_\rho \varphi \partial_\lambda \varphi + V(\varphi )\right] \,,
\end{equation}
and its equation of motion is
\begin{equation} \label{eq:eom-lag}
\frac{\partial }{\partial \varphi} \left( \sqrt{-g}{\cal L}_{\varphi}\right) = -\frac{g^{\mu \nu}}{\sqrt{-g}}\partial_\mu \left( \sqrt{-g}\partial_\nu \varphi \right) - \frac{\partial V}{\partial \varphi} = 0\,.
\end{equation}
In an FLRW Universe (see the line element in \Eq{eq:FLRW-metric}) with the spatial curvature $K=0$ (as it is suppressed during inflation) and a homogeneous scalar field $\varphi$, \Eq{eq:eom-lag} becomes
\begin{equation} \label{eq:KG-varphi}
\frac{\dd^2 \varphi}{\dd t^2} + 3H\frac{\dd \varphi}{\dd t} + \frac{\partial V}{\partial \varphi} = 0\,,
\end{equation}
where the $H^2=\rho_\varphi /(3\Mp^2)$ is constrained to the energy density of the scalar field $\rho_\varphi$ via \Eq{eq:friedmann}. One can always treat the scalar field as a perfect fluid due to there being no anisotropic stress,\footnote{Since there can only be one degree of freedom.} hence we can obtain the energy density of $\varphi$ 
\begin{equation}
\rho_\varphi = \frac{1}{2}\left( \frac{\dd \varphi}{\dd t}\right)^2 + V(\varphi ) \,,
\end{equation}
and the pressure of $\varphi$
\begin{equation}
\ {\sf P}_\varphi = \frac{1}{2}\left( \frac{\dd \varphi}{\dd t}\right)^2 - V(\varphi ) \,.
\end{equation}
Hence the equation of state for $\varphi$ is
\begin{equation} \label{eq:eosvarphi}
\ w_\varphi \equiv \frac{{\sf P}_\varphi}{\rho_\varphi} = \frac{\left( \frac{\dd \varphi}{\dd t}\right)^2 - 2V(\varphi )}{\left( \frac{\dd \varphi}{\dd t}\right)^2 + 2V(\varphi )} \,.
\end{equation}

Inflation requires an accelerated expansion of the Universe, so the condition on the scale factor for this epoch is
\begin{equation} \label{eq:accel-req}
\frac{\dd^2 a}{\dd t^2} > 0\,. 
\end{equation}
In this regime it will also prove convenient to define some new parameters
\begin{equation} \label{eq:slowrollp}
\epsilon_{i+1} = \frac{\dd{\ln}\vert \epsilon_i\vert}{\dd N}\,,
\end{equation}
where $\epsilon_0 \equiv 1/H$, $\dd N \equiv \dd \ln a = H\dd t$ and $N$ is known as the number of `\efold{s}'. In rewriting \Eq{eq:accel-req} in terms of $H$ and \Eq{eq:slowrollp} one finds
\begin{equation}
\frac{1}{a}\frac{\dd^2 a}{\dd t^2} = H^2 + \frac{\dd H}{\dd t} = H^2(1-\epsilon_1 ) > 0\,,
\end{equation}
thus inflation corresponds to $\epsilon_1 < 1$. Notice that matching \Eq{eq:raychaud} with \Eq{eq:accel-req} also is equivalent to the condition
\begin{equation}
\ w_\varphi < -\frac{1}{3}\,.
\end{equation}
Hereafter, we shall use the term `slow-roll' for dynamics which satisfy $\vert \epsilon_i \vert \ll 1 \,, \, \forall i$. This regime is interesting due to its known attractor behaviour, limiting the arbitrariness required in setting the initial conditions to the inflationary epoch.   

By comparison with \Eq{eq:eosvarphi} we see that this condition on the equation of state requires the field to be dominated by its potential energy $V(\varphi ) > \dd^2 \varphi / (\dd t)^2$ and so, using \Eq{eq:eosvarphi}, typical models of inflation have $w_{\varphi}\simeq -1$ corresponding to (or close to) a pure de Sitter spacetime where $\dd H / \dd t \simeq 0$ and $a(t)\propto \ee^{Ht}$.\footnote{Note that this is exactly the same as a spacetime dominated by a cosmological constant $a(t)\propto \ee^{\sqrt{\frac{\Lambda}{3}} t}$.} Slow-roll inflation achieves exactly this feat by considering the gradual roll of a scalar field towards its potential minimum (where an initial condition has to be set by some mechanism) while the slope of the potential is typically gentle enough that the second derivative in time of \Eq{eq:KG-varphi} is never important. Due to this fact, \Eq{eq:KG-varphi} reduces to
\begin{equation} \label{eq:classical-slow-roll-KG}
\frac{\dd \varphi}{\dd N} = -\frac{1}{3H^2}\frac{\partial V}{\partial \varphi} \,,
\end{equation}
defining what are known as `classical' slow-roll dynamics.\footnote{Reasons for this distinction from `quantum' dynamics will become clear in later sections.} In this limit, we must also assume that
\begin{equation} \label{eq:friedmann-slow-roll}
\ H^2 = \frac{V}{3\Mp^3}\,,
\end{equation}
such that \Eq{eq:classical-slow-roll-KG} can be rewritten as
\begin{equation} \label{eq:classical-slow-roll-KG-2}
\frac{\dd \varphi}{\dd N} = -\frac{\Mp^2}{V}\frac{\partial V}{\partial \varphi} \,.
\end{equation}

The number of \efold{s} $N$ serves as a useful parameter to characterise the length of time that inflation takes place. Between $N_X=N(t_X)$ and $N_Y=N(t_Y)$, \Eq{eq:classical-slow-roll-KG-2} can be manipulated to give
\begin{equation}
\ N_Y - N_X = \int^{N_Y}_{N_X} \dd \tilde{N} = \Mp^2 \int^{\varphi_X}_{\varphi_Y} \dd \tilde{\varphi} \, V(\tilde{\varphi})\, \frac{\partial \tilde{\varphi}}{\partial V}  \label{eq:numefolds-formula} \,,
\end{equation}
where we have integrated between $\varphi_Y = \varphi (t_Y)$ and  $\varphi_X = \varphi (t_X)$ during some slow-roll phase of the homogeneous field $\varphi$. Note that for single field slow-roll inflation it is simple to show, using Eqs.\eqref{eq:friedmann-slow-roll}, \eqref{eq:classical-slow-roll-KG-2} and \eqref{eq:slowrollp}, that
\begin{equation} \label{eq:slow-roll-param1-V}
\epsilon_1 = \frac{1}{2}\left( \frac{\dd \varphi}{\dd N}\right)^2 \simeq \frac{\Mp^2}{2}\left( \frac{\partial {\ln}V}{\partial \varphi}\right)^{2} \,.
\end{equation}

\subsection{\textsf{Sourcing cosmological perturbations}} \label{sec:sourcing-cos-pert}

To begin with, let us break the homogeneity assumption of $\varphi$ by splitting it up into a homogenous part and a small fluctuation $\delta \varphi$ like so
\begin{equation} \label{eq:mean-field-app}
\varphi (x_i,\eta ) = \varphi (\eta ) + \delta \varphi (x_i,\eta ) \,,
\end{equation}
where we are now using conformal time $\eta$ as our time variable. This expansion should be considered in conjunction with the line element for scalar metric perturbations to linear order~\cite{Peter:1208401,Malik:2008im}
\begin{align}
&\dd s^2 = \nonumber \\
& a^2(\eta ) \left\{ -(1+2A)\dd \eta^2 + 2\frac{\partial B}{\partial x^i}\dd x^i\dd\eta + \left[ (1-2\psi ) \delta_{ij} + 2\frac{\partial^2E}{\partial x^i \partial x^j}\right] \dd x^i \dd x^j \right\} \label{eq:perturbed-scalar-fLrW} \,,
\end{align}
which follows from a flat FLRW background. Using \Eq{eq:perturbed-scalar-fLrW} and \Eq{eq:mean-field-app} it can be shown that the scalar sector of matter and gravitational fluctuating degrees of freedom can be described entirely by the following gauge-invariant quantity~\cite{Peter:1208401,Malik:2008im}
\begin{align}
\ Q 
\equiv  a \left( \delta \varphi + \psi \frac{\dd \varphi}{\dd N}\right) \label{eq:ms-variable} \,,
\end{align}
where $Q$ is known as the Mukhanov-Sasaki variable~\cite{Mukhanov:1988jd,1984PThPS..78....1K}. We will avoid discussing \Eq{eq:ms-variable} in too much detail here, however, let us simply note that gauge freedom permits the fixing of variables within $Q$ to eliminate unphysical degrees of freedom.\footnote{Choose $E=\psi = 0$ for the spatially flat gauge, $B=E=0$ for the Newtonian gauge, $\delta \varphi =0$ for comoving gauge and $B=A=0$ for synchronous gauge~\cite{Malik:2008im}.}

Let us now expand \Eq{eq:efes-action-with-varphi} to second order in $Q$ to give~\cite{Brandenberger:2003vk}\footnote{No linear order terms in $Q$ can exist since they have to vanish in order to extremise the action.}
\begin{equation}
S_2 = \int \dd^4 x  \frac{1}{2} \left[ \left( \frac{\partial Q}{\partial \eta}  \right)^2 - \delta^{ij}\frac{\partial Q}{\partial x^i}\frac{\partial Q}{\partial x^j} + \frac{1}{a\sqrt{\epsilon_1}}\frac{\dd^2 (a\sqrt{\epsilon_1})}{\dd \eta^2 } Q^2 \right] \label{eq:action-scalar-perturbed-2ndorder} \,,
\end{equation}
where we have used \Eq{eq:slow-roll-param1-V} and the action, upon variation with respect to $Q$, and a Fourier transform (such that $\nabla^2 \mapsto k^2$) gives the following equation of motion\footnote{This is known as the `Mukhanov-Sasaki' equation~\cite{Mukhanov:1988jd,1984PThPS..78....1K}.}
\begin{equation} \label{eq:eom-scalar-pert}
\frac{\partial^2Q_k}{\partial \eta^2} + \left[ k^2 - \frac{1}{a\sqrt{\epsilon_1}}\frac{\dd^2 (a\sqrt{\epsilon_1})}{\dd \eta^2 } \right] Q_k = 0\,,
\end{equation}
where $\nabla \equiv (\partial / \partial x^i) (\partial / \partial x_i)$ is the (comoving) spatial Laplacian. Note that \Eq{eq:eom-scalar-pert} now characterises the dynamics of the entire scalar sector on both super ($k < aH$) and sub-horizon ($k > aH$) scales. 

Note that $Q$ and its canonical momentum conjugate 
\begin{equation}
\Pi \equiv \frac{\delta S_2}{\delta (\partial_\eta Q)} = \frac{\partial Q}{\partial \eta}\,,
\end{equation}
can be expanded in a Fourier basis in terms of its mode functions such that
\begin{gather} \label{eq:modfuncdelphi}
\ Q = \int \frac{\dd^3k}{(2\pi )^{\frac{3}{2}}} \left[  \hat{a}_k Q_k(\eta )\ee^{-ik_j x^j} + \hat{a}^\dagger_k Q_k^*(\eta )\ee^{ik_j x^j } \right]  \\
\Pi = \int \frac{\dd^3k}{(2\pi )^{\frac{3}{2}}} \left[  \hat{a}_k\Pi_k(\eta )\ee^{-ik_j x^j} + \hat{a}^\dagger_k \Pi_k^*(\eta )\ee^{ik_j x^j } \right] \label{eq:modfuncPi} \,, 
\end{gather}
where $a_k^\dagger$ and $a_k$ are the creation and annihilation operators, respectively. These are normalised according to the standard commutation relations
\begin{equation}
\ [\hat{a}_{\tilde{k}},\hat{a}^\dagger_k] = \delta^3 (\tilde{k}_i-k_i)\,, \quad [\hat{a}^\dagger_{\tilde{k}},\hat{a}^\dagger_k] = [\hat{a}_{\tilde{k}},\hat{a}_k] = 0\,,
\end{equation}
which in conjuction with satisfying the equal-time commutation relations of the field operators of \Eq{eq:modfuncdelphi} and \Eq{eq:modfuncPi} (in order for the theory to be causal)
\begin{gather}
\ [\hat{Q}(\tilde{x}_i,\eta ),\hat{\Pi}(x_i,\eta )] = i\delta^3 (\tilde{x}_i-x_i) \\
\ [\hat{Q}(\tilde{x}_i,\eta ),\hat{Q}(x_i,\eta )] = [\hat{\Pi}(\tilde{x}_i,\eta ),\hat{\Pi}(x_i,\eta )] = 0\,,
\end{gather}
give rise to the following Wronskian normalisation 
\begin{equation} \label{eq:Wronskiannorm}
\ Q_k\frac{\dd Q_k^*}{\dd \eta} - \frac{\dd Q_k}{\dd \eta} Q_k^* = i\,.
\end{equation}

Substituting into \Eq{eq:eom-scalar-pert} the leading order slow-roll expansion for $a\sqrt{\epsilon_1}$ at a point $t_X$ in time,\footnote{To leading order in slow roll expansion about $a_X \equiv a(t_X)$, the change in scale factor can be expressed as 
\begin{equation*}
\ln \left( \frac{a}{a_X}\right) = N-N_X = \int^{\eta}_{\eta_X}{\cal H}(\tilde{\eta})\dd \tilde{\eta} \simeq \int^{\eta_X}_{\eta} (1+\epsilon_{1,X})\dd{\ln }\tilde{\eta} =(1+\epsilon_{1,X}) {\ln }\left( \frac{\eta_X}{\eta}\right)\,, 
\end{equation*}
where ${\cal H}\equiv \dd{\ln}a/\dd \eta$. Equivalently, one finds that the first slow roll parameter varies according to
\begin{equation*}
\epsilon_1 - \epsilon_{1,X} = \left. \epsilon_{1,X}\frac{\dd{\ln}\epsilon_1}{\dd N}\right\vert_{t_X}(N-N_X) \simeq - \epsilon_{1,X}\epsilon_{2,X}(1+\epsilon_{1,X}) {\ln }\left( \frac{\eta}{\eta_X}\right) \simeq - \epsilon_{1,X}\epsilon_{2,X} {\ln }\left( \frac{\eta}{\eta_X}\right)\,.
\end{equation*}} one finds
\begin{equation}
\frac{\dd^2Q_k}{\dd \eta^2} + \left[ k^2 - \frac{1}{\eta^2}\left( 2+3\epsilon_{1,X} + \frac{3}{2} \epsilon_{2,X}\right) \right] Q_k = 0 \label{eq:MS-non-gauge-inv}\,.
\end{equation}
\Eq{eq:MS-non-gauge-inv} has the following solution to leading-order in the slow roll (such that $\epsilon_{1,X}$ and $\epsilon_{2,X}$ are constant)
\begin{equation}
\ Q_k(\eta ) = \sqrt{-\eta} \left[ W_1 H^{(1)}_\nu (-k\eta ) + W_2 H^{(2)}_\nu (-k\eta )\right] \label{eq:unnorm-solution-hankel} \,,
\end{equation}
where: $H^{(1)}_\nu (-k\eta )$ and $H^{(2)}_\nu (-k\eta )$ are Hankel functions of the first and second kind, respectively; both $W_1$ and $W_2$ here are constants to be set by initial conditions; and we have defined\footnote{Note that due to the expansion in \Eq{eq:action-scalar-perturbed-2ndorder}, one can gain more physical intuition by using \Eq{eq:classical-slow-roll-KG-2} and \Eq{eq:slowrollp} to rewrite \Eq{eq:muHankeldef} as 
\begin{equation*}
\nu \simeq \frac{3}{2} + \epsilon_1 + \frac{V}{3H^2}\left( \frac{\partial{\ln}V}{\partial \varphi}\right)^2 - \frac{1}{3H^2}\frac{\partial^2 V}{\partial \varphi^2}\,,
\end{equation*}
which holds more generically for test fields as well (fields whose energy density is so sub-dominant that, effectively, $H\neq H(\varphi )$).}
\begin{equation}
\nu \equiv \frac{3}{2}\sqrt{1+\frac{4}{3}\epsilon_{1,X}+\frac{2}{3}\epsilon_{2,X}} \simeq \frac{3}{2} + \epsilon_{1,X} + \frac{1}{2}\epsilon_{2,X} \label{eq:muHankeldef}\,. 
\end{equation}

A subtle, yet deep issue arises when na\"{i}vely attempting to set the initial conditions $W_1$ and $W_2$ of \Eq{eq:unnorm-solution-hankel}. Notice that the mode functions which satisfy \Eq{eq:MS-non-gauge-inv} will have a time-dependent frequency. Due to this fact, it becomes problematic to define the vacuum state unambiguously. Consider that the set of mode functions for which the Hamiltonian, constructed out of Eqs.~\eqref{eq:modfuncdelphi},~\eqref{eq:modfuncPi} and~\eqref{eq:Wronskiannorm}, is minimised (to find the ground state) at one point in time $\eta$ will not be the same set of mode functions to minimise the Hamiltonian at a later time $\eta + \delta \eta$. The solution to this problem of ambiguity in the ground state lies in noticing that the sub-Hubble limit $\vert k\eta \vert \simeq \vert k/(aH) \vert \gg 1$ of \Eq{eq:MS-non-gauge-inv} removes this time dependence. Hence we may asymptotically define a ground state that is identical to that in Minkowski space known as the Bunch-Davies vacuum
\begin{equation}
\ Q_k \xrightarrow[\vert k\eta \vert \gg 1]{} \frac{1}{\sqrt{2k}}\ee^{-ik\eta} \,,
\end{equation}
and hence by comparison to the sub-Hubble limit of \Eq{eq:unnorm-solution-hankel} (up to an irrelevant phase factor of $\exp [-i\pi (1 + 2\nu )/4]$ which the power spectrum cannot observe) we see that the necessary initial conditions to set for the Bunch-Davies vacuum are
\begin{equation} \label{eq:BD-init-cond}
\ W_1 = \frac{\sqrt{\pi}}{2}\,, \qquad W_2 = 0\,.
\end{equation}
Now that we are able to set the conditions in \Eq{eq:BD-init-cond}, our solution which asymptotically matches the Bunch-Davies vacuum is
\begin{equation}
\ Q_k(\eta ) = \frac{\sqrt{-\pi \eta}}{2} H^{(1)}_\nu (-k\eta ) \label{eq:Hankelmodfunc}\,.
\end{equation}

In slow roll $\epsilon_{1,X},\epsilon_{2,X} < 1$ and hence we can approximate the amplitude-squared of \Eq{eq:Hankelmodfunc} in the super-horizon limit, i.e, the limit where $\vert k\eta \vert = \vert k/(aH)\vert \ll 1$, as
\begin{align}
\ Q_k^*(\eta )Q_k(\eta ) \xrightarrow[\vert k\eta \vert \ll 1]{} \,\,\, &\frac{\Gamma^2 (\nu )}{2\pi k} \left( \frac{-k\eta}{2}\right)^{-2\nu + 1} \nonumber \\
\simeq \,\, &\Gamma^2 \left( \frac{3}{2} + \epsilon_{1,X} + \frac{1}{2}\epsilon_{2,X}\right)\frac{1}{2\pi k} \left( \frac{-k\eta}{2}\right)^{-2 - 2\epsilon_{1,X} - \epsilon_{2,X}} \nonumber \\
\simeq \,\, &\frac{1}{2k^3\eta^2} = \frac{(aH)^2}{2k^3} \qquad ({\rm where} \,\, \epsilon_{1,X},\epsilon_{2,X}\rightarrow 0)  \label{eq:modfuncamp-approx}\,. 
\end{align}
In order to calculate the variance of $Q$ itself, the integral in the Fourier basis of \Eq{eq:modfuncdelphi} gives rise to an additional factor of $2\pi^2/k^3$ in \Eq{eq:modfuncamp-approx}, hence we may define the power spectrum ${\cal P}_{Q}$ which quantifies the variance of field fluctuations through
\begin{equation}
\left\langle  Q_k^*(\eta )Q_{\tilde{k}}(\eta ) \right\rangle = \frac{k^3}{2\pi^2} {\cal P}_{Q}(k,\eta )\delta^3 (k_j+\tilde{k}_j) \label{eq:powers-def}\,,
\end{equation}
where $\langle \cdot \rangle$ in this expression is to be understood as an ensemble average (computed from a quantum average) over the field fluctuations. By comparison of \Eq{eq:modfuncamp-approx} with \Eq{eq:powers-def}, we arrive at
\begin{equation} \label{eq:scale-inv-spec}
\frac{1}{a^2} {\cal P}_{Q}(k,\eta ) = \frac{H^2}{(2\pi )^2}\,,
\end{equation}
on super-horizon scales, indicating a scale-invariant spectrum. 

The variable $Q$ in \Eq{eq:ms-variable} can also be directly related to the comoving curvature peturbation $\zeta$ (which can be shown to be constant super-Hubble scales as long as the fluctuations are adiabatic~\cite{Bassett:2005xm,Mukhanov:991646,Malik:2008im,Lyth:2003im,weinberg2008cosmology,Lyth:2009zz}, making it extremely useful for translating the curvature perturbation to later epochs) by fixing $\delta \varphi = 0$ such that 
\begin{equation}
\frac{Q}{a} = \psi \frac{\dd \varphi}{\dd N} = \zeta \frac{\dd \varphi}{\dd N} \,.
\end{equation}
Therefore, the power spectrum of $\zeta$ that is sourced by the field $\varphi$ is
\begin{equation} \label{eq:Pzeta:class-intro}
\ {\cal P}_\zeta (k_*) = \left( \frac{\dd N}{\dd \varphi} \right)^2 \left( \frac{H}{2\pi }\right)^2 = \frac{V^3}{12 \pi^2\Mp^6} \left. \left( \frac{\partial \varphi}{\partial V}\right)^{2} \right\vert_{k_*} \,,
\end{equation}
which is typically evaluated at some pivot scale $k_* = 0.05{\rm Mpc}^{-1}$ of the comoving wave vector. Varying \Eq{eq:Pzeta:class-intro} with respect to $k$ up to second order, we find a new pair of parameters which can be constrained from CMB data
\begin{align}
\nS - 1 &\equiv \left. \frac{\dd{\ln}{\cal P}_{\zeta}}{\dd{\ln}k} \right\vert_{k_*} \simeq - 2\epsilon_1 - \epsilon_2 \label{eq:spectral-ind} \\
\alphaS &\equiv \left. \frac{\dd^2{\ln}{\cal P}_\zeta }{\dd ({\ln}k)^2} \right\vert_{k_*} \simeq - 2\epsilon_1 \epsilon_2 - \epsilon_2 \epsilon_3 \label{eq:alphaS-intro} \,,
\end{align}
which are the spectral index $\nS$ and running of the spectral index $\alphaS$, respectively. Notice that the last equalities in both expressions are valid only for single-field models to leading order in slow roll. 

The fluctuations in the spacetime metric can be decomposed into more than just the scalar degree of freedom that we have studied so far.  In fact it is known that, due to the conservation of angular momentum, vector perturbations decay during inflation. In contrast,  one can expand the tensor degrees of freedom --- two tensor helicities, $h_+$ and $h_-$, are available\footnote{This is due to the constraint that the full tensor degree of freedom $h_{ij}$ arising from the tensor-perturbed metric, with line element
\begin{equation*}
\dd s^2 = a^2(\eta ) \left[ -\dd \eta^2 + (\delta_{ij} + 2h_{ij}) \dd x^i \dd x^j\right] \,,
\end{equation*}
must be transverse $\partial_ih_{ij}=0$ and trace-free $h^i{}_{i}=0$.} --- out of the full action up to second order to find an equivalent expression to \Eq{eq:action-scalar-perturbed-2ndorder}.
Varying 
this expression with respect to the metric, we arrive at the equations of motion for each polarisation of the tensor perturbations which are the same as for the massless scalar
\begin{equation} \label{eq:eom-tensors}
\frac{\dd^2 h_{\pm}}{\dd \eta^2} + 2{\cal H}\frac{\dd  h_{\pm}}{\dd \eta} + k^2h_{\pm} = 0\,,
\end{equation}
where ${\cal H}\equiv \dd{\ln}a/\dd \eta$. In order to compute the same vacuum fluctuations as in the scalar case, we must normalise $h_{\pm}$ in the same way such that the newly defined tensor perturbation is
\begin{equation}
\gamma_{\pm} \equiv \sqrt{2}\Mp h_{\pm} \,.
\end{equation}
Finally, using the same reasoning as for the scalars, \Eq{eq:eom-tensors} and accounting for the two separate polarisations, we compute the power spectrum of tensor perturbations as\footnote{This becomes clear from its definition $\langle 2h_{ij}(k,\eta )2h^{ij}(\tilde{k},\eta ) \rangle = [k^3/(2\pi^2)] {\cal P}_{h}(k,\eta ) \delta^3(k_j+\tilde{k}_j)$.}
\begin{equation}
{\cal P}_h(k_*) = \frac{8}{\Mp^2} \left( \frac{H}{2\pi}\right)^2 = \left. \frac{2V}{3\pi^2\Mp^4} \right\vert_{\varphi = \varphi (k_*)} \label{eq:tensors-power-spect} \,,
\end{equation}
where, as before, we used the slow roll equation~\eqref{eq:friedmann-slow-roll} to compute the second equality. Using \Eq{eq:tensors-power-spect} and \Eq{eq:Pzeta:class-intro} we can define the tensor-to-scalar ratio
\begin{equation}
\ r \equiv \frac{{\cal P}_{h}(k_*)}{{\cal P}_{\zeta}(k_*)} \simeq 16\epsilon_1\,, \label{eq:tens-scalar-ratio}
\end{equation}
which is used to compare inflationary models to CMB data, and where we applied slow roll to obtain the last equality. 

In this section, we obtained the observables $\AmpS \equiv \vert {\cal P}_\zeta \vert$, $\nS$, $\alphaS$ and $r$ all from classical inflationary field dynamics. Let us now take a quick example of a popular potential $V$ from which we can compute the observables. The Starobinksy potential~\cite{Starobinsky:1980te} is a plateau inflationary model with
\begin{equation}\label{eq:starobinsky-pot-example}
\ V(\phi)\propto \left( 1-\ee^{-\sqrt{\frac{2}{3}}\frac{\varphi}{\Mp}}\right)^2\,.
\end{equation}
The slow-roll parameters for this model, which may be calculated from \Eq{eq:slowrollp} and \Eq{eq:slow-roll-param1-V}, are~\cite{Martin:2013tda}
\begin{align}
\epsilon_1 &= \frac{4}{3}\left( 1-\ee^{\sqrt{\frac{2}{3}}\frac{\varphi}{\Mp}}\right)^{-2} \\
\epsilon_2 &= \frac{2}{3}\left[ \sinh \left( \frac{\varphi}{\sqrt{6}\Mp} \right) \right]^{-2}  \\
\epsilon_3 &= \frac{2}{3}\coth \left( \frac{\varphi}{\sqrt{6}\Mp} \right) \left[ \coth \left( \frac{\varphi}{\sqrt{6}\Mp} \right) -1 \right] \,.
\end{align}
At a value of, e.g., 60 \efold{s} before the end of inflation,\footnote{In many models, 50-60 is the typical number of \efold{s} before the end of inflation at which the observable perturbations crossed the Hubble radius~\cite{Liddle:2003as}. } \Eq{eq:numefolds-formula} gives us a value of $\phi \simeq 5.453\Mp$~\cite{Martin:2013tda}. From Eqs.~\eqref{eq:spectral-ind}, \eqref{eq:alphaS-intro}, \eqref{eq:tens-scalar-ratio} and this value we find that, to leading order in slow roll, $\nS = 0.968$, $\alphaS = -0.0005$ and $r = 0.003$.

\subsection{\textsf{Resumming divergences}} \label{sec:intro-stochastic-approach}

So far the extent to which fluctuations of the quantum field $\varphi$ have been taken into account is in describing how cosmological perturbations are sourced from its vacuum fluctuations with a Bunch-Davies initial condition. What we shall consider now is a consequence of these fluctuations leaving the Hubble radius on large scales and accumulating in the Infra-Red (IR) limit. To begin with, let us rewrite \Eq{eq:KG-varphi} for a massless test field $\varphi = \varphi (x_i,\eta )$ in terms of conformal time
\begin{equation} \label{eq:massless-inhomo-field}
\frac{\partial^2 \varphi}{\partial \eta^2} + 2{\cal H}\frac{\partial \varphi}{\partial \eta} - \nabla^2 \varphi = 0\,,
\end{equation}
where we are now including the inhomogeneity of the field explicitly such that the comoving spatial Laplacian $\nabla^2$ is non-vanishing. Expanding the field of \Eq{eq:massless-inhomo-field} in a Fourier basis, similarly to \Eq{eq:modfuncdelphi}, we find 
\begin{equation} \label{eq:fourier-freefield}
\varphi (x_i,\eta ) = \int \frac{\dd^3k}{(2\pi )^{\frac{3}{2}}} \left[  \hat{a}_k \varphi_k(\eta )\ee^{-ik_j x^j} + \hat{a}^\dagger_k \varphi_k^*(\eta )\ee^{ik_j x^j } \right] \,,
\end{equation}
and hence the equation of motion that $\varphi_k(\eta )$ satisfies
is simply 
\begin{equation} \label{eq:massless-inhomo-field-fourier}
\frac{\dd^2 \varphi_k}{\dd \eta^2} + 2{\cal H}\frac{\dd \varphi_k}{\dd \eta} + k^2 \varphi_k = 0\,,
\end{equation}
where the Bunch-Davies solution to this equation is equivalent to the massless limit of \Eq{eq:Hankelmodfunc}
\begin{align}
\varphi_k (\eta ) = \frac{\sqrt{-\pi \eta}}{2a} H^{(1)}_{\frac{3}{2}} (-k\eta ) &= \frac{1}{\sqrt{2k}a}\left( \frac{i}{k\eta}-1 \right) \ee^{-ik\eta} \\
& = \frac{H}{\sqrt{2k}}\left( \eta -\frac{i}{k} \right) \ee^{-ik\eta} \label{eq:pure-de-sitter-modfunc} \,,
\end{align}
where \Eq{eq:pure-de-sitter-modfunc} is obtained by using the fact that, in de Sitter, $H$ is constant and so $a=-1/(\eta H)$. Returning to the $\varphi (x_i,\eta )$ form of the field, we are now ready to calculate an expectation value (in the quantum sense) between spatially-separated points $x_i$ and $\tilde{x}_i$. Denoting the vacuum state with $\vert 0\rangle$,\footnote{This is defined such that $a_k\vert 0\rangle =0$.} the two-point function is~\cite{Prokopec:2007ak}
\begin{align}
\langle 0 \vert \varphi (x_i,\eta )\varphi (\tilde{x}_i,\tilde{\eta} ) \vert 0\rangle &= \int \frac{\dd^3k}{(2\pi )^3} \ee^{ik_j(x^j-\tilde{x}^j)}\varphi_k(\eta ) \varphi^*_k(\tilde{\eta }) \nonumber \\
&= \int \frac{k^2\dd k}{2\pi^2} \frac{\sin \left( k\vert x-\tilde{x}\vert \right)}{k\vert x-\tilde{x}\vert}\varphi_k(\eta ) \varphi^*_k(\tilde{\eta }) \label{eq:IRdivergences}\,,
\end{align}
where we have obtained the second equality by an angular integral over $k_j/k = k_j/(\vert k_j \vert )$. Substituting \Eq{eq:pure-de-sitter-modfunc} into \Eq{eq:IRdivergences} and expanding about the super-horizon limit $\vert k \eta \vert, \vert k \tilde{\eta} \vert \ll 1$, we obtain the following $k$-behaviour in the indefinite form of the integral\footnote{Note that this result coincides with integrating the result from \Eq{eq:modfuncamp-approx} in the spatially flat gauge, where $\varphi_k = Q_k/a$.}
\begin{equation}
\langle 0 \vert \varphi (x_i,\eta )\varphi (\tilde{x}_i,\tilde{\eta} ) \vert 0\rangle \xrightarrow[\vert k_\Lambda\eta\vert , \vert k_\Lambda \tilde{\eta}\vert \ll 1]{} \frac{H^2}{4\pi^2}\left( \ln k - \frac{k^2 \vert x-\tilde{x}\vert^2}{12} \right) \label{eq:log-div} \,,
\end{equation}
where we have implicitly also used the fact that the spatial separation must satisfy $\vert x-\tilde{x} \vert < \vert \eta -\tilde{\eta} \vert$ for the correlator to be causal. This example demonstrates that the correlation functions of quantum fields in an inflationary spacetime exhibit logarithmic divergences in the IR limit.\footnote{Taking $k\rightarrow 0$, there is a divergence, but in practice there is a cutoff because $k<H$ is impossible as an initial condition~\cite{Prokopec:2007ak}.}

In addition to the divergence of \Eq{eq:log-div}, we will now show that an additional problem emerges when one wishes to consider interactions~\cite{Prokopec:2007ak, Burgess:2010dd}. Using \Eq{eq:eom-scalar-pert} in the spatially flat gauge (where $\varphi = Q/a$), one can write the full equation of motion for an inhomogenous test field where the potential $V$ is now included (and hence the interaction terms within it)
\begin{equation}
\frac{\partial^2 \varphi}{\partial \eta^2} + 2{\cal H}\frac{\partial \varphi}{\partial \eta} - \nabla^2 \varphi + a^2\frac{\partial V}{\partial \varphi} =0 \label{eq:KG-inhomo-1-IR}\,,
\end{equation}
and perturbatively expand $\varphi (x_i,\eta ) = \varphi^{(0)}(x_i,\eta ) + \varphi^{(1)}(x_i,\eta ) +\dots$ such that $\varphi^{(0)}(x_i,\eta )$ is the free field, $\varphi^{(1)}(x_i,\eta )$ follows \Eq{eq:KG-inhomo-1-IR} where $\varphi^{(0)}$ is the source to any interactions and so on to higher order.\footnote{Note that this expansion is separate from \Eq{eq:mean-field-app} since the former is a mean field expansion and the latter is performed for the expansion of cosmological perturbations.} In such a picture, one uses the inhomogenous solution to the free field \Eq{eq:KG-inhomo-1-IR}  
\begin{equation}
\left( \frac{\partial^2}{\partial \eta^2} + 2{\cal H}\frac{\partial }{\partial \eta} - \nabla^2  + a^2m^2 \right) {\sf G} (x_i,\eta ;\tilde{x}_i,\tilde{\eta}) = \frac{1}{a^3}\delta^3(x_i-\tilde{x}_i) \label{eq:greens-func-def-KG}\,,
\end{equation}
where $m^2 \equiv \partial^2 V(\varphi^{(0)}) /(\partial \varphi^{(0)})^2$ here is the mass of the free field potential (and hence does not depend on the field itself) and ${\sf G} (x_i,\eta ;\tilde{x}_i,\tilde{\eta}) = (i/a^2) \Theta (\eta - \tilde{\eta})[\varphi^{(0)}(x_i,\eta ),\varphi^{(0)}(\tilde{x}_i,\tilde{\eta} )]$ is the retarded Greens function, which we can calculate using \Eq{eq:fourier-freefield} and \Eq{eq:pure-de-sitter-modfunc}. 

We now collect all remaining terms of the potential (beyond free field) using the following method. Using ${\sf G}$ to construct the Yang-Feldman equation~\cite{PhysRev.79.972}, one integrates the interactions of the field in \Eq{eq:KG-inhomo-1-IR} up to all orders in the expansion
\begin{equation}
\varphi (x_i,\eta ) =  \varphi^{(0)} (x_i,\eta ) - \int \dd \tilde{\eta} \int \dd^3\tilde{x} \sqrt{-g}\,{\sf G} (x_i,\eta ;\tilde{x}_i,\tilde{\eta}) \, a^2(\tilde{\eta} )\frac{\partial \tilde{V}}{\partial \varphi^{(0)}} + {\cal O}({\sf G}^2) \label{eq:YFequation}\,,
\end{equation}
where $\tilde{V}=V[\varphi (\tilde{x}_i,\tilde{\eta})]$ is now the full potential. Note here that the vertex integration contributes a factor of $\dd \eta \, \dd^3x \sqrt{-g}\,  {\sf G}\, a^2 \sim \dd a/a$, causing a (rather catastrophic) break down in the perturbative expansion after some critical timescale~\cite{Burgess:2010dd} and hence originating an additional IR logarithm that must be removed.

It transpires that the first of these divergences may be removed through the application of a cutoff. Notice that, because the observable perturbations are super-horizon during inflation, one can choose to place a cutoff in \Eq{eq:fourier-freefield} on the modes up to a Fourier coarse-graining scale $k_\Lambda = \sigma_\Lambda aH$ (where $\sigma_\Lambda \ll 1$) such that our new field has the UV modes integrated out like so
\begin{equation} 
\bar{\varphi} (x_i,\eta ) = \int \frac{\dd^3k}{(2\pi )^{\frac{3}{2}}} \Theta [k_\Lambda (\eta ) - k] \left[  \hat{a}_k \varphi_k(\eta )\ee^{-ik_j x^j} + \hat{a}^\dagger_k \varphi_k^*(\eta )\ee^{ik_j x^j } \right] \label{eq:fourier-coarsegrainedfield} \,.
\end{equation}
However, if we were to substitute the remaining modes (by simply flipping the $\Theta [k_\Lambda (\eta ) -k] \leftrightarrow \Theta [k-k_\Lambda (\eta ) ]$ in \Eq{eq:fourier-coarsegrainedfield}) into \Eq{eq:pure-de-sitter-modfunc} and compute the two-point function, we would find that \Eq{eq:log-div} is rendered finite since the integrand is predominantly oscillatory in that mode range. This means that first divergence we identified in the two-point function has been removed!

Let us now compute the commutator of $\bar{\varphi}$ using the free field mode functions 
\begin{align}
\ & [\bar{\varphi} (x_i,\eta ),\bar{\varphi} (\tilde{x}_i,\tilde{\eta} )] = \nonumber \\
&\int^{k_\Lambda} \frac{\dd k}{k}\frac{\sin \left( k\vert x-\tilde{x}\vert \right)}{k\vert x-\tilde{x}\vert}\left( \frac{H^2}{4\pi^2}\right) \left\{ (k\eta +i)(k\tilde{\eta } -i) \left[ a_k^\dagger a_k \ee^{ik(\eta -\tilde{\eta})} - a_k a_k^\dagger \ee^{ik(\tilde{\eta}-\eta )} \right] \right. \nonumber \\
& \qquad \qquad \qquad \qquad \qquad \left. + (k\tilde{\eta } +i)(k\eta -i) \left[ a_k a_k^\dagger \ee^{ik(\tilde{\eta}-\eta )} - a_k^\dagger a_k \ee^{ik(\eta - \tilde{\eta})} \right] \right\} \label{eq:vanishing-IR-commutator}\,.
\end{align}
The emergence of solely classical fluctuations of the field, i.e. those for which the commutator becomes much smaller than the corresponding anticommutator, can be immediately seen in the super-horizon limit $\vert k\eta \vert , \vert k\tilde{\eta}\vert \ll 1$. This feature exists for $\varphi$ in the same way --- since the only difference would be the removal of the $k_\Lambda$ regulator in the upper limit of the integral --- and is one example of the quantum-to-classical transition~\cite{Polarski:1995jg,Lesgourgues:1996jc} that explains why cosmological perturbations with a quantum source are observed as classical.\footnote{In more detail, this is thought to be a consequence of the unique two-mode squeezed quantum state~\cite{PhysRevLett.59.2555,
PhysRevD.46.1440,PhysRevD.42.3413} that fields find themselves in during inflation. In addition, one must also study the transition without taking the super-horizon limit. }

So far, we have demonstrated that by a redefinition of the field $\varphi (x_i,\eta )\rightarrow \bar{\varphi}(x_i,\eta )$, which leaves the observables unchanged through the application of a regularisation procedure that cannot affect the observables in the IR, one can successfully remove the divergence in the free field correlation functions. Furthermore, we have shown that the new IR field $\bar{\varphi}$ has a vanishing commutator which implies that it may be described as a classical, stochastic field. An additional problem still appears to persist, however. In order to describe an interacting field, one can attempt to use $\bar{\varphi}$ in \Eq{eq:YFequation}. Perturbation theory breaks down after a finite timescale in the vertex integration itself, which contributes a secular growth factor of $\sim \dd a/a$, and this problem has not yet been resolved. For practical applications of \Eq{eq:fourier-coarsegrainedfield}, this remaining problem suggests that a non-perturbative solution is required in order to take into account of the time evolution of the system.   

Let us return to \Eq{eq:classical-slow-roll-KG} for the dynamics of a field during slow-roll inflation. This equation is still valid for an inhomogeneous field $\varphi (x_i,\eta )$ in the super-horizon limit (where the gradient term $\propto \nabla^2$ may safely be neglected). Combining \Eq{eq:fourier-coarsegrainedfield} and \Eq{eq:classical-slow-roll-KG}, we find\footnote{This is easily confused with a similar type of expansion as \Eq{eq:mean-field-app}. This is true \emph{only} instantaneously since \Eq{eq:langevin} represents the slow-roll expansion evaluated at each new moment in time (rather than the expansion performed about, e.g., the end of inflation). This optimisation of the perturbative expansion is similar to a renormalisation group flow --- an observation we will make later. }
\begin{equation}
\frac{\partial \bar{\varphi}}{\partial N} = -\frac{1}{3H^2}\frac{\partial V}{\partial \bar{\varphi}} + f(x_i,\eta ) \label{eq:langevin} \,,
\end{equation}
where we remind the reader that $N=N(\eta )$ is the number of \efold{s} and we have defined a new term~\cite{Starobinsky:1994bd}\footnote{The identity $\dd \Theta [f(x)] / \dd x = \delta (x) (\dd f / \dd x)$ has been used here as well as 
\begin{equation*}
\frac{\dd k_\Lambda }{ \dd N } = \frac{1}{aH}\frac{\dd k_\Lambda }{ \dd \eta } = \frac{k_\Lambda}{aH} \left( \frac{\dd{\ln}a}{\dd \eta}  +  \frac{\dd{\ln}H}{\dd \eta} \right) = k_\Lambda \left( \frac{\dd{\ln}a}{\dd N}  +  \frac{\dd{\ln}H}{\dd N} \right) = k_\Lambda (1-\epsilon_1)\,.
\end{equation*}}
\begin{align}
\ & f(x_i,\eta ) \equiv \nonumber \\
& (1-\epsilon_1) k_\Lambda (\eta ) \int \frac{\dd^3k}{(2\pi )^{\frac{3}{2}}} \delta [k-k_\Lambda (\eta )]  \left[  \hat{a}_k \varphi_k(\eta )\ee^{-ik_j x^j} + \hat{a}^\dagger_k \varphi_k^*(\eta )\ee^{ik_j x^j } \right] \label{eq:def-white-noise1}\,,
\end{align}
which is valid in slow roll, where $\epsilon_1 = -\dd{\ln}H/\dd N$. Evaluating the two-point function of this new $f(x_i,\eta )$ term, under the assumption of massless mode functions, we find 
\begin{align}
\ & \langle 0 \vert f(x_i,\eta ) f(\tilde{x}_i,\tilde{\eta} ) \vert 0\rangle = \nonumber \\
& \qquad \quad (1-\epsilon_1) \frac{k_\Lambda^3(\eta )}{2\pi^2} \frac{\delta (\eta - \tilde{\eta})}{aH} \frac{\sin \left[ k_\Lambda (\eta ) \vert x-\tilde{x}\vert \right]}{k_\Lambda (\eta ) \vert x-\tilde{x}\vert} \varphi_{k_\Lambda}(\eta )\varphi_{k_\Lambda}^*(\eta ) \,, \label{eq:def-white-noise1-two-point}
\end{align}
where we have used the fact that $\delta [k-k_\Lambda (\eta ) ] \delta [k-k_\Lambda (\tilde{\eta} ) ] = (\dd \tilde{\eta} / \dd k_{\Lambda})\delta (\eta -\tilde{\eta})$. Note that to give the temporal correlation in terms of \efold{s}, one simply relates $\delta (\eta -\tilde{\eta})/(aH) = \delta (N-\tilde{N})$. In the super-horizon limit \Eq{eq:def-white-noise1-two-point} thus informs us that $f(x_i,\eta )$ becomes a white noise with an amplitude of $H^2/(4\pi^2)$ in \Eq{eq:langevin}.

In light of this new development, one is correct in the interpretation of \Eq{eq:langevin} as a Langevin equation --- a stochastic differential equation. If one were to evolve it under many realisations and integrate over time, the result would be that a Probability Density Function (PDF) could be constructed over the values that the field could take over a specified interval and given an appropriate initial condition. Note that this is a non-perturbative resummation which transcends the need for an expansion of the form in \Eq{eq:YFequation}, as long as the slow roll is satisfied.\footnote{In fact, this method can be applied to more general situations than slow roll. Applying this technique to the full phase space requires a second noise (and accompanying coupled Langevin equation) for the conjugate momentum~\cite{Grain:2017dqa}. It transpires that slow roll is still an attractor, however, and since we shall predominately consider test fields on a slow-roll background \Eq{eq:langevin} will be adequate for our needs.} This is due to the fact that the backreaction from small quantum fluctuations is inherently included into the background evolution described by \Eq{eq:langevin}, thus optimising the perturbative expansion at each new scale in time --- a cosmological analog to (but not exactly the same as~\cite{Woodard:2008yt}) the renormalisation group flow~\cite{Tsamis:2005hd}. Let us also note that massless mode functions were used to evaluate the white noise in \Eq{eq:langevin}, hence if the massless assumption ($m \ll H$) were no longer correct then it would invalidate this current approach. Applying \Eq{eq:langevin} to non-perturbatively calculate the IR behaviour of light (effectively massless) fields in an inflationary background is known as the stochastic inflation formalism~\cite{Starobinsky:1982ee,Starobinsky:1986fx,Starobinsky:1994bd}. 

If one considers how the background energy density is affected by the evolution of $\bar{\varphi}$, there are two distinct possibilities: it is an inflaton (or `non-test field') meaning that inflation proceeds with $H$ being contributed to by $\bar{\varphi}$; or, it is a `test field' which is sub-dominant to the overall energy density of the Universe during inflation and thus one can effectively treat $H$ as independent of $\bar{\varphi}$. In the former case it has been shown that in order to correctly reproduce the results from Quantum Field Theory (QFT) on curved spacetime, one must use $N$ as the time variable in \Eq{eq:langevin}. Many works have considered this issue~\cite{Finelli:2008zg, Finelli:2010sh, Finelli:2011gd, Vennin:2015hra} and incorporated the quantum diffusion given by \Eq{eq:langevin} directly into the inflationary dynamics, with interesting results. For example, the power spectrum~(\ref{eq:Pzeta:class-intro}) becomes~\cite{Vennin:2015hra, Assadullahi:2016gkk, Vennin:2016wnk}
\begin{align}
\calP_\zeta(k) =& 2
\left\lbrace \int_{\bar{\varphi}_*}^{\infty}\frac{\mathrm{d} A}{M_{{}_\mathrm{Pl}}}\frac{24\pi^2\Mp^4}{V( A)}\exp\left[\frac{24\pi^2\Mp^4}{V(A)}-\frac{24\pi^2\Mp^4}{V\left(\bar{\varphi}_*\right)}\right] \right\rbrace^{-1}
\nonumber \\
&
\times  \int_{\bar{\varphi}_*}^{\infty}\frac{\mathrm{d} A}{M_{{}_\mathrm{Pl}}}\left\lbrace\int_{A}^{\infty} \frac{\mathrm{d} B}{M_{{}_\mathrm{Pl}}} \frac{24\pi^2\Mp^4}{V(B)}\exp\left[\frac{24\pi^2\Mp^4}{V(B)}-\frac{24\pi^2\Mp^4}{V(A)}\right] \right\rbrace^2 \label{eq:Pzeta:sto-intro}\,,
\end{align}   
where $\bar{\varphi}_* = \bar{\varphi} (k_*)$ and one can see that the backreaction onto the inflationary dynamics, caused by these divergences, leads to the sensitivity of the power spectrum (among other observables~\cite{Assadullahi:2016gkk, Vennin:2016wnk}) to the entire inflationary domain.\footnote{A modified form of the separate Universe approach~\cite{Wands:2000dp} (and in \Sec{sec:sourcing-cos-pert}) has been employed to obtain the perturbations here.} Though this is a fascinating area of current research, in this thesis we shall focus primarily on the latter situation where $\bar{\varphi}$ is a test field.

The noise amplitude, calculated in \Eq{eq:def-white-noise1-two-point}, is $H^2/(4\pi^2)$ in the super-horizon limit, hence the corresponding Fokker-Planck equation to \Eq{eq:langevin} is
\begin{equation} \label{FP}
\frac{\partial }{\partial N}P\left[ \bar{\varphi} (x_i,\eta ) \right] =  \frac{\partial }{\partial \bar{\varphi}} \left\{ \frac{1}{3H^2}\frac{\partial V}{\partial \bar{\varphi}} P\left[ \bar{\varphi} (x_i,\eta ) \right]\right\} + \frac{\partial^2}{\partial \bar{\varphi}^2} \left\{ \frac{H^2}{8\pi^2} P\left[ \bar{\varphi} (x_i,\eta ) \right] \right\} \,,
\end{equation}
where we have defined $P\left[ \bar{\varphi} (x_i,\eta ) \right]$ as the one-point PDF.\footnote{There is a subtlety in defining the diffusion term ($\sim \partial^2/\partial \bar{\varphi}^2$) in this equation with the interpretation of stochastic process (either It\^{o} or Stratonovich). Here we choose It\^{o} as one can show that this exceeds the accuracy of the approximation one makes in the stochastic formalism in its current implementation~\cite{Vennin:2015hra}.} Notice that \Eq{FP} is similar to a continuity equation so that one may define a probability current $J$ as follows
\begin{equation}
\frac{\partial }{\partial N}P[\bar{\varphi}(x_i,\eta )] = - \frac{\partial }{\partial \bar{\varphi}}J[\bar{\varphi}(x_i,\eta )]\,, 
\end{equation}
where $J$ itself can be deduced as
\begin{equation} \label{eq:probcurr-test-field}
\ J[\bar{\varphi}(x_i,\eta )] = -\frac{1}{3H^2}\frac{\partial V}{\partial \bar{\varphi}}P[\bar{\varphi}(x_i,\eta )] - \frac{H^2}{8\pi^2}\frac{\partial}{\partial \bar{\varphi}}P[\bar{\varphi}(x_i,\eta )] \,.
\end{equation}
In the limit where $J=0$,\footnote{In later chapters we shall demonstrate why $J=0$ is an interesting limit. Suffice it here to state that when there is only one field, integrability at infinity enforces $J=0 \,, \forall \bar{\varphi}$. } and $\bar{\varphi}$ is a test field, the equilibrium distribution of this Fokker-Planck equation is
\begin{equation} \label{eq:equilib-spectator-dist}
\ P(\bar{\varphi} , N) \propto \exp \left( -\frac{8\pi^2V(\bar{\varphi })}{3H^4}\right) \,,
\end{equation}
and we shall return to discussing the stochastic formalism in more detail in later chapters, though predominately in \Chap{sec:infra-red-divergences}.

\section{\textsf{Perturbative reheating}} \label{sec:pert-reheat-intro}

Inflation itself leaves the Universe empty of SM particles. So far, the discussion of inflation has been confined to the processes that take place throughout its duration. A crucially important phase after inflation which is required to set cosmological initial conditions correctly is reheating. Reheating is the process by which the Universe fills with SM particles and it typically achieved through the thermalisation of the inflaton. The most thorough non-perturbative calculations of this process to date typically are performed by a lattice simulation~\cite{Amin:2014eta}. In this section however, we shall follow the perturbative arguments in Refs.~\cite{Kofman:1994rk, Kofman:1997yn, Kofman:1997ga, Bassett:2005xm} to attain a brief, overall picture for this process.   

\begin{figure}[h]
\begin{center}
\includegraphics[width=9cm]{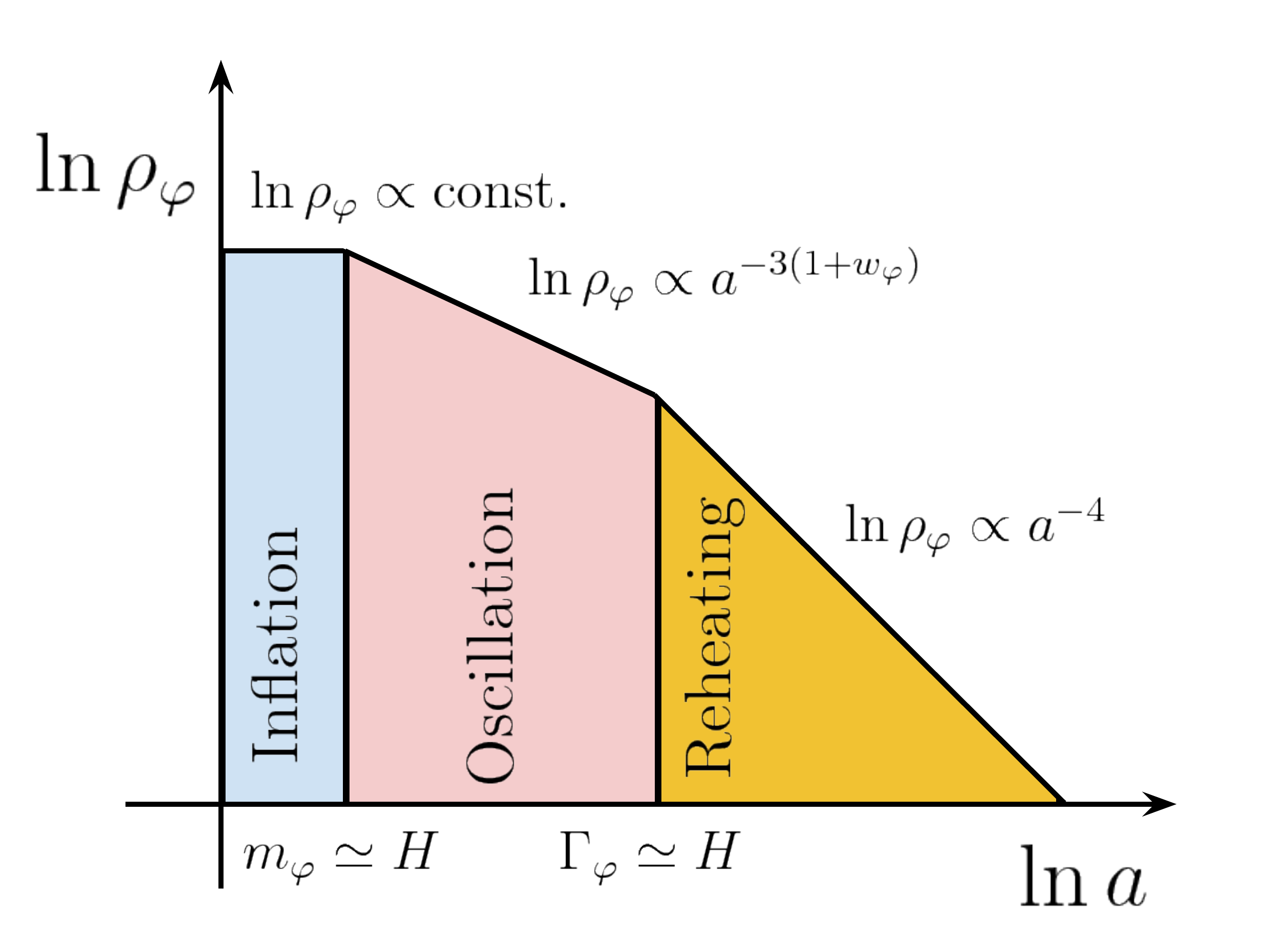}
\caption[Simple decay of the inflaton]{The basic decay of the inflaton given in \Eq{eq:inflaton-decay-boltz}.}
\label{fig:inflaton-simple-decay}
\end{center}
\end{figure}

In \Sec{sec:past} we noted that the continuity equation derived from the Einstein field equations (\Eq{eq:continuity}) could be verified by integrating \Eq{eq:boltz-rel} in the collisionless limit over $\dd^3p$ and combining with Eqs.~\eqref{eq:liouvilleop}, \eqref{eq:en-dens-bos} and \eqref{eq:pressure-bos}.\footnote{We can see that this is straightforward by using the chain rule
\begin{equation*}
\ - \hat{L}[f] = E\frac{\partial f}{\partial t} - H\vert p_i\vert^2 \frac{\partial f}{\partial E} = \frac{\partial (Ef)}{\partial t} - \frac{\partial{\ln} E}{\partial t}Ef - H\vert p_i\vert^2 \frac{\partial f}{\partial E} \,,
\end{equation*}
where after a $\dd^3p$ integration, the expression becomes $\dd \rho / \dd t + 3H\rho + 3H{\sf P}$.} In perturbative reheating one approximates the thermalisation of the inflaton as a decay process (such as a trilinear interaction) modeled by the following integrated Boltzmann equation
\begin{equation} \label{eq:inflaton-decay-boltz}
\frac{\dd \rho_\varphi}{\dd t} + 3(1+w_\varphi )H\rho_\varphi = -\Gamma_\varphi \rho_\varphi \,,
\end{equation}
where the term on the RHS is an approximate form for the collision operator and $\Gamma_\phi$ is the decay rate. The mechanism is such that when the Hubble rate drops to and below the decay rate $H=\Gamma_\varphi$, the thermalisation occurs and the inflaton field decays into SM particles. By considering the coherent oscillations of scalar fields in a cosmological background one can deduce, e.g., that if $\varphi$ oscillates about a quartic minimum $\propto \varphi^4$ then $w_\varphi \simeq 1/3$ and if it oscillates about a quadratic minimum $\propto \varphi^2$ then $w_\varphi \simeq 0$~\cite{Turner:1983he}. For a potential minimum with the shape $\propto \varphi^q$, one expects $w_{\varphi} = (q-2)/(q+2)$~\cite{PhysRevD.28.1243}.

A term such as the one on the RHS can be estimated through the physics of decay associated with $\varphi$. In particular, if the decay of $\varphi$ were gravitationally mediated, one would expect a relation of the form~\cite{Harigaya:2013vwa,Enqvist:2013gwf}
\bea
\label{eq:Mstar-before}
\Gamma_\varphi \simeq \frac{m_\varphi^3}{\Mp^2}\,,
\eea
which is, in practice, the smallest decay rate expected in the early Universe and hence it is essentially a lower bound on all possible decay rates.

The standard post-inflationary phenomenology is thus as follows: inflation terminates due to slow-roll violation $\epsilon_1 =1$; shortly after, the mass of the inflaton becomes of the same order as the Hubble rate $m_\varphi \simeq H$, it dynamically unfreezes and begins to coherently oscillate; and finally, after some time, the Hubble rate lowers to the same order as the decay rate of the inflaton $H\simeq \Gamma_\varphi$ and the field thermalises. This sequence of events is depicted in \Fig{fig:inflaton-simple-decay}.  

At the time of decay, assuming that the products of the process are in equilibrium with the thermal bath of SM particles, we can use \Eq{eq:en-dens-bos} to relate energy densities $\rho_\urad$ contained in radiation fluids to temperatures through
\bea
T=\left(\frac{30\rho_\urad}{\pi^2 g_*}\right)^{\frac{1}{4}}\, ,
\label{eq:T:rho}
\eea
where $g_*$ is the effective number of degrees of freedom 
\bea
\ g_* = \sum_{\rm B} \nu_{\rm B}\left( \frac{T_{\rm B}}{T}\right)^4 + \frac{7}{8}\sum_{\rm F} \nu_{\rm F}\left( \frac{T_{\rm F}}{T}\right)^4 \, ,
\label{eq:T:rho:effreldof}
\eea
which one calculates through a rescaling to account for both Bosonic and Fermionic degrees of freedom: $\nu_{\rm B}$ and $\nu_{\rm F}$, respectively.

\subsection{\textsf{The curvaton mechanism}} \label{sec:sourcing-cos-pert-curvaton}

We will now apply the tools developed in the previous sections to compute the observables of the curvaton model~\cite{Lyth:2001nq, Enqvist:2001zp, Moroi:2001ct}. This is a two-field model where the generic potential is of the form\footnote{Note that the curvaton itself is not required to specifically have a quadratic potential, though in the original realisation of the model this is the case~\cite{Lyth:2001nq, Enqvist:2001zp, Moroi:2001ct} as this proves useful to the reheatic kinematics.}
\bea
V\left(\varphi ,\sigma\right) = U\left(\varphi\right) + \frac{1}{2}m_{\sigma}^2\sigma^2 \label{eq:pot:gen:curvaton}\,.
\eea
During inflation, additional light (masses smaller than the Hubble rate $m<H$) test (energetically sub-dominant such that $H$ is independent of them) fields, such as $\sigma$, can fluctuate in an orthogonal direction to the inflaton perturbations (also known as adiabatic). These are known as isocurvature perturbations and can be observed directly as relic fluctuations in the relative number density of a given particle species~\cite{1984JETPL..40.1333L}. In the case of the curvaton model, one typically assumes that these relic number density variations have fully thermalised and reached thermodynamic equilibrium with the background radiation. When this happens, the perturbations of $\sigma$ can be shown to contribute only to the adiabatic perturbations~\cite{Lyth:2002my, Weinberg:2004kf, Smith:2015bln}.\footnote{Some curvaton models leave non-adiabatic fluctuations even after thermalisation of the decay products, e.g., in the presence of
a conserved quantum number, like baryon number (see~\cite{Lyth:2003ip}).} A curvaton then provides a mechanism to source the observed primordial density perturbations in the CMB independently of the inflaton. 

After inflation, the inflaton field energy density $\rho_{\varphi}$ decays into radiation and the energy density contained in the curvaton field, $\rho_\sigma$, may grow relative to the background energy density, until it also decays into radiation. If isocurvature perturbations do not persist but instead fully thermalise into an adiabatic perturbation when inflation ends, the total adiabatic power spectrum is given by the sum of 
the power spectra,
\bea
\label{ps}
{\mathcal{P}}^{\mathrm{total}}_{\zeta}(k_*)={\mathcal{P}}^{\varphi}_{\zeta}(k_*)+ {\mathcal{P}}^{\sigma}_{\zeta}(k_*) \, ,
\eea
where in the case of observational interest that $\sigma (k_*) \ll\Mp$ (reminding the reader that $k_* = 0.05\, \mathrm{Mpc}^{-1}$) 
\bea 
\label{eq:powerspecs}
\mathcal{P}^{\varphi}_{\zeta} \simeq \left. \frac{1}{2\epsilon_{1}}\left(\frac{H}{2\pi \Mp}\right)^2 \right\vert_{k_*}
 \;\; \mathrm{and} \;\;
\mathcal{P}^{\sigma}_{\zeta} \simeq \left. \rdec^2\left(\frac{H}{3\pi \sigma}\right)^2 \right\vert_{k_*} \,.
\eea
Here we have calculated the amplitude of the perturbation coming from $\sigma$ in \Eq{eq:powerspecs} by perturbing to linear order with an isocurvature fluctuation $\zeta_x-\zeta_{\rm tot}$ from a uniform density hypersurface, such that
\begin{equation}
\rho_{\rm tot} = \rho_{\varphi}\ee^{-4(\zeta_\varphi - \zeta_{\rm tot})} + \rho_{\sigma}\ee^{-3(\zeta_\varphi - \zeta_{\rm tot})} \,\, \Rightarrow \,\, \zeta_{\rm tot} = \zeta_\varphi + \rdec \zeta_{\sigma} \label{eq:first-order-hypersurface-curvaton} \,,
\end{equation}
and estimating\footnote{Note that $\zeta$ is a gauge-invariant quantity and so we have been able to compute this in the spatially flat gauge where $\zeta = \delta \rho /[3(\rho + {\sf P})]$~\cite{Lyth:2003im}. There is energy conservation of each species (see \Eq{eq:continuity}) on super-horizon scales as they evolve along their own FLRW comoving worldlines. }
\begin{equation}
\ {\cal P}^\sigma_\zeta (k) = \rdec^2\frac{(\delta \ln \rho_\sigma )^2}{9(1+w_\sigma )^2} \simeq \frac{4}{9}\rdec^2\left( \frac{\delta \sigma}{\sigma} \right)^2 \,,
\end{equation}
where $\rdec \equiv 3\rho_\sigma /(3\rho_\sigma + 4\rho_\varphi )$. Note that here we have used the fact that the equation of state of the curvaton during its oscillations will be $w_\sigma \simeq 0$ due to its quadratic minimum and, in order to obtain \Eq{eq:first-order-hypersurface-curvaton}, we have assumed the sudden-decay approximation for the curvaton~\cite{Malik:2006pm, Sasaki:2006kq}. Note also that $\rdec$ can vary from zero to unity in the case that $\sigma$ dominates the background energy density at the time it decays. 

The spectral index $\nS$ and tensor-to-scalar ratio $r$ of this model, following our definitions in Eqs.~\eqref{eq:spectral-ind} and~\eqref{eq:tens-scalar-ratio}, are given to leading order in slow roll by~\cite{Wands:2002bn}
\begin{align}
\label{eq:nsr:slowroll}
\nS-1 & =   \lambda \left(-2\epsilon_{1}+2\eta_{\sigma}\right) +\left(1-\lambda\right)\left(-6\epsilon_{1}+2 \eta_{\varphi}\right) \\
r & = 16\epsilon_{1} \left(1- \lambda\right)\,,
\end{align}
where here we have defined $\eta_\varphi \equiv (3H^2)^{-1}\partial^2V/(\partial \varphi^2)$ and $\eta_\sigma \equiv (3H^2)^{-1}\partial^2V/(\partial \sigma^2)$ while $\lambda$ denotes the fraction of the total perturbations originating from $\sigma$,
\bea
\label{eq:lambda:def}
\lambda\equiv\frac{{\mathcal P}^{\sigma}_{\zeta}}{{\mathcal P}^{\rm total}_{\zeta}} \, .
\eea
Note that \Eq{eq:lambda:def} may be evaluated by inserting \Eq{eq:powerspecs}. When the primordial density perturbation is entirely due to curvaton field fluctuations then the original curvaton model~\cite{Lyth:2001nq, Enqvist:2001zp, Moroi:2001ct} is realised. 

Another way to detect the curvaton is through primordial non-linearity of the density perturbations, of which the key observable is the local non-Gaussianity of the bispectrum, parametrised by $\fnl$ through the relation~\cite{Wands:2010af}
\begin{equation}
\Phi (x_i) = \phi (x_i) +\fnl \left[ \phi^2 (x_i) - \langle \phi^2 \rangle \right] + \dots \,,
\end{equation}
where $\Phi$ is the spatially varying metric potential ($\Phi = (3/5)\zeta$ during matter domination) and $\phi$ is a single Gaussian random field.

In curvaton models, the value of $\fnl$ can be approximately related to
\bea
  \label{eq:fnl}
\fnl \simeq \lambda^2 \left(\frac{5}{4\rdec}-\frac53-\frac{5\rdec}{6} \right)\,,
\eea
where we have assumed sudden-decay approximation for the curvaton here as well~\cite{Ichikawa:2008iq}. Note that this formula follows naturally if one perturbs on constant density hypersurfaces to second order in $\zeta$ such that~\cite{Sasaki:2006kq}
\begin{gather}
\ 16\left[ \zeta^{(1)}_{\rm tot}\right]^2 (\rho_{\rm tot}-\rho_\sigma ) - 4 \zeta^{(2)}_{\rm tot}(\rho_{\rm tot}-\rho_\sigma ) = - 9\left[ \zeta_\sigma^{(1)} - \zeta_{\rm tot}^{(1)}\right]^2 \rho_\sigma - 3\left[ \zeta_\sigma^{(2)} - \zeta_{\rm tot}^{(2)}\right] \rho_\sigma \nonumber \\
\Rightarrow \frac{\zeta^{(2)}_{\rm tot}}{\left[ \zeta^{(1)}_{\rm tot}\right]^2} \simeq \lambda^2 \left( \frac{3}{2\rdec} -2 -\rdec \right)  \,,
\end{gather}
where we have used \Eq{eq:first-order-hypersurface-curvaton} and $\zeta^{(2)}_\sigma \simeq -(3/2)[\zeta^{(1)}_\sigma ]^2$~\cite{Sasaki:2006kq}.

\chapter{\textsf{Statistical introduction}}
\label{sec:statistical-intro} \niceline {\vskip+1ex} 

\begin{center}
\fbox{\parbox[c]{13cm}{\vspace{1mm}{\textsf{\textbf{Abstract.}}} In this chapter we will very briefly review some topics in Bayesian statistics~\cite{Cox:1946,Jeffreys:1961,deFinetti:1974}, dealing with the mathematical formulation of inference and model selection. 
In addition, some useful concepts in classical information theory~\cite{kullback1951} will be covered as well as a short review of the fundamentals for Bayesian experimental design~\cite{Chaloner1995aa} in order to prepare for its application in \Chap{sec:future-prospects}.\vspace{1mm}}}
\end{center}

\section{\textsf{Bayesian inference}} \label{sec:bayes-inference-sec}

The robustness of the scientific method relies upon a continual comparison between theory and experiment. Rigorous statistical analysis is thus a cornerstone of any scientific result, where there still exits a lively debate over the optimal method.\footnote{Though the debate between methods is philosophical in nature, it is important to still acknowledge that the perspective taken in this thesis will be largely that of a `Bayesian point of view', and hence we will avoid addressing these fundamental questions in favour of a more direct technical application of the formalism itself.}

In probability theory, one can denote the probability of an event $A$ occurring by ${\sf p}(A)$. If one has another event $B$, upon which $A$ may or may not rely, then one may construct: the probability of $B$ occurring ${\sf p}(B)$; the probability of $A$ occurring given that $B$ has occurred ${\sf p}(A\vert B)$ (and its converse); and the joint probability of both $A$ and $B$ occurring, ${\sf p}(A, B)$. The essential concept of Bayesian statistics originates from considering the following identity between conditional probabilities of $A$ and $B$ and their joint probability
\begin{equation} \label{eq:probAB}
\ {\sf p} (A\vert B) {\sf p} (B) = {\sf p} (B\vert A) {\sf p} (A) = {\sf p} (A, B) \,. 
\end{equation}
Adapting \Eq{eq:probAB}, we immediately find Bayes' theorem
\begin{equation} \label{eq:bayes-theorem}
\ {\sf p} (A\vert B) = \frac{{\sf p} (B \vert 
A) {\sf p} (A)}{{\sf p} (B)} \,.
\end{equation}
\Eq{eq:bayes-theorem} informs us on the correct procedure that one must take in updating knowledge about $A$ with $B$. Hence, it is \Eq{eq:bayes-theorem} which forms the basis upon which all Bayesian reasoning is founded.

Statistical inference in the Bayesian paradigm falls naturally out of \Eq{eq:bayes-theorem}. If one wishes to update knowledge of a parameter $\theta$ with data ${\cal D}$ to obtain a \emph{posterior} distribution over it $p(\theta \vert {\cal D})$, \Eq{eq:bayes-theorem} tells us to multiply the \emph{likelihood} function over a collection of data ${\cal L}({\cal D}\vert \theta )$ to some given \emph{prior} information $\pi (\theta )$, such that
\begin{equation} \label{eq:bayes-theorem-ppiL}
\ p (\theta \vert {\cal D}) \propto {\cal L}({\cal D}\vert \theta ) \, \pi (\theta ) \,.
\end{equation}
Let us illustrate the Bayesian update of $\pi$ into $p$ using the following simple example: consider a Gaussian prior 
\begin{equation}
\pi (\theta ) = \frac{1}{\sqrt{2\pi}\sigma_\pi}\exp \left[ -\frac{(\mu -\theta )^2}{2\sigma^2_\pi} \right] \label{eq:pr-pi}\,,
\end{equation}
and likelihood function 
\begin{equation}
\ {\cal L} ({\cal D}\vert \theta ) = \frac{1}{\sqrt{2\pi}\sigma_{\cal L}}\exp \left[ -\frac{(\mu -\theta )^2}{2\sigma^2_{\cal L}} \right]  \label{eq:gauss-lik-func-examp}\,,
\end{equation}
which share the same mean $\mu$ but have different standard deviations $\sigma_\pi$ and $\sigma_{\cal L}$, respectively. The posterior distribution which corresponds to these distributions can be calculated using \Eq{eq:bayes-theorem-ppiL} (ignoring the normalisation), where one finds
\begin{equation}
\ p\left(\theta \vert\mathcal{D}\right) \propto \exp \left[ -(\mu - \theta )^2\left( \frac{1}{2\sigma_{\pi}^2} + \frac{1}{2\sigma_{{\cal L}}^2} \right) \right] \label{eq:post-bayes-update-analy}\,.
\end{equation}
Comparing \Eq{eq:post-bayes-update-analy} with \Eq{eq:pr-pi}, we see that the prior standard deviation has been updated by the data using Bayes' theorem $\sigma_\pi \rightarrow ( \sigma_\pi^{-2} + \sigma_{\cal L}^{-2})^{-1/2}$. From this example, we see that the net results will always increase the precision over $\theta$ for finite $\sigma_{\cal L}$. 

If we were to go a step further and assume that a model $\mathcal{M}$ had a defined set of parameters $\theta$, the posterior probability $p$ of its parameters $\theta$ would be expressed as
\bea
\label{eq:posterior:def:before}
p\left(\theta \vert\mathcal{D},\mathcal{M}\right)=\frac{\mathcal{L}
\left(\mathcal{D}\vert\theta ,\mathcal{M}\right)
\pi\left(\theta \vert \mathcal{M}\right)}{\mathcal{E}\left(\mathcal{D}\vert\mathcal{M} \right) } \, ,
\eea
where $\mathcal{L}(\mathcal{D}\vert\theta ,\mathcal{M})$ is the likelihood and represents the probability of observing the data $\mathcal{D}$ assuming the model $\mathcal{M}$ is true and $\theta$ are the actual values of its parameters and $\pi (\theta \vert \mathcal{M} )$ is the prior distribution on the parameters $\theta$. Notice that, in contrast to \Eq{eq:bayes-theorem-ppiL}, we have now specifically defined $\mathcal{E}\left(\mathcal{D}\vert\mathcal{M} \right)$ as the normalisation constant called the Bayesian evidence, which we shall discuss further in \Sec{sec:bayes-model-selection}.

\Eq{eq:posterior:def:before} shows that $\mathcal{L}$ is an important quantity to construct when a statistical inference is to be performed. It is possible to conduct an inference on parameters with very little information about this quantity,\footnote{We refer the reader to the many reviews on the topic, e.g., Refs.~\cite{doi:10.1093-sysbio-syw077, blum2013, 2011arXiv1101.0955M}.} however in this thesis we shall primarily focus on situations where the likelihood function is well known and parameterised in an optimal way, e.g., such as that of \Ref{Ringeval:2013lea}. The dimensionality of $\theta$ is often an important indication of what methodology to use --- splitting ${\cal L}$ into two approximately categories, either: 
\begin{enumerate}
\item{The number of dimensions is low enough such that one can perform Importance (or Rejection) sampling~\cite{2003itil.book.....M,gelman2013bayesian}. We will make use of this technique combined with Nested sampling in \Chap{sec:statistical-comp-importancesamp}, where more detail can be found in Appendix~\ref{sec:statistical-comp-importancesamp}.}
\item{The number of dimensions is too high, in which case one may select from a number of sampling techniques, e.g., Metropolis-Hastings, Gibbs and Hamiltonian Monte Carlo sampling~\cite{2003itil.book.....M,gelman2013bayesian}. }
\end{enumerate}

\section{\textsf{The Kullback-Leibler divergence}} \label{sec:info-theory-tools}

Information theory can provide a powerful insight into statistical inference. In particular, it is quite common to find quantities which are reparameterisation invariant and hence extremely useful for robust analysis. The relative (or conditional) entropy between the prior $\pi(\theta \vert {\cal M})$ and posterior $p(\theta \vert {\cal D},{\cal M})$ distributions on some parameter $\theta$ is called the Kullback-Leibler divergence, and is defined for a 1-dimensional $\theta$-space as  
\bea
\label{eq:app:dkl}
\dkl\left(p \vert\vert \pi\right) \equiv \int {p}\left(\theta \vert {\cal D},{\cal M} \right) \log_2 \left[\frac{{p}\left(\theta \vert {\cal D},{\cal M}\right)}{\pi \left(\theta \vert {\cal M}\right)} \right] \dd \theta\, ,
\eea
where we have chosen a base of $2$ such that $\dkl$ is measured in bits and the integration limits are those specified by the domain of $\theta$. This is a measure of the amount of information provided by the data about the parameter $\theta$. Since it uses a logarithmic score function, it is a well-behaved measure of information~\cite{bernardo:2008}. Note that \Eq{eq:app:dkl} can easily be generalised to an arbitrary number of parameter dimensions, but we shall here keep $\theta$ as 1-dimensional for simplicity.

The $\dkl$ is indeed invariant under a generic reparameterisation $\theta\rightarrow \theta^\prime$. This is because the prior and posterior on $\theta^\prime$ can be calculated according to
\bea
\pi (\theta \vert {\cal M} ){\rm d}\theta = \bar{\pi}(\theta' \vert {\cal M}){\rm d}\theta'\, ,\quad\quad
p (\theta \vert {\cal D},{\cal M}){\rm d}\theta = \bar{p}(\theta' \vert {\cal D},{\cal M}){\rm d}\theta' \, ,
\eea
and hence one can determine that
\bea
\bar{{p}}(\theta' \vert {\cal D},{\cal M}) \log_2 \left[ \frac{\bar{p}(\theta' \vert {\cal D},{\cal M})}{\bar{\pi} (\theta'\vert {\cal M})} \right] {\rm d}\theta' = {p}(\theta \vert {\cal D},{\cal M}) \log_2 \left[ \frac{{p}(\theta \vert {\cal D},{\cal M})}{\pi (\theta \vert {\cal M})} \right] {\rm d}\theta \, .
\eea
Another very important property of the Kullback-Leibler divergence is that it is always positive, due to Gibbs' inequality which states that for two continuous normalised distributions $\pi(\theta)$ and $p(\theta)$, one has\footnote{One may also show this from Jensen's inequality, due to the fact that the logarithm is a concave function.}
\bea
\int p(\theta \vert {\cal D},{\cal M})\log_2 p(\theta \vert {\cal D},{\cal M}) \dd\theta \geq \int p(\theta \vert {\cal D},{\cal M})\log_2 \pi(\theta \vert {\cal M}) \dd\theta\, .
\eea

In order to gain some immediate insight into how $\dkl$ is affected by the shape of the prior and posterior distributions, let us compute its value in the case where both distributions are 1-dimensional Gaussians, with mean values $\mu_\pi$ and $\mu_p$ respectively, and with standard deviations of $\sigma_\pi$ and $\sigma_p$, respectively. Their distributions should take the form 
\begin{align}
\pi (\theta \vert {\cal M}) &= \frac{1}{\sqrt{2\pi}\sigma_\pi}\exp \left[ -\frac{(\mu_\pi -\theta )^2}{2\sigma^2_\pi} \right] \label{eq:prior-gaussian-dis} \,, \\ 
\ p(\theta \vert {\cal D},{\cal M}) &= \frac{1}{\sqrt{2\pi}\sigma_p}\exp \left[ -\frac{(\mu_p-\theta )^2}{2\sigma^2_p} \right]\,.
\end{align}
Defining $\delta\mu \equiv \mu_p-\mu_\pi$, one obtains
\bea
\label{eq:DKL:gaussian}
\dkl = \frac{1}{2\ln 2}\left[ \frac{(\delta \mu )^2}{\sigma_{\pi}^2} + \frac{\sigma_{p}^2}{\sigma_{\pi}^2}+ 2\ln \left( \frac{\sigma_{\pi}}{\sigma_{p}}\right) -1\right]\,,
\eea
where the first term accounts for the update in the preferred value and can be understood as follows: if the change in the preferred value is large compared to the uncertainty level of the prior, then non-trivial information is gained and the value of $\dkl$ is large. In contrast, the other terms depend only on the ratio $\sigma_p/\sigma_\pi$, and therefore yield a contribution to $\dkl$ which increases when $\sigma_p/\sigma_\pi$ decreases, corresponding to improved measurements of $\theta$.

Let us also define a quantity which we dub the `information density' $\delta \dkl$, which one can view as the information gained in each bin $\dd\theta$ of the parameter $\theta$, such that
\bea
\int {p}\left(\theta \vert {\cal D},{\cal M}\right) \log_2 \left[\frac{{p}\left(\theta \vert {\cal D},{\cal M}\right)}{\pi \left(\theta \vert {\cal M}\right)} \right] \dd \theta \equiv \int \delta \dkl\left(\theta\right)\dd \theta \, .
\label{eq:DKL}
\eea 
Contrary to $\dkl$, this quantity is parameterisation dependent, but it indicates where information is mostly gained and lost. We shall use both $\dkl$ and $\delta \dkl$ in later chapters.

\section{\textsf{Bayesian model selection}} \label{sec:bayes-model-selection}

Let us now consider that a higher-dimensional $\theta$ contains information about an additional parameter $\theta_a$ (or many parameters) that we do not want to study, one should marginalise out $\theta_a$ (or all of the parameters) 
\bea \label{eq:marginalisation}
\ p\left(\theta \vert\mathcal{D},\mathcal{M}\right) = \int p(\theta , \theta_a\vert\mathcal{D},\mathcal{M})\dd\theta_a \, ,
\eea
where now the integration limits correspond to the domain of $\theta_a$. Marginalisation is a generic feature of probability distributions and considering where it is present within \Eq{eq:bayes-theorem} will yield us a tool which is key for Bayesian model selection. In \Eq{eq:posterior:def:before} we can use \Eq{eq:marginalisation} to marginalise out $\theta$, leaving
\bea
\label{eq:evidence:def:before}
\mathcal{E}\left(\mathcal{D}\vert\mathcal{M} \right) 
= \int \mathcal{L}
\left(\mathcal{D}\vert\theta ,\mathcal{M}\right)
\pi\left(\theta \vert \mathcal{M}\right) \dd\theta\, ,
\eea
which is often known as the marginal likelihood or the Bayesian evidence~\cite{jeffprob,Trotta:2017wnx}. 

The Bayesian evidence is a full integration over the parameter space of $\theta$ (or its analogue in an arbitrary number of dimensions), and hence contains all of the marginal information about the fit of the model to the data which is therefore reparameterisation-invariant.\footnote{Note that this is obvious since ${\cal E}$ has no explicit dependence on $\theta$ since it has been integrated out. An equivalent statement is $\mathcal{L}
\left(\mathcal{D}\vert\theta ,\mathcal{M}\right)
\pi\left(\theta \vert \mathcal{M}\right) \dd\theta = \mathcal{L}
\left(\mathcal{D}\vert\theta' ,\mathcal{M}\right)
\pi\left(\theta' \vert \mathcal{M}\right) \dd\theta'$.} Motivated by this basic property of ${\cal E}$, we can use it in a Bayesian equivalent to a classic maximum likelihood ratio test to compare two models ${\cal M}_\alpha$ and ${\cal M}_\beta$
\begin{equation} \label{eq:BayesFactor-intro}
\ {\rm B}_{\alpha \beta}  = \frac{{\cal E}_{\alpha} ({\cal D}\vert {\cal M}_\alpha )}{{\cal E}_{\beta} ({\cal D}\vert {\cal M}_\beta )}\,,
\end{equation}
which is known as the Bayes factor. The benefit of using \Eq{eq:BayesFactor-intro} to compare between the relative fitting performance of models to data is that it manifestly penalises against too much model structure. A very simple example of this is to once again consider a Gaussian distribution for the likelihood of the same form as \Eq{eq:gauss-lik-func-examp}, but with a mean set to $\mu = 0$, and a prior constructed from a finite-domain Dirac comb 
\begin{equation}
\pi (\theta \vert {\cal M}) = \frac{1}{N}\sum_{n=0}^{N} \delta (\theta -n) \label{eq:Dirac-comb-prior}\,,
\end{equation}
with $N$ denoting the number of delta functions, hence acting as a crude metric for model structure. Notice also that the normalisation factor of $1/N$ in \Eq{eq:Dirac-comb-prior} is necessary for the prior to be normalised to $1$. Using \Eq{eq:evidence:def:before} one can show that the evidence in this example becomes
\begin{equation}
\ {\cal E}({\cal D}\vert {\cal M}) = \frac{1}{\sqrt{2\pi}\sigma_{\cal L}N} \sum_{n=0}^{N} \exp \left[ -\frac{n^2}{2\sigma^2_{\cal L}} \right] \label{eq:evidence-ex-analytic-fun}\,.
\end{equation}
\Eq{eq:evidence-ex-analytic-fun} thus demonstrates how increasing the structure of a model, i.e., increasing $N$ decreases ${\cal E}$, the evidence for model decreases as a penalty for overfitting the data.   

Before we move on to the next section, we shall mention here briefly that, although there is no universally derivable threshold for the Bayes' factor ${\rm B}_{\alpha \beta}$ to take the value of such that it indicates a `ruling-out' of ${\cal M}_{\alpha}$ with respect to ${\cal M}_{\beta}$, a useful guideline is provided by the Jeffreys threshold~\cite{jeffprob,2008arXiv0804.3173R}. This essentially suggests that $\vert \ln {\rm B}_{\alpha \beta} \vert \simeq 5$ is a reasonable criterion to use.

\section{\textsf{Choice of prior}}
\label{sec:priors-intro}

In \Sec{sec:bayes-inference-sec} we highlighted the importance of accurate computation for the likelihood function ${\cal L}$ for rigorous statistical inference. We will now discuss how best to choose a prior distribution $\pi (\theta \vert {\cal M})$ such that the scientific question one seeks to answer through the inference is well posed. Priors may be constructed from either subjective theoretical prejudice (here described as `informative') or derived using general methodologies (here described as `non-informative'). 

Taking the non-informative viewpoint, an optimal prior from the perspective of the likelihood is the `Jeffreys prior'~\cite{1946RSPSA.186..453J}. This prior is $\pi (\theta ) \propto \sqrt{\det {\cal F}_{ij}}$, where ${\cal F}_{ij}$ is the Fisher information matrix. ${\cal F}_{ij}$ of a general distribution $P (\theta ; Y)$, equipped with a set of hyperparameters $Y=\{ y_i\}$, is defined as\footnote{Notice that Taylor expanding either $\dkl$ from \Sec{sec:info-theory-tools} about their minimum values ($Y_0$ where $A(\theta ;Y_0)=B(\theta ;Y_0)$) with respect to the shape parameters in $Y$, we find that both quantities vanish at first order in the expansion leaving terms $\propto {\cal F}_{ij}\Delta y_i\Delta y_j$.}
\begin{equation}
\ {\cal F}_{ij} \equiv \int  P(\theta ;Y) \, \frac{\partial{\ln}P}{\partial y_i}\frac{\partial {\ln}P}{\partial y_j} \dd \theta = \left\langle \frac{\partial{\ln}P}{\partial y_i}\frac{\partial {\ln}P}{\partial y_j} \right\rangle_P \,.
\end{equation}
In the case of constructing the Jeffreys prior out of ${\cal F}_{ij}$, one makes the choice $P = {\cal L}$. The key property of this prior is that it is invariant under a reparameterisation of the likelihood~\cite{1946RSPSA.186..453J}, and hence it may be used to motivate a choice of logarithmic prior for scale parameters. Such a choice for scale parameters may also be motivated in other ways, as we shall discuss below.
 
Continuing in our discussion of non-informative priors is a similar notion to the eigenfunction of ${\cal L}$. Such priors are known as `conjugate priors' which have the property that the family of probability distribution of the posterior $p$, computed through \Eq{eq:bayes-theorem}, is ensured to be the same as $\pi$ up to a variation in hyperparameters of that family~\cite{2003itil.book.....M}. For example, \Eq{eq:pr-pi} demonstrates that a Gaussian prior with known $\mu$ is conjugate to a Gaussian posterior distribution. 

Symmetry can also be used to motivate a non-informative prior choice. If a prior is invariant under location transformations
\begin{equation}
\pi (\theta ) \rightarrow \pi (\theta + a)\,,
\end{equation}
with a domain of $\theta_{\rm max}>\theta > \theta_{\rm min}$, then it is known that the measure choice which leaves the prior volume invariant will be $\pi (\theta ) \dd \theta \propto \Theta (\theta - \theta_{\rm min}) \Theta (\theta_{\rm max} - \theta ) \dd \theta$ which may be shown by the solution to the corresponding differential equation. Similarly, if the prior is invariant under scale transformations of the form
\begin{equation}
\pi (\theta ) \rightarrow \pi (a\theta ) \,,
\end{equation}
then the prior volume is left invariant if one chooses the logarithmic prior measure $\pi (\theta ) \dd \theta \propto \dd{\ln}\theta$. These are both very simple examples of Haar measures~\cite{haar1933massbegriff}, which generalise this concept to measures which are invariant under left and right actions from an arbitrary group $G$.

In this thesis we will also make use of informative priors, which we derive through theoretically-motivated calculations --- see, e.g., Chapters~\ref{sec:curvaton-reheating} and~\ref{sec:isocurvature-fields}.

\section{\textsf{Bayesian experimental design}}
\label{sec:experimental-design-principles}

In Bayesian analysis, one can forecast a future observation $d \in {\cal D}_{\rm fut}$, given the current data ${\cal D}_{\rm cur}$ using the posterior distribution over $\theta$ given ${\cal D}_{\rm cur}$. One achieves this by the following marginalisation
\begin{equation}
\ p (d \vert {\cal D}_{\rm cur}) = \int p(d\vert \theta , {\cal D}_{\rm cur}) \, p(\theta \vert {\cal D}_{\rm cur}) \dd \theta \label{eq:forecast-margin} \,.
\end{equation}
In the same vein, one may forecast any $\theta$-dependent quantity, say $U(\theta )$, by inserting it in place of $p(d\vert \theta , {\cal D}_{\rm cur})$ in \Eq{eq:forecast-margin}. The quantity one thus constructs is an expectation value $\langle U\rangle$. If one now identifies $U$ with a utility function~\cite{Chaloner1995aa,Trotta:2010ug,Leclercq:2015rta}, which attributes a value to each possible realisation forecast by the posterior, then the expected performance of a given probabilistic process will be given by
\begin{equation}
\langle U \rangle = \int U(\theta ) \, p(\theta \vert {\cal D}_{\rm cur}) \dd \theta \label{eq:expected-ut-intro} \,,
\end{equation}
where $U(\theta )$ can be, e.g., the information gain $\dkl$ between the current and future posteriors. \Eq{eq:expected-ut-intro} will be adapted to forecast the performance of astronomical experiments in \Chap{sec:future-prospects}. We shall leave further development of these concepts until then.

\chapter{\textsf{Curvaton reheating}}
\label{sec:curvaton-reheating} \niceline {\vskip+1ex} 

\begin{center}
\fbox{\parbox[c]{13cm}{\vspace{1mm}{\textsf{\textbf{Abstract.}}} In this chapter we will study the situation where inflation is driven by a single scalar inflaton field, but an extra light (relative to the inflationary Hubble scale) scalar field can also contribute to the total amount of curvature perturbations. This field is essentially a `curvaton' --- which we introduced in \Sec{sec:sourcing-cos-pert-curvaton} --- and is assumed to be subdominant during inflation but can store a substantial part of the energy budget of the Universe during reheating. We demonstrate that when an additional field exists, and contributes to the curvature perturbation, it leads to a substantial gain in information about the precise temperature of the Universe at reheating~\cite{Hardwick:2016whe}. \vspace{1mm}}}
\end{center}

\section{\textsf{Introduction}}
\label{sec:intro}
How inflation ends and is connected to the subsequent hot Big-Bang phase through the reheating era is still poorly constrained. The main reason is that at linear order, in absence of entropic perturbations, curvature perturbations are preserved on large scales~\cite{Lukash:1980iv, Bardeen:1983qw}, hence their statistical properties at recombination time carry limited direct information about the microphysics at play during the reheating epoch. 

Nevertheless, the amount of expansion between the end of inflation and the onset of the radiation epoch determines the amount of expansion between the Hubble crossing time of the physical scales probed in the CMB and the end of inflation~\cite{Martin:2006rs, Martin:2010kz, Easther:2011yq, Dai:2014jja, Rehagen:2015zma, Domcke:2015iaa}. As a consequence, the kinematic properties of reheating set the time frame during which the fluctuations probed in cosmological experiments emerge, hence defining the location of the observational window along the inflationary potential. If inflation is realised with a single slowly-rolling field for instance, this effect can be used to extract constraints on a certain combination of the averaged equation-of-state parameter during reheating and the reheating temperature, the so-called ``reheating parameter'', yielding an information gain of about 1 bit on the reheating history~\cite{Martin:2014nya, Martin:2016oyk}. 

Since the reheating parameter is related to quantities such as the effective potential of the inflationary fields during reheating and the couplings between these fields and their decay products, this provides an indirect probe into the fundamental microphysical parameters of reheating~\cite{Drewes:2015coa}. Deriving such a relationship for concrete reheating models is therefore an important, although often laborious, task. Let us also notice that since the dependence of inflationary predictions on the reheating history is now of the same order as the accuracy of the data itself, different prescriptions for the reheating dynamics give rise to substantially different results regarding which inflationary models are preferred by the data~\cite{Martin:2014nya, Martin:2014rqa}. Therefore, improving our understanding of reheating has become crucial to derive meaningful constraints on inflation itself.
\begin{figure}
\begin{center}
\includegraphics[width=10cm]{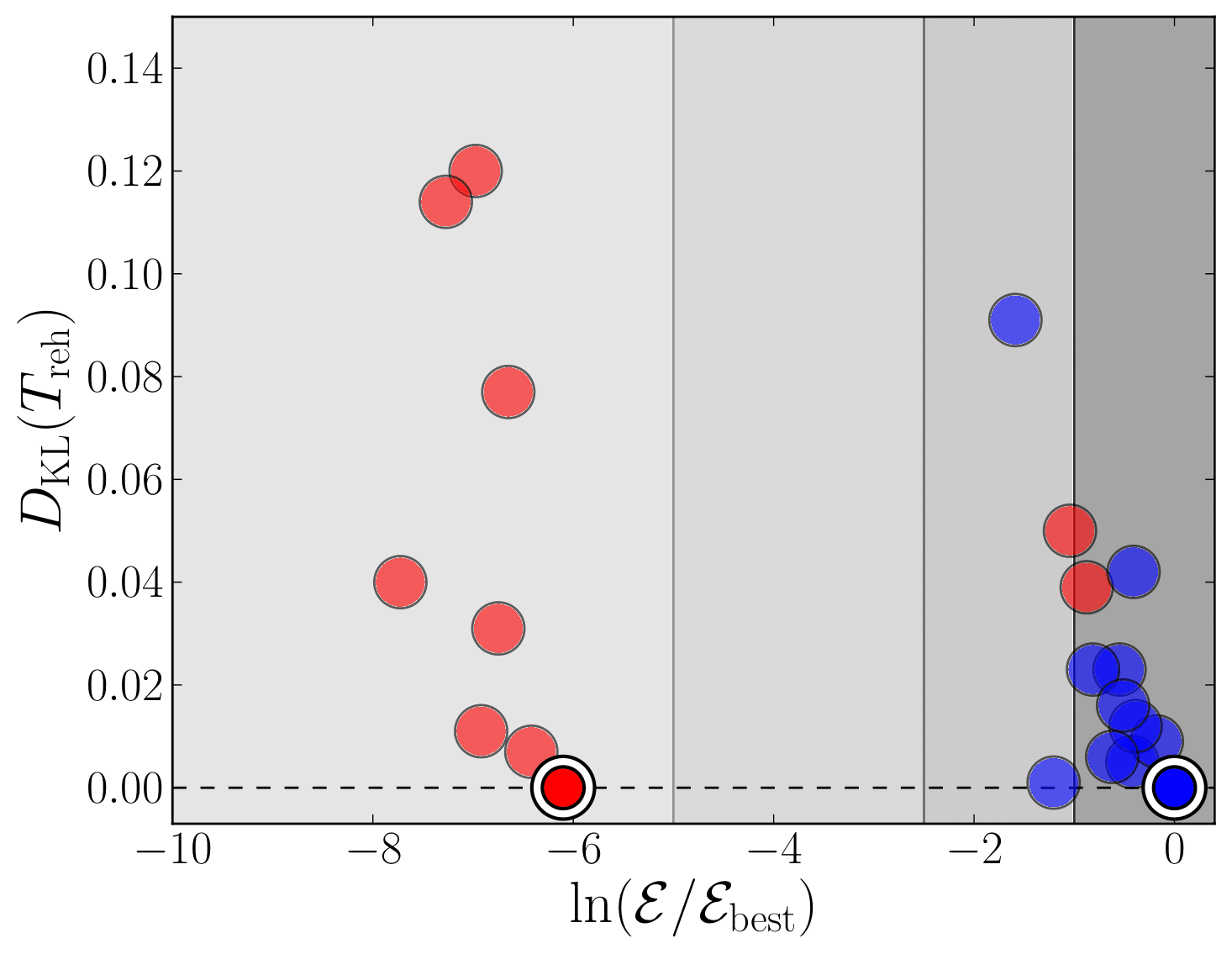}
\caption[Information gain on the reheating temperature]{Relative information gain between prior and posterior distributions over the reheating temperature $T_\ureh$, plotted against the Bayes factor normalised to the best model (single-field Higgs inflation). The white circled disks stand for purely single-field models (blue: Higgs inflation $\mathrm{HI}$, red: quartic inflation $\mathrm{LFI_4}$) and the other disks are the equivalent models with an additional light scalar field in the different reheating scenarios (blue: $\mathrm{MC}_i\mathrm{HI}$, red: $\mathrm{MC}_i\mathrm{LFI}_4$, for $i=1\cdots 10$). From left to right, the grey shading darkens, denoting the respective evidence ratios: strongly disfavoured, moderately disfavoured, weakly disfavoured and favoured. The two red disks that lie in the favoured region are reheating scenarios 5 and 8 for quartic inflation. A logarithmically flat prior has been used on the vev of the additional light scalar field at the end of inflation, see \Sec{sec:Bayesian}.}
\label{fig:mainresult}
\end{center}
\end{figure}

We will follow this line of research and study the situation where inflation is driven by a single scalar inflaton field $\phi$, but an extra light (relative to the inflationary Hubble scale) scalar field $\sigma$ can also contribute to the total amount of curvature perturbations. This additional field $\sigma$ is assumed to be subdominant during inflation but can store a substantial part of the energy budget of the Universe during reheating. In the limit where it is entirely responsible for the observed primordial curvature perturbations, the class of models this describes is essentially the curvaton scenario of~Refs.\cite{Linde:1996gt, Enqvist:2001zp, Lyth:2001nq, Moroi:2001ct, Bartolo:2002vf} and \Sec{sec:sourcing-cos-pert-curvaton}. Here however, we address the generic setup where both $\phi$ and $\sigma$ can a priori contribute to curvature perturbations~\cite{Dimopoulos:2003az, Langlois:2004nn, Lazarides:2004we, Moroi:2005np}. The reasons why we focus on these scenarios are threefold. First, from a theoretical perspective, most physical setups that have been proposed to embed inflation contain extra scalar fields that can play a role either during inflation or afterwards. This is notably the case in string theory models where extra light scalar degrees of freedom are usually considered~\cite{Turok:1987pg, Damour:1995pd, Kachru:2003sx, Krause:2007jk, Baumann:2014nda}. Second, from an observational point of view, these scenarios predict levels of non-Gaussianities that may lie within the reach of the next generation of cosmological surveys~\cite{Cooray:2006km, Pillepich:2006fj, Alvarez:2014vva, Munoz:2015eqa}. Their observational status is therefore likely to evolve in the coming years, which is why it is important to improve our understanding of these models. Third, at the practical level, these scenarios are interesting since the reheating parameter is an explicit function of the decay rates of both fields, the mass of the light field $\sigma$ and its vev at the end of inflation. This means that the same parameters determine the direct imprint of $\sigma$ on the statistics of curvature perturbations and the reheating kinematic effect on the location of the observational window along the inflaton potential. The associated increased sensitivity of the data to these parameters should allow us to better constrain them.

These scenarios have recently been brought into the full domain of Bayesian analysis in \Refs{Hardwick:2015tma, Vennin:2015vfa, Vennin:2015egh}. In this chapter, we make use of the Bayesian inference techniques developed in these works to derive constraints on the inflationary energy scale and the reheating temperatures, and quantify the gain in information about these quantities from current observations.

In \Sec{sec:Modeling}, we present in greater details the scenarios at hand and explain how information on reheating can be extracted using Bayesian inference. In \Sec{sec:results-creh}, we provide our main results and analyse their implications for the physics of reheating and the amount of information that has been gained. In \Sec{sec:Discussion}, we extend the discussion by considering the role played by the inflationary energy scale in plateau potentials, the impact of gravitino overproduction bounds and the constraints on decay rates. We present our conclusions in \Sec{sec:Conclusions} and then end the chapter with several appendices. In \App{sec:DKL}, we present the Kullback-Leibler divergence as a tool to quantify information gain. In \App{Sec:IndividualScenarios}, we present our results for individual reheating scenarios. In \App{Sec:DKLDensity} finally, we discuss information gain densities.
\section{\textsf{Method}}
\label{sec:Modeling}
The method we employ here combines the analytical work of \Ref{Vennin:2015vfa} with the numerical tools developed in \Refs{Ringeval:2013lea, Martin:2013nzq, Vennin:2015egh}. In this section, we describe its main aspects and explain the use of Bayesian inference techniques and information gain quantification to analyse constraints on the parameters of reheating. 
\subsection{\textsf{Curvaton and reheating}}
\label{sec:introscenarios}
As explained in \Sec{sec:intro}, we study the case where inflation is driven by a single field $\phi$ slowly rolling down its potential $U(\phi)$, and an extra light scalar field $\sigma$ (with mass $m_\sigma$ smaller than the inflationary Hubble scale) is present both during inflation and reheating. We therefore consider potentials of the type given in \Eq{eq:pot:gen:curvaton}. 

\begin{figure}
\begin{center}
\includegraphics[width=13cm]{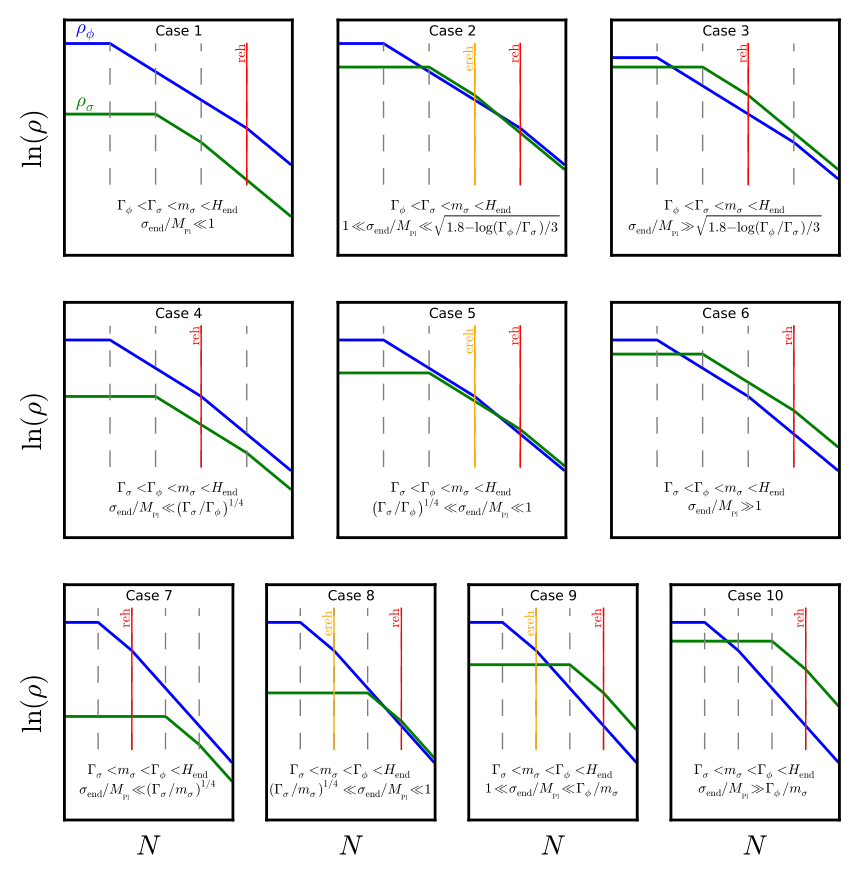}
\caption[Curvatonic reheating scenarios]{Different possible reheating scenarios, depending on the values taken by $\Gamma_\sigma$, $m_\sigma$, $\Gamma_\phi$, $H_\uend$ and $\sigma_\uend$. Cases 1, 2 and 3 correspond to $\Gamma_\phi<\Gamma_\sigma<m_\sigma<H_\uend$; cases 4, 5 and 6 correspond to $\Gamma_\sigma<\Gamma_\phi<m_\sigma<H_\uend$; cases 7, 8, 9 and 10 correspond to $\Gamma_\sigma<m_\sigma<\Gamma_\phi<H_\uend$. Within each row, different cases are distinguished by $\sigma_\uend/\Mp$ which controls when $\sigma$ dominates the total energy density (the precise values for $\sigma_\uend$ at the limit between the different scenarios are given in \Ref{Vennin:2015vfa}). The blue curves stand for the energy density of $\phi$ while the green ones are for $\sigma$. The time at which the total energy density corresponds to the reheating temperature $T_{\rm reh}$ is marked out in red for each case, and early reheating temperature $T_{\rm ereh}$ are denoted in orange when two disconnected radiation phases exist.}
\label{fig:cases}
\end{center}
\end{figure}

We remind the reader that this extra field $\sigma$ is taken to be subdominant at the level of the background energy density during the whole inflationary epoch. Both fields are assumed to be slowly rolling during inflation, and eventually decay into radiation fluids with decay rates\footnote{
Here, $\Gamma_\phi$ (respectively $\Gamma_\sigma$) are effective values for which assuming instantaneous decay at $H=\Gamma_\phi$ (respectively $H=\Gamma_\sigma$) provides a good description of the full decay dynamics. 
} respectively denoted $\Gamma_\phi$ and $\Gamma_\sigma$, during reheating. While we require that $\phi$ becomes massive at the end of inflation, we do not make any assumption as to the ordering of the three events: $\sigma$ becomes massive, $\phi$ decays and $\sigma$ decays. Nor do we restrict the epochs during which $\sigma$ can dominate the energy content of the Universe. This leaves us with 10 possible cases (including situations where $\sigma$ drives a secondary phase of inflation~\cite{Langlois:2004nn, Moroi:2005kz, Ichikawa:2008iq, Dimopoulos:2011gb}), depending on the vev of $\sigma$ at the end of inflation $\sigma_\uend$. These ten ``reheating scenarios'' are listed and detailed in \Ref{Vennin:2015vfa} but are sketched in \Fig{fig:cases}. The usual curvaton scenario corresponds to case number 8 but one can see that a much wider class of models is covered by the present analysis.

In this section, we also assume that all particles are in full thermal equilibrium after $\phi$ and $\sigma$ decay. Therefore, there are no residual isocurvature modes~\cite{Lyth:2002my, Weinberg:2004kf}, that would otherwise give rise to additional constraints. Such constraints depend on the specific processes of decay and thermalisation~\cite{Langlois:2004nn, Lemoine:2006sc, Langlois:2008vk, Lemoine:2008qj, Smith:2015bln}. Thermal equilibrium also allows us to relate energy densities $\rho_\urad$ contained in radiation fluids to temperatures through \Eq{eq:T:rho}. When this expression is evaluated at the onset of the Big-Bang radiation epoch, it yields the ``reheating temperature'' $T_\ureh$. In reheating scenarios 1, 2, 4 and 7 (see \Fig{fig:cases}), this corresponds to the temperature of the thermalised decay products of $\phi$, while for scenarios 3, 5, 6, 8, 9 and 10, this corresponds to the decay products of $\sigma$. However, it can also happen that a transient radiation epoch takes place during reheating (as in reheating scenarios 2, 5, 8 and 9), in which case the energy density of the Universe at the beginning of this first radiation phase is called ``early reheating temperature'' and is noted $T_\uereh$. In reheating scenarios 5, 8 and 9, this corresponds to the decay products of $\phi$, while in scenario 2, this corresponds to the decay products of $\sigma$.

In \Ref{Vennin:2015vfa}, the $\delta N$ formalism~\cite{Starobinsky:1986fxa, Salopek:1990jq, Sasaki:1995aw, Sasaki:1998ug, Wands:2000dp, Lyth:2004gb, Lyth:2005fi} and the sudden decay approximation~\cite{Malik:2006pm,Sasaki:2006kq} were employed to relate observables of the models considered here to variations in the energy densities of both fields at the decay time of the last field. This allows one to calculate all relevant physical quantities by only keeping track of the background energy densities. Analytical expressions have been derived for all $10$ reheating scenarios, that have been implemented in the publicly available \texttt{ASPIC} library~\cite{aspic}. For a given inflaton potential, and from the values of $\Gamma_\phi$, $\Gamma_\sigma$, $m_\sigma$ and $\sigma_\uend$, this code returns the value of the first three slow-roll parameters (or equivalently at second order in the slow-roll approximation, of the scalar spectral index $\nS$ and its running, and of the tensor-to-scalar ratio $r$) and of the local-type non-Gaussianity parameter $f_{\mathrm{NL}}$. In \Ref{Vennin:2015egh}, this has been interfaced with the ``effective likelihood via slow-roll reparametrisation'' of \Ref{Ringeval:2013lea}, and Bayesian constraints were derived for the models that we consider here. The results presented in this chapter are obtained from this numerical pipeline, where the Planck 2015 $TT$ data are combined with the high-$\ell$ $C_\ell^{TE}+C_\ell^{EE}$ likelihood and the low-$\ell$ temperature plus polarisation likelihood (PlanckTT,TE,EE+lowTEB in the notations of \Ref{Aghanim:2015xee}, see table~1 there), together with the BICEP2-Keck/Planck likelihood described in \Ref{Ade:2015tva}.

An important result of \Ref{Vennin:2015egh} is that the models favoured by the data are of two types: either the inflaton has a ``plateau potential'' (\ie is a monotonically increasing function of $\phi$ that asymptotes a constant positive value at infinity) and the reheating scenario can be any of the 10 cases listed in \Fig{fig:cases}, or the inflaton has a ``quartic potential'' (\ie is proportional to $\phi^4$) and reheating occurs in scenario 5 or 8. For this reason, we restrict the following analysis to these two kinds of potential. As an example of a plateau potential, we consider the one of Higgs inflation ($\mathrm{HI}$)
\bea
U\left(\phi\right)=M^4\left(1-e^{-\sqrt{\frac{2}{3}}\frac{\phi}{\Mp}}\right)^2\, ,
\label{eq:pot:hi}
\eea
which also matches the Starobinsky model~\cite{Starobinsky:1980te} (in \Sec{sec:KMIII}, another plateau potential is studied, ``K\"ahler moduli II inflation'', to investigate the role played by the inflationary energy scale in plateau models). The other potential we consider is the one of quartic inflation ($\mathrm{LFI}_4$)
\bea
\label{eq:pot:quartic}
U\left(\phi\right)=M^4\left(\frac{\phi}{\Mp}\right)^4\, .
\eea 
Here, ``$\mathrm{HI}$'' and ``$\mathrm{LFI}_4$'' refer to the terminology of \Ref{Martin:2013tda} and stand for the purely single-field versions of these models. When the prefix ``$\mathrm{MC}$'' is appended (for ``Massive Curvaton''), the index following the prefix refers to the reheating scenario number. For example, $\mathrm{MC}_5\mathrm{LFI}_4$ corresponds to the case where the inflaton potential is of the quartic type, and where the reheating scenario is of the fifth kind.

In \Fig{fig:mainresult}, some of the results of \Ref{Vennin:2015egh} have been summarised for the $\mathrm{HI}$ models (blue disks, where the white circled disk stands for the single-field version of the model and the other disks represent the $10$ reheating scenarios) and the $\mathrm{LFI}_4$ models (red disks). On the horizontal axis, the Bayesian evidence is displayed. One can see that for Higgs inflation, adding a light scalar field slightly decreases the Bayesian evidence of the model but at a level which is inconclusive for most reheating scenarios (and never more than weakly disfavoured). For quartic inflation, the single-field version of the model is strongly disfavoured and so are most of the reheating scenarios when a light scalar field is added. Two exceptions are to be noted however, namely cases 5 and 8, which lie in the favoured region. On the vertical axis, the information gained on $T_\ureh$ is displayed, as will be defined and analysed in \Sec{sec:InformationGain}.
\subsection{\textsf{Inverse problem for reheating parameters}}
\label{sec:InversionProblem}
As mentioned in \Sec{sec:intro}, a specific feature of the models considered in this section is that the same parameters determine the expansion history during reheating as well as the contribution from the additional light scalar field to the total curvature perturbations. This is responsible for a high level of interdependency between these parameters, that plays an important role in shaping the constraints we obtain in \Sec{sec:results-creh}. For this reason, it is important to first better understand their origin.

The number of \efolds $\Delta N_*$ elapsed between the Hubble exit time of the CMB pivot scale $k_{{}_\mathrm{P}}$ and the end of inflation is given by~\cite{Martin:2006rs, Martin:2010kz, Easther:2011yq}
\bea
\Delta N_* = \frac{1-3\bar{w}_\ureh}{12\left(1+\bar{w}_\ureh\right)}\ln\left(\frac{\rho_\ureh}{\rho_\uend}\right)
+\frac{1}{4}\ln\left(\frac{\rho_*}{9\Mp^4}\frac{\rho_*}{\rho_\uend}\right)
-\ln\left(\frac{k_{{}_\mathrm{P}}/a_\mathrm{now}}{\tilde{\rho}_{\gamma,\,\mathrm{now}}^{1/4}}\right)\, ,
\label{eq:DeltaNstar}
\eea
which can be calculated by stitching together separate epochs with known equations of state. In this expression, $\bar{w}_\ureh=\int_\ureh w(N)\dd N/N_\ureh$ is the averaged equation of state parameter during reheating, $\rho_\ureh$ is the energy density of the Universe at the end of reheating, $\rho_*$ is the energy density calculated $\Delta N_*$ \efolds before the end of inflation (all the quantities with a subscript ``*'' are evaluated at that time), $a_\mathrm{now}$ is the present value of the scale factor, and $\tilde{\rho}_{\gamma,\,\mathrm{now}}$ is the the energy density of radiation today rescaled by the number of relativistic degrees of freedom. Taking the pivot scale $k_{{}_\mathrm{P}}/a_\mathrm{now}$ to be $0.05\, \mathrm{Mpc}^{-1}$ and $\tilde{\rho}_{\gamma,\,\mathrm{now}}$ to its measured value, the last term is $N_0\equiv-\ln(k_{{}_\mathrm{P}}/a_\mathrm{now}/\tilde{\rho}_{\gamma,\,\mathrm{now}}^{1/4})\simeq 61.76$.

Let us first illustrate the use of \Eq{eq:DeltaNstar} to constrain reheating in the simple case of single-field quartic inflation, where the potential is given by \Eq{eq:pot:quartic} and there is no additional light scalar field $\sigma$. As mentioned above, we require that $\phi$ becomes massive at the end of inflation, so that in this case, one simply has $\bar{w}_\ureh=0$. Inflation ends by slow-roll violation at $\phi_\uend=2\sqrt{2}\Mp$, so that $\rho_\uend=3 U(\phi_\uend)/2 = 96 M^4$. On the other hand, the slow-roll trajectory is given by $\phi_*^2/\Mp^2 = 8(\Delta N_*+1)$, so that $\rho_*=U(\phi_*)= 64 M^4(\Delta N_*+1)^2$. For this reason, $\Delta N_*$ also appears in the right hand side of \Eq{eq:DeltaNstar} and this formula should be viewed as an implicit equation for $\Delta N_*$. In fact, this is all the more true since $M^4$ also implicitly depends on $\Delta N_*$. Indeed, this mass scale can be fixed by requiring that the correct scalar power spectrum amplitude $A_{{}_\mathrm{S}}=(M/\Mp)^4(\phi_*/\Mp)^6/(192\pi^2) $ is obtained (where $A_{{}_\mathrm{S}}$ has been evaluated at leading order in slow roll in quartic inflation). Making use of \Eq{eq:T:rho} to express $\rho_\ureh$ in terms of $T_\ureh$, one then obtains
\begin{align}
\left.\Delta N_*\right\vert_{\mathrm{LFI}_4} &= 
\frac{1}{12}\ln\left(\frac{512}{135}g_*\right)
+\frac{1}{2}\ln\left[\frac{64\pi}{3}\left(1+\Delta N_*\right)^3\right] \nonumber \\
& \qquad \qquad +\frac{1}{3}\ln\left(\sqrt{A_{{}_\mathrm{S}}}\frac{T_\ureh}{\Mp}\right)
+N_0
\, .
\end{align}
This equation can be inverted using the $-1$ branch of the Lambert function $W$, and one finds
\bea
\left.\Delta N_*\right\vert_{\mathrm{LFI}_4} &=& -1 -\frac{3}{2}W_{-1}\left[-\frac{5^{1/18} \ee^{-\frac{2}{3}(1+N_0)}}{2^{3/2} 3^{1/2} \pi^{1/3} g_*^{1/18} A_{{}_\mathrm{S}}^{1/9}}\left(\frac{\Mp}{T_\ureh}\right)^{2/9}\right]
\\ 
& \simeq & -1-\frac{3}{2}W_{-1}\left[-4.11\times 10^{-14} \left(\frac{\mathrm{MeV}}{T_\ureh}\right)^{2/9}\right] \\
&\simeq &
45.23+\frac{1}{3}\ln\left(\frac{T_\ureh}{\mathrm{MeV}}\right)\, , 
\eea
where in the second equality, we have used $A_{{}_\mathrm{S}}\simeq 2.2\times 10^{-9}$~\cite{Ade:2013sjv}, $g_*\simeq 106.75$ (which is calculated from the SM effective degrees of freedom above the EW scale\footnote{There are 28 Bosonic (2 photon helicities, 3 massive gauge Bosons each with 3 spins, 1 Higgs Boson and 8 gluons each with 2 spins) and 90 Fermionic (12 quarks each with 3 colours and 2 spins, 6 charged leptons each with 2 spins and 6 neutrinos) degrees of freedom, giving $g_* = 28 + (7/8)90 = 106.75$. } --- see \Fig{fig:coshist-illus}) and the value given above for $N_0$, and the last expression corresponds to the limit $\Delta N_*\gg 1$. This makes explicit the dependence of $\Delta N_*$ on the reheating temperature $T_\ureh$. Since observable quantities such as the scalar spectral index $\nS$ or the tensor-to-scalar ratio $r$ depend on $\Delta N_*$ through $\phi_*$, this means that the reheating temperature is directly constrained by CMB measurements,
\bea
\label{eq:lfi4:nsr:Treh}
\left.\nS\right\vert_{\mathrm{LFI}_4} \simeq 1-\frac{3}{46.23+\frac{1}{3}\ln\left(\frac{T_\ureh}{\mathrm{MeV}}\right)}\, ,\quad
\left. r\right\vert_{\mathrm{LFI}_4} \simeq \frac{16}{46.23+\frac{1}{3}\ln\left(\frac{T_\ureh}{\mathrm{MeV}}\right)}\, .
\eea 
From these expressions, it is clear that observational constraints on $\nS$ and $r$ directly translate into constraints on the reheating temperature $T_\ureh$. As this simple calculation shows, this is the consequence of many interdependencies between the parameters of the problem. 

When a light scalar field is added, these dependencies are substantially more complicated. For instance, the averaged equation of state parameter $\bar{w}_\ureh$ does not vanish anymore but is a non-trivial function of $\rho_\uend$, $\Gamma_\phi$, $\Gamma_\sigma$, $m_\sigma$ and $\sigma_\uend$, that is different for each of the 10 reheating scenarios of \Fig{fig:cases} (this function is given in Appendix~B of \Ref{Vennin:2015vfa}). Then, the mass scale of the potential $M^4$ is not simply related to the amplitude of the scalar power spectrum since $A_{{}_\mathrm{S}}$ also receives a contribution from the light scalar field $\sigma$, and this contribution depends on $\rho_\uend$, $\Gamma_\phi$, $\Gamma_\sigma$, $m_\sigma$ and $\sigma_\uend$. As a result, the dependency of observable quantities on these parameters is much more complicated than the one obtained for a purely single-field model, and the constraints one can infer on the reheating temperatures for instance are a priori much less trivial. The goal of this chapter is precisely to derive these constraints.
\subsection{\textsf{Bayesian inference and prior choices}}
\label{sec:Bayesian}
Starting from the data sets $\mathcal{D}$ mentioned in \Sec{sec:introscenarios}, our goal is to derive observational constraints on the energy scale of inflation $\rho_\uend$ and the reheating temperatures $T_\ureh$ and $T_\uereh$. This can be done using Bayesian inference techniques~\cite{Cox:1946,Jeffreys:1961,deFinetti:1974,Trotta:2005ar,Trotta:2008qt}. Following \Eq{eq:posterior:def:before}, we assume a model $\mathcal{M}_i$, where the posterior probability $p$ of its parameters $\theta_{ij}$ (labeled by $j$) is expressed as
\bea
\label{eq:posterior:def}
p\left(\theta_{ij}\vert\mathcal{D},\mathcal{M}_i\right)=\frac{\mathcal{L}
\left(\mathcal{D}\vert\theta_{ij},\mathcal{M}_i\right)
\pi\left(\theta_{ij}\vert \mathcal{M}_i\right)}{\mathcal{E}\left(\mathcal{D}\vert\mathcal{M}_i \right) } \, .
\eea
In this expression, $\mathcal{L}(\mathcal{D}\vert\theta_{ij},\mathcal{M}_i)$ is the likelihood and represents the probability of observing the data $\mathcal{D}$ assuming the model $\mathcal{M}_i$ is true and $\theta_{ij}$ are the actual values of its parameters, $\pi (\theta_{ij}\vert \mathcal{M}_i )$ is the prior distribution on the parameters $\theta_{ij}$, and $\mathcal{E}\left(\mathcal{D}\vert\mathcal{M}_i \right)$ is a normalisation constant called the Bayesian evidence, which using \Eq{eq:evidence:def:before}, is
\bea
\label{eq:evidence:def}
\mathcal{E}\left(\mathcal{D}\vert\mathcal{M}_i \right) 
= \int\dd\theta_{ij}\mathcal{L}
\left(\mathcal{D}\vert\theta_{ij},\mathcal{M}_i\right)
\pi\left(\theta_{ij}\vert \mathcal{M}_i\right)\, .
\eea
The Bayesian evidence of the models considered in this section have been computed in \Ref{Vennin:2015vfa} and here, we are interested in the posterior distributions $p$ for the energy scale of inflation and the reheating temperatures. Notice that these quantities are not necessarily ``fundamental'' parameters that we start from but can be derived from them. For example, as stressed in \Sec{sec:InversionProblem}, $\rho_\uend$ is a complicated function of the parameters $\lbrace \theta_V \rbrace$ characterising the inflaton potential, $\Gamma_\phi$, $\Gamma_\sigma$, $m_\sigma$ and $\sigma_\uend$. In this case, for a derived parameter $\theta_d$ that can be expressed as $\theta_d=f_i(\theta_{ij})$, one marginalises the distribution obtained in \Eq{eq:posterior:def} according to \Eq{eq:marginalisation} such that
\bea 
p\left(\theta_d\vert\mathcal{D},\mathcal{M}_i\right) = \int_{f(\theta_{ij})=\theta_d}p(\theta_{ij}\vert\mathcal{D},\mathcal{M}_i)\dd\theta_{ij}\, .
\eea

In this method, the priors are important quantities as they encode physical information one has ``a priori'' on the values of the parameters that describe the models. For the parameters of the potential $\lbrace \theta_V \rbrace$, we use the same priors as the ones proposed in \Ref{Martin:2013nzq}, based on \Ref{Martin:2013tda}. Because the extra field $\sigma$ is supposed to be still light at the end of inflation, its mass $m_\sigma$ must be smaller than the Hubble scale at the end of inflation, $H_\uend$. The same condition applies to the two decay rates, $\Gamma_\phi,\ \Gamma_\sigma<H_\uend$, since both fields decay after inflation. On the other hand, we want the Universe to have fully reheated before Big Bang Nucleosynthesis (BBN), which means that the two decay rates are also bounded from below by $H_{\mathrm{BBN}}\simeq (10\,\MeV)^2/\Mp$. The same lower bound applies to $m_\sigma$ since, assuming perturbative decay, $m_\sigma>\Gamma_\sigma$. Between these two values, the order of magnitude of $m_\sigma$ and of the two decay rates is a priori unknown, which is why a logarithmically flat prior (or ``Jeffreys prior'') is chosen:
\begin{align}
\ln H_{\mathrm{BBN}} < \ln \Gamma_\phi,\,\ln\Gamma_\sigma,\,\ln m_\sigma < \ln H_\uend\, .
\label{eq:prior:massscales}
\end{align}
The relative orderings identified in \Fig{fig:cases} then determine which of the 10 reheating scenarios is realised for a given set of parameters. For $\sigma_\uend$, two different priors are considered. The first one, denoted $\pilog$, is logarithmic and consists in assuming that the order of magnitude of $\sigma_\uend$ is unknown,
\begin{align}
\label{eq:sigmaend:LogPrior}
\ln\sigma_\uend^\umin < \ln\sigma_\uend < \ln \sigma_\uend^\umax \, .
\end{align}
Here, $\sigma_\uend^\umin$ and $\sigma_\uend^\umax$ are the boundary values given for each reheating case in \Fig{fig:cases}. For cases 1, 4 and 7, the lower bound is taken to be $\sigma_\uend^\umin=H_\uend/(2\pi)$, corresponding to the minimal quantum dispersion of the field, and for cases 3, 6 and 10, the upper bound $\sigma_\uend^\umax$ is set by the condition that the extra phase of inflation driven by $\sigma$ is sufficiently short so that the pivot scale $k_{{}_\mathrm{P}}$ exits the Hubble radius during the first phase of inflation, driven by $\phi$. The second prior relies on the equilibrium distribution of a light spectator field in a Sitter space-time with Hubble scale $H_\uend$~\cite{Starobinsky:1986fxa, Enqvist:2012}, 
\bea 
\pisto \left(\sigma_\uend\right) \propto \exp\left(-\frac{4\pi^2 m_\sigma^2\sigma_\uend^2}{3H_\uend^4}\right)\, ,
\label{eq:sigmaend:GaussianPrior}
\eea
which is the same as \Eq{eq:equilib-spectator-dist} as will be referred to as the ``stochastic'' prior on $\sigma_\uend$. A few words of caution regarding the use of this prior are in order here. In practice, the timescale of equilibration can be very large for small values of $m_\sigma$, and the initial conditions for spectator fields are not necessarily erased during inflation~\cite{Enqvist:2012}. Also note that in non-plateau models, the time variation of $H$, even in the slow-roll regime, is such that the distribution~(\ref{eq:sigmaend:GaussianPrior}) is not an equilibrium solution anymore, even approximatively. Moreover, since $H_\uend$ depends on $\sigma_\uend$ itself (see the discussion of \Sec{sec:InversionProblem}), \Eq{eq:sigmaend:GaussianPrior} is not a simple Gaussian function of $\sigma_\uend$. This is why the use of \Eq{eq:sigmaend:GaussianPrior} should only be seen as a way to study the effects of picking a specific preferred scale for $\sigma_\uend$. In practice, we therefore implement this prior by simply rejecting realisations for which the argument of the exponential function in \Eq{eq:sigmaend:GaussianPrior} is smaller than $1/10$ or larger than $10$ (we have checked that when changing these arbitrary values to, say, $1/100$ and $100$, very similar results are obtained). In what follows, the inclusion of these two priors for $\sigma_\uend$ allow us to examine prior dependency of the reheating constraints.
\section{\textsf{Results and analysis}}
\label{sec:results-creh}
Let us now present our main results. In \Sec{sec:contraints}, we display and analyse the constraints obtained on the energy scale of inflation $\rho_\uend$ and the two reheating temperatures $T_\ureh$ and $T_\ureh$. In \Sec{sec:InformationGain}, we quantify how much information has been gained about these quantities.
\subsection{\textsf{Constraints on inflationary energy and reheating temperatures}}
\label{sec:contraints}
The posteriors on $\rho_\uend$, $T_\ureh$ and $T_\uereh$ for all 10 individual reheating scenarios (see \Fig{fig:cases}) are given in \App{Sec:IndividualScenarios}. In this section, for the sake of conciseness, as well as to allow direct comparison with purely single-field models, only the constraints averaged over the reheating scenarios are shown. Such distributions can be computed in the following manner. For the purpose of illustration, let us consider two toy models $\mathcal{M}_1$ and $\mathcal{M}_2$, that both depend on the same parameter $\theta$. In model $\mathcal{M}_1$, $\theta$ is assumed to lie within the range $[a,b]$ with a flat prior distribution, while in model $\mathcal{M}_2$, $\theta$ lies within the range $[b,c]$ with a flat prior distribution too. The model $\mathcal{M}_{1+2}$ is defined to be the ``union'' of $\mathcal{M}_1$ and $\mathcal{M}_2$, where $\theta$ lies in $[a,c]$ with a flat prior distribution, so that $\mathcal{M}_1$ and $\mathcal{M}_2$ are simply sub-models of $\mathcal{M}_{1+2}$ (in the same manner as all 10 reheating scenarios $\mathrm{MC}_i\mathrm{XXI}$, for $1\leq i \leq 10$ and some inflaton potential $\mathrm{XXI}$, are submodels of $\mathrm{MCXXI}$). From \Eq{eq:posterior:def}, one can see that
\begin{align}
p\left(\theta\vert\mathcal{D},\mathcal{M}_{1+2}\right) &= 
\frac{\pi\left(\theta\vert \mathcal{M}_{1+2}\right)}
{\mathcal{E}\left(\mathcal{D}\vert \mathcal{M}_{1+2}\right)}
\left[
\frac{\mathcal{E}\left(\mathcal{D}\vert \mathcal{M}_{1}\right)}{\pi\left(\theta\vert \mathcal{M}_{1}\right)}p\left(\theta\vert\mathcal{D},\mathcal{M}_1\right) \right. \nonumber \\
& \left. \qquad \qquad \qquad \qquad \qquad  +
\frac{\mathcal{E}\left(\mathcal{D}\vert \mathcal{M}_{2}\right)}{\pi\left(\theta\vert \mathcal{M}_{2}\right)}p\left(\theta\vert\mathcal{D},\mathcal{M}_2\right)
\right]\, .
\label{eq:posterior:mixed}
\end{align}
In this expression, the Bayesian evidence of $\mathcal{M}_{1+2}$ can be evaluated with \Eq{eq:evidence:def}, which gives rise to
\bea
\mathcal{E}\left(\mathcal{D}\vert \mathcal{M}_{1+2}\right) = \frac{b-a}{c-a}\mathcal{E}\left(\mathcal{D}\vert \mathcal{M}_{1}\right) + \frac{c-b}{c-a}\mathcal{E}\left(\mathcal{D}\vert \mathcal{M}_{2}\right)\, .
\label{eq:evid:combined}
\eea
By combining \Eqs{eq:posterior:mixed} and~(\ref{eq:evid:combined}), the posterior distribution of the parameter $\theta$ within model $\mathcal{M}_{1+2}$ can be written as
\begin{align}
p\left(\theta\vert\mathcal{D},\mathcal{M}_{1+2}\right) &=
\frac{\mathcal{E}\left(\mathcal{D}\vert \mathcal{M}_{1}\right) (b-a) p\left(\theta\vert\mathcal{D},\mathcal{M}_{1}\right)}
{\mathcal{E}\left(\mathcal{D}\vert \mathcal{M}_{1}\right) (b-a)
+
\mathcal{E}\left(\mathcal{D}\vert \mathcal{M}_{2}\right) (b-a)}\nonumber \\
& \qquad \qquad +\frac{\mathcal{E}\left(\mathcal{D}\vert \mathcal{M}_{2}\right) (b-a) p\left(\theta\vert\mathcal{D},\mathcal{M}_{2}\right)}
{\mathcal{E}\left(\mathcal{D}\vert \mathcal{M}_{1}\right) (b-a)
+
\mathcal{E}\left(\mathcal{D}\vert \mathcal{M}_{2}\right) (b-a)}
\, .
\label{eq:posterior:averaged}
\end{align}
In other words, it is given by the averaged sum of the posterior distributions within each sub-model, weighted by the product of the Bayesian evidence and the fractional prior volume of the sub-models. These fractional prior volumes can be viewed as priors for the sub-models themselves. In particular, one can check that \Eq{eq:posterior:averaged} is correctly normalised. 

The above formula can easily be generalised for arbitrary priors and arbitrary number of sub-models. In practice, the Bayesian evidence and fractional prior volumes of all 10 reheating scenarios are given in \Ref{Vennin:2015egh} for the inflaton potentials considered here, and we compute posterior distributions averaged over reheating scenarios adopting this approach. They correspond to the constraints one would obtain starting from the priors~(\ref{eq:prior:massscales}), without the ordering conditions of \Fig{fig:cases}, and simply computing observables according to the reheating scenario in which each sampled point falls.
\subsubsection{\textsf{Energy density at the end of inflation}}
\label{sec:result:rhoend}

\begin{figure}
\figpilogsto
\begin{center}
\includegraphics[width=7cm]{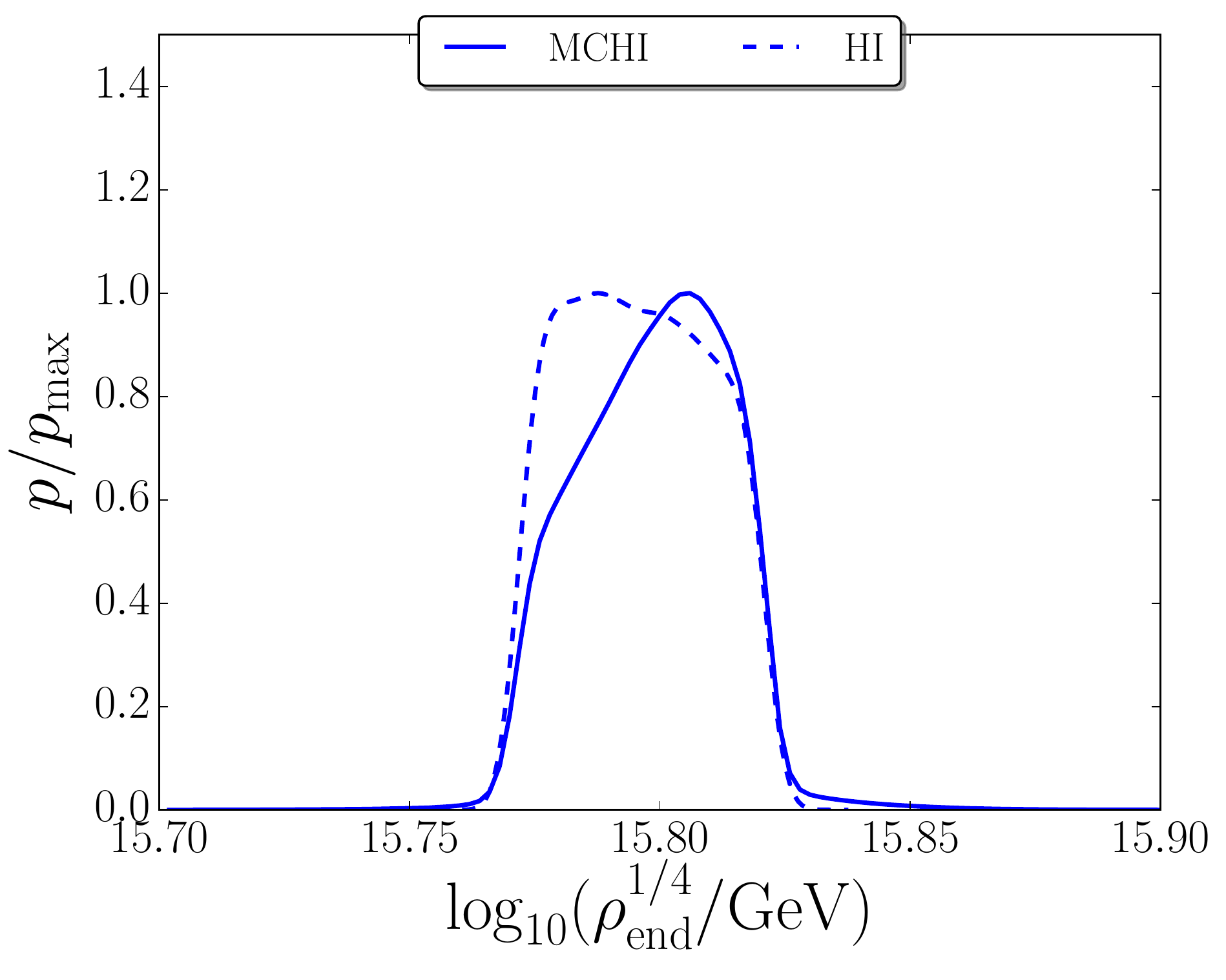}
\includegraphics[width=7cm]{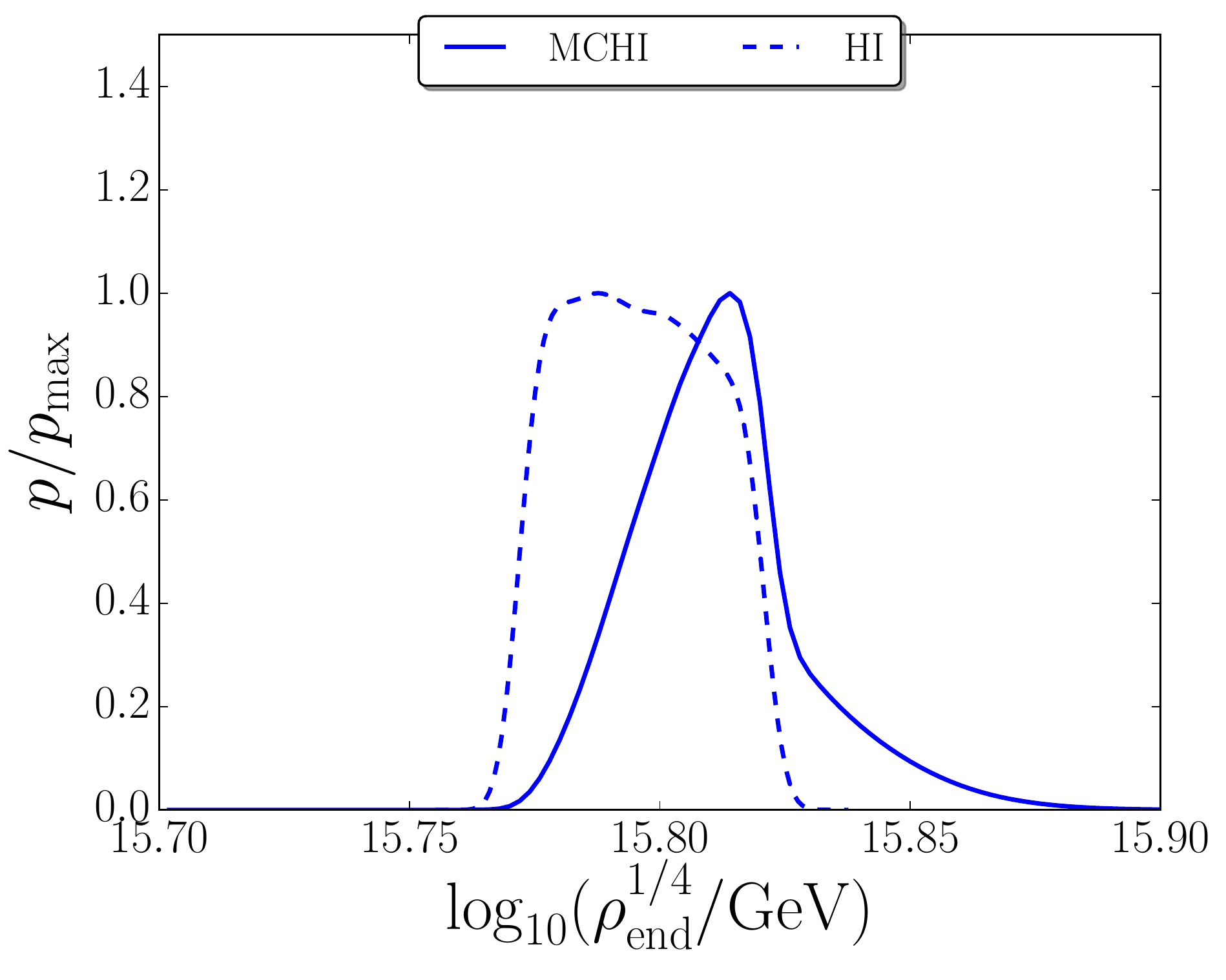}
\includegraphics[width=7cm]{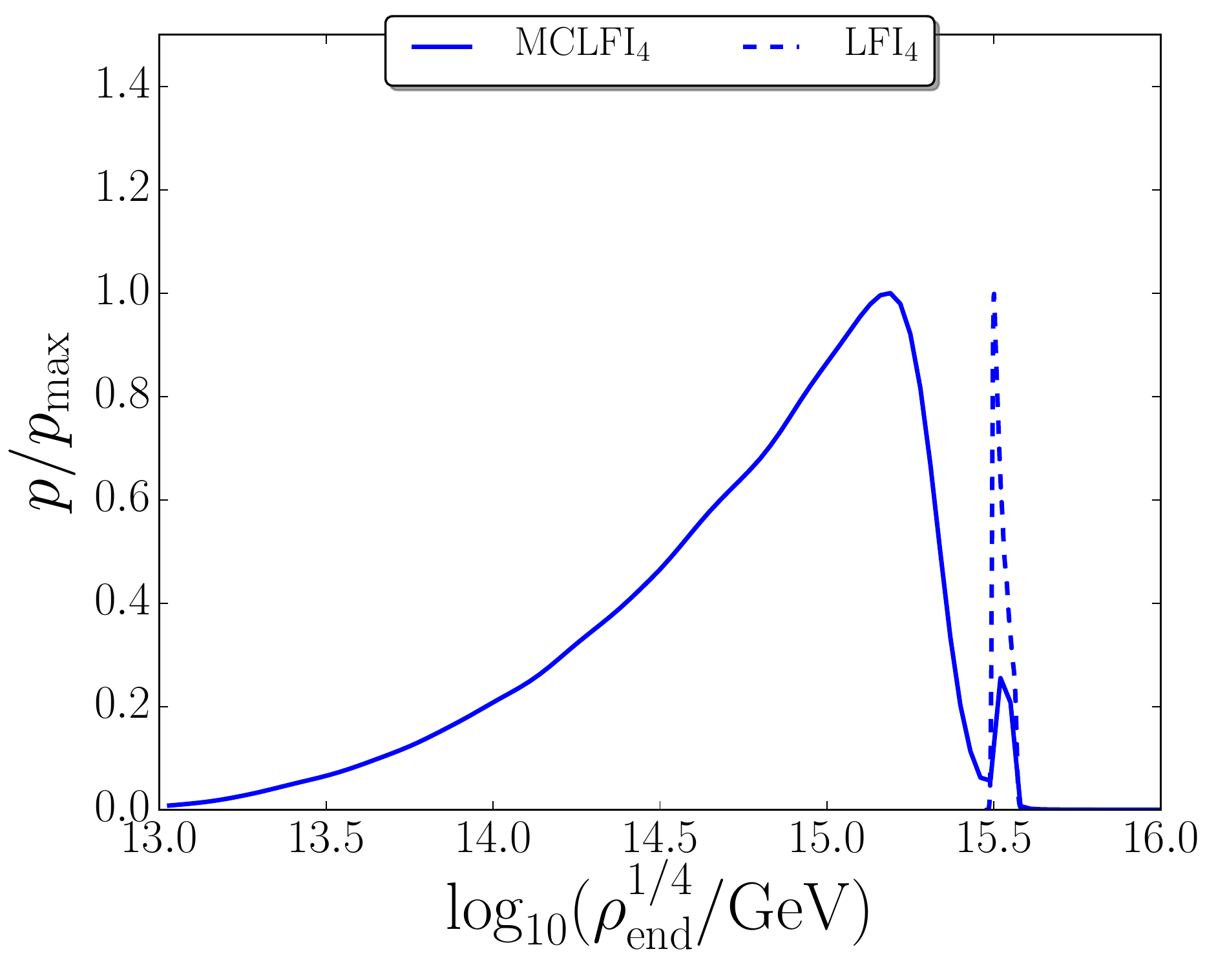}
\includegraphics[width=7cm]{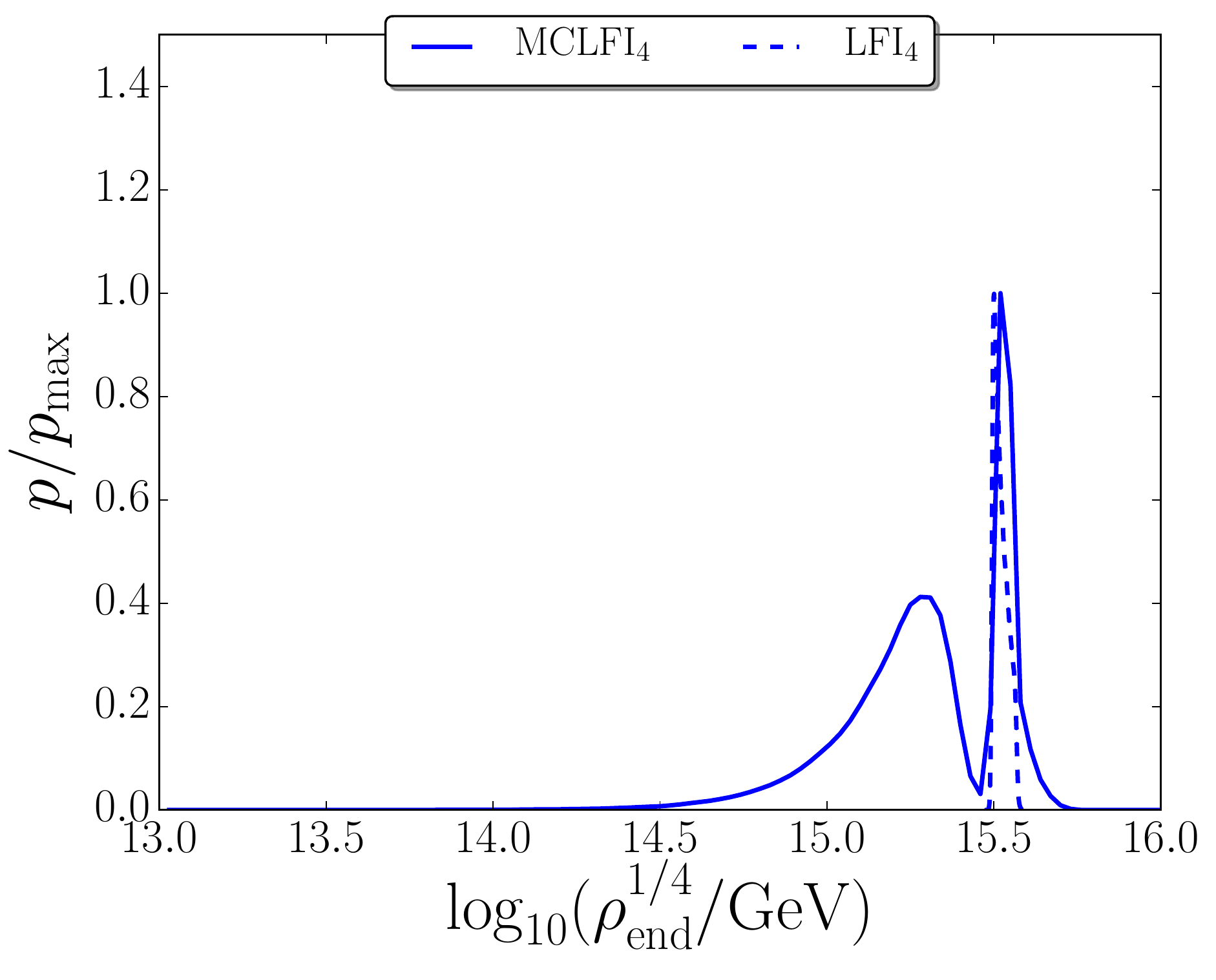}
\caption[Posterior distributions on the energy density at the end of inflation]{Posterior distributions on the energy density at the end of inflation with the plateau potential~(\ref{eq:pot:hi}) of Higgs inflation (top panels) and the quartic potential~(\ref{eq:pot:quartic}) (bottom panels). The left panels correspond to the logarithmically flat prior~(\ref{eq:sigmaend:LogPrior}) $\pilog$ on $\sigma_\uend$, and the right panels stand for the stochastic prior~(\ref{eq:sigmaend:GaussianPrior}) $\pisto$ derived from the equilibrium distribution of a light scalar field in a de Sitter space-time with Hubble scale $H_\uend$. The dashed lines correspond to the single-field versions of the models, while the solid lines stand for the averaged posterior distributions over all $10$ reheating scenarios.}
\label{fig:post:rhoend:averaged}
\end{center}
\end{figure}

In \Fig{fig:post:rhoend:averaged}, the posterior distributions on $\rho_\uend$, the energy density at the end of inflation, is displayed. If the inflaton potential is of the plateau type (Higgs inflation, top panels), the difference between the purely single-field result and the one with an extra light scalar field, averaged over all $10$ reheating scenarios, is very small. One can check that this is also the case at the level of the individual posterior distributions for the different reheating scenarios in \Fig{fig:post:rhoend:individual} of \App{sec:app:rhoend:individual}. This is consistent with the generic robustness of plateau models under the introduction of extra light scalar fields noticed in \Ref{Vennin:2015egh}. In particular, the range of values allowed for $\rho_\uend$ is remarkably narrow. The stochastic prior tends to favour slightly larger values of the energy density. This is because this prior samples larger values of $\sigma_\uend$, hence larger contributions of $\sigma$ to the total curvature power spectrum~\cite{Vennin:2015egh}, hence bluer values of $\nS$. This effect can be compensated for by increasing $\Delta N_*$, hence $\rho_\uend$ [see \Eq{eq:DeltaNstar}], which decreases $\nS$ back into the data's sweet spot~\cite{Vennin:2015vfa}.

The situation is quite different for the quartic potential (bottom panels). In this case, the single-field version of the model provides a very poor fit to the data due to values of the tensor-to-scalar ratio $r$ that are too large. When a light scalar field is introduced, $r$ is typically decreased, and so is $\rho_\uend$. In scenarios where the amount of non-Gaussianities remains small, \ie scenarios 5 and 8, this explains why lower values of $\rho_\uend$ are favoured, see \Fig{fig:post:rhoend:individual}. In other cases, $\fnl$ increases when $r$ decreases, and the trade-off between both effects leads to bimodal posterior distributions. Since scenarios 5 and 8 are favoured however (see \Fig{fig:mainresult}), the clear preference is for lower values of $\rho_\uend$. If a stochastic prior on $\sigma_\uend$ is used, the maximum of the distribution is switched back to the single-field prediction, but all reheating scenarios are moderately or strongly disfavoured in this case anyway~\cite{Vennin:2015egh}.
\subsubsection{\textsf{Reheating temperature}}
\label{sec:result:Treh}
\begin{figure}
\figpilogsto
\begin{center}
\includegraphics[width=7cm]{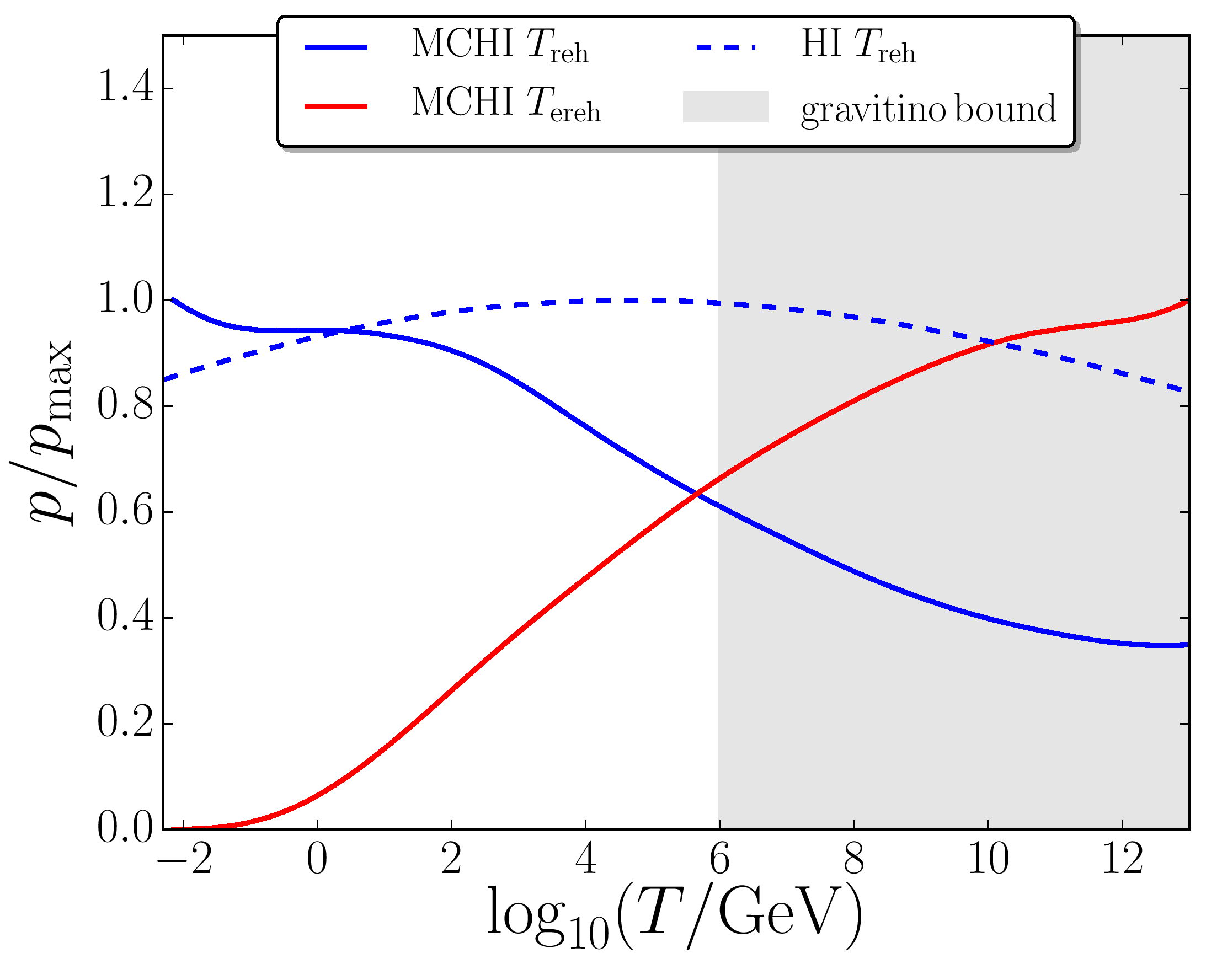}
\includegraphics[width=7cm]{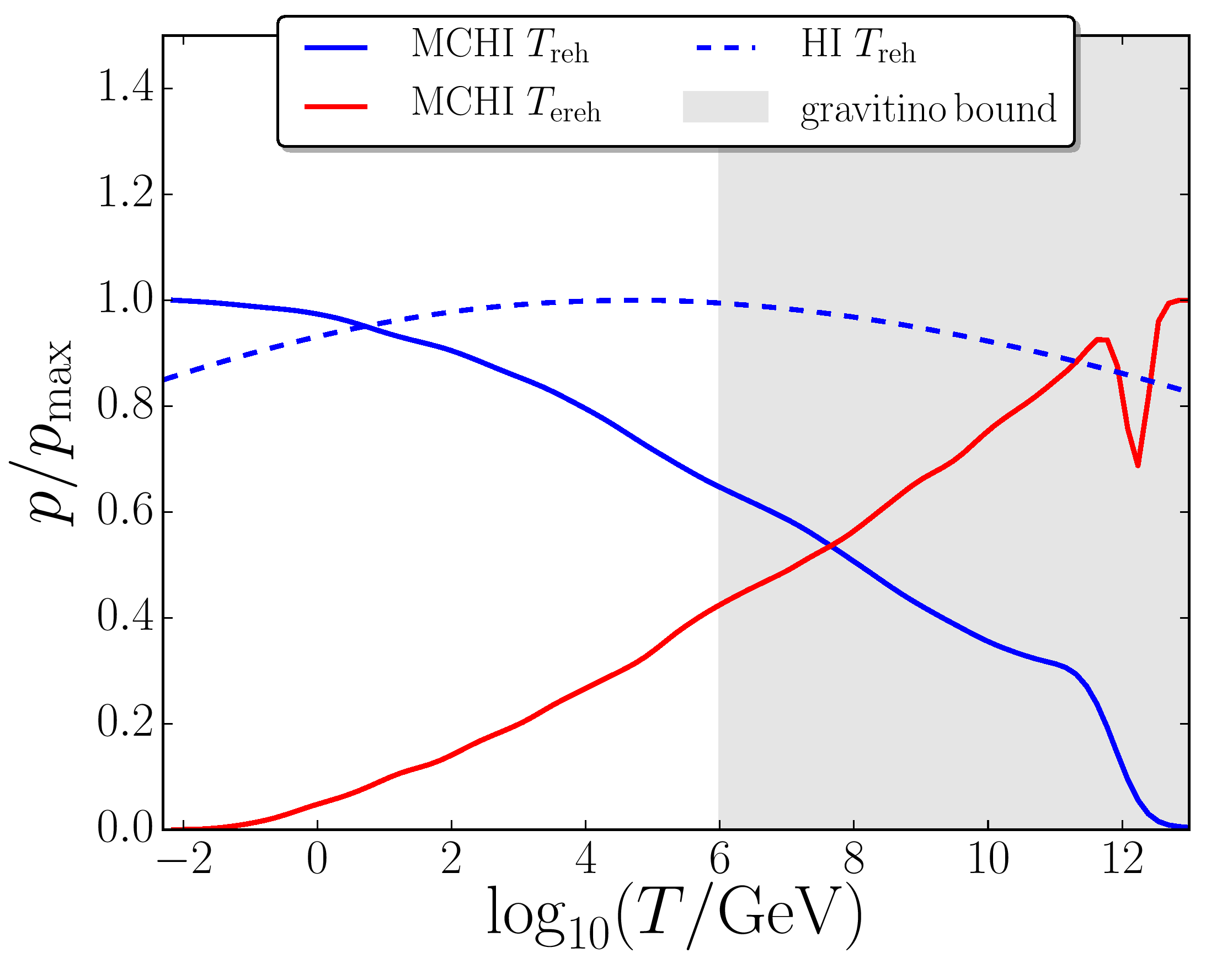}
\includegraphics[width=7cm]{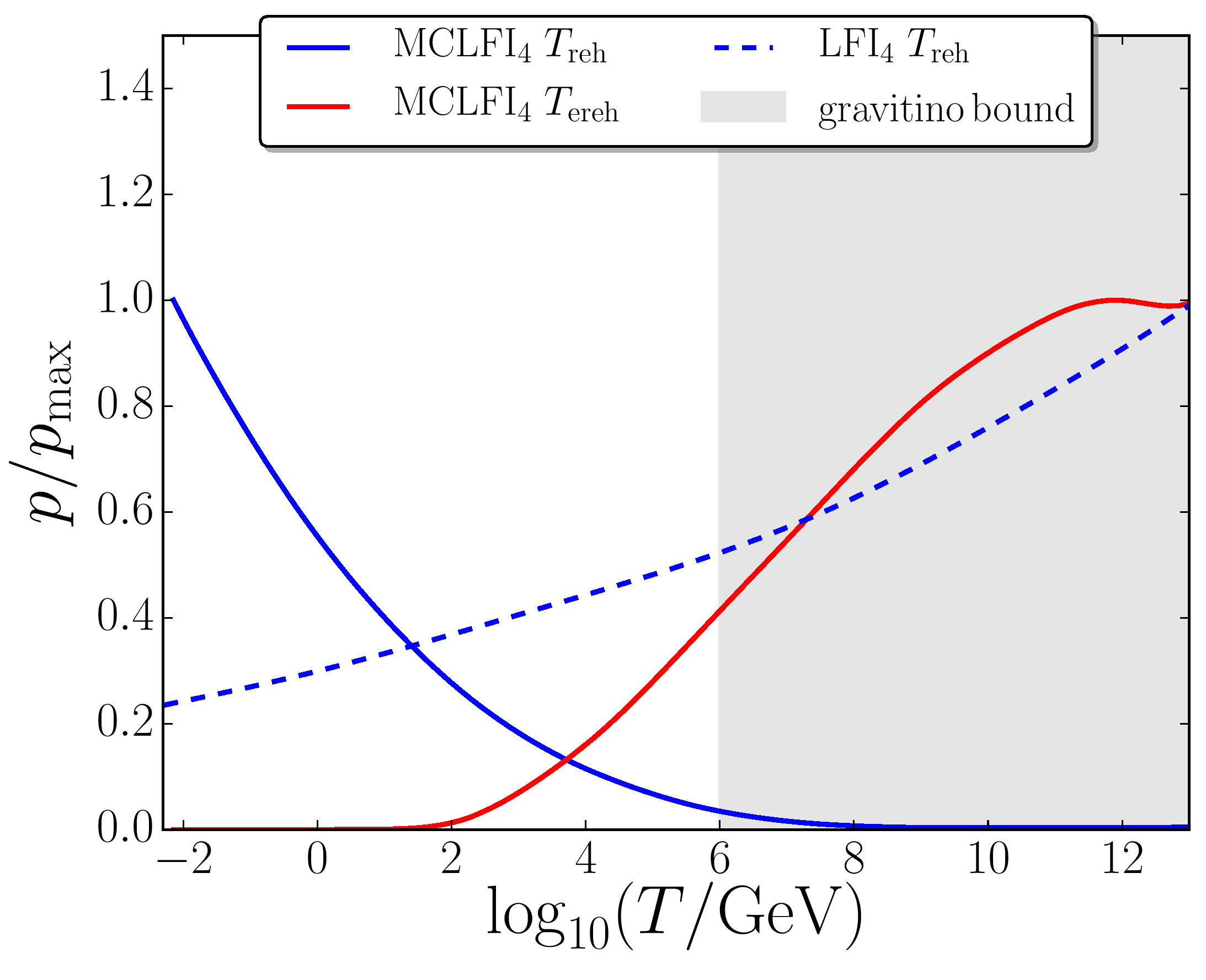}
\includegraphics[width=7cm]{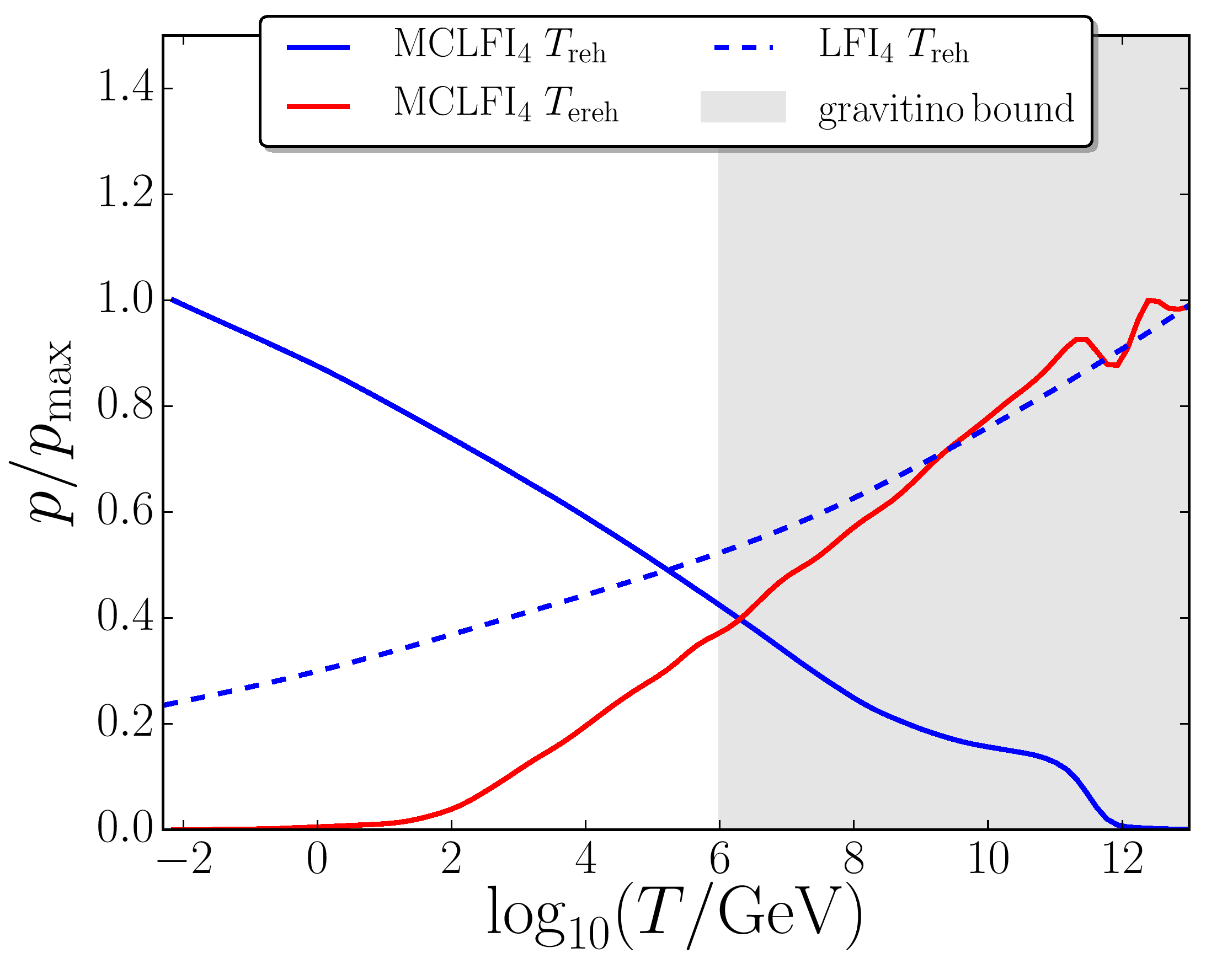}
\caption[Posteriors on the reheating (and early) temperature]{Posterior distributions on the reheating temperature $T_\ureh$ and early reheating temperature $T_\uereh$ with the plateau potential~(\ref{eq:pot:hi}) of Higgs inflation (top panels) and the quartic potential~(\ref{eq:pot:quartic}) (bottom panels). The left panels correspond to the logarithmically flat prior~(\ref{eq:sigmaend:LogPrior}) $\pilog$ on $\sigma_\uend$, and the right panels stand for the stochastic prior~(\ref{eq:sigmaend:GaussianPrior}) $\pisto$ derived from the equilibrium distribution of a light scalar field in a de Sitter space-time with Hubble scale $H_\uend$. The dashed blue lines correspond to the single-field constraints on $T_\ureh$, while the solid lines stand for the averaged posterior distributions on $T_\ureh$ (blue) and $T_\uereh$ (red) when an extra light scalar field is added. The grey shaded region corresponds to reheating temperatures that would be excluded by gravitino production, see \Sec{sec:gravitino}.}
\label{fig:post:Trehs:averaged}
\end{center}
\end{figure}

In \Fig{fig:post:Trehs:averaged}, the posterior distributions on the reheating temperature $T_\ureh$ are displayed. In the single-field version of the plateau model of Higgs inflation, the reheating temperature is rather unconstrained. This is because all reheating temperatures can accommodate the data equally well for this model (at least when $\bar{w}_\ureh=0$, see \Ref{Martin:2016oyk} otherwise). When a light scalar field is introduced however, a slight preference is found for lower reheating temperatures. Looking at \Fig{fig:post:Treh:individual} of \App{sec:app:Treh:individual}, one can see that in the case of the logarithmic prior on $\sigma_\uend$, this trend is mostly due to reheating scenarios 1, 2, 5, 6, 8 and 9, for which $T_\ureh$ is bounded from above. For scenarios 3, 4, and 10 however, the distributions have a maximum around the scale $T_\ureh\sim 10^4\, \mathrm{GeV}$, and for scenario 7, larger values of $T_\ureh$ are even preferred. A similar dichotomy is observed with the stochastic prior on $\sigma_\uend$ where scenarios 1, 2, 5 and 6 prefer smaller values of $T_\ureh$, scenarios 3, 7, 9 and 10 prefer larger values of $T_\ureh$, and scenarios 4 and 8 leave $T_\ureh$ unconstrained. When averaging over the 10 reheating scenarios, the resulting distributions show preference for lower values of $T_\ureh$, but because of these opposite individual behaviours, the constraint is not very strong.

For the single-field version of quartic inflation, larger values of the reheating temperature are preferred since they lead to smaller values for the tensor-to-scalar ratio $r$ as well as larger values of $\nS$ that are in better agreement with the data, as shown explicitly in \Eq{eq:lfi4:nsr:Treh}. When a light scalar field is introduced, one can note in \Fig{fig:post:Treh:individual} that the same variety of individual behaviours of the 10 reheating scenarios is obtained as with Higgs inflation. However, since scenarios 5 and 8 strongly dominate the averaged posterior distribution due to their large Bayesian evidence, and since they both show preference for lower values of $T_\ureh$, better constraints are obtained from the averaged posterior distribution than with a plateau potential. In practice, an upper bound on the reheating temperature can be derived,
\bea
\left.T_\ureh\right\vert_{\mathrm{MCLFI}_4} < 5\times 10^4\,\GeV \, (95\%\,\mathrm{C.L.})\, .
\label{eq:LFI4:constraint:Treh}
\eea
This value has been obtained with the logarithmic prior $\pilog$ on $\sigma_\uend$. With the stochastic prior, the constraint would be much weaker, but one should remember that this prior is not well motivated in that case and that $\mathrm{MCLFI}_4$ is strongly disfavoured~\cite{Vennin:2015egh} when $\pisto$ is used anyway. 
\subsubsection{\textsf{Early reheating temperature}}
\label{sec:result:Tereh}
The weighted posterior distributions on the early reheating temperature $T_{\rm ereh}$ are displayed as the solid red lines in \Fig{fig:post:Trehs:averaged}. Obviously, these distributions are averaged over the scenarios for which $T_\ureh$ is defined only, that is to say cases 2, 5, 8 and 9, and the individual posteriors are given in \Fig{fig:post:Tereh:individual} in \App{sec:app:Tereh:individual} for these scenarios. Contrary to the reheating temperature discussed in \Sec{sec:result:Treh}, one can see that larger values are preferred and that lower bounds on $T_\uereh$ can be obtained,
\begin{align}
\left.T_\uereh\right\vert_{\mathrm{MCHI}} &> 251\,\GeV \, (95\%\,\mathrm{C.L.}) \nonumber \\
\left.T_\uereh\right\vert_{\mathrm{MCLFI}_4} &> 10^5\,\GeV \,(95\%\,\mathrm{C.L.})\, ,
\end{align}
with a logarithmic flat prior on $\sigma_\uend$. In this case, from \Fig{fig:post:Tereh:individual}, one can see that the constraint mostly comes from scenarios 8 and 9, while the posterior distribution for scenarios 2 and 5 has a maximum around $10^7\, \mathrm{GeV}$ for Higgs inflation and $10^9\, \mathrm{GeV}$ for quartic inflation. If one uses the stochastic prior instead, one obtains
\begin{align}
\left.T_\uereh\right\vert_{\mathrm{MCHI}} &> 501\,\GeV \, (95\%\,\mathrm{C.L.}) \nonumber \\
\left.T_\uereh\right\vert_{\mathrm{MCLFI}_4} &> 2.5\times 10^4\,\GeV \, (95\%\,\mathrm{C.L.})\, .
\end{align}
In this case, one can check in \Fig{fig:post:Tereh:individual} that all reheating scenarios favour large values for $T_\uereh$.
\subsection{\textsf{Information gain}}
\label{sec:InformationGain}
\begin{figure}[t]
\begin{center}
\includegraphics[width=9cm]{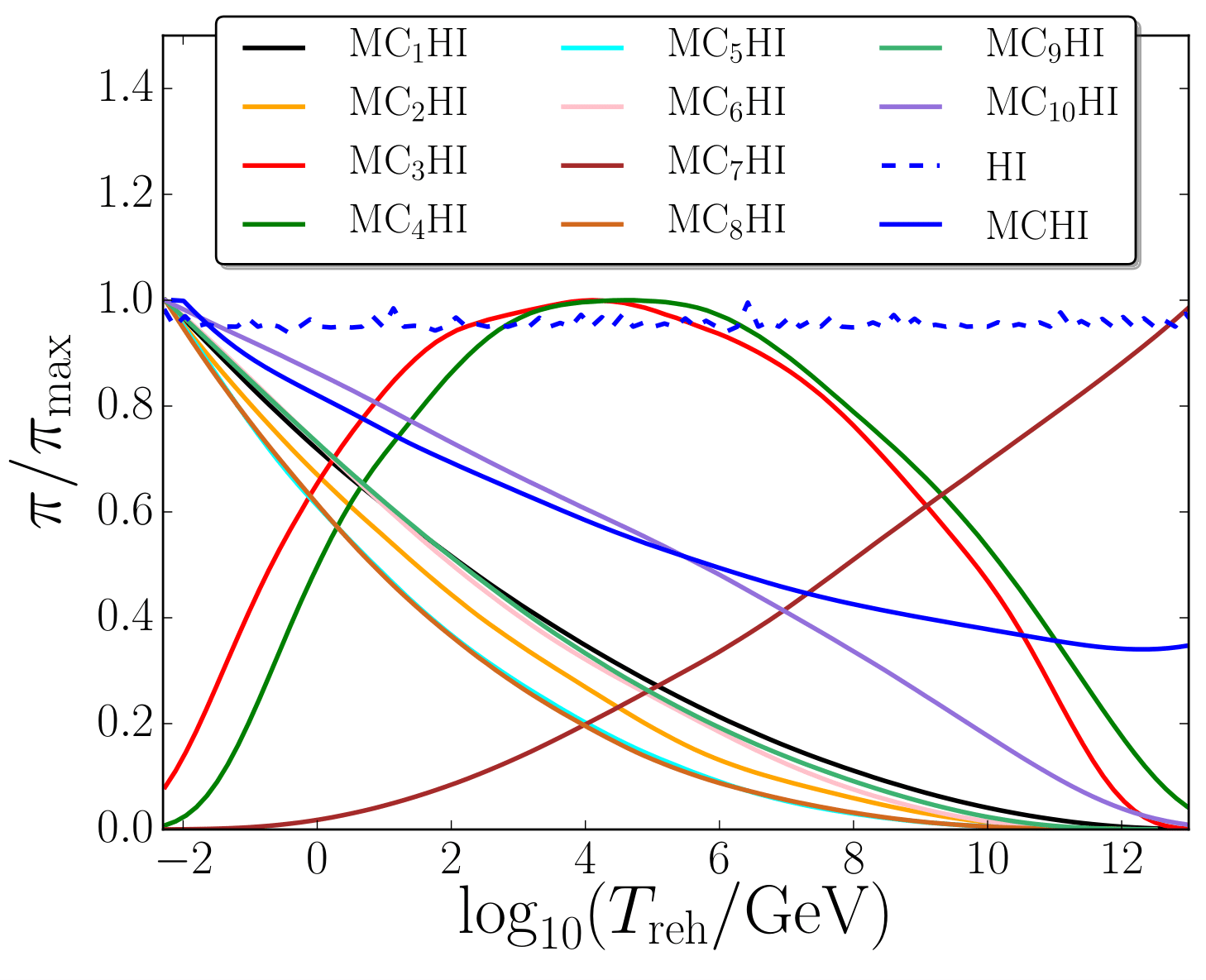}
\caption[Induced prior on the reheating temperature for the Higgs potential]{Induced prior on $\log_{10}(T_{\rm reh}/\mathrm{GeV})$ for the Higgs inflation potential and using the logarithmic prior~(\ref{eq:sigmaend:LogPrior}) on $\sigma_{\rm end}$. The dashed blue line stands for the single-field version of the model for which the prior is flat, the solid coloured lines correspond to the 10 reheating scenarios when a light scalar field is added and the solid blue line is the averaged prior distribution over all reheating scenarios.}
\label{fig:prior:Treh}
\end{center}
\end{figure}
In \Sec{sec:contraints}, the posterior distributions on $\rho_\uend$, $T_\ureh$ and $T_\uereh$ have been displayed and it was shown that, compared to single-field models, different constraints are obtained when a light scalar field is included. In \Sec{sec:InversionProblem}, we explained that both situations are indeed qualitatively different, since in the latter case the same parameters define both the contribution from $\sigma$ to the curvature power spectrum and the kinematic properties of reheating that determine the location of the observational window along the inflaton potential. This leads to an increased interdependence between these parameters and observations, which yields more information about these quantities. This is why in this section, we quantify the information gain on reheating parameters to quantitatively describe this effect.

A first remark is that since the induced priors on $\rho_\uend$, $T_\ureh$ and $T_\uereh$ are not logarithmically flat, information gain cannot be simply assessed by measuring how the distributions of \Sec{sec:contraints} are peaked, or more generally deviate from a flat profile. For example, in \Fig{fig:prior:Treh}, the induced priors\footnote{
In practice, induced priors are reconstructed using a fiducial, constant likelihood in our Bayesian inference code (so that the posteriors we extract correspond to the actual induced priors), where only the value of $A_{{}_\mathrm{S}}$ is used to normalise the mass scale $M^4$ appearing in the inflaton potentials. This is because $A_{{}_\mathrm{S}}$ is so accurately measured that it effectively reduces the support of the posterior to a hypersurface in parameter space, and distributions are considered along this hypersurface only.} on $\log T_\ureh$ are  displayed in the case of Higgs inflation, for the single-field model (dashed blue line), for the 10 reheating scenarios (coloured solid lines), and when averaged over all reheating scenarios (solid blue line). The prior is exactly flat in the single-field case since $T_\ureh$ is directly related to $\Gamma_\phi$ in this case, over which the logarithmically flat prior $ \ln H_{\mathrm{BBN}} < \ln \Gamma_\phi < \ln H_\uend$ is chosen. When a light scalar field is added however, $T_\ureh$ is either related to $\Gamma_\phi$ (in cases 1, 2, 4 and 7) or to $\Gamma_\sigma$ (in cases 3, 5, 6, 8, 9 and 10). Since the ordering conditions of \Fig{fig:cases} are further imposed on top of the logarithmically flat priors for these quantities, the non-flat induced priors of \Fig{fig:prior:Treh} are obtained. 

This is why the posterior distributions are not sufficient to estimate the information gain, but one needs to compute the relative information between the prior and posterior distributions. This can be done using the Kullback-Leibler divergence~\cite{kullback1951} $\dkl$ between the prior $\pi (\theta )$ and the posterior $p(\theta )$ of some parameter $\theta$ (here, for display convenience, the notations of \Sec{sec:Bayesian} are simplified, $p(\theta)\equiv p(\theta\vert\mathcal{D},\mathcal{M}_i)$, etc.), as introduced in \Sec{sec:info-theory-tools}.\footnote{The application of $\dkl$ here between the derived marginalised priors and posteriors over the reheating temperatures and energy densities can be contrasted with the more conventional application of $\dkl$ over the entire space of parameters. In this latter application, the reparameterisation invariance becomes manifest. }

In this section, only the integrated Kullback-Leibler divergences are discussed. The numbers obtained for all models previously discussed are given in table~\ref{table:DKL} in \App{Sec:DKLDensity}. In table~\ref{table:DKL:summary}, the results are summarised and the divergence obtained in the single-field versions of the models are compared with the ones obtained from the averaged distributions over all 10 reheating scenarios. The averaged posterior distribution has been defined in \Sec{sec:contraints}, and the averaged prior distribution is simply the averaged sum of all prior distributions weighted by the fractional prior volume of the sub-models. Let us note that this divergence cannot be obtained by a simple weighted summation over each individual value. For instance, in table~\ref{table:DKL}, one can check that the divergence between averaged distribution can be larger than all individual divergences, as further discussed in \Sec{sec:InformationGain:Treh}.
\begin{table}[t]
\centering
\begin{tabular}{|l||*{3}{c|}}\hline
\backslashbox{Model}{$\dkl$}
&\makebox[3em]{$\rho_\uend$}&\makebox[3em]{$T_\ureh$}&\makebox[3em]{$T_\uereh$}
\\\hline\hline
$\mathrm{HI}$ & 1.370 & 0.004 & -\\\hline
$\mathrm{MCHI}(\pilog)$  & 0.114 & 0.005 & 0.018 \\\hline
$\mathrm{MCHI}(\pisto)$ & 0.224 & 0.006 & 0.014 \\\hline
$\mathrm{LFI}_4$ & 1.171 & 0.108 & - \\\hline
$\mathrm{MCLFI}_4(\pilog)$  & 3.104 & 0.656 & 0.181 \\\hline
$\mathrm{MCLFI}_4(\pisto)$ & 4.780 & 0.111 & 0.281 \\\hline
\end{tabular}
\caption[Information gain on energy densities and reheating temperatures]{Kullback-Leibler divergences $\dkl$ (quoted in binary bits) on $\rho_\uend$, $T_\ureh$ and $T_\uereh$ for Higgs inflation and quartic large field inflation. The result is given for the single-field versions of the models and for the model-averaged priors and posteriors over the 10 reheating scenarios, when a logarithmically flat prior $\pilog$ on $\sigma_\uend$ is used, and with the stochastic prior $\pisto$ of \Eq{eq:sigmaend:GaussianPrior} as well. Note that the early reheating temperature $T_\ureh$ is not defined for single-field models, which is why no value is displayed.} 
\label{table:DKL:summary}
\end{table}
\subsubsection{\textsf{Energy density at the end of inflation}}
\label{sec:InformationGain:rhoend}
In table~\ref{table:DKL:summary}, one can see that more than one bit of information is gained on $\rho_\uend$ for the two single-field models considered here, $\mathrm{HI}$ and $\mathrm{LFI}_4$. The main reason is that, since these single-field potentials have no free parameters (apart from the overall mass scale $M^4$), as shown in \Sec{sec:InversionProblem}, $\rho_\uend$ is entirely fixed by $A_{{}_\mathrm{S}}$, up to a small dependence on $T_\ureh$. In this case, the support of both the priors and the posteriors on $\rho_\uend$ are very narrow, and even a small difference between their preferred values is enough to yield a large Kullback-Leibler divergence, see the discussion around \Eq{eq:DKL:gaussian} in \App{sec:DKL}. However, as soon as another free parameter is introduced in the inflaton potential for instance, this effect disappears as will be explicitly checked in \Sec{sec:KMIII}. Therefore, these large values of $\dkl$ for $\mathrm{HI}$ and $\mathrm{LFI}_4$ are mostly a consequence of the very sharp measurement on $A_{{}_\mathrm{S}}$.

When a light scalar field is added, a few tenths of bits of information on $\rho_\uend$ are typically gained with the plateau potential of Higgs inflation. This number can be larger for individual reheating scenarios, see for instance $\mathrm{MC}_{3}\mathrm{HI}$ and $\mathrm{MC}_{10}\mathrm{HI}$ in table~\ref{table:DKL} where, depending on the prior chosen for $\sigma_\uend$, one gains between one and two bits of information. The situation is particularly interesting for quartic inflation, where the by far favoured reheating scenarios are 5 and 8 (see \Fig{fig:mainresult}). For these models, one typically obtains one bit of information with the logarithmic prior on $\sigma_\uend$ and $3.5$ bits with the stochastic prior, see table~\ref{table:DKL}. This is because, as explained in \Sec{sec:result:rhoend}, the data favours regions of parameter space where $\sigma$ provides the main contribution to curvature perturbations and $\rho_\uend$ is smaller than its single-field counterpart, yielding non-trivial information about the energy density at the end of inflation. The divergence between the averaged distributions displayed in table~\ref{table:DKL:summary} is even larger, the additional information coming from the update in the relative degrees of belief between the different reheating scenarios, namely the fact that the data strongly favours scenarios 5 and 8. 

\subsubsection{\textsf{Reheating temperature}}
\label{sec:InformationGain:Treh}
For the reheating temperature, very little information is gained with the single-field versions of the models. One may wonder whether this is consistent with \Ref{Martin:2016oyk}, where it is found that almost one bit of information is obtained on the reheating parameter of single-field models, on average. This is in fact the case since, in \Ref{Martin:2016oyk}, $\bar{w}_{\ureh}$ is allowed to vary between $-1/3$ and $1$. In \Eq{eq:DeltaNstar}, one can see that the dependence of $\Delta N_*$ on $T_\ureh$ is maximal when $\bar{w}_{\ureh}=-1/3$ [that is to say, the multiplying factor $(1-3\bar{w}_{\ureh})/(1+\bar{w}_{\ureh})$ between $\rho_\ureh$ and $\Delta N_*$ is maximal when $\bar{w}_{\ureh}=-1/3$], which explains why most of the information measured in \Ref{Martin:2016oyk} is gained close to $\bar{w}_{\ureh}=-1/3$. In the present section however, one imposes $\bar{w}_{\ureh}=0$ in the single-field models, to allow fair comparison with the situation where an extra light scalar field is introduced where it is assumed that the inflaton is massive between the end of inflation and its decay.

For the plateau potential of Higgs inflation, although more information on $T_\ureh$ is gained once an extra light scalar field is introduced, the Kullback-Leibler divergences remain small. With a quartic potential however, $0.66$ bits of information are obtained with the logarithmic prior on $\sigma_\uend$, which is a sizeable value. Looking at table~\ref{table:DKL}, one can see that it is in fact much more than any individual reheating scenario for the quartic potential. This means that these $0.66$ bits of information mostly correspond to the selection of scenarios 5 and 8 amongst all 10 possible reheating scenarios, similarly to what was discussed in \Sec{sec:InformationGain:rhoend} for $\rho_\uend$.

The values of the individual Kullback-Leibler divergences on $T_\ureh$ are also shown in \Fig{fig:mainresult}, together with the Bayesian evidence of the models they correspond to.
\subsubsection{\textsf{Early reheating temperature}}
\label{sec:InformationGain:Tereh}
The early reheating temperature is defined only for scenarios 2, 5, 8 and 9. One obtains small information gains with plateau potentials, and depending on the prior one uses on $\sigma_\uend$, $0.2$ or $0.3$ bits with the quartic potential. 

In summary, one finds that more information about reheating can be extracted from the data in models where an extra light scalar field is added than in purely single-field setups. In particular, the Kullback-Leibler divergences on the reheating temperatures can be substantial if the inflaton potential is quartic, and are more modest for a plateau potential.
\section{\textsf{Discussion}}
\label{sec:Discussion}
In \Sec{sec:results-creh}, constraints were derived on the energy scale of inflation, the reheating temperature and the early reheating temperature. In this section, we extend the discussion in a few directions to investigate the physical implications of the constraints we obtained.
\subsection{\textsf{Inflationary energy scale in plateau models}}
\label{sec:KMIII}

\begin{figure}
\figpilogsto
\begin{center}
\includegraphics[width=7cm]{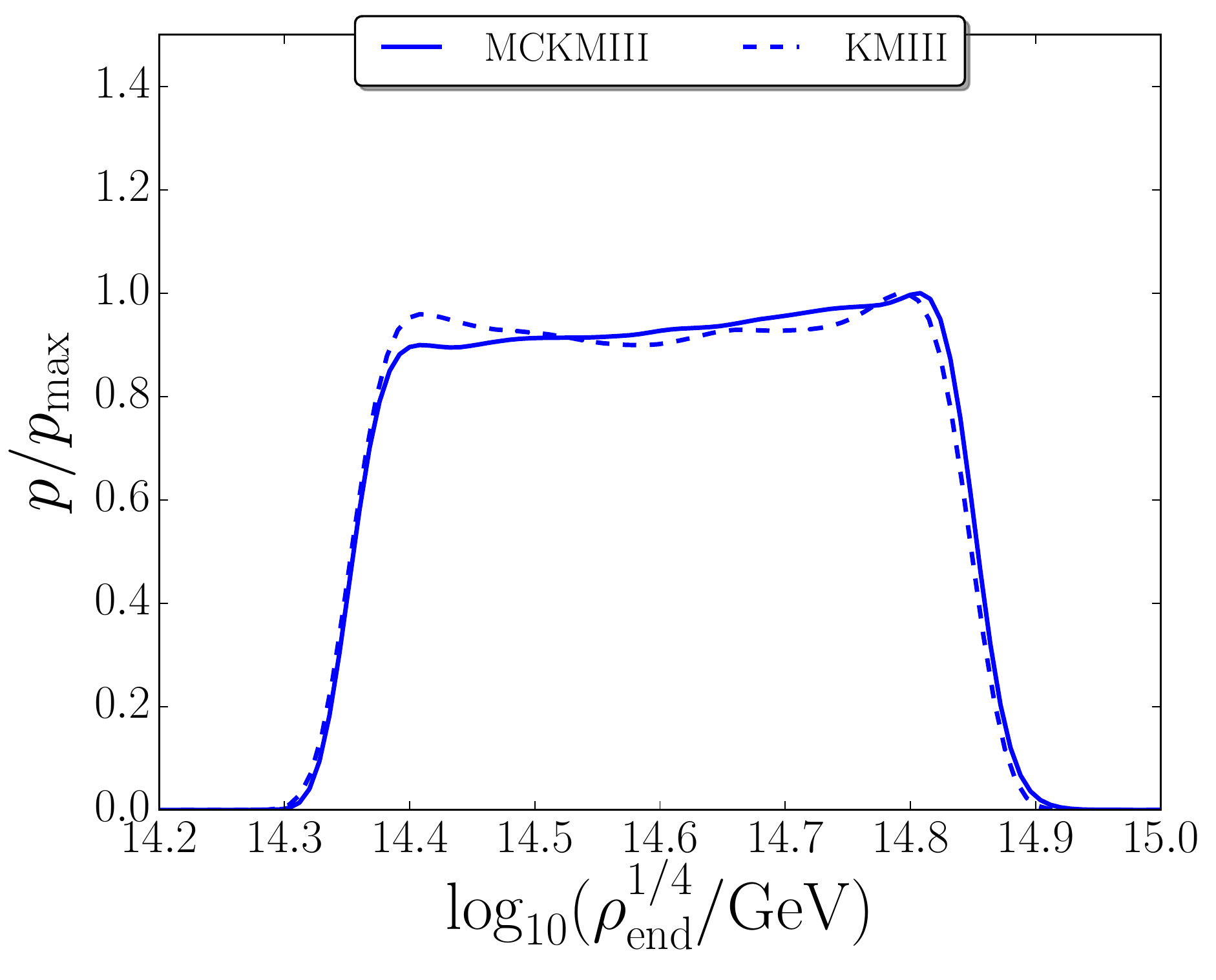}
\includegraphics[width=7cm]{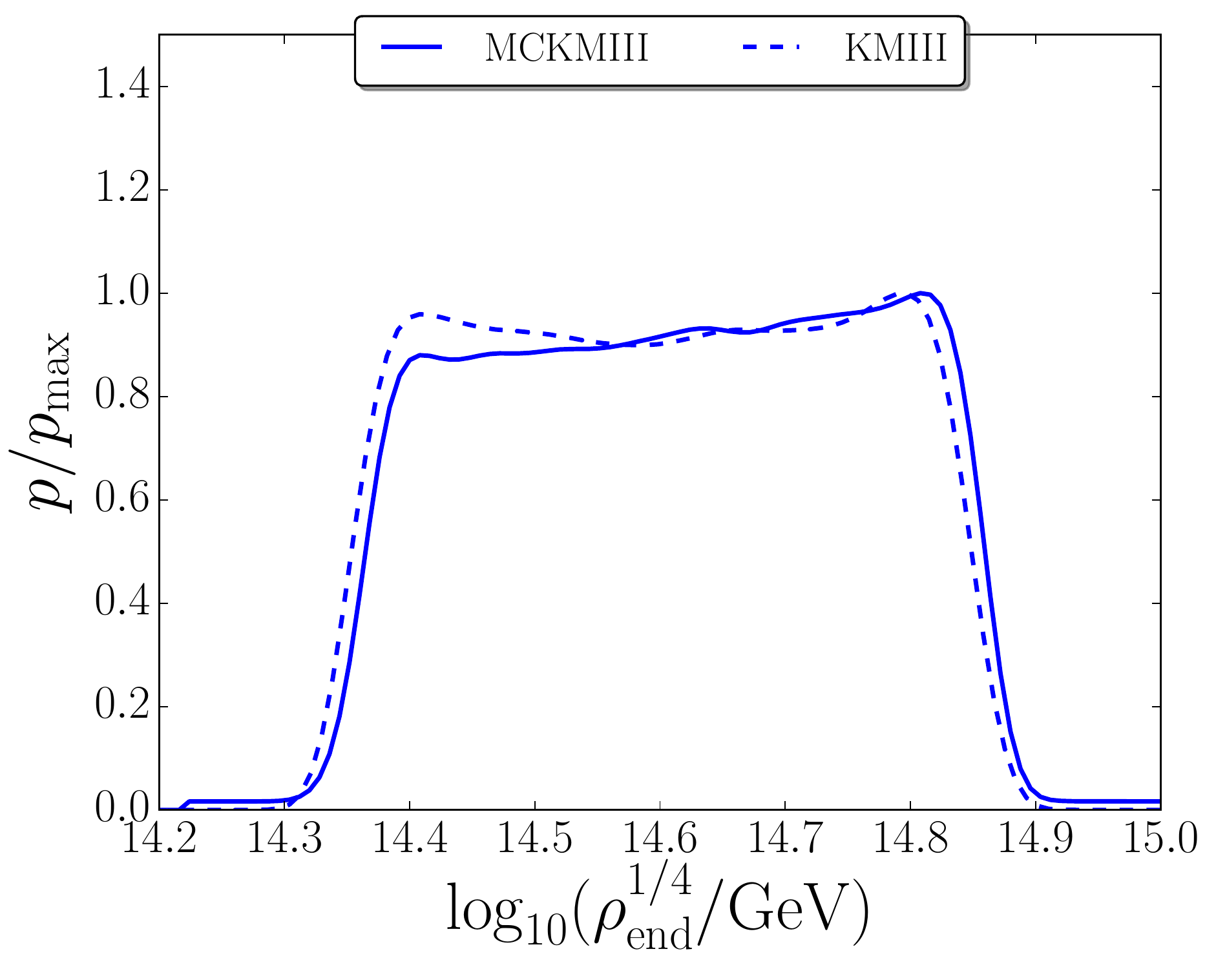}
\includegraphics[width=6.9cm]{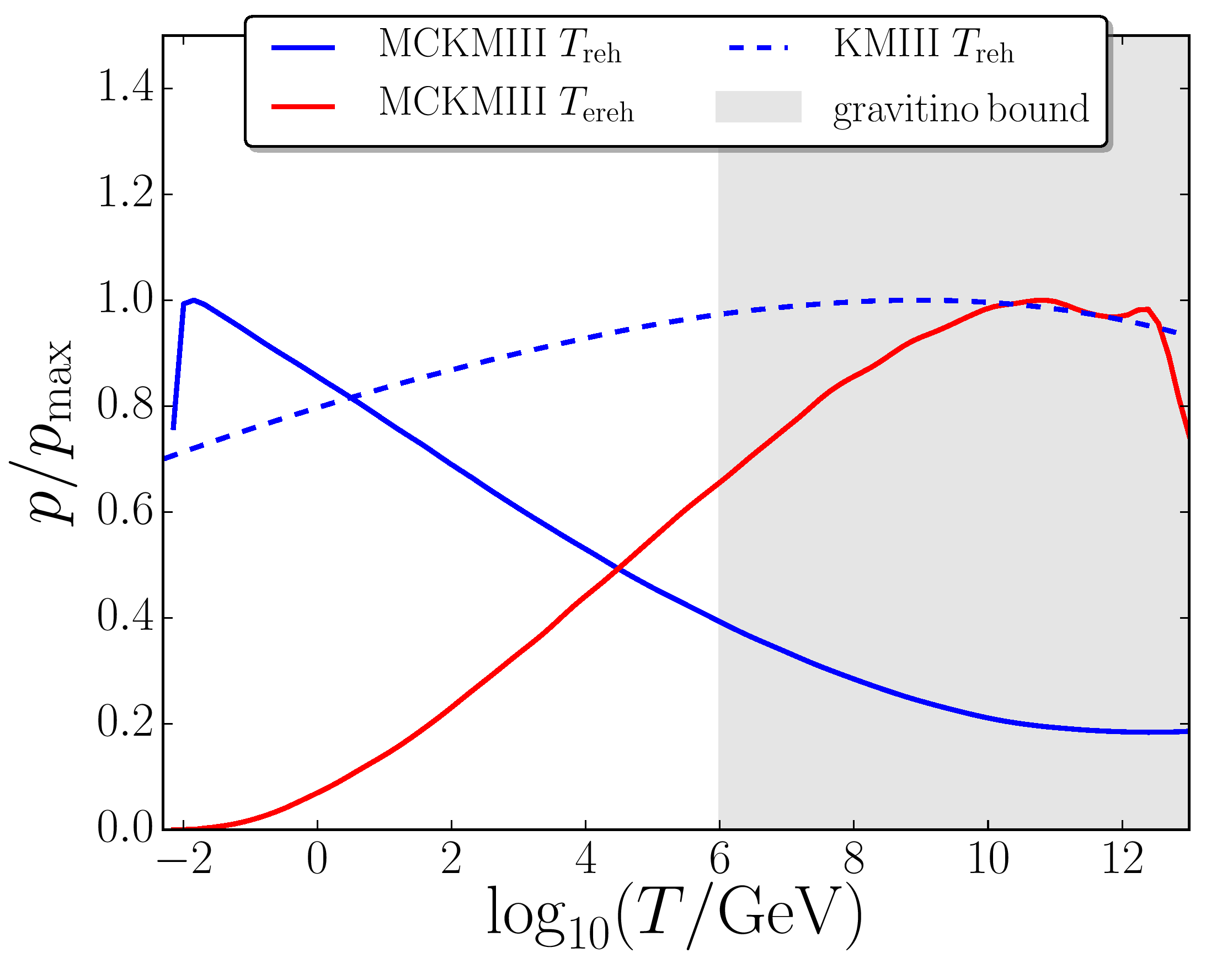}
\includegraphics[width=6.9cm]{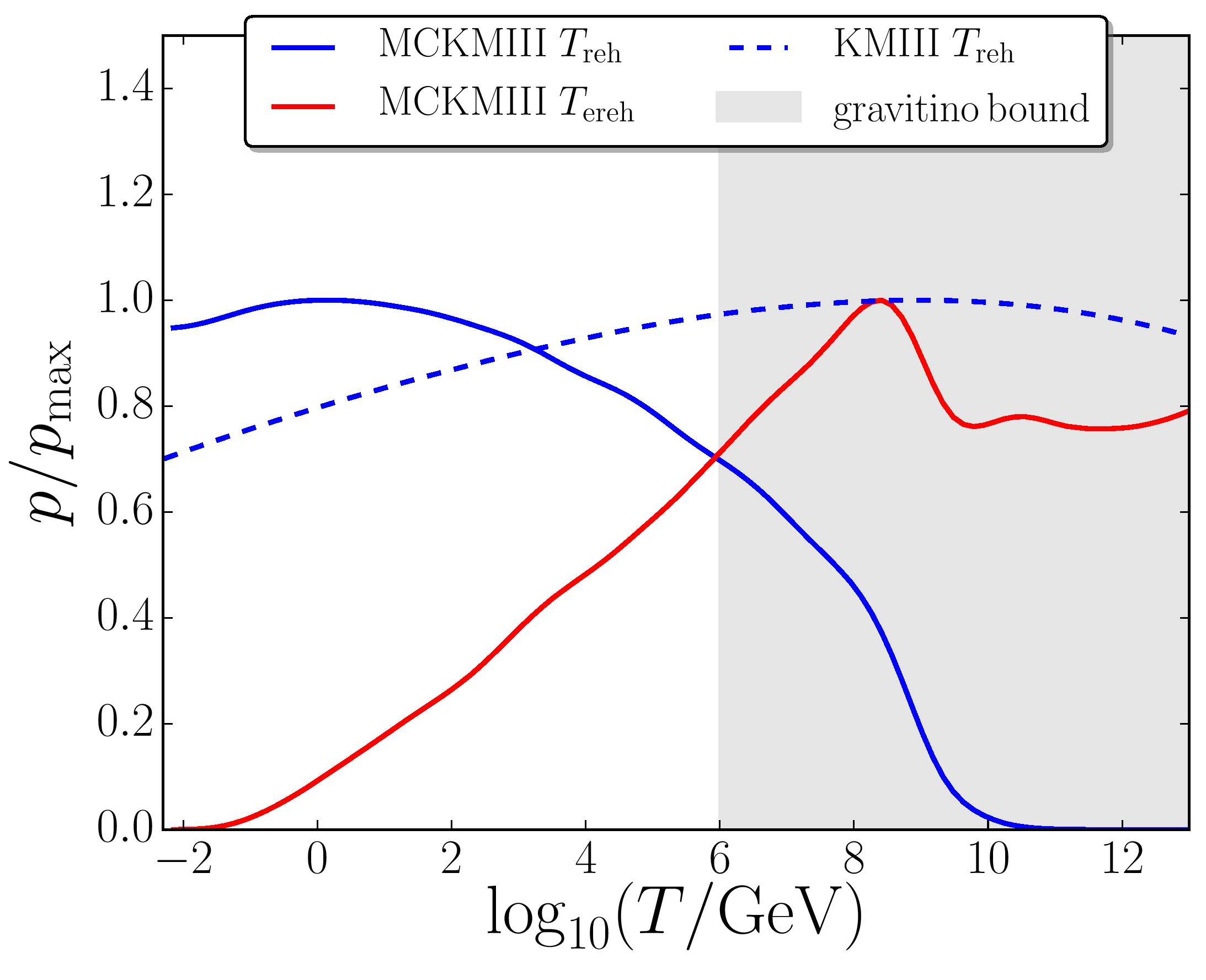}
\caption[Posteriors on the energy density and reheating temperatures]{Posterior distribution on the energy density at the end of inflation (top panels) and the reheating temperatures (bottom panels) in the K\"ahler moduli II potential~(\ref{eq:pot:KMIII})  of inflation. The left panels correspond to the logarithmically flat prior~(\ref{eq:sigmaend:LogPrior}) $\pilog$ on $\sigma_\uend$, and the right panels stand for the stochastic prior~(\ref{eq:sigmaend:GaussianPrior}) $\pisto$ derived from the equilibrium distribution of a light scalar field in a de Sitter space-time with Hubble scale $H_\uend$. The dashed lines correspond to the single-field version of the model, while the solid lines stand for the averaged posterior distributions over all $10$ reheating scenarios.}
\label{fig:KMIII}
\end{center}
\end{figure}

As explained in \Sec{sec:intro}, the Bayesian model comparison program applied to the scenarios discussed in the present chapter show that~\cite{Vennin:2015egh} the models favoured by the data are of two types: either plateau potentials, in any of the 10 reheating scenarios, or quartic potentials in scenarios 5 and 8. Quartic potentials are rather uniquely defined but several versions of plateau inflation have been proposed in the literature. So far, the potential of Higgs inflation (or equivalently the Starobinsky model) has been used to study these models. As noticed in \Fig{fig:post:rhoend:averaged} and further commented on in \Sec{sec:InformationGain:rhoend}, this leads to very sharp constraints on $\rho_\uend$, which, in the absence of any other free parameter in the potential, is mostly fixed by $A_{{}_\mathrm{S}}$. However, plateau potentials exist where inflation can be realised at different energies. To study how the conclusions drawn above are dependent on the specific shape (and energy scale) of the plateau potential considered, in this section, we include another plateau potential in our analysis, K\"ahler moduli II inflation (KMIII in the terminology of \Ref{Martin:2013tda}),
\bea
U(\phi)&= M^4\left[1-\alpha\left(\frac{\phi}{\Mp}\right)^{4/3}\ee^{-\beta\left(\frac{\phi}{\Mp}\right)^{4/3}}\right]\, .
\label{eq:pot:KMIII}
\eea
The posterior constraints on $\rho_\uend$, $T_\ureh$ and $T_\uereh$ are shown in \Fig{fig:KMIII}, and the individual reheating scenarios are displayed in \App{Sec:IndividualScenarios}. 

Compared to \Fig{fig:post:rhoend:averaged}, one can see that inflation proceeds at lower energy, with a wider range of allowed energy scales due to the presence of the free parameters $\alpha$ and $\beta$ in \Eq{eq:pot:KMIII}. This leads to a much smaller Kullback-Leibler divergence on $\rho_\uend$ than in the case of single-field Higgs or quartic inflation, see table~\ref{table:DKL}. However, one still notices that the $\rho_\uend$ posteriors when an extra light scalar field is added are very close to the single-field constraints. For the reheating temperatures, the same remarks apply as in \Secs{sec:result:Treh} and~\ref{sec:result:Tereh} for Higgs inflation. In particular, small reheating temperatures and large early reheating temperatures are preferred. Therefore, apart from the large value of $\dkl$ for $\rho_\uend$, the results obtained above for Higgs inflation seem to characterise plateau potentials in general.
\subsection{\textsf{Gravitino overproduction bounds}}
\label{sec:gravitino}
Reheating affects cosmology in different ways. First, as explained in \Sec{sec:InversionProblem}, it contributes to the expansion history through its averaged equation-of-state parameter and its energy density at completion. This is the effect we used to constrain reheating in single-field models. Second, it may produce additional features (such as gravitational waves, magnetic fields, topological defects, baryon asymmetries or dark matter, etc.), and enhance the contribution from light scalar fields (that are otherwise spectator fields during inflation) to curvature perturbations. This is the case of the scenarios considered in the present section and this additional effect is the one we have used to constrain reheating in these setups. Third, it affects the subsequent thermal history of the Universe, since it determines the temperature at the onset of the radiation dominated epoch. 

To illustrate how this last effect can be important to constrain reheating, in this section, we consider gravitinos, the gauge fermion supersymmetric partners of the graviton of supergravity theories. Gravitinos are produced from scatterings in the hot plasma during reheating, and their abundance is directly related to the magnitude of the reheating temperature~\cite{Terada:2014uia}. Their lifetime depends on their mass $m_{3/2}$, and if they survive long enough, their decay products can produce spectral distortions of the CMB. Combining current constraints on CMB spectral distortions and BBN, upper bounds can be derived on $T_\ureh$. In \Ref{Dimastrogiovanni:2015wvk}, it is found that, with $m_{3/2} \sim {\cal O}(100\,\GeV)$, one typically obtains the most stringent constraint $T_\ureh<10^{6}\,\GeV$.\footnote{In full generality, combining these constraints in a rigorous analysis would require deriving a likelihood function that takes into account correlations between this data on smaller scale fluctuations with those on the larger scale fluctuations from CMB experiments. The constraint also assumes that local supersymmetry is indeed the correct extension to the standard model of particles.} 

This value is shown in \Fig{fig:post:Trehs:averaged} and the bottom panels of \Fig{fig:KMIII} where the posterior distributions on $T_\ureh$ and $T_\uereh$ are displayed. One can see that it excludes a large set of possible temperatures. However, scenarios where an extra light scalar field is added seem to more easily evade the gravitino overproduction bound than their single-field counterpart. For instance, in quartic inflation with an additional light field, the reheating temperature is typically smaller than $10^6 \, \GeV$ [see the bottom left panel of \Fig{fig:post:Trehs:averaged} and \Eq{eq:LFI4:constraint:Treh}], which is not the case of the single-field versions of Higgs inflation, quartic inflation or even K\"ahler moduli III inflation in \Fig{fig:KMIII}. On the other hand, since large early reheating temperatures are preferred in general, the gravitino problem might be worsened if gravitinos are generated from the decay products of the first decaying field in scenarios 2, 5, 8 and 9. 

Interestingly, this also shows that if gravitinos exist, they provide a powerful indirect way to further constrain the models discussed in this section. In particular, gravitino production bounds seem to yield less additional constraints for quartic models than for plateau models (with an extra light scalar field in both cases). If they were explicitly included in the set of observations, they would therefore probably lead to a slight preference of the former against the later. 
\subsection{\textsf{Decay mediation scale}}
\label{sec:DecayMediationScale}

\begin{figure}
\figpilogsto
\begin{center}
\includegraphics[width=7cm]{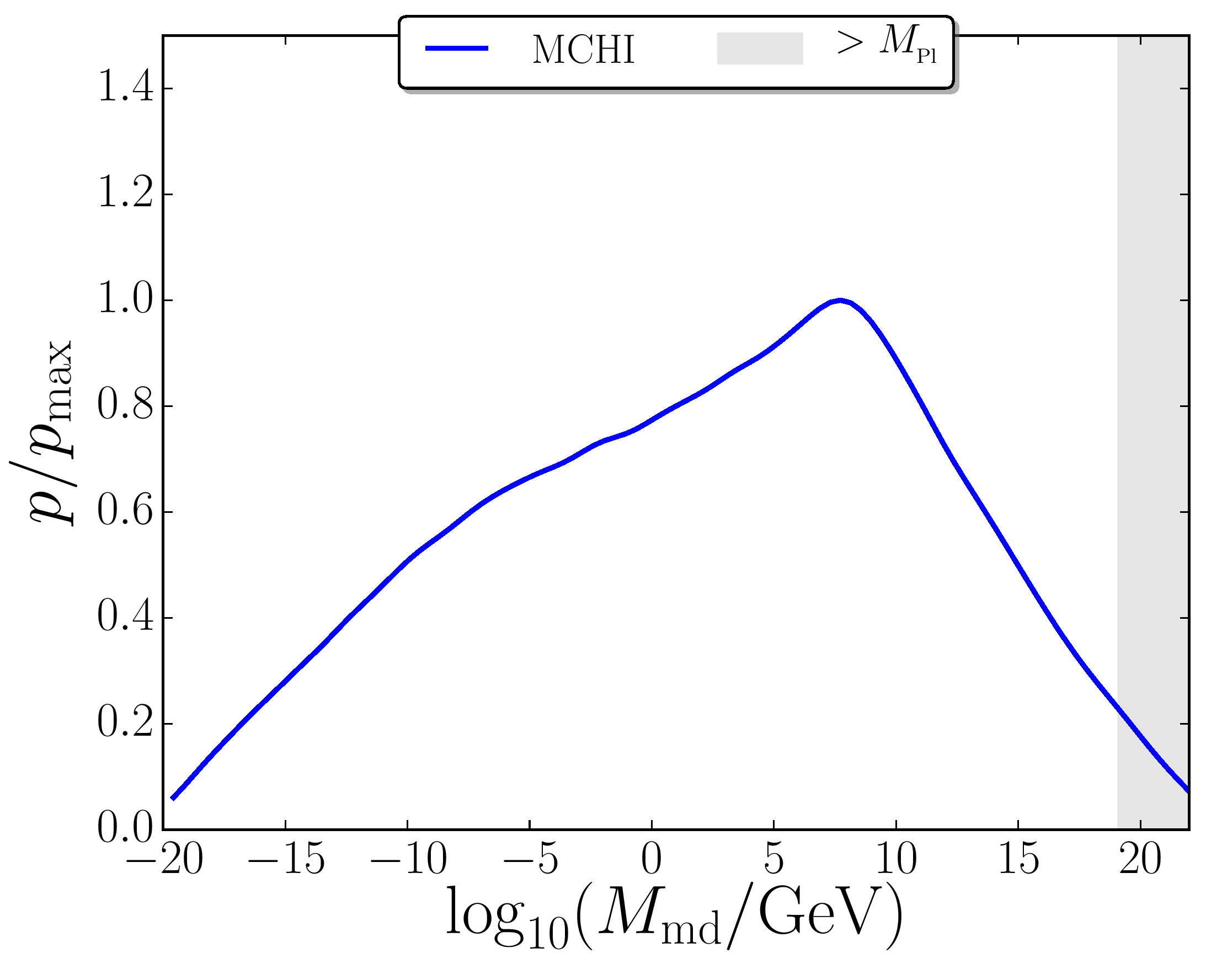}
\includegraphics[width=7cm]{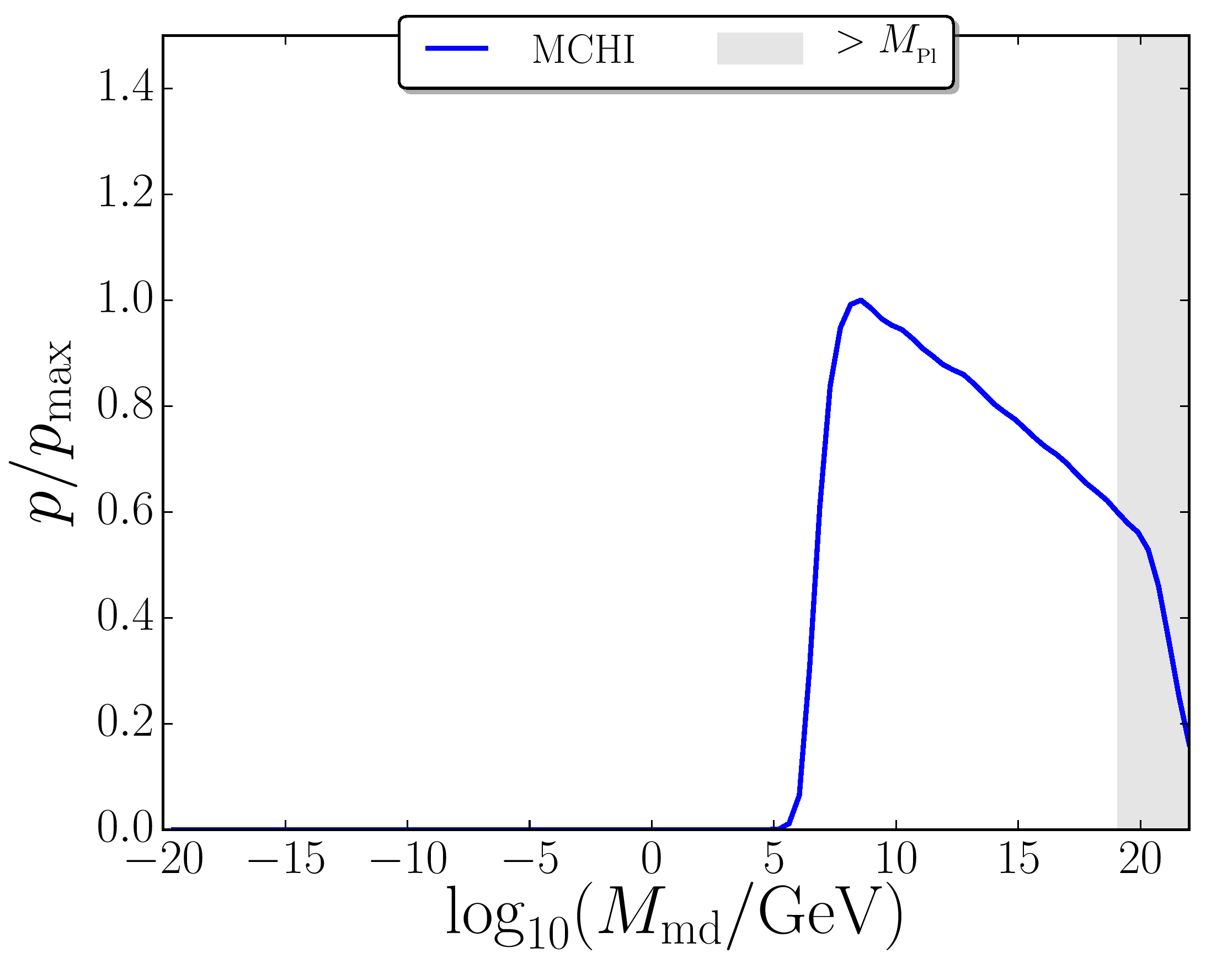}
\includegraphics[width=7cm]{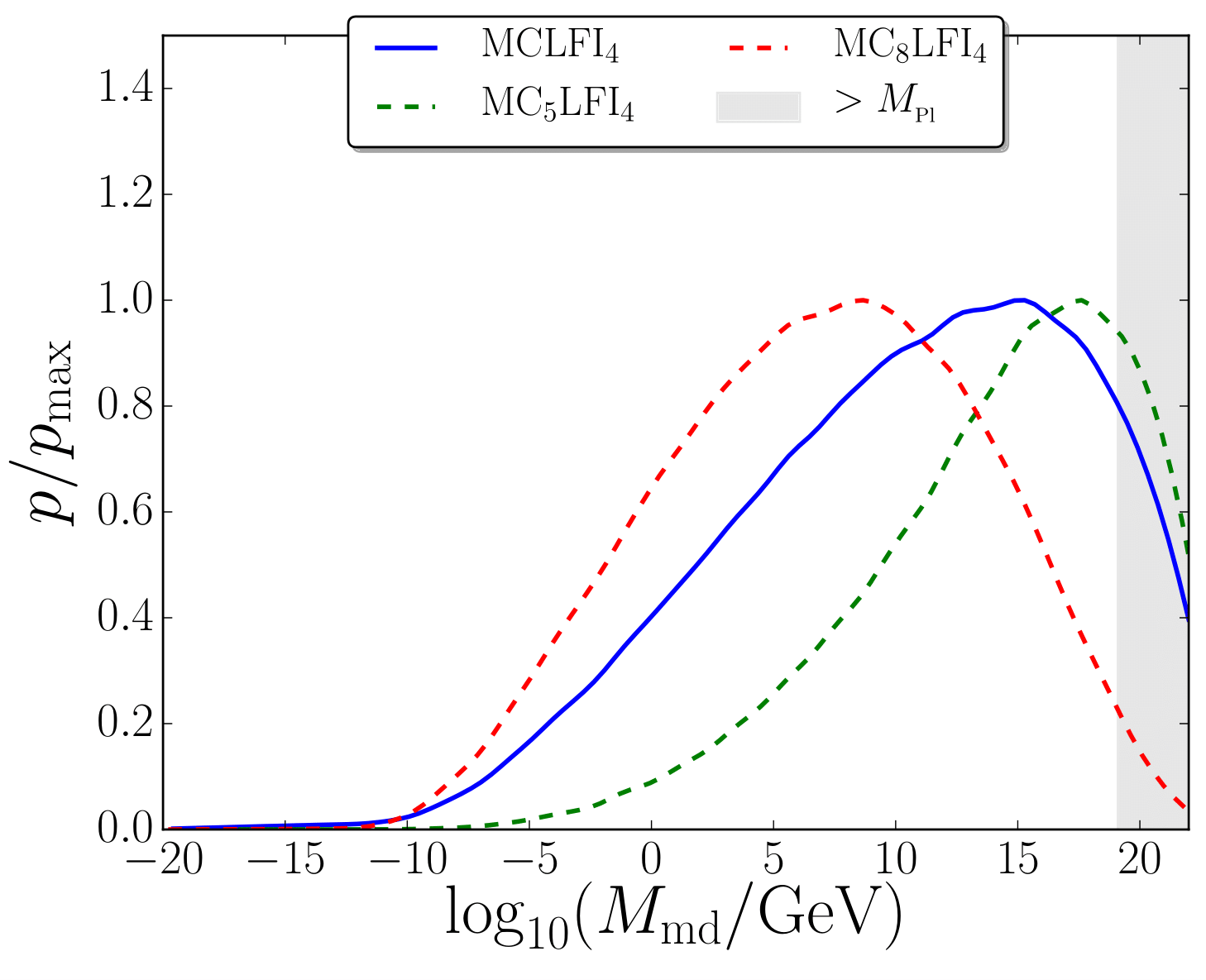}
\includegraphics[width=7cm]{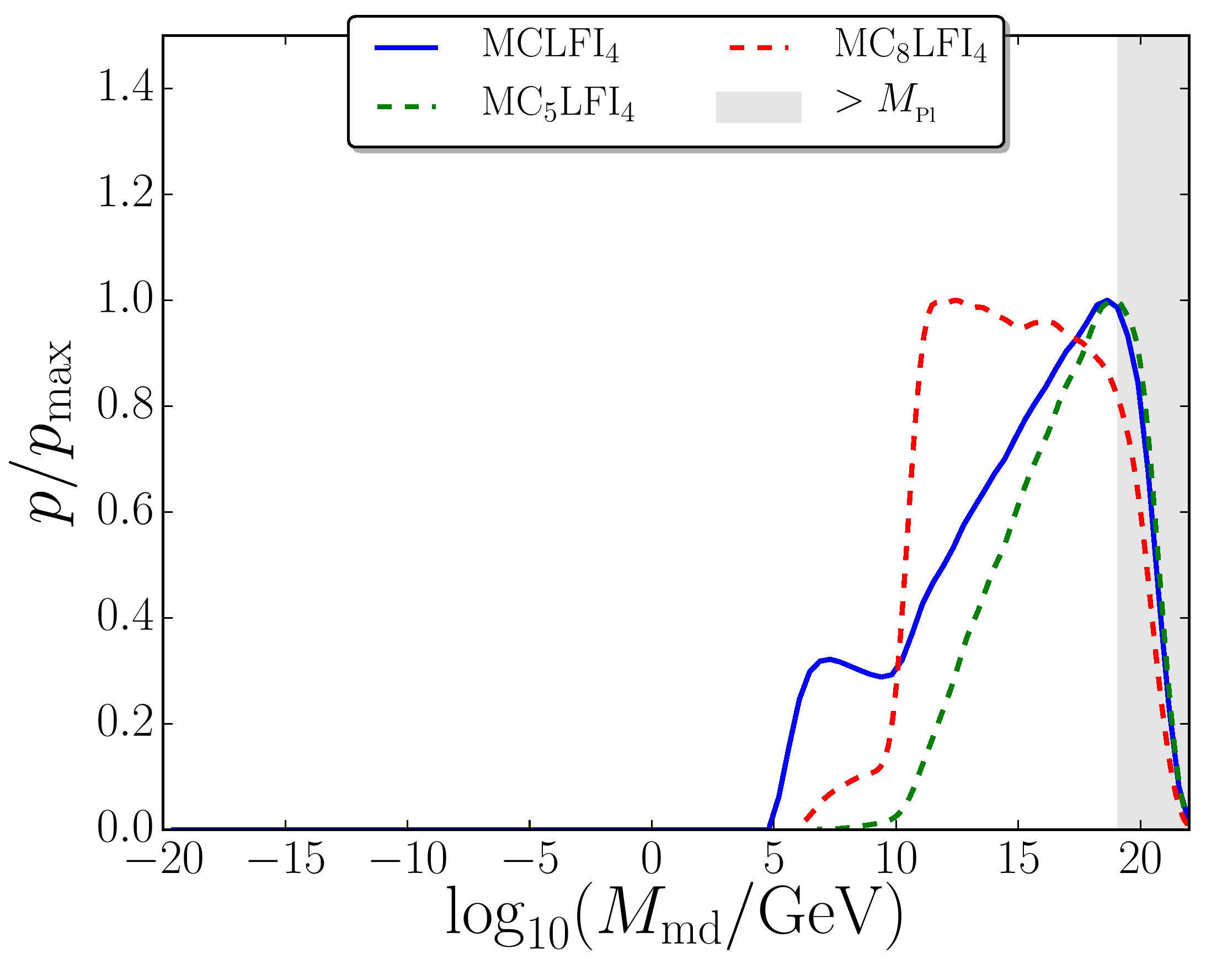}
\caption[Posterior distributions of the mass mediation scale]{Posterior distributions of the mass mediation scale $\Mmd$ defined for $\sigma$ in \Eq{eq:Mstar}, for Higgs inflation (top panels) and quartic inflation (bottom panels), when the logarithmically flat prior (left panels) and the stochastic prior (right prior) are used on $\sigma_\uend$. The distributions are averaged over all 10 reheating scenarios, and for quartic inflation, the individual posteriors for scenarios 5 and 8 are shown since these are the only non-ruled out scenarios. The region corresponding to decay rates that are smaller than gravitational mediation, $\Mmd>\Mp$, is shaded in grey.}
\label{fig:MediationScale}
\end{center}
\end{figure}

So far, the decay rate of the additional scalar field $\sigma$, $\Gamma_\sigma$, and its mass $m_\sigma$, have been assumed to be independent (up to the ordering conditions of \Fig{fig:cases}). However, these scales may be related by the physics of the decay of $\sigma$, and in this section we study the implications of the results we obtained on such processes. More specifically, we consider the case where spontaneous decay of $\sigma$ by dimension $5$ operators is mediated by some scale $\Mmd$. The decay rate and the mass are then given by \Eq{eq:Mstar-before} with $\Mp$ replaced by the new scale such that
\begin{equation} \label{eq:Mstar}
\Gamma_\sigma \simeq \frac{m_\sigma^3}{\Mmd^2}\,. 
\end{equation}
Let us study which values of $\Mmd$ are typically predicted by the scenarios considered in this section. In \Fig{fig:MediationScale}, the averaged (over reheating scenarios) posterior distributions for $\Mmd$ are displayed. For electroweak suppressed decay for instance, one should have $\Mmd\sim 100\, \mathrm{GeV}$. Although such values are well within the distributions when a logarithmically flat prior on $\sigma_\uend$ is used,  higher mediation scales are typically preferred, which is in agreement with the standard curvaton picture where  gravitationally mediated decay~\cite{Enqvist:2013gwf} is assumed. 

One can also see that the large-field quartic models favour slightly higher mediation scales than the plateau potential of Higgs inflation (that has a very similar posterior on $\Mmd$ as K\"ahler moduli II inflation introduced in \Sec{sec:KMIII}, which is why this other plateau potential is not displayed here). This is due to the fact that the most likely scenarios for the quartic potential, cases 5 and 8, yield large values compared to the other scenarios, while for plateau potentials all reheating cases contribute to the distributions plotted in \Fig{fig:MediationScale}. For this reason, individual cases 5 and 8 are also displayed (green and red dashed lines respectively) for the quartic potential in \Fig{fig:MediationScale}. Interestingly, while these two scenarios are indistinguishable with respect to all criteria discussed so far, they give slightly different preferred values for $\Mmd$, which suggests that requiring specific mediated decay scales may be a way to distinguish between these cases.

When the stochastic prior on $\sigma_\uend$ is used, one notices that mediated decay cannot happen for $\Mmd$ below $10^5\,\GeV$. This is in sharp contrast with the result obtained with a logarithmically flat prior on $\sigma_\uend$ and can be understood as follows. When $\sigma_\uend$ is super-Planckian, $\sigma$ drives a second phase of inflation (cases 6, 9 and 10 in \Fig{fig:cases}), the duration of which is roughly given by $\sigma_\uend^2/(4\Mp^2)$ in numbers of $e$-folds. Therefore, $\sigma_\uend$ cannot be much larger than, say, $20\Mp$. Therefore the stochastic prior on $\sigma_\uend$, which implies that $m_\sigma\sim H_\uend^2/\sigma_\uend$ [see \Eq{eq:sigmaend:GaussianPrior}], yields a lower bound on $m_\sigma$, that does not exist when a logarithmically flat prior on $\sigma_\uend$ is used. This explains why higher values of $m_\sigma$, hence of $\Mmd$, are obtained with the stochastic prior.
\section{\textsf{Conclusion}}
\label{sec:Conclusions}
In this section, we have presented the first systematic observational constraints on reheating in scenarios where inflation is driven by a single scalar inflaton field $\phi$, but an extra light scalar field $\sigma$ can also contribute to the total amount of curvature perturbations. Following the results of \Ref{Vennin:2015egh}, the analysis was performed in the two classes of models that are favoured by the data, where the inflationary potential is either of the plateau or the quartic type.  

Bayesian inference techniques were employed to derive posterior constraints on the energy density at the end of inflation $\rho_\uend$, and the temperature of the Universe $T_\ureh$  (and $T_\uereh$) at the onset of the radiation dominated epoch(s). If inflation is realised with a plateau potential, it was found that the constraints on $\rho_\uend$ are scarcely altered by the introduction of a light scalar field (compared to the purely single-field case), in agreement with the strong robustness of these models under the introduction of an additional scalar field noted in \Ref{Vennin:2015egh}. For a quartic inflationary potential however, it was found that lower values of $\rho_\uend$ are favoured with an extra light scalar field. Indeed, quartic inflation predicts a value of the tensor-to-scalar ratio that is not too large only when the extra field provides the dominant contribution to curvature perturbations, in which case $\rho_\uend$ is smaller than in the single-field scenario. 

For the reheating temperature, plateau potentials yield constraints on $T_\ureh$ that depend on the reheating scenario (these scenarios are listed in \Fig{fig:cases} and the constraints are given in \Fig{fig:post:Treh:individual}). For quartic inflation, the only favoured reheating scenarios are 5 and 8, and both show a preference for lower reheating temperatures than with the single-field counterpart of the model. When a logarithmically flat prior on the vev of the extra light field at the end of inflation is used, one obtains the averaged $95\,\%\ \mathrm{CL}$ upper bound $T_\ureh<5\times 10^{4}\,\GeV$. In reheating scenarios 2, 5, 8 and 9, the Universe undergoes a transient early radiation dominated epoch during a two-stage reheating process and the constraints on the temperature at its onset, $T_\uereh$, were also derived. Contrary to $T_\ureh$, lower bounds can be derived on $T_\uereh$, typically larger than $\sim 10^2\,\GeV$ for a plateau potential and larger than $\sim 10^5\,\GeV$ for the quartic potential.

In general, it was observed that tighter constraints on reheating are derived with an additional light scalar field than without, in agreement with the results of \Ref{Vennin:2015egh} where Bayesian complexity was used to quantify the number of unconstrained parameters. Indeed, when the extra field is present, the same parameters define both its contribution to curvature perturbations and to the expansion history of reheating that determines the location of the observational window along the inflationary potential. More information about reheating can therefore be gained in scenarios with an additional scalar field, compared to the single-field case where only the later effect allows one to constrain reheating from observations. This information gain was quantified by computing the Kullback-Leibler divergence between the prior and posterior distributions of $\rho_\uend$, $T_\ureh$ and $T_\uereh$. Even if the information gain remains modest when the inflationary potential is of the plateau type, it becomes substantial in quartic inflation (where, for instance, more than 3 bits of information are gained on the energy density at the end of inflation).

Since the process of reheating determines the temperature of the Universe at the onset of the radiation dominated epoch, it affects its subsequent thermal history. The constraints we derived thus have implications for post-inflationary physics. For instance, we have considered gravitino overproduction bounds and shown that since models with an additional scalar field predict lower reheating temperatures, they evade those bounds more easily than their single-field counterpart. This is particularly true if the inflationary potential is of the quartic type, so that if gravitino bounds were explicitly included in the set of observations used to constrain the models, they would probably lead to a slight preference of quartic inflation with an extra light scalar field (in reheating scenarios 5 and 8) over all other models, including the single-field plateau ones. 

The sensitivity to the microphysics of reheating has also been demonstrated with the mass mediation scale of the extra scalar field decay, on which constraints have been derived. Notably, it was found that reheating scenarios 5 and 8 in quartic inflation, otherwise indistinguishable with respect to all other criteria discussed in this chapter, give slightly different preferred values for this mass scale.

In this analysis, the crucial role played by the prior on the vev of the extra light scalar field at the end of inflation, $\sigma_\uend$, has also been highlighted. Even though the main conclusions quoted above are robust under changes of priors on $\sigma_\uend$, the detailed constraints on reheating and the relative parameter space volume associated with the 10 reheating scenarios depend on the assumptions one makes about its value. In particular, for quartic inflation, which, in reheating scenarios 5 and 8, is one of the most favoured models, if $\sigma_\uend$ is set by the quantum diffusion effects during inflation, one finds that the Gaussian distribution~(\ref{eq:sigmaend:GaussianPrior}) is not an equilibrium solution of the stochastic dynamics of $\sigma$. In fact, there is no equilibrium solution in this case, and the typical value acquired by the additional scalar field at the end of inflation both depends on its value at the onset of inflation and on the total duration of inflation. This may be relevant to the question~\cite{Starobinsky:1986fxa, Linde:2005yw, Enqvist:2012, Burgess:2015ajz} whether observations can give access to scales beyond the classical horizon, and we plan to study this question further in later chapters.

\newpage

\begin{subappendices}
\section{\textsf{Kullback-Leibler divergence}}
\label{sec:DKL}
We introduced the Kullback-Leibler divergence in \Sec{sec:info-theory-tools} as a quantity which computes the information gain between prior and posterior distributions.

In table~\ref{table:DKL}, the Kullback-Leibler divergences on the energy scale of inflation and the reheating temperatures are given for the three potentials considered in this section (Higgs inflation, quartic inflation and K\"ahler moduli II inflation), for the single-field  versions of the model as well as for all 10 reheating scenarios, where the divergence between the averaged priors and posteriors are also given. The left tables were obtained with a logarithmically flat prior on $\sigma_\uend$, and the right priors with the stochastic prior~(\ref{eq:sigmaend:GaussianPrior}).

\newpage
\begin{table}[!h]
  \small
  \centering
  \resizebox{4.2cm}{!}{ 
\begin{tabular}{ | l | l | l | l |}
\hline
  $\pi_{\log} (\sigma_{\rm end})$   & \multicolumn{2}{l}{\qquad \qquad $\dkl$} & \\ \hline
    Model & $\rho_{\rm end}$ & $T_{\rm reh}$ & $T_{\rm ereh}$   \\ \hline \hline
    HI & 1.370 & 0.004 & - \\ \hline
    MCHI & 0.114 & 0.005 & 0.018 \\ \hline \hline
    ${\rm MC}_1$HI & 0.107 & 0.005 & -\\ \hline
    ${\rm MC}_2$HI & 0.009 & 0.009 & 0.001 \\ \hline
    ${\rm MC}_3$HI & 1.059 & 0.001 & -\\ \hline
    ${\rm MC}_4$HI & 0.061 & 0.042 & -\\ \hline
    ${\rm MC}_5$HI & 0.504 & 0.023 & 0.039 \\ \hline
    ${\rm MC}_6$HI & 0.687 & 0.023 & -\\ \hline
    ${\rm MC}_7$HI & 0.280 & 0.012 & - \\ \hline
    ${\rm MC}_8$HI & 0.587 & 0.016 & 0.015 \\ \hline
    ${\rm MC}_9$HI & 0.548 & 0.006 & 0.001 \\ \hline
    ${\rm MC}_{10}$HI & 1.539 & 0.091 & -\\ \hline \hline
        ${\rm LFI}_4$ & 1.171 & 0.108 & - \\ \hline
    MC${\rm LFI}_4$ & 3.104 & 0.656 & 0.181 \\ \hline \hline
    ${\rm MC}_1{\rm LFI}_4$ & 0.000 & 0.120 & -\\ \hline
    ${\rm MC}_2{\rm LFI}_4$ & 0.080 & 0.077 & 0.019 \\ \hline
    ${\rm MC}_3{\rm LFI}_4$ & 0.971 & 0.039 & -\\ \hline
    ${\rm MC}_4{\rm LFI}_4$  & 0.190 & 0.011 & -\\ \hline 
    ${\rm MC}_5{\rm LFI}_4$  & 0.911 & 0.039 & 0.125 \\ \hline
    ${\rm MC}_6{\rm LFI}_4$ & 0.425 & 0.114 & -\\ \hline
    ${\rm MC}_7{\rm LFI}_4$ & 0.317 & 0.007 & - \\ \hline
    ${\rm MC}_8{\rm LFI}_4$  & 1.093 & 0.050 & 0.044 \\ \hline
    ${\rm MC}_9{\rm LFI}_4$  & 0.719 & 0.031 & 0.044 \\ \hline
    ${\rm MC}_{10}{\rm LFI}_4$ & 1.195 & 0.223 & -\\ \hline \hline
       KMIII & 0.083 & 0.008 & - \\ \hline
    MCKMIII & 0.121 & 0.015 & 0.010 \\ \hline \hline
    ${\rm MC}_1$KMIII & 0.092 & 0.021 & -\\ \hline
    ${\rm MC}_2$KMIII & 0.000 & 0.102 & 0.006 \\ \hline
    ${\rm MC}_3$KMIII & 0.072 & 0.022 & -\\ \hline
    ${\rm MC}_4$KMIII & 0.089 & 0.002 & -\\ \hline
    ${\rm MC}_5$KMIII & 0.095 & 0.003 & 0.002 \\ \hline
    ${\rm MC}_6$KMIII & 2.584 & 0.125 & -\\ \hline
    ${\rm MC}_7$KMIII & 0.095 & 0.000 & - \\ \hline
    ${\rm MC}_8$KMIII & 0.095 & 0.002 & 0.000 \\ \hline
    ${\rm MC}_9$KMIII & 0.000 & 0.012 & 0.011 \\ \hline
    ${\rm MC}_{10}$KMIII & n.c. & n.c. & -\\ \hline
    \end{tabular}
    }
      \resizebox{4.2cm}{!}{
\begin{tabular}{ | l | l | l | l |}
\hline
   $\pisto (\sigma_{\rm end})$  & \multicolumn{2}{l}{\qquad \qquad $\dkl$} & \\ \hline
    Model & $\rho_{\rm end}$ & $T_{\rm reh}$ & $T_{\rm ereh}$   \\ \hline \hline
    HI & 1.370 & 0.004 & - \\ \hline
    MCHI & 0.224 & 0.006 & 0.014 \\ \hline \hline
    ${\rm MC}_1$HI & 0.060 & 0.004 & -\\ \hline
    ${\rm MC}_2$HI & 0.058 & 0.006 & 0.000 \\ \hline
    ${\rm MC}_3$HI & 1.077 & 0.007 & -\\ \hline
    ${\rm MC}_4$HI & 0.087 & 0.001 & -\\ \hline
    ${\rm MC}_5$HI & 0.015 & 0.002 & 0.000 \\ \hline
    ${\rm MC}_6$HI & 0.800 & 0.032 & -\\ \hline
    ${\rm MC}_7$HI & - & - & - \\ \hline
    ${\rm MC}_8$HI & 0.046 & 0.002 & 0.000 \\ \hline
    ${\rm MC}_9$HI & 0.606 & 0.028 & 0.015 \\ \hline
    ${\rm MC}_{10}$HI & 2.069 & 0.130 & -\\ \hline \hline
     ${\rm LFI}_4$ & 1.171 & 0.108 & - \\ \hline
    MC${\rm LFI}_4$ & 4.780 & 0.111 & 0.281 \\ \hline \hline
    ${\rm MC}_1{\rm LFI}_4$ & 0.176 & 0.110 & -\\ \hline
    ${\rm MC}_2{\rm LFI}_4$ & 0.167& 0.088 & 0.010 \\ \hline
    ${\rm MC}_3{\rm LFI}_4$ & 1.337 & 0.049 & -\\ \hline
    ${\rm MC}_4{\rm LFI}_4$  & 0.141 & 0.105 & -\\ \hline
    ${\rm MC}_5{\rm LFI}_4$  & 3.499 & 0.051 & 0.097 \\ \hline
    ${\rm MC}_6{\rm LFI}_4$ & 1.197 & 0.117 & -\\ \hline
    ${\rm MC}_7{\rm LFI}_4$ & - & - & - \\ \hline
    ${\rm MC}_8{\rm LFI}_4$  & 3.695 & 0.157 & 0.028 \\ \hline
    ${\rm MC}_9{\rm LFI}_4$  & 1.035 & 0.174 & 0.016 \\ \hline
    ${\rm MC}_{10}{\rm LFI}_4$ & 1.528 & 0.175 & -\\ \hline \hline
        KMIII & 0.083 & 0.008 & - \\ \hline
    MCKMIII & 0.162 & 0.011 & 0.010 \\ \hline \hline
    ${\rm MC}_1$KMIII & 0.098 & 0.016 & -\\ \hline
    ${\rm MC}_2$KMIII & n.c. & n.c. & n.c. \\ \hline
    ${\rm MC}_3$KMIII & n.c. & 0.021 & -\\ \hline
    ${\rm MC}_4$KMIII & 0.099 & 0.006 & -\\ \hline
    ${\rm MC}_5$KMIII & 0.095 & 0.011 & 0.001 \\ \hline
    ${\rm MC}_6$KMIII & n.c. & n.c. & -\\ \hline
    ${\rm MC}_7$KMIII & - & - & - \\ \hline
    ${\rm MC}_8$KMIII & 0.079 & 0.004 & 0.001 \\ \hline
    ${\rm MC}_9$KMIII & n.c. & n.c. & n.c. \\ \hline
    ${\rm MC}_{10}$KMIII & n.c. & 0.133 & -\\ \hline
    \end{tabular}
    } 
\caption[Information gain on energy densities and reheating temperatures]{Kullback-Leibler divergences $\dkl$ on $\rho_\uend$, $T_\ureh$ and $T_\uereh$ for Higgs (top row), quartic large field (middle row) and K\"{a}hler moduli II (bottom row) inflation. The result is given for the single-field versions of the model and for the 10 reheating scenarios of \Fig{fig:cases} as well. The divergence between the averaged (over reheating scenarios) priors and posteriors is also displayed. The left tables were obtained with a logarithmically flat prior on $\sigma_\uend$, and the right tables with the stochastic prior~(\ref{eq:sigmaend:GaussianPrior}). Note that the early reheating temperature $T_\ureh$ is defined only for scenarios 2, 5, 8 and 9, and that scenario 7 cannot be sampled when a stochastic prior is used. For some of the K\"{a}hler moduli II cases, denoted n.c. (for ``not converged''), numerically robust results could not be obtained.}
\label{table:DKL}
\end{table}

\newpage
\section{\textsf{Individual reheating scenarios constraints}}
\label{Sec:IndividualScenarios}
In this appendix, we display the posterior constraints on $\rho_\uend$, $T_\ureh$ and $T_\uereh$, for the individual 10 reheating scenarios of \Fig{fig:cases}, for the three potentials considered in this section (Higgs inflation, quartic inflation and K\"ahler moduli II inflation) and when the logarithmically flat prior or the stochastic prior~(\ref{eq:sigmaend:GaussianPrior}) on $\sigma_\uend$ are used. For the K\"{a}hler moduli II cases denoted ``n.c.'' in table~\ref{table:DKL}, well-converged distributions could not be inferred due to the numerical difficulty in sampling these scenarios.
%
%
\subsection{\textsf{Energy density at the end of inflation}}
\label{sec:app:rhoend:individual}
\begin{figure}
\figpilogsto
\begin{center}
\includegraphics[width=7cm]{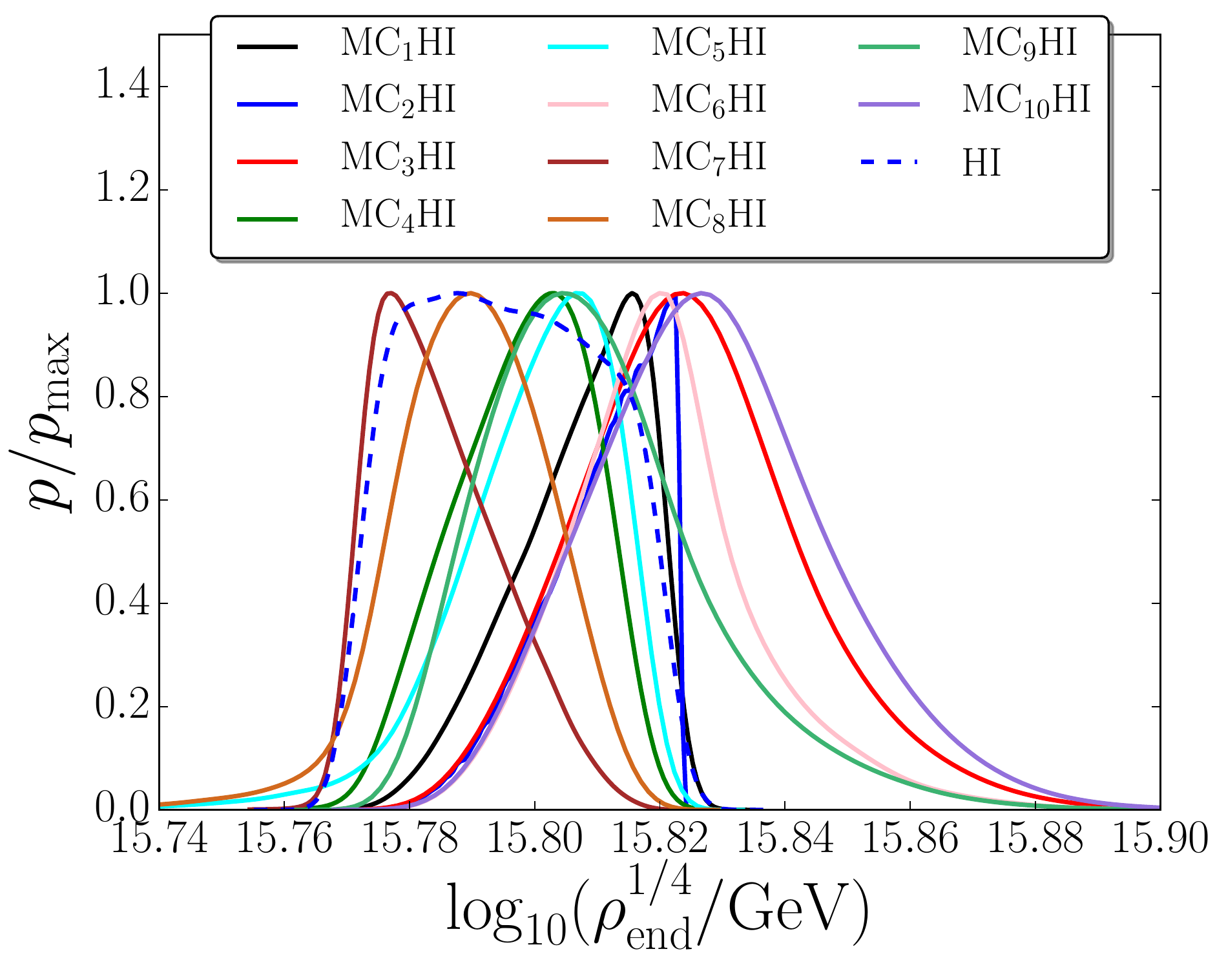}
\includegraphics[width=7cm]{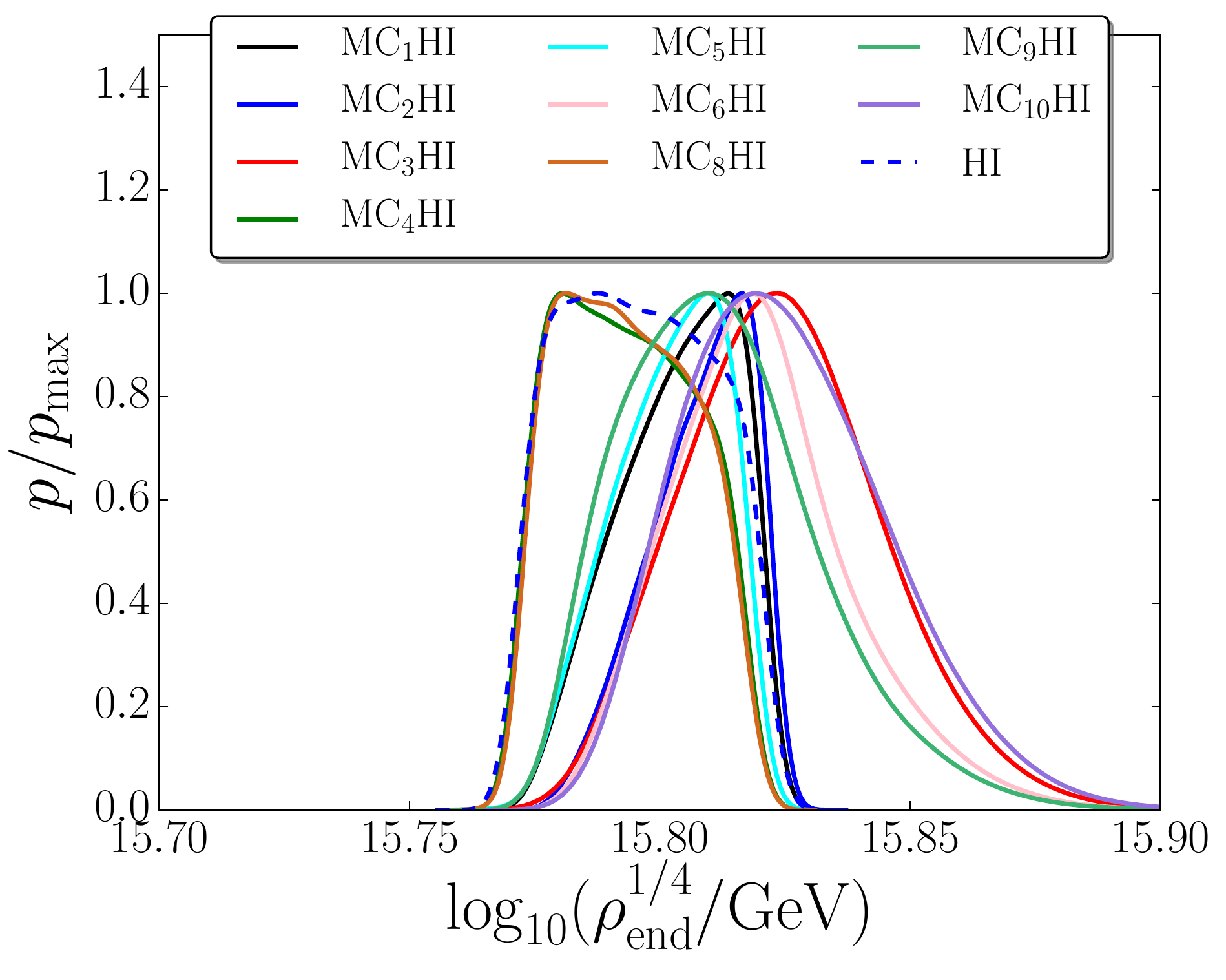}
\includegraphics[width=7cm]{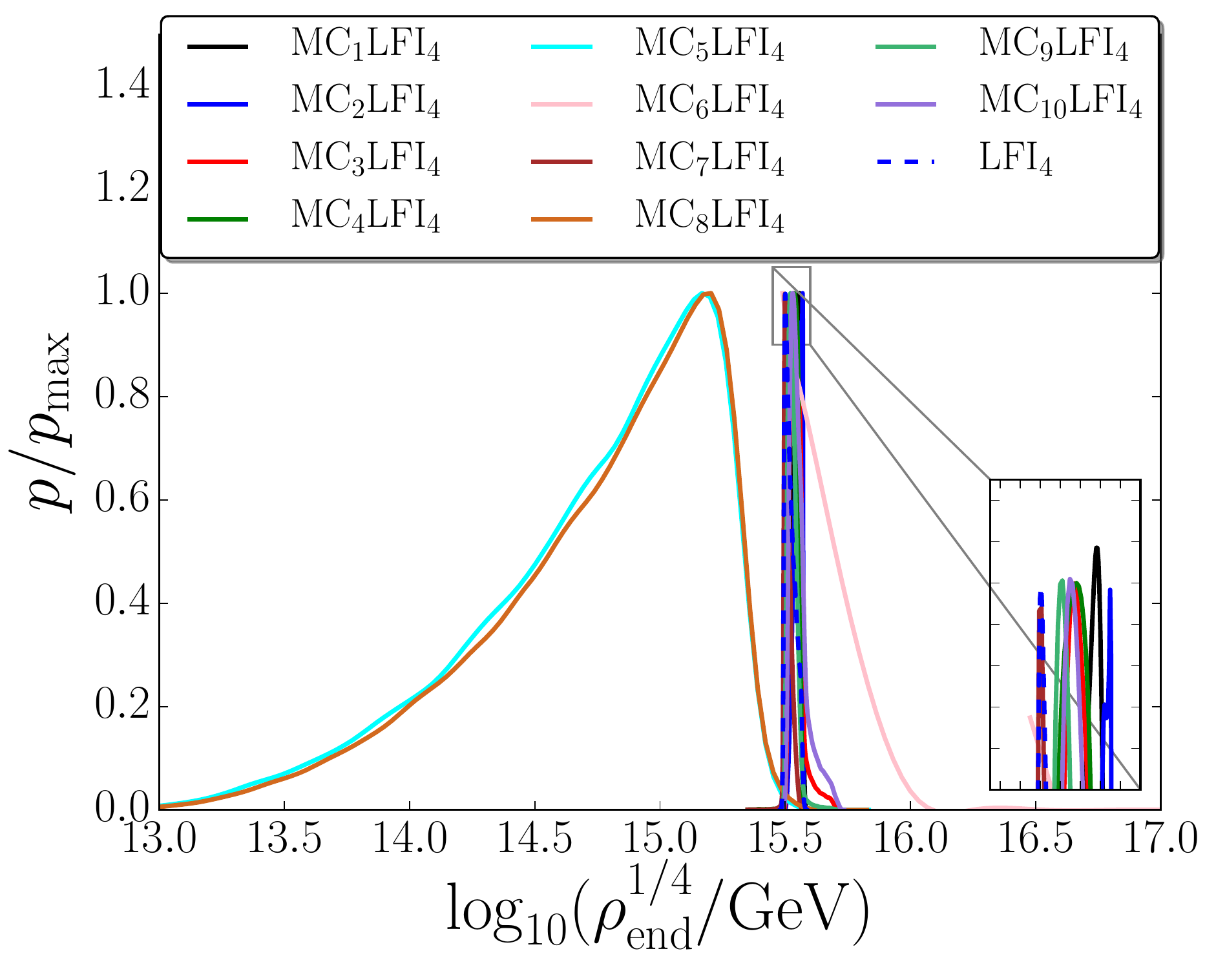}
\includegraphics[width=7cm]{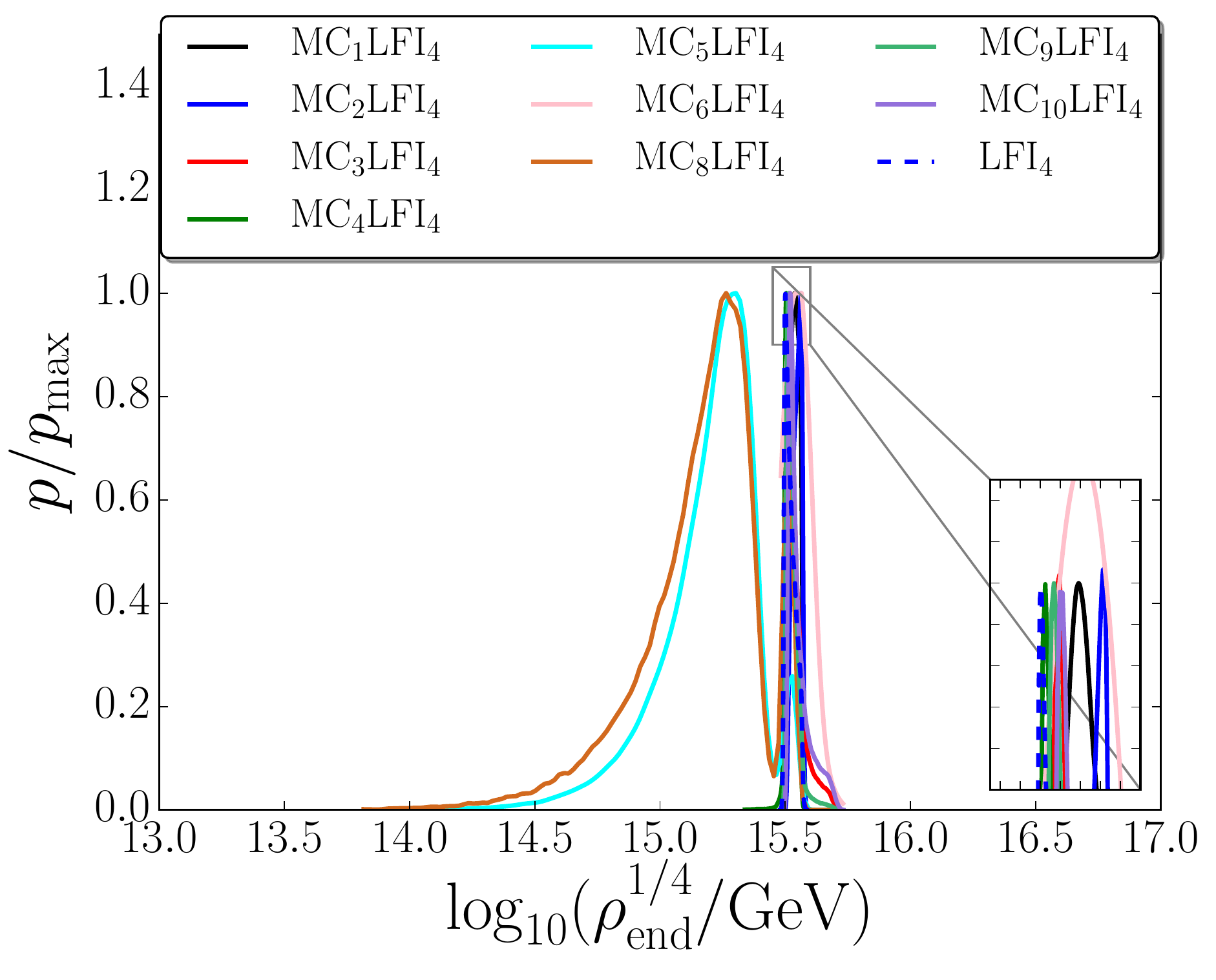}
\includegraphics[width=7cm]{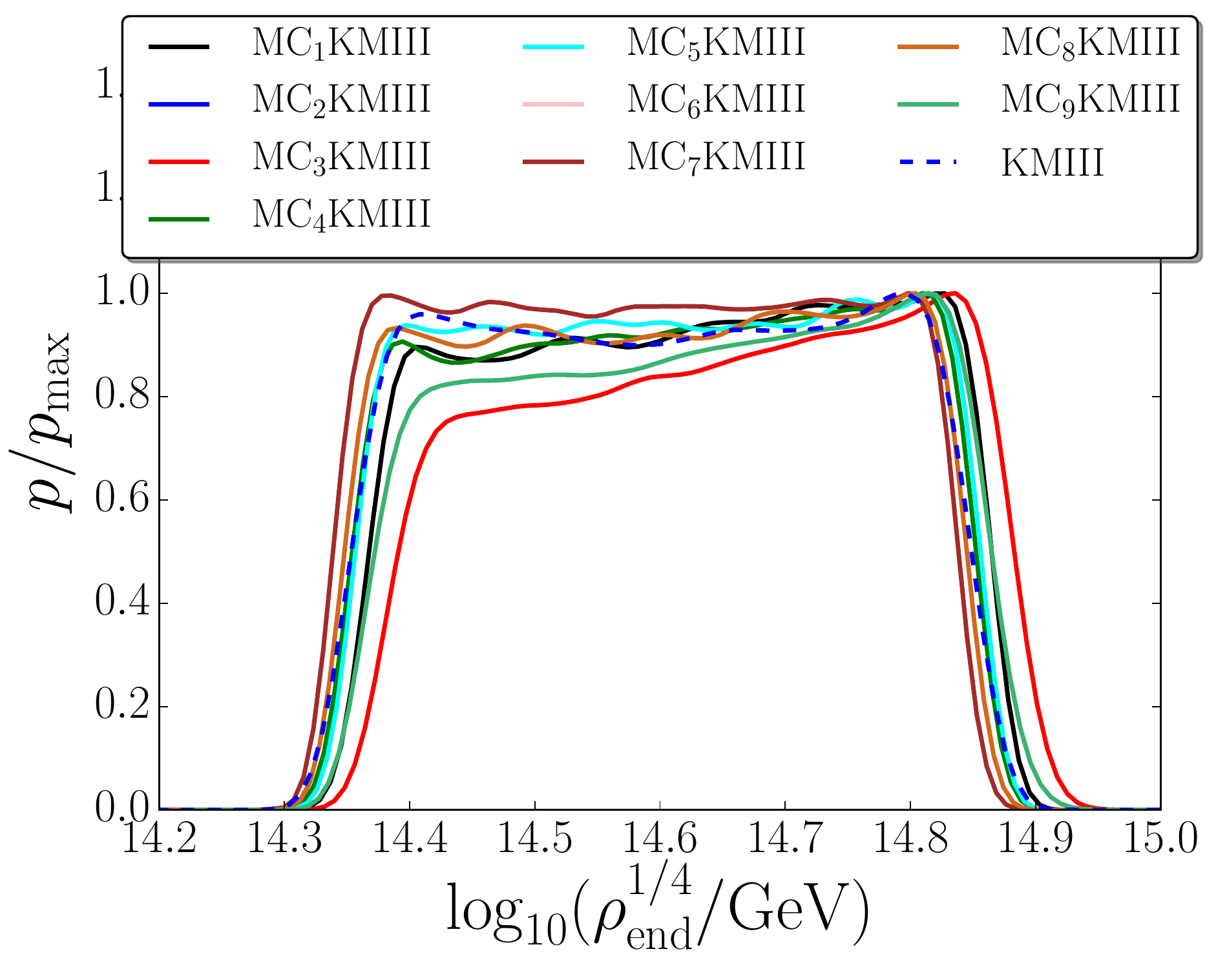}
\includegraphics[width=7cm]{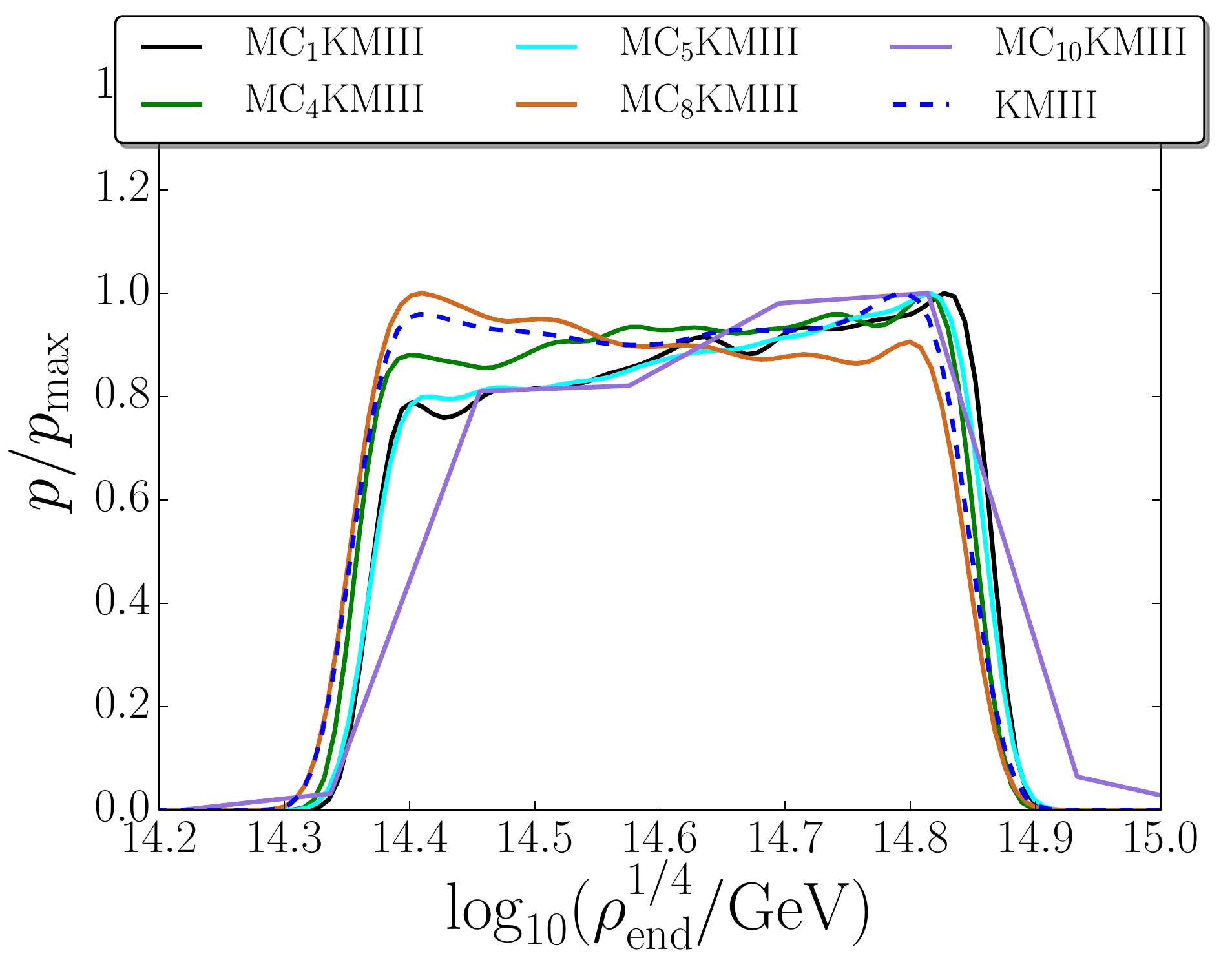}
\end{center}
\caption[Posterior distributions on the energy density at the end of inflation]{}\label{fig:post:rhoend:individual}
\end{figure}
\begin{figure}
\contcaption{Posterior distributions on the energy density at the end of inflation with the plateau potential~(\ref{eq:pot:hi}) of Higgs inflation (top panels), the quartic potential~(\ref{eq:pot:quartic}) (middle panels), and the plateau potential~(\ref{eq:pot:KMIII}) of K\"ahler moduli inflation II (bottom panels). The left panels correspond to the logarithmically flat prior~(\ref{eq:sigmaend:LogPrior}) on $\sigma_\uend$, and the right panels stand for the stochastic prior~(\ref{eq:sigmaend:GaussianPrior}) derived from the equilibrium distribution of a light scalar field in a de Sitter space-time with Hubble scale $H_\uend$. The dashed blue lines correspond to the single-field versions of the models, while the solid coloured lines stand for the $10$ reheating scenarios of \Fig{fig:cases} when an extra light scalar field is present.}
\end{figure}
\newpage 
\subsection{\textsf{Reheating Temperature}}
\label{sec:app:Treh:individual}
\begin{figure}
\figpilogsto
\begin{center}
\includegraphics[width=7cm]{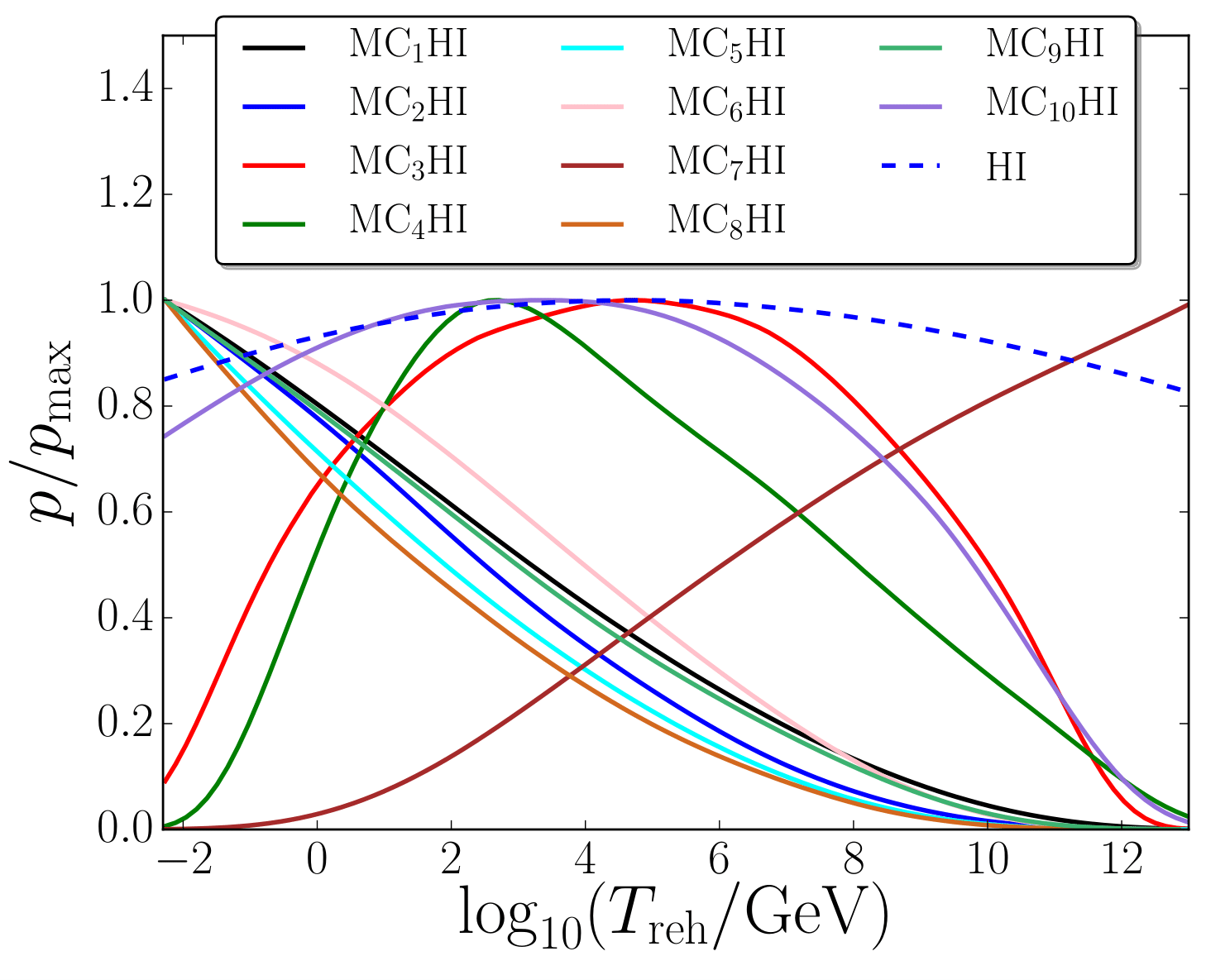}
\includegraphics[width=7cm]{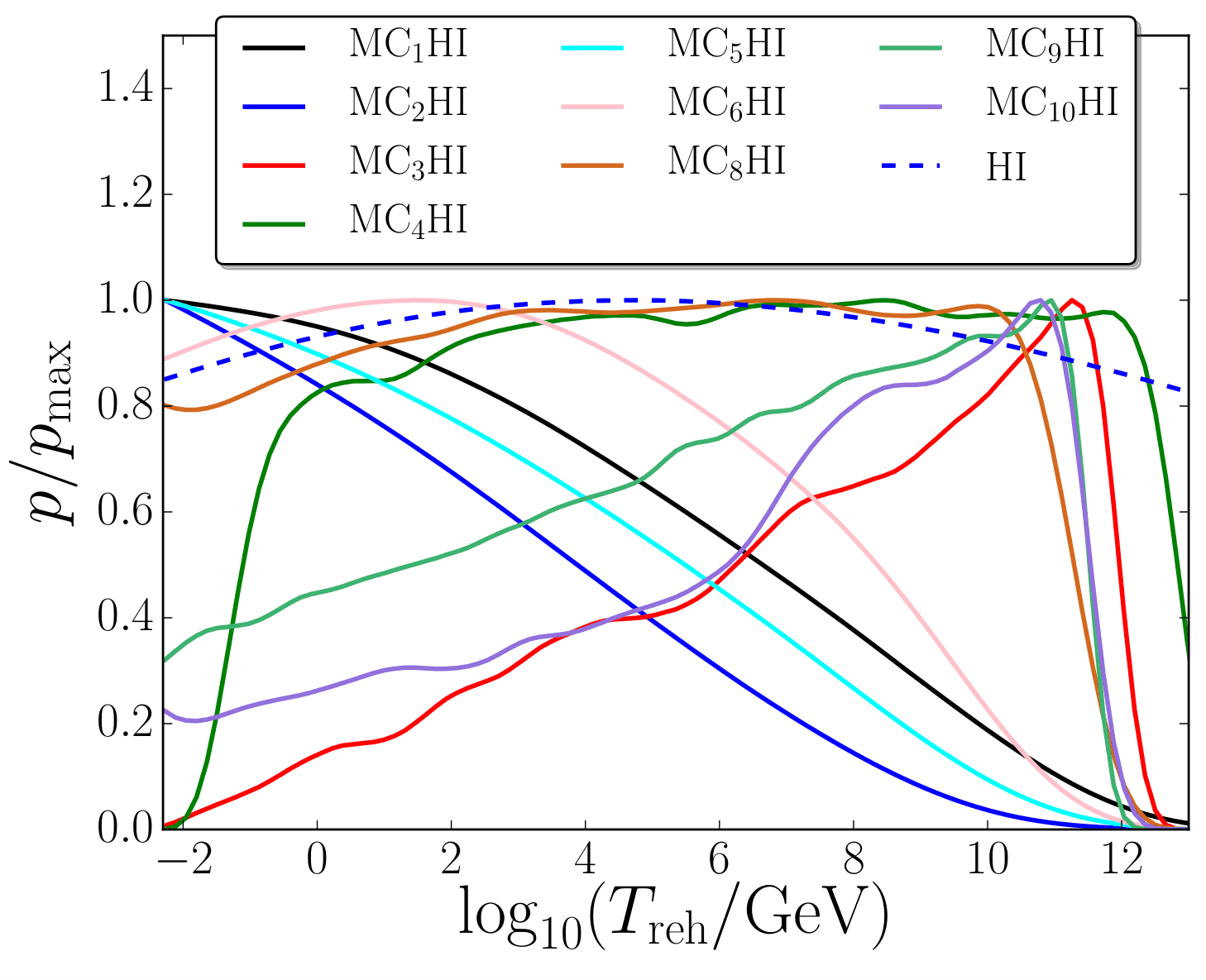}
\includegraphics[width=7cm]{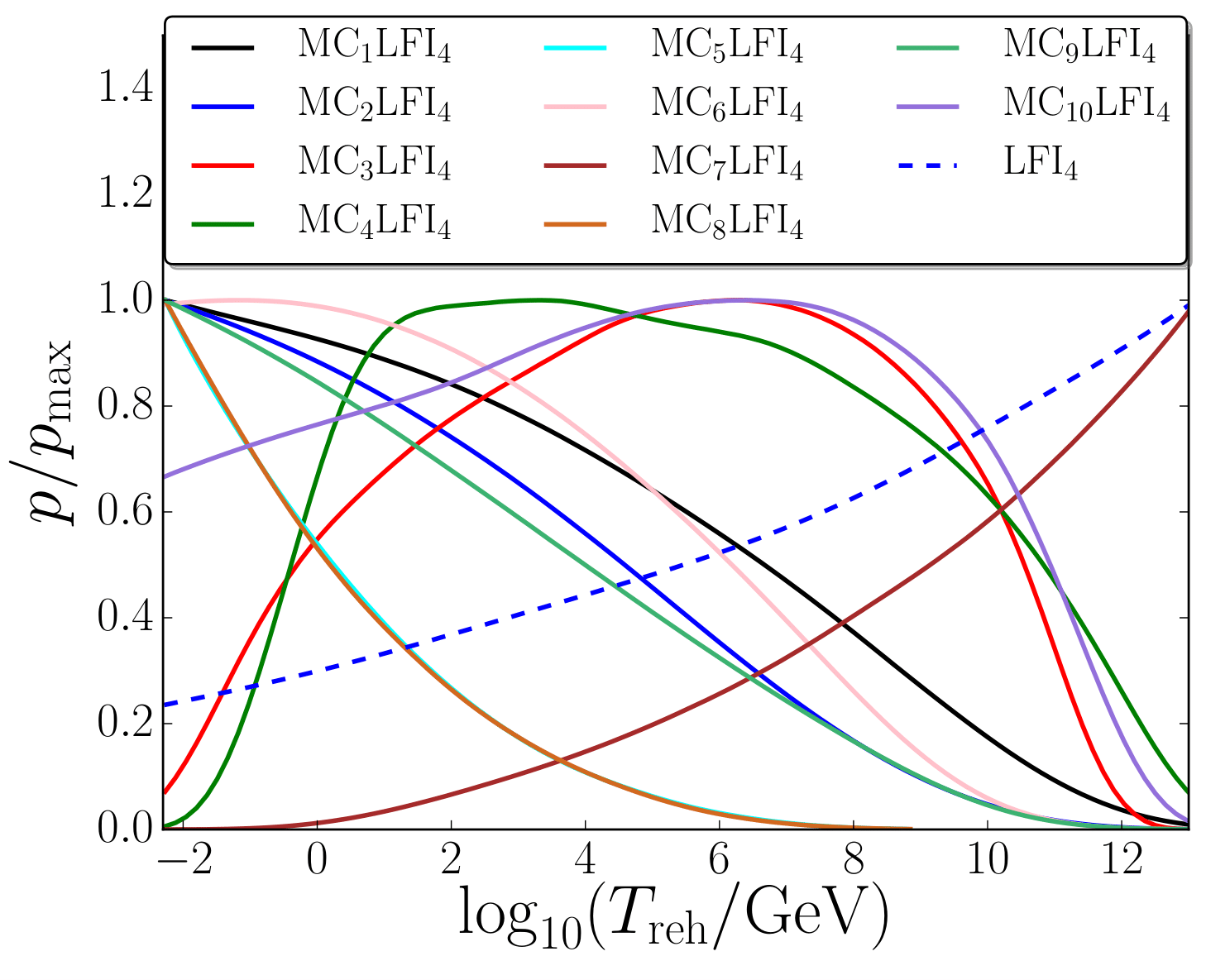}
\includegraphics[width=7cm]{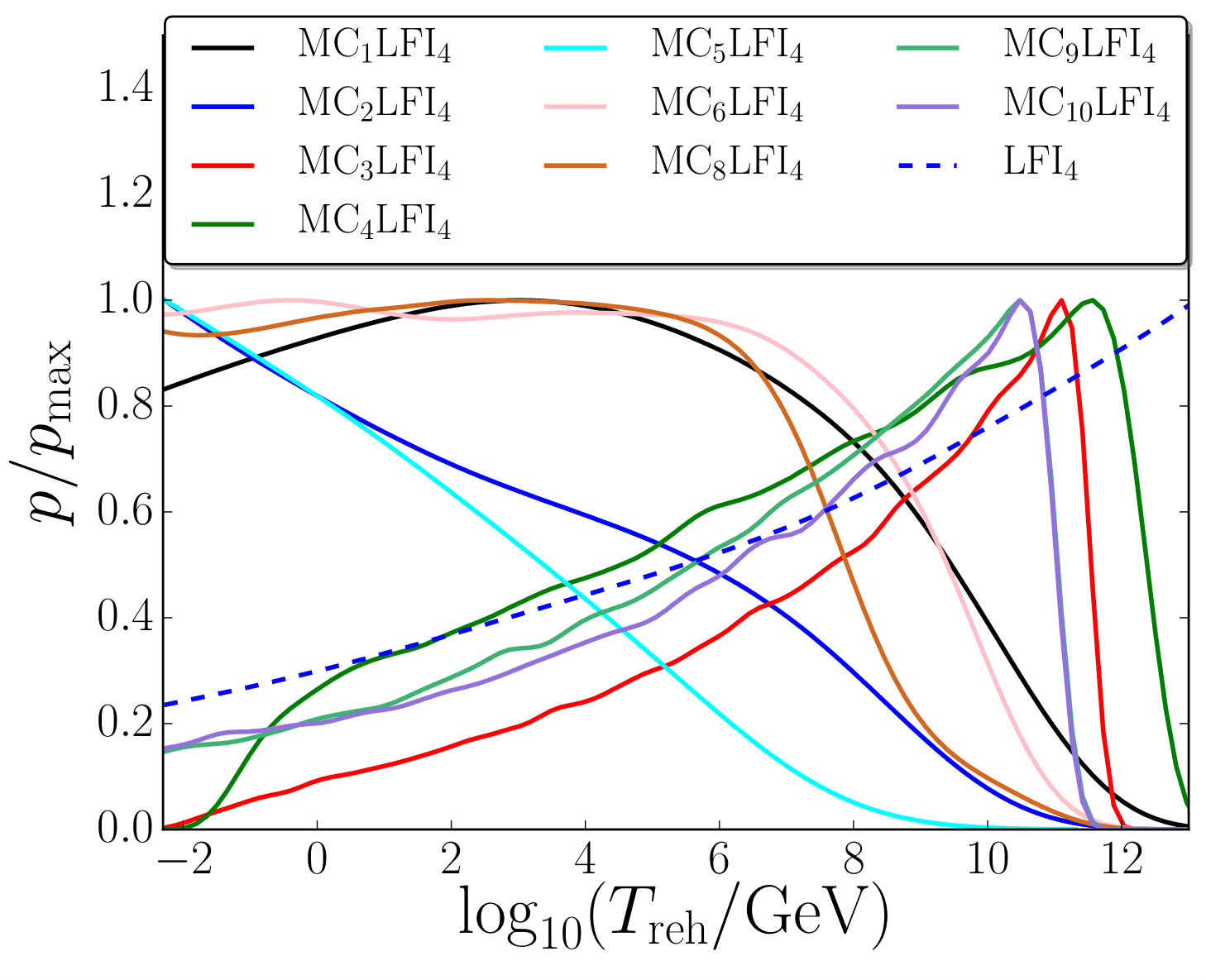}
\includegraphics[width=7.2cm]{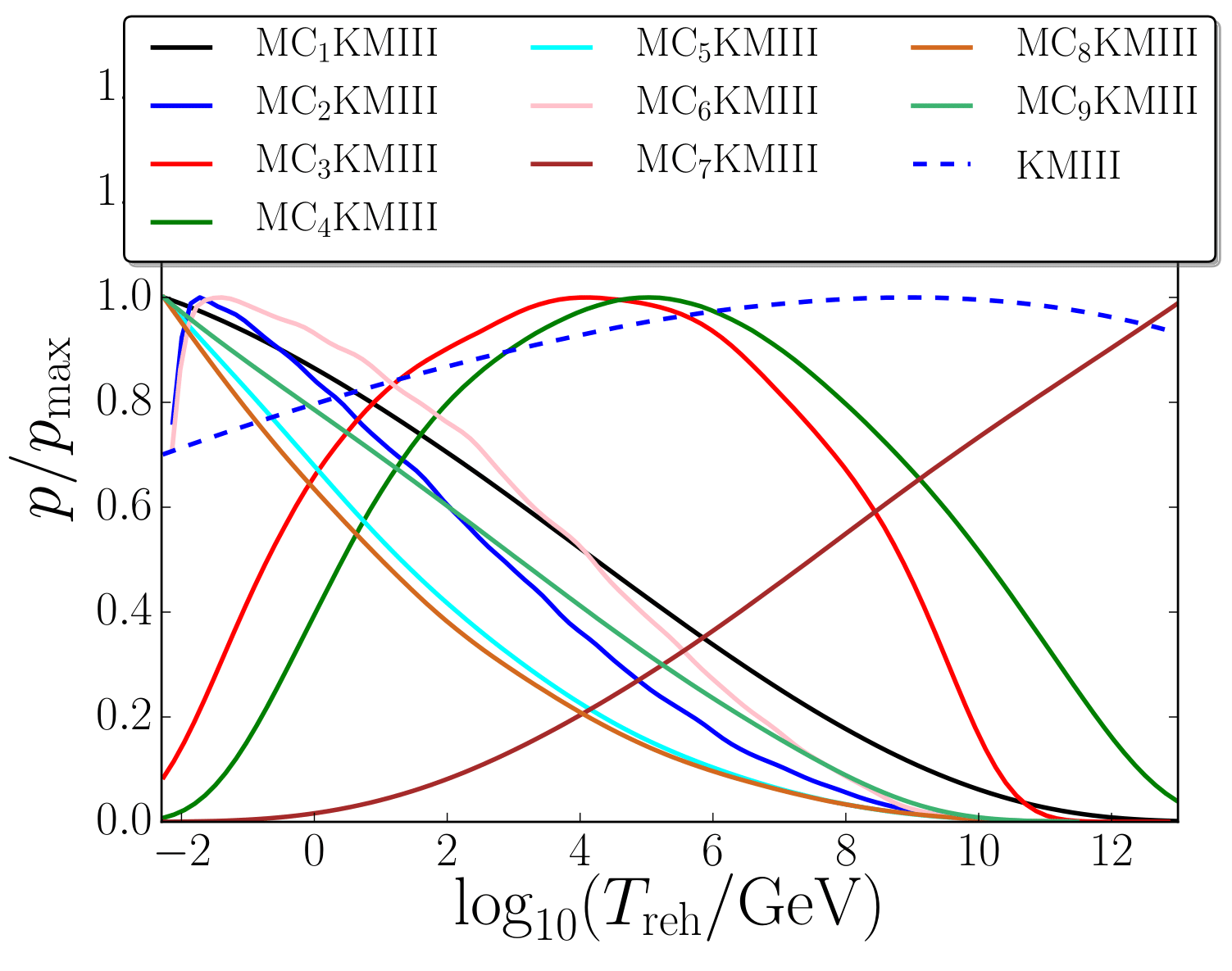}
\includegraphics[width=7.2cm]{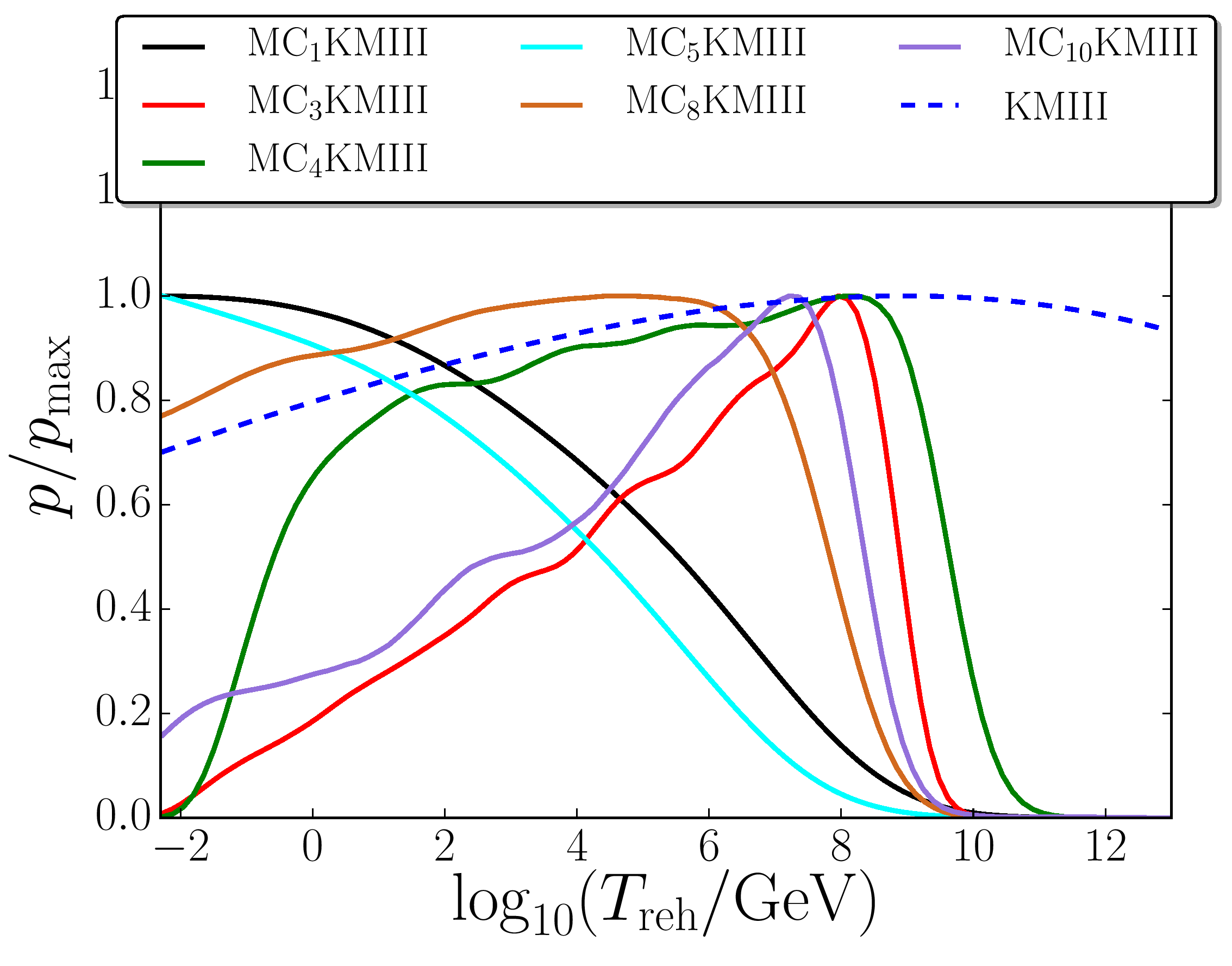}
\caption[Posterior distributions on the reheating temperature]{}
\label{fig:post:Treh:individual}
\end{center}
\end{figure}
\begin{figure}
\contcaption{Posterior distributions on the reheating temperature $T_\ureh$ with the plateau potential~(\ref{eq:pot:hi}) of Higgs inflation (top panels), the quartic potential~(\ref{eq:pot:quartic}) (middle panels), and the plateau potential~(\ref{eq:pot:KMIII}) of K\"ahler moduli inflation II (bottom panels). The left panels correspond to the logarithmically flat prior~(\ref{eq:sigmaend:LogPrior}) on $\sigma_\uend$, and the right panels stand for the stochastic prior~(\ref{eq:sigmaend:GaussianPrior}) derived from the equilibrium distribution of a light scalar field in a de Sitter space-time with Hubble scale $H_\uend$. The dashed blue lines correspond to the single-field versions of the models, while the solid coloured lines stand for the $10$ reheating scenarios of \Fig{fig:cases} when an extra light scalar field is present.}
\end{figure}

\newpage
\subsection{\textsf{Early reheating temperature}}
\label{sec:app:Tereh:individual}
\begin{figure}
\figpilogsto
\begin{center}
\includegraphics[width=7cm]{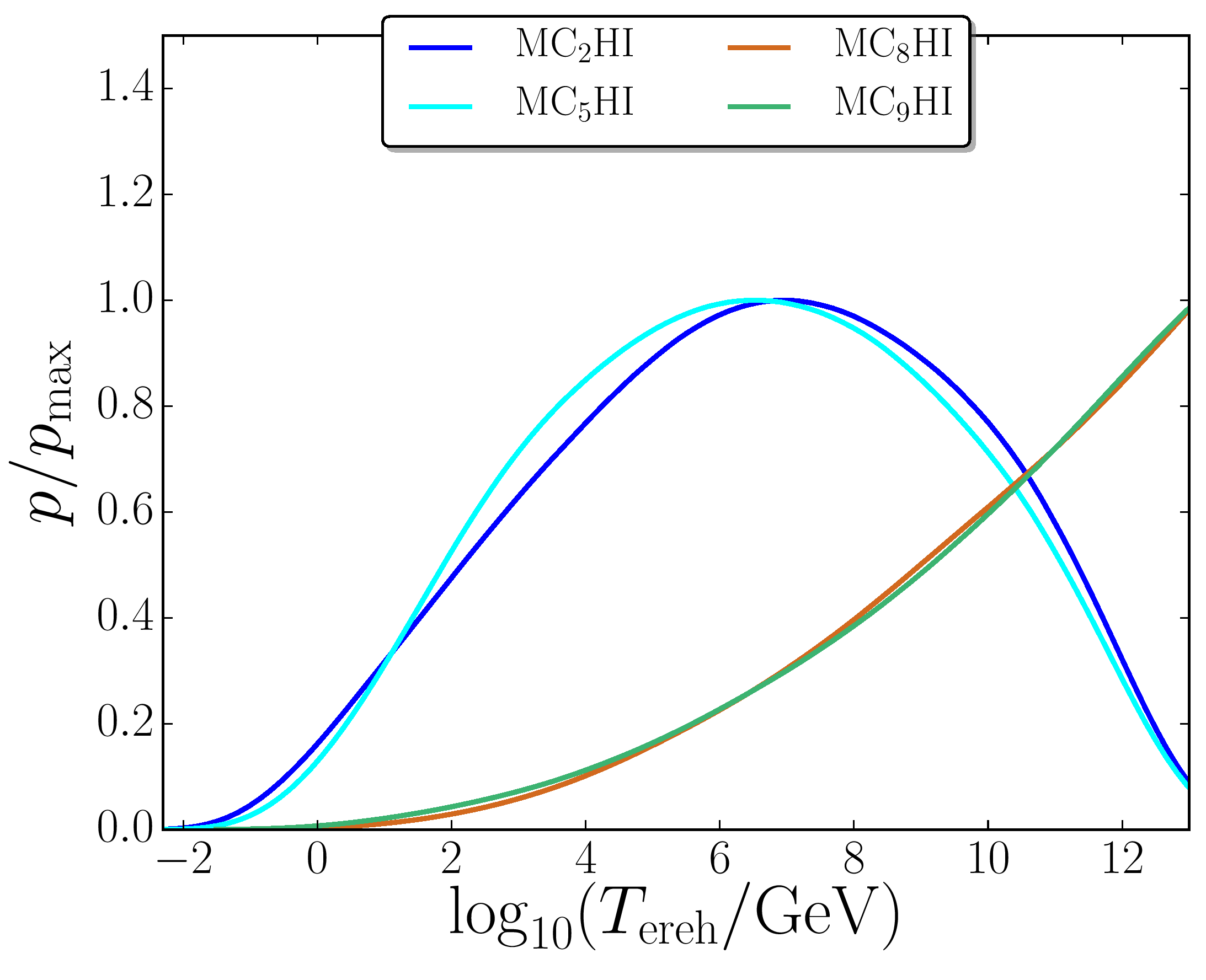}
\includegraphics[width=7cm]{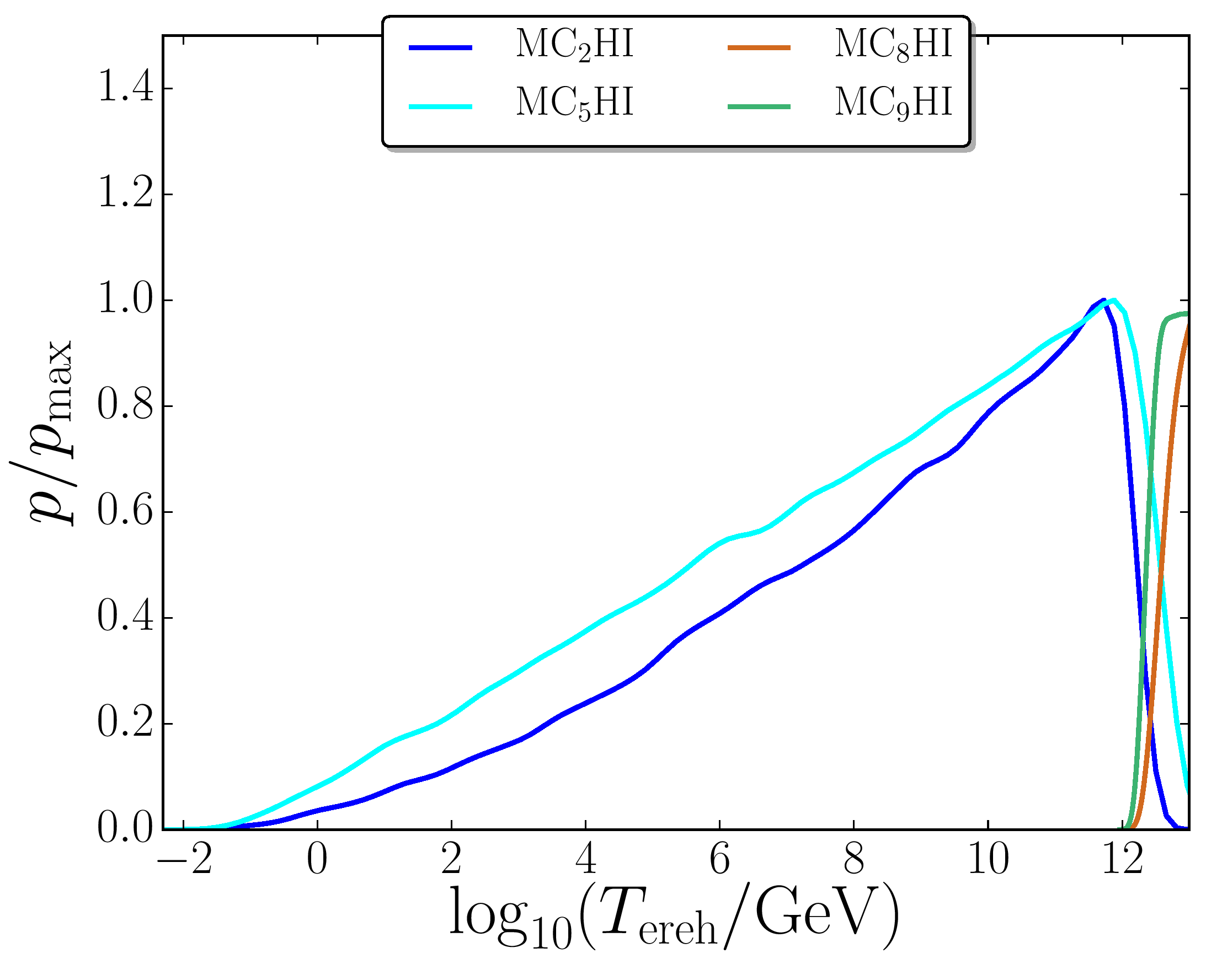}
\includegraphics[width=7cm]{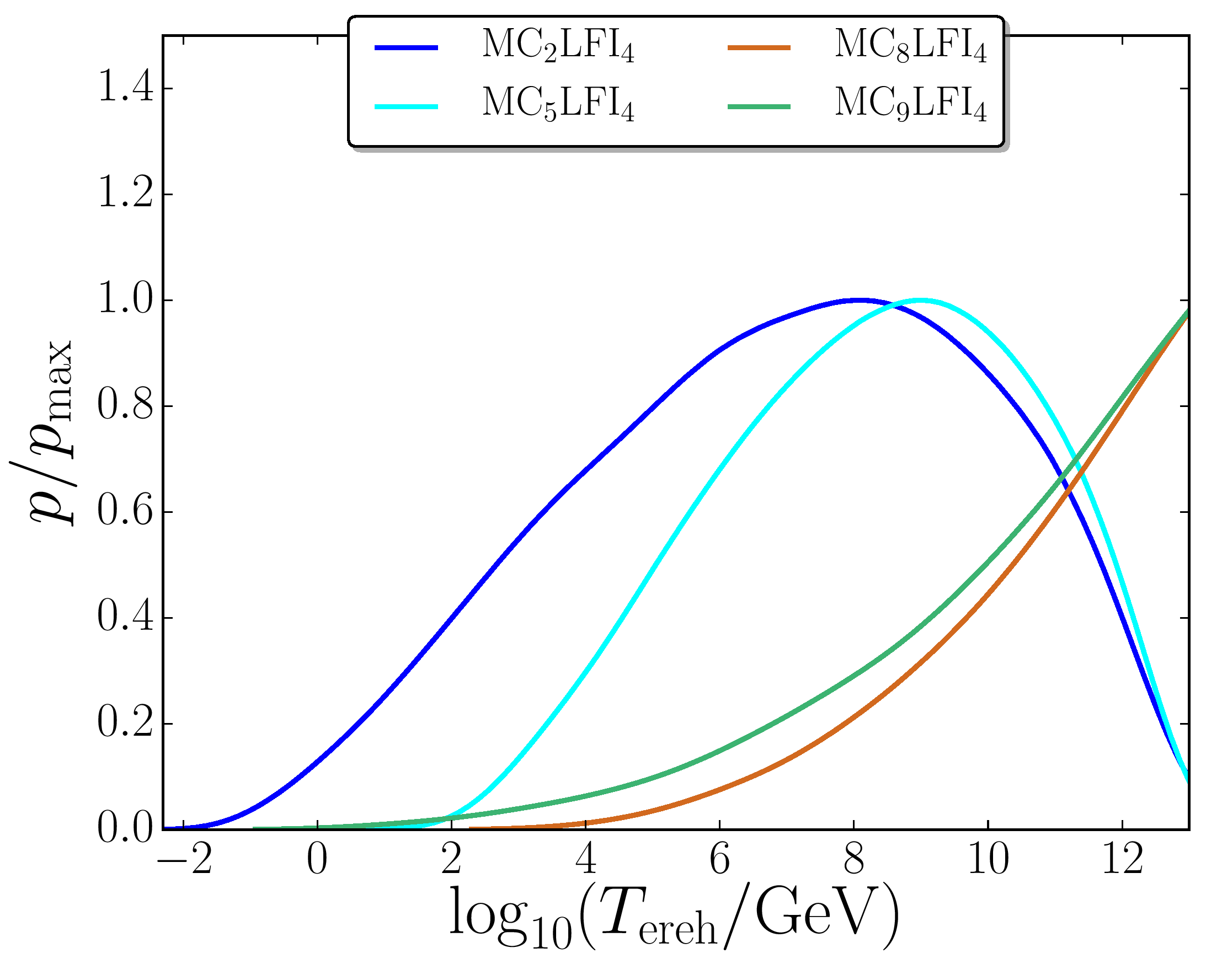}
\includegraphics[width=7cm]{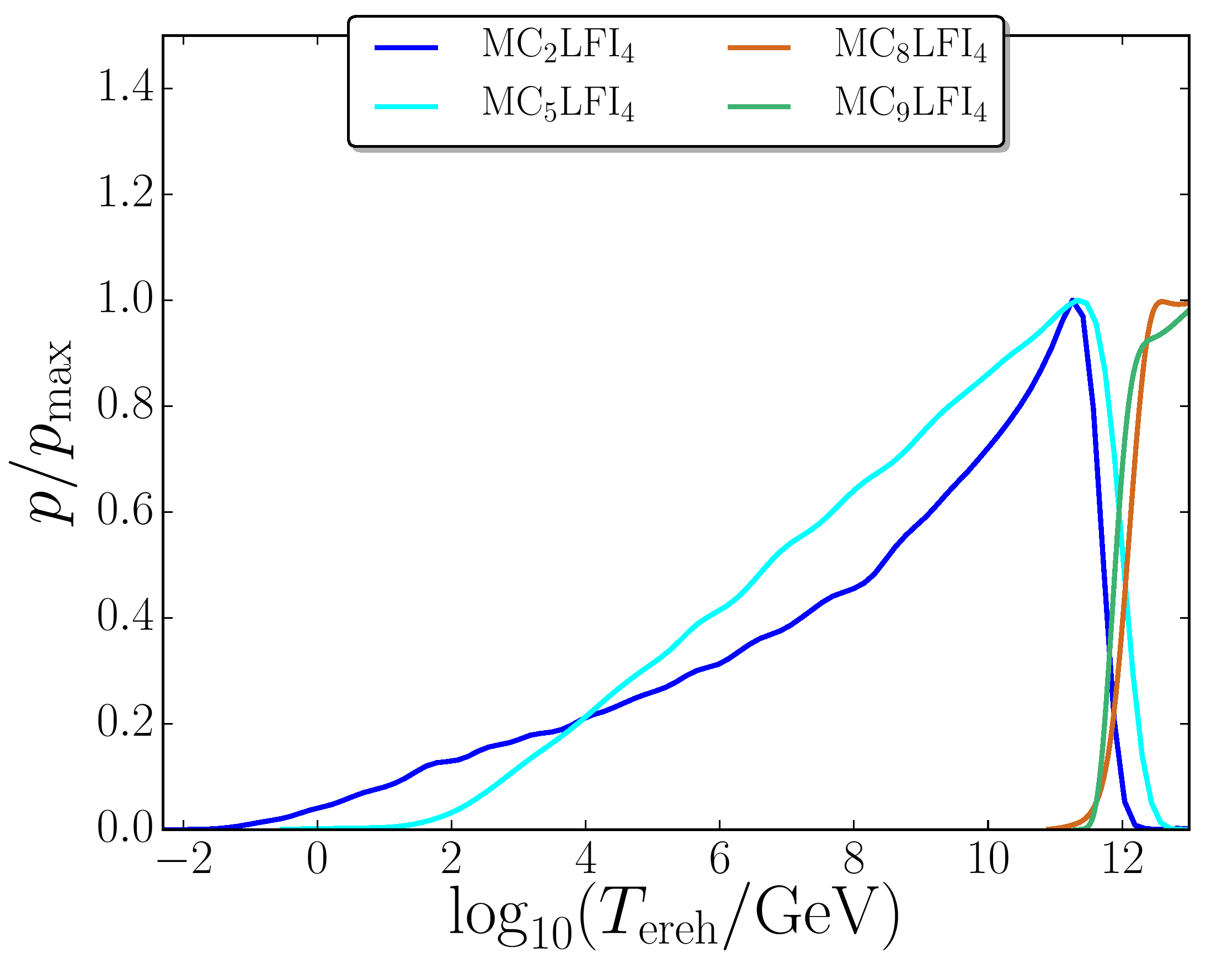}
\includegraphics[width=7cm]{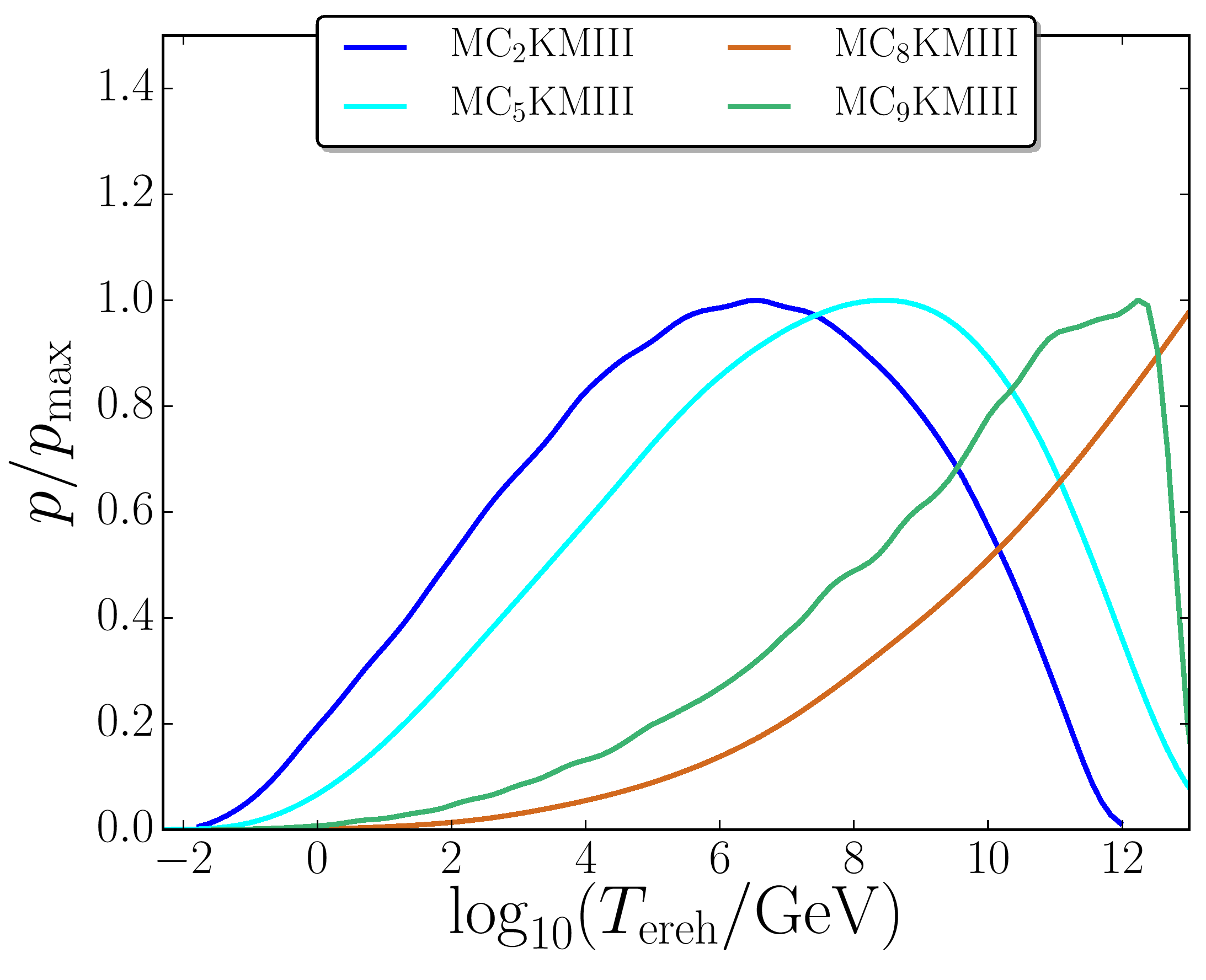}
\includegraphics[width=7cm]{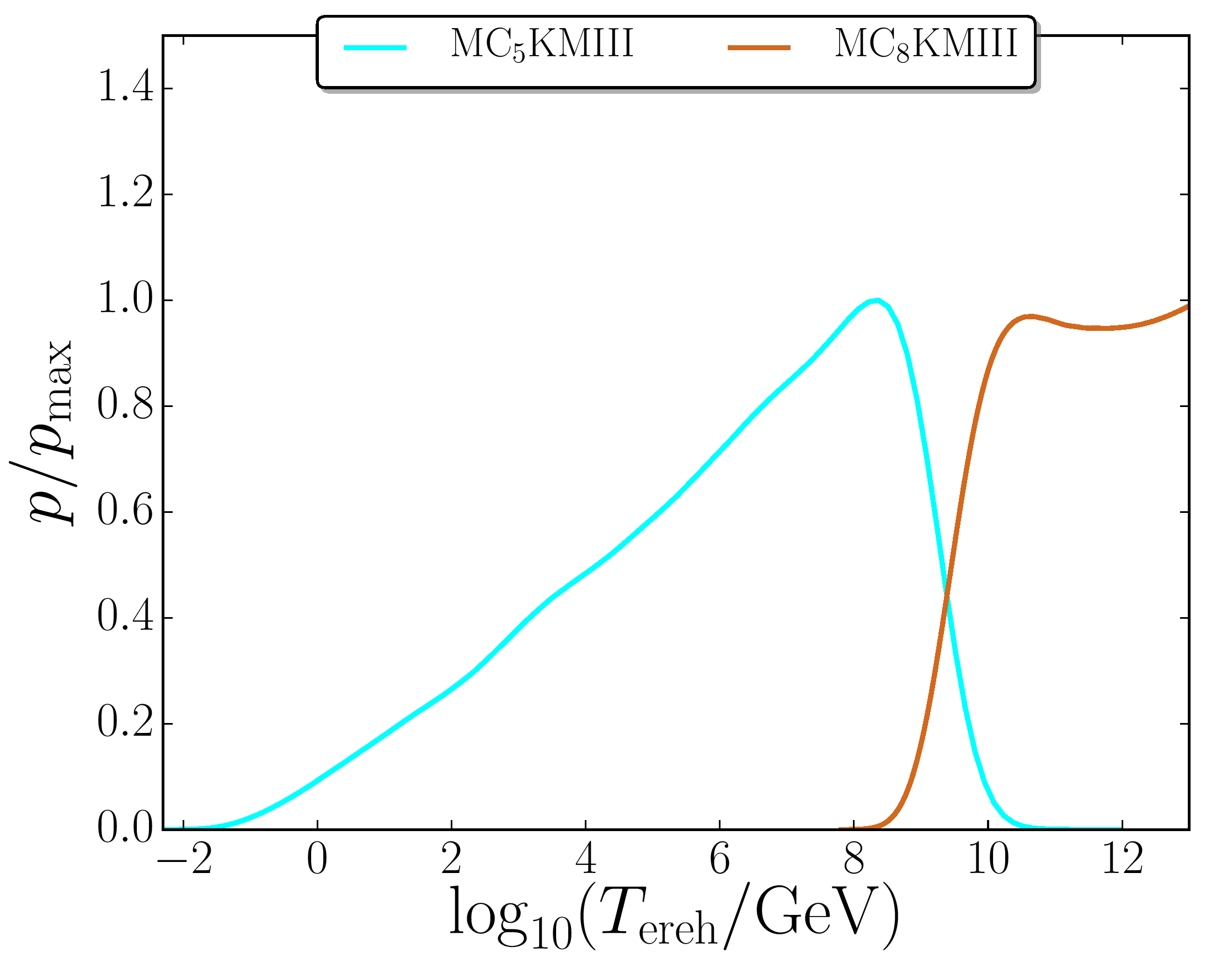}
\caption[Posterior distributions on the early reheating temperature]{}
\label{fig:post:Tereh:individual}
\end{center}
\end{figure}
\begin{figure}
\contcaption{Posterior distributions on the early reheating temperature $T_\uereh$ with the plateau potential~(\ref{eq:pot:hi}) of Higgs inflation (top panels), the quartic potential~(\ref{eq:pot:quartic}) (middle panels), and the plateau potential~(\ref{eq:pot:KMIII}) of K\"ahler moduli inflation II (bottom panels). The left panels correspond to the logarithmically flat prior~(\ref{eq:sigmaend:LogPrior}) on $\sigma_\uend$, and the right panels stand for the stochastic prior~(\ref{eq:sigmaend:GaussianPrior}) derived from the equilibrium distribution of a light scalar field in a de Sitter space-time with Hubble scale $H_\uend$. The dashed blue lines correspond to the single-field versions of the models, while the solid coloured lines stand for the $10$ reheating scenarios of \Fig{fig:cases} when an extra light scalar field is present.}
\end{figure}

\newpage
\section{\textsf{Information density}}
\label{Sec:DKLDensity} 
%
%
\subsection{\textsf{Energy density at the end of inflation}}
\label{sec:app:DKL:rhoend}
\begin{figure}
\figpilogsto
\begin{center}
\includegraphics[width=7cm]{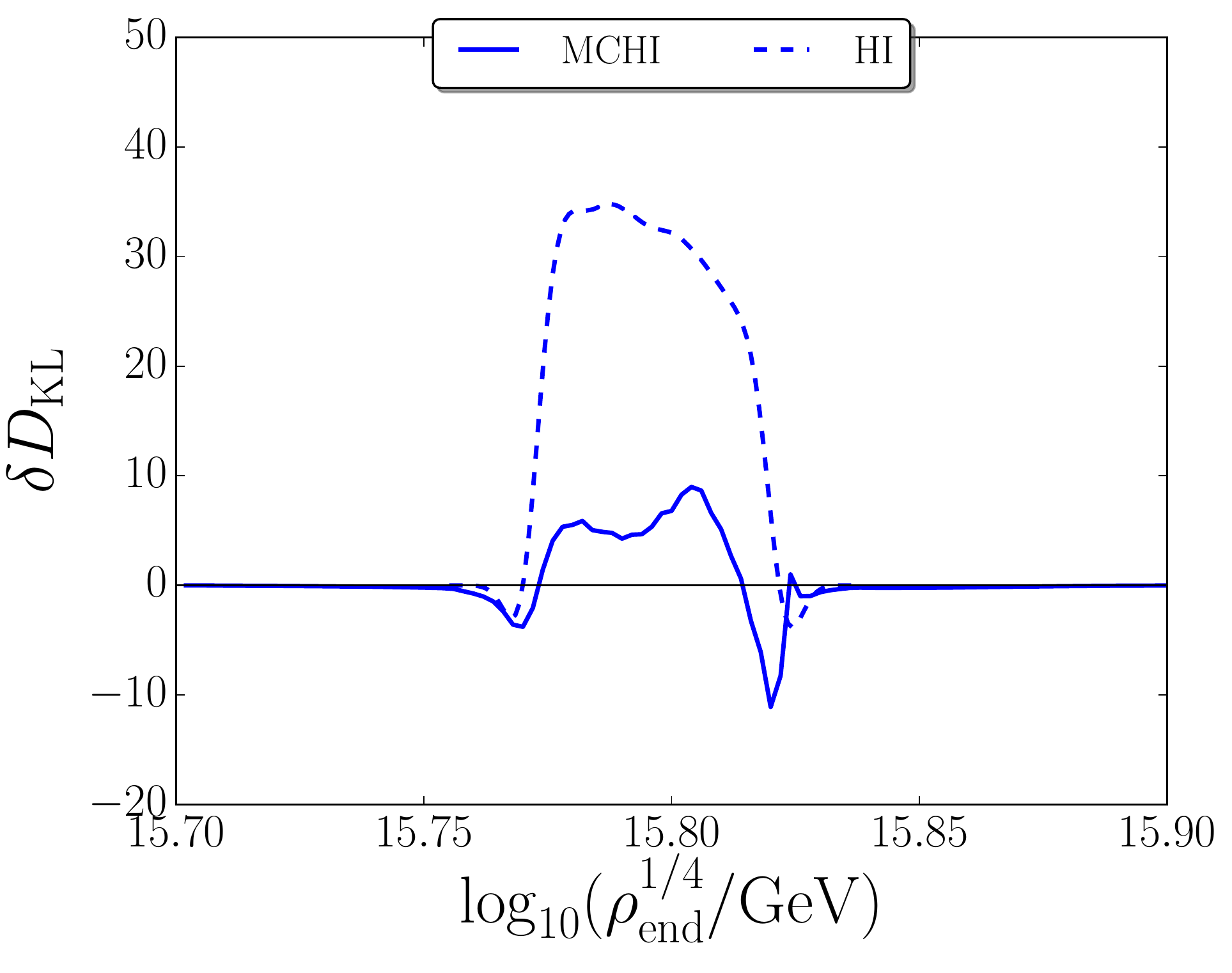}
\includegraphics[width=7cm]{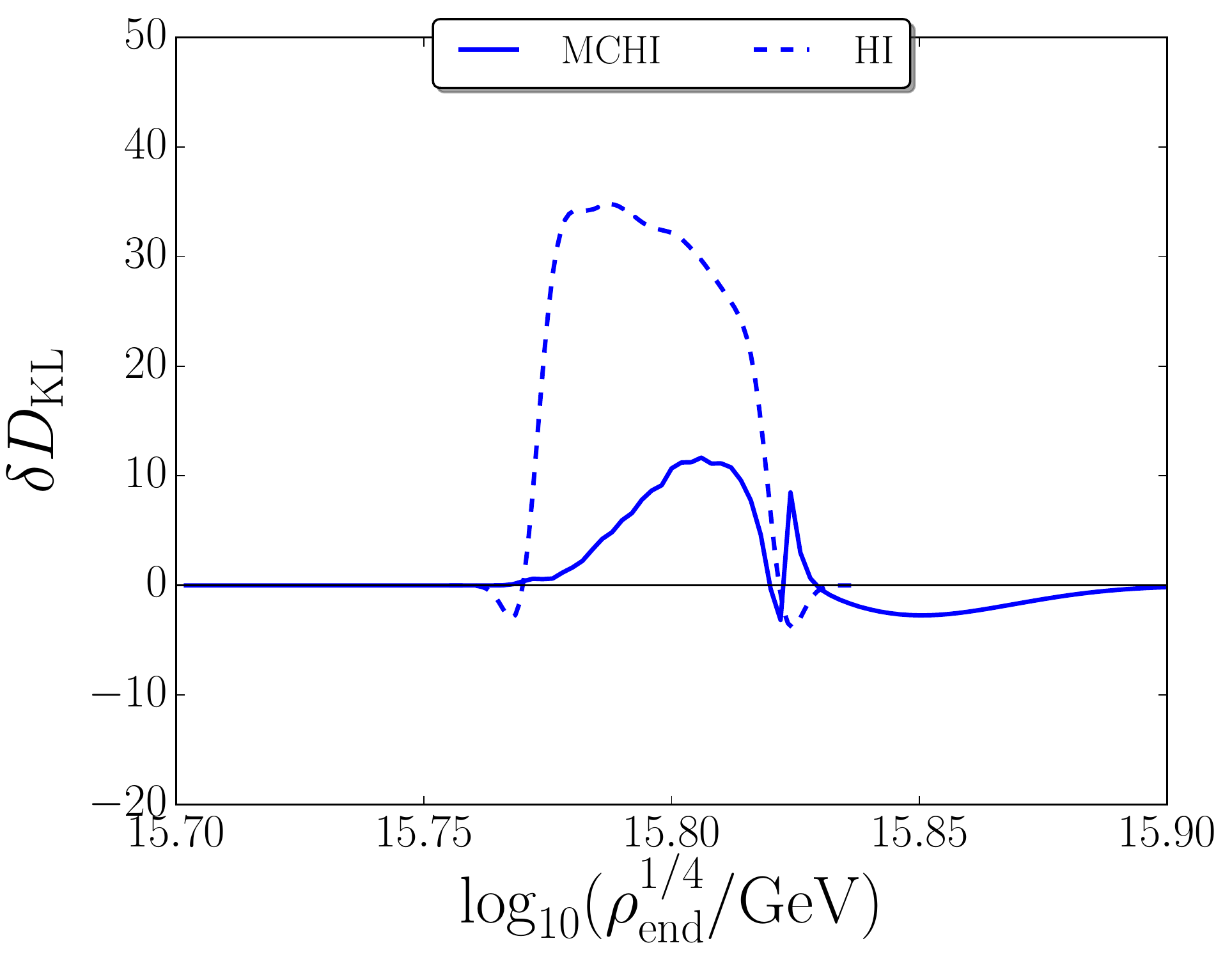}
\includegraphics[width=7cm]{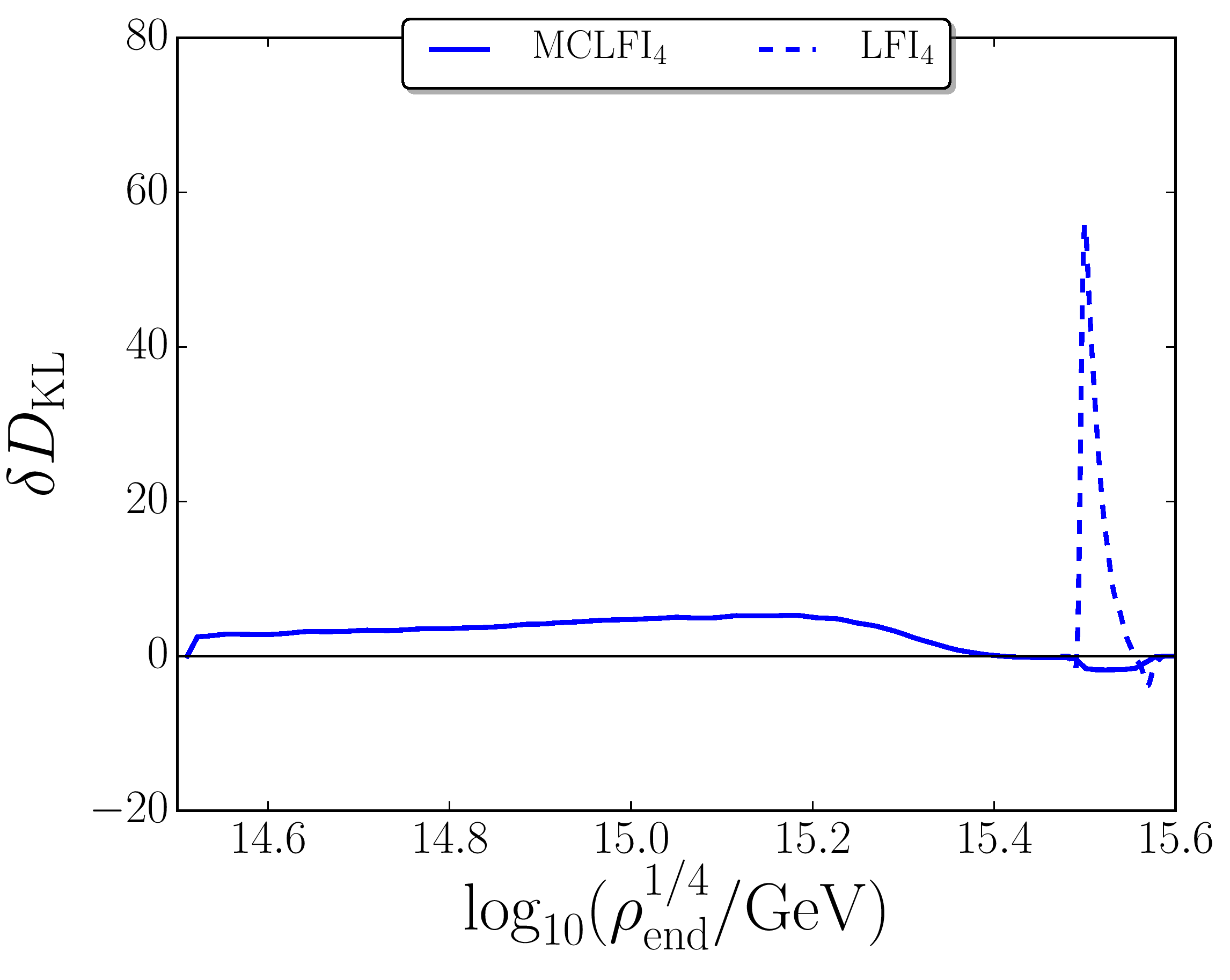}
\includegraphics[width=7cm]{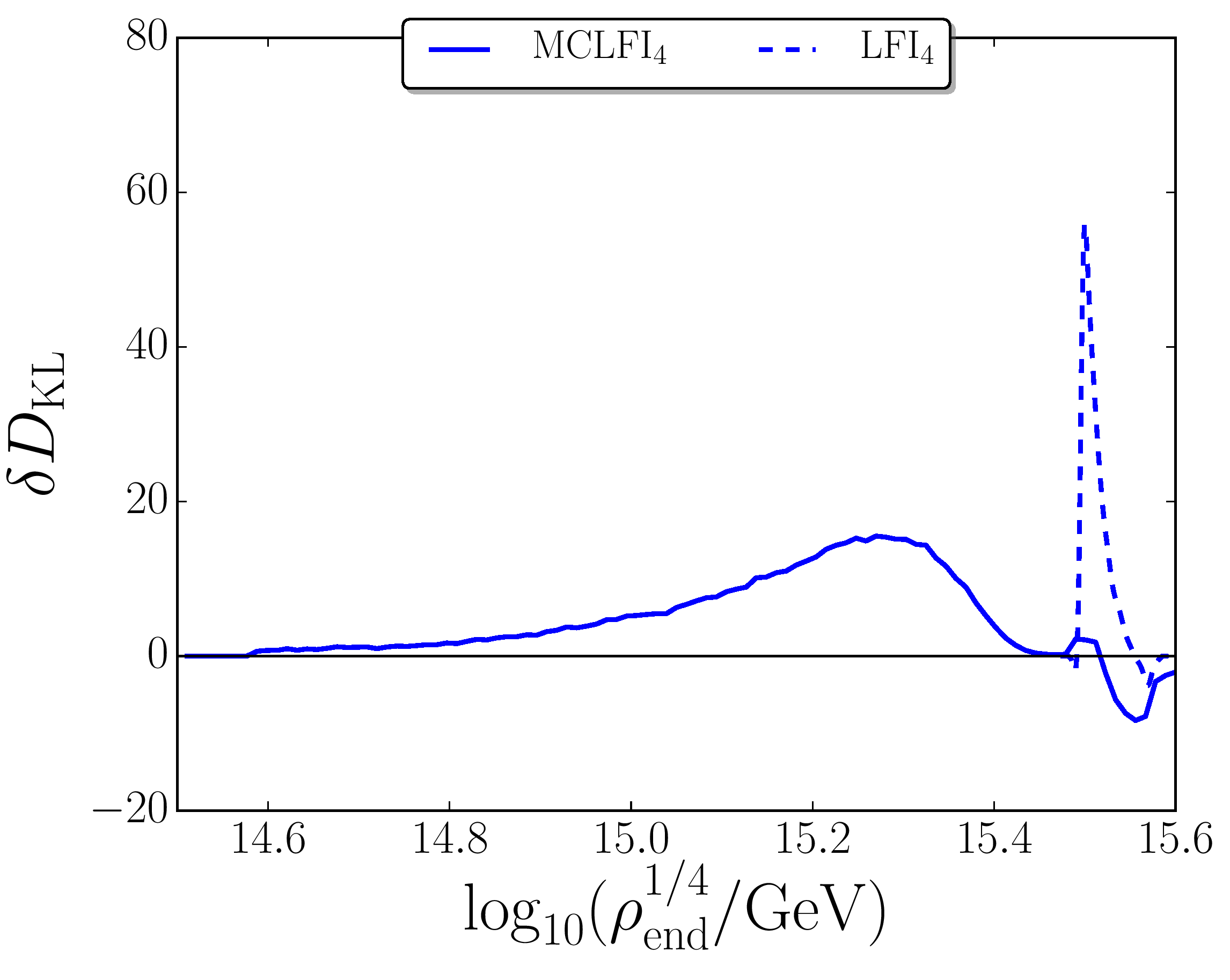}
\includegraphics[width=7cm]{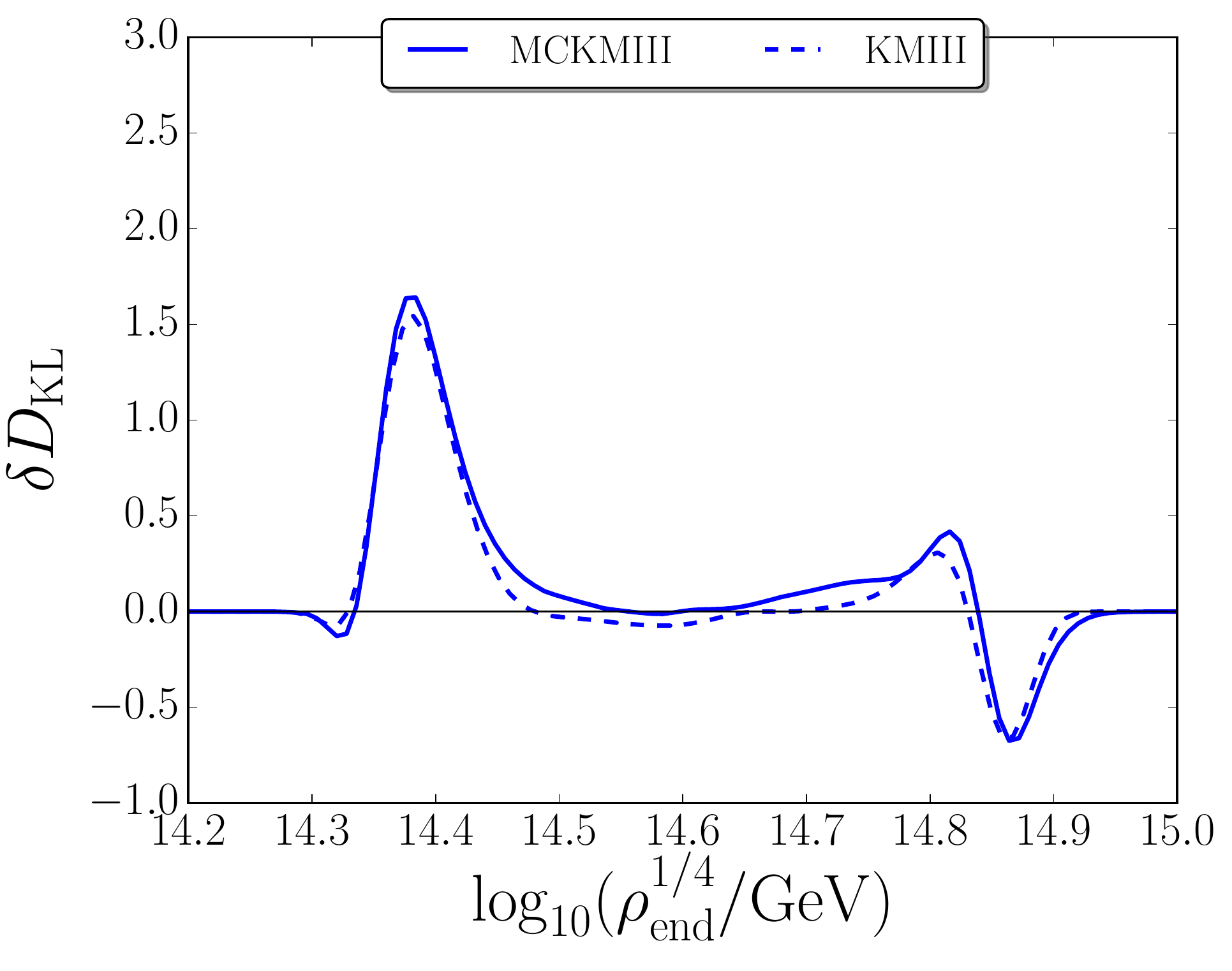}
\includegraphics[width=7cm]{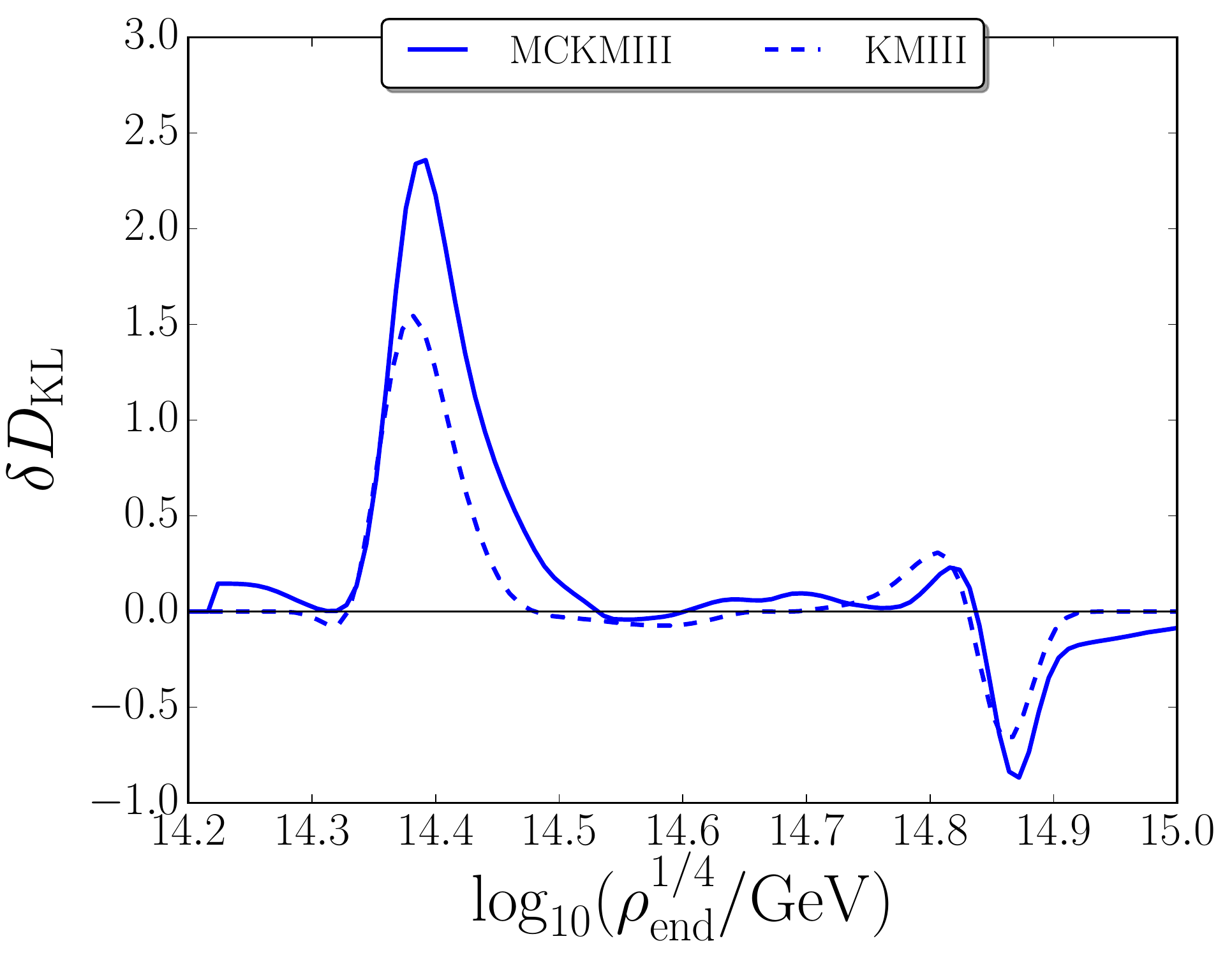}
\caption[Information density over the averaged energy density]{}
\label{fig:DKL:rhoend:averaged}
\end{center}
\end{figure}
\begin{figure}
\contcaption{Information density on $\rho_\uend$ for Higgs inflation (top panels), quartic inflation (middle panels) and K\"ahler moduli inflation II (bottom panels). The left panels correspond to the logarithmically flat prior~(\ref{eq:sigmaend:LogPrior}) on $\sigma_\uend$, and the right panels stand for the stochastic prior~(\ref{eq:sigmaend:GaussianPrior}). The dashed lines correspond to the single-field versions of the models, while the solid lines are derived from the averaged distributions over all $10$ reheating scenarios. 
}
\end{figure}

\newpage
\begin{figure}
\figpilogsto
\begin{center}
\includegraphics[width=7cm]{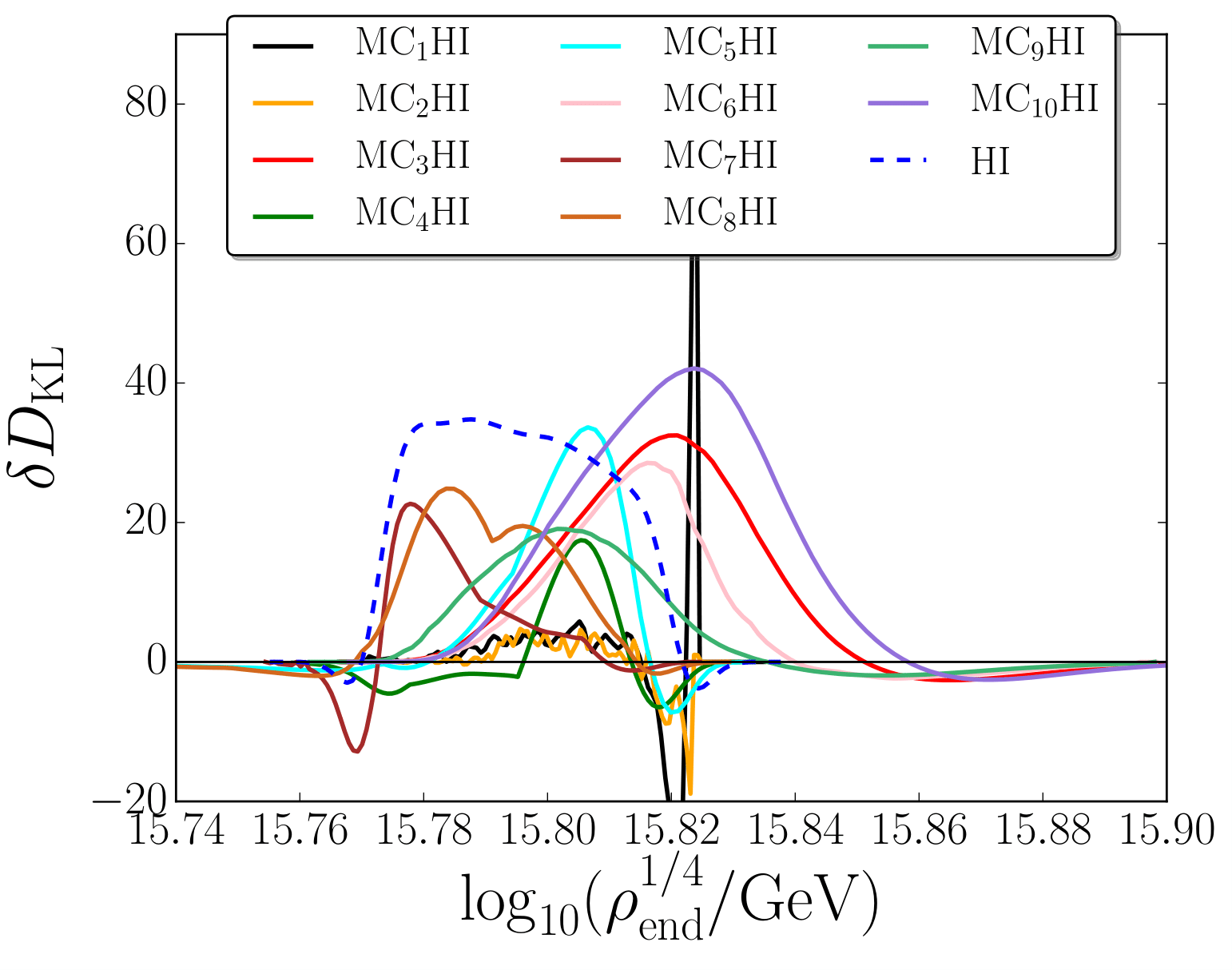}
\includegraphics[width=7cm]{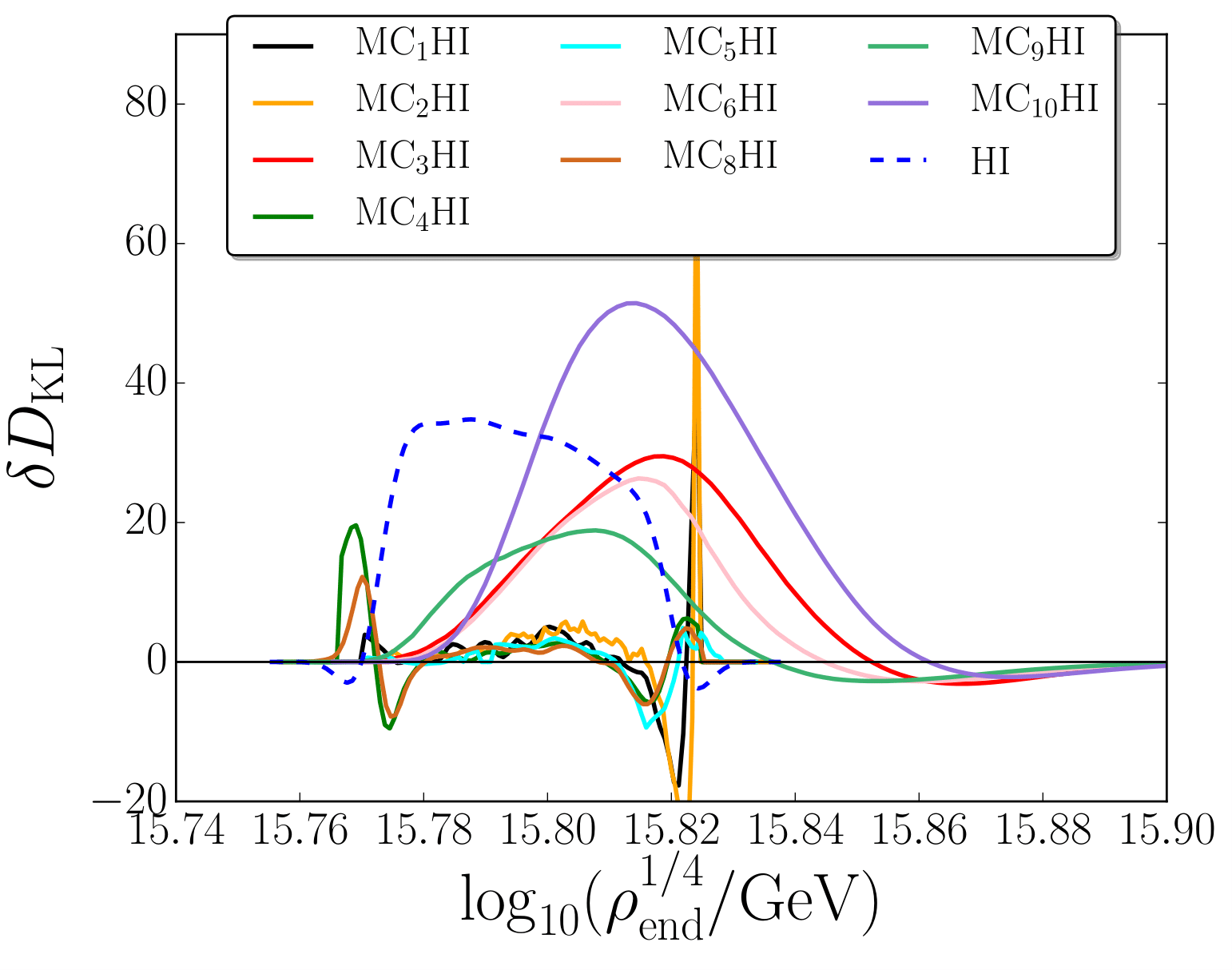}
\includegraphics[width=7cm]{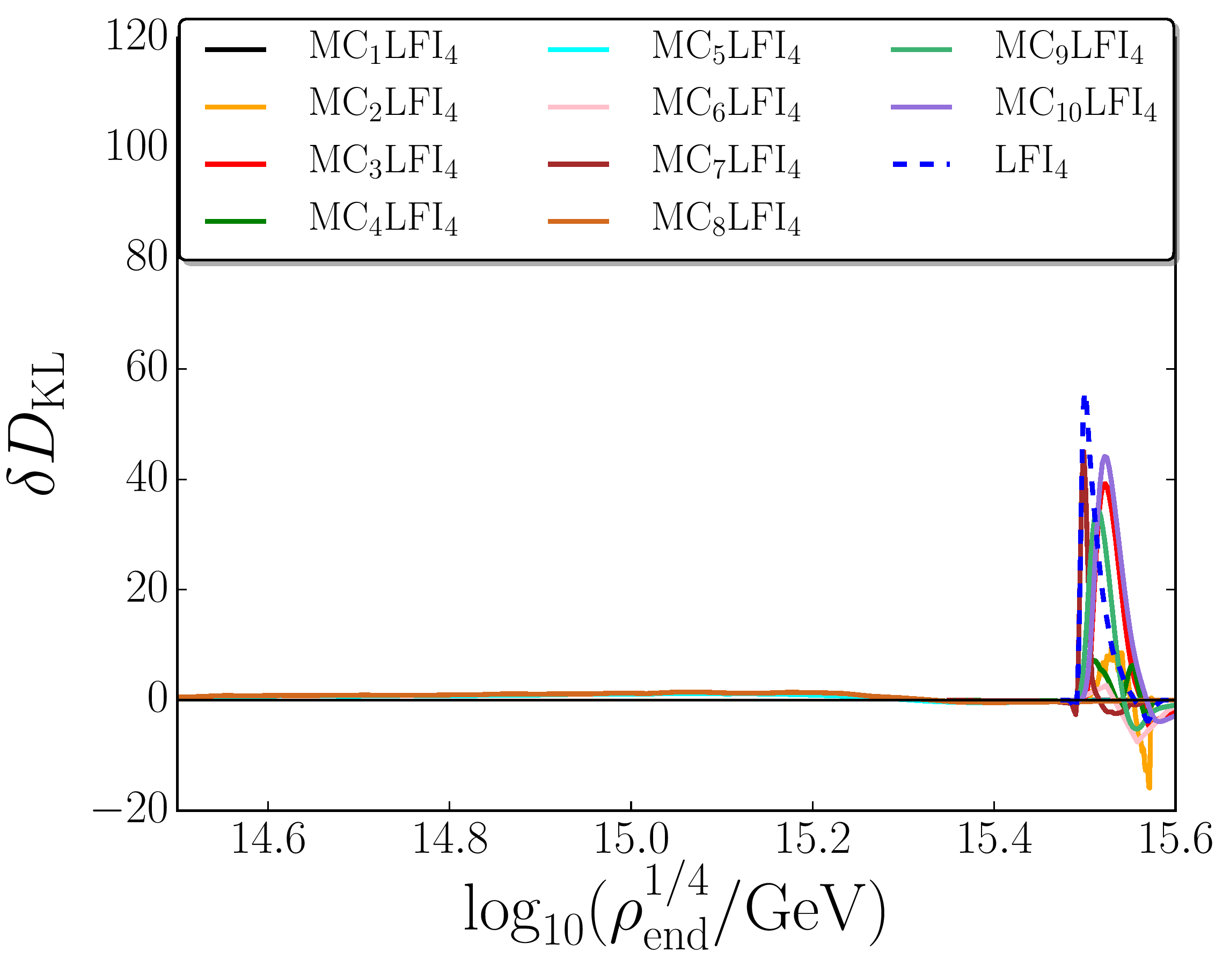}
\includegraphics[width=7cm]{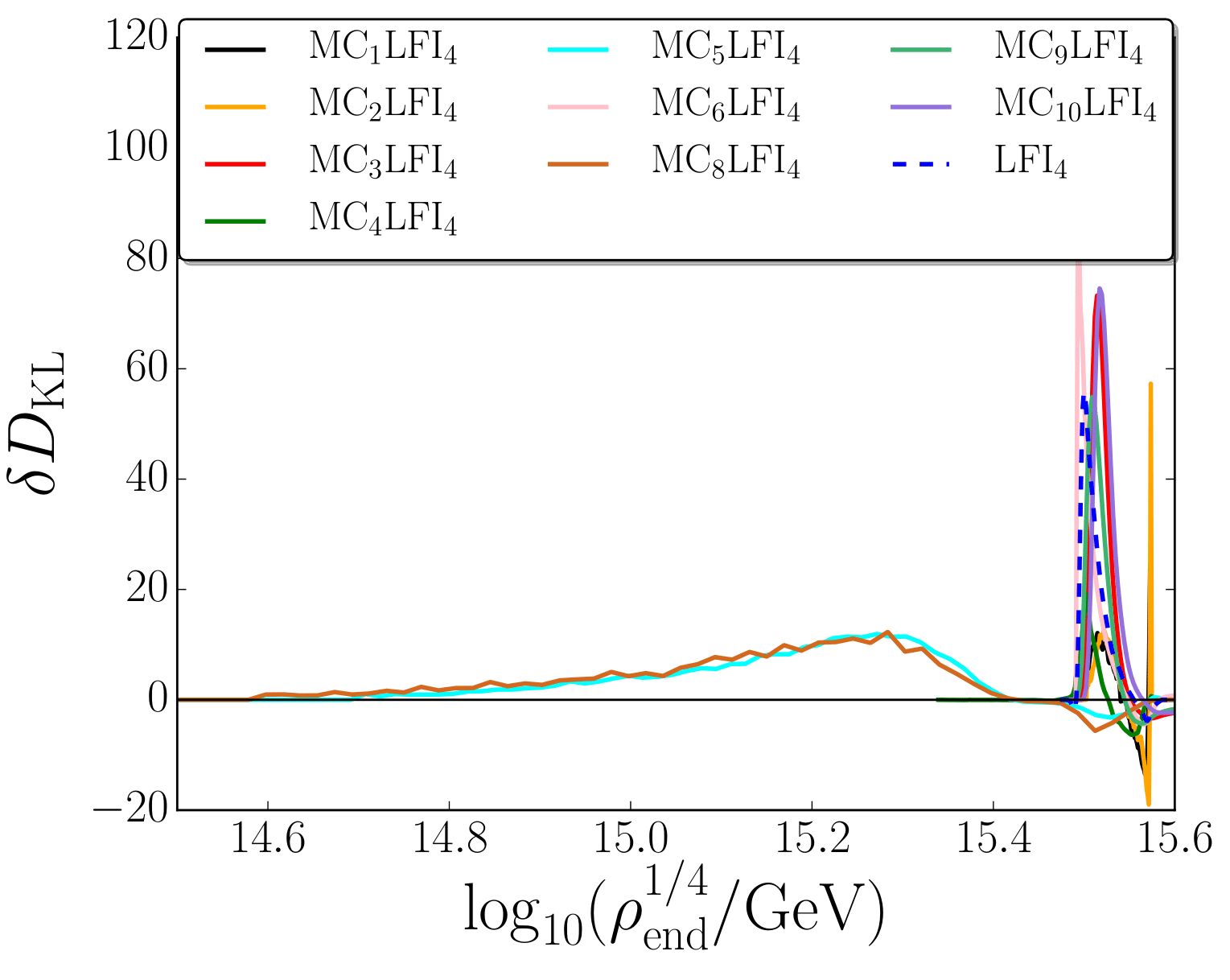}
\includegraphics[width=7.1cm]{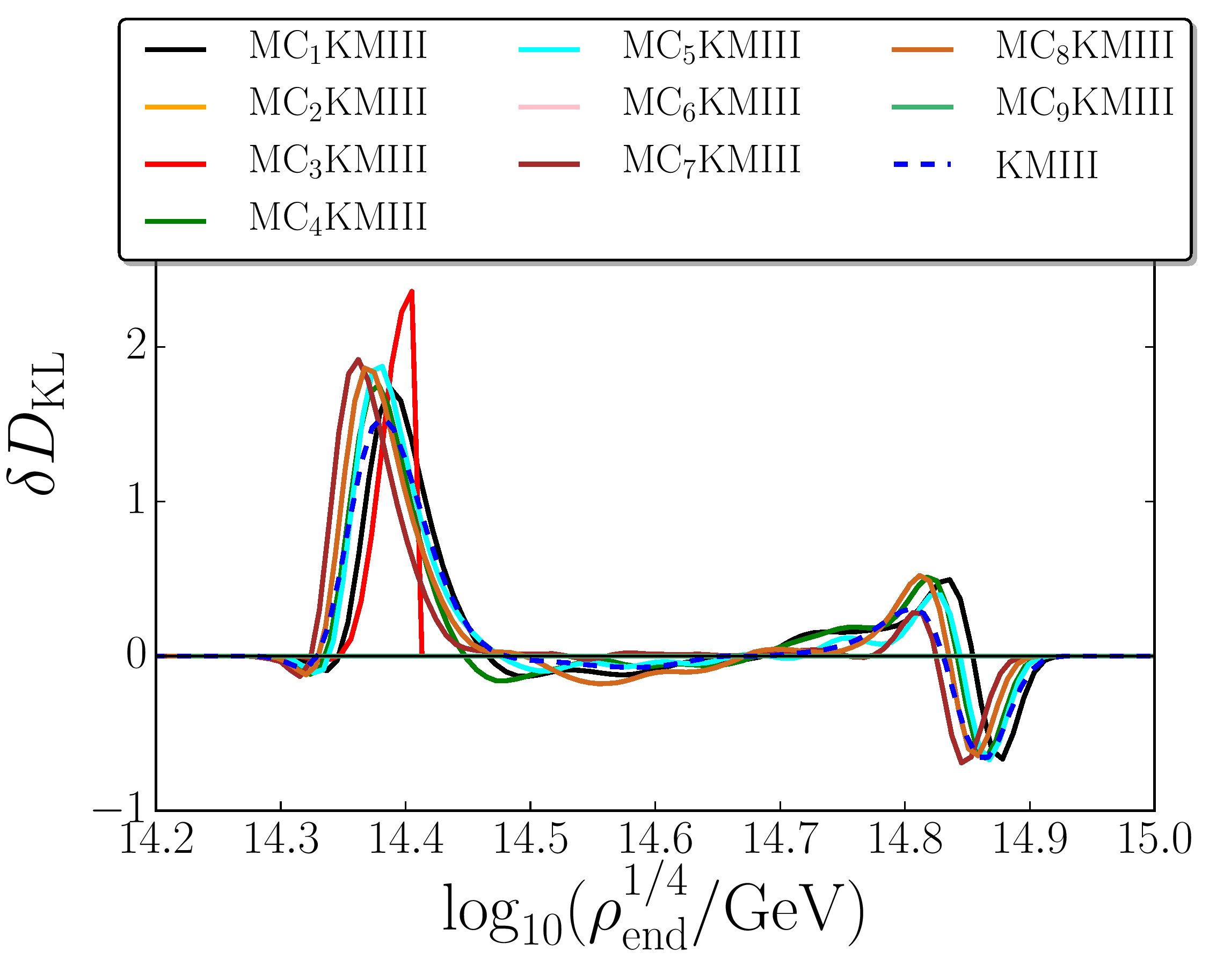}
\includegraphics[width=7cm]{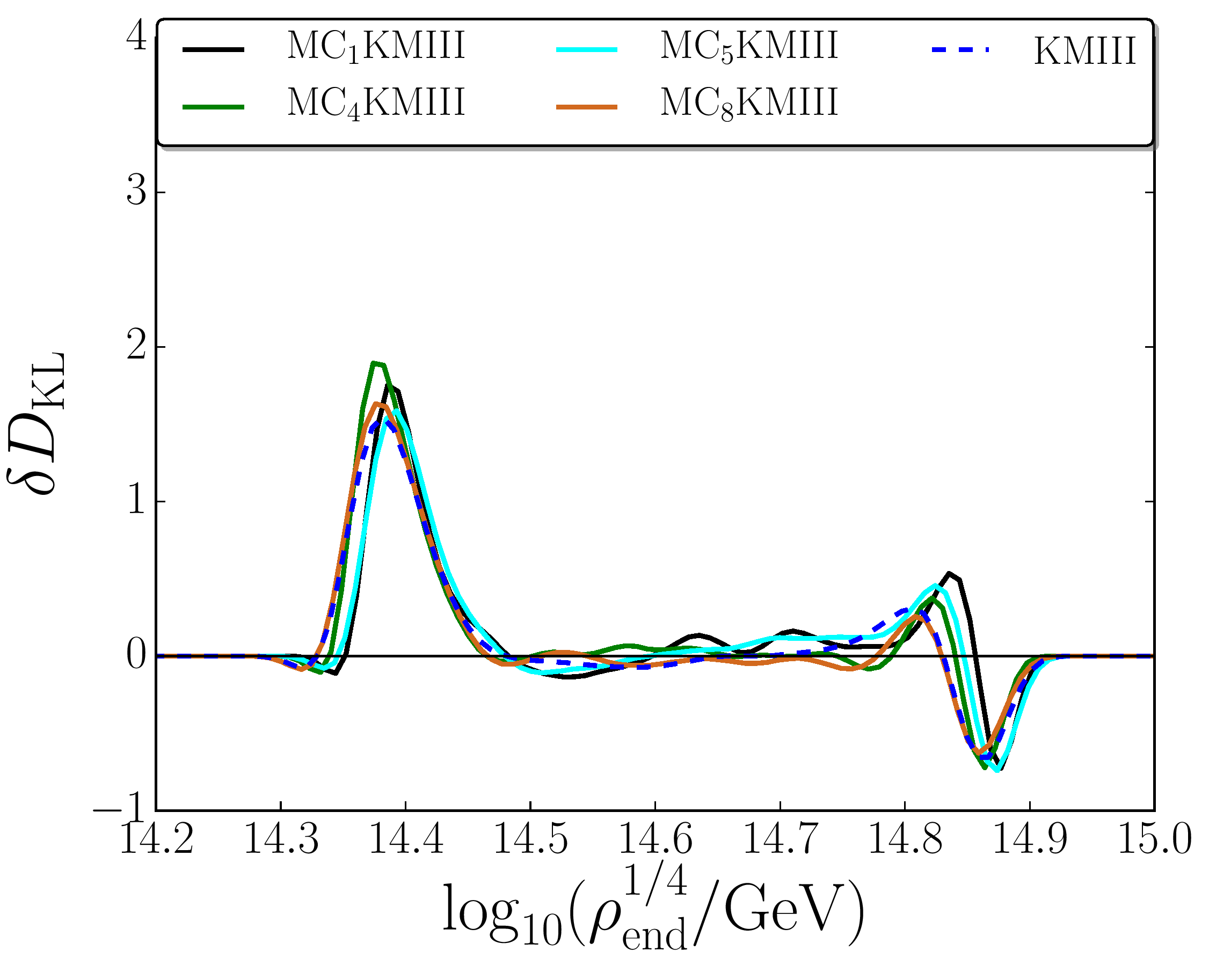}
\caption[Information density on the energy density at the end of inflation]{}
\label{fig:DKL:rhoend:individual}
\end{center}
\end{figure}
\begin{figure}
\contcaption{Information density on $\rho_\uend$ for Higgs inflation (top panels), quartic inflation (middle panels) and K\"ahler moduli II inflation (bottom panels). The left panels correspond to the logarithmically flat prior~(\ref{eq:sigmaend:LogPrior}) on $\sigma_\uend$, and the right panels stand for the stochastic prior~(\ref{eq:sigmaend:GaussianPrior}) derived from the equilibrium distribution of a light scalar field in a de Sitter space-time with Hubble scale $H_\uend$. The dashed blue lines correspond to the single-field versions of the models, while the solid coloured lines stand for the $10$ reheating scenarios. 
}
\end{figure}
\newpage
\subsection{\textsf{Reheating temperature}}
\label{sec:app:DKL:Treh}
\begin{figure}
\figpilogsto
\begin{center}
\includegraphics[width=7cm]{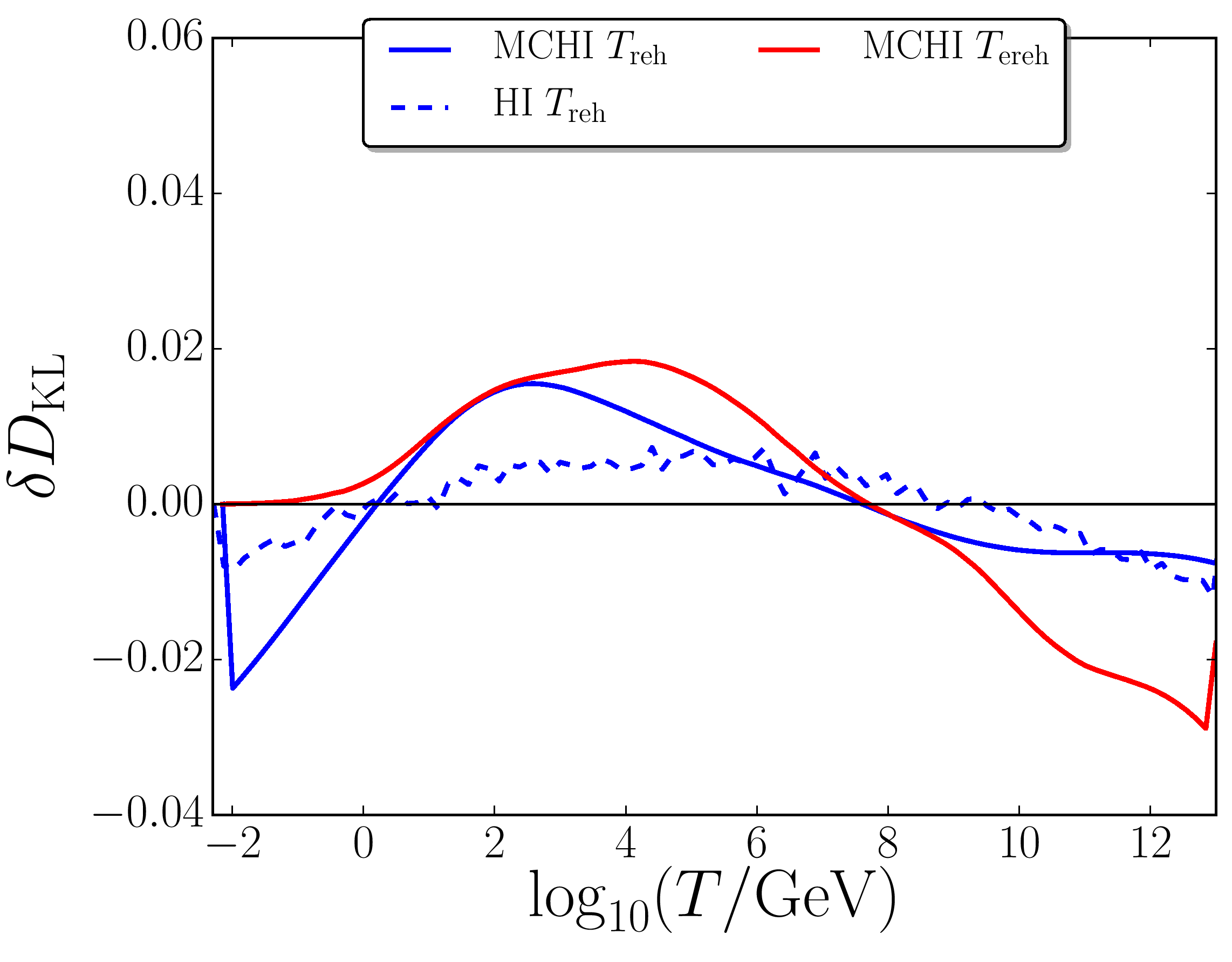}
\includegraphics[width=7cm]{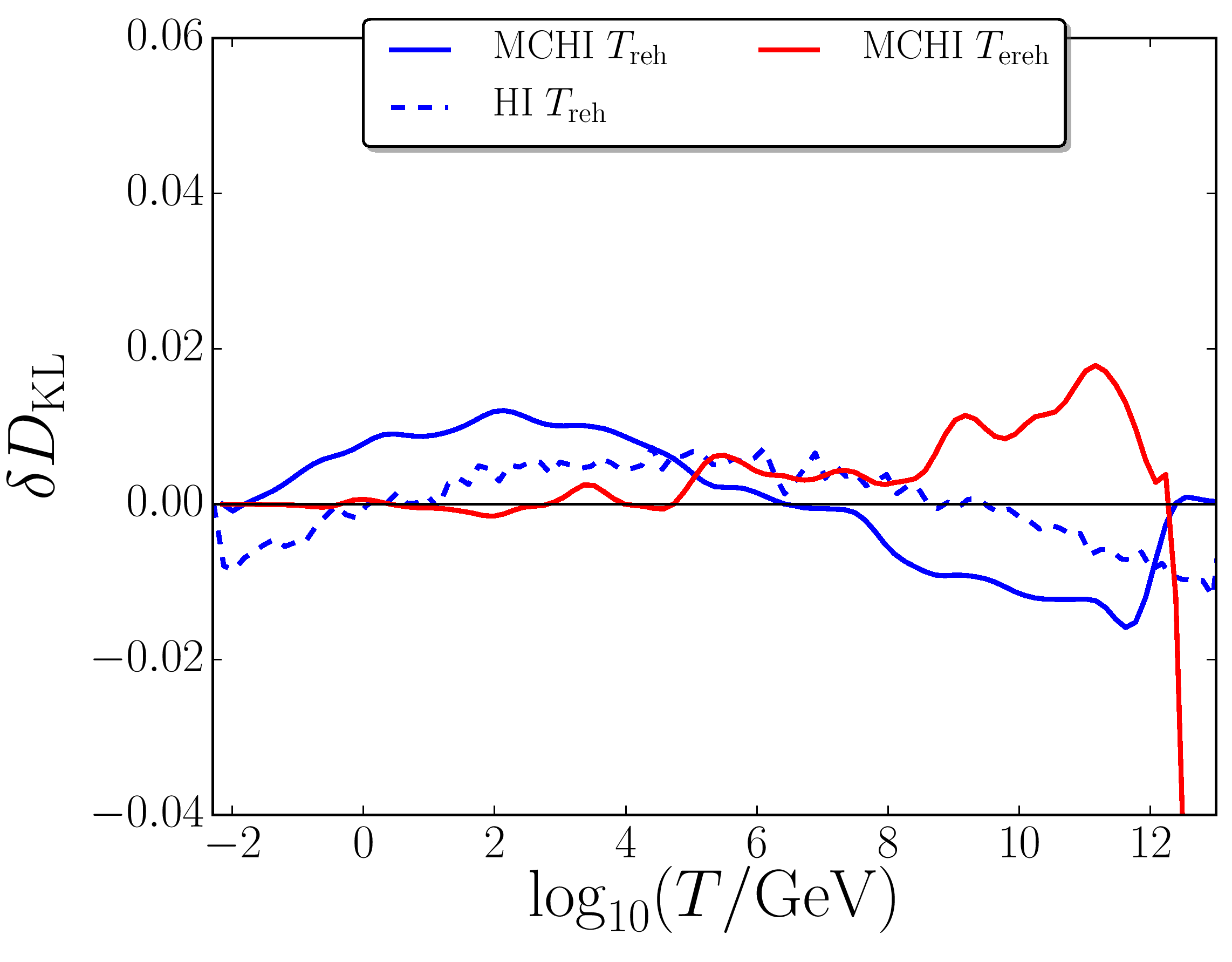}
\includegraphics[width=7cm]{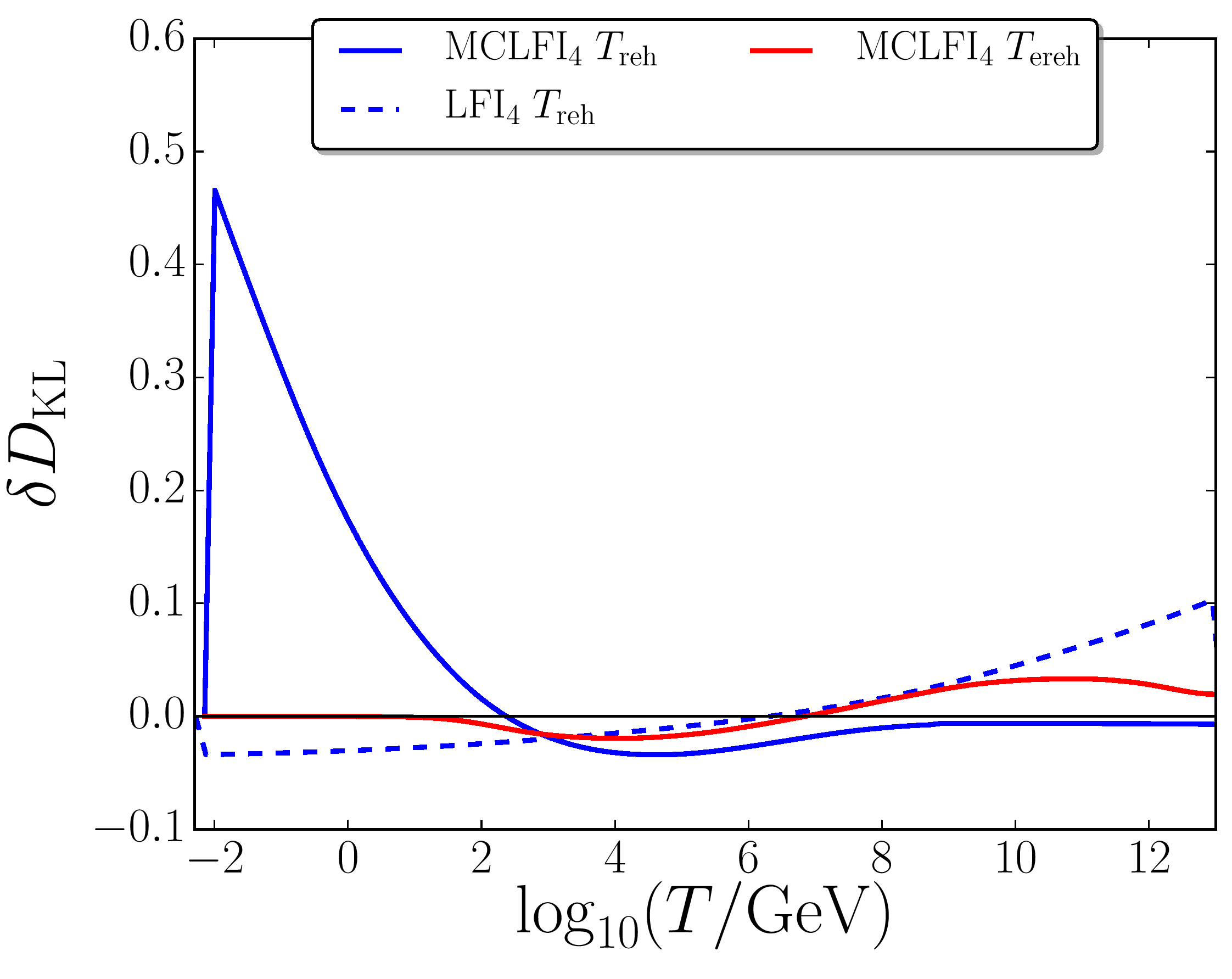}
\includegraphics[width=7cm]{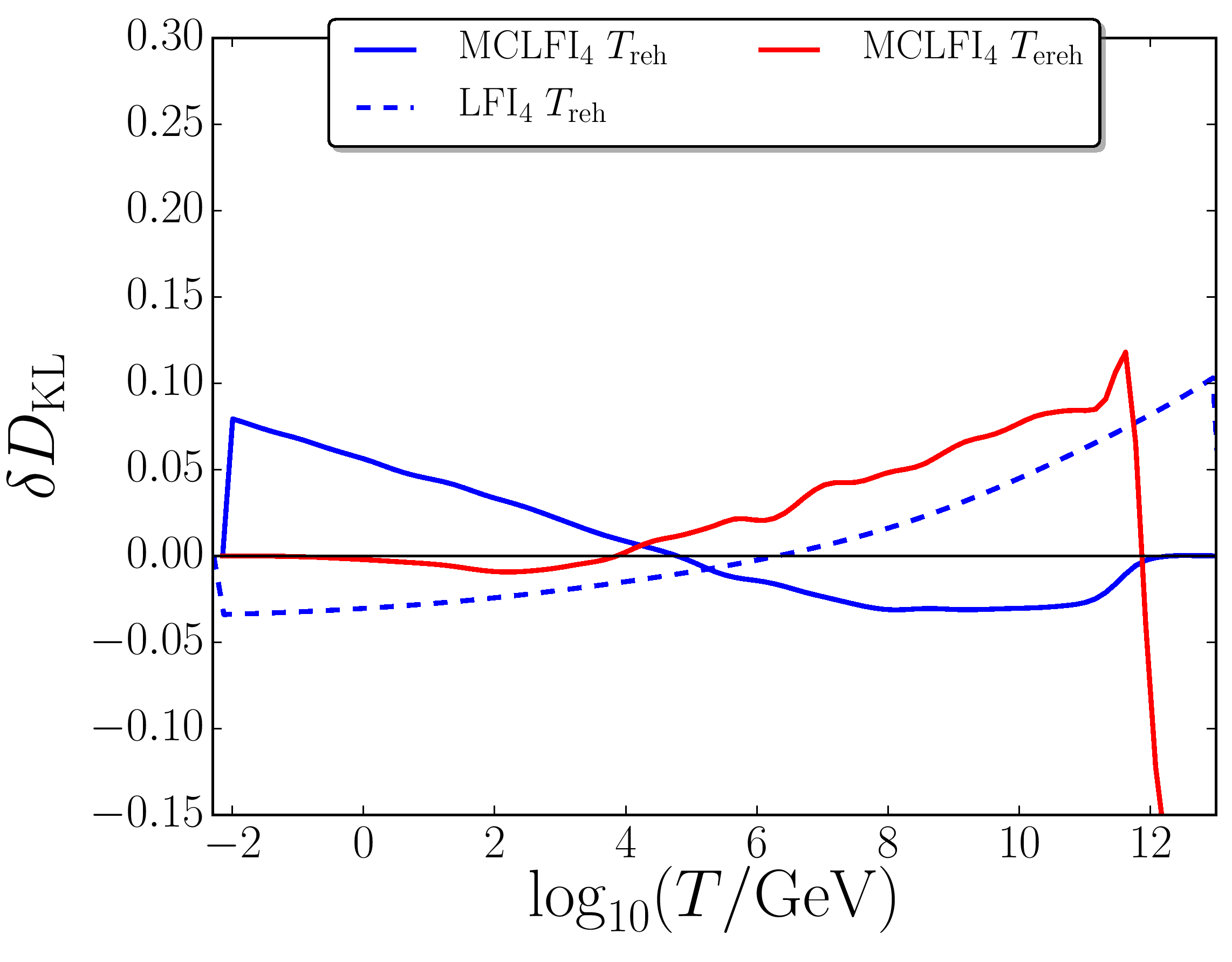}
\includegraphics[width=7cm]{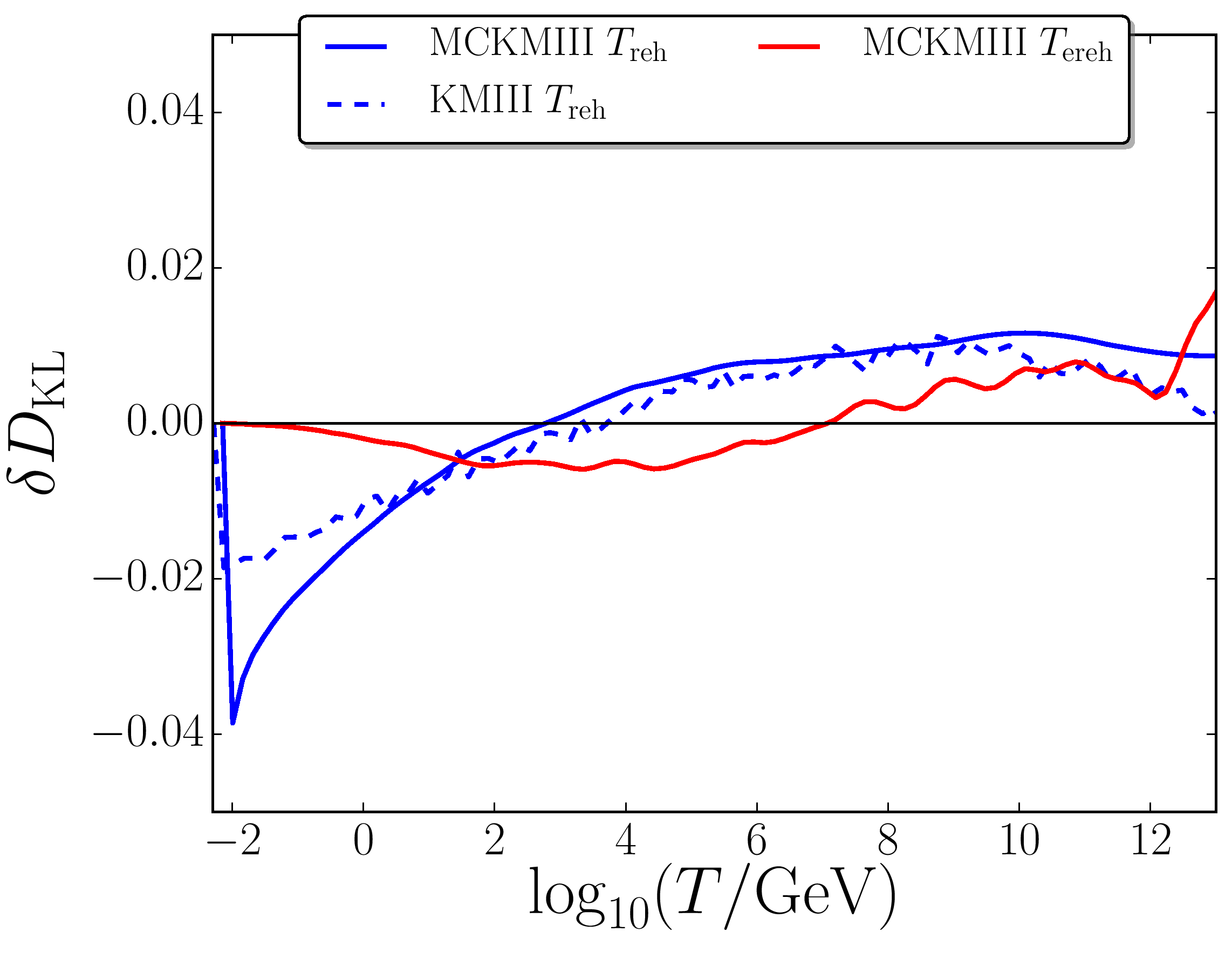}
\includegraphics[width=7cm]{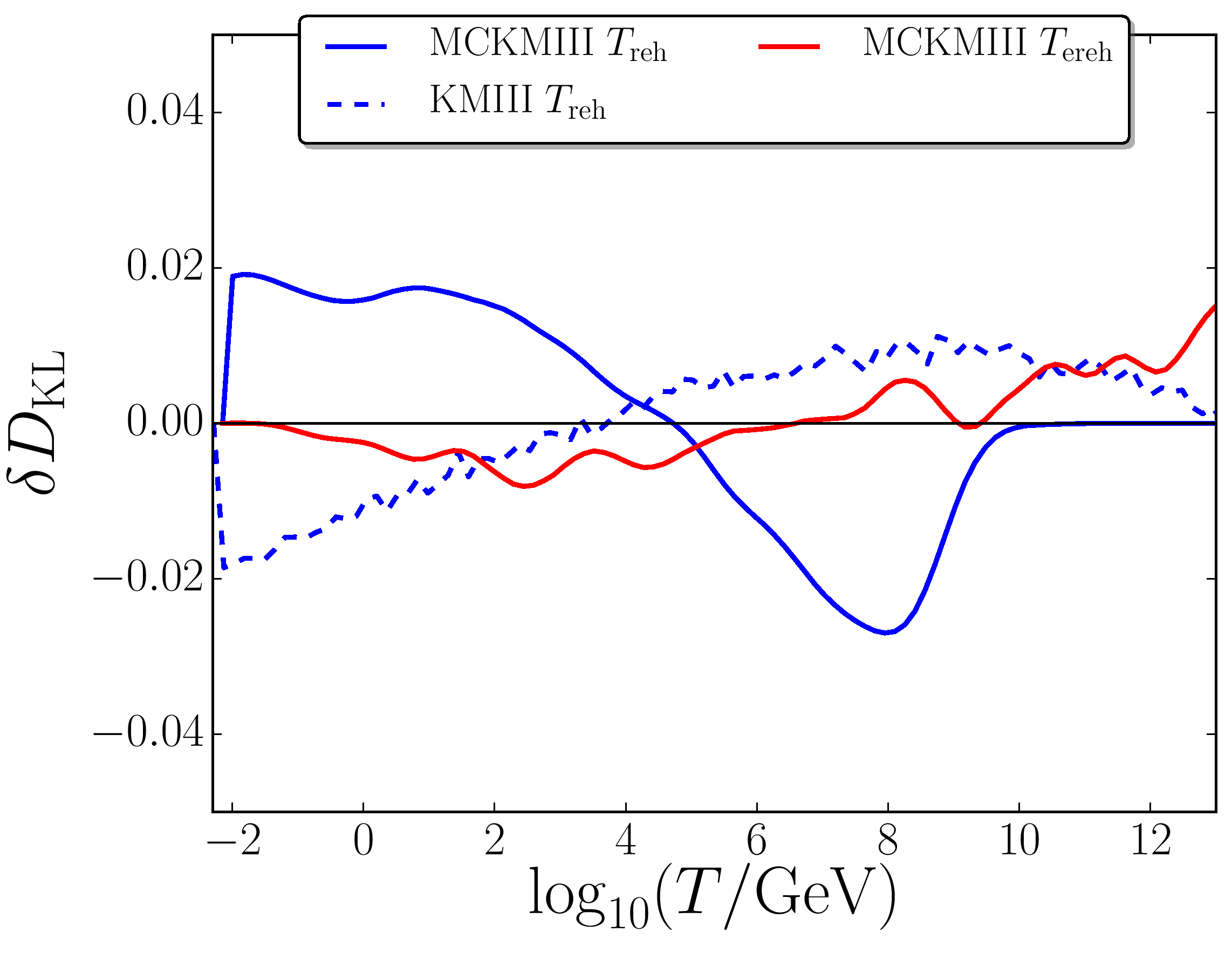}
\caption[Information density on the averaged reheating temperature]{}
\label{fig:DKL:Treh:averaged}
\end{center}
\end{figure}
\begin{figure}
\contcaption{Information density on $T_\ureh$ and $T_\uereh$ for Higgs inflation (top panels), quartic inflation (middle panels) and K\"ahler moduli II inflation (bottom panels). The left panels correspond to the logarithmically flat prior~(\ref{eq:sigmaend:LogPrior}) on $\sigma_\uend$, and the right panels stand for the stochastic prior~(\ref{eq:sigmaend:GaussianPrior}) derived from the equilibrium distribution of a light scalar field in a de Sitter space-time with Hubble scale $H_\uend$. The dashed blue lines correspond to the single-field versions of the models, while the solid lines are derived from the averaged distributions on $T_\ureh$ (blue) and $T_\uereh$ (red), when an extra light scalar field is added. 
}
\end{figure}

\newpage
\begin{figure}
\figpilogsto
\begin{center}
\includegraphics[width=7cm]{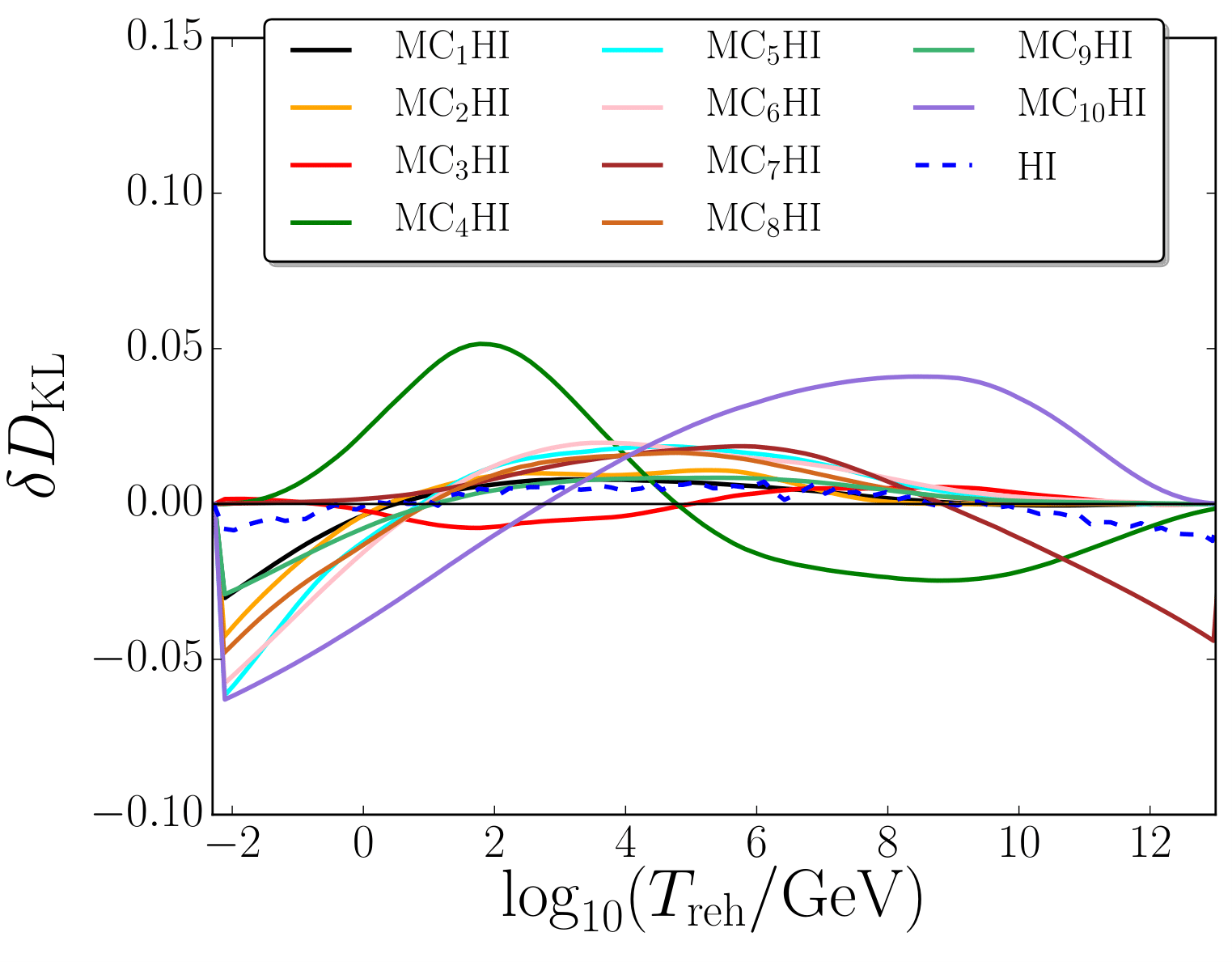}
\includegraphics[width=7cm]{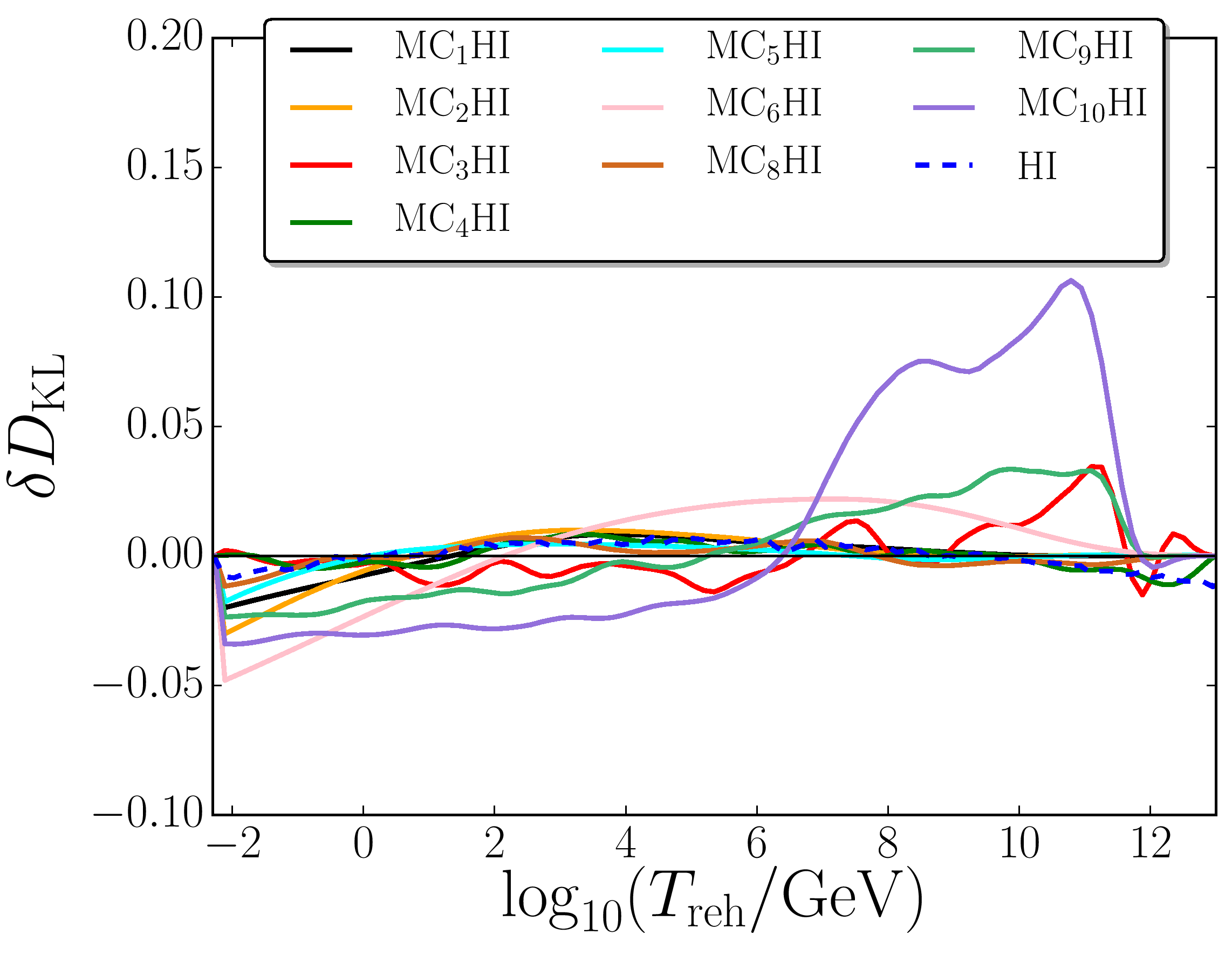}
\includegraphics[width=7cm]{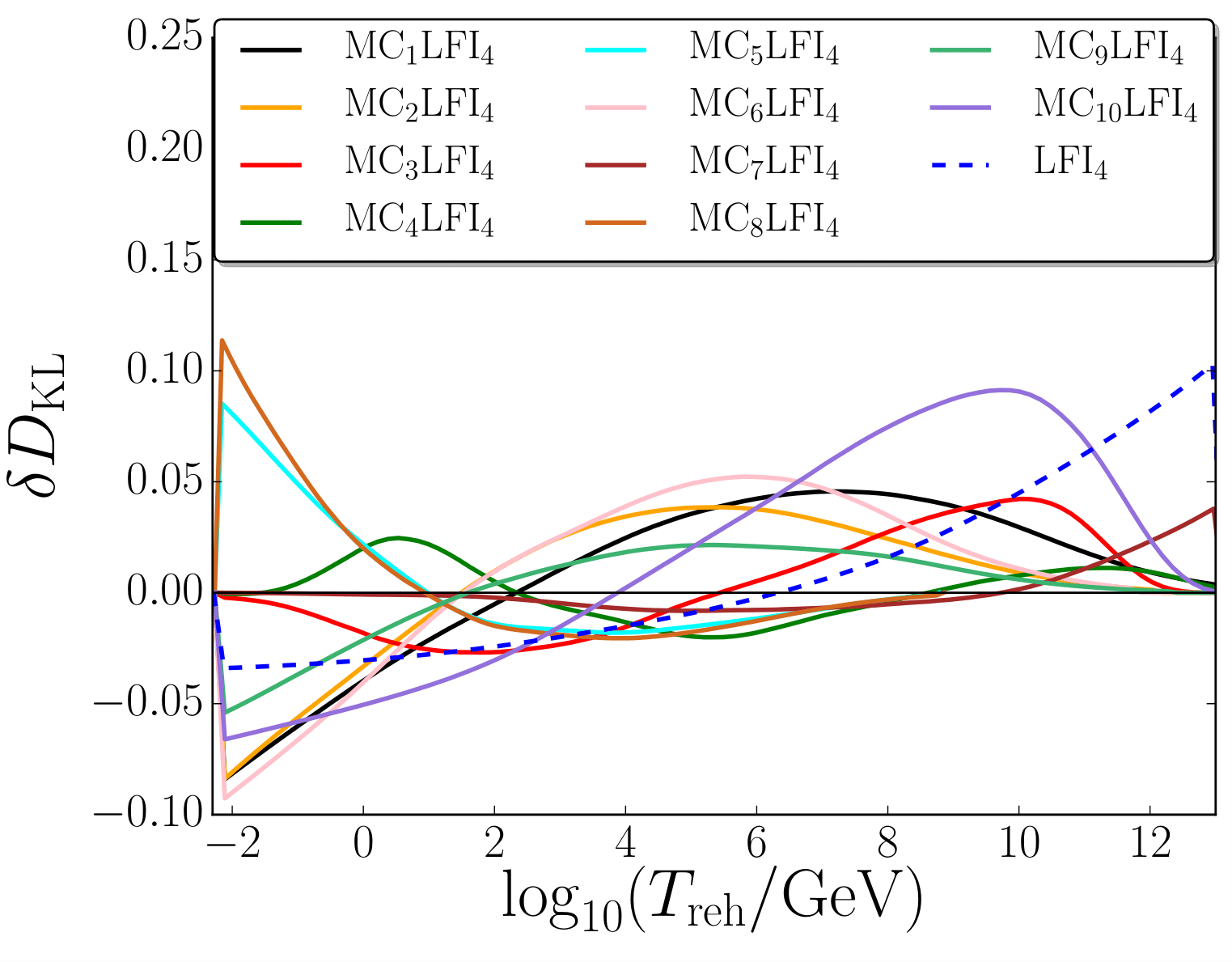}
\includegraphics[width=7cm]{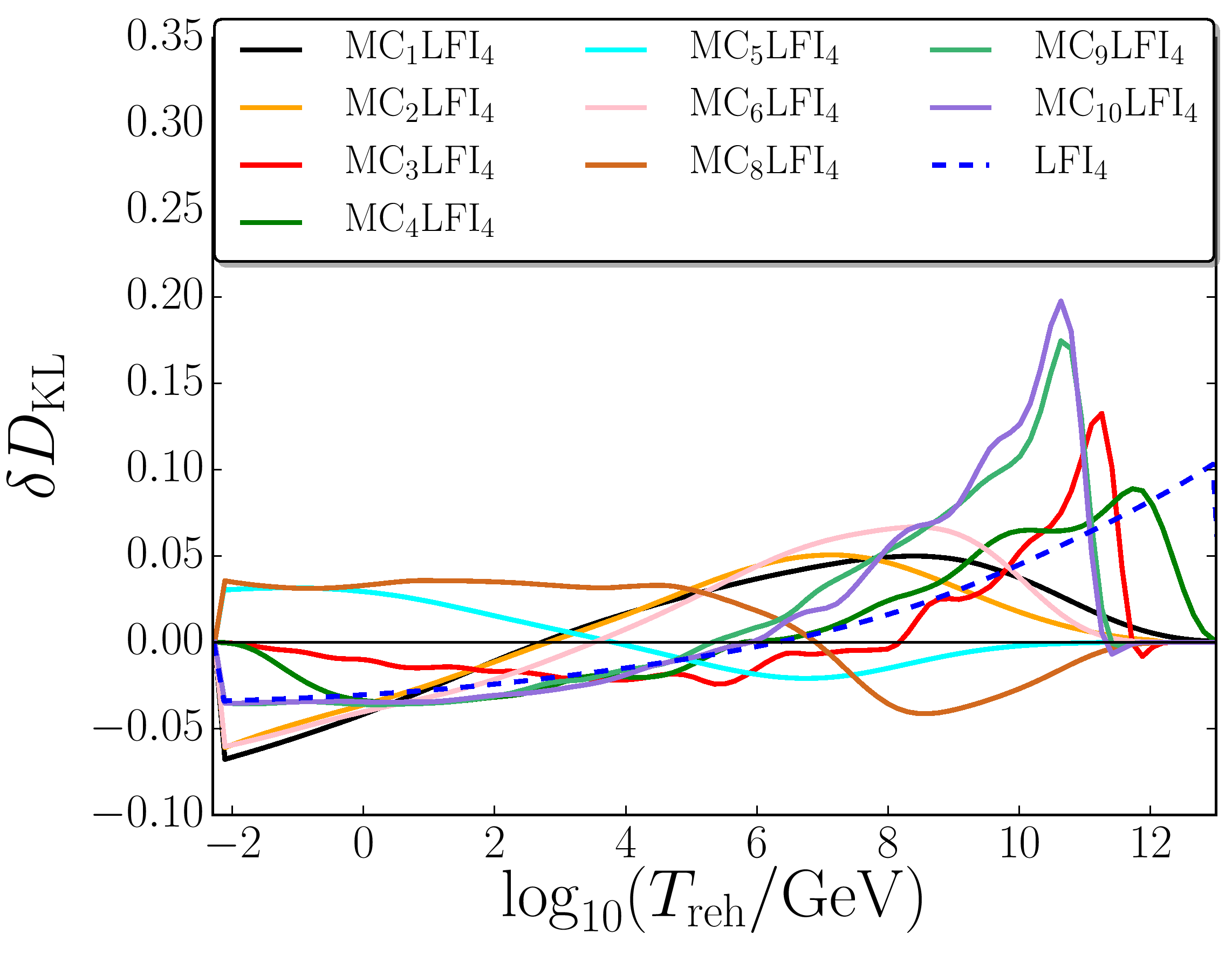}
\includegraphics[width=7.2cm]{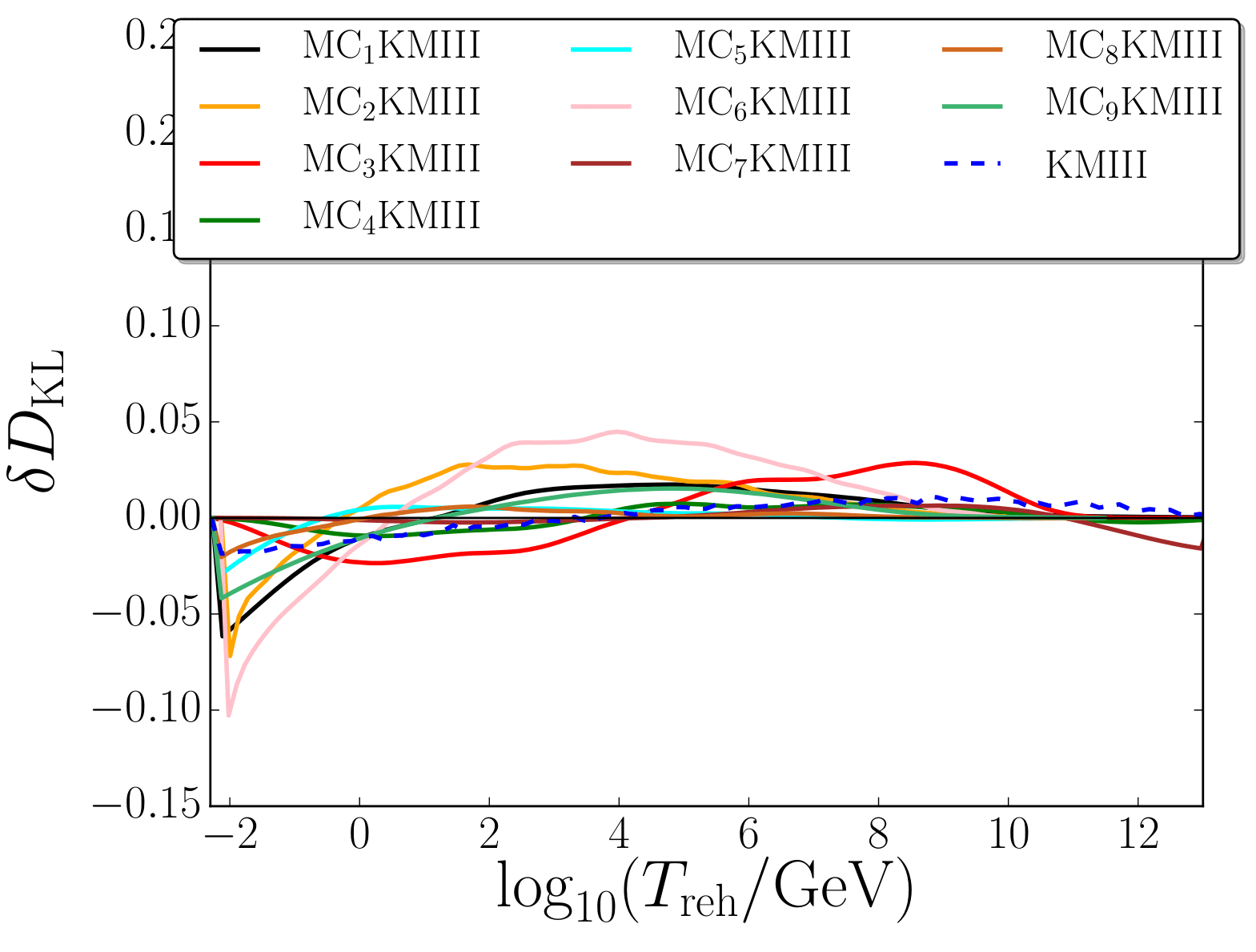}
\includegraphics[width=7.2cm]{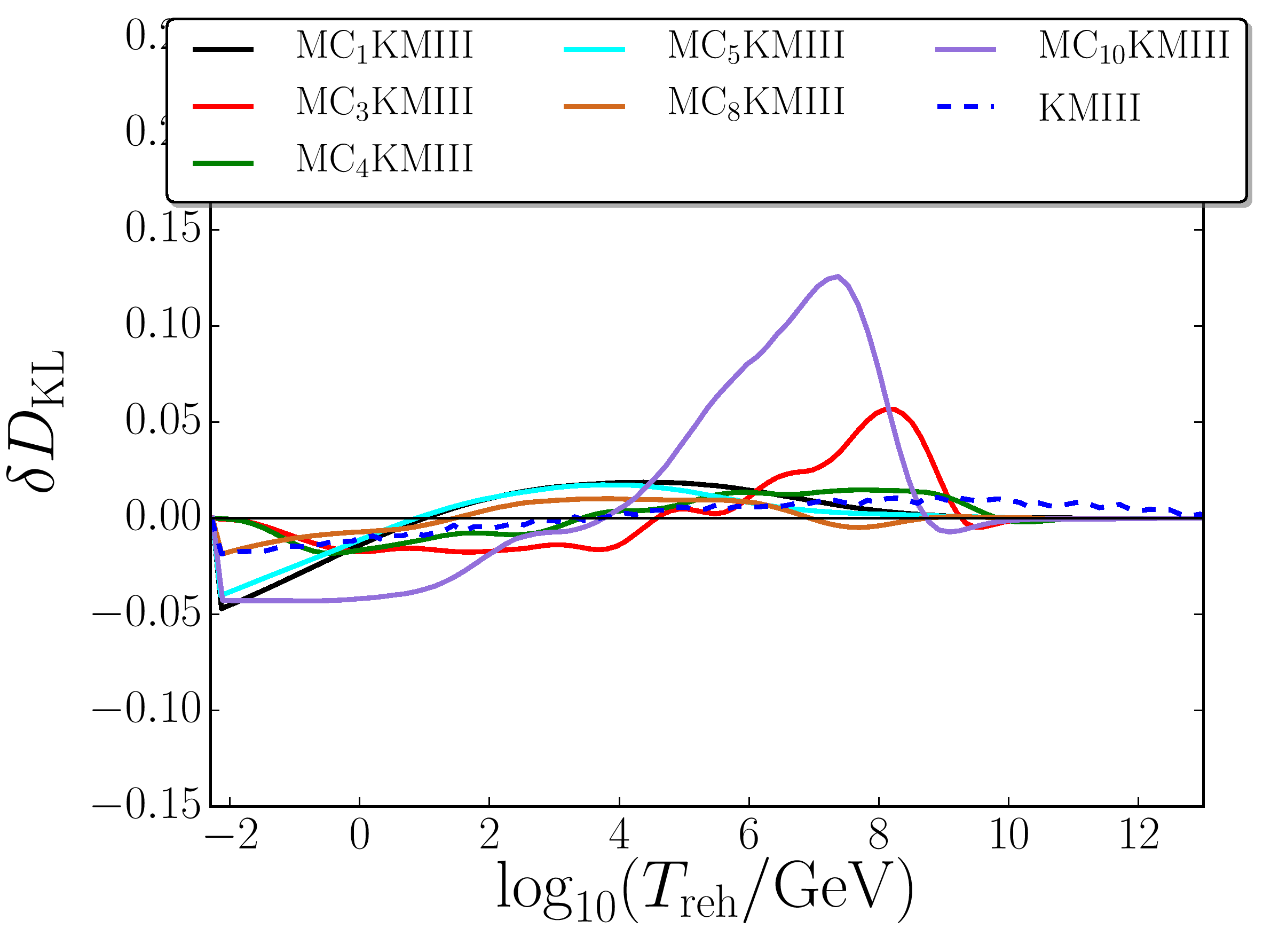}
\caption[Information density on the reheating temperature]{}
\label{fig:DKL:Treh:individual}
\end{center}
\end{figure}
\begin{figure}
\contcaption{Information density on $T_\ureh$ for Higgs inflation (top panels), quartic inflation (middle panels) and K\"ahler moduli II inflation (bottom panels). The left panels correspond to the logarithmically flat prior~(\ref{eq:sigmaend:LogPrior}) on $\sigma_\uend$, and the right panels stand for the stochastic prior~(\ref{eq:sigmaend:GaussianPrior}) derived from the equilibrium distribution of a light scalar field in a de Sitter space-time with Hubble scale $H_\uend$. The dashed blue lines correspond to the single-field versions of the models, while the solid coloured lines stand for the 10 reheating scenarios. 
}
\end{figure}

\newpage
\subsection{\textsf{Early reheating temperature}}
\label{sec:app:DKL:Tereh}
\begin{figure}
\figpilogsto
\begin{center}
\includegraphics[width=7cm]{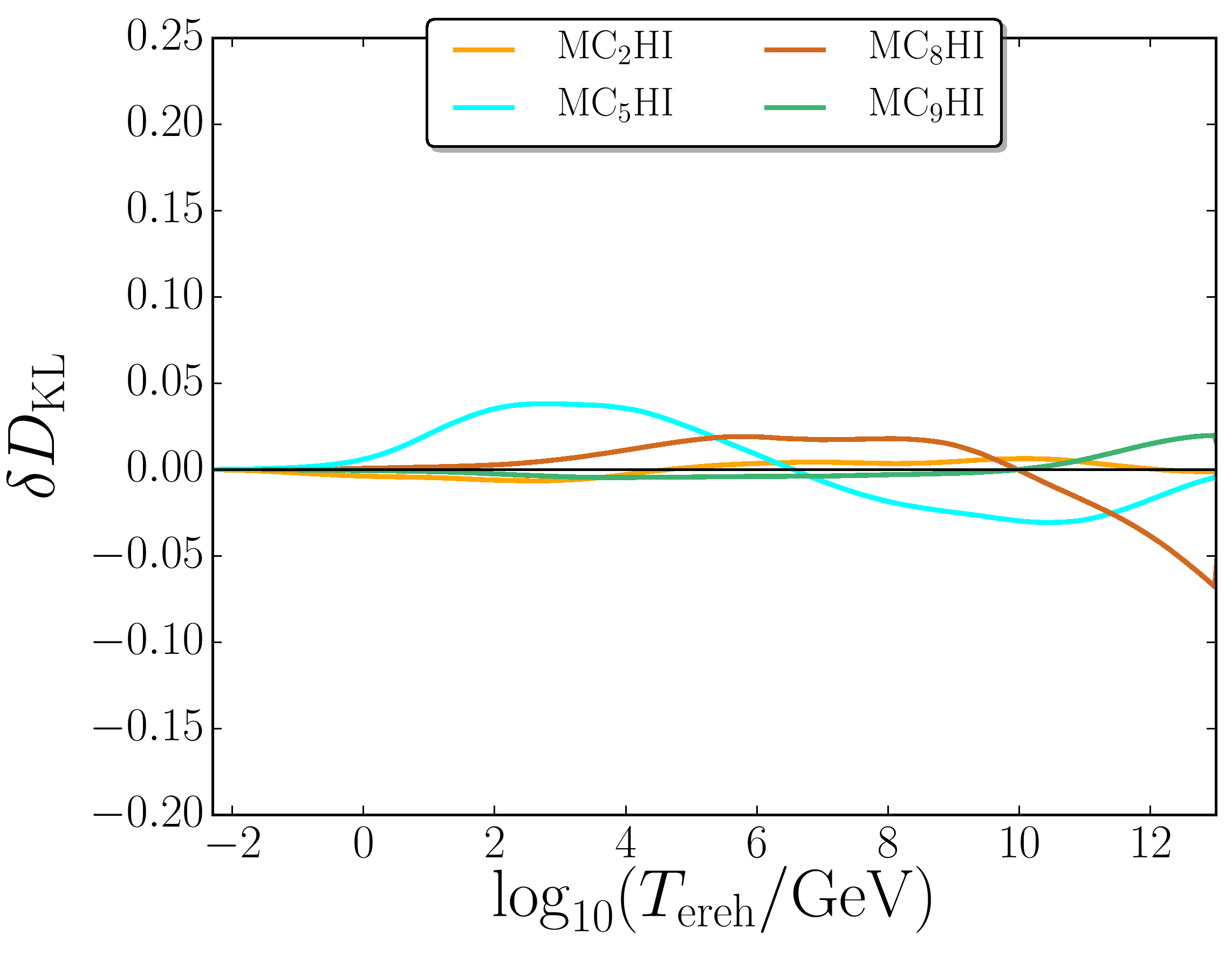}
\includegraphics[width=7cm]{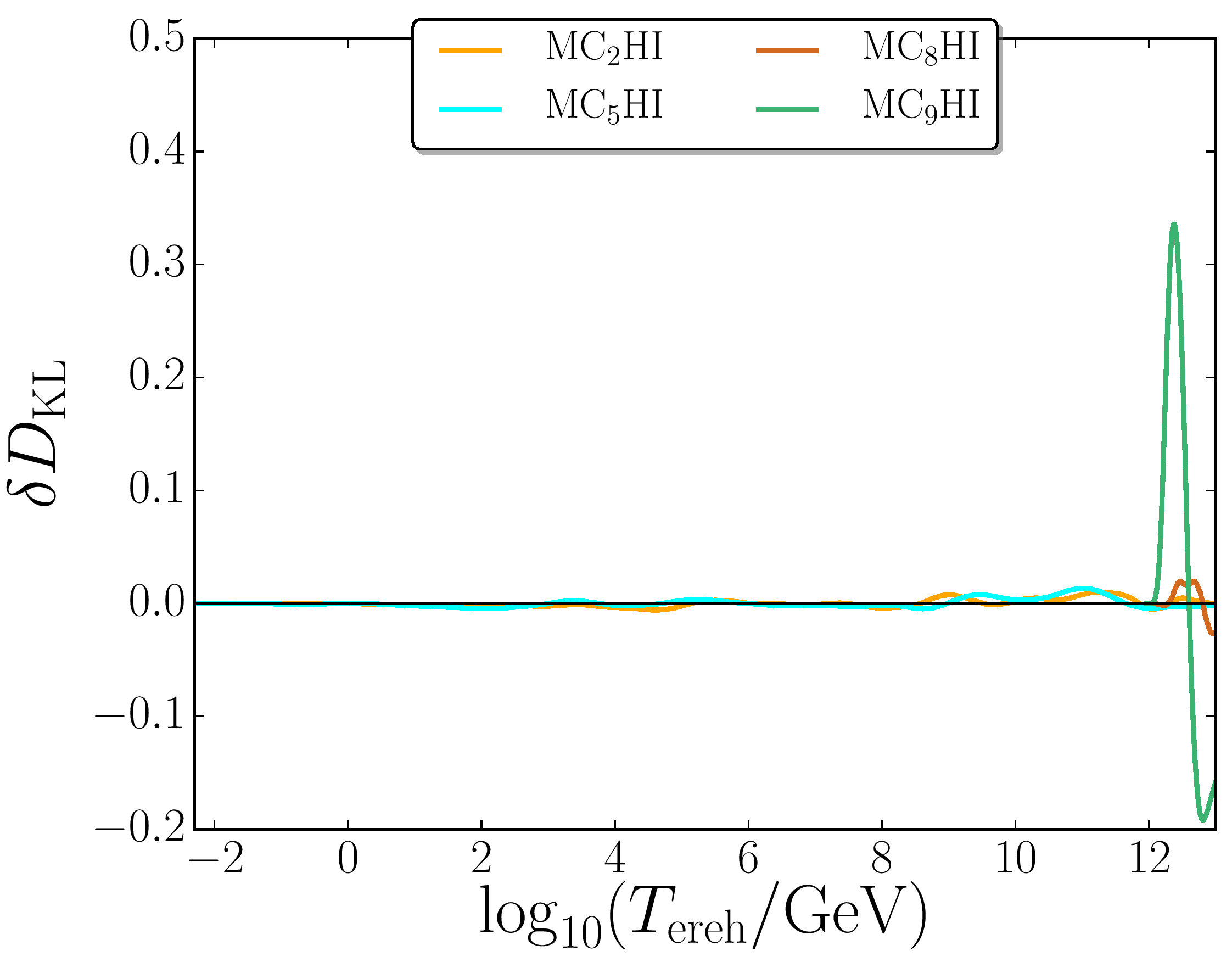}
\includegraphics[width=7cm]{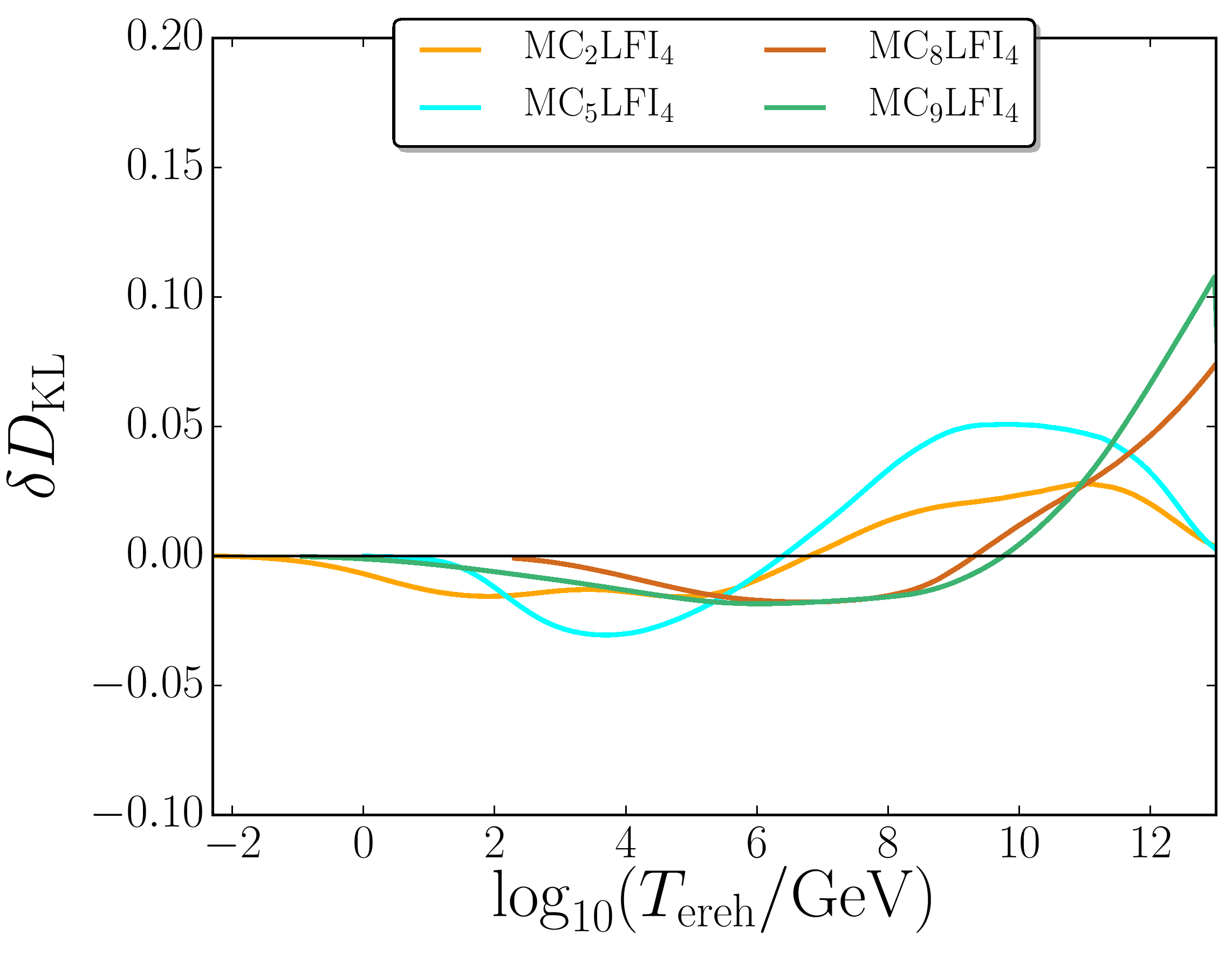}
\includegraphics[width=7cm]{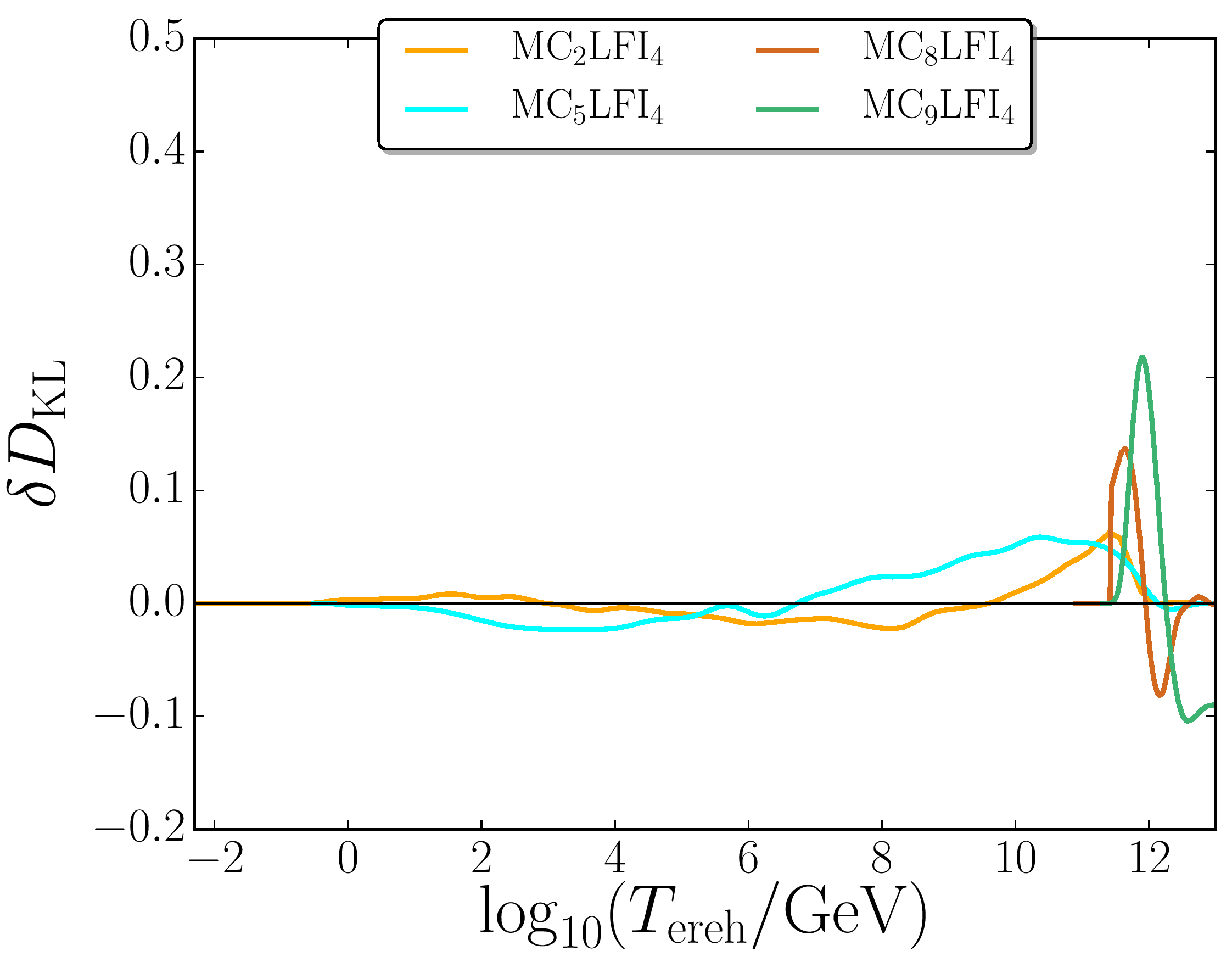}
\includegraphics[width=7cm]{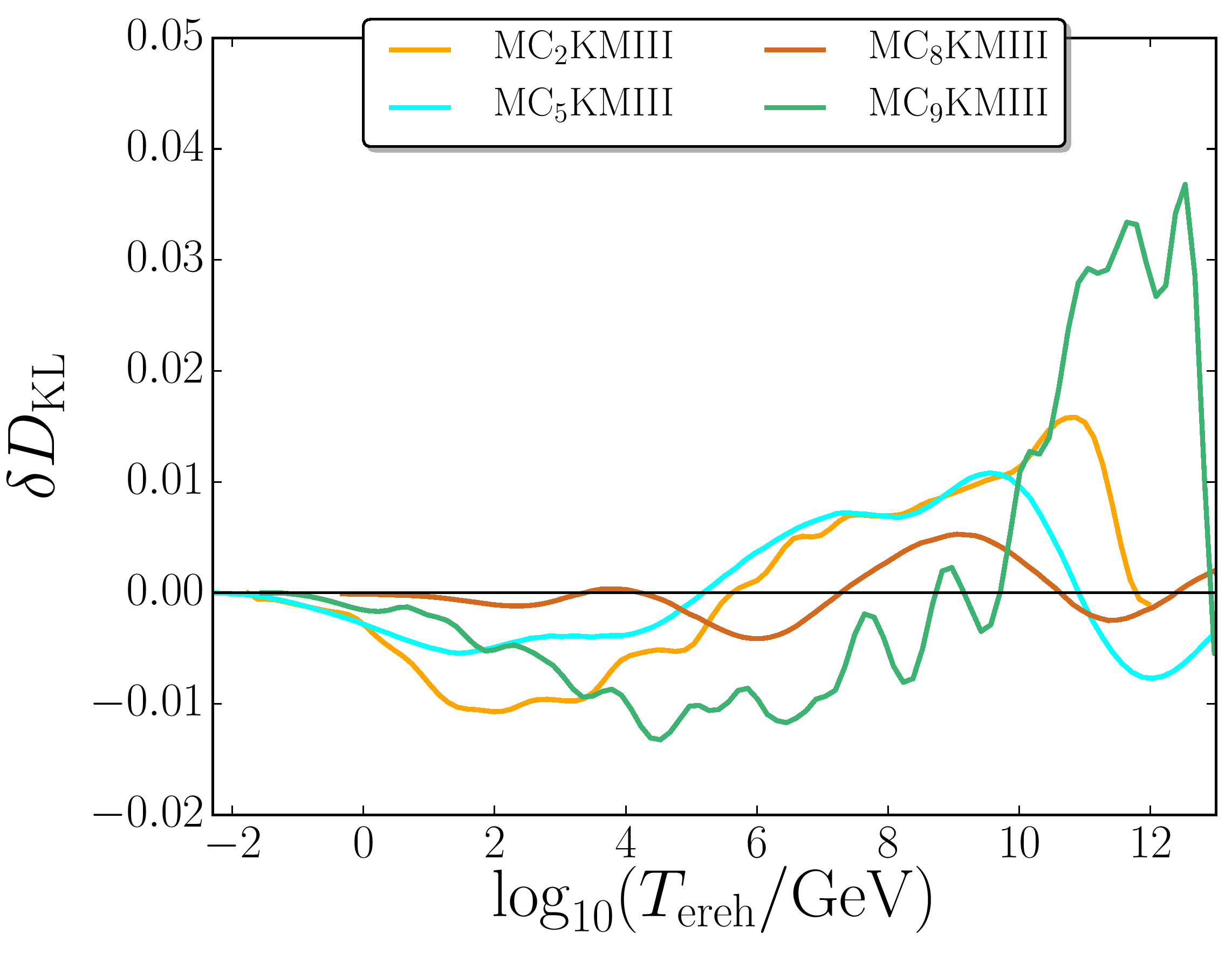}
\includegraphics[width=7cm]{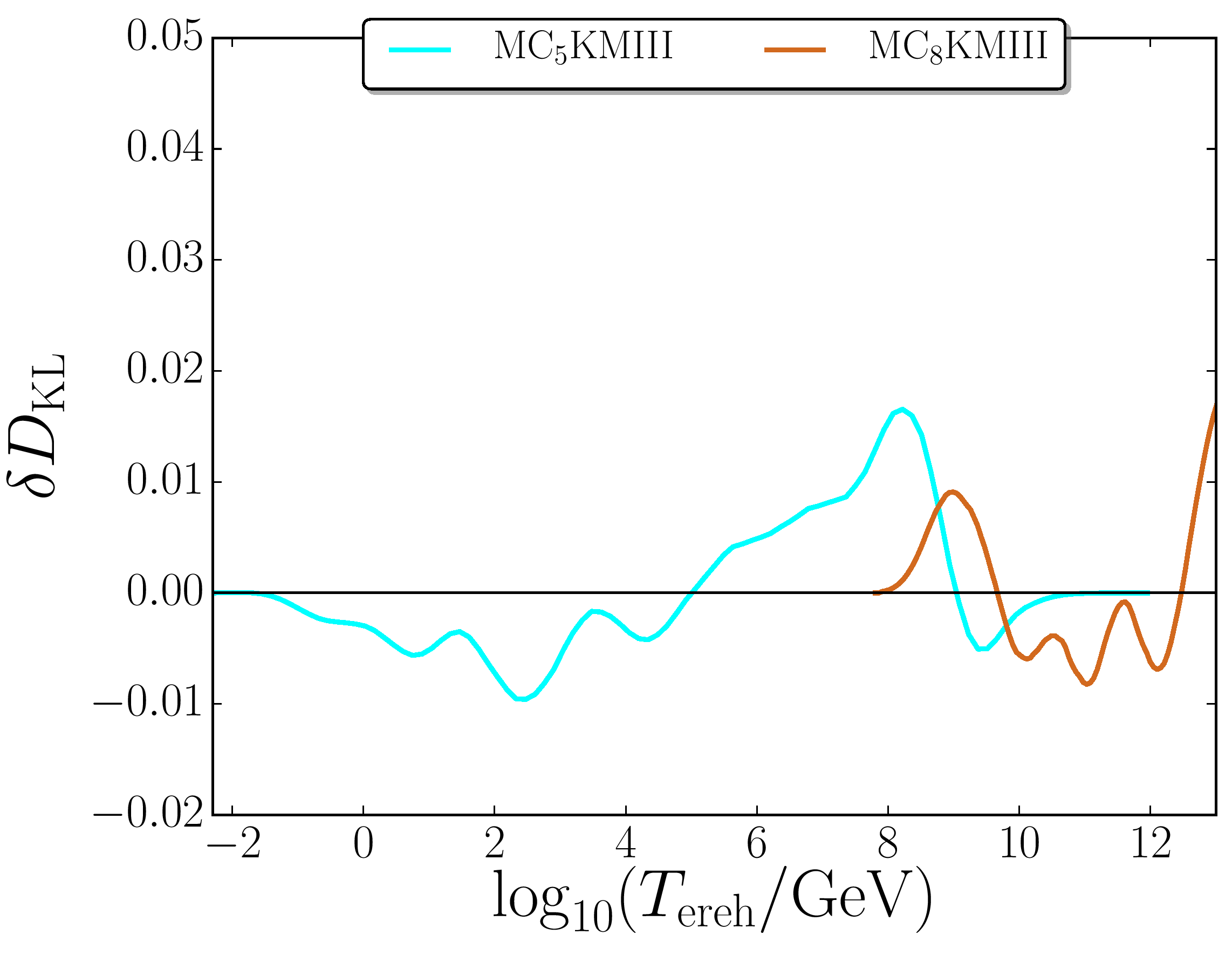}
\caption[Information density on the averaged early reheating temperature]{}
\label{fig:DKL:Tereh:averaged}
\end{center}
\end{figure}
\begin{figure}
\contcaption{Information density on $T_\uereh$ for Higgs inflation (top panels), quartic inflation (middle panels) and K\"ahler moduli II inflation (bottom panels). The left panels correspond to the logarithmically flat prior~(\ref{eq:sigmaend:LogPrior}) on $\sigma_\uend$, and the right panels stand for the stochastic prior~(\ref{eq:sigmaend:GaussianPrior}) derived from the equilibrium distribution of a light scalar field in a de Sitter space-time with Hubble scale $H_\uend$. 
}
\end{figure}
\end{subappendices}

\chapter{\textsf{Spectator field condensates}}
\label{sec:infra-red-divergences} \niceline {\vskip+1ex} 

\begin{center}
\fbox{\parbox[c]{13cm}{\vspace{1mm}{\textsf{\textbf{Abstract.}}} In this chapter we will study the dynamics of light (sub-Hubble mass) test (energetically sub-dominant) fields --- also dubbed `spectator fields' --- in an inflationary background. We have already shown in \Sec{sec:intro-stochastic-approach} that the dynamics of such fields may be accurately described by a stochastic approach. Here we shall focus on implementing this formalism to compute the typical variance acquired by these fields (effectively a condensate) up to the end of inflation: for different spectator field potentials; in different slow-roll inflationary backgrounds; and for multiple coupled spectators. In this review we combine work from Refs.~\cite{Hardwick:2017fjo,Hardwick:2018sck}, more recent work on non-minimally coupled fields and introduce the publicly available code, \href{https://sites.google.com/view/nfield-py}{\texttt{nfield}}, (which now supports multiple test and non-test fields during inflation) as a new computational tool. Motivated originally by the requirement to set the initial conditions for the curvaton in the previous chapter: the results from this chapter are crucial to setting the initial conditions for many other models of post-inflationary physics, including the majority discussed in this thesis. The results from this chapter are thus applicable to a great variety of models for the early Universe. \vspace{1mm}}}
\end{center}

\section{\textsf{Introduction}}
\label{sec:Introduction}
\label{sec:SpectatorExamples}
From a theoretical point of view, inflation takes place in a regime that is far beyond the reach of terrestrial particle accelerators, and the physical details of how the inflaton is connected with the standard model of particle physics and its extensions are still unclear. In particular, most physical setups that have been proposed to embed inflation contain extra scalar fields. This is notably the case in string theory models where many extra light moduli fields may be present~\cite{Turok:1987pg, Damour:1995pd, Kachru:2003sx, Kofman:2003nx, Krause:2007jk, Baumann:2014nda}.

Even if such fields are purely spectators during inflation (\ie masses smaller than the Hubble rate and contribute a negligible amount to the total energy density of the Universe), as we have shown already in \Chap{sec:curvaton-reheating}, they can still play an important dynamical role afterwards. The details of their post-inflationary contribution typically depend on the field displacement they acquire during inflation. In this context, if inflation provides initial conditions for cosmological perturbations, it should also be seen as a mechanism that generates a distribution of initial field displacements for light degrees of freedom. In this chapter, we investigate what possibilities this second channel offers to probe the physics of inflation. In practice, we study how the field value acquired by light scalar spectator fields at the end of inflation depends on the  inflaton field potential, on the spectator field potential and on the initial distribution of spectator field values.

As an illustration of post-inflationary physical processes for which the field value acquired by spectator fields during inflation plays an important role, we may consider the curvaton scenario of \Sec{sec:sourcing-cos-pert-curvaton}, \Chap{sec:curvaton-reheating} and, originally, of Refs.~\cite{Linde:1996gt, Enqvist:2001zp, Lyth:2001nq, Moroi:2001ct}. We are reminded that the curvaton density perturbation is given by $\delta \rho_\sigma/\rho_\sigma\sim \delta\sigma/\sigma$, where $\rho_\sigma$ denotes the energy density contained in $\sigma$, and the effect of this perturbation on the total density perturbation of the Universe is reduced by the relative energy density of the curvaton field to the total energy density. The curvaton field, like every light scalar field, is perturbed at Hubble radius exit by an amount $\delta\sigma\sim H_*\lesssim 10^{-6} \Mp$, where $H_*$ is the Hubble parameter evaluated at the time of Hubble radius crossing during inflation and $\Mp$ is the reduced Planck mass. If the curvaton perturbations produce the entire observed primordial density perturbation with amplitude $10^{-5}$, the average field value in our Hubble patch, $\sigma$, is of order $\sigma\sim 10^5 H_* $. An important question is therefore whether such a field value can naturally be given to the curvaton during inflation. In the limit of low energy scale inflation in particular, this implies that $\sigma \ll \Mp$. 

The requirement for a very sub-Planckian spectator field value in models where an initially isocurvature field perturbation is later converted into the observed adiabatic curvature perturbation is common but not completely generic, and may be intuitively understood by realising that if the spectator field fluctuations are negligible compared to the background value (\ie $\delta\sigma< 10^{-5} \sigma$), then it is difficult to make the primordial density perturbation have a significant dependence on $\delta\sigma$ if the background value is not very sub-Planckian. This is discussed in the conclusions of \Ref{Byrnes:2008zz}, which shows that it typically also applies to scenarios such as modulated reheating \cite{Dvali:2003em,Kofman:2003nx}. The dark energy model proposed in \Ref{Ringeval:2010hf} also requires sub-Planckian spectator fields during inflation, and the new results we derive on the field value distribution of a spectator field with a quartic potential may have implications for the stability of the Higgs vacuum during inflation as well, see \eg \Refs{Espinosa:2007qp,Herranen:2014cua,Kearney:2015vba}.

This naturally raises the question of whether having a sub-Planckian spectator field value represents a fine tuning of the initial conditions or not. Provided that inflation lasts long enough, we address this question here by calculating the stochastically generated distribution of spectator field values. We will show cases in which sub-Planckian field values are natural, and others in which super-Planckian field values are preferred.

If the spectator field value is driven to become significantly super-Planckian, it can drive a second period of inflation, which may have observable effects even if the inflaton field perturbations dominate, because the observable scales exit the Hubble radius at a different time during the first period of inflation, when the inflaton is traversing a different part of the potential~\cite{Vennin:2015vfa, Vennin:2015egh}. In some cases, we will show that the spectator field value may naturally become so large that it drives more than 60 \efolds of inflation. In this case we would not observe the initial period of inflation at all,  but its existence remains important for generating the initial conditions for the second, observable period of inflation.

If no isocurvature perturbations persist after reheating, the linear perturbations from the inflaton and spectator field are likely to be observationally degenerate. Non-linear perturbations, especially the coupling between primordial long- and short-wavelength perturbations, help to break this degeneracy. We will not study non Gaussianity in this chapter, but highlight that the results calculated here help to motivate a prior distribution for the initial spectator field value, which is a crucial ingredient of model comparison between single- and multiple-field models of inflation \cite{Hardwick:2015tma,Vennin:2015vfa, Vennin:2015egh, dePutter:2016trg}.

\subsection{\textsf{Stochastic single spectator}}
\label{sec:StochasticInflation}

As we have seen in \Sec{sec:intro-stochastic-approach}, in the stochastic framework, the short wavelength fluctuations behave as a classical noise acting on the dynamics of the super-Hubble scales as they cross the coarse-graining scale.  The coarse-grained fields can thus be described by a stochastic classical theory, following Langevin equations
\bea
\label{eq:Langevin}
\frac{\dd\sigma}{\dd N}=-\frac{1}{3H^2}\frac{\partial V}{\partial \sigma}+\frac{H}{2\pi}\xi (N)\, .
\eea
In this expression, $\sigma$ denotes a coarse-grained field with potential $V(\sigma)$. The time variable $N\equiv \ln(a)$ has been used but the choice of the time variable is irrelevant for test fields~\cite{Finelli:2008zg, Finelli:2010sh, Finelli:2011gd, Vennin:2015hra}. We are also reminded that $\xi$ is a Gaussian white noise with vanishing mean and unit variance such that $\langle \xi(N) \rangle = 0$ and $\langle \xi(N_1)\xi(N_2)\rangle = \delta(N_1-N_2)$, where $\langle\cdot\rangle$ denotes ensemble average. The Langevin equation~(\ref{eq:Langevin}) is valid for a light test field with $\partial^2V/\partial \sigma^2\ll H$.  In the It\^o interpretation, it gives rise to a Fokker-Planck equation for the probability density $P(\sigma,N)$ of the coarse-grained field $\sigma$ at time $N$~\cite{Starobinsky:1986fx, Vilenkin:1999kd}
\bea
\label{eq:FP}
\frac{\partial P(\sigma,N)}{\partial N} = \frac{\partial}{\partial\sigma}\left[\frac{1}{3H^2}\frac{\partial V}{\partial \sigma} P(\sigma,N)\right]+\frac{H^2}{8\pi^2}\frac{\partial^2}{\partial\sigma^2}\left[ P(\sigma,N)\right]\,,
\eea
which is the same as we found in \Sec{sec:intro-stochastic-approach}. As in \Eq{eq:probcurr-test-field}, this equation can be written as $\partial P/\partial N = -\partial J/\partial \sigma$, where $J\equiv - P/(3H^2)\partial V/\partial \sigma -H^2/(8\pi^2)\partial P/\partial \sigma$ is the probability current.

When $H$ is constant, a stationary (equilibrium) solution $P_\mathrm{stat}$ to \Eq{eq:FP} can be found, however we need to demonstrate that $J$ vanishes in order to identify this solution with \Eq{eq:equilib-spectator-dist}. Since  $P_\mathrm{stat}$ does not depend on time, the probability current does not depend on $\sigma$ (or on time either). Therefore, if $J$ vanishes at the boundaries of the field domain, it vanishes everywhere. So we can find that the solution now matches \Eq{eq:equilib-spectator-dist} like so
\begin{equation}
P_\mathrm{stat}(\sigma)\propto \exp\left[-\frac{8\pi^2V(\sigma )}{3H^4} \right]\, ,
\label{eq:Pstat}
\end{equation}
where the overall integration constant is fixed by requiring that the distribution is normalised, $\int P(\sigma)\dd\sigma = 1$. In the following, the solution~(\ref{eq:Pstat}) will be referred to as the ``de Sitter equilibrium''. For instance, if the spectator field has a quadratic potential $V(\sigma)=m^2\sigma^2/2$, the de Sitter equilibrium is a Gaussian with standard deviation $\sqrt{\langle \sigma^2 \rangle} \sim H^2/m$. In this case, it will be shown in \Sec{sec:plateau_quad_spec} that this equilibrium solution is in fact an attractor of \Eq{eq:Pstat}, that is reached over a time scale $N_\mathrm{relax}\sim H^2/m^2$. Therefore, provided inflation lasts more than  $N_\mathrm{relax}$ $e$-folds, the typical field displacement is of order $H^2/m$ at the end of inflation in this case~\cite{Enqvist:2012xn}.
\subsection{\textsf{Limitations of the adiabatic approximation}}
\label{sec:validity-stat}
In the absence of more general results prior to this chapter, the de Sitter results derived in \Sec{sec:StochasticInflation} have been commonly used and/or assumed to still apply to more realistic slow-roll backgrounds, see \eg \Refs{Enqvist:2012xn,Enqvist:2013kaa,Herranen:2014cua,Hardwick:2015tma,Vennin:2015egh,Figueroa:2016dsc}. The reason is that $H$ varies slowly during slow-roll inflation, which thus does not deviate much from de Sitter. This is why in practice, \Eq{eq:Pstat} is often used to estimate the field value acquired by spectator fields during inflation. However, one can already see why this ``adiabatic'' approximation, which assumes that one can simply replace $H$ by $H(N)$ in \Eq{eq:Pstat} and track the local equilibrium at every time, is not always valid. Indeed,   the time scale over which $H$ varies by a substantial amount in slow-roll inflation is given by $N_H=1/\epsilon_1$, which can be deduced from \Eq{eq:slowrollp}. During inflation, $\epsilon_1\ll 1$, so that $N_H\gg 1$. However, in order to see whether a spectator field tracks the de Sitter equilibrium, one should not compare $N_H$ to $1$, but to $N_\mathrm{relax}$, the number of \efolds required by the spectator field to relax towards the equilibrium. In other words, only if the adiabatic condition
\bea
\label{eq:adiabatic:condition}
N_\mathrm{relax} \ll N_H=\frac{1}{\epsilon_1} \,,
\eea
holds can $H$ be considered as a constant over the time required by the spectator field to relax to the equilibrium, and only in this case can the stationary distribution~(\ref{eq:Pstat}) be used.

If the inflaton potential is of the plateau type and asymptotes to a constant as the field value asymptotes to infinity, one typically has~\cite{Roest:2013fha, Martin:2016iqo} $\epsilon_1\simeq \order{1}/(N_\uend-N)^2$ in the limit where $N_\uend-N\gg 1$, where $N_\uend$ denotes the number of \efolds at the end of inflation where $\epsilon_1\simeq 1$. This leads to
\bea
H\simeq H_\mathrm{plateau} \exp\left[{\frac{\order{1}}{N-N_\uend-1}}\right]\, ,
\eea
where $H_\mathrm{plateau}$ is the asymptotic value of $H$ at large-field value, hence $N_H\simeq \order{1} (N_\uend-N)^2$, meaning that $H$ cannot change by more than a factor of order one throughout the entire inflationary phase. For instance, if one considers the Starobinsky potential of \Eq{eq:starobinsky-pot-example}, one finds $\epsilon_1\simeq 3/[4(N_\uend-N)^2]$ and $H_\uend/H_\mathrm{plateau}\simeq 0.53$. In this case, the de Sitter equilibrium~(\ref{eq:Pstat}), $\langle V(\sigma) \rangle \sim H^4$, only changes by a relatively small fraction and therefore provides a useful estimate for the order of magnitude of spectator field displacements at the end of inflation [using either $H=H_\mathrm{plateau}$ or $H=H_\uend$ in \Eq{eq:Pstat}]. Note that the same can be true for hilltop potentials where $H$ also asymptotes a constant in the infinite past.

In the context of single-field inflation however, plateau potentials are known to provide a good fit to the data only in the last $\sim \! 50$ \efolds of inflation. The shape of the inflaton potential is not constrained beyond this range and is typically expected to receive corrections when the field varies by more than the Planck scale. In multiple-field inflation, observations allow the inflaton potential to be of the large-field type all the way down to the end of inflation~\cite{Vennin:2015egh}. Therefore we also consider monomial inflaton potentials $V(\phi)\propto \phi^p$ with $p>0$. In these models, one has
\bea
\label{eq:Hubble}
H(N)=H_\uend\left[1+\frac{4}{p}\left(N_\uend-N\right)\right]^{\frac{p}{4}}\, .
\eea
If $p>1$, this corresponds to convex inflaton potentials (meaning $V''>0$), while this describes concave inflaton potentials ($V''<0$) for $p<1$, and the de Sitter case is recovered in the limit $p\rightarrow 0$. From \Eq{eq:Hubble}, one has $\epsilon_1=(H_\uend/H)^{4/p}$, so that $N_H=(H/H_\uend)^{4/p}$. If the spectator field has a quadratic potential for instance, as mentioned above, it will be shown in \Sec{sec:plateau_quad_spec} that $N_\mathrm{relax}\sim H^2/m^2$. In this case, the adiabatic condition~(\ref{eq:adiabatic:condition}) reads $(H/H_\uend)^{2/p-1}\gg H_\uend/m$. If $p\geq 2$, one can see that this can never be realised since $H_\uend>m$ and $H>H_\uend$. If $p <2$, the adiabatic condition is satisfied when $H$ is sufficiently large, that is to say at early enough times when $N_\uend-N>p[(H_\uend/m)^{4/(2-p)}-1]/4$. If $m/H_\uend \sim 0.01$ for instance, this number of \efolds is larger than $\sim 400$ as soon as $p>0.1$ (and larger than $\sim 10^7$ for $p>1$), which means that even in this case, the adiabatic regime lies far away from the observable last $50$ \efolds of inflation. One concludes that in most cases, the de Sitter equilibrium solution does not provide a reliable estimate of the field value acquired by spectator fields during inflation. In the following, we therefore study the dynamics of such fields beyond the adiabatic approximation.
\section{\textsf{Quadratic spectator}}
\label{sec:quad_spec}
In this section, we consider a quadratic spectator field, for which
\bea
V(\sigma)=\frac{m^2}{2}\sigma^2\, .
\eea
In this case, the Langevin equation~(\ref{eq:Langevin}) is linear, which allows one to solve it analytically. In \App{sec:quadmoments}, we explain how to calculate the first two statistical moments of the spectator field $\sigma$. The first moment is given by
\bea
\label{eq:meansigma}
\langle\sigma\left(N\right)\rangle = \langle\sigma\left(N_0\right)\rangle\exp\left[-\frac{m^2}{3}\int_{N_0}^N\frac{\dd{N}^\prime}{H^2({N}^\prime)}\right]\,,
\eea
which corresponds to the classical solution of \Eq{eq:Langevin} in the absence of quantum diffusion, and where we have set $\langle \sigma \rangle = \langle \sigma(N_0) \rangle $ at the initial time $N_0$. For the second moment, one obtains
\begin{align}
\left\langle \sigma^2(N)\right\rangle =& \left\langle \sigma^2(N_0)\right\rangle \exp\left[-\frac{2m^2}{3}\int_{N_0}^{N}\frac{\dd N^{\prime}}{H^2(N^{\prime})}\right]
\nonumber \\ &
+ \int_{N_0}^N \dd N^\prime \frac{H^2(N^\prime)}{4\pi^2} \exp\left[\frac{2m^2}{3}\int_N^{N^\prime}\frac{\dd N^{\prime\prime}}{H^2(N^{\prime\prime})}\right] \label{eq:meansigmasquare:final}\, .
\end{align}
In this expression, the structure of the first term in the right-hand side is similar to the first moment~(\ref{eq:meansigma}) while the second term is due to quantum diffusion, so that the variance of the distribution $\langle \sigma^2 \rangle-\langle\sigma\rangle^2$ is given by the same formula as the second moment (\ie one can replace $\langle \sigma^2\rangle$ by $\langle \sigma^2 \rangle-\langle\sigma\rangle^2$ in \Eq{eq:meansigmasquare:final} and the formula is still valid).

One can also show that the Fokker-Planck equation~(\ref{eq:FP}) admits Gaussian solutions,
\bea
\label{eq:ansatz:Gaussian}
P\left(\sigma,N\right)=\frac{1}{\sqrt{2\pi\left\langle \sigma^2(N) \right\rangle}}\exp\left\lbrace-\frac{\left[\sigma-\left\langle\sigma(N)\right\rangle\right]^2}{2\left\langle \sigma^2 \right\rangle}\right\rbrace\, ,
\eea
where $\langle \sigma(N) \rangle$ and  $\langle \sigma^2(N) \rangle$ are given by \Eqs{eq:meansigma} and~(\ref{eq:meansigmasquare:final}) respectively. However, let us stress that \Eqs{eq:meansigma} and~(\ref{eq:meansigmasquare:final}) are valid for any (\ie not only Gaussian) probability distributions.
\subsection{\textsf{Plateau inflation}}
\label{sec:plateau_quad_spec}
As explained in \Sec{sec:validity-stat}, if the inflaton potential is of the plateau type, $H$ can be approximated by a constant. In this case, the mean coarse-grained field~(\ref{eq:meansigma}) is given by
\bea
\left\langle \sigma(N) \right\rangle =
\left\langle \sigma(N_0) \right\rangle
\exp\left[-\frac{m^2}{3H^2}\left(N-N_0\right)\right]
\, .
\eea
It follows the classical trajectory as already pointed out below \Eq{eq:meansigma}, and becomes small when $N-N_0\gg H^2/m^2$. For the second moment, \Eq{eq:meansigmasquare:final} gives rise to
\bea
\left\langle \sigma^2(N) \right\rangle = \left[\left\langle \sigma^2(N_0) \right\rangle - \frac{3H^4}{8\pi^2m^2}\right]
\exp\left[-\frac{2m^2}{3H^2}\left(N-N_0\right)\right] + \frac{3H^4}{8\pi^2m^2} \label{eq:plateau-quadratic-evolve}\, .
\eea
When $N-N_0\gg H^2/m^2$, it approaches the constant value $\langle \sigma^2 \rangle = 3H^4/(8\pi^2 m^2)$. One can check that this asymptotic value corresponds to the de Sitter equilibrium in \Eq{eq:Pstat}. Moreover, one can see that the typical relaxation time that is required to reach the attractor is given by
\bea
\label{eq:quadratic:Nrelax}
N_\mathrm{relax}=\frac{H^2}{m^2}\, ,
\eea
which corresponds to the value reported in \Sec{sec:StochasticInflation}.
\subsection{\textsf{Monomial inflation}}
\label{sec:monomial_quad_spec}
If the inflaton potential is monomial and of the form $V(\phi)\propto \phi^p$, the Hubble factor is given by \Eq{eq:Hubble}. Substituting this expression for $H(N)$ into \Eq{eq:meansigma}, one obtains (for $p\neq 2$)
\bea
\label{eq:meansigma:quadratic:monomial}
\left\langle \sigma(H)\right\rangle = \left\langle \sigma(H_0)\right\rangle \exp\left\lbrace\frac{\mu}{2} \left[\left(\frac{H}{H_\uend}\right)^{\frac{4}{p}-2}-\left(\frac{H_0}{H_\uend}\right)^{\frac{4}{p}-2}\right]\right\rbrace\, ,
\eea
where $H_0$ is the value of $H$ at an initial time $N_0$, and we have defined
\bea
\mu\equiv \frac{m^2}{3H_\uend^2}\frac{p}{2-p}\, .
\eea
In \Eq{eq:meansigma:quadratic:monomial}, time is parametrised by $H$ instead of $N$ for convenience but the two are directly related through \Eq{eq:Hubble}. For the second moment (or for the variance), by substituting \Eq{eq:Hubble} into \Eq{eq:meansigmasquare:final}, one obtains
\begin{align}
&\left\langle \sigma^2(H)\right\rangle = \left\langle \sigma^2(H_0)\right\rangle \ee^{ \mu \left[\left(\frac{H}{H_\uend}\right)^{\frac{4}{p}-2}-\left(\frac{H_0}{H_\uend}\right)^{\frac{4}{p}-2}\right] } +\frac{p H_\uend^2 \mu^{\frac{p+2}{p-2}}}{8\pi^2(p-2)}
\ee^{\mu \left(\frac{H}{H_\uend}\right)^{\frac{4}{p}-2}}
 \nonumber \\ & \qquad \qquad 
\times \left\lbrace \Gamma \left[ \frac{2+p}{2-p}, \mu\left(\frac{H_0}{H_\uend}\right)^{\frac{4}{p}-2}\right] -\Gamma \left[ \frac{2+p}{2-p}, \mu\left(\frac{H}{H_\uend}\right)^{\frac{4}{p}-2}\right] \right\rbrace \,,
\label{eq:meansigma2:quadratic:generic}
\end{align}
where $\Gamma$ denotes the incomplete Gamma function. One can note that both \Eqs{eq:meansigma:quadratic:monomial} and~(\ref{eq:meansigma2:quadratic:generic}) can be expressed as functions $\mu (H/H_\uend)^{4/p-2}$ only, which is directly proportional to the ratio $N_H/N_\mathrm{relax}$. As noted in \Sec{sec:validity-stat}, for $p\geq 2$ this ratio is always small, while for $p<2$, it is large unless $H$ is sufficiently large. The two cases $p\geq2$ and $p<2$ must therefore be treated distinctly.
\subsubsection{\textsf{Case where $p \geq 2$}}
\label{sec:quadratic:pgt2}
If $p>2$, one has  $N_H\ll N_\mathrm{relax}$ and the quantity $\mu (H/H_\uend)^{4/p-2}$ in \Eqs{eq:meansigma:quadratic:monomial} and~(\ref{eq:meansigma2:quadratic:generic}) is always much smaller than one. This implies that the argument of the exponential in \Eq{eq:meansigma:quadratic:monomial} can be neglected, and $\langle \sigma(H) \rangle\simeq \langle \sigma(H_0) \rangle$ stays constant. Therefore, the distribution remains centred at the initial value. Note that the case $p=2$ is singular and gives rise to
\bea
\label{eq:meansigma:quadratic:peq2}
\left\langle \sigma(H)\right\rangle = \left\langle \sigma(H_0) \right\rangle \left(\frac{H}{H_0}\right)^{\frac{m^2}{3H_\uend^2}}\, ,
\eea
which also yields $\langle \sigma(H) \rangle\simeq \langle \sigma(H_0) \rangle$ unless $H_0/H_\uend\gg \exp(3H_\uend^2/m^2)$.

For the second moment, the second arguments of the incomplete Gamma functions in \Eq{eq:meansigma2:quadratic:generic} are always much smaller than one and in this limit, one finds
\bea
\label{eq:meansigma2:quadratic:generic:pGT2}
\left\langle \sigma^2(H)\right\rangle \simeq  \left\langle \sigma^2(H_0)\right\rangle+\frac{H_\uend^2 p}{8\pi^2(p+2)}\left[\left(\frac{H_0}{H_\uend} \right)^{2+\frac{4}{p}}-\left(\frac{H}{H_\uend} \right)^{2+\frac{4}{p}}\right]\, .
\eea
In this expression, one can see that $\langle\sigma^2\rangle$ can only increase as time proceeds, in a way that does not depend on the mass (as long as it is sub-Hubble). The result is therefore the same as if one set the mass to zero, and corresponds to a free diffusion process. This is consistent with the fact that $\langle\sigma\rangle$ stays constant in this case. If $p=2$, \Eq{eq:meansigma2:quadratic:generic} is singular and one has
\begin{align}
\left\langle \sigma^2(H)\right\rangle = & \left\langle \sigma^2(H_0)\right\rangle  \left(\frac{H}{H_0}\right)^{\frac{2m^2}{3H_\uend^2}}
+\frac{H_\uend^2}{8\pi^2\left(2-\frac{m^2}{3H_\uend^2}\right)}\left(\frac{H}{H_\uend}\right)^{\frac{2m^2}{3H_\uend^2}}
 \nonumber \\ & \qquad \qquad \qquad
\times\left[\left(\frac{H_0}{H_\uend}\right)^{4-\frac{2m^2}{3H_\uend^2}}-\left(\frac{H}{H_\uend}\right)^{4-\frac{2m^2}{3H_\uend^2}}\right]
\label{eq:meansigma2:quadratic:peq2} \, .
\end{align}
In this case, it was also shown in \Sec{sec:validity-stat} that $N_H\ll N_\mathrm{relax}$ so there is no adiabatic regime either. Unless $H_0/H_\uend\gg \exp(3H_\uend^2/m^2)$, in the limit $m\ll H_\uend$, \Eq{eq:meansigma2:quadratic:peq2} coincides with \Eq{eq:meansigma2:quadratic:generic:pGT2} evaluated at $p=2$ so in practice the latter formula can be used for all values of $p\geq 2$.

An important feature of \Eq{eq:meansigma2:quadratic:generic:pGT2} is that it strongly depends on the initial conditions $\langle \sigma(H_0)\rangle$ and $H_0$. This is because there is no adiabatic regime in this case and hence no attractor that would erase initial conditions. As a consequence, the typical spectator field displacement at the end of inflation cannot be determined without specifying initial conditions.

One should also note that the present analysis relies on the assumption that the inflaton is not experiencing large stochastic diffusion, which allows us to use \Eq{eq:Hubble}. This is in fact the case if $H\ll H_\mathrm{eternal}$, where
\bea
\label{eq:Hei:def}
H_\mathrm{eternal} \equiv H_\uend \left( \frac{\Mp}{H_\uend}2\pi \sqrt{2} \right)^{\frac{p}{2+p}} \,,
\eea
is the scale above which a regime of so-called ``eternal inflation'' takes place.\footnote
{More precisely, $H_\mathrm{eternal}$ is defined~\cite{Winitzki:2008zz} as the scale above which, over the typical time scale of an $e$-fold, the mean quantum diffusion received by the inflaton field, $H/(2\pi)$, is larger than the classical drift, $\sqrt{2\epsilon_1}\Mp$. Since $\epsilon_1=(H_\uend/H)^{4/p}$ in monomial inflation~(\ref{eq:Hubble}), this condition gives rise to $H>H_\mathrm{eternal}$ where $H_\mathrm{eternal}$ is given by \Eq{eq:Hei:def}.}
For this reason, $H_\mathrm{eternal}$ is the largest value one can use for $H_0$ in order for the calculation to be valid. Setting $H_0=H_\mathrm{eternal}$, and substituting \Eq{eq:Hei:def} into \Eq{eq:meansigma2:quadratic:generic:pGT2}, one obtains at the end of inflation
\bea
\label{eq:quadratic:sigmaend:pgt2}
\left\langle \sigma^2_\uend \right\rangle \simeq \left\langle \sigma^2_\mathrm{eternal} \right\rangle + \frac{p}{p+2}\Mp^2 \, .
\eea
This expression is displayed in the left panel of \Fig{fig:quadratic:summary}. It means that the field value of the spectator field is at least of the order of the Planck mass at the end of inflation. If one assumes the de Sitter equilibrium distribution~(\ref{eq:Pstat}) at the end of eternal inflation for instance, $\langle \sigma^2_\mathrm{eternal} \rangle = 3H_\mathrm{eternal}^4/(8\pi^2 m^2)$, even much larger field displacements are obtained at the end of inflation.
\subsubsection{\textsf{Case where $p < 2$}}
\label{sec:quadratic:plt2}
\begin{figure}
\begin{center}
\includegraphics[width=6.269cm]{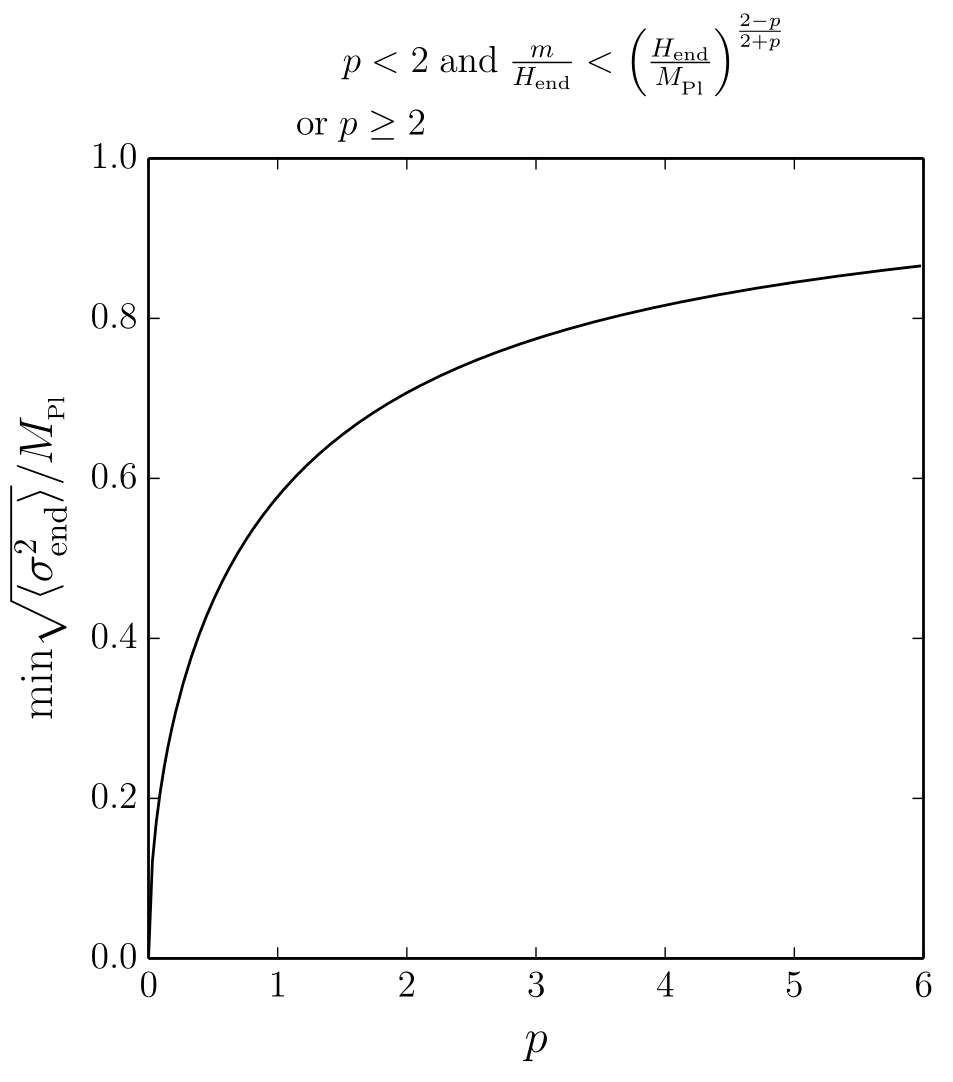}
\includegraphics[width=6.566cm]{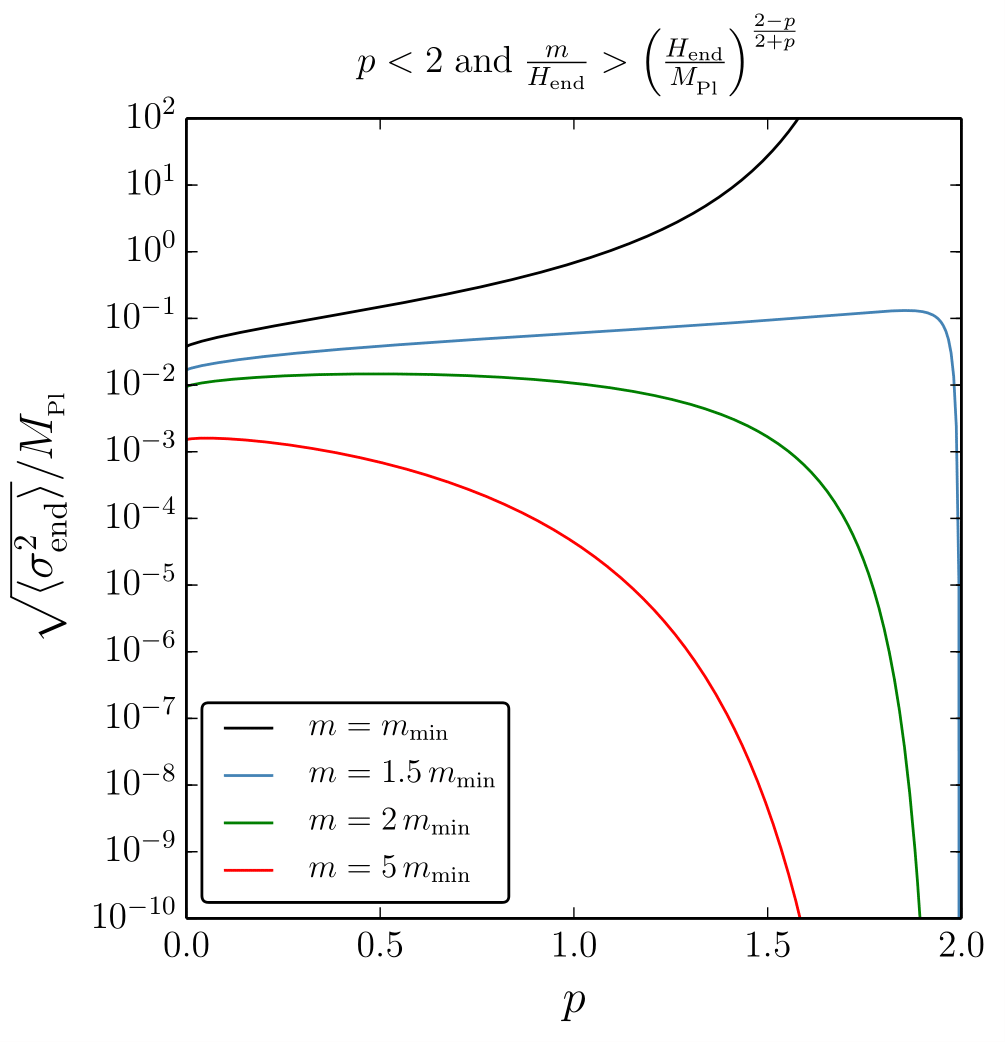}
\caption[Quadratic spectator typical field displacement]{The typical field displacement $\sqrt{\langle\sigma_\uend^2 \rangle}$ acquired by a test field $\sigma$ with quadratic potential $V(\sigma)=m^2\sigma^2/2$ at the end of an inflationary phase driven by an inflaton potential $V(\phi)\propto\phi^p$. In the left panel, the cases $p\geq 2$ and $p<2$ with $m/H_\uend<(H_\uend/\Mp)^{(2-p)/(2+p)}$ are displayed, where the minimum value of $\langle\sigma_\uend^2 \rangle$ is given if one initially sets $\langle\sigma^2 \rangle=0$ at the time $H=H_\mathrm{eternal}$ when stochastic corrections to the inflaton dynamics stop being large. This corresponds to \Eq{eq:quadratic:sigmaend:pgt2} and shows that spectator fields are typically at least close to super-Planckian at the end of inflation in these cases. In the right panel, the case $p<2$ with $m/H_\uend>(H_\uend/\Mp)^{(2-p)/(2+p)}$ is displayed, where there is an early adiabatic regime that allows the dependence on initial condition to be erased. The typical field displacement is given by \Eq{eq:quadratic:sigmaend:plt2}, which is expressed as a function of $p$ and $m/m_\umin$ in \Eq{eq:meansigma2:quadratic:plt2:adiabatic:HadiabHeternal}, where $m_\umin$ is the lower bound on $m$ associated to the condition $m/H_\uend>(H_\uend/\Mp)^{(2-p)/(2+p)}$. One can check that, as soon as  $m \gtrsim 1.5\, m_\umin$,  $\sqrt{\langle\sigma_\uend^2 \rangle}$  is always sub-Planckian in this case.
\label{fig:quadratic:summary}}
\end{center}
\end{figure}
If $p<2$, whether the ratio $N_H/N_\mathrm{relax}$ is small or large depends on the value of $H$. More precisely, if $H \gg H_\mathrm{adiab}$, where
\bea
\label{eq:Hadiab:def}
H_\mathrm{adiab} \equiv H_\uend \left(\frac{H_\uend}{m}\right)^{\frac{p}{2-p}}\, ,
\eea
one is in the adiabatic regime and $N_H\gg N_\mathrm{relax}$. As soon as $H$ drops below $H_\mathrm{adiab}$ however, one leaves the adiabatic regime. In order to set initial conditions during the adiabatic regime, it should apply after the eternal inflationary phase during which our calculation does not apply, which implies that $H_\mathrm{adiab}<H_\mathrm{eternal}$. Making use of \Eqs{eq:Hei:def} and~(\ref{eq:Hadiab:def}), this condition gives rise to
\bea
\label{eq:quadratic:adiabstart:condition}
\frac{m}{H_\uend}>\left(\frac{H_\uend}{\Mp}\right)^{\frac{2-p}{2+p}}\, .
\eea
Let us distinguish the two cases where this relation is and is not satisfied.
\paragraph{\textsf{Starting out in the adiabatic regime}}\mbox{} \\
If \Eq{eq:quadratic:adiabstart:condition} is satisfied, one can set initial conditions for the spectator field $\sigma$ in the adiabatic regime while being outside the eternal inflationary phase, that is to say one can take $H_\mathrm{adiab}<H_0<H_\mathrm{eternal}$. From \Eq{eq:meansigma:quadratic:monomial}, this implies that $\langle \sigma_\uend \rangle \ll \langle \sigma_0\rangle$ and the distribution becomes centred around smaller field values as time proceeds. Regarding the width of the distribution, two regimes of interest need to be considered.

At early time, \ie when $H\gg H_\mathrm{adiab}$, the incomplete Gamma functions in \Eq{eq:meansigma2:quadratic:generic} can be expanded in the large second argument limit and one obtains
\begin{align}
\langle \sigma^2(H) \rangle \simeq \left[\langle \sigma^2(H_0) \rangle - \frac{3H_0^4}{8\pi^2 m^2}\right] \ee^{\mu\left[\left(\frac{H}{H_\uend}\right)^{\frac{4}{p}-2}-\left(\frac{H_0}{H_\uend}\right)^{\frac{4}{p}-2}\right]} + \frac{3H^4}{8\pi^2m^2} \label{eq:meansigma2:quadratic:plt2:adiabatic} \, .
\end{align}
In this expression, one can see that as soon as $H$ decreases from $H_0$, the first term is exponentially suppressed and one obtains $\langle \sigma^2 \rangle \simeq 3H^4/(8\pi m^2)$, which corresponds to the de Sitter equilibrium formula\footnote
{More precisely, in a de Sitter universe where $H$ is constant and equal to the instantaneous value $H(N)$ for a given $N$ in the case at hand, the asymptotic value reached by $\langle\sigma^2\rangle$ at late time is the same as the instantaneous value  $\langle\sigma^2(N)\rangle$ obtained from \Eq{eq:meansigma2:quadratic:plt2:adiabatic}. In this sense, the time evolution of $H$ can be neglected and this corresponds, by definition, to an adiabatic regime.}
and confirms that one is in the adiabatic regime. This also shows that the de Sitter equilibrium is an attractor of the stochastic dynamics in this case, and that it is reached within a number of \efolds $\sim H_0^2/m^2$, which exactly corresponds to $N_\mathrm{relax}$ given in \Eq{eq:quadratic:Nrelax} when $H=H_0$.

At later times, \ie when $H\ll H_\mathrm{adiab}$, one leaves the adiabatic regime and while the first incomplete Gamma function in \Eq{eq:meansigma2:quadratic:generic} can still be expanded in the large second argument limit, the second one must be expanded in the small second argument limit and this gives rise to
\bea
\label{eq:quadratic:sigmaend:plt2}
\left\langle \sigma^2(H) \right\rangle \simeq  \frac{H_\uend^2}{8\pi^2}\frac{p}{2-p}\Gamma\left(\frac{2+p}{2-p}\right)\left(\frac{3H_\uend^2}{m^2}\frac{2-p}{p}\right)^{\frac{2+p}{2-p}}\, .
\eea
Interestingly, this expression does not depend on $H$, meaning that $\langle \sigma^2 \rangle$ stays constant as soon as one leaves the adiabatic regime (and obviously stops tracking the adiabatic solution). One can also check that in this expression, the limit $p\rightarrow 0$ gives rise to $\langle \sigma^2_\uend\rangle \simeq 3H_\uend^4/(8\pi^2m^2)$, that is to say the de Sitter equilibrium formula.

An important consequence of this result is that in the case $p<2$ and if $m>m_\umin$, where $m_\umin$ corresponds to the lower bound on $m$ given by \Eq{eq:quadratic:adiabstart:condition}, even if the end of inflation lies far outside the adiabatic regime, the existence of an early adiabatic phase allows initial conditions to be erased. At the end of inflation, the field value of the spectator field only depends on $m$, $H_\uend$ and $p$. This is in contrast with the case $p\geq 2$ where there is no adiabatic regime, even at early time, and initial conditions remain important even at the end of inflation. A second important consequence is that the typical field displacement is always sub-Planckian at the end of inflation in this case. Indeed, substituting the expression given for $m_\umin$ by \Eq{eq:quadratic:adiabstart:condition} into \Eq{eq:quadratic:sigmaend:plt2}, one obtains
\bea
\label{eq:meansigma2:quadratic:plt2:adiabatic:HadiabHeternal}
\frac{\left\langle\sigma^2_\uend\right\rangle}{\Mp^2}=\frac{1}{8\pi^2}\left(\frac{p}{2-p}\right)^{\frac{2p}{p-2}}\Gamma\left(\frac{2+p}{2-p}\right)\left(3 \frac{m_\umin^2}{m^2}\right)^{\frac{2+p}{2-p}}\, .
\eea
This expression is displayed in the right panel of \Fig{fig:quadratic:summary} for a few values of $m/m_\umin$. One can see that as soon as $m\gtrsim 1.5\, m_\umin$, the spectator field is always sub-Planckian at the end of inflation.
\paragraph{\textsf{Starting out away from the adiabatic regime}}\mbox{} \\
If the condition~(\ref{eq:quadratic:adiabstart:condition}) is not satisfied, the adiabatic regime cannot be used to erase initial conditions dependence. If both $H_0$ and $H$ are much smaller than $H_\mathrm{adiab}$, the incomplete Gamma functions in \Eq{eq:meansigma2:quadratic:generic} can be expanded in the small second argument limit and one obtains \Eq{eq:meansigma2:quadratic:generic:pGT2} again. When $H$ becomes small compared to $H_0$, $\langle\sigma^2\rangle$ reaches a constant and the distribution remains frozen until the end of inflation. Letting $H_0=H_\mathrm{eternal}$ as in \Sec{sec:quadratic:pgt2}, this gives rise to \Eq{eq:quadratic:sigmaend:pgt2} and one concludes that, in this case, the spectator field acquires a super-Planckian field value at the end of inflation.
\mbox{}\\
\mbox{}\\
\indent The situation is summarised in the first line of table~\ref{table:summary} in \Sec{sec:conclusions}. If $p<2$ and $m/H_\uend>(H_\uend/\Mp)^{(2-p)/(2+p)}$, quadratic spectator fields acquire sub-Planckian field values at the end of inflation, while if $p\geq 2$ or if $p<2$ with $m/H_\uend<(H_\uend/\Mp)^{(2-p)/(2+p)}$, they are typically super-Planckian.
\subsection{\textsf{Can a spectator field drive a second phase of inflation?}}
\begin{figure}
\begin{center}
\includegraphics[width=6.5cm]{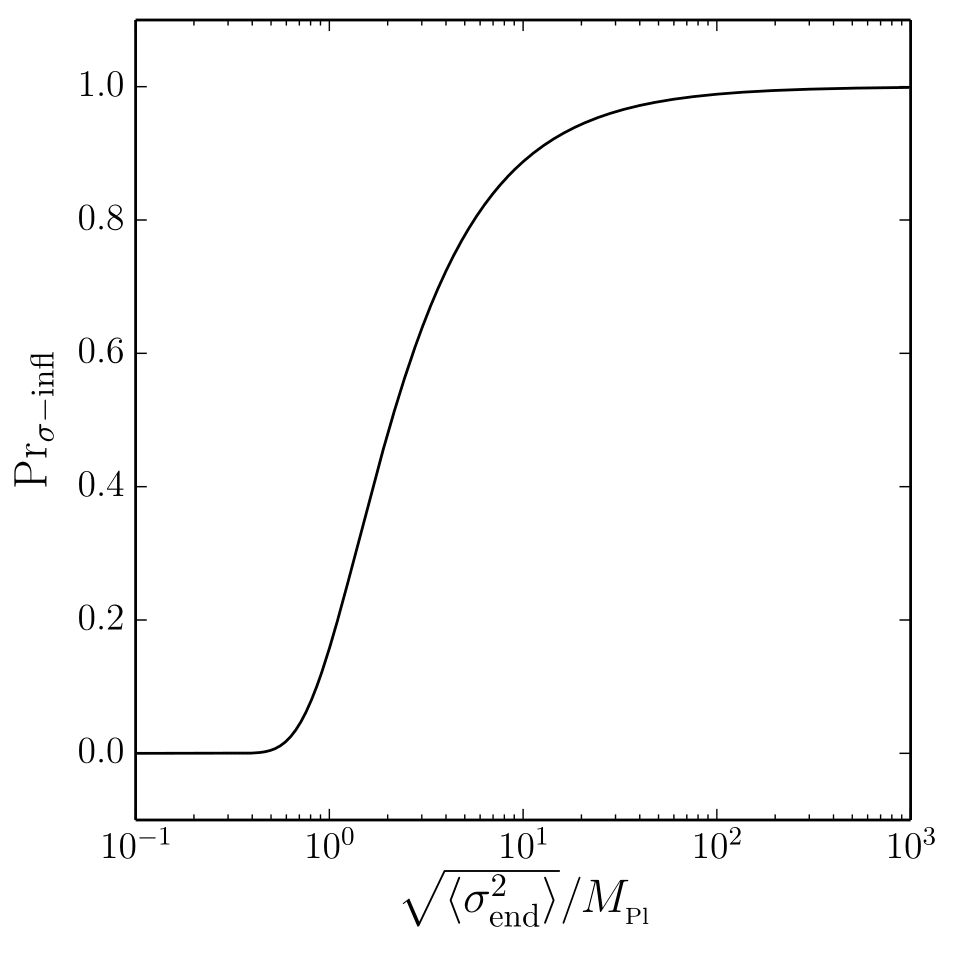}
\includegraphics[width=6.5cm]{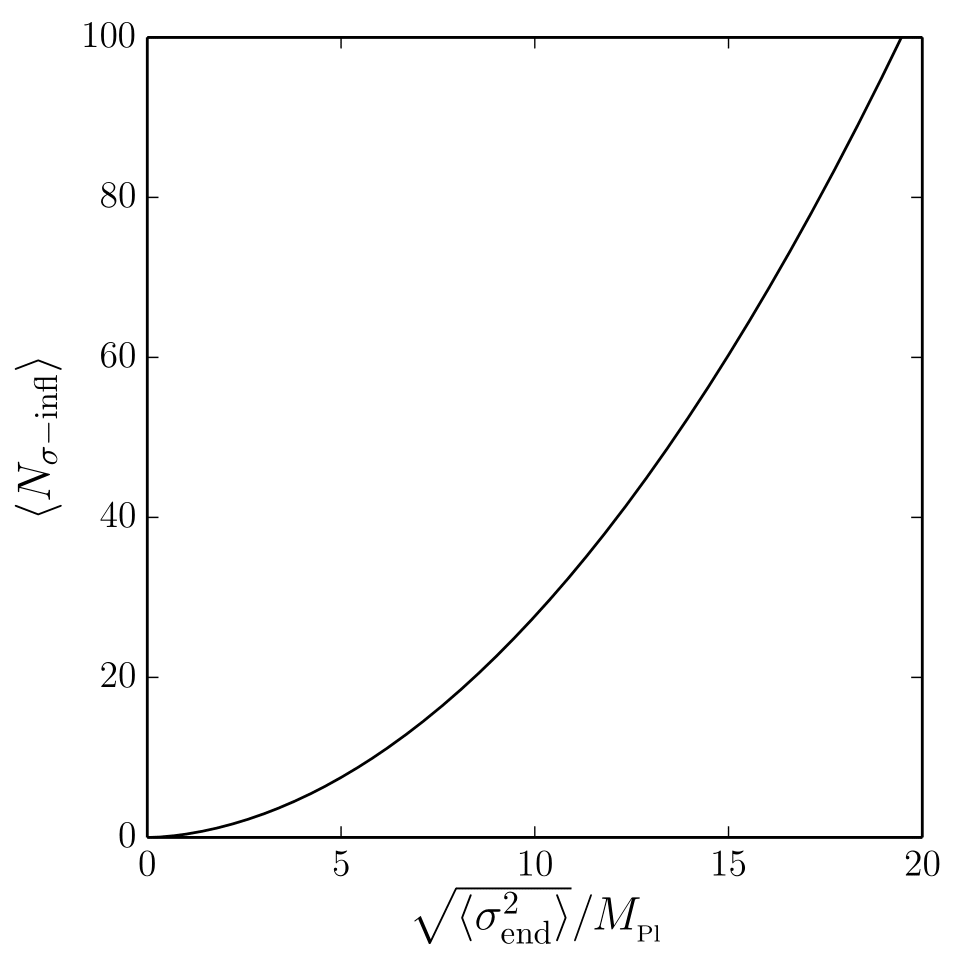}
\caption[Probability of second inflationary phase]{A quadratic spectator field $\sigma$ can trigger a second phase of inflation if $\vert \sigma_\uend \vert>\sqrt{2}\Mp$. Assuming a centred Gaussian distribution with variance $\langle\sigma_\uend^2\rangle$, the left panel displays the probability for such a condition to be satisfied, while the mean number of \efolds realised in the second phase of inflation is given in the right panel.
\label{fig:quadratic:SecondInflation}}
\end{center}
\end{figure}
If inflation is driven by a monomial potential $V\propto\phi^p$ with $p\geq2$, in \Sec{sec:monomial_quad_spec} it was shown that quadratic spectator fields typically acquire super-Planckian field values at the end of inflation. This can have important consequences as discussed in \Sec{sec:SpectatorExamples}, amongst which is the ability for the spectator field to drive a second phase of inflation. This can happen if $\vert \sigma_\uend \vert>\sqrt{2}\Mp$, and the probability associated to this condition is given by
\begin{align}
\mathrm{Pr}_{\sigma\text{-}\mathrm{infl}}= \int_{\left\vert\sigma\right\vert > \sqrt{2}\Mp}
P (\sigma,N_\uend)\dd \sigma = \erfc \left( \frac{\Mp}{\sqrt{\langle \sigma_\uend^2 \rangle}} \right)\, .
\end{align}
In the second expression, we have assumed that the probability distribution of the spectator field value at the end of inflation is a Gaussian with vanishing mean and variance $\langle\sigma_\uend^2\rangle$, and $\erfc$ denotes the complementary error function. This probability is displayed in the left panel of \Fig{fig:quadratic:SecondInflation}. If a second phase of inflation starts driven by the quadratic potential with initial field value $\sigma_\uend$, then the number of \efolds realised is given by $\sigma_\uend^2/(4\Mp^2)-1/2$. The mean duration of this additional inflationary period can thus be calculated according to
\begin{align}
\langle N_{\sigma\text{-}\mathrm{infl}} \rangle&=   \frac{1}{\mathrm{Pr}\left(\sigma\text{-}\mathrm{infl}\right)}\displaystyle\int_{\left\vert\sigma\right\vert > \sqrt{2}\Mp} \left(\frac{\sigma^2}{4\Mp^2}-\frac{1}{2}\right) P (\sigma,N_\uend)\dd \sigma
\nonumber \\ &=
\left( \frac{\langle \sigma_\uend^2\rangle}{4\Mp^2} - \frac{1}{2}\right) +
\frac{\sqrt{\langle \sigma_\uend^2 \rangle}}{2\sqrt{\pi}\Mp}
\frac{\exp \left( -\frac{\Mp^2}{\langle \sigma_\uend^2\rangle}\right)}{\erfc \left( \frac{\Mp}{\sqrt{\langle \sigma_\uend^2 \rangle}} \right)} \, ,
\end{align}
where in the second expression, again, we have assumed that the probability distribution of the spectator field value at the end of inflation is a centred Gaussian. This mean number of \efolds is shown in the right panel of \Fig{fig:quadratic:SecondInflation}. When $\langle \sigma_\uend^2 \rangle$ is super-Planckian, one has a non-negligible probability of a second phase of inflation. For instance, with $\sqrt{\langle \sigma_\uend^2 \rangle }=5\Mp$, one finds $\mathrm{Pr}\left(\sigma\text{-}\mathrm{infl}\right)\simeq 0.77$ and $\langle N_{\sigma\text{-}\mathrm{infl}}\rangle = 7.5$.

\section{\textsf{Quartic spectator}}
\label{sec:quart_spec}
In \Sec{sec:quad_spec}, it was shown that quadratic spectator fields with potential $V(\sigma)=m^2\sigma^2/2$ typically acquire super-Planckian field displacements at the end of inflation if the inflaton potential is of the form $V(\phi)\propto\phi^p$ with $p\geq 2$ at large-field values or with $p<2$ and $m/H_\uend<(H_\uend/\Mp)^{(2-p)/(2+p)}$. In this section, we investigate whether these super-Planckian field values can be tamed by making the spectator field potential steeper at large-field values. In practice, we consider a quartic spectator field,
\bea
\label{eq:quartic:potential}
V(\sigma)=\lambda \sigma^4\, ,
\eea
where $\lambda$ is a dimensionless constant. Contrary to the quadratic case in \Sec{sec:quad_spec}, the Langevin equation~(\ref{eq:Langevin}) is not linear for quartic spectators and cannot be solved analytically. Numerical solutions are therefore presented in this section, where a large number (typically $10^5$ or $10^6$) of realisations of \Eq{eq:Langevin} are generated with a fourth order Runge-Kutta method, over which moments of the spectator field value are calculated at fixed times. These results have been checked with independent numerical solutions of the Fokker-Planck equation~(\ref{eq:FP}).
\subsection{\textsf{Plateau inflation}}
\label{sec:quartic:plateau}
As explained in \Sec{sec:validity-stat}, if the inflaton potential is of the plateau type, $H$ can be approximated by a constant and the spectator field value reaches the de Sitter equilibrium~(\ref{eq:Pstat}) where the typical field displacement, for the quartic spectator potential~(\ref{eq:quartic:potential}), is given by
\bea
\label{eq:quartic:deSitter}
\left\langle\sigma^2 \right\rangle = \frac{\Gamma\left(\frac{3}{4}\right)}{\Gamma\left(\frac{1}{4}\right)}\sqrt{\frac{3}{2\lambda}} \frac{H^2}{2\pi}\, .
\eea
The relaxation time required to reach this asymptotic value can be assessed as follows. Since the equilibrium~(\ref{eq:Pstat}) is of the form $P(\sigma)\propto e^{-\alpha \sigma^4}$, with $\alpha=8\pi^2\lambda/(3H^4)$, let us assume that the time evolving distribution for $\sigma$ is more generally given by
\bea
\label{eq:quartic:ansatz}
P(\sigma,N)=
\frac{2 \alpha^{1/4}(N)}{\Gamma\left(\frac{1}{4}\right)}\exp\left[-\alpha(N)\sigma^4\right]\, ,
\eea
where $\alpha(N)$ is a free function of time and the prefactor is set so that the distribution remains normalised, and track the stochastic dynamics with this ansatz. By substituting \Eq{eq:quartic:ansatz} into \Eq{eq:FP}, an ordinary differential equation for $\alpha(N)$ is derived in \App{sec:nonlin_drift}, that reads
\begin{equation}
\label{eq:quartic:quarticappr:alpha:eom}
\frac{\dd \alpha}{\dd N} =  \frac{\Gamma \left( \frac{1}{4} \right)}{2\Gamma \left( \frac{3}{4} \right)} \left( \frac{\lambda}{H^2}\alpha^{1/2} - \frac{3H^2}{8\pi^2} \alpha^{3/2}\right) \,.
\end{equation}
If $H$ is a constant, this equation can be solved analytically and the solution is given by \Eq{eq:solution:alpha}. Since \Eq{eq:quartic:ansatz} gives rise to $\langle \sigma^2 \rangle= \alpha^{-1/2} \Gamma(3/4)/\Gamma(1/4)$, one obtains for the second moment
\begin{align}
& \left\langle \sigma^2(N) \right\rangle = \nonumber \\
& \quad \dfrac{\dfrac{\Gamma\left(\frac{3}{4}\right)}{\Gamma\left(\frac{1}{4}\right)}\sqrt{\frac{3H^4}{8\pi^2\lambda}} }{{\rm tanh}\left\lbrace \sqrt{\dfrac{3\lambda}{2}} \dfrac{\Gamma\left(\frac{1}{4}\right)}{8\pi \Gamma\left(\frac{3}{4}\right)} (N-N_0) +{\rm atanh} \left[\sqrt{\dfrac{3H^4 }{8\pi^2\lambda}}\dfrac{\Gamma\left(\frac{3}{4}\right)}{\Gamma\left(\frac{1}{4}\right)\left\langle \sigma^2(N_0)\right\rangle}\right]  \right\rbrace} \label{eq:quartic:plateau:Appr} \,.
\end{align}
In the late time limit, one recovers the de Sitter equilibrium value~(\ref{eq:quartic:deSitter}). Let us stress however that \Eq{eq:quartic:plateau:Appr} is not an exact solution to \Eq{eq:FP} but only provides an approximation under the ansatz~(\ref{eq:quartic:ansatz}). This approximation will be shown to be reasonably accurate in \Sec{sec:quartic:monomial}, but for now, expanding $\tanh(x)\simeq 1-2 \ee^{-2x}$ when $x\gg 1$ at late time, it provides an estimate of the relaxation time as
\bea
\label{eq:quartic:Nrelax}
N_\mathrm{relax} =  \frac{1}{\sqrt{\lambda}}\, .
\eea
It is interesting to notice that this expression is consistent with the numerical exploration of \Ref{Enqvist:2012xn}, see Eq.~(2.12) of this reference.
\subsection{\textsf{Monomial inflation}}
\label{sec:quartic:monomial}
If the inflaton potential is monomial and of the form $V(\phi)\propto \phi^p$, the Hubble factor is given by \Eq{eq:Hubble} and varies over time scales of order $N_H=(H/H_\uend)^{4/p}$ as explained in \Sec{sec:validity-stat}. Making use of \Eq{eq:quartic:Nrelax}, the adiabatic condition $N_H\gg N_\mathrm{relax}$ then requires $H\gg H_\mathrm{adiab}$, where
\bea
\label{eq:quartic:Hadiab}
H_\mathrm{adiab} \equiv \lambda^{-p/8} H_\uend\, .
\eea
A fundamental difference with the quadratic spectator is that in the quartic case, for all values of $p$, there always exists an adiabatic regime at early times. However, it is not guaranteed  that this regime is consistent with the classical inflaton solution~(\ref{eq:Hubble}), \ie extends beyond the eternal inflationary phase. This is the case only if $H_\mathrm{adiab}<H_\mathrm{eternal}$, where $H_\mathrm{eternal}$ is given in \Eq{eq:Hei:def}, that is to say if $\lambda$ is large enough,
\bea
\label{eq:quartic:adiabatic:condition}
\lambda>\left(\frac{H_\uend}{\Mp}\right)^{\frac{8}{p+2}}\, .
\eea
Let us distinguish the case where this condition is satisfied and one can use the stationary solution~(\ref{eq:Pstat}) to describe the distribution in the adiabatic regime independently of initial conditions, and the case where this is not possible.
\begin{figure}
\begin{center}
\includegraphics[width=9cm]{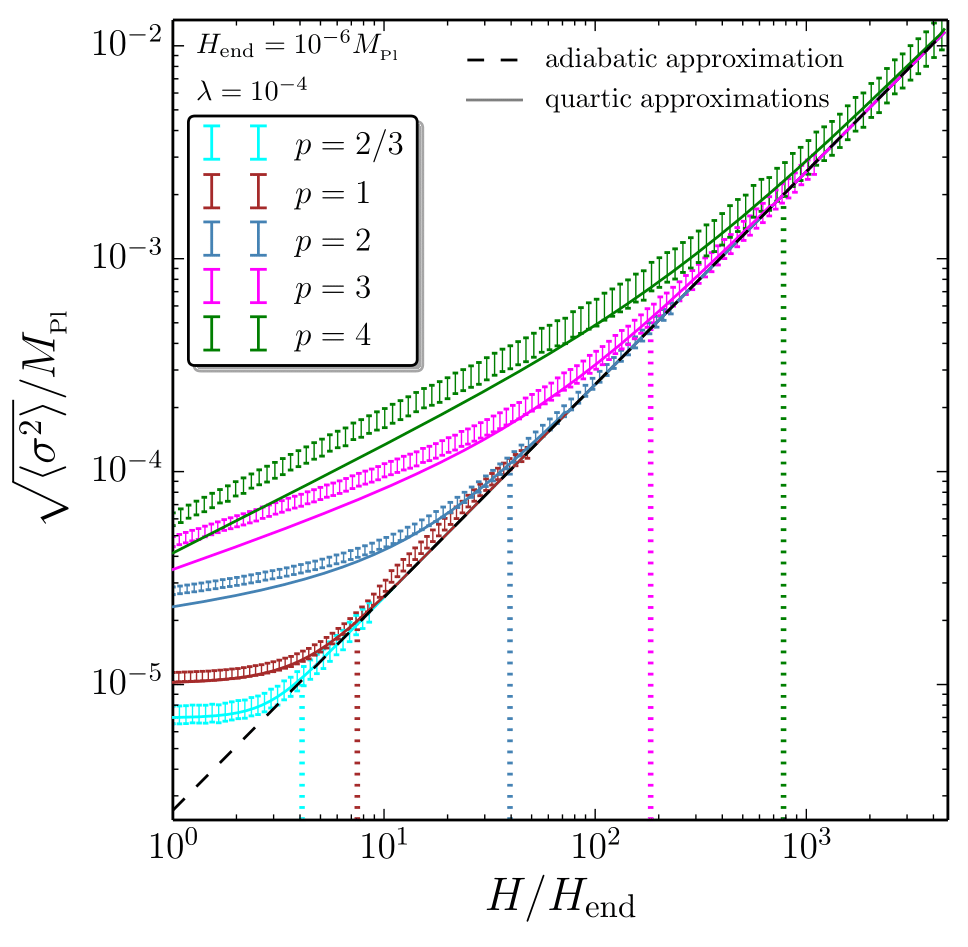}
\caption[Quartic spectator typical field displacement]{Standard deviation $\sqrt{\langle \sigma^2 \rangle}$ of the distribution of a quartic spectator field with potential $V(\sigma)=\lambda\sigma^4$ with $\lambda=10^{-4}$, as a function of time parametrised by the Hubble scale $H$ (time flows from the right to the left). The inflaton potential is of the monomial type $V(\phi)\propto\phi^p$, with $H_\uend=10^{-6}\Mp$. The coloured \textsc{i} symbols correspond to numerical solutions of the Langevin equation where $10^{5}$ realisations of \Eq{eq:Langevin} have been produced for the values of $p$ displayed in the legend. The centres of the vertical bars correspond to ensemble averages of $\sigma^2$ while the heights of the bars are statistical noise estimates (due to having a finite number of realisations only) obtained from the jackknife resampling technique. The realisations are initially drawn according to the adiabatic distribution~(\ref{eq:Pstat}). The black dashed line corresponds to this adiabatic value~(\ref{eq:quartic:deSitter}) for $\langle\sigma^2\rangle$. The coloured dashed vertical lines denote the value of $H$ such that the argument of the Bessel functions in \Eq{eq:analytic_quart} equals one, which corresponds to $H_\mathrm{adiab}$ given by \Eq{eq:quartic:Hadiab} up to an order one prefactor. One can see that when $H$ drops below $H_\mathrm{adiab}$, the numerical solutions depart from the de Sitter equilibrium, denoting the end of the adiabatic regime. Finally, the coloured solid lines correspond to the quartic approximation~(\ref{eq:analytic_quart}).\label{fig:quartic:sigmaend:time}}
\end{center}
\end{figure}

\subsubsection{\textsf{Starting out in the adiabatic regime}}
If the condition~(\ref{eq:quartic:adiabatic:condition}) is satisfied, one can set initial conditions for the spectator field  $\sigma$ in the adiabatic regime after the eternal inflationary phase. In \Fig{fig:quartic:sigmaend:time}, we present the results of a numerical integration of the Langevin equation~(\ref{eq:Langevin}) in this case (with the values used for $H_\uend$ and $\lambda$, one can check that \Eq{eq:quartic:adiabatic:condition} is satisfied up to $p=10$). The values of $H_\mathrm{adiab}$ given by \Eq{eq:quartic:Hadiab} are denoted by the vertical coloured dashed lines. When $H\gg H_\mathrm{adiab}$, the numerical results follow the de Sitter stationary solution~(\ref{eq:quartic:deSitter}) represented by the black dashed line. When $H$ drops below $H_\mathrm{adiab}$, this is not the case anymore, and the distributions are wider at the end of inflation than the adiabatic approximation would naively suggest.

In this regime, the behaviour of $\langle \sigma^2\rangle$ can in fact still be tracked analytically by making use of the quartic ansatz~(\ref{eq:quartic:ansatz}) introduced in \Sec{sec:quartic:plateau}. Indeed, in the case where $H$ is given by \Eq{eq:Hubble}, one can cast \Eq{eq:quartic:quarticappr:alpha:eom} into a Ricatti equation and in \App{sec:nonlin_drift} it is shown that its solution reads
\bea
 \label{eq:analytic_quart}
\langle \sigma^2 (H)\rangle = \frac{\Gamma\left(\frac{3}{4}\right)}{\Gamma\left(\frac{1}{4}\right)}\sqrt{\frac{3}{2\lambda}} \frac{H^2}{2\pi} \dfrac{K_{\frac{p}{4}+\frac{1}{2}}\left[\frac{p}{4\pi}\sqrt{\frac{\lambda}{6}}\frac{\Gamma \left( \frac{1}{4}\right)}{\Gamma \left( \frac{3}{4}\right)} \left( \frac{H}{H_\uend}\right)^{4/p}\right]}{K_{\frac{p}{4}-\frac{1}{2}}\left[\frac{p}{4\pi}\sqrt{\frac{\lambda}{6}}\frac{\Gamma \left( \frac{1}{4}\right)}{\Gamma \left( \frac{3}{4}\right)} \left( \frac{H}{H_\uend}\right)^{4/p}\right] } \,.
\eea
In this expression, $K$ is a modified Bessel function of the second kind. One can note that the argument of the Bessel functions is directly proportional to $N_H/N_\mathrm{relax}$, confirming that this ratio controls the departure from the adiabatic solution~(\ref{eq:quartic:deSitter}). At early times when $N_H \gg N_\mathrm{relax}$, or equivalently $H\gg H_\mathrm{adiab}$, one can expand the Bessel functions in the large argument limit, $K_\alpha(x)\simeq \sqrt{\pi/(2x)}\ee^{-x}$, and one recovers the adiabatic approximation~(\ref{eq:quartic:deSitter}). The formula~(\ref{eq:analytic_quart}) is displayed in \Fig{fig:quartic:sigmaend:time} with the solid coloured lines. One can see that even when $H<H_\mathrm{adiab}$, it still provides a reasonable approximation to the numerical solutions. One can also notice that the lower $p$ is, the better this quartic approximation. At the end of inflation, $N_H/N_\mathrm{relax} = \sqrt{\lambda} \ll 1$, so the Bessel functions can be expanded in the small argument limit, which depends on the sign of the index of the Bessel function.\footnote
{In the limit $x\ll 1$, if $\alpha<0$, $K_\alpha(x)\simeq \Gamma(-\alpha) 2^{-1-\alpha}x^\alpha$, if $\alpha>0$, $K_\alpha(x)\simeq \Gamma(\alpha) 2^{\alpha-1} x^{-\alpha}$ and if $\alpha=0$, $K_\alpha(x)\simeq \ln(2/x)-\gamma$, where $\gamma\simeq 0.577$ is the Euler constant~\cite{Abramovitz:1970aa}.}
Because the index of the Bessel function in the denominator of \Eq{eq:analytic_quart} is proportional to $p-2$, this leads to different results whether $p$ is smaller or larger than $2$, namely
\bea
\left\langle \sigma_\uend^2\right\rangle \simeq
\left\lbrace
\begin{array}{lcc}
 \dfrac{\Gamma\left(\frac{1}{2}+\frac{p}{4}\right)}{\Gamma\left(\frac{1}{2}-\frac{p}{4}\right)}
\left[\sqrt{\dfrac{3}{2}}\dfrac{\Gamma\left(\frac{3}{4}\right)}{\Gamma\left(\frac{1}{4}\right)}\right]^{1+\frac{p}{2}}
\dfrac{\left(\frac{16\pi}{p}\right)^{\frac{p}{2}}}{2\pi}
\dfrac{H_\uend^2}{\lambda^{\frac{1}{2}+\frac{p}{4}}} & \quad\quad\quad & \mathrm{if}\ p<2\\
& & \\
 \dfrac{6 }{\bar{\gamma}-\ln(\lambda)} \dfrac{\Gamma^2\left(\frac{3}{4}\right)}{\Gamma^2\left(\frac{1}{4}\right)} \dfrac{H_\uend^2}{\lambda}& \quad\quad\quad & \mathrm{if}\ p=2\\
& & \\
 \left(3-\frac{6}{p}\right)\dfrac{\Gamma^2\left(\frac{3}{4}\right)}{\Gamma^2\left(\frac{1}{4}\right)}\dfrac{H_\uend^2}{\lambda}& \quad\quad\quad &\mathrm{if}\ p>2
\end{array}
\right.\quad ,
\label{eq:quartic:sigmaend:adiab}
\eea
where we have defined $\bar{\gamma}\equiv 2\ln[4\pi\sqrt{6}\Gamma(3/4)/\Gamma(1/4)]-2\gamma\simeq 3.53$, where $\gamma$ is the Euler constant. Ignoring the overall constants of order one, if $p\geq 2$, one finds $\langle\sigma_\uend^2\rangle \sim H_\uend^2/\lambda$, and if $p<2$, $\langle\sigma_\uend^2\rangle \sim H_\uend^2/\lambda^{1/2+p/4}$. This needs to be compared to the de Sitter case~(\ref{eq:quartic:deSitter}) where $\langle\sigma_\uend^2\rangle \sim H_\uend^2/\sqrt{\lambda}$. In monomial inflation,  $\langle\sigma_\uend^2\rangle$ is therefore larger than in plateau inflation for the same value of $H_\uend$, by a factor $\lambda^{-p/4}$ if $p<2$ and $\lambda^{-1/2}$ if $p\geq 2$. One should also note that the condition~(\ref{eq:quartic:adiabatic:condition}) for the adiabatic regime to extend beyond the eternal inflationary phase can be substituted into \Eq{eq:quartic:sigmaend:adiab} and gives rise to $\sqrt{\langle\sigma_\uend^2\rangle}/\Mp \ll (H_\uend/\Mp)^{(p-2)/(p+2)}$ if $p\geq 2$ and $\sqrt{\langle\sigma_\uend^2\rangle}/\Mp \ll 1$ if $p < 2$. In both cases, the spectator field displacement at the end of inflation is therefore sub-Planckian.
\subsubsection{\textsf{Starting out away from the adiabatic regime}}
\begin{figure}
\begin{center}
\includegraphics[width=7cm]{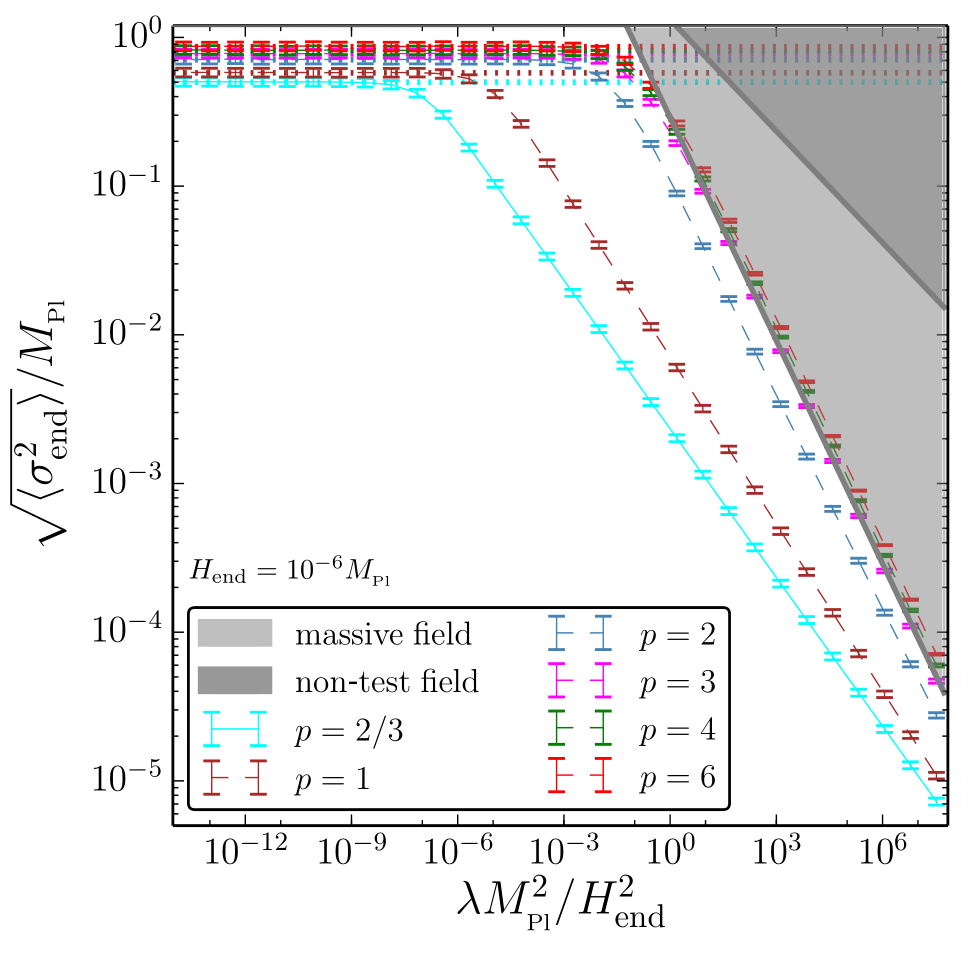}
\includegraphics[width=7cm]{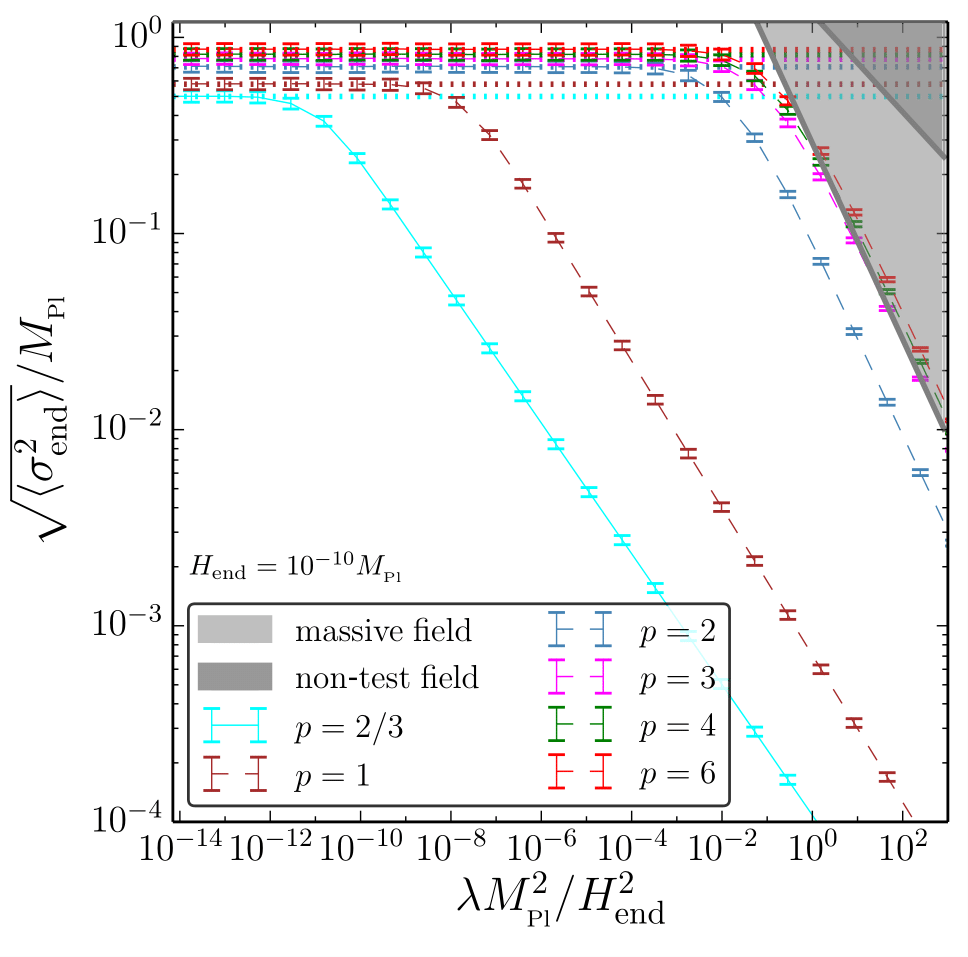}
\caption[Quartic spectator field displacement at the end of inflation]{Field displacement $\sqrt{\langle \sigma^2 \rangle}$ acquired by a quartic spectator field with potential $V(\sigma)=\lambda\sigma^4$ at the end of inflation, as a function of $\lambda \Mp^2/H_\uend^2$, for $H_\uend=10^{-6}\Mp$ (left panel) and  $H_\uend=10^{-10}\Mp$ (right panel).  The inflaton potential is of the monomial type $V(\phi)\propto\phi^p$. The coloured \textsc{i} symbols correspond to numerical solutions of the Langevin equation where $10^{5}$ realisations of \Eq{eq:Langevin} have been produced for the values of $p$ displayed in the legend. The centres of the vertical bars correspond to ensemble averages of $\sigma^2$ while the heights of the bars are statistical noise estimates (due to having a finite number of realisations only) obtained from the jackknife resampling technique. All realisations are initiated with $\sigma=0$ at $H=H_\mathrm{eternal}$. The horizontal dashed lines correspond to \Eq{eq:quadratic:sigmaend:pgt2} to which the numerical results asymptote in the limit $\lambda\rightarrow 0$. The pale grey region corresponds to $m_\ueff>H$ where the spectator field is not light and our calculation does not apply, and the dark region stands for $\lambda \sigma_\uend^4>3\Mp^2H_\uend^2$ where $\sigma$ cannot be considered as a spectator field anymore.
\label{fig:quartic:sigmaend:nonadiabatic}}
\end{center}
\end{figure}
If the condition~(\ref{eq:quartic:adiabatic:condition}) is not satisfied, the adiabatic regime lies entirely within the eternal inflationary phase and cannot be used to erase initial conditions. In this case, the spectator field displacement at the end of inflation is thus strongly dependent on initial conditions at the start of the classical inflaton evolution. In this section, we derive a lower bound on $\langle\sigma_\uend^2\rangle$, assuming that it vanishes when $H=H_\mathrm{eternal}$ and solving the subsequent stochastic dynamics numerically. The result is presented in \Fig{fig:quartic:sigmaend:nonadiabatic} where $\langle\sigma_\uend^2\rangle$ is displayed as a function of $\lambda\Mp^2/H_\uend^2$ for $H_\uend=10^{-6}\Mp$ (left panel) and for $H_\uend=10^{-10}\Mp$ (left panel). The two cases $p\geq 2$ and $p<2$ must be treated separately.
\paragraph{\textsf{Case where $p\geq 2$}}\mbox{}\\
If $p\geq 2$, it was shown in \Sec{sec:quadratic:pgt2} that a light quadratic spectator field always acquires a super-Planckian field value at the end of inflation. The mean effective mass of the quartic spectator field is given by
 \bea
 \label{eq:quartic:effectivemass}
m_\ueff^2 = 12 \lambda \left\langle \sigma^2 \right\rangle \, ,
 \eea
and is smaller than $H_\uend$ for $\sqrt{\langle\sigma_\uend^2\rangle}\sim \Mp$ if $\lambda<H_\uend^2/\Mp^2$. This explains why, in \Fig{fig:quartic:sigmaend:nonadiabatic}, in the regime $\lambda<H_\uend^2/\Mp^2$, one recovers \Eq{eq:quadratic:sigmaend:pgt2} that is displayed with the horizontal coloured lines, and which shows that the spectator field acquires a super-Planckian field value in this case. Otherwise, if $H_\uend^2/\Mp^2<\lambda<(H_\uend/\Mp)^{8/(p+2)}$ [the upper bound coming from breaking the inequality~(\ref{eq:quartic:adiabatic:condition})], one can see in \Fig{fig:quartic:sigmaend:nonadiabatic} that the field displacement can be made sub-Planckian, but that its effective mass becomes of order $H$.\footnote{Strictly speaking, the present calculation does not apply when the effective mass of the spectator field is of order $H$ or larger. However, if the effects of the mass were taken into account, the amplitude of the noise term in \Eq{eq:Langevin} would not be $H/(2\pi)$ but would become smaller as $m_\ueff$ approaches $H$. This would result in a smaller value for $\langle\sigma^2\rangle$, hence for $m_\ueff$, and therefore a larger noise amplitude. One can expect the two effects to compensate for a value of $m_\ueff$ around $H$.} In this regime, the spectator field cannot be considered as light anymore.
\paragraph{\textsf{Case where $p< 2$}}\mbox{}\\
If $p< 2$, it was shown in \Sec{sec:quadratic:plt2} that a quadratic spectator field acquires a super-Planckian field value at the end of inflation if its mass is smaller than $H_\uend (H_\uend/\Mp)^{(2-p)/(2+p)}$, see \Eq{eq:quadratic:adiabstart:condition}. When evaluated at the Planck scale, the effective mass~(\ref{eq:quartic:effectivemass}) of the quartic spectator field is smaller than this threshold when $\lambda < (H_\uend/\Mp)^{8/(2+p)}$, which exactly corresponds to breaking the inequality~(\ref{eq:quartic:adiabatic:condition}). One can check in \Fig{fig:quartic:sigmaend:nonadiabatic} that when $\lambda < (H_\uend/\Mp)^{8/(2+p)}$, one does indeed recover \Eq{eq:quadratic:sigmaend:pgt2} which is displayed with the horizontal dashed coloured lines. One concludes that in this case, the spectator field always acquires a field value at least of order the Planck mass at the end of inflation.
\mbox{}\\
\mbox{}\\
\indent The situation is summarised in the second line of table~\ref{table:summary} in \Sec{sec:conclusions}. If $\lambda>(H_\uend/\Mp)^{8/(p+2)}$, the spectator field is sub-Planckian at the end of inflation. Otherwise, if $p\geq 2$, either the spectator field is super-Planckian or not light at the end of inflation, and if $p<2$, it is always super-Planckian. Considering the quadratic spectator discussed in \Sec{sec:quad_spec} where it was shown that super-Planckian field displacements are usually generated at the end of inflation, one thus concludes that an additional self-interacting term $\lambda\sigma^4$ in the potential can render the field value sub-Planckian if $\lambda$ is large enough, namely if $\lambda>(H_\uend/\Mp)^{8/(p+2)}$. One can check that for such a value of $\lambda$, if $V(\sigma)=m^2\sigma^2/2+\lambda\sigma^4$ with $m<H_\uend$, the quartic term always dominates over the quadratic one when $\sigma\sim\Mp$, which is consistent.
\section{\textsf{Axionic spectator}}
\label{sec:axion_spec}
In \Sec{sec:quad_spec}, it was shown that quadratic spectator fields with potential $V(\sigma)=m^2\sigma^2/2$ typically acquire super-Planckian field displacements at the end of inflation if the inflaton potential is of the form $V(\phi)\propto\phi^p$ with $p\geq 2$ at large-field value or with $p<2$ and $m/H_\uend<(H_\uend/\Mp)^{(2-p)/(2+p)}$. In \Sec{sec:quart_spec}, we discussed how adding a quartic self-interaction term in the potential could help to tame these super-Planckian values. In this section, we investigate another possibility, which consists in making the field space compact and of sub-Planckian extent. This is typically the case for axionic fields, with periodic potentials of the type
\bea
\label{eq:axionpot}
V(\sigma) = \Lambda^4\left[ 1- \cos \left( \frac{\sigma}{f} \right)\right]\,.
\eea
In this expression, $\Lambda$ and $f$ are two mass scales that must satisfy $\Lambda^2<fH_\uend$ in order for the curvature of the potential to remain smaller than the Hubble scale throughout inflation, \ie for the axionic field to remain light, which we will assume in the following.
\subsection{\textsf{Plateau inflation}}
\label{sec:axionic:plateau}
As explained in \Sec{sec:validity-stat}, if the inflaton potential is of the plateau type, $H$ can be approximated by a constant and the spectator field value reaches the de Sitter equilibrium~(\ref{eq:Pstat}). If $H\gg \Lambda$, such a distribution is approximately flat, in which case $\langle\sigma^2\rangle\simeq \pi^2 f^2/3$ if $\sigma$ is restricted to one period of the potential~(\ref{eq:axionpot}). In this regime, the classical drift due to the potential gradient in \Eq{eq:Langevin} can be neglected and the spectator field experiences a free diffusion process. The relaxation time is therefore the time it takes to randomise $\sigma$ over the period of the potential and is given by $N_\mathrm{relax}\simeq (\pi^2f/H)^2$. In the opposite limit when $H\ll \Lambda$, the distribution is localised close to the minimum of the potential where it can be approximated by a quadratic function $V(\sigma)\simeq m^2\sigma^2/2$ with mass $m^2=\Lambda^4/f^2$. In this case, according to \Sec{sec:plateau_quad_spec}, one has $\langle \sigma^2 \rangle = 3H^4f^2/(8\pi^2\Lambda^4)$, and the relaxation time is of order $N_\mathrm{relax}=H^2/m^2\simeq H^2f^2/\Lambda^4$.
\begin{figure}
\begin{center}
\includegraphics[width=9cm]{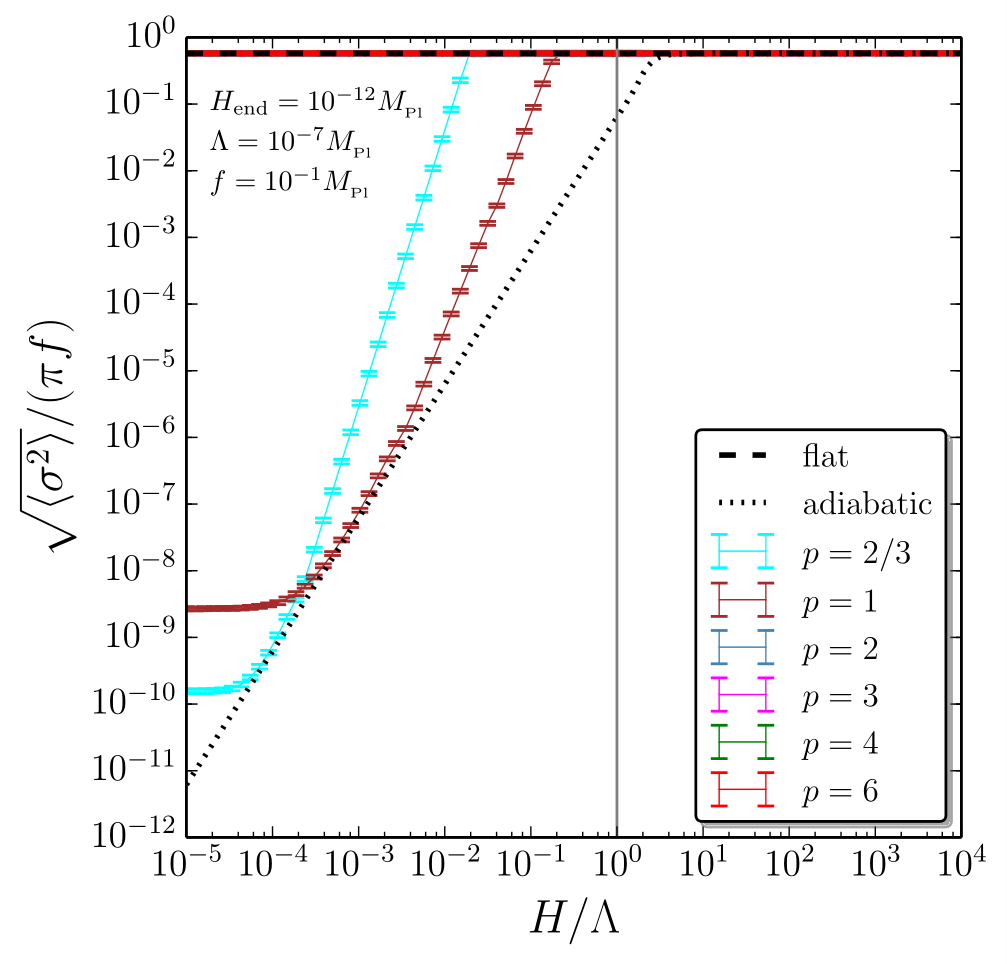}
\caption[Axionic spectator typical field displacement]{Standard deviation $\sqrt{\langle \sigma^2 \rangle}$ of the distribution of an axionic spectator field with potential $V(\sigma)=\Lambda^4[1-\cos(\sigma/f)]$ with $\Lambda=10^{-7}\Mp$ and $f=10^{-1}\Mp$, as a function of time parametrised by the Hubble scale $H$ (time flows from the right to the left). The inflaton potential is of the monomial type $V(\phi)\propto\phi^p$, with $H_\uend=10^{-12}\Mp$. The coloured \textsc{i} symbols correspond to numerical solutions of the Langevin equation where $10^{6}$ realisations of \Eq{eq:Langevin} have been produced for the values of $p$ displayed in the legend. The centres of the vertical bars correspond to ensemble averages of $\sigma^2$ while the heights of the bars are statistical noise estimates (due to having a finite number of realisations only) obtained from the jackknife resampling technique. The realisations are initially drawn according to a flat distribution when $H/\Lambda=10^4$. The black dashed line  corresponds to the standard deviation of a distribution that is flat over one period of the potential, $\langle \sigma^2\rangle=\pi^2f^2/3$.  The black dotted line corresponds to the adiabatic solution~(\ref{eq:Pstat}), which remains flat when $H$ is larger than $\Lambda$, represented by the grey vertical line. When $p\geq 2$ the distributions remain flat until the end of inflation (and one cannot distinguish the different values of $p$ that are superimposed). When $p<2$, the distributions narrow down once $H\ll \Lambda$ since the parameters have been chosen to satisfy \Eq{eq:axionic:adiabatic:condition}.
\label{fig:axionic:sigmaend:time}}
\end{center}
\end{figure}

\subsection{\textsf{Monomial inflation}}
\label{sec:axionic:monomial}

If inflation is realised by a monomial potential $V(\phi)\propto \phi^p$, there is always an epoch when $H>\Lambda$ in the past and during which the spectator field distribution is made flat within a number of \efolds of order $N_\mathrm{relax}\simeq (\pi^2f/H)^2$. Therefore, contrary to the quadratic and to the quartic spectators, the field displacement of an axionic spectator at the end of inflation is always independent of initial conditions, provided that inflation lasts long enough. If $\Lambda<H_\uend$, the distribution remains flat until the end of inflation. In the opposite case, when $H$ drops below $\Lambda$, the subsequent dynamics of $\sigma$ depends on whether $p\geq 2$ or $p<2$.
\subsubsection{\textsf{Case where $p\geq 2$}}
\label{sec:axionic:pgt2}
If $p\geq 2$, in \Sec{sec:quad_spec} it was shown that the evolution of a quadratic field with mass $m<H_\uend$ is effectively described by a free-diffusion process where the potential drift can be neglected. For an axionic spectator, the potential is always flatter than its quadratic expansion around its minimum and can therefore also be neglected. As a consequence, the distribution remains flat until the end of inflation and one finds $\langle\sigma_\uend^2\rangle\simeq \pi^2 f^2/3$.
\subsubsection{\textsf{Case where $p< 2$}}
\label{sec:axionic:plt2}
If $p<2$, in \Sec{sec:quad_spec} it was shown that the distribution of a quadratic field with mass $m<H_\uend$ tracks the adiabatic equilibrium until $H=H_\mathrm{adiab}$, where $H_\mathrm{adiab}$ is given by \Eq{eq:Hadiab:def}, and remains frozen afterwards. This implies that an axionic spectator distribution narrows down from a flat profile if $H_\mathrm{adiab}<\Lambda$, which gives rise to
\bea
\label{eq:axionic:adiabatic:condition}
\frac{\Lambda}{H_\uend}>\left(\frac{f}{H_\uend}\right)^{\frac{p}{p+2}}\, .
\eea
Notice that for this condition to be compatible with the light-field prescription given below \Eq{eq:axionpot}, one must have $H_\uend<f$ for $p<2$ (which makes sense, otherwise the distribution would be randomised over one \efold~even towards the end of inflation). In this case, $\langle \sigma^2\rangle$ settles down to $3 H_\mathrm{adiab}^4/(8\pi^2 m^2)$, which gives rise to
\bea
\label{eq:axionic:adiabatic:sigmaend}
\sqrt{\left\langle \sigma_\uend^2 \right\rangle} \simeq \sqrt{\frac{3}{2}}\frac{H_\uend}{2\pi}\left(\frac{H_\uend f}{\Lambda^2}\right)^{\frac{2+p}{2-p}}\, .
\eea
If \Eq{eq:axionic:adiabatic:condition} is not satisfied however, the field distribution remains flat until the end of inflation and one has $\langle\sigma_\uend^2\rangle\simeq \pi^2 f^2/3$.
\mbox{}\\
\mbox{}\\
\indent
In order to check the validity of these considerations, in \Fig{fig:axionic:sigmaend:time} we present numerical solutions of the Langevin equation~(\ref{eq:Langevin}). When $p\geq2$, one can check that the distributions remain flat until the end of inflation. The values of the parameters $\Lambda$, $f$ and $H_\uend$ have been chosen to satisfy \Eq{eq:axionic:adiabatic:condition}, which explains why for $p<2$, the distributions narrow down once $H$ drops below $\Lambda$ (otherwise, we have checked that even when $p<2$, the distributions remain flat). However, one can see that when the distributions start moving away from the flat configuration, they do not exactly follow the adiabatic solution displayed with the black dotted line, even though $H>H_\mathrm{adiab}$. This is because in the above discussion, we have approximated the axionic potential with its quadratic expansion around its minimum, which is not strictly valid at the stage where the distribution is still flat and sensitive to the full potential shape. Nonetheless, the distributions converge towards the adiabatic profile at later time and the final value of $\langle \sigma^2 \rangle$ is well described by \Eq{eq:axionic:adiabatic:sigmaend}.

The situation is summarised in the third line of table~\ref{table:summary} in \Sec{sec:conclusions}. If $H_\uend > \Lambda$, $p\geq 2$, or $p<2$ with $\Lambda<H_\uend (f/H_\uend)^{p/(p+2)}$, the distribution of the axionic spectator remains flat until the end of inflation and $\sqrt{\smash[b]{\langle\sigma_\uend^2}\rangle }= \pi f/\sqrt{3}$. Only if $p<2$ with $\Lambda>H_\uend (f/H_\uend)^{p/(p+2)}$ does the distribution narrow down and $\sqrt{\smash[b]{\langle\sigma_\uend^2\rangle}} \simeq H_\uend(H_\uend f /\Lambda^2)^{(2+p)/(2-p)} $. In all cases, if $f$ is sub-Planckian, the typical field displacement obviously remains sub-Planckian as well.

\section{\textsf{Non-minimally coupled spectator}}
\label{sec:nonmin_coup_spec}

In this section we extend the calculations made in \Sec{sec:quad_spec} to include the existence of a non-minimal coupling to gravity. During inflation, scalar fields can radiatively generate a non-minimal coupling between themselves and the background~\cite{Callan:1970ze, Freedman:1974gs}. One then may expect such fields to appear with an effective potential of the form
\begin{equation} \label{eq:potential-nm-spec}
\ V(\sigma ) = \frac{1}{2}( m^2 + \xi R )\sigma^2 \,,
\end{equation}
where $\xi$ is the non-minimal coupling strength and, during inflation, the Ricci scalar is $R=6(2-\epsilon_1)H^2$ ($H$ being the Hubble parameter and here we will assume that $\epsilon_1$ is negligible during inflation) and, hence, so long as $\vert \xi \vert \ll 1/12$ the spectator can remain light and acquire a non-zero variance during inflation. We also note here that it has recently been checked that neglecting metric fluctuations of the background is consistent with the known behaviour of a spectator field in the Einstein frame~\cite{Markkanen:2017dlc}. Thus, at this level, we may confidently ignore differences between Jordan and Einstein frames. 

As we have already discussed in \Sec{sec:quad_spec} the limit where the non-minimal coupling is negligible, we shall now take the opposite limit $\xi R \gg m^2$ such that \Eq{eq:potential-nm-spec} during inflation becomes
\begin{equation} \label{eq:unbounded-pot}
\ V(\sigma ) = 6\xi H^2 \sigma^2 \,.
\end{equation}
Given \Eq{eq:unbounded-pot} and following the same reasoning as was used to obtain \Eq{eq:meansigmasquare:final}, one may find the variance with the implicit solution
\begin{align}
\left\langle \sigma^2\right\rangle &= \left\langle \sigma^2(N_0)\right\rangle \ee^{-8\xi \left( N-N_0\right)} + \int_{N_0}^N \dd N^\prime \frac{H^2(N^\prime )}{4\pi^2} \ee^{ -8\xi \left( N-N_0\right) } \label{eq:meansigmasquare:final:nmcoupling}\,.
\end{align}
We note here that, as in the minimally-coupled quadratic case, the Fokker-Planck equation (\Eq{eq:FP}) for this case admits stationary Gaussian solutions. The variance computed from \Eq{eq:meansigmasquare:final:nmcoupling} is hence sufficient to characterise the entire stationary (and near-stationary) PDF. Upon complete departure from equilibrium, however, the PDF can deviate from Gaussianity.

\subsection{\textsf{Plateau inflation}}

In a plateau inflationary background, $H$ does not evolve in time and one immediately finds the following explicit solution to \Eq{eq:meansigmasquare:final:nmcoupling} 
\begin{equation} \label{eq:stat-sol-nm}
\left\langle \sigma^2 \right\rangle =  \left[ \left\langle \sigma^2(N_0) \right\rangle - \frac{H^2}{32\pi^2 \xi } \right] \ee^{- 8\xi \left( N-N_0\right)} +  \frac{H^2}{32\pi^2 \xi } \,,
\end{equation}
hence in the limit where $N-N_0 \gg 1/\xi$ we find that $\left\langle \sigma^2 \right\rangle = H^2/(32\pi^2\xi)$. From \Eq{eq:stat-sol-nm} we may also read off the relaxation timescale
\begin{equation} \label{eq:Nrelax-nm-spec}
\ N_{\rm relax} = \frac{1}{\xi }\,.
\end{equation}

\subsection{\textsf{Monomial inflation}}

Going beyond a plateau inflationary background, if $N_{\rm relax}<N_H$ (where we remind the reader that $N_H=(H/H_\uend )^{4/p}$ in a monomial background with power $p$, i.e., $V(\phi ) \propto \phi^p$) the system may relax after each successive time step, and hence we may use the solution quoted in \Eq{eq:stat-sol-nm}. If $N_{\rm relax}>N_H$, however, we must understand how the system changes in time. 

Using $H(N)$ in a monomial background, as given in \Eq{eq:Hubble}, and rewriting \Eq{eq:meansigmasquare:final:nmcoupling} in terms of $H$ as the time variable, the solution to \Eq{eq:meansigmasquare:final:nmcoupling} is
\begin{align} \label{eq:general:variance:nm-coupling}
\left\langle \sigma^2(H) \right\rangle &= \left\langle \sigma^2(H_0)\right\rangle \ee^{ 2p\xi \left[ \left( \frac{H}{H_\uend}\right)^{\frac{4}{p}}-\left( \frac{H_0}{H_\uend}\right)^{\frac{4}{p}} \right] } + \frac{H_\uend^2}{4\pi^2} \dfrac{ \ee^{2p\xi\left( \frac{H}{H_\uend} \right)^{\frac{4}{p}}}}{2^{\frac{p}{2}+3}p^{\frac{p}{2}}\xi^{\frac{p}{2}+1}} \nonumber \\
& \quad \times \left\{ \Gamma \left[ \frac{p}{2}+1 ; 2p\xi\left(\frac{H}{H_\uend}\right)^{\frac{4}{p}}\right] - \Gamma \left[ \frac{p}{2}+1 ; 2p\xi \left( \frac{H_0}{H_\uend}\right)^{\frac{4}{p}} \right] \right\} \,.
\end{align}
We can see from \Eq{eq:general:variance:nm-coupling} that the non-minimally coupled spectator follows the same qualitative behaviour as in the quartic case: at early times there is always an adiabatic regime, however this can be so early as to be beyond the self-reproducing regime of inflation. Hence, using \Eq{eq:Nrelax-nm-spec}, the condition analogous to \Eq{eq:quartic:adiabatic:condition} that ensures an adiabatic initial condition is
\begin{equation} \label{eq:adiab-cond-nm}
\xi > \left( \frac{H_\uend}{\Mp}\right)^{\frac{4}{p+2}} \,.
\end{equation}
Given that \Eq{eq:adiab-cond-nm} holds and that the field is light ($\xi \ll 1/12$, as discussed before), we can expand \Eq{eq:general:variance:nm-coupling} in the late-time limit to find the variance. Taking the $H\rightarrow H_\uend$ limit, the initial variance $\left\langle \sigma^2(H_0)\right\rangle$ has been washed away and the second incomplete Gamma function in \Eq{eq:general:variance:nm-coupling} becomes negligible, leaving us with a small second-argument expansion of the first incomplete Gamma function\footnote{The small argument limit of the (upper) incomplete Gamma function can be obtained by rewriting it as a combination of the Gamma function $\Gamma [a]$ in the first argument and the lower incomplete Gamma function $\gamma [a;x]$
\begin{equation*}
\lim_{x\rightarrow 0} \left\{ \Gamma [a;x]  \right\} = \lim_{x\rightarrow 0} \left\{ \Gamma [a] - \gamma [a;x]  \right\} \rightarrow \Gamma [a] - \frac{x^a}{a} \,.
\end{equation*}}
\begin{equation} \label{eq:nonminimal-variance-extras}
\left\langle \sigma^2_\uend \right\rangle \simeq \frac{H^2_\uend}{8\pi^2\xi^{\frac{p}{2}+1}} \frac{\Gamma \left( \frac{p}{2}+1\right)}{2^{\frac{p}{2}+2}p^{\frac{p}{2}}}  \,.
\end{equation}
This expression coincides with stationary limit of \Eq{eq:stat-sol-nm} when $p\rightarrow 0$, which is consistent with the expectation that, in the same limit, monomial inflation approaches a plateau.

The overall picture for the non-minimally coupled spectator in a monomial background is as follows. For values of $\xi < (H_\uend/\Mp )^{\frac{4}{p+2}}$, the spectator will never follow the de Sitter equilibrium distribution and hence will acquire an initial variance that will be unchanged from the point at which it becomes light during inflation. For values of $( H_\uend/\Mp )^{\frac{4}{p+2}} < \xi < 1/12$, the spectator will follow the de Sitter equilibrium distribution at early times, followed by a transition away from this solution at the point when $N_{\rm relax} = N_H \, \Leftrightarrow \, H = \xi^{-\frac{p}{4}}H_\uend $.

\section{\textsf{Information retention from initial conditions}}
\label{sec:sigmaendpriors}
When calculating the field value acquired by spectator fields at the end of inflation, we have found situations in which initial conditions are erased by the existence of an adiabatic regime at early times, and situations in which this is not the case. In this section, building from \Sec{sec:info-theory-tools}, we propose to quantify this memory effect using information theory in order to better describe the amount of information about early time physics (potentially pre-inflationary) available in the final field displacements of spectator fields. As in \Eq{eq:app:dkl}, the relative information between two distributions $P_1(\sigma)$ and $P_2(\sigma)$ can be measured using the Kullback-Leibler divergence~\cite{kullback1951} $\dkl$,
\bea
\dkl\left(P_1 \vert\vert P_2 \right) \equiv \int^{\infty}_{-\infty} {P_1}\left(\sigma \right) \log_2 \left[\frac{{P_1}\left(\sigma \right)}{P_2 \left(\sigma \right)} \right] \dd \sigma\, .
\label{eq:DKL-spectator}
\eea
It is invariant under any reparametrisation $\sigma^\prime = f(\sigma)$, and since it uses a logarithmic score function as in the Shannon's entropy, it is a well-behaved measure of information~\cite{bernardo:2008}. Considering two initial distributions separated by an amount of information $\delta\dkl^0$, giving rise to two final distributions separated by $\delta\dkl^\uend$, we define the information retention criterion by
\bea
\label{eq:calI:def}
\mathcal{I}\equiv\frac{\delta\dkl^\uend}{\delta\dkl^0}\, .
\eea
When $\mathcal{I}<1$, the initial information is contracted by the dynamics of the distributions. This is typically the case when there is an attractor, or an adiabatic regime, which tends to erase the initial conditions dependence of final states. When $\mathcal{I}>1$, the initial information is amplified and the final state is sensitive to initial conditions. Values of $\mathcal{I}\gg 1$ might signal the presence of chaotic dynamics in which case initial conditions are difficult to infer. For this reason, $\mathcal{I}=\order{1}$ represents an optimal situation in terms of initial conditions reconstruction. In practice, $\mathcal{I}$ depends both on the initial (or final) state around which the infinitesimal variation is performed, and on the direction in the space of distributions along which it is performed.
\begin{figure}
\centering
\includegraphics[width=7cm]{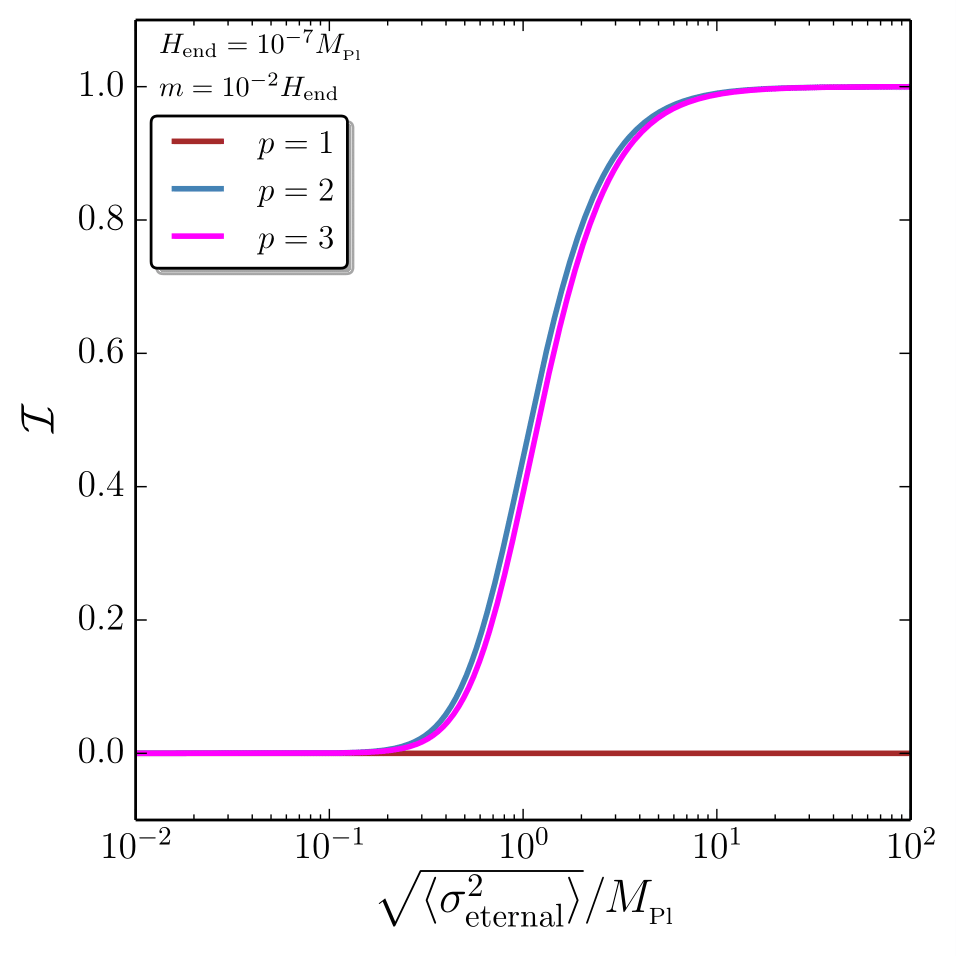}
\includegraphics[width=7cm]{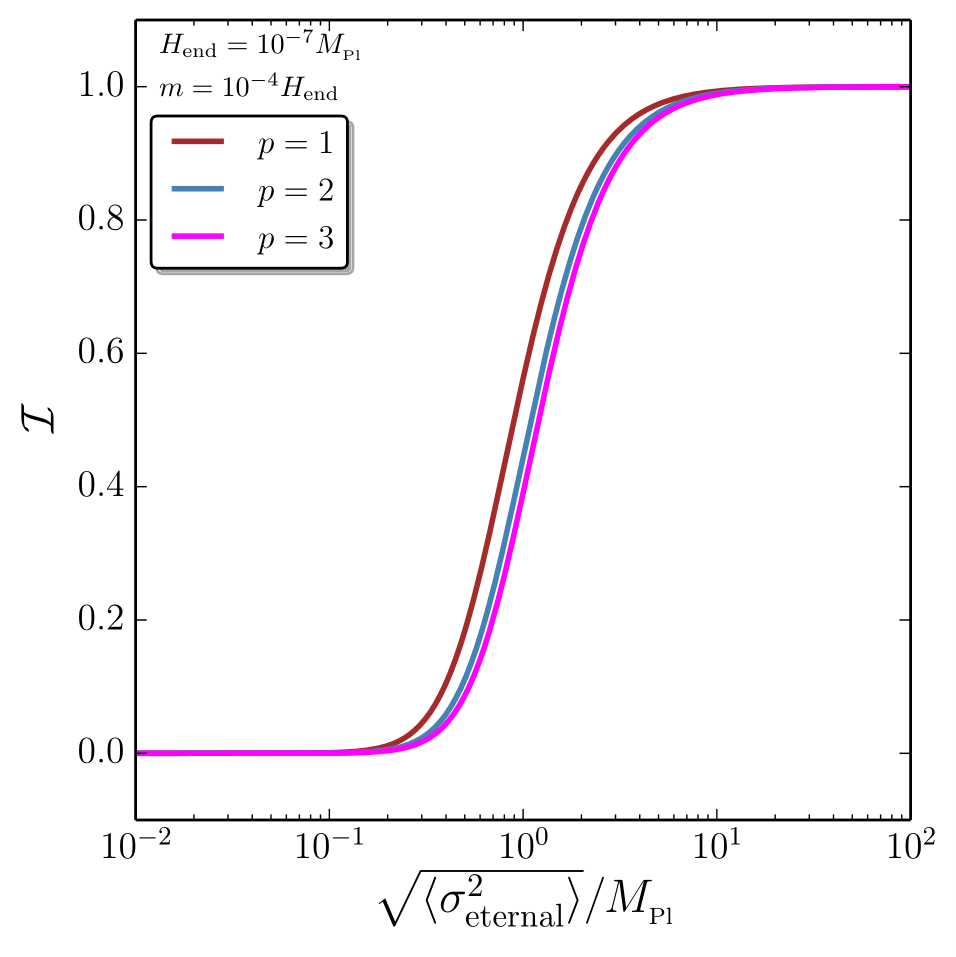}
\caption[Spectator information retention]{\label{fig:quadratic_information} Information retention criterion~(\ref{eq:calI:def}) as a function of the initial standard deviation $\sqrt{\langle\sigma_\mathrm{eternal}^2\rangle}$ for a quadratic spectator field with potential $V(\sigma)=m^2\sigma^2/2$, if inflation is driven by a monomial potential $V(\phi)\propto\phi^p$. Initial conditions are set at $H_0=H_\mathrm{eternal}$ where the inflaton exits the eternal inflationary regime. In both panels, $H_\uend=10^{-7}\Mp$, and $m=10^{-2} H_\uend$ in the left panel and $m=10^{-4}H_\uend$ in the right panel. Different colours represent different values of $p$. If $p\geq 2$, initial conditions are not erased but provide a subdominant contribution to the final distribution if the field displacement is initially sub-Planckian. This is why, if $\sqrt{\smash[b]{\langle \sigma^2_\mathrm{eternal}\rangle}} \ll \Mp$, $\mathcal{I}\simeq 0$, while if $\sqrt{\smash[b]{\langle \sigma^2_\mathrm{eternal}\rangle}} \gg \Mp$, $\mathcal{I}\simeq 1$. In the left panel, the condition~(\ref{eq:quadratic:adiabstart:condition}) is satisfied for $p=1$, so initial conditions are erased ($\mathcal{I}\simeq 0$), while in the right panel, the condition~(\ref{eq:quadratic:adiabstart:condition}) is not satisfied for $p=1$ which therefore behaves as the cases $p\geq 2$.}
\end{figure}

For concreteness, let us restrict the analysis to the space of symmetric Gaussian distributions, fully characterised by a single parameter, $\langle \sigma^2\rangle$. In this case, \Eq{eq:DKL:gaussian} gives rise to the equivalent expression in this case
\bea
\label{eq:DKL:Gaussian}
\dkl\left(P_1 \vert\vert P_2 \right) = \frac{1}{2\ln 2}\left[\frac{\left\langle \sigma_2^2\right\rangle}{\left\langle \sigma_1^2\right\rangle}-\ln\left(\frac{\left\langle \sigma_2^2\right\rangle}{\left\langle \sigma_1^2\right\rangle}\right)-1\right]\, ,
\eea
where $\langle\sigma_1^2\rangle$ (respectively $\langle\sigma_2^2\rangle$) is the variance of $P_1$ (respectively $P_2$). One then has $\delta\dkl  = (\delta\langle\sigma^2\rangle/\langle\sigma^2\rangle)^2/(4\ln 2)$, which gives rise to\footnote
{The same expression is obtained if one uses the Jensen-Shannon divergence as a measure of the relative information between two distributions,
\bea
\label{eq:jsd}
\djs\left(P_1 \vert\vert P_2 \right) =  \frac{1}{2}\dkl \left(P_1\left\vert\left\vert \frac{P_1+P_2}{2}\right.\right.\right) + \frac{1}{2}\dkl \left(P_2\left\vert\left\vert \frac{P_1+P_2}{2} \right.\right.\right)  \,,
\eea
which is a symmetrised and smoothed version of the Kullback-Leibler divergence. The Jensen-Shannon divergence between two Gaussian distributions cannot be expressed in a closed form comparable to \Eq{eq:DKL:Gaussian}. However, in the limit where the two Gaussian distributions have variances $\langle\sigma^2\rangle$ and $\langle\sigma^2\rangle+\delta \langle\sigma^2\rangle$ infinitesimally close one to the other, one can expand the integrands of \Eq{eq:jsd} at quadratic order in $\delta \langle\sigma^2\rangle$ and obtain $\delta \djs = (\delta\langle\sigma^2\rangle/\langle\sigma^2\rangle)^2/(16\ln 2)= \delta\dkl/4$. As a consequence, $\delta \djs^\uend/\delta\djs^0 = \delta \dkl^\uend/\delta\dkl^0$ and the same information retention criterion is obtained.
}
\bea
\label{eq:info-ret:Gaussian}
\mathcal{I} = \left(  \frac{\partial\ln \langle \sigma^2_\uend\rangle}{\partial  \ln\langle \sigma^2_{0}\rangle} \right)^2\,.
\eea
In practice, the functional relationship between $\langle\sigma_0^2\rangle$ and $\langle\sigma_\uend^2\rangle$ depends on the details of the stochastic dynamics followed by $\sigma$. When $\langle \sigma_\uend^2\rangle$ is independent of $\langle \sigma_0^2\rangle$ for instance, initial conditions are irrelevant to determine the final state and $\mathcal{I}=0$.

For quadratic spectator fields, in \Sec{sec:quad_spec} it was shown that the distributions remain Gaussian if they were so initially, and the relationship~(\ref{eq:meansigmasquare:final}) between $\langle\sigma_0^2\rangle$ and $\langle\sigma_\uend^2\rangle$ was derived. The formula~(\ref{eq:info-ret:Gaussian}) can therefore directly be evaluated, and it is displayed in \Fig{fig:quadratic_information} in the case where inflation is driven by a monomial potential $V\propto\phi^p$ and initial conditions are taken at the time when the inflaton exits the eternal inflationary epoch. When $p\geq 2$, there is no adiabatic regime and therefore no erasure of initial conditions. Since quantum diffusion contributes a field displacement of order the Planck mass, if the initial field value is much smaller than the Planck mass, it provides a negligible contribution to the final field value and one has $\mathcal{I}\simeq 0$. If it is much larger than the Planck mass it provides the dominant contribution to the final field value and  $\mathcal{I}\simeq 1$. In the left panel, the value of $m$ has been chosen so that the condition~(\ref{eq:quadratic:adiabstart:condition}) is satisfied for $p=1$. In this case, initial conditions are erased during the adiabatic regime and one has $\mathcal{I}\simeq 0$. In the right panel, the value chosen for $m$ is such that \Eq{eq:quadratic:adiabstart:condition} is not satisfied and the situation for $p=1$ is similar to the cases $p\geq2$.

For quartic spectator fields, in \Sec{sec:quart_spec} it was shown that either the condition~(\ref{eq:quartic:adiabatic:condition}) is satisfied and initial conditions are erased during an early adiabatic phase, leading to $\mathcal{I}\simeq 0$; or if the condition~(\ref{eq:quartic:adiabatic:condition}) is not satisfied, the dynamics of the spectator field is described by a free diffusion process and the situation is the same as in the right panel of  \Fig{fig:quadratic_information}.

For axionic spectator fields finally, in  \Sec{sec:axion_spec}, initial conditions were shown to always be erased at early times, yielding $\mathcal{I}\simeq 0$.

The amount of information one can recover about the initial state from the final one therefore depends both on the potential of the spectator field and on the inflationary background. Let us stress that in some situations, initial conditions are not erased ($\mathcal{I}\simeq 1$). This suggests that, if observations yield non-trivial constraints on spectator field values at the end of inflation in our local patch, one may be able to infer a non-trivial probability distribution on its field value at much earlier time, for instance when one leaves the regime of eternal inflation. This might be relevant to the question~\cite{Linde:2005yw} of whether observations can give access to scales beyond the observational horizon.

\section{\textsf{Multiple spectator condensates from inflation}}
\label{sec:intro-multispec}
Consider now the evolution of multiple spectator fields in the inflationary background. As we have shown in \Sec{sec:intro-stochastic-approach}, the quantum correction to the classical field dynamics can thus be well-described as a stochastic system of drift and diffusion captured by the following Langevin equation for an indexed field $\sigma_i$ appearing in a multi-field potential $V$
\begin{equation} \label{eq:langevin-multi}
\frac{\dd \sigma_i}{\dd N} = -\frac{1}{3H^2}\frac{\partial V}{\partial \sigma_i} + \frac{H}{2\pi}\xi_i (N) \,,
\end{equation}
where $\xi_i$ is a Gaussian white noise term (without cross-correlation) with a unit amplitude ensemble-average $\left\langle \xi_i (N)\xi_j (N') \right\rangle = \delta_{ij}\delta (N-N')$. In all equations throughout the remainder of this chapter, we will use the indices $i,j,k =\{ 1,2,\dots ,n_{\rm f}\}$, where $n_{\rm f}$ is the number of spectator fields.

We note here that the noise term in \Eq{eq:langevin-multi} originates from the effectively massless and uncoupled mode functions derived from the vacuum solutions to the field in a quasi-de Sitter background. Should the effective mass $\partial^2V/\partial \sigma_i^2$ of the field $\sigma_i$ exceed the Hubble rate, then this formalism is no longer valid and other methods must be developed~\cite{Bunch:1978yq, Birrell:1982ix, Markkanen:2016aes}. Hence, it seems natural here to consider the evolution of light fields up until the threshold where their effective mass is equal to the Hubble rate, and beyond which we shall refer to the condensate as having `collapsed' to the Hubble scale and the effective mass has also saturated to $H$. We shall return to this point in \Sec{sec:comp-analytic-arg} where we, e.g. evaluate the critical couplings required to achieve this saturation.

We stress here another point raised in \Sec{sec:intro-stochastic-approach} which is that, for interacting fields in de Sitter spacetimes, another critical value is known to exist which signals the breakdown of the semi-classical approximation. As discussed in \Sec{sec:sourcing-cos-pert}, in the mean-field approximation, one separates a classical `mean' background field from perturbatively small quantum fluctuations. For quartic scalar fields, in \Ref{Burgess:2010dd}, it was shown that a breakdown in this peturbative expansion occurs in the regime where the bare mass is less than $\lambda H^2/(4\pi^2)$ which cannot be removed by reorganising the perturbative expansion to include a running effective mass. We stress here that non-peturbative methods of resummation, such as those of this chapter, are potentially unaffected by such a bound. This is due to the fact that the backreaction from small quantum fluctuations is inherently included into the background evolution described by \Eq{eq:langevin-multi}, thus optimising the perturbative expansion at each new scale in time --- a cosmological analog to (but not exactly the same as~\cite{Woodard:2008yt}) the Renormalisation Group flow~\cite{Tsamis:2005hd}. 

The corresponding multi-field Fokker-Planck equation to \Eq{eq:langevin-multi} is
\begin{equation} \label{eq:dist-multi}
\frac{\partial }{\partial N}P(\sigma_i ,N) = \frac{1}{3H^2} \sum^{n_{\rm f}}_{j=1}\frac{\partial }{\partial \sigma_j} \left[ \frac{\partial V}{\partial \sigma_j} P(\sigma_i ,N)\right] + \frac{H^2}{8\pi^2}\sum^{n_{\rm f}}_{j=1}\frac{\partial^2}{\partial \sigma_j^2} P(\sigma_i ,N)\,,
\end{equation}
where we have implicitly made use of the test field condition $\partial H /\partial \sigma_i = 0$ and defined $P(\sigma_i ,N)$ as the probability distribution function over field values at a given $N$, when normalised. Thus, the evolution of modes as they accumulate outside of the horizon typically yields an $n_{\rm f}$-dimensional distribution of field displacements throughout the inflationary phase $P(\sigma_i ,N)$. It has recently been remarked~\cite{Pinol:2018euk} that, when more than one field is present, one must use the Stratonovich interpretation of the stochastic process which maintains general covariance over the field space at the cost of introducing spurious frame dependencies into the noise term of \Eq{eq:dist-multi} --- which has been obtained from the It\^{o} interpretation. Due to the fact that we are considering test fields, however, the backreaction onto $H$ from all of the $\sigma_i$ fields is negligible. In this case, one can likely remove these without any loss of information about the physical system because, as is further remarked in \Ref{Pinol:2018euk} by analogy with non-linear sigma models, the Riemann curvature of field space only enters the mass matrix. This is equivalent to test fields developing a preferred set of field space coordinates due to their potential gradients only entering into the drift term.

\Eq{eq:dist-multi} may also be written essentially as a continuity equation~\cite{doi:10.1002/bimj.4710280614,Assadullahi:2016gkk}
\begin{equation}
\frac{\partial P}{\partial N} + \sum^{n_{\rm f}}_{i=1}\frac{\partial J_{i}}{\partial \sigma_i} = 0\,,
\end{equation}
where $J_i$ is the probability current and the right hand side of the equation must vanish for probability conservation. By inspection of \Eq{eq:dist-multi}, one may verify that in this case
\begin{equation}~\label{eq:probcurr-multi}
\ J_i = -\frac{1}{3H^2}  \frac{\partial V}{\partial \sigma_i} P(\sigma_i ,N) - \frac{H^2}{8\pi^2}\frac{\partial}{\partial \sigma_i} P(\sigma_i ,N) \,.
\end{equation}

In de Sitter-like inflation the Hubble parameter is effectively constant in time, hence there is a stationary\footnote{$\partial P/\partial N = 0$ in this context.} solution to \Eq{eq:dist-multi}, $P_{\rm stat}$, corresponding to a vanishing divergence $\nabla \cdot \boldsymbol{J} = 0$ of the probability current --- an incompressible flow of the vector field with components $J_i$. Where $n_{\rm f}=1$ in an unbounded field domain\footnote{In the case of a bounded field domain, probability conservation at the specified boundary implies that $J_i=0$ directly.} one can show that in order for the distribution to have a finite normalisation $P(\sigma_1, N) \dd \sigma_1 \rightarrow 0$ (and hence $J_1\rightarrow 0$) as $\sigma_1 \rightarrow \infty$. Furthermore, given $n_{\rm f}=1$, one can also show that in the stationary limit, the vanishing divergence of $J_1$ simply reduces to $\partial J_1 / \partial \sigma_1 = 0$, and $J_1$ must therefore vanish $\forall \sigma_1$. Hence, the left hand side of \Eq{eq:probcurr-multi} may always be set to zero and the well-known exponential solution to \Eq{eq:dist-multi} for the stationary probability distribution is obtained~\cite{Starobinsky:1986fx} $P_{\rm stat} (\sigma_1 ) \propto \exp \left[ -8\pi^2V(\sigma_1)/3H^4\right]$.

For unbounded $V$ with arbitrary $n_{\rm f}$, it is still natural to consider a boundary condition where $J_i=0$ as $\sigma_i\rightarrow \infty$ to restrict unphysical possibilities, and this may even in practice occur at a set finite scale $\Lambda$ that denotes the chosen cutoff of the theory. However, one can no longer generally state that $J_i$ vanishes everywhere throughout the $n_{\rm f}$-field domain since any class of incompressible vector $J_i$ flows are permitted. Because $\partial J_i / \partial \sigma_i = 0$ is still possible, it is true that one stationary solution to \Eq{eq:dist-multi} is
\begin{equation} \label{eq:exp-stat-dist}
\ P_{\rm stat}(\sigma_i ) \propto \exp \left[ -\frac{8\pi^2V(\sigma_i)}{3H^4}\right] \,,
\end{equation}
but it is no longer unique, and one must use either use further analytical arguments or full numerical solutions for verification.

For any $n_{\rm f}$, the stationary distribution $P_{\rm stat}$ is in practice only reached after some equilibration timescale $N_{\rm eq}$. The timescale $N_{\rm eq}$ is defined as the number of \efold{s} it takes for $P(\sigma_i,N)= \prod^{n_{\rm f}}_{i=1}\delta (\sigma_i)$ --- an $n_{\rm f}$-dimensional Dirac delta function\footnote{We note that, for the symmetric potentials about the origin used in this chapter, this is of course equivalent to the more general definition of a Dirac function at the global minimum, $\prod^{n_{\rm f}}_{i=1}\delta (\sigma_i - \sigma^{\rm min}_i)$.} --- to relax to $P(\sigma_i,N)=P_{\rm stat}$. Hence, $N_{\rm eq}$ can be thought of as the time it takes for the effective condensate to grow to its maximal value in every field dimension. It is also important to note here that the definition of $N_{\rm eq}$ used in this chapter relies on the inflationary background being de Sitter-like. In slow-roll backgrounds where $H$ varies more substantially, such as those permitted by a monomial $U(\phi ) \propto \phi^p$ inflationary potential, this timescale will have to be recomputed~\cite{Hardwick:2017fjo}.

\section{\textsf{Vanishing probability current with symmetric potentials}} \label{sec:proof-symmetric}

In the previous section, we stated that the exponential form (\Eq{eq:exp-stat-dist}) of the stationary solution to \Eq{eq:dist-multi} may no longer be stable when any divergence-free (incompressible) probability currents are potentially allowed. For any choice of $n_{\rm f}>1$, only the divergence of the current must vanish for a stationary solution, which leaves the possibility of a curl in the vector field $\nabla \times \boldsymbol{J}$. Because $\boldsymbol{J}\cdot \hat{\boldsymbol{{\rm e}}}$ vanishes, where $\hat{\boldsymbol{{\rm e}}}$ is the normal to the boundary, the total integral of the curl over the domain of the fields $\sigma_i \in \Sigma$ vanishes according to Stokes' theorem
\begin{equation} \label{eq:stokes}
\int_{\Sigma} (\nabla \times \boldsymbol{J})_i \,\, \dd^{n_{\rm f}}\sigma_i = 0\,,
\end{equation}
however there are still an infinite number of functions for $\nabla \times \boldsymbol{J}$ that can satisfy this criterion. Examining \Eq{eq:probcurr-multi}, and using the general properties of the totally antisymmetric symbol $\epsilon_{ijk}$, one can show that
\begin{equation}
\ ( \nabla \times \boldsymbol{J} )_i  = -\sum^{n_{\rm f}}_{j = 1}\sum^{n_{\rm f}}_{k = 1}\epsilon_{ijk}\frac{1}{3H^2}\frac{\partial V}{\partial \sigma_k}\frac{\partial P}{\partial \sigma_j} \label{eq:curl-J} \,.
\end{equation}
 
Our first remark is that \Eq{eq:curl-J} vanishes at the extrema of $V$ and $P$ (a fact that we numerically verify for a given potential in \Sec{sec:non-vanish}) but not necessarily everywhere in the domain of $\sigma_i$. Secondly, for all choices of potential and initial distribution, if the gradients of $V$ and $P$ align, i.e. $\partial V/\partial \sigma_i \propto \partial P/\partial \sigma_i$, then \Eq{eq:curl-J} vanishes and hence $J_i=0$ must be true at this point. If one takes a derivative of \Eq{eq:exp-stat-dist}, it is clear that the stationary solution that we have quoted satisfies this criterion.

Without an alternative ansatz to compute $P$, it is difficult to make any general claims about stationary solutions to \Eq{eq:dist-multi}, even when $V$ is symmetric\footnote{Indeed, even with symmetric $V$ and $P$ (the latter can be proved to follow from a symmetric initial condition), if $n_{\rm f}=2$ it can be shown that
\begin{equation} \label{eq:arb-func}
\ ( \nabla \times \boldsymbol{J} )_3 = f(\sigma_1,\sigma_2) - f(\sigma_2,\sigma_1) \,,
\end{equation}
where $f(\sigma_1,\sigma_2)$ is an arbitrary function of both variables. \Eq{eq:arb-func} trivially satisfies the integral constraint from Stokes' theorem (\Eq{eq:stokes}) and hence we are left with no further determination of its exact form without working through an explicit example. Note, however, that \Eq{eq:arb-func} gives $( \nabla \times \boldsymbol{J} )_3(\sigma_1,\sigma_2)=-( \nabla \times \boldsymbol{J} )_3(\sigma_2,\sigma_1)$ and hence, if one can also demonstrate that $\boldsymbol{J}(\sigma_1,\sigma_2)=\boldsymbol{J}(\sigma_2,\sigma_1)$ for symmetric potentials, it must be true that $( \nabla \times \boldsymbol{J} )_3=0$. }. However, we conjecture that when $V$ is symmetric, the solution for $P$ --- which we assume has a been evolved from a symmetric initial condition --- typically has a gradient which aligns with $V$ and hence \Eq{eq:exp-stat-dist} is a stable stationary solution to \Eq{eq:dist-multi}. We have verified numerically that \Eq{eq:exp-stat-dist} provides a stable solution to the late-time dynamics with symmetric potentials in \Sec{sec:comp-analytic-arg}. Note also that in asymmetric potentials\footnote{For example, some of the potentials we introduce and discuss in \Sec{sec:dec-coup}.} we can no longer assume that the $J_i$ components vanish everywhere, and \Eq{eq:exp-stat-dist} is no longer the stationary solution. In such instances, one can also turn to numerical methods.

\section{\textsf{Computation and analytic arguments}}
\label{sec:comp-analytic-arg}

The general problem for arbitrary $V$ defined by Eqs. \eqref{eq:langevin-multi} and \eqref{eq:dist-multi} cannot be solved analytically, and so in this section we shall make our computations for the condensates formed from multi-field spectator potentials combining both analytic and numerical methods. The details of our numerical implementation can be found in Appendix \ref{sec:Num-implement}, where we briefly outline our development of a new publicly available python code, \href{https://sites.google.com/view/nfield-py}{\texttt{nfield}}.

In light of our discussion in \Sec{sec:proof-symmetric}, we cannot always expect to use moments of the distribution in \Eq{eq:exp-stat-dist} to reliably evaluate the stationary variance for asymmetric potentials. However, we shall not need this distribution to hold true in order to still gain an insight from some approximations. 

Consider a general multi-field interacting spectator potential. In the limit of small field displacements, one can typically perform a Taylor expansion about the minimum of a potential which defines an effective mass in each orthogonal field dimension\footnote{Where we have already implicitly performed any necessary rotations in field space such that $\sigma_i\sigma_j$ cross-terms vanish.} as
\begin{equation} \label{eq:effm}
\ M_i^2 \equiv \frac{\partial^2 V}{\partial \sigma_i^2}\,.
\end{equation}
Hence, a generic multi-field potential can be approximated by 
\begin{equation} \label{eq:Veffm}
\ V \simeq \frac{1}{2} \sum^{n_{\rm f}}_{j=1} M_j^2 \sigma_j^2\,,
\end{equation}
where one may account for interactions (both self and with other fields) through the typical values that one finds for $M_i$. For example, quartic self-interacting terms where $M_i\propto \sigma_i$ may be approximately written as $M_i\propto \sqrt{\left\langle \sigma_i^2\right\rangle}$. As another example, consider the situation where $M_i\propto \sigma_k^2$ due to interaction terms, then the effective mass becomes approximately $M_i\propto \left\langle \sigma_k^2\right\rangle$. This approximation will prove sufficient to calculate the desired quantities in \Sec{sec:crit-coup}. 

Using \Eq{eq:Veffm}, one can derive a second-moment evolution equation from \Eq{eq:langevin-multi} of the form~\cite{Starobinsky:1994bd}
\begin{equation} \label{eq:approx-2om}
\frac{\dd \left\langle \sigma^2_i\right\rangle}{\dd N} \simeq -\frac{2M_i^2}{3H^2}\left\langle \sigma_i^2\right\rangle + \frac{H^2}{4\pi^2} \,,
\end{equation}
and hence the stationary\footnote{$\dd \left\langle \sigma_i^2 \right\rangle/\dd N =0$ in this context, hence this need only be `stationary' in the $i$th field dimension. } variance can be immediately derived\footnote{Notice that this equation is indeed consistent with inserting \Eq{eq:Veffm} into \Eq{eq:exp-stat-dist} and taking the second-order moment.}~\cite{Starobinsky:1994bd}
\begin{equation} \label{eq:approx-2om-stat}
\left\langle \sigma_i^2\right\rangle \vert_{\rm stat} = \frac{3H^4}{8\pi^2M_i^2}\,.
\end{equation}
Note that in the limit where \Eq{eq:Veffm} is no longer an approximation, such as for a quadratic non-interacting spectator, then \Eq{eq:approx-2om} and \Eq{eq:approx-2om-stat} are precise equations with $M_i$ corresponding to the bare mass.

\begin{figure}
\begin{center}
\includegraphics[width=14cm]{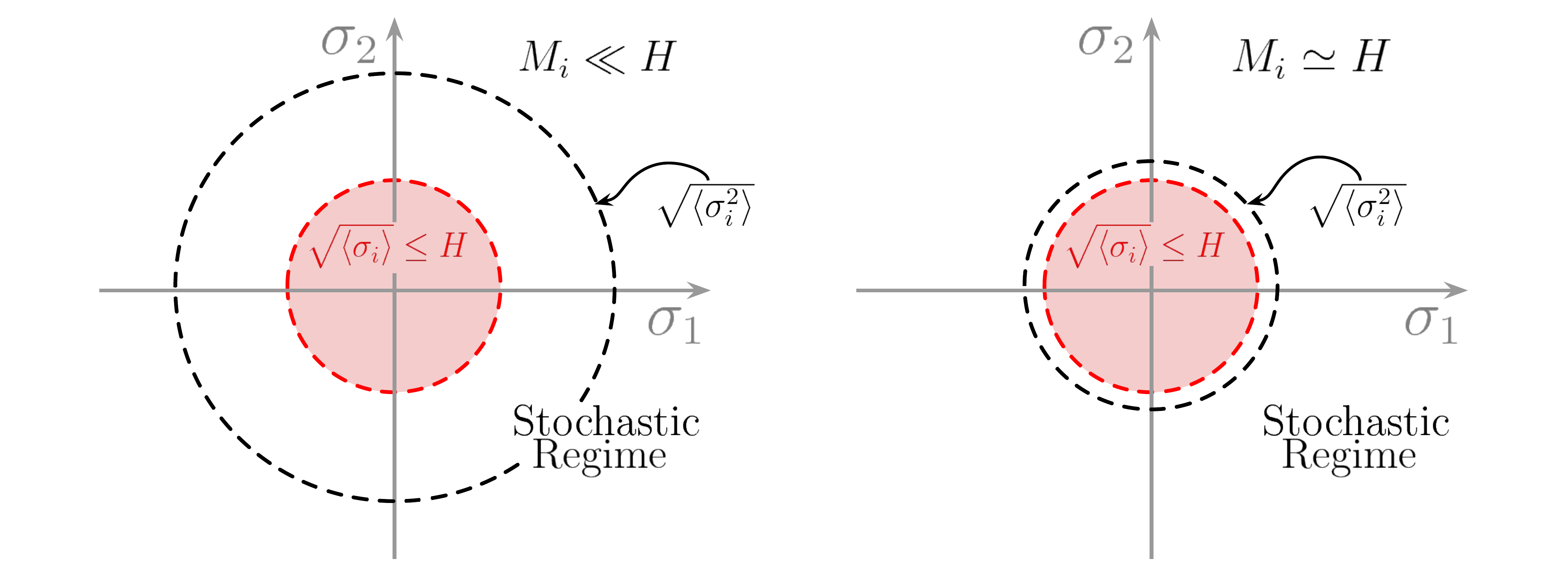}
\caption[Saturation of the multi-field variance]{~\label{fig:saturation-diagram} Schematic diagrams of the condensate $\sqrt{\left\langle \sigma_i^2 \right\rangle}$ (dashed black circles) in a symmetric two-field potential. When $M_i\simeq H$ the condensate collapses to the Hubble rate, corresponding to the dashed red circle. The shaded red region corresponds to situations where $M_i>H$ and there is conjectured saturation at the Hubble scale $\sqrt{\left \langle \sigma_i^2 \right\rangle } \simeq H$. The entire region in field space outside of the red shaded area can be considered where the stochastic formalism is valid, with the caveat that it is possible for $M_i>H$ to also occur for a field at large displacements, e.g. quartic self-interacting fields will have $M_i \propto \sigma_i$. }
\end{center}
\end{figure}

In \Eq{eq:approx-2om-stat}, the inverse-proportionality between the effective mass and the variance indicates that there is a critical value for $M_i \simeq H$ above which the stationary condensate collapses to the Hubble rate $\sqrt{\left\langle \sigma_i^2\right\rangle} = H$. Taking a two-field example for illustration, we have plotted a schematic diagram of the physical situation in \Fig{fig:saturation-diagram} for a symmetric potential. In the left panel, the condensate (dashed black circle) is relatively large because the effective mass $M_i \ll H$. In the right panel, the condensate collapses to the value of the Hubble rate (dashed red circle) because $M_i\simeq H$. For the shaded red region the suppression $\left\langle \sigma_i^2\right\rangle \propto 1/M_i^2$ in the variance from \Eq{eq:approx-2om-stat} is no longer valid and the stochastic approach can no longer be used. This is because when $M_i>H$ the mode functions which source the fluctuations of the field can no longer be accurately described by the simple form of noise correlator in the definition of \Eq{eq:langevin-multi}.

Other calculations do exist for situations considering a constant super-Hubble mass $M_i>H$~\cite{Bunch:1978yq, Birrell:1982ix, Markkanen:2016aes} where, in these instances, the variance is known to experience further suppression. However, the assumption of constant $M_i$ is one we cannot make for the potentials studied in this section. We anticipate that a similar suppression occurs but leave the verification of this to future work. Even if this is not always true (e.g. for many non-interacting quadratic spectators), the `saturation' value is still of interest since it characterises the fundamental domain of validity for the stochastic formalism. Hence, in this regard, we shall leave the calculation of possible condensates in the $M_i>H$ regime to future work, and therefore we will focus our efforts on the regime where the stochastic formalism is valid.

\subsection{\textsf{The critical coupling}} \label{sec:crit-coup}

In this subsection, we will demonstrate that when one generalises the formation of spectator field condensates to many coupled fields, a critical value for the coupling appears, above which the equilibrium variances of all fields have collapsed to the Hubble scale and effective mass of each field has saturated to $H$. To show this we will consider a simplified potential that will allow us to calculate this critical coupling both analytically and numerically, for verification.

Now consider the multi-field spectator potential 
\begin{equation} \label{eq:VA}
\ V_{\rm A} = \frac{1}{2} g\sum_{i\neq j} \sigma_i^2\sigma^2_j\,.
\end{equation}
Mindful of the approximation made with $M_i$ in \Eq{eq:Veffm}, one thus expects that incrementally strengthening the interaction between spectator fields $\propto g \sigma_i^2\sigma_j^2$ can lead to the eventual saturation of the condensate value at the Hubble scale due to the effective mass of each field being progressively larger, and we therefore anticipate a critical value for the inter-field coupling $g_{\rm crit}$ to exist, for a given $n_{\rm f}$, above which the stationary condensate collapses to $\sqrt{\left\langle \sigma_i^2\right\rangle} \simeq H$.

By inspection between Eqs. \eqref{eq:Veffm} and \eqref{eq:VA}, the typical value of the effective mass in the $i$th field dimension corresponds to $M_i^2 \simeq g\sum_{k\neq i}\left\langle  \sigma_k^2\right\rangle \simeq g (n_{\rm f}-1)\left\langle  \sigma_i^2\right\rangle $, where in the second equality we have assumed that the distribution (using \Eq{eq:VA} as the potential) has reached stationarity $P(\sigma_i,N) = P_{\rm stat}$ and, hence, due to symmetry $\left\langle \sigma^2_i \right\rangle = \left\langle \sigma^2_k \right\rangle \,\, \forall \,\, i,k$. Because $\left\langle \sigma^2_i \right\rangle = \left\langle \sigma^2_i \right\rangle \vert_{\rm stat}$, given in \Eq{eq:approx-2om-stat}, we can now obtain an approximate relation for the critical coupling\footnote{Note that because \Eq{eq:VA} is a symmetric potential --- i.e. $V(\sigma_{j}) = V(\sigma_{{\rm perm}(j)})$ for any permutation of field indices ${\rm perm}(j)$ --- our discussion in \Sec{sec:proof-symmetric} indicates the stability of \Eq{eq:exp-stat-dist} in this situation. Hence, another way to compute \Eq{eq:gcrit-variance-analytic} would be to take the second moment of \Eq{eq:exp-stat-dist}.}
\begin{align}
\left\langle  \sigma_i^2\right\rangle &= \frac{3H^4}{8\pi^2 M_i^2} \simeq \frac{3H^4}{8\pi^2 g (n_{\rm f}-1)\left\langle  \sigma_i^2\right\rangle } \nonumber \\
&\Rightarrow \left\langle  \sigma_i^2\right\rangle \simeq \sqrt{ \frac{3H^4}{8\pi^2 g (n_{\rm f}-1)}} \label{eq:gcrit-variance-analytic} \\
&\Rightarrow g_{\rm crit} \simeq \frac{3}{8\pi^2 (n_{\rm f}-1)}  \label{eq:gcrit-analytic}\,,
\end{align}
where we have found $g_{\rm crit}$ by setting $\left\langle \sigma_i^2\right\rangle = H^2$ in \Eq{eq:gcrit-variance-analytic}.

\begin{figure}
\begin{center}
\includegraphics[width=7cm]{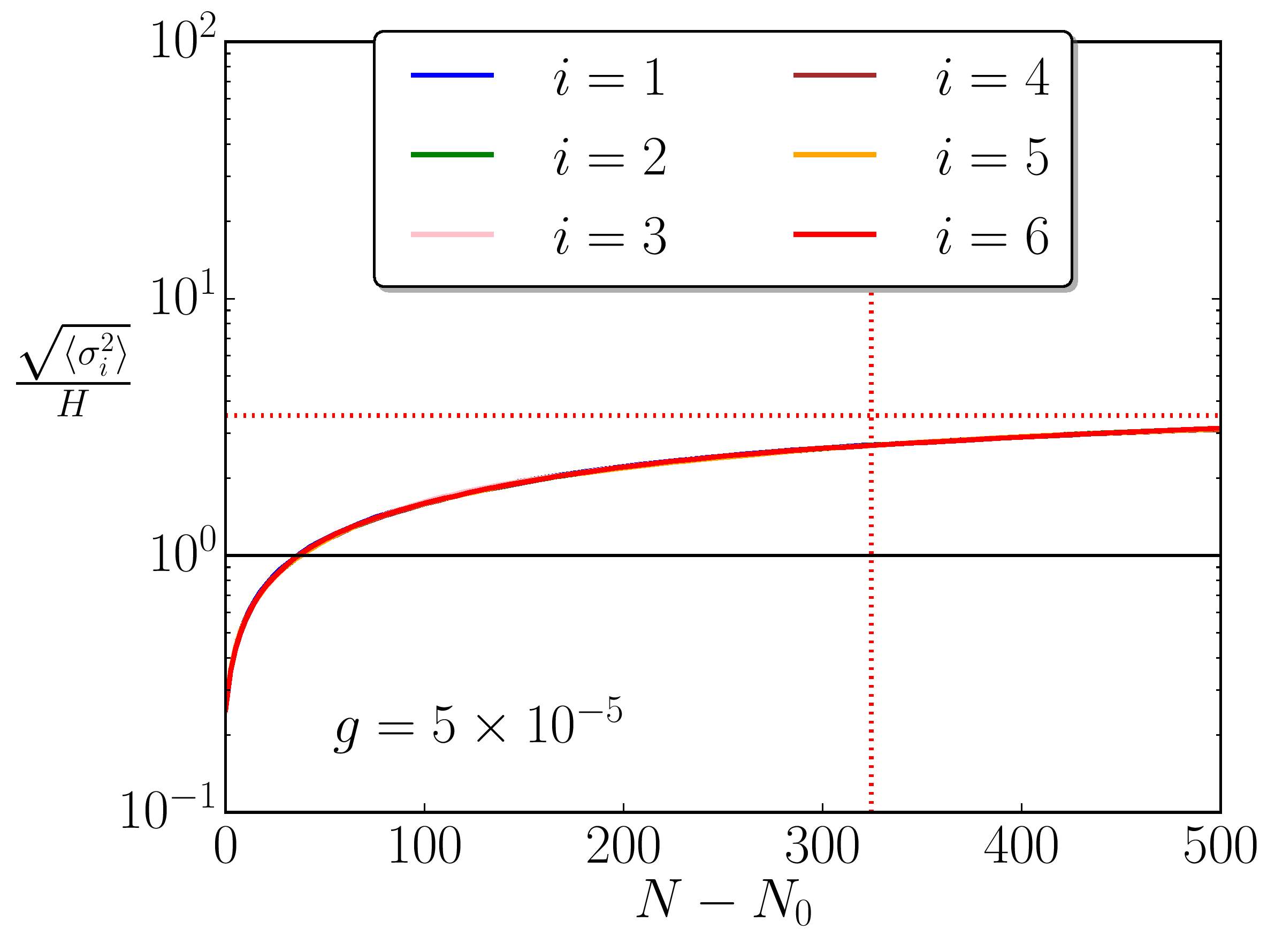}
\includegraphics[width=7cm]{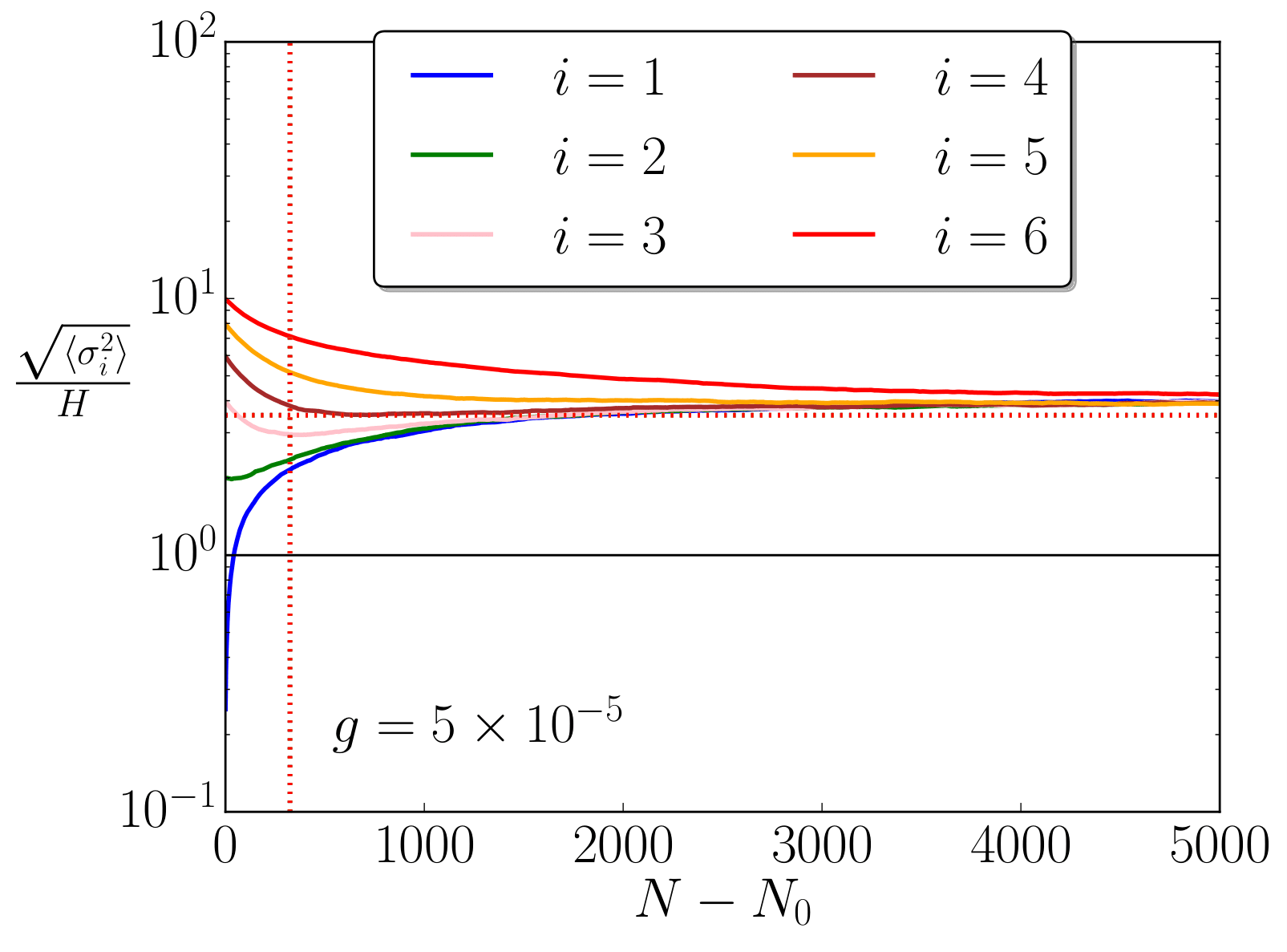}
\caption[Time evolution of many spectators]{~\label{fig:VA-variance-plot} The numerically evaluated (solid lines) time evolution of the variance for each spectator field in the case of the $V_{\rm A}$ potential (\Eq{eq:VA}) with example value $n_{\rm f}=6$. The variance of all fields is initialised at $\left\langle \sigma_i^2\right\rangle = 0$ in the left panel and we have chosen a range of initial conditions for the fields in the right panel to indicate the robustness of the late time stationary behaviour. All of the field variances overlap due to the symmetry of the potential. Dotted horizontal and vertical lines represent the stationary variance (\Eq{eq:gcrit-variance-analytic}) and equilibration timescale (\Eq{eq:Neqi}), respectively. The number of realisations used is $10^4$. }
\end{center}
\end{figure}

\begin{figure}
\begin{center}
\includegraphics[width=10cm]{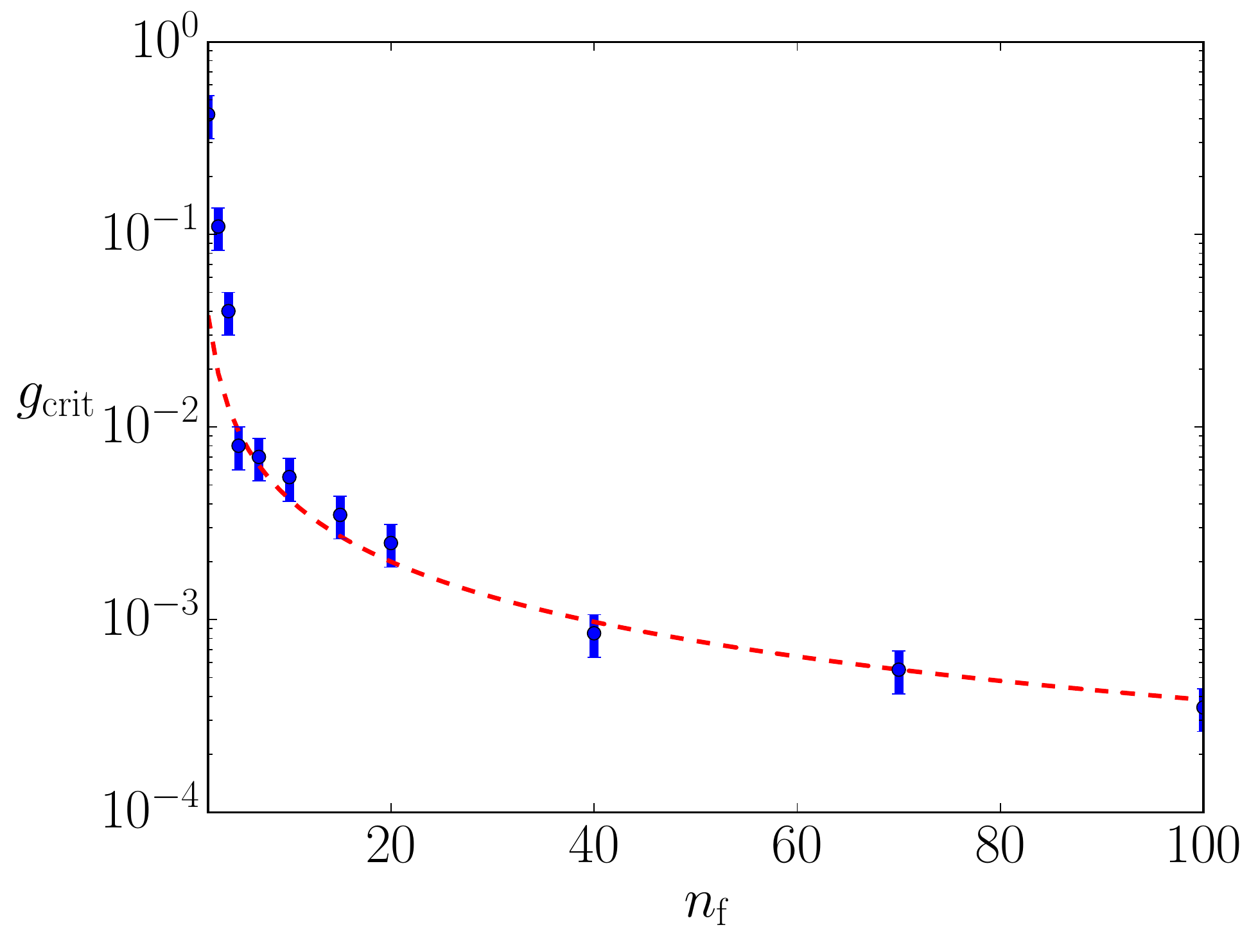}
\caption[Critical coupling for multiple fields]{~\label{fig:gcrit-plot} The numerically evaluated (blue data points with error bars related to both statistical and numerical uncertainty from having a finite number of realisations and a finite stepsize in numerically finding $g_{\rm crit}$, respectively) value of $g_{\rm crit}$ as a function of the number of fields $n_{\rm f}$ for potential $V_{\rm A}$ (see \Eq{eq:VA}). The line clearly matches the analytically derived relation in \Eq{eq:gcrit-analytic} (dashed red line) very well. The number of realisations used is $10^4$ for each point. Due to how rapidly $g_{\rm crit}$ varies with the number of fields for $n_{\rm f}\lesssim 5$, we are likely underestimating our error in this region, and hence these points may appear slightly inconsistent. }
\end{center}
\end{figure}

For illustration, we plot the time evolution for variances, averaging over multiple Langevin realisations (realisations of \Eq{eq:langevin-multi}), of an example where $n_{\rm f}=6$ in \Fig{fig:VA-variance-plot} and \Eq{eq:gcrit-variance-analytic} is shown to be a good description of the stationary values against the numerically evaluated variances (all identical to each other due to the symmetry). The approximate form of \Eq{eq:gcrit-analytic} must also be verified numerically, and hence we plot in \Fig{fig:gcrit-plot} the comparison between numerical and analytic approaches to obtain the functional relationship between $g_{\rm crit}(n_{\rm f})$. Due to the apparently excellent agreement between the two calculations in \Fig{fig:gcrit-plot} we can be confident in \Eq{eq:gcrit-analytic} as a reliable formula to extrapolate to large $n_{\rm f}$.

We further note that one may derive the equilibration timescale for each field dimension $N_{{\rm eq},i}$ for the $V_{\rm A}$ potential, and this is approximately be given by
\begin{equation} \label{eq:Neqi}
\ N_{{\rm eq},i} \simeq \frac{H^2}{M_i^2} \simeq \sqrt{\frac{8\pi^2}{3g(n_{\rm f}-1)}} \,.
\end{equation}
The first relation of \Eq{eq:Neqi} can be derived from \Eq{eq:approx-2om} (see also Refs.~\cite{Starobinsky:1986fx,Enqvist:2012xn,Hardwick:2017fjo}), where it is also natural to consider the `steepness' of the effective potential in \Eq{eq:Veffm} to control the rate of equilibration. Note that \Eq{eq:Neqi} has been derived by assuming the stationary variance, however, because $N_{\rm eq,i}$ is precisely the time it takes to relax to the stationary limit, this assumption is not strictly valid and requires comparison with full numerical solutions. Interestingly, in the example with $n_{\rm f}=6$ plotted in \Fig{fig:VA-variance-plot}, \Eq{eq:Neqi} appears to perform well regardless of its less trustworthy origin.

\subsection{\textsf{The decoupling limit}} \label{sec:dec-coup}

We will now investigate another limit of the inter-field coupling, which can also be analytically estimated for some specific potentials and the numerical verification will also serve to showcase further applications of the \href{https://sites.google.com/view/nfield-py}{\texttt{nfield}} code.

Consider two further examples of interacting spectator potentials
\begin{align}
\ V_{\rm B} &= \frac{1}{2}m^2\left( \sigma^2_1 + \alpha \sigma^2_2\right) + \frac{1}{2}g \sigma_1^2\sigma^2_2 \label{eq:VBpot} \\
\ V_{\rm C} &= \frac{1}{4}\lambda \left( \sigma^4_1 + \alpha \sigma^4_2\right) + \frac{1}{2}g \sigma_1^2\sigma^2_2\,. \label{eq:VCpot}
\end{align}
$V_{\rm B}$ a generalisation from $V_{\rm A}$ by introducing additional masses $m$ and $\sqrt{\alpha} m$, with a hierarchy parameter $\alpha$, but we have now specified that $n_{\rm f}=2$ to capture the essential phenomenology. $V_{\rm C}$ is another generalisation from $V_{\rm A}$ to include self-interactions. We note here that, in each case, decoupling the system in the limit where $g \rightarrow 0$ will yield the well-known formulae~\cite{Starobinsky:1986fx,Enqvist:2012xn,Hardwick:2017fjo} for the stationary variance of each non-interacting field (see \Eq{eq:approx-2om-stat})
\begin{align}
\ {\rm Decoupled} \,\, V_{\rm B} \Rightarrow \left\langle \sigma_1^2 \right\rangle \vert_{\rm stat} &= \frac{3H^4}{8\pi^2 m^2} \label{eq:stationary_VB_variances_1} \\
\left\langle \sigma_2^2 \right\rangle \vert_{\rm stat} &= \frac{3H^4}{8\pi^2 \alpha m^2} \label{eq:stationary_VB_variances_2} \\
\ {\rm Decoupled} \,\, V_{\rm C} \Rightarrow \left\langle \sigma_1^2 \right\rangle \vert_{\rm stat} &= \frac{\Gamma \left( \frac{3}{4}\right)}{\Gamma \left( \frac{1}{4}\right)}\sqrt{\frac{3H^4}{2\pi^2 \lambda}} \label{eq:stationary_VC_variances_1} \\
\left\langle \sigma_2^2 \right\rangle \vert_{\rm stat} &= \frac{\Gamma \left( \frac{3}{4}\right)}{\Gamma \left( \frac{1}{4}\right)}\sqrt{\frac{3H^4}{2\pi^2 \alpha \lambda}} \label{eq:stationary_VC_variances_2} \,.
\end{align}
In this same limit one can also obtain the respective equilibration timescales
\begin{align}
\ {\rm Decoupled} \,\, V_{\rm B} \Rightarrow N_{{\rm eq},1} \vert_{\rm stat} &\simeq \frac{H^2}{M_1^2} \simeq \frac{H^2}{m^2} \label{eq:stationary_VB_Neq_1} \\
\ N_{{\rm eq},2} \vert_{\rm stat} &\simeq \frac{H^2}{M_2^2} \simeq \frac{H^2}{\alpha m^2} \label{eq:stationary_VB_Neq_2} \\
\ {\rm Decoupled} \,\, V_{\rm C} \Rightarrow N_{{\rm eq},1} \vert_{\rm stat} &\simeq \frac{H^2}{M_1^2} \propto \frac{1}{\sqrt{\lambda}} \label{eq:stationary_VC_Neq_1} \\
\ N_{{\rm eq},2} \vert_{\rm stat} &\simeq \frac{H^2}{M_2^2} \propto \frac{1}{\sqrt{ \alpha \lambda}} \label{eq:stationary_VC_Neq_2} \,,
\end{align}
where we recall that the effective masses $M_i$ are defined in \Eq{eq:effm}. We note there that, as in \Eq{eq:Neqi}, these timescales have been derived using the stationary form of the variances which is not strictly valid, hence they must be checked for validity against the numerical implementation to ensure that they are still accurate.

A `decoupling' value of $g=g_{\rm dec}$ can be derived analytically from these stationary variances by obtaining the value of $g$ above which the main contribution to the effective mass is from the coupling term $\propto g$ and not from the bare mass or self-interaction. Looking at the effective mass of either of the fields in each potential, one can hence show that in the stationary limit
\begin{align}
\ V_{\rm B} &\Rightarrow M^2_1 \simeq m^2 + g \left\langle \sigma_2^2 \right\rangle\vert_{\rm stat} \Rightarrow g_{\rm dec} \simeq \frac{8\pi^2}{3}\alpha \left( \frac{m}{H}\right)^4  \label{eq:gdecVB} \\
\ V_{\rm C} &\Rightarrow M^2_1 \simeq 3\lambda \left\langle \sigma_1^2 \right\rangle \vert_{\rm stat} + g \left\langle \sigma_2^2 \right\rangle\vert_{\rm stat} \Rightarrow g_{\rm dec} \simeq 3 \lambda \sqrt{\alpha} \,. \label{eq:gdecVC}
\end{align}
If one were to re-derive \Eq{eq:gdecVB} and \Eq{eq:gdecVC} by replacing $M_1 \Longleftrightarrow M_2$, it is trivial to show that the same formulae are obtained.

\begin{figure}
\begin{center}
\includegraphics[width=7cm]{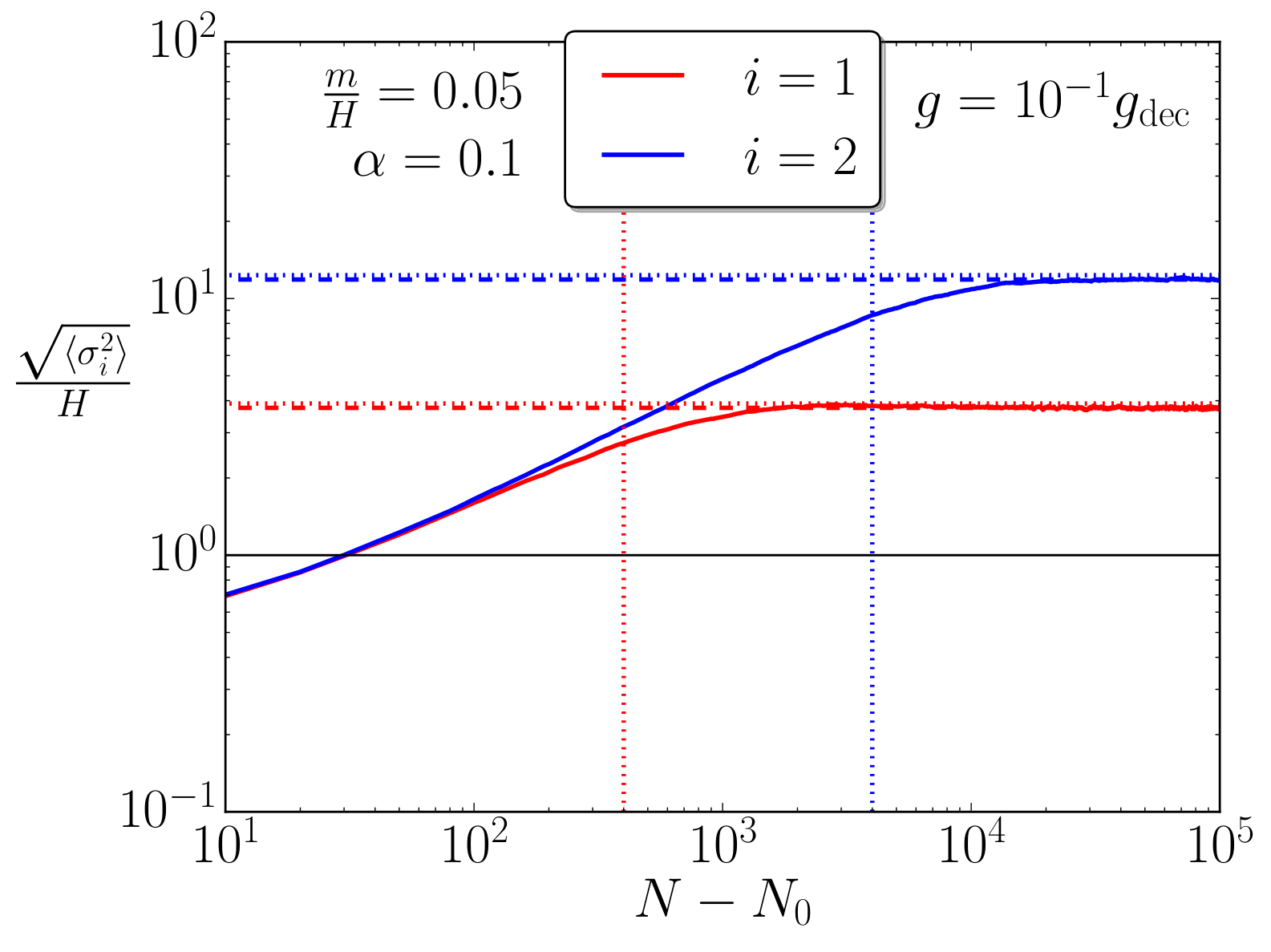}
\includegraphics[width=7cm]{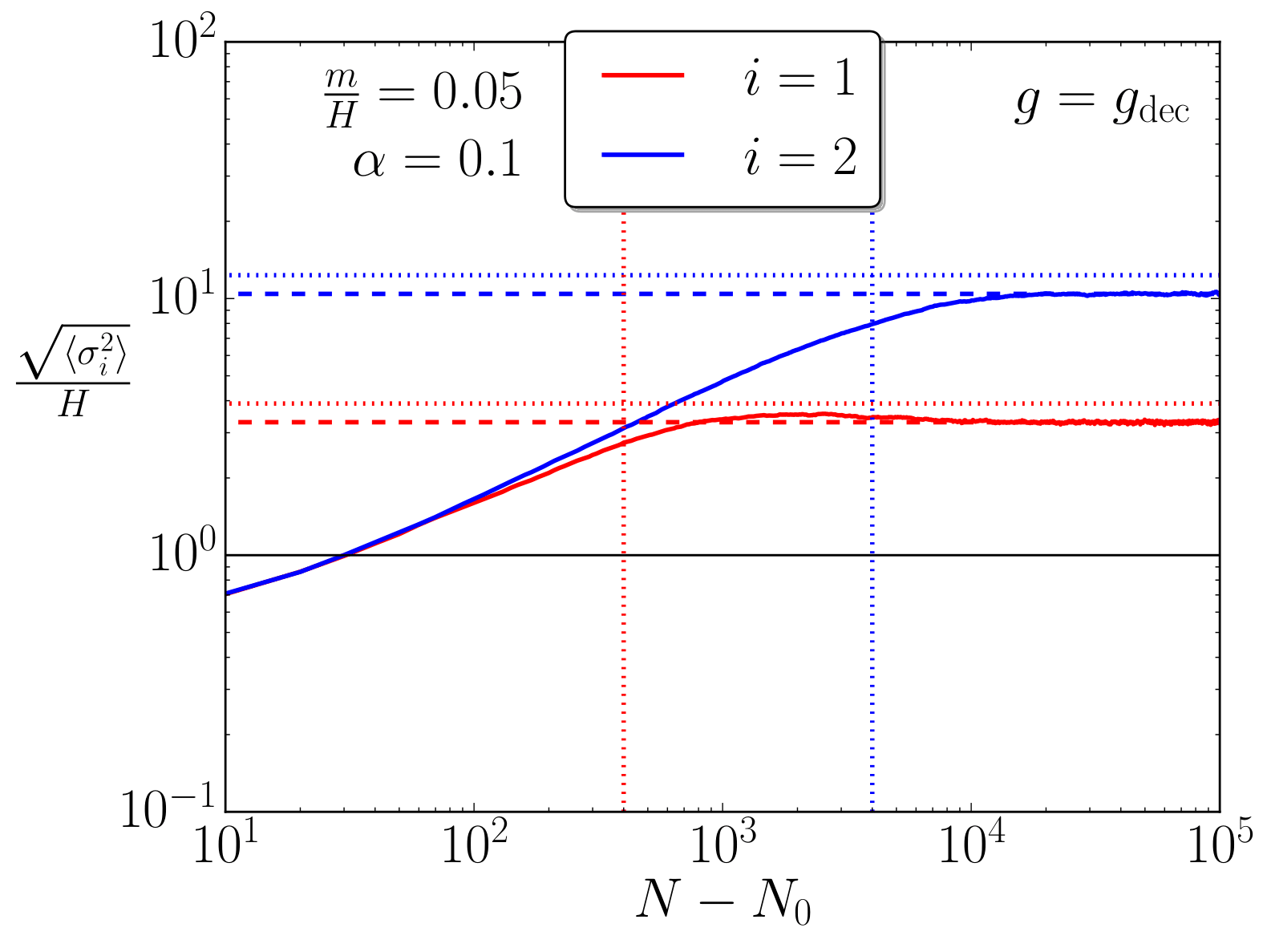} \\
\includegraphics[width=7cm]{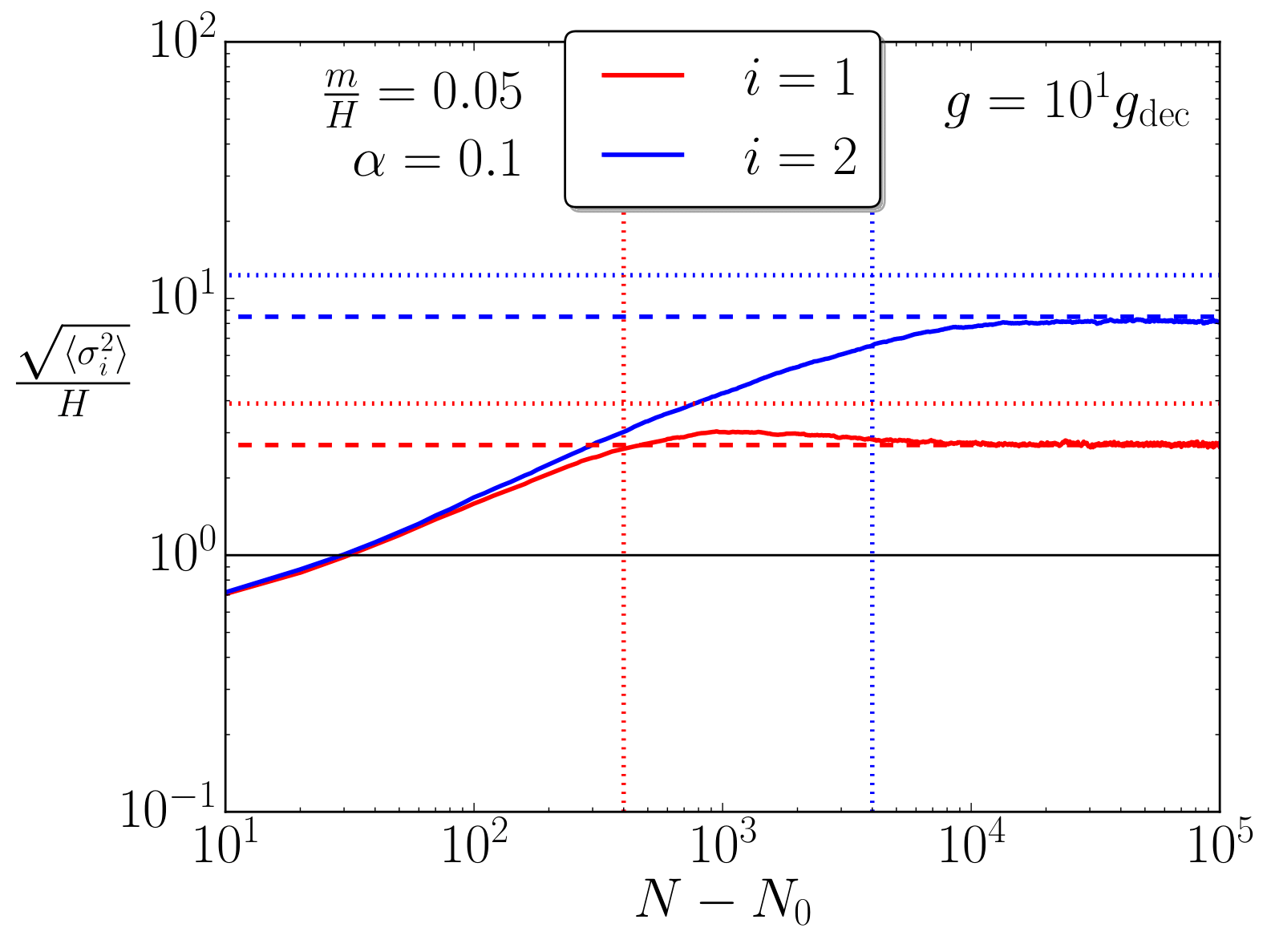}
\includegraphics[width=7cm]{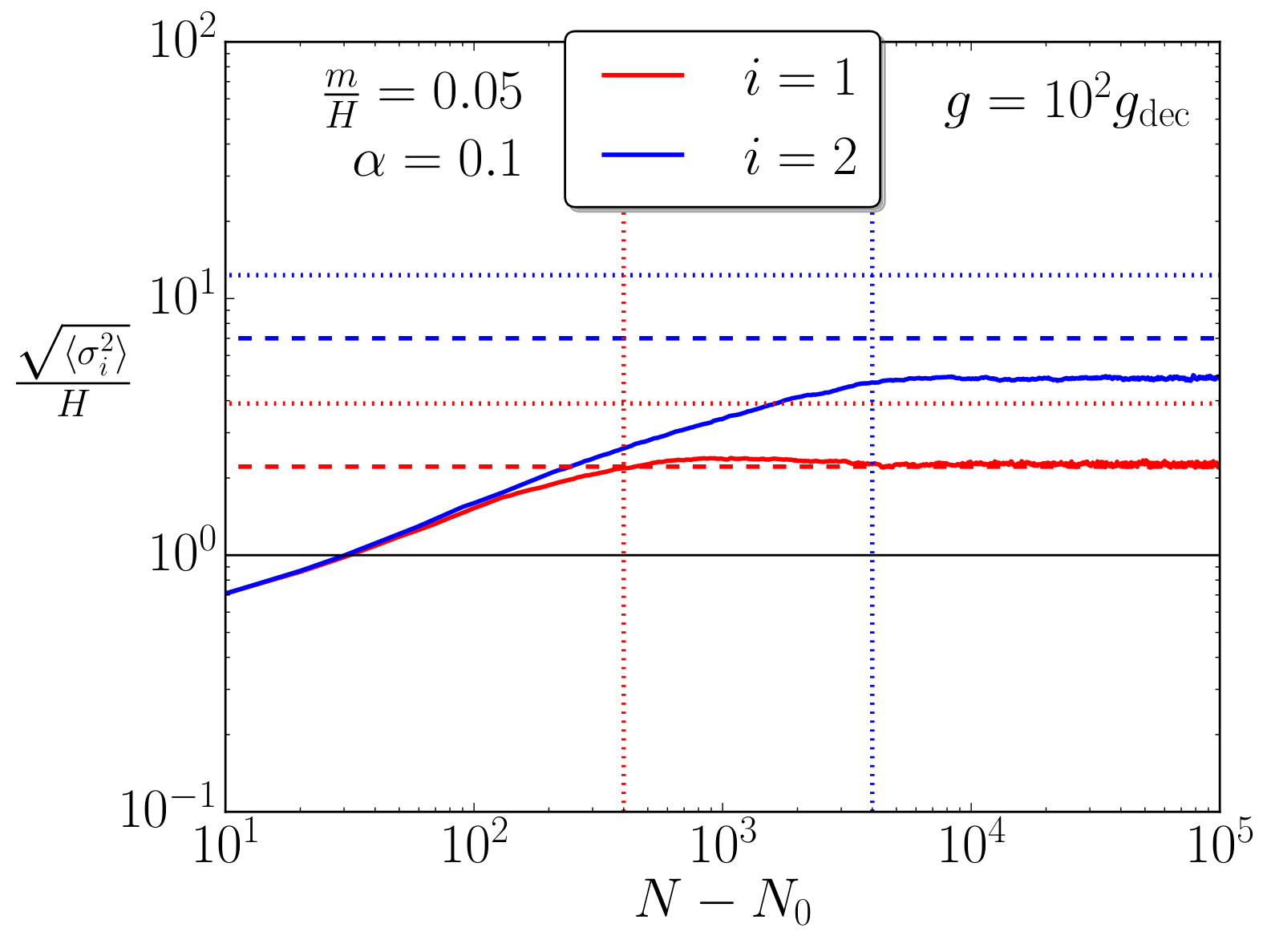}
\caption[Time evolution of coupled quadratic spectator fields]{~\label{fig:quadratic-plots} The numerically evaluated (solid lines) time evolution of the variance for each spectator field in the case of the $V_{\rm B}$ potential (see \Eq{eq:VBpot}), where variance of both fields is initialised at $\left\langle \sigma_i^2\right\rangle = 0$ and $10^4$ realisations of \Eq{eq:langevin-multi} were used in \href{https://sites.google.com/view/nfield-py}{\texttt{nfield}}. Dotted horizontal and dotted vertical lines represent the stationary variances (Eqs. \eqref{eq:stationary_VB_variances_1} and \eqref{eq:stationary_VB_variances_2}) and equilibration timescales (Eqs. \eqref{eq:stationary_VB_Neq_1} and \eqref{eq:stationary_VB_Neq_2}), respectively, computed in the decoupled limit $g\leq g_{\rm dec}$. The dashed horizontal lines use an alternative method to derive the stationary variance by numerically evaluating the second moment of \Eq{eq:exp-stat-dist}. The number of realisations used in each case is $10^4$. }
\end{center}
\end{figure}

\begin{figure}
\begin{center}
\includegraphics[width=6.5cm]{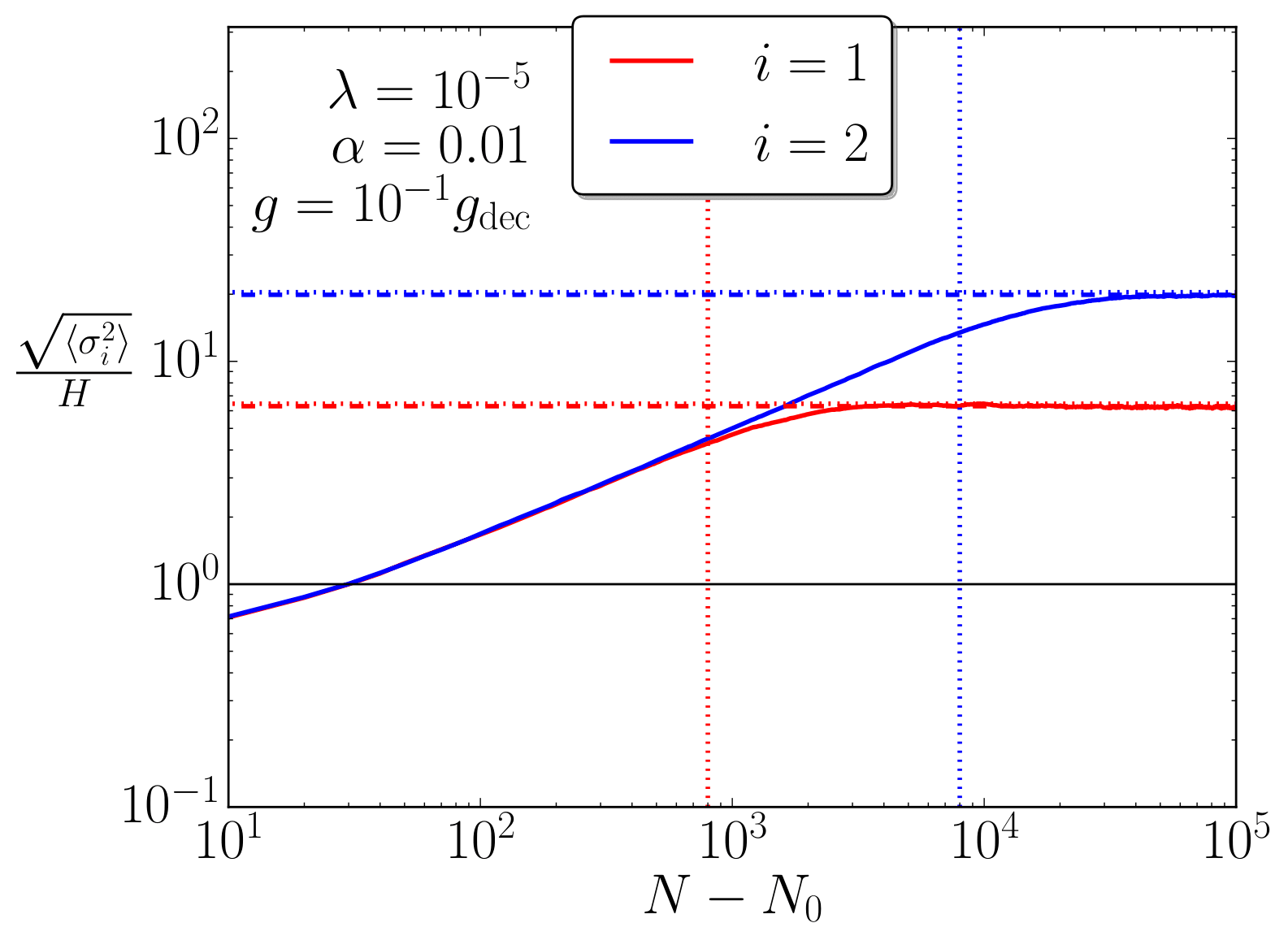}
\includegraphics[width=6.5cm]{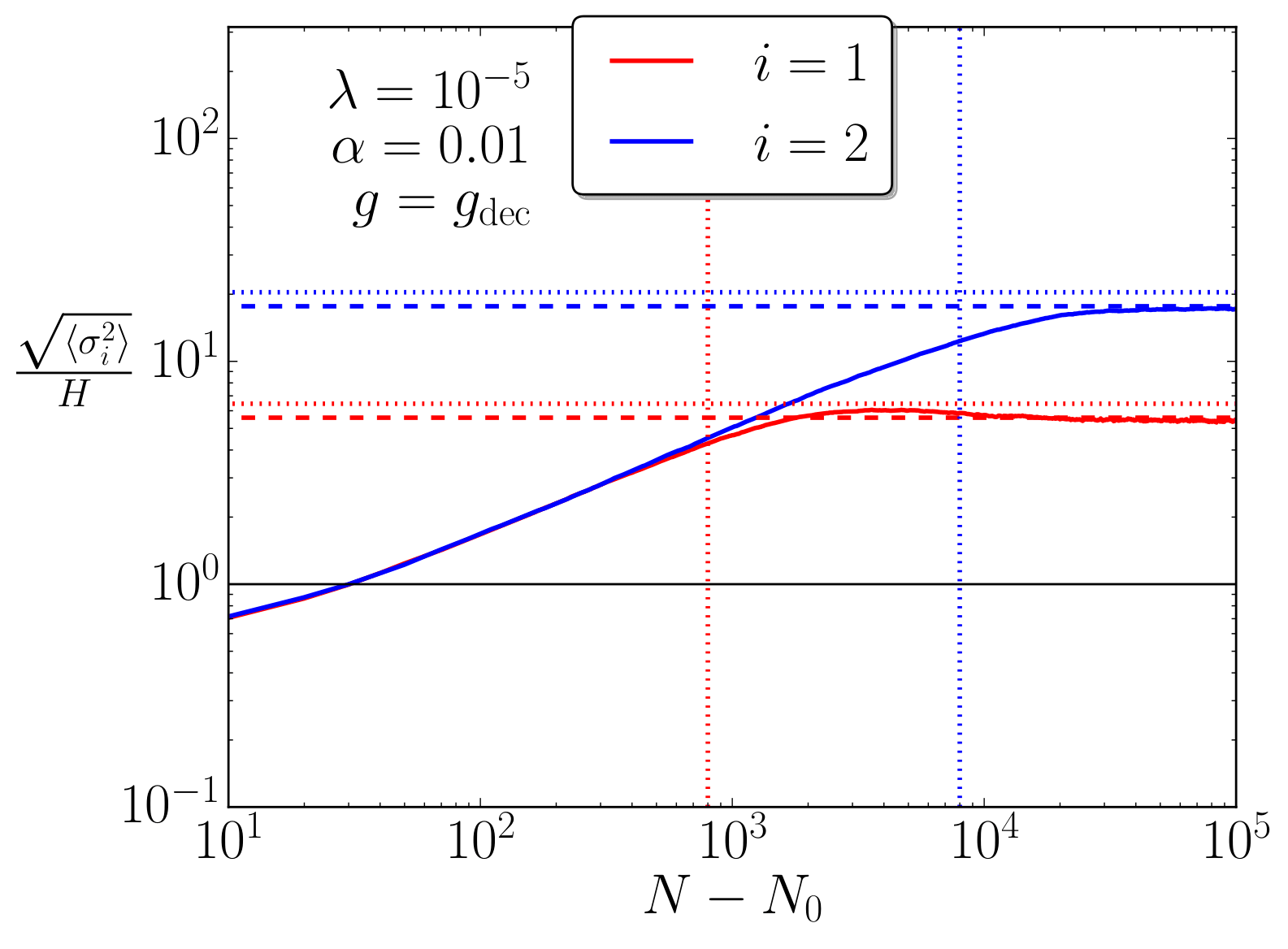} \\
\includegraphics[width=6.5cm]{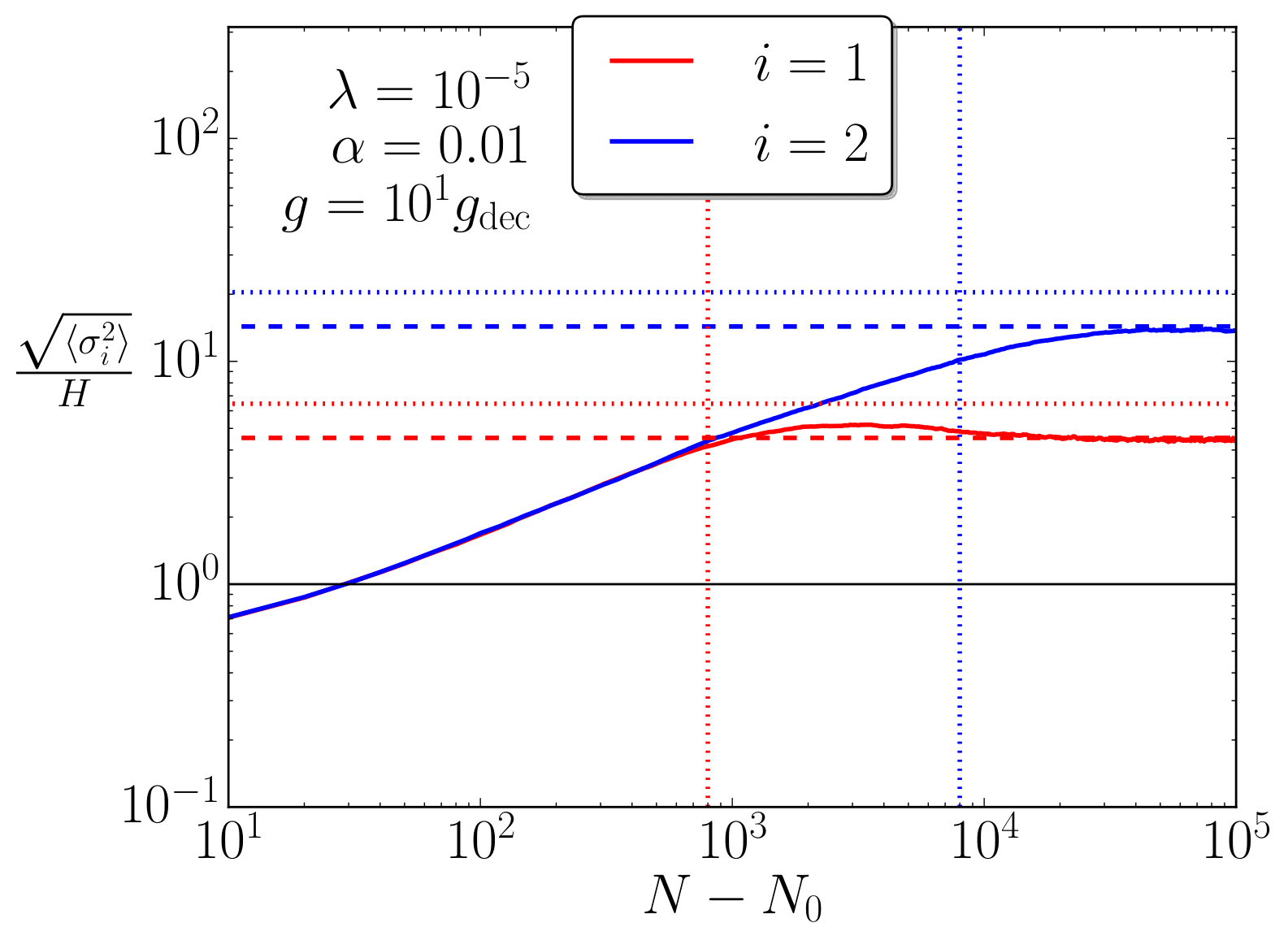}
\includegraphics[width=6.5cm]{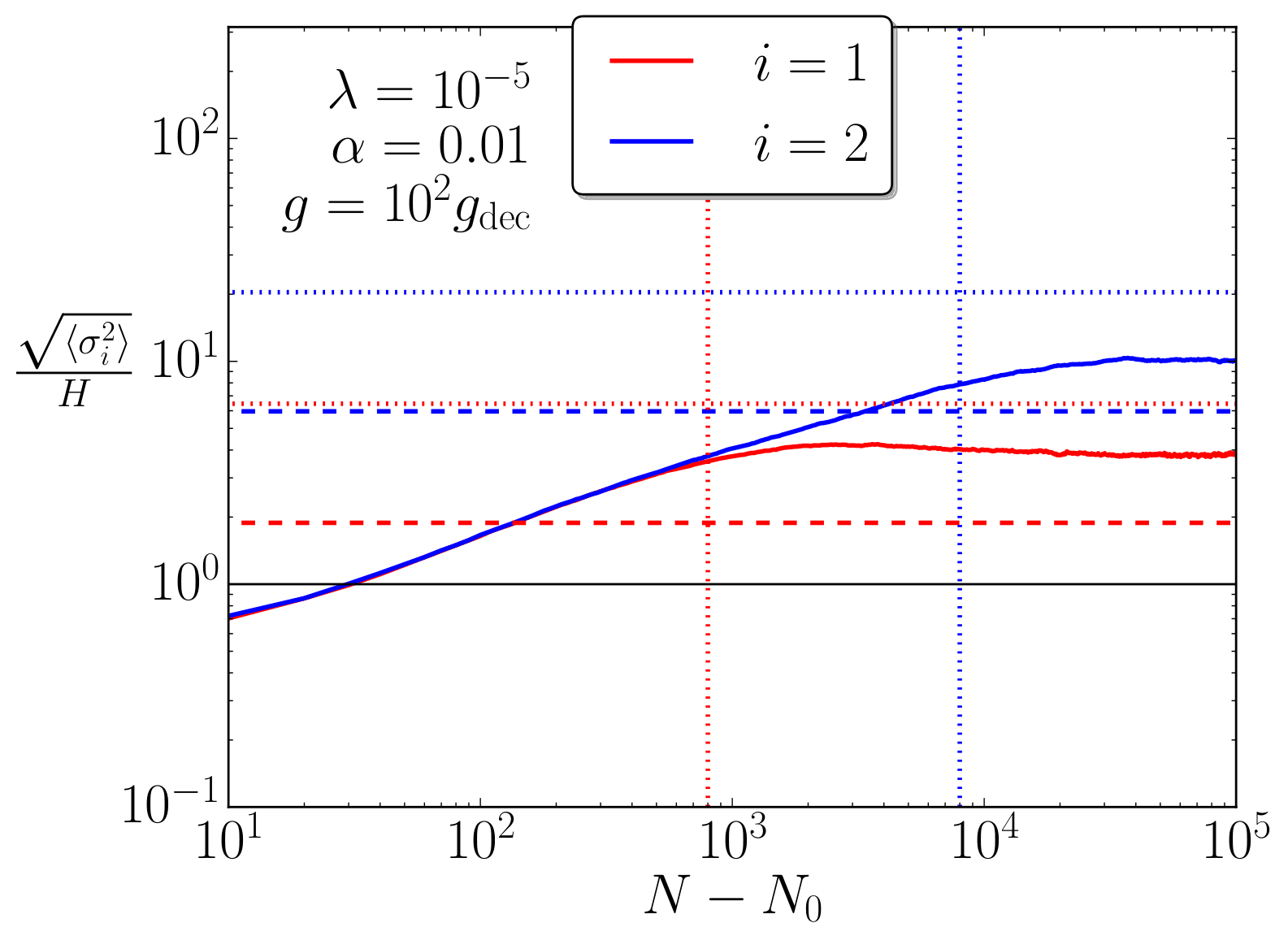}
\caption[Time evolution of coupled quartic spectator fields]{~\label{fig:quartic-plots} The numerically evaluated (solid lines) time evolution of the variance for each spectator field in the case of the $V_{\rm C}$ potential (see \Eq{eq:VCpot}), where variance of both fields is initialised at $\left\langle \sigma_i^2\right\rangle = 0$ and $10^4$ realisations of \Eq{eq:langevin-multi} were used in \href{https://sites.google.com/view/nfield-py}{\texttt{nfield}}. Dotted horizontal and dotted vertical lines represent the stationary variances (Eqs. \eqref{eq:stationary_VC_variances_1} and \eqref{eq:stationary_VC_variances_2}) and equilibration timescales (Eqs. \eqref{eq:stationary_VC_Neq_1} and \eqref{eq:stationary_VC_Neq_2}), respectively, computed in the decoupled limit $g\leq g_{\rm dec}$. The dashed horizontal lines use an alternative method to derive the stationary variance by numerically evaluating the second moment of \Eq{eq:exp-stat-dist}. The number of realisations used in each case is $10^4$. }
\end{center}
\end{figure}

\begin{figure}
\begin{center}
\includegraphics[width=7cm]{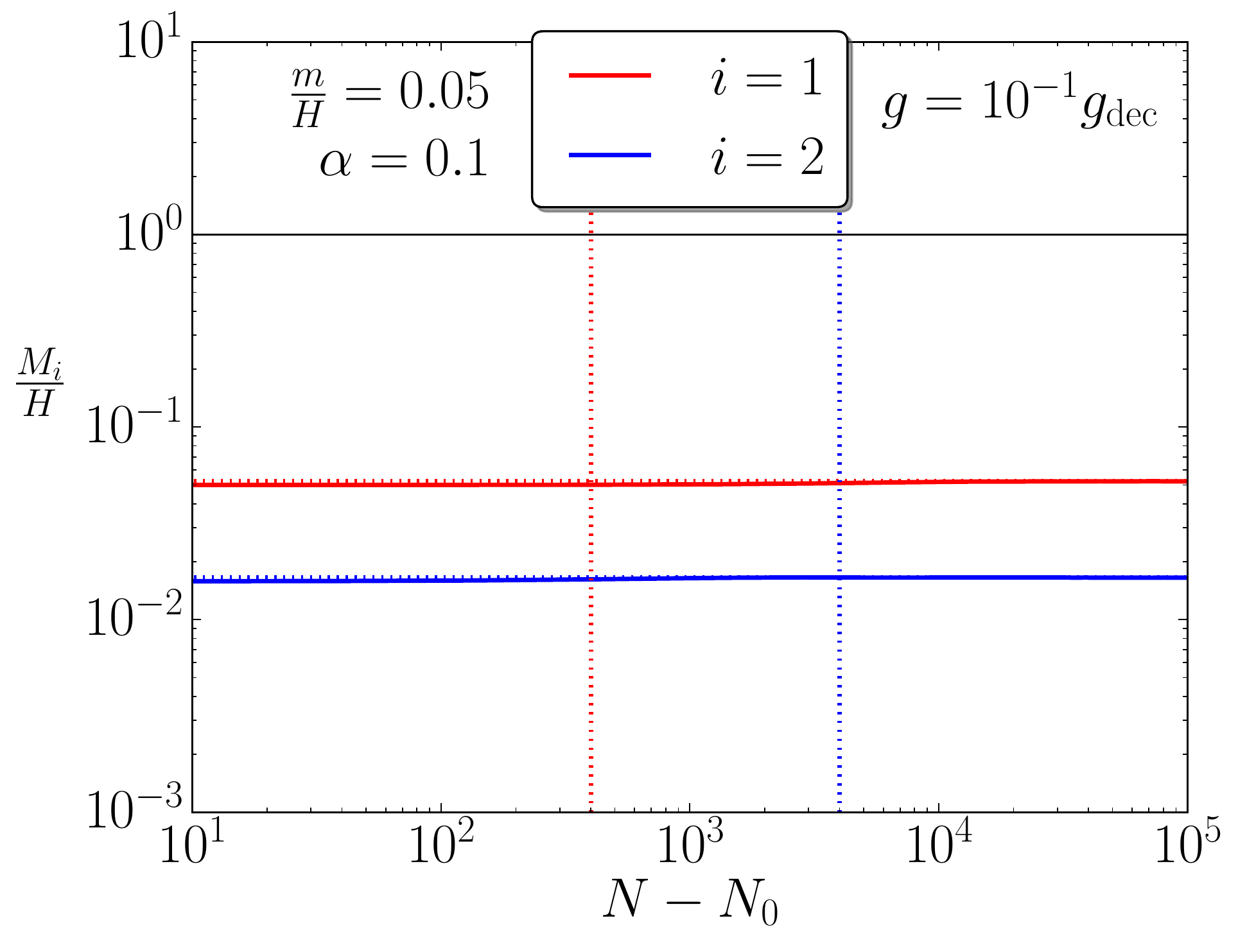}
\includegraphics[width=7cm]{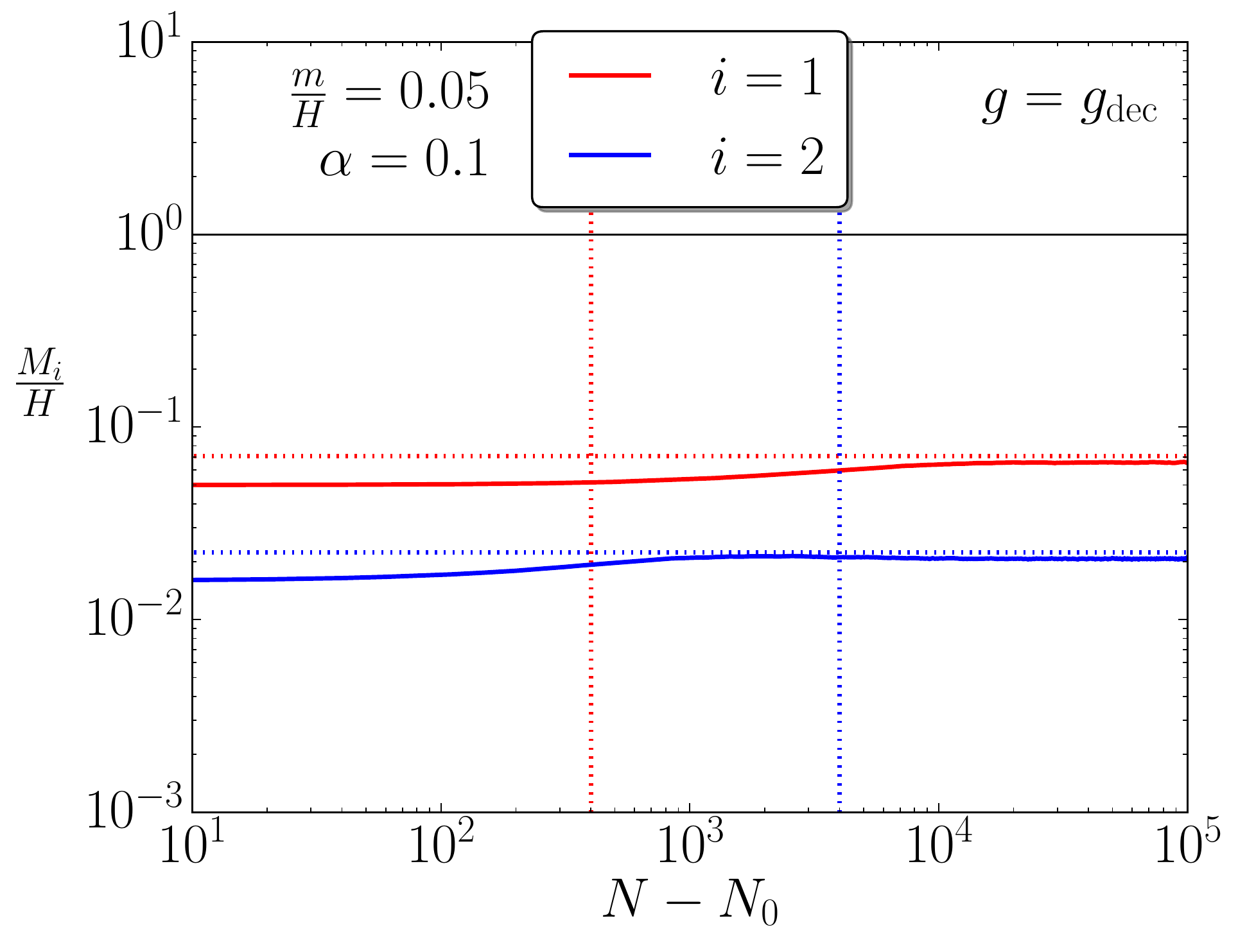} \\
\includegraphics[width=7cm]{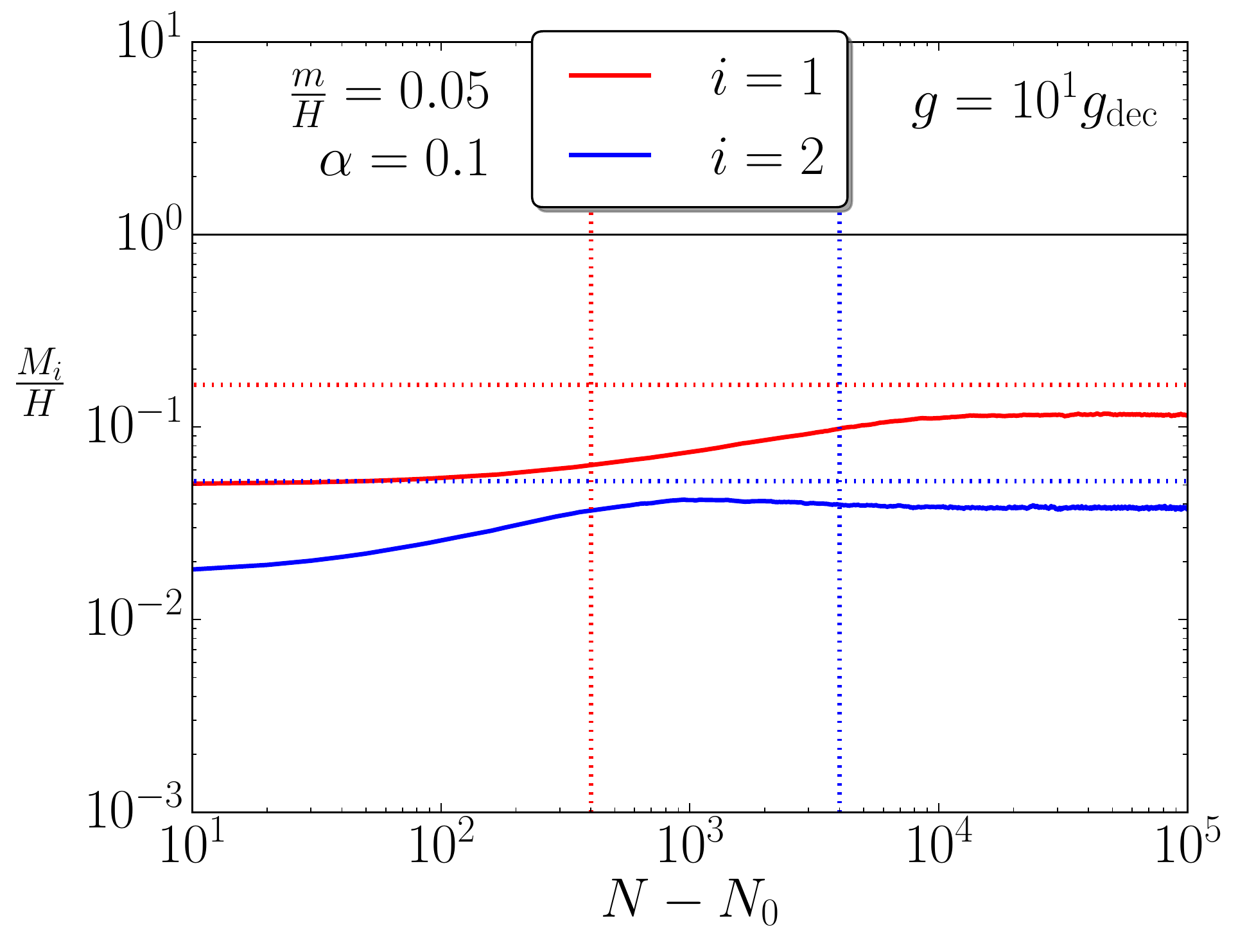}
\includegraphics[width=7cm]{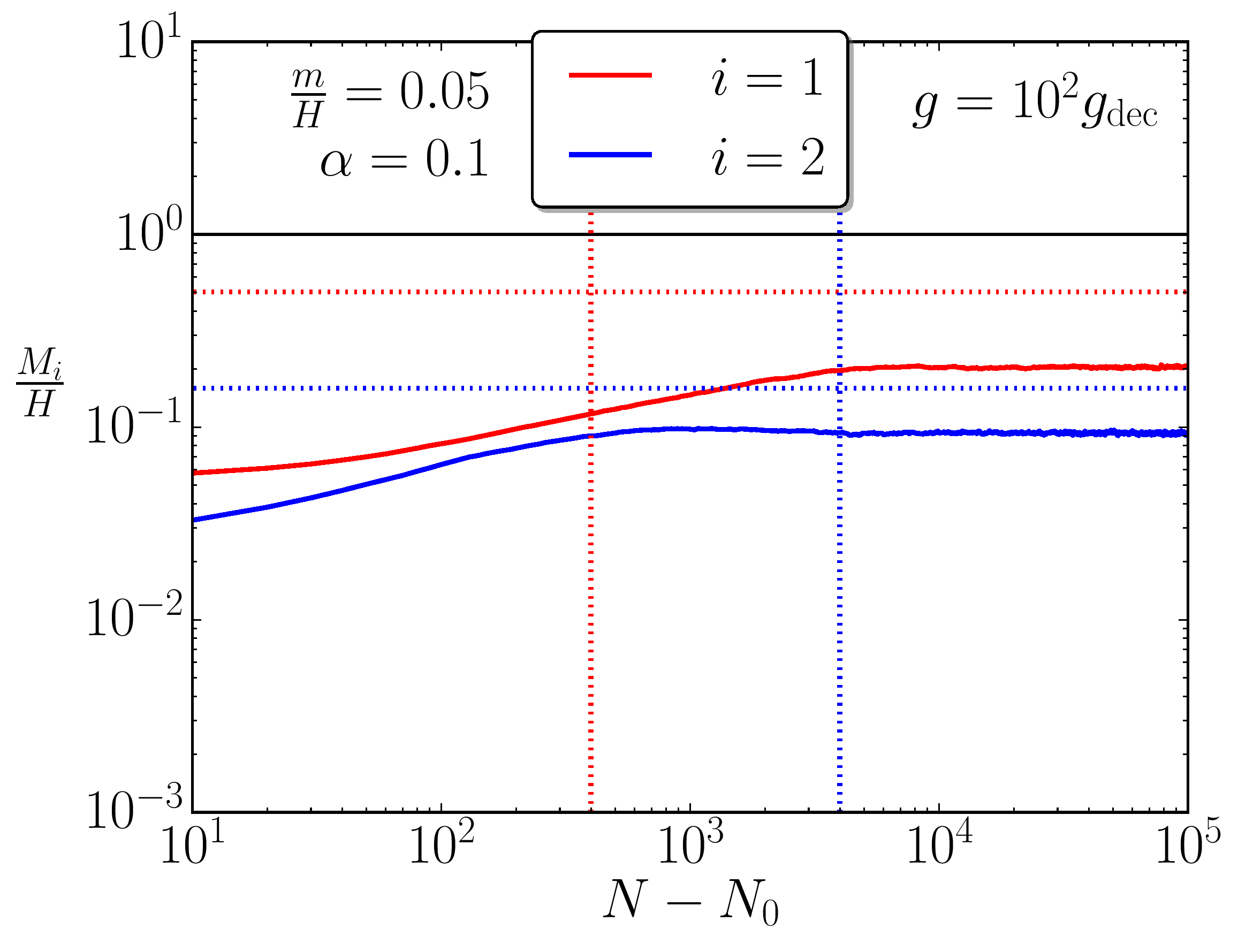}
\caption[The time evolution of the effective mass for coupled quadratic fields]{~\label{fig:meff-quadratic-plots} The numerically evaluated (solid lines) time evolution of the effective mass (see \Eq{eq:effm}) for each spectator field in \Fig{fig:quadratic-plots}. Dotted horizontal and dotted vertical lines represent $M_i$ derived using the stationary variances (Eqs. \eqref{eq:stationary_VB_variances_1} and \eqref{eq:stationary_VB_variances_2}) and the equilibration timescales (Eqs. \eqref{eq:stationary_VB_Neq_1} and \eqref{eq:stationary_VB_Neq_2}), respectively. }
\end{center}
\end{figure}

\begin{figure}
\begin{center}
\includegraphics[width=7cm]{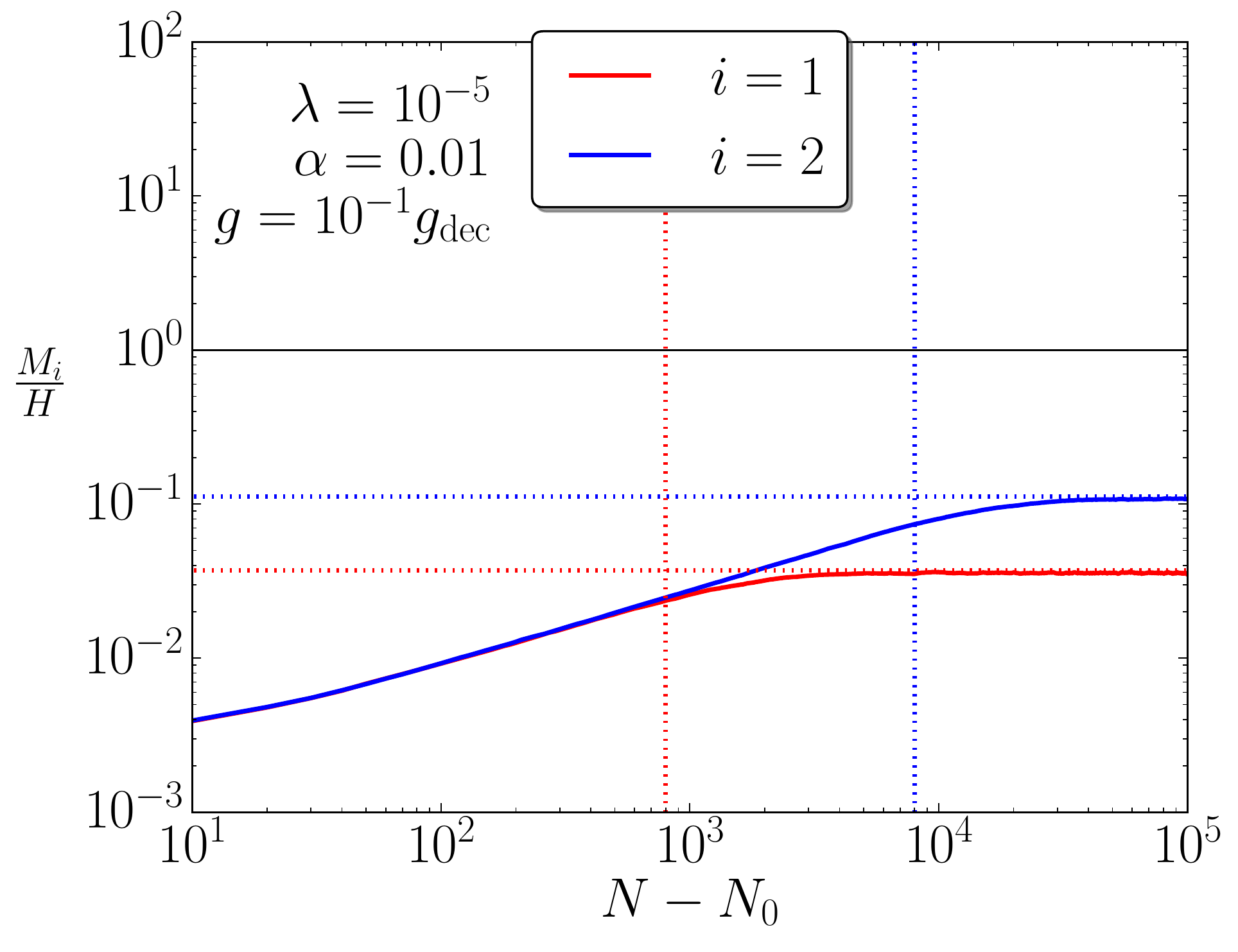}
\includegraphics[width=7cm]{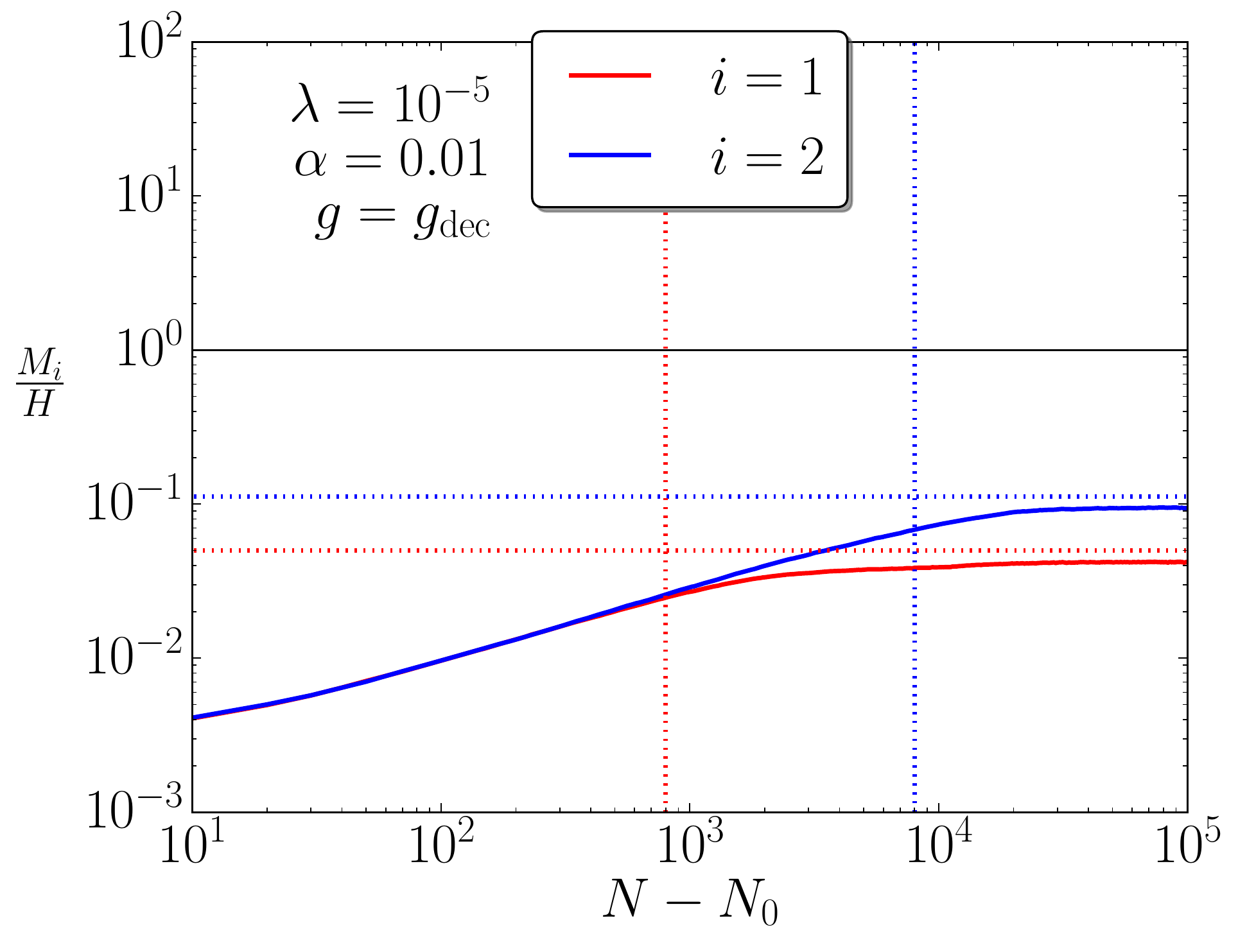} \\
\includegraphics[width=7cm]{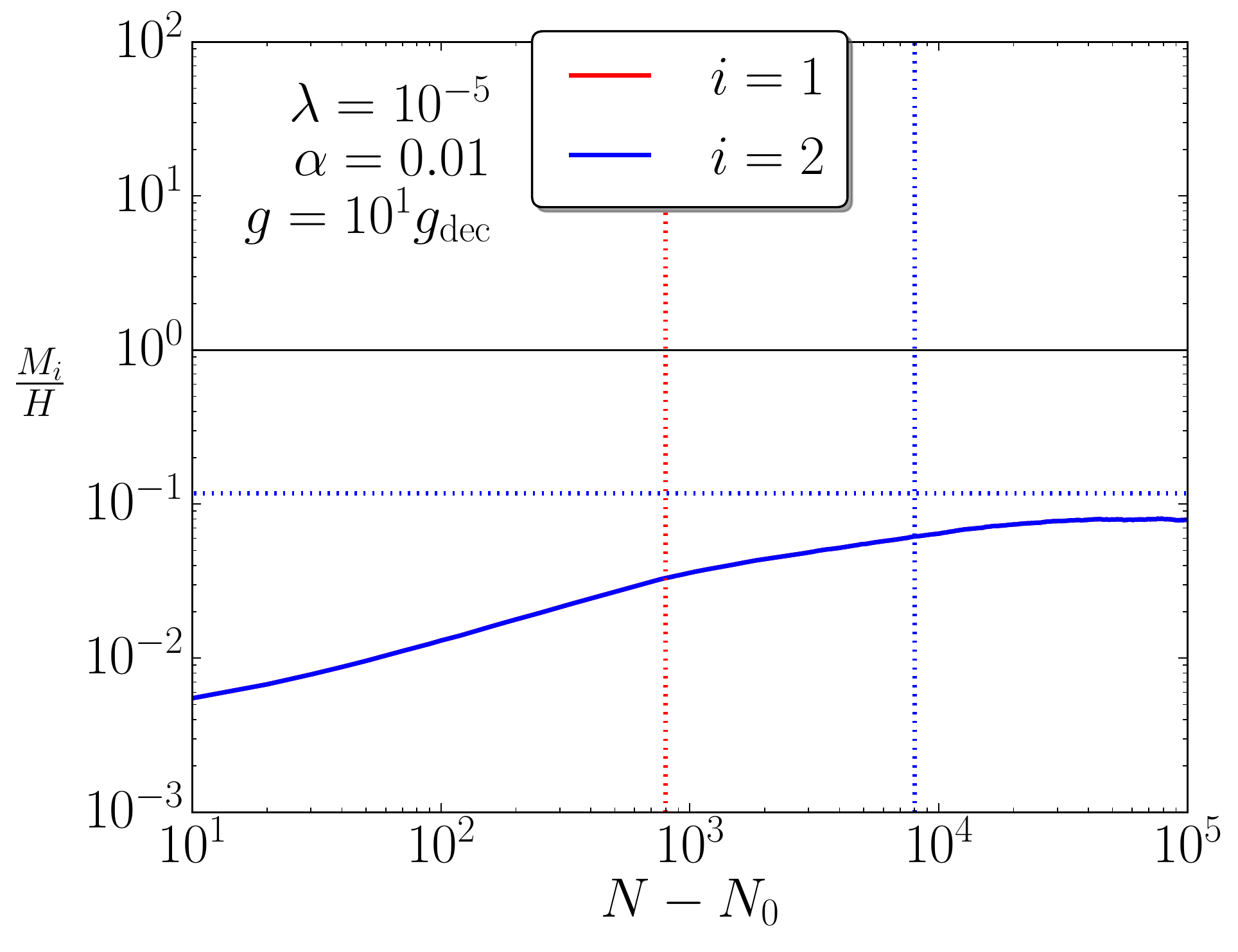}
\includegraphics[width=7cm]{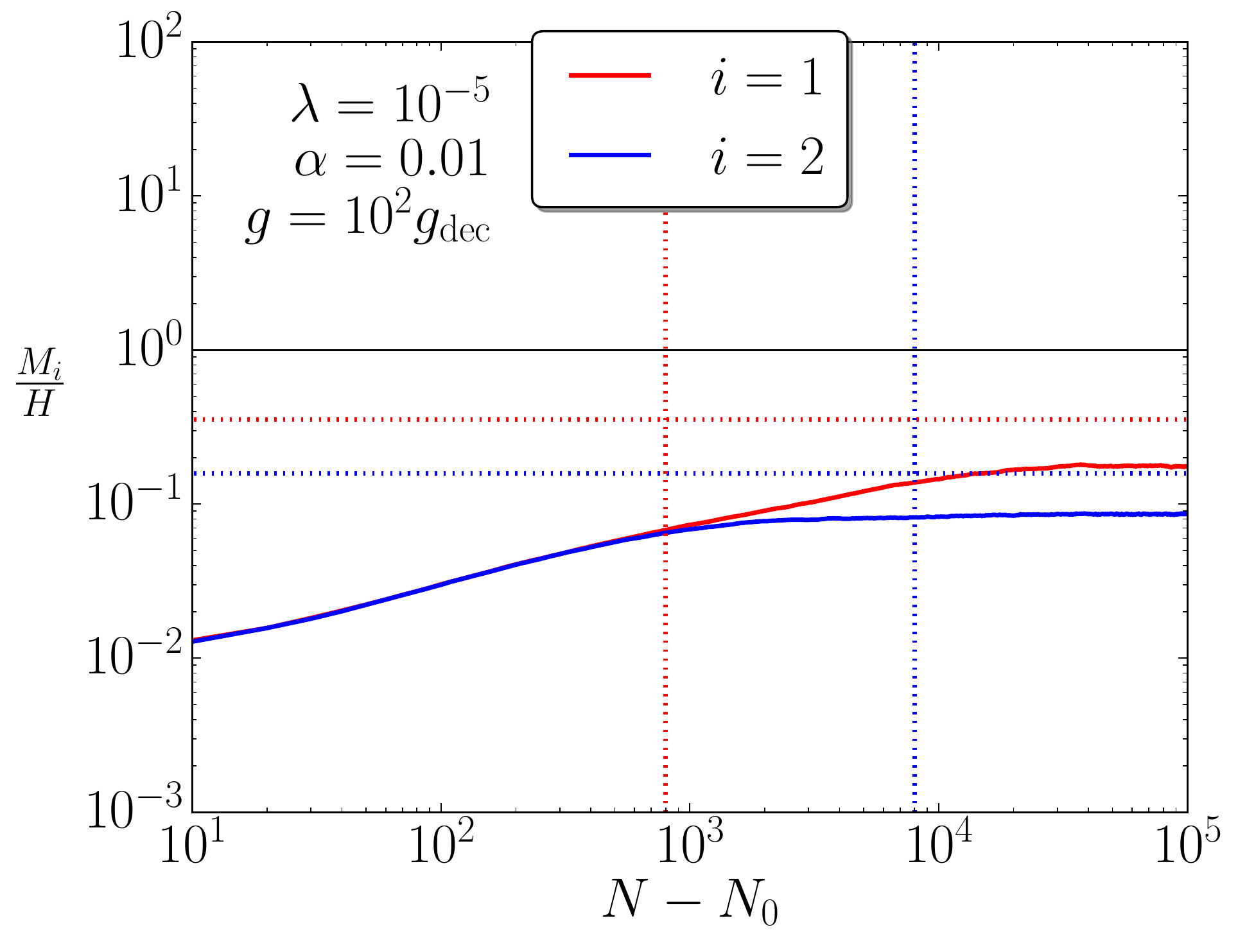}
\caption[The time evolution of the effective mass for coupled quartic fields]{~\label{fig:meff-quartic-plots} The numerically evaluated (solid lines) time evolution of the effective mass (see \Eq{eq:effm}) for each spectator field in \Fig{fig:quartic-plots}. Dotted horizontal and dotted vertical lines represent $M_i$ derived using the stationary variances (Eqs. \eqref{eq:stationary_VC_variances_1} and \eqref{eq:stationary_VC_variances_2}) and the equilibration timescales (Eqs. \eqref{eq:stationary_VC_Neq_1} and \eqref{eq:stationary_VC_Neq_2}), respectively. }
\end{center}
\end{figure}

If $g > g_{\rm dec}$ and $\alpha \neq 1$ however, then neither the equations above, nor the symmetry of $P(\sigma_i,N)$, can be exploited for analytic calculations and hence one must rely upon the numerically evaluated solution in order to study the system. In \Fig{fig:quadratic-plots} and \Fig{fig:quartic-plots} we plot these numerical solutions (and their corresponding effective masses in \Fig{fig:meff-quadratic-plots} and \Fig{fig:meff-quartic-plots}) given some specific values of $( m \, {\rm or} \, \lambda , \alpha , g )$ for both fields in the quadratic and quartic potentials, respectively. The variances are all initialised with $\left\langle \sigma_i^2\right\rangle = 0$ and hence the number of \efold{s} it takes for each solution to reach the effectively decoupled stationary values (dotted horizontal lines in the relevant colour using Eqs. \eqref{eq:stationary_VB_variances_1}, \eqref{eq:stationary_VB_variances_2}, \eqref{eq:stationary_VC_variances_1} and \eqref{eq:stationary_VC_variances_2}) is well-approximated by the analytic relaxation timescales derived in Eqs. \eqref{eq:stationary_VB_Neq_1}, \eqref{eq:stationary_VB_Neq_2}, \eqref{eq:stationary_VC_Neq_1} and \eqref{eq:stationary_VC_Neq_2} (depicted with vertical dotted lines in the relevant colour) in cases where $g\leq g_{\rm dec}$ (the top row plots of both sets of Figs.). In all plots, one can also clearly see the strong deviation from the decoupled predictions with larger values of $g$, which highlights the importance for a numerical solution from \href{https://sites.google.com/view/nfield-py}{\texttt{nfield}} in this large regime of parameter values to obtain the correct equilibrium as well as out-of-equilibrium behaviour.

\begin{figure}
\begin{center}
\includegraphics[width=9cm]{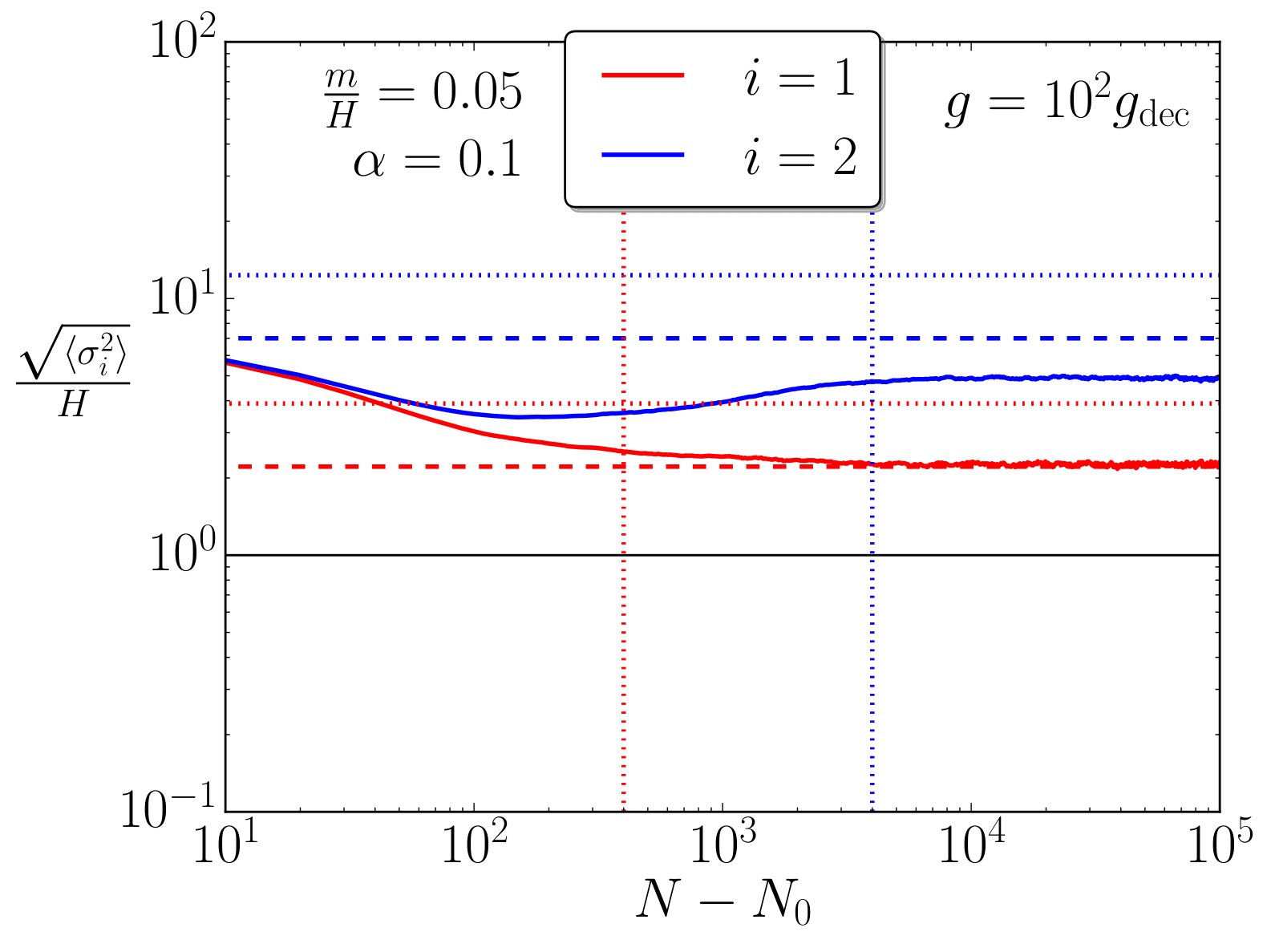}
\caption[Initial condition independence of coupled quadratic spectator fields]{~\label{fig:varyinitcond-VB-variance} An illustrative re-plotting of the numerical variance evolution, with an initial condition much closer to the analytic stationary variance derived from the second moment of \Eq{eq:exp-stat-dist}, for the case in the bottom right-hand corner of \Fig{fig:quadratic-plots}.  }
\end{center}
\end{figure}

\subsection{\textsf{Non-vanishing probability currents}} \label{sec:non-vanish}

In \Fig{fig:quadratic-plots} there is also an important anomaly which appears to be repeated in \Fig{fig:quartic-plots}. In both figures we have also provided (dashed horizontal lines) an alternative calculation for the stationary variance using the numerically calculated second moment of \Eq{eq:exp-stat-dist}. There is generally excellent agreement  between this solution and the one obtained from the many realisations of \Eq{eq:langevin-multi} in \href{https://sites.google.com/view/nfield-py}{\texttt{nfield}} for $g\leq 10g_{\rm dec}$, however these no longer agree precisely when $g = 10^2g_{\rm dec}$ in both sets of plots. This deviation has been checked for numerical robustness by increasing the number of Langevin realisations to $10^5$ and altering the initial conditions --- see \Fig{fig:varyinitcond-VB-variance} for illustration.

\begin{figure}
\begin{center}
\includegraphics[width=9cm]{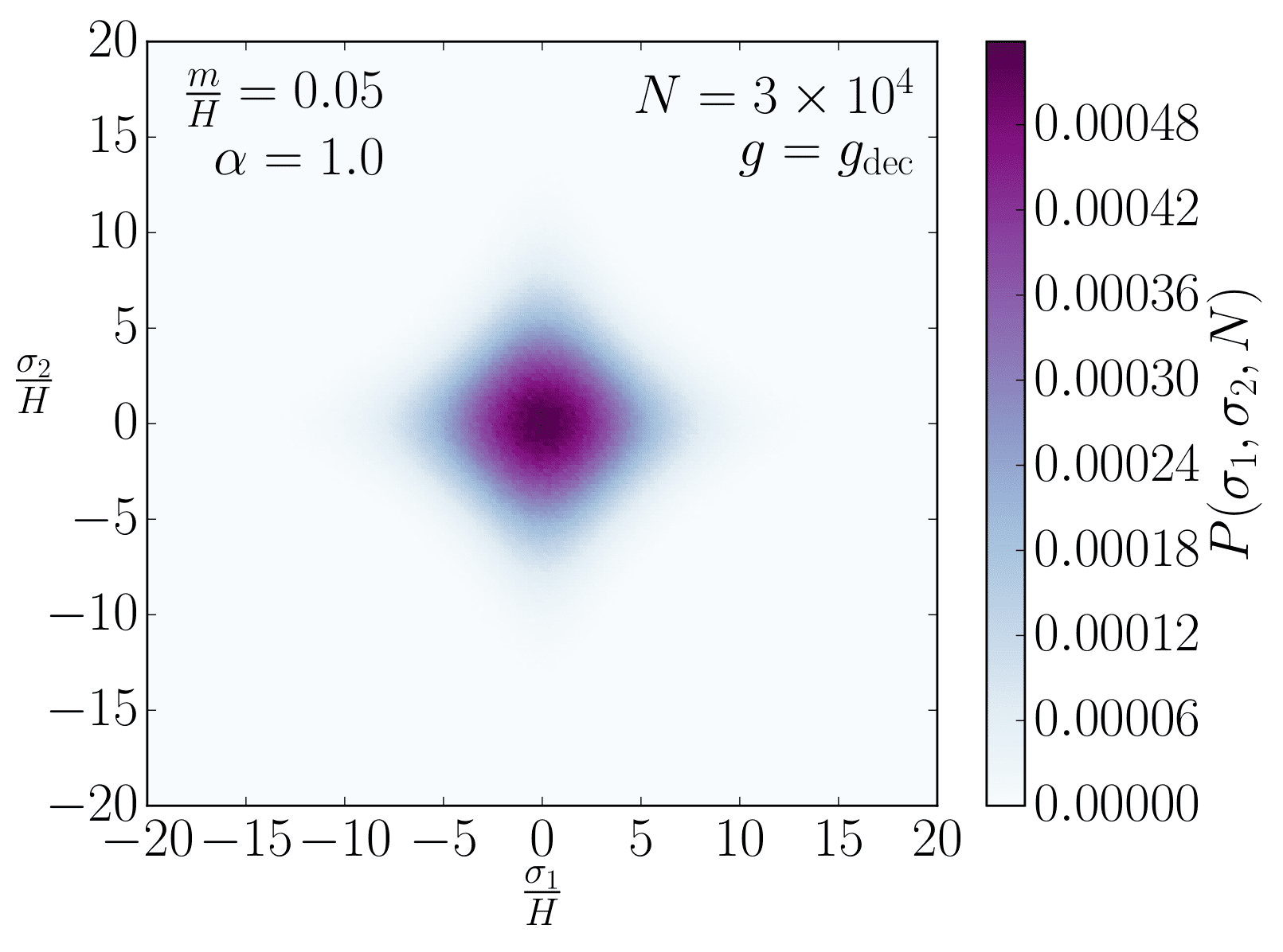} \\
\includegraphics[width=7cm]{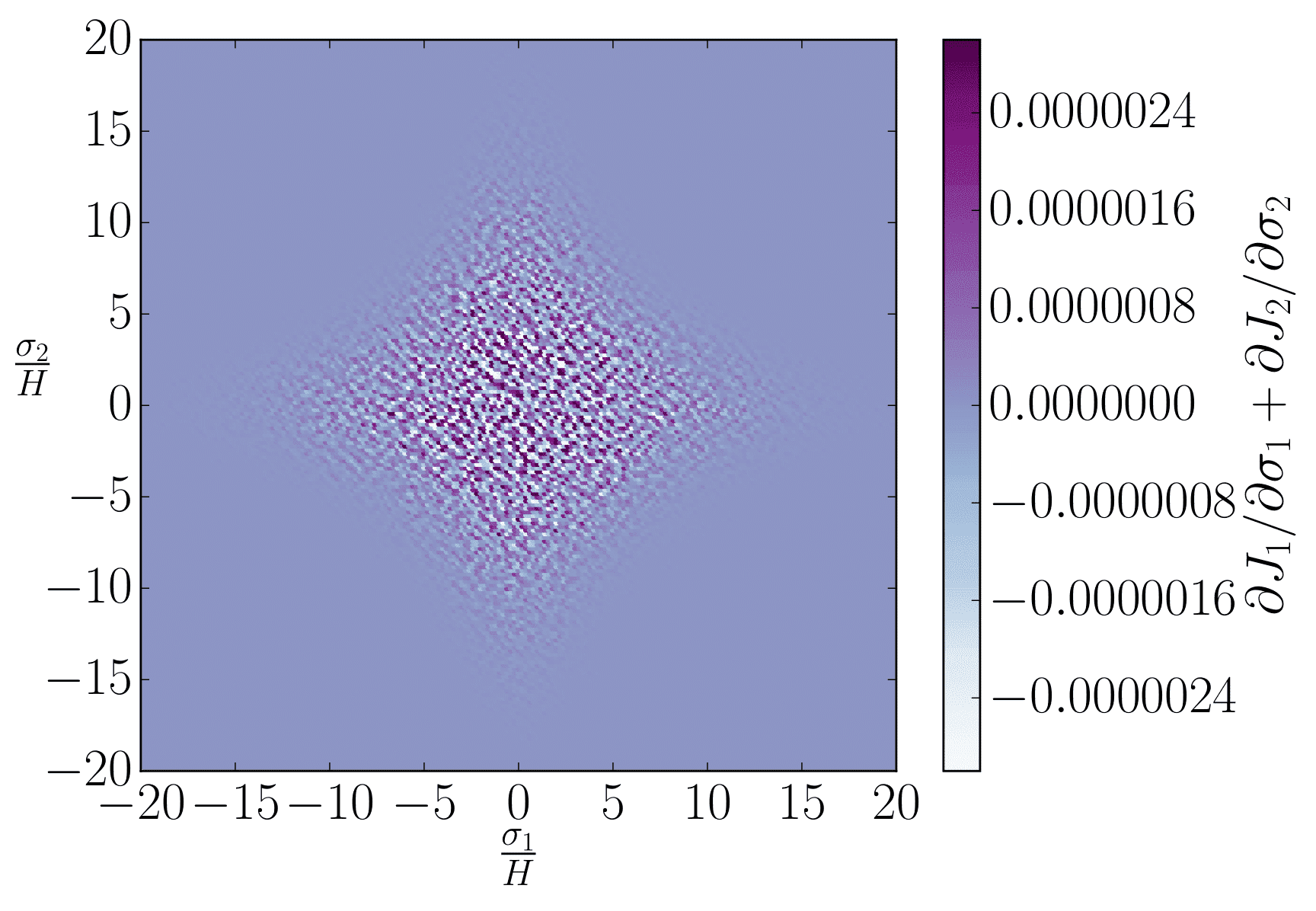}  \includegraphics[width=7cm]{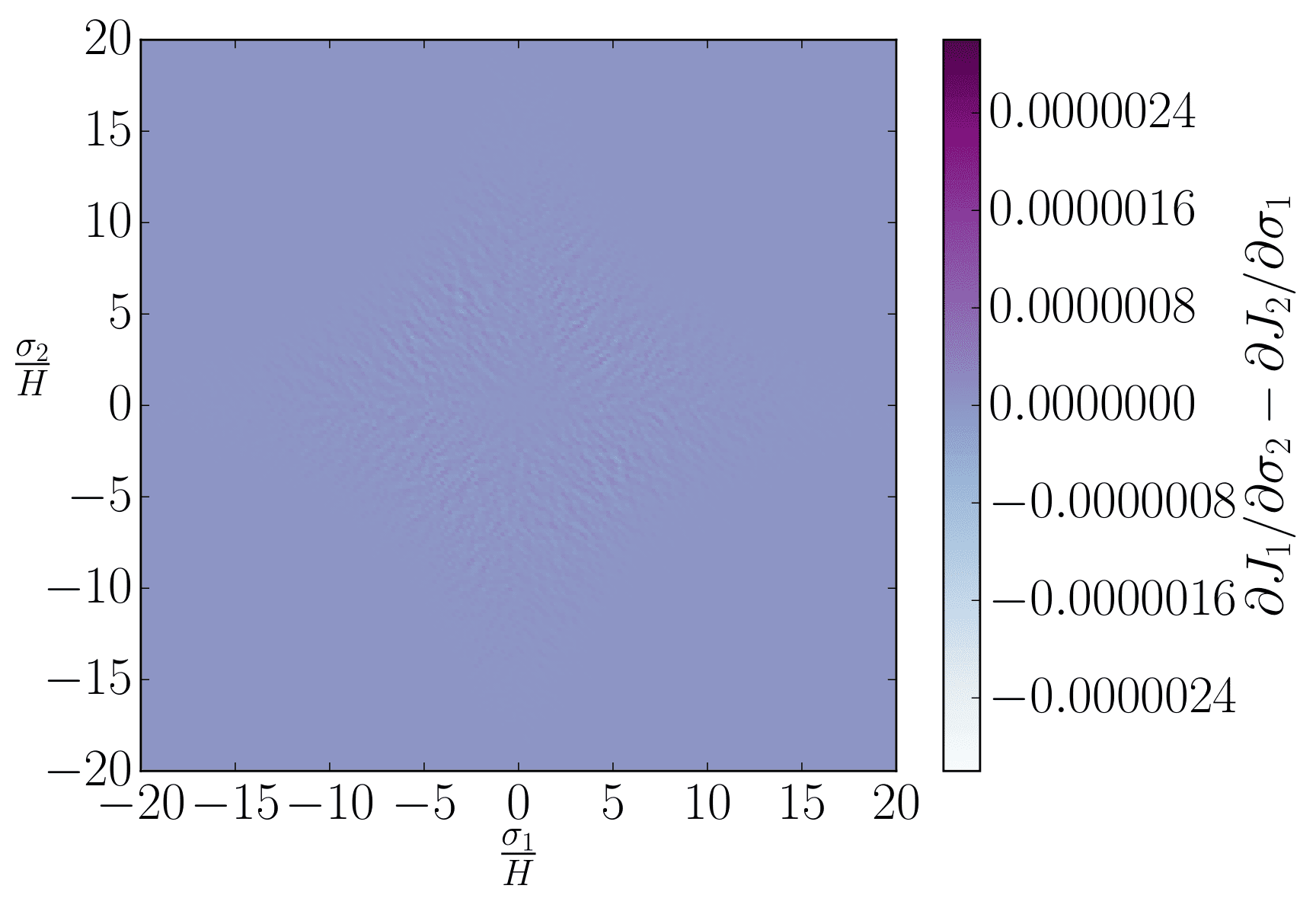}
\caption[Probability density function for coupled quadratic spectators]{~\label{fig:pdf-cur-plot1} Using specific parameter choices of the $V_{\rm B}$ potential (see \Eq{eq:VBpot}) we plot the binned stationary probability density $P_{\rm stat}$ (on top), probability current divergence $\nabla \cdot \boldsymbol{J}$ (bottom left) and the only non-zero component of the probability current curl $\nabla \times \boldsymbol{J}$ (bottom right). These have all been numerically obtained from $10^7$ realisations of \Eq{eq:langevin-multi} in the \href{https://sites.google.com/view/nfield-py}{\texttt{nfield}} code. }
\end{center}
\end{figure}

\begin{figure}[t]
\begin{center}
\includegraphics[width=7cm]{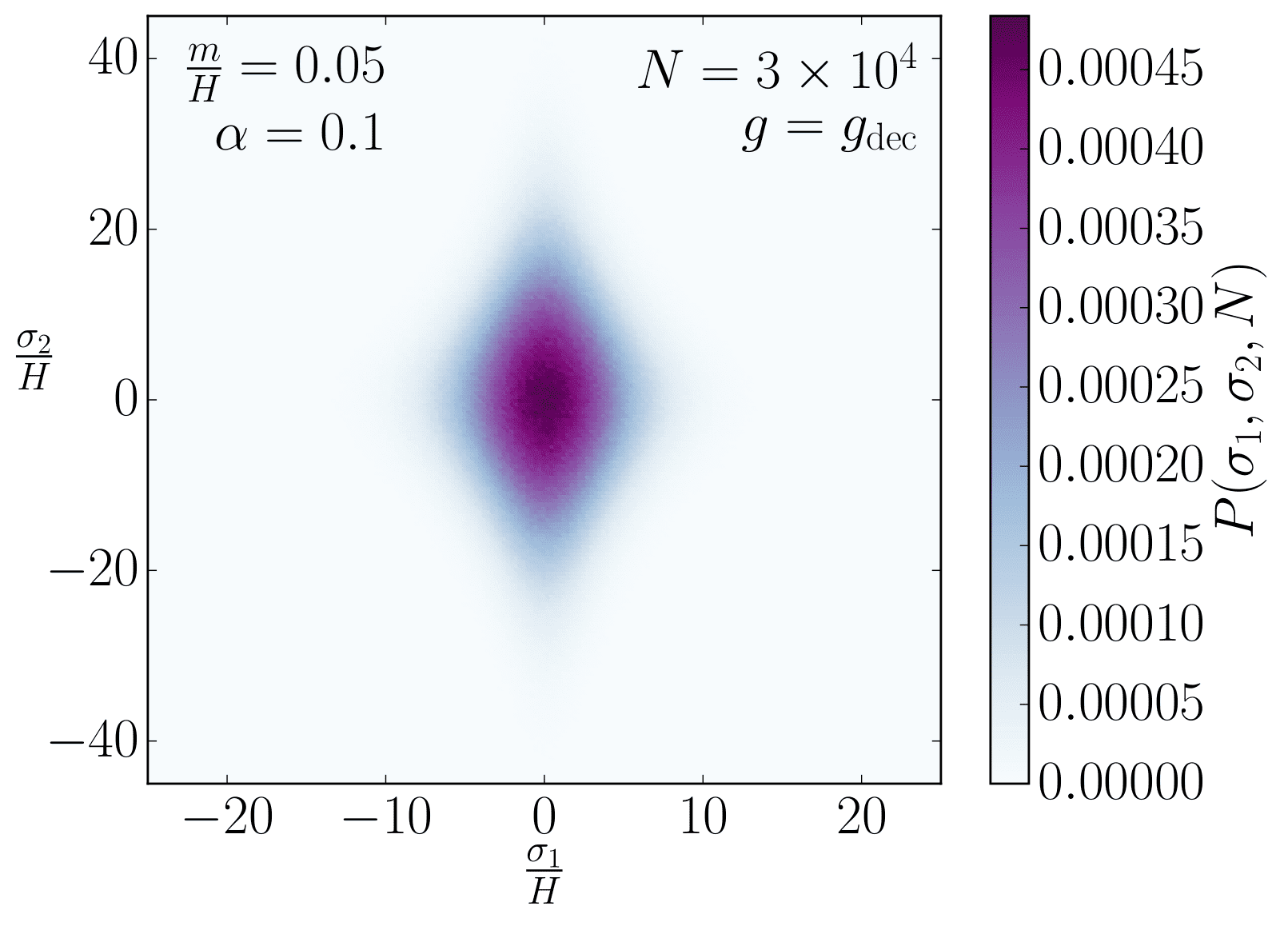}
\includegraphics[width=7cm]{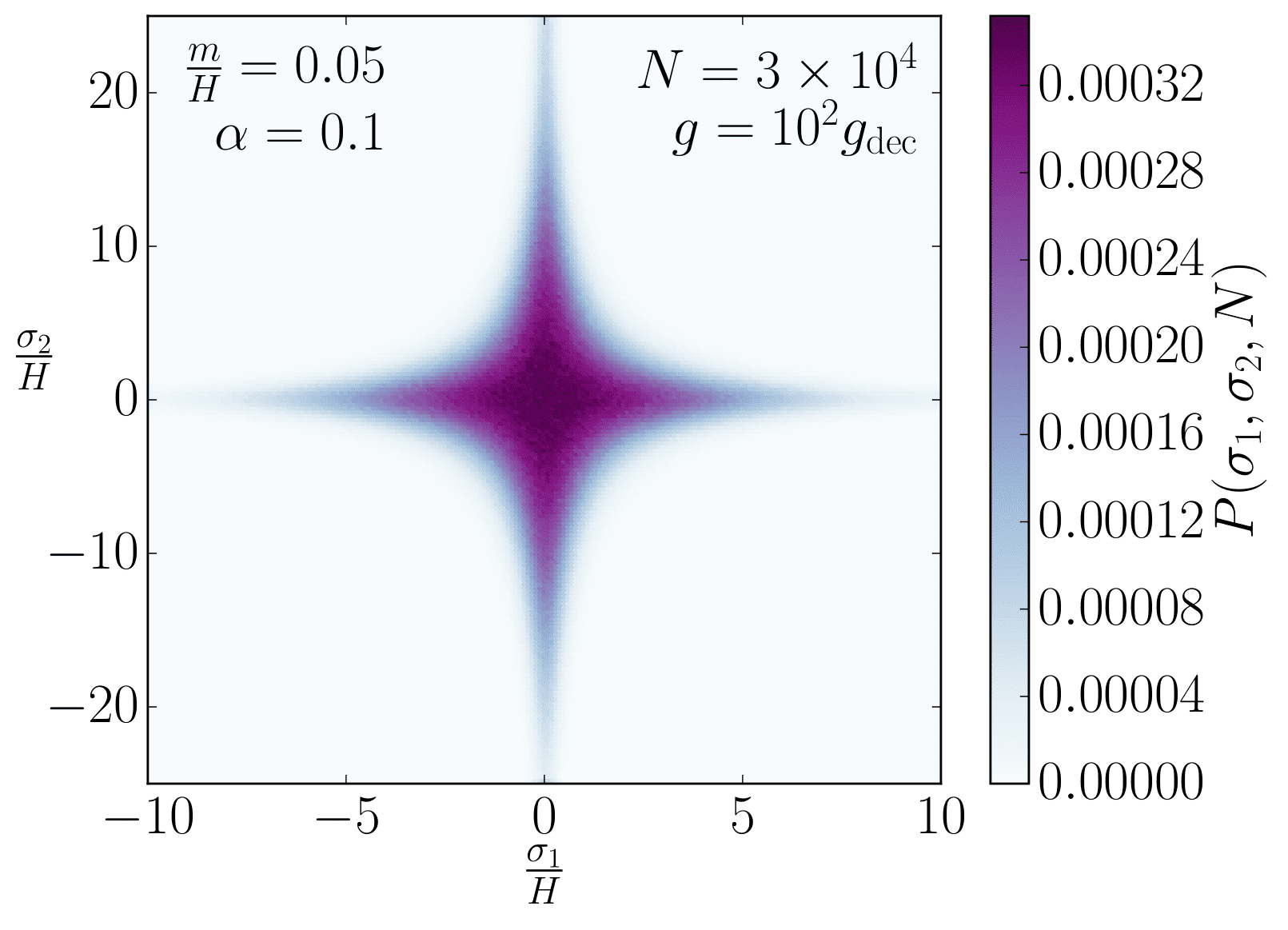}  \\
\includegraphics[width=7cm]{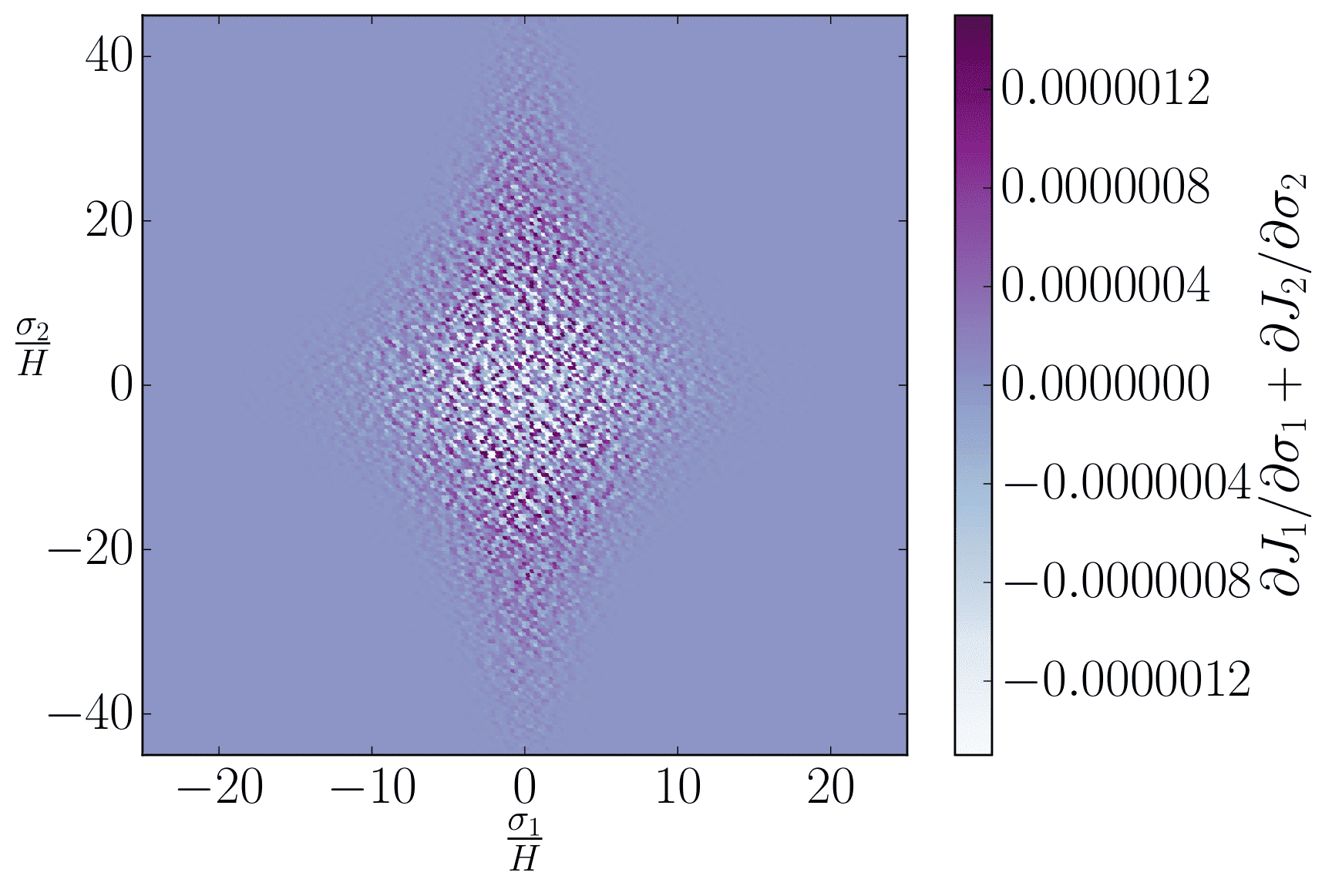}
\includegraphics[width=7cm]{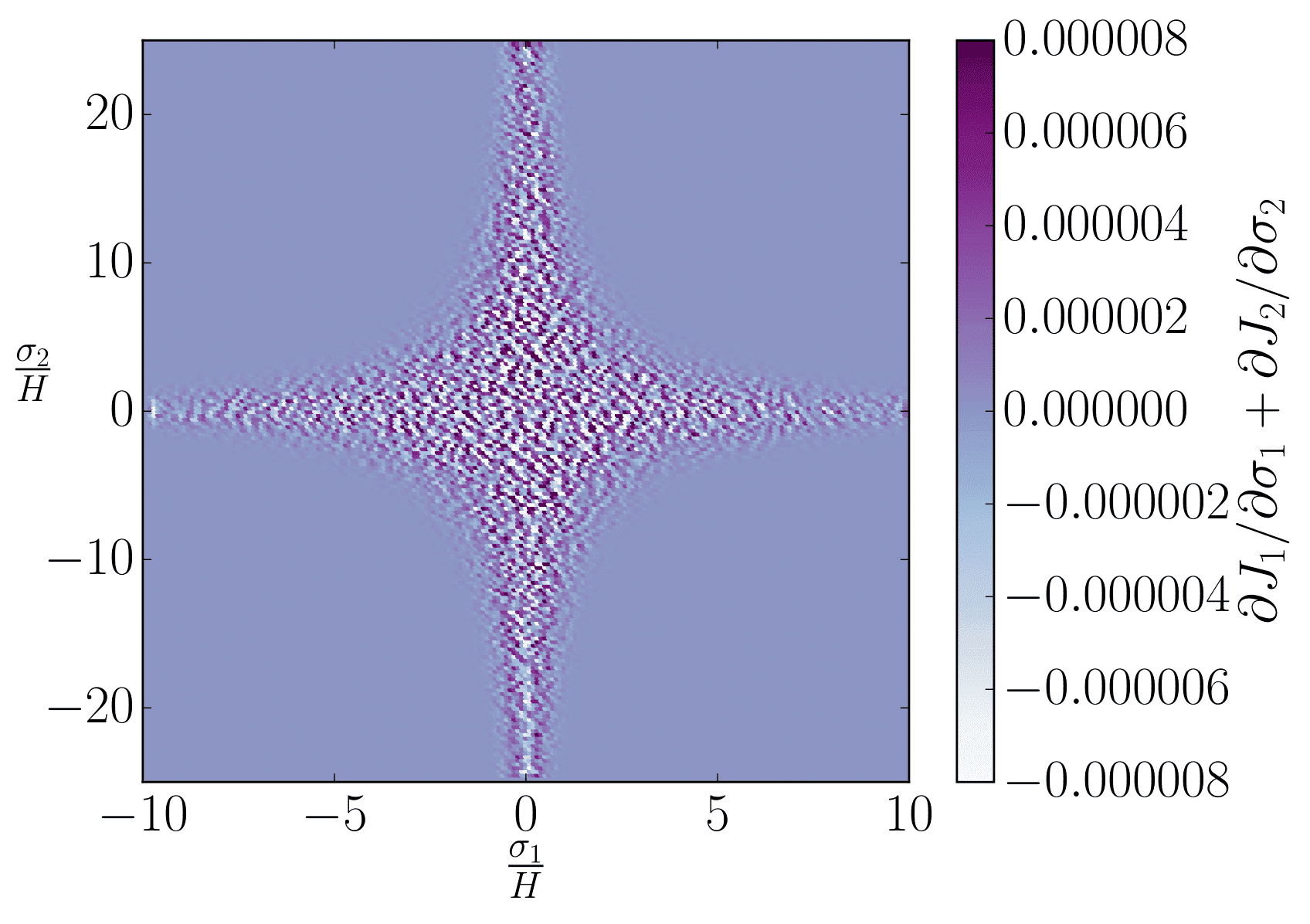} \\
\includegraphics[width=7cm]{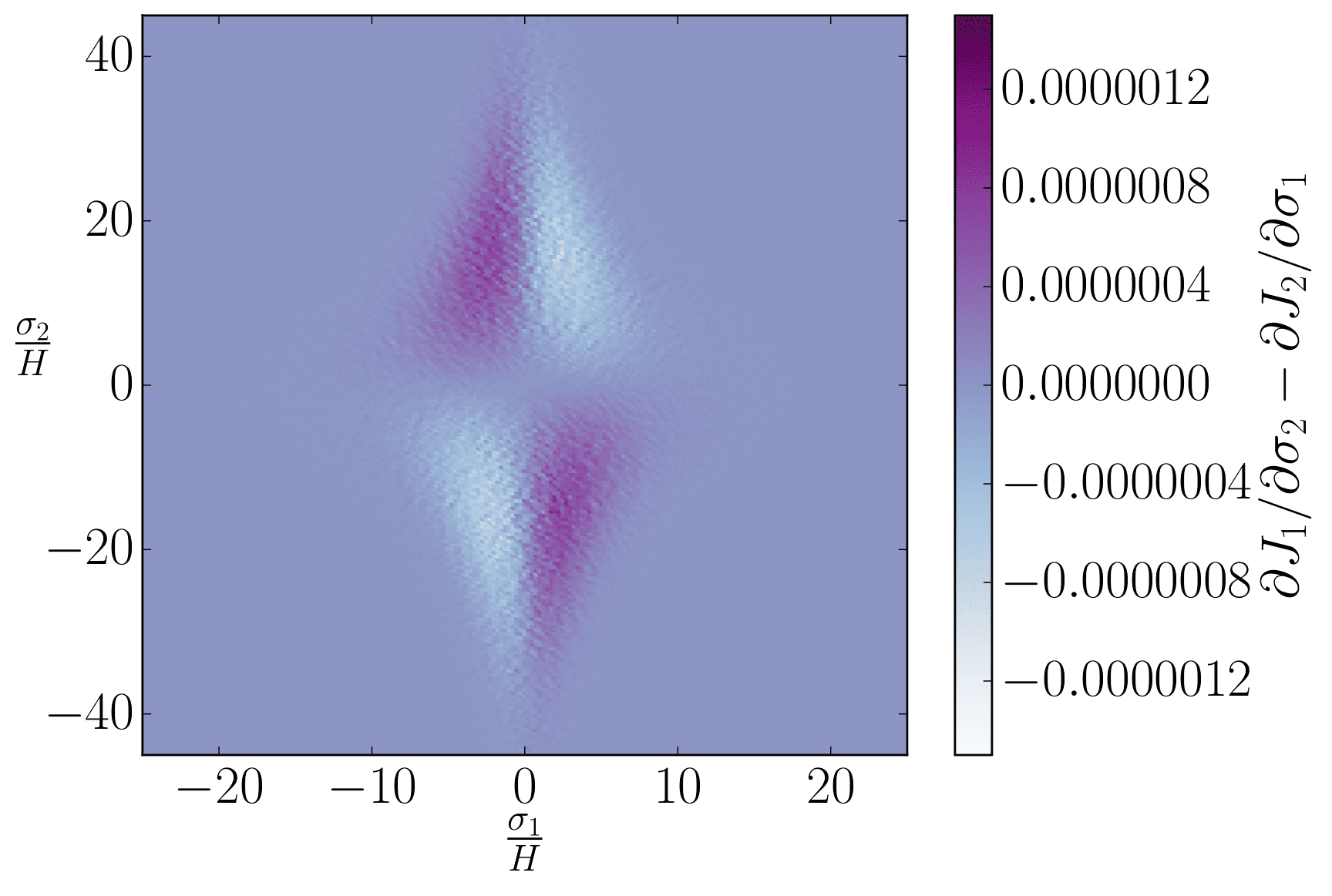}
\includegraphics[width=7cm]{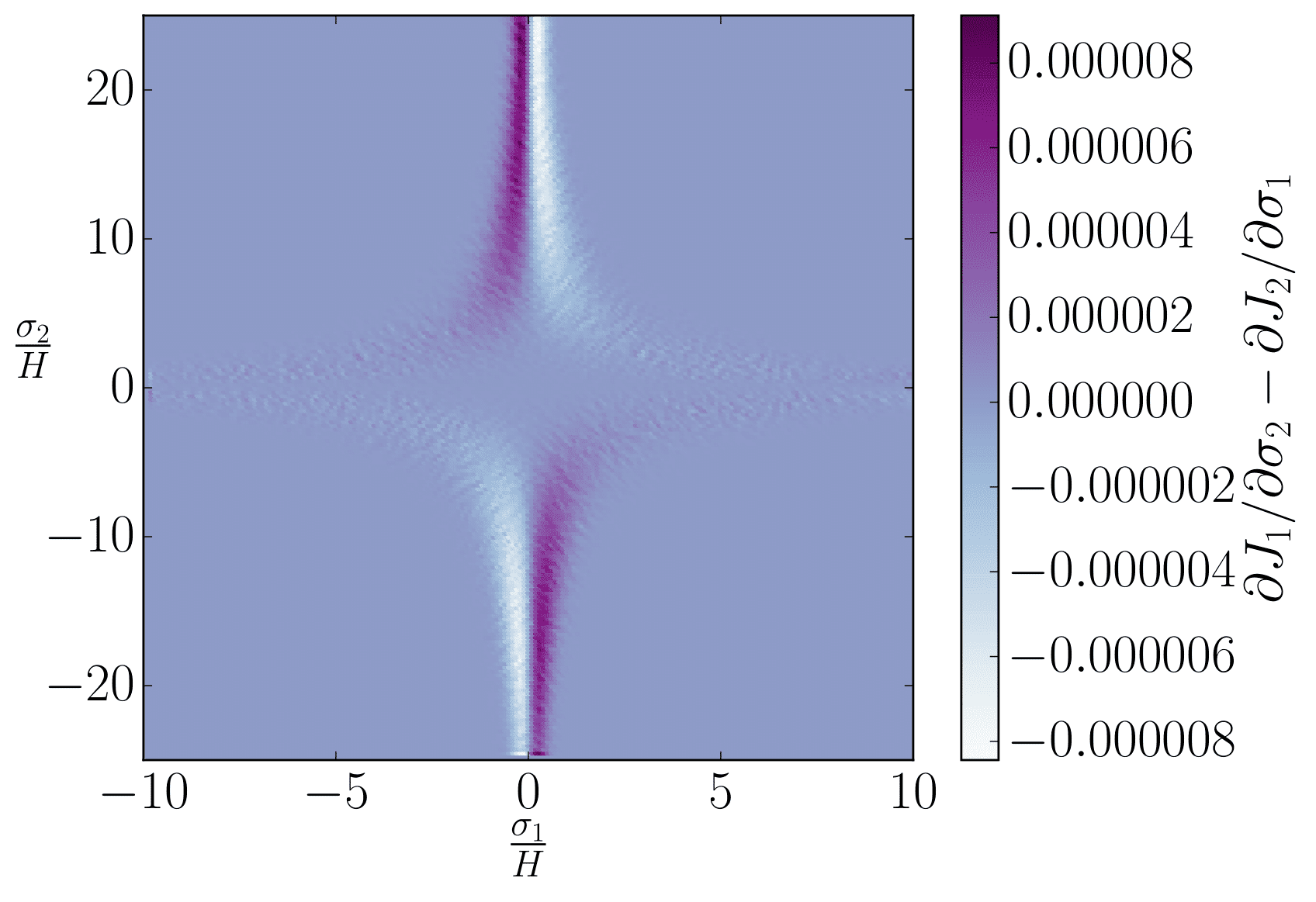}
\caption[Probability current for coupled quadratic spectators]{~\label{fig:pdf-cur-plot2} Using specific parameter choices of the $V_{\rm B}$ potential (see \Eq{eq:VBpot}) we plot the binned stationary probability density $P_{\rm stat}$ (top row), probability current divergence $\nabla \cdot \boldsymbol{J}$ (middle row) and the only non-zero component of the probability current curl $\nabla \times \boldsymbol{J}$ (bottom row). These have all been numerically obtained from $10^7$ realisations of \Eq{eq:langevin-multi} in the \href{https://sites.google.com/view/nfield-py}{\texttt{nfield}} code. }
\end{center}
\end{figure}

We are left with the interesting conclusion that for a sufficiently large coupling $g$, and an asymmetric potential induced by the mass hierarchy parameter $\alpha < 1$, \Eq{eq:exp-stat-dist} is no longer sufficient to describe the stationary probability distribution. In \Sec{sec:proof-symmetric} we conjectured that \Eq{eq:exp-stat-dist} is the stationary solution for symmetric potentials (here when $\alpha =1$). However, when $\alpha <1$, since only the divergence of the probability current $\nabla \cdot \boldsymbol{J}$ must vanish and \emph{not} its curl $\nabla \times \boldsymbol{J}$, \Eq{eq:exp-stat-dist} is no longer the true stationary solution and hence the solution must be elucidated through full numerical evaluation of either \Eq{eq:langevin-multi} or \Eq{eq:dist-multi}. We have plotted $\nabla \times \boldsymbol{J}$ for different choices of parameter in \Fig{fig:pdf-cur-plot1} and \Fig{fig:pdf-cur-plot2}, where one can see in particular that the only component of $\nabla \times \boldsymbol{J}$ is much larger when $g$ is increased for the $\alpha =0.1$ cases plotted in \Fig{fig:pdf-cur-plot2}. As a further numerical check, we have verified that when the symmetry of the potential is restored ($\alpha = 1$) in \Fig{fig:pdf-cur-plot1}, the curl vanishes up to some numerical noise.

Note that $\nabla \times \boldsymbol{J}$ also vanishes at the origin in both \Fig{fig:pdf-cur-plot1} and \Fig{fig:pdf-cur-plot2}. This is confirmed by the analytic expression in \Eq{eq:curl-J}, in which the curl is indeed vanishing at the extrema $\partial V/\partial \sigma_i=\partial P/\partial \sigma_i=0$.

\section{\textsf{Conclusions}}
\label{sec:conclusions}
The typical field value acquired by spectator fields during inflation is an important parameter of many post-inflationary physical processes. Often, in slow-roll inflationary backgrounds, it is estimated using the stochastic equilibrium solution in de Sitter space-times~(\ref{eq:Pstat}), since slow  roll is parametrically close to de Sitter. However, slow roll only implies that the Hubble scale $H$ varies over time scales larger than one \efold. Since the relaxation time of a spectator field distribution towards the de Sitter equilibrium is typically much larger than an \efold, this does not guarantee that the spectator distribution adiabatically tracks the de Sitter solution. In practice, we have found that when the inflaton potential is monomial everywhere, the de Sitter approximation is never a reliable estimate of the spectator typical field value at the end of inflation. Instead, spectator fields acquire field displacements that depend on the details of both the spectator potential and the inflationary background. These results are summarised in table~\ref{table:summary}.

In some cases, the existence of an adiabatic regime at early times leads to an erasure of initial conditions and the spectator field distribution is fully determined by the microphysical parameters of the model. When this is the case, we have shown that spectator fields always acquire sub-Planckian field values at the end of inflation (even when the spectator is dominated by a non-minimal coupling term). However, it can also happen that adiabatic regimes either do not exist or take place at a stage where quantum corrections to the inflaton dynamics are large and our calculation does not apply. In such cases, a dependence on the initial conditions is unavoidable if the inflaton dynamics are indeed dominated by quantum corrections, which we have quantified in the context of information theory. This suggests that observations might have the potential to give access to scales beyond the observable horizon, through processes that are integrated over the whole inflationary period, such as spectator field displacements.

In general, we have found that light spectator fields acquire much larger field displacements during inflation than the de Sitter approximation suggests, which has important consequences. As an illustration, let us mention one of the curvaton models which is favoured by observations, where inflation is driven by a quartic potential in the presence of a quadratic spectator field, the curvaton, that later dominates the energy budget of the Universe and provides the main source of cosmological perturbations. In order for this model to provide a good fit to the data, the field value of the curvaton at the end of inflation should lie in the range~\cite{Vennin:2015vfa, Vennin:2015egh} $\Gamma_\sigma/\Gamma_\phi \ll \sigma_\uend/\Mp \ll 1$, where $\Gamma_\phi$ and $\Gamma_\sigma$ are the decay rates of the inflaton and of the curvaton, respectively. In this case however, we have found that if inflation starts from the eternal inflation regime, then the curvaton typically acquires a super-Planckian field value at the end of inflation, which challenges this model, at least in its simplest form. As shown in this section, possible solutions could be to add either quartic coupling or non-minimal coupling terms to the curvaton potential, or to consider axionic curvaton potentials. Whether the model is still in agreement with the data in this case is an important question that we plan to study in a future work.

\newgeometry{top=3.1cm}
\begin{landscape}
\begin{table}
\begin{center}
\includegraphics[height=8cm]{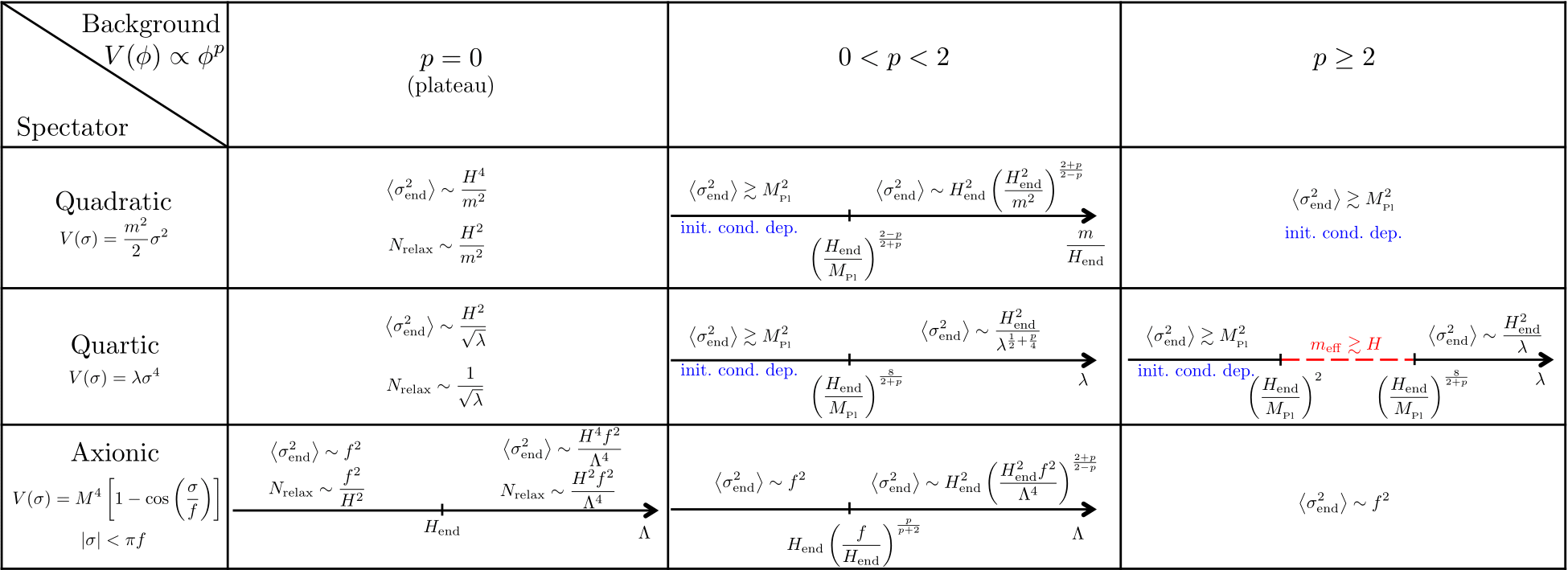}
\caption[Summary of variances]{Summary of the results obtained in this chapter. The stochastic dynamics of spectator scalar fields with quadratic, quartic and axionic potentials have been studied in inflationary backgrounds driven by plateau and monomial potentials. In each case, the typical field displacement $\langle \sigma_\uend^2\rangle$ acquired by spectator fields at the end of inflation is given in this table. When inflation is realised with a plateau potential, the de Sitter equilibrium is reached within a number of \efolds $N_\mathrm{relax}$ also given in the table. If the inflaton potential receives monomial corrections at large-field values, the de Sitter approximation is never a reliable estimate of the spectator typical field value, and the result depends on the details of both the spectator potential and the inflationary background. In some cases, the lack of adiabatic attractors at early time also introduces initial conditions dependence (denoted by ``init. cond. dep''.).    }
\label{table:summary}
\end{center}
\end{table}
\end{landscape}
\restoregeometry

In this chapter we have also demonstrated the usefulness of numerical solutions in order to evaluate the variances of multiple light coupled fields during inflation. In doing so we have identified a lower limit $g_{\rm dec}$ in some example two-field potentials on the coupling $g$, for interactions of the form $\propto g \sigma_1^2\sigma_2^2$, below which the fields may be considered as effectively decoupled and the standard formulae for stationary variances may be used. We have further verified that for choices of $g\geq g_{\rm dec}$, the analytic decoupling approximation for the variances breaks down. In such situations, the solutions from either evaluating the moments of \Eq{eq:exp-stat-dist} (Eqs. \eqref{eq:VBpot} and \eqref{eq:VCpot}, stable in the stationary limit when the potential is either symmetric $\alpha =1$ or decoupled $g<g_{\rm dec}$) or full numerical solutions (for all potentials and generic initial conditions) are the methods to obtain correct values.

In \Sec{sec:proof-symmetric} we have given a general argument as to why it is possible for \Eq{eq:exp-stat-dist} to still remain stable for some symmetric potentials due to vanishing of the probability current everywhere in the domain. Conversely, we have shown that by breaking the symmetry in the potential (e.g. $\alpha \neq 1$ in Eqs. \eqref{eq:VBpot} and \eqref{eq:VCpot}) the form of \Eq{eq:exp-stat-dist} may no longer be stable as a solution to the stationary behaviour of the multi-spectator system. We have supported these conclusions with the numerically obtained figures provided in \Sec{sec:dec-coup} and Figs. \ref{fig:pdf-cur-plot1} and \ref{fig:pdf-cur-plot2}.

A simple generalisation for future work may be to check how this limit changes as the number of coupled fields $n_{\rm f}$ is increased, where we anticipate that because increasing $n_{\rm f}$ typically increases the contribution from the coupling to the effective mass of each field, the lower limit on $g=g_{\rm dec}$ should decrease in order to compensate. Due to the complexity of such a system, a numerical scheme such as the one we have developed in this chapter\footnote{One can go to the following repository to access the code: \href{https://github.com/umbralcalc/nfield}{https://github.com/umbralcalc/nfield}.} (\href{https://sites.google.com/view/nfield-py}{\texttt{nfield}}) will likely be required for such an extension.

By considering an arbitrary $n_{\rm f}$ in the symmetric potential of \Eq{eq:VA} we have also discovered a critical value for $g=g_{\rm crit}$ that varies $\propto 1/n_{\rm f}$ (see \Eq{eq:gcrit-analytic} for a more precise form) above which the formation of stationary spectator condensates collapses to the Hubble rate. For values of $g>g_{\rm crit}$, we cannot yet precisely say that the formation of stationary condensates in such a potential is suppressed (as it is when increasing $g$ up to this point) because this phenomenon results from the effective mass $M_i$ of each field reaching ${\cal O}(1) H$. At this point the mode functions which source the fluctuations of the field can no longer be accurately described by the simple form of noise correlator in the definition of \Eq{eq:langevin-multi}, and hence the stochastic formalism cannot be exactly trusted when $M_i>H$. It is known~\cite{Bunch:1978yq, Birrell:1982ix, Markkanen:2016aes}, however, that the suppression may be further enhanced when $M_i > H$ --- assuming that $M_i$ is constant --- and so we anticipate that further (perhaps fully QFT-theoretic) computations to include a field-dependent effective mass in future work may support our current conjecture beyond this point.

\newpage

\begin{subappendices}

\section{\textsf{Statistical moments of quadratic spectators}}
\label{sec:quadmoments} 
In this section, we derive the first two statistical moments of quadratic spectator fields, for which $V(\sigma)=m^2\sigma^2/2$. If the initial distribution is Gaussian, it remains so throughout the entire evolution so these two moments fully characterise the distribution at any time. Otherwise, higher-order moments can be derived along the same lines.

The first moment can be obtained by taking the stochastic average of \Eq{eq:Langevin}, which gives rise to
\bea
\label{eq:ode:meansigma}
\frac{\dd\langle \sigma \rangle }{\dd N}=-\frac{m^2}{3H^2}\langle \sigma \rangle\, .
\eea
In this expression, the fact that $\sigma$ is a test field plays an important role since it implies that $H$ does not depend on $\sigma$ and is thus a classical (\ie non-stochastic) quantity. Interestingly, \Eq{eq:ode:meansigma} is the same as \Eq{eq:Langevin} in the absence of quantum diffusion, which is why $\langle\sigma\rangle$ follows the classical dynamics
\bea
\label{eq:meansigma:App}
\langle\sigma\left(N\right)\rangle = \langle\sigma\left(N_0\right)\rangle\exp\left[-\frac{m^2}{3}\int_{N_0}^N\frac{\dd{N}^\prime}{H^2({N}^\prime)}\right]\,,
\eea
where $\langle\sigma\left(N_0\right)\rangle$ is the value of $\langle \sigma \rangle$ at the initial time $N_0$.

The second moment can be obtained by multiplying \Eq{eq:Langevin} by $\sigma$ and taking the stochastic average, which leads to
\bea
\label{eq:meansigma2:eom}
\frac{1}{2}\frac{\dd\langle\sigma^2\rangle}{\dd N} = -\frac{m^2}{3H^2}\langle\sigma^2\rangle+\frac{H}{2\pi}\langle \sigma\xi\rangle,
\eea
where $\langle \sigma\xi\rangle$ needs to be calculated separately. This can be done by noticing that a formal solution to \Eq{eq:Langevin} is given by
\bea
\sigma=\int_{A}^N\dd N^\prime \frac{H(N^\prime)}{2\pi}\xi(N^\prime)\exp\left[\int_N^{N^\prime}\frac{m^2}{3H^2(N^{\prime\prime})}\dd N^{\prime\prime}\right]\, ,
\eea
where $A$ is an integration constant. This gives rise to
\begin{align}
\left\langle\sigma(N)\xi(N)\right\rangle& =\int_{A}^N\dd N^\prime \frac{H(N^\prime)}{2\pi}\left\langle\xi(N)\xi(N^\prime)\right\rangle\exp\left[\int_N^{N^\prime}\frac{m^2}{3H^2(N^{\prime\prime})}\dd N^{\prime\prime}\right] \nonumber \\
& =\int_{A}^N\dd N^\prime \frac{H(N^\prime)}{2\pi}\delta(N-N^\prime)\exp\left[\int_N^{N^\prime}\frac{m^2}{3H^2(N^{\prime\prime})}\dd N^{\prime\prime}\right] \nonumber \\
&=\frac{1}{2} \frac{H(N)}{2\pi}\, ,
\end{align}
where the factor $1/2$ comes from the fact that the delta function is centred at one of the boundaries of the integral (recall that $\int_{x_0}^{x_1}f(x)\delta(x-x_0)=f(x_0)/2$). One can then write \Eq{eq:meansigma2:eom} as
\bea
\label{eq:ode:meansigmasquare}
\frac{1}{2}\frac{\dd\langle\sigma^2\rangle}{\dd N} = -\frac{m^2}{3H^2}\langle\sigma^2\rangle+\frac{H^2}{8\pi^2}.
\eea
This equation can be solved and one obtains
\bea
\label{eq:meansigmasquare}
\left\langle \sigma^2(N)\right\rangle = \int_B^N \frac{\dd N^\prime}{4\pi^2} H^2(N^\prime)\exp\left[\frac{2m^2}{3}\int_N^{N^\prime}\frac{\dd N^{\prime\prime}}{H^2(N^{\prime\prime})}\right]\, .
\eea
In this expression, $B$ is an integration constant that can be solved requiring that $\langle \sigma^2\rangle =\langle \sigma^2(N_0)\rangle $ at the initial time $N_0$. This gives rise to
\begin{align}
\label{eq:meansigmasquare:final:App}
\left\langle \sigma^2(N)\right\rangle =& \left\langle \sigma^2(N_0)\right\rangle \exp\left[-\frac{2m^2}{3}\int_{N_0}^{N}\frac{\dd N^{\prime}}{H^2(N^{\prime})}\right]
\nonumber \\ & 
\qquad \qquad + \int_{N_0}^N \dd N^\prime \frac{H^2(N^\prime)}{4\pi^2} \exp\left[\frac{2m^2}{3}\int_N^{N^\prime}\frac{\dd N^{\prime\prime}}{H^2(N^{\prime\prime})}\right]\, .
\end{align}
In this expression, the structure of the first term is similar to the first moment~(\ref{eq:meansigma:App}), so that the variance of the distribution $\langle \sigma^2 \rangle-\langle\sigma\rangle^2$ evolves according to the same formula as the second moment (\ie one can replace $\langle \sigma^2\rangle$ by $\langle \sigma^2 \rangle-\langle\sigma\rangle^2$ in \Eq{eq:meansigmasquare:final:App} and the formula is still valid).
\section{\textsf{Adiabatic solution for quartic spectators}}
\label{sec:nonlin_drift}
For quartic spectator fields, the Langevin equation is not linear anymore and cannot be solved analytically. In this section we provide a solution using the ansatz
\bea
\label{eq:quartic:ansatz:App}
P(\sigma,N)=
\frac{2 \alpha^{1/4}(N)}{\Gamma\left(\frac{1}{4}\right)}\exp\left[-\alpha(N)\sigma^4\right]\, .
\eea
This ansatz is satisfied by the de Sitter equilibrium~(\ref{eq:Pstat}), so we expect the solution to be valid at least in the adiabatic regime and potentially beyond. By plugging \Eq{eq:quartic:ansatz:App} into \Eq{eq:FP}, one obtains
\begin{align}
  \left( \frac{1}{4\alpha} - \sigma^4 \right) \frac{\dd \alpha}{\dd N} P\left(\sigma,N\right) =& \left( \frac{4\lambda}{H^2} - \frac{3H^2\alpha}{2\pi^2} \right) \sigma^2 P\left(\sigma,N\right) \nonumber \\
  & + \left( \frac{2H^2\alpha^2}{\pi^2} - \frac{16\lambda\alpha}{3 H^2} \right) \sigma^6 P\left(\sigma,N\right) \label{eq:quartic:App:interm2} \,.
\end{align}
Multiplying this equation by $\sigma^2$ and integrating over $\sigma$, this gives rise to
\begin{align}
  \left( \frac{\left\langle\sigma^2\right\rangle}{4\alpha} - \left\langle\sigma^6\right\rangle \right) \frac{\dd \alpha}{\dd N}  =& \left( \frac{4\lambda}{H^2} - \frac{3H^2}{2\pi^2}\alpha \right) \left\langle\sigma^4\right\rangle \nonumber \\
  & + \left( \frac{2H^2}{\pi^2} \alpha^2 - \frac{16\lambda}{3 H^2}\alpha \right) \left\langle\sigma^8\right\rangle P\left(\sigma,N\right) \label{eq:quartic:App:interm} \,.
\end{align}
From the ansatz~(\ref{eq:quartic:ansatz:App}), the moments $\langle \sigma^2 \rangle$, $\langle \sigma^4 \rangle$, $\langle \sigma^6 \rangle$ and $\langle \sigma^8 \rangle$ are directly related to $\alpha$, through
\bea
\label{eq:moments}
\langle \sigma^2 \rangle = \frac{\Gamma \left( \frac{3}{4}\right) }{\alpha^{1/2}\Gamma \left( \frac{1}{4}\right)}\,,  \quad
\langle \sigma^4 \rangle = \frac{1}{4\alpha}\,, \quad
\langle \sigma^6 \rangle = \frac{3\Gamma \left( \frac{3}{4}\right) }{4\alpha^{3/2}\Gamma \left( \frac{1}{4}\right)} \,,  \quad
\langle \sigma^8 \rangle = \frac{5}{16\alpha^2}\,.
\eea
By substituting these expressions into \Eq{eq:quartic:App:interm}, one obtains
\begin{align}
\label{eq:quartic:App:interm3}
\frac{\dd \alpha}{\dd N} &=  \frac{\Gamma \left( \frac{1}{4} \right)}{2\Gamma \left( \frac{3}{4} \right)} \left( \frac{\lambda}{H^2}\alpha^{1/2} - \frac{3H^2}{8\pi^2} \alpha^{3/2}\right) \,.
\end{align}
Notice that if one had directly integrated \Eq{eq:quartic:App:interm2} over $\sigma$ and substituted \Eq{eq:moments}, one would have obtained a trivial relationship, which is why we first multiplied \Eq{eq:quartic:App:interm2} by $\sigma^2$ before integrating over $\sigma$.

If the inflaton potential is of the plateau type and  $H$ can be approximated by a constant, this equation can be solved and one finds
\bea
\label{eq:solution:alpha}
\alpha (N) =\frac{8\pi^2\lambda}{3H^4}\tanh^2\left\lbrace \sqrt{\frac{3\lambda}{2}}\dfrac{\Gamma\left(\frac{1}{4}\right)}{8\pi\Gamma\left(\frac{3}{4}\right)}\left(N-N_0\right)+\mathrm{atanh}\left[\sqrt{\frac{3H^4\alpha\left(N_0\right)}{8\pi^2\lambda}}\right] \right\rbrace ,
\eea
which gives rise to \Eq{eq:quartic:plateau:Appr} for the second moment $\langle \sigma^2 \rangle$.

If the inflaton potential is monomial, the function $H(N)$ is given by \Eq{eq:Hubble} and although an analytical solution still exists, it is less straightforward to derive. The first step consists of writing \Eq{eq:quartic:App:interm3} in terms of an equation for $\langle\sigma^2\rangle$ using \Eq{eq:moments},
\begin{equation}
\frac{\dd \langle \sigma^2 \rangle}{\dd N} = -\frac{2}{3}\left[\frac{\Gamma \left( \frac{1}{4} \right)}{\Gamma \left( \frac{3}{4} \right)}\right]^2\frac{\lambda \langle \sigma^2\rangle^2}{H^2} + \frac{H^2}{4\pi^2}  \, .
\end{equation}
The next step is to use $x\equiv H/H_\uend$ as a time variable, which gives rise to
\begin{equation}
\label{eq:Ricatti}
\frac{\dd \langle \sigma^2 \rangle}{\dd x} = \frac{2}{3}\left[\frac{\Gamma \left( \frac{1}{4} \right)}{\Gamma \left( \frac{3}{4} \right)}\right]^2\frac{\lambda x^{4/p-3}}{H_\uend^2}\langle \sigma^2\rangle^2 - \frac{H_\uend^2x^{4/p+1}}{4\pi^2}  \,.
\end{equation}
This equation is of the Ricatti type and can be transformed into a second-order linear differential equation making use of the change of variables
\begin{equation}
\label{eq:gtrans}
\langle \sigma^2 \rangle= -\frac{3}{2}\frac{H_\uend^2}{\lambda}\left[ \frac{\Gamma \left( \frac{3}{4} \right)}{\Gamma \left( \frac{1}{4} \right)}\right]^2 x^{3-\frac{4}{p}} \frac{1}{f(x)}\frac{\dd f}{\dd x}\, .
\end{equation}
By plugging \Eq{eq:gtrans} into \Eq{eq:Ricatti}, one obtains
\begin{equation}
\frac{\dd^2 f}{\dd x^2} + \left( 3- \frac{4}{p}\right)\frac{1}{x} \frac{\dd f}{\dd x} - \frac{\lambda}{6\pi^2}\left[ \frac{\Gamma \left( \frac{1}{4} \right)}{\Gamma \left( \frac{3}{4} \right)}\right]^2x^{8/p-2}f = 0\,.
\end{equation}
This equation can be solved in terms of modified Bessel functions of the first kind $I$. Making use of \Eq{eq:gtrans}, the solution one obtains gives rise to
\begin{align}
\label{eq:quartic:App:interm10}
\langle \sigma^2 (x)\rangle =& \frac{x^{2-4/p}}{2A}\left( 2 - \frac{4}{p} \right) - \sqrt{\frac{B}{A}}\frac{x^2}{2} 
\nonumber \\ &
\times \left\lbrace \frac{  I_{-\frac{p}{4}-\frac{1}{2}}(W) + I_{-\frac{p}{4}+\frac{3}{2}}(W)  + C \left[ I_{\frac{p}{4}+\frac{1}{2}}(W) + I_{\frac{p}{4}-\frac{3}{2}}(W) \right]}{I_{\frac{p}{4}-\frac{1}{2}}(W) + C I_{-\frac{p}{4}+\frac{1}{2}}(W)} \right\rbrace \,,
\end{align}
where we have defined
\bea
A = \frac{2}{3}\left[ \frac{\Gamma \left( \frac{1}{4} \right)}{\Gamma \left( \frac{3}{4} \right)}\right]^2 \frac{\lambda}{H_\uend^2}\,, \qquad \ B = \frac{H_\uend^2}{4\pi^2}\,, \qquad \ W = \frac{p}{4}\sqrt{AB} x^{4/p}\, ,
\eea
where $C$ is an integration constant that can be set as follows: In the asymptotic past, $W\gg 1$ and the Bessel functions can be expanded in this limit, $I_\alpha (W) \simeq \ee^W/\sqrt{2\pi W}$. Unless $C=-1$, the term inside square brackets in \Eq{eq:quartic:App:interm10} goes to $1$ and one finds $\langle\sigma^2\rangle\simeq -\sqrt{B/A}x^2/2 <0$ which would not be consistent. As a consequence, $C=-1$ is the only choice that allows the solution~(\ref{eq:quartic:App:interm10}) to be defined over the entire inflationary period. Setting $C=-1$, \Eq{eq:quartic:App:interm10} can be simplified and one obtains
\bea
\langle \sigma^2 (N)\rangle =\frac{\Gamma\left(\frac{3}{4}\right)}{\Gamma\left(\frac{1}{4}\right)}\sqrt{\frac{3}{2\lambda}} \frac{H^2}{2\pi} \frac{K_{\frac{p}{4}+\frac{1}{2}}(W)}{K_{\frac{p}{4}-\frac{1}{2}}(W) } \,,
\eea
where $K$ is the modified Bessel function of the second kind.

\section{\textsf{Numerical implementation}}~\label{sec:Num-implement}
{\noindent Few analytic solutions to either \Eq{eq:langevin-multi} or \Eq{eq:dist-multi} for $n_{\rm f} \geq 2$ are known to exist, except in the stationary limit of various cases, as given in \Eq{eq:exp-stat-dist}. A robust method for numerical evaluation of a coupled system of Langevin equations of the form in \Eq{eq:langevin-multi} is the modified Improved Euler scheme, introduced in \Ref{2012arXiv1210.0933R}, where it is also proven to exhibit strong first-order convergence. Due to the more complicated potentials studied, a relatively simple implementation of this scheme was developed for the numerical solutions obtained in this section.}

The code is written in the python language and achieves runtimes of $\sim$ 5-10 minutes on a standard netbook laptop for $10^4$ realisations of with any potential up to $n_{\rm f} \sim 10$ for $10^5$ \efold{s}. For increased performance, e.g. $n_{\rm f} \sim {\cal O}(100)$ or more, then it is advised to use a computer cluster. The code has also been made publicly available at the following repository: \href{https://github.com/umbralcalc/nfield}{https://github.com/umbralcalc/nfield}. The repository also contains an example script with 5 fields to help the user get started.

In \Fig{fig:pdf-cur-plot1} and \Fig{fig:pdf-cur-plot2} we have plotted some binned realisations of \Eq{eq:langevin-multi} that are used in the code. These plots can also serve as a useful tool to test for numerical convergence, e.g. to check that no arbitrary asymmetry has appeared or if the divergence of the probability current has not vanished due to elevated numerical noise. In such instances, the code may simply be rerun with more realisations to ensure convergence. Even though $10^7$ realisations were used for these plots, numerical noise (and noise from a finite number of samples) still appears for those values of $\boldsymbol{\nabla} \cdot \boldsymbol{J}$ and $\boldsymbol{\nabla} \times \boldsymbol{J}$ which are meant to vanish. Up to this noise amplitude, however, a strong signal can still be seen in $\boldsymbol{\nabla} \times \boldsymbol{J}$ for the $g=10^2g_{\rm dec}$ potential in \Fig{fig:pdf-cur-plot2}, and we leave further improvements to these visualisations for future work.

\end{subappendices}

\chapter{\textsf{Probing inflation with extra fields}}
\label{sec:isocurvature-fields} \niceline {\vskip+1ex} 

\begin{center}
\fbox{\parbox[c]{13cm}{\vspace{1mm}{\textsf{\textbf{Abstract.}}} In this chapter we shall use the initial conditions derived from \Chap{sec:infra-red-divergences} to compute observable predictions. In detail, we argue that spectator field condensates represent a sensitive probe of the entire inflationary potential~\cite{Hardwick:2017qcw} and demonstrate this through two explicit examples of post-inflationary physics: freeze-in dark matter, which is shown to constrain the energy scale of inflation~\cite{Enqvist:2017kzh}; and the curvaton model, which can constrain the number of inflationary \efold{s}~\cite{Torrado:2017qtr}. \vspace{1mm}}}
\end{center}

\section{\textsf{`A quantum window'}}

If inflation is driven by a single scalar field $\phi$ with potential $V(\phi)$, the power spectrum of curvature perturbations $\zeta$ at scale $k$, given in \Eq{eq:Pzeta:class-intro}, is
\begin{equation}
\label{eq:Pzeta:class}
\calP_\zeta(k) \simeq \frac{V^3}{12 \pi^2\Mp^6} \left. \left( \frac{\partial \phi}{\partial V}\right)^{2} \right\vert_{\phi = \phi_*(k)}\,,
\end{equation}
where $\phi_*(k)$ is the value of $\phi$ when $a/k$ exits the Hubble radius. The range of scales probed \eg in the CMB then translates into a time interval during inflation of length $N\sim 7$, measured by the number of \efolds $N$. If one includes the large-scale structure of our Universe, this window is extended but cannot exceed the last $\sim 60$ \efolds of inflation. But can we ever learn about larger scales, hence earlier times?

As discussed at length in \Sec{sec:intro-stochastic-approach} and \Chap{sec:infra-red-divergences}, during inflation, the coarse-grained fields (\ie scales larger than the Hubble radius) are constantly sourced by the small-wavelength quantum fluctuations as they cross the Hubble radius. This quantum backreaction on the dynamics of the Universe can be modeled through the stochastic inflation formalism~\cite{Starobinsky:1986fx}. The system then explores parts of the potential that would be inaccessible under the classical dynamics. For example, the power spectrum~(\ref{eq:Pzeta:class}) is now computed by \Eq{eq:Pzeta:sto-intro} such that
\begin{align}
\calP_\zeta(k) =& 2
\left\lbrace \int_{\phi_*}^{\infty}\frac{\mathrm{d} A}{M_{{}_\mathrm{Pl}}}\frac{24\pi^2\Mp^4}{V( A)}\exp\left[\frac{24\pi^2\Mp^4}{V(A)}-\frac{24\pi^2\Mp^4}{V\left(\phi_*\right)}\right] \right\rbrace^{-1}
\nonumber \\
&
\times  \int_{\phi_*}^{\infty}\frac{\mathrm{d} A}{M_{{}_\mathrm{Pl}}}\left\lbrace\int_{A}^{\infty} \frac{\mathrm{d} B}{M_{{}_\mathrm{Pl}}} \frac{24\pi^2\Mp^4}{V(B)}\exp\left[\frac{24\pi^2\Mp^4}{V(B)}-\frac{24\pi^2\Mp^4}{V(A)}\right] \right\rbrace^2 \label{eq:Pzeta:sto}\,.
\end{align}
Contrary to \Eq{eq:Pzeta:class}, this expression does not only depend on the potential evaluated at $\phi_*(k)$, but relies on the properties of the potential in the entire inflationary domain. For this reason, even the limited range of scales probed in the CMB may contain imprints from early features of the inflationary dynamics and in this sense, quantum diffusion in an expanding background greatly extends the observational window. In practice, when $V\ll \Mp^4$, \Eq{eq:Pzeta:sto} is well approximated by \Eq{eq:Pzeta:class} so the dependence on the potential function outside the standard observational window is usually Planck suppressed. This is however not the case when several fields drive inflation~\cite{Assadullahi:2016gkk, Vennin:2016wnk, Kawasaki:2015ppx}, or in very flat regions of the potential that can drive the dynamics at smaller (but still accessible~\cite{Pattison:2017mbe}) scales than the ones probed in the CMB.

\begin{figure}
\begin{center}
\includegraphics[width=14cm]{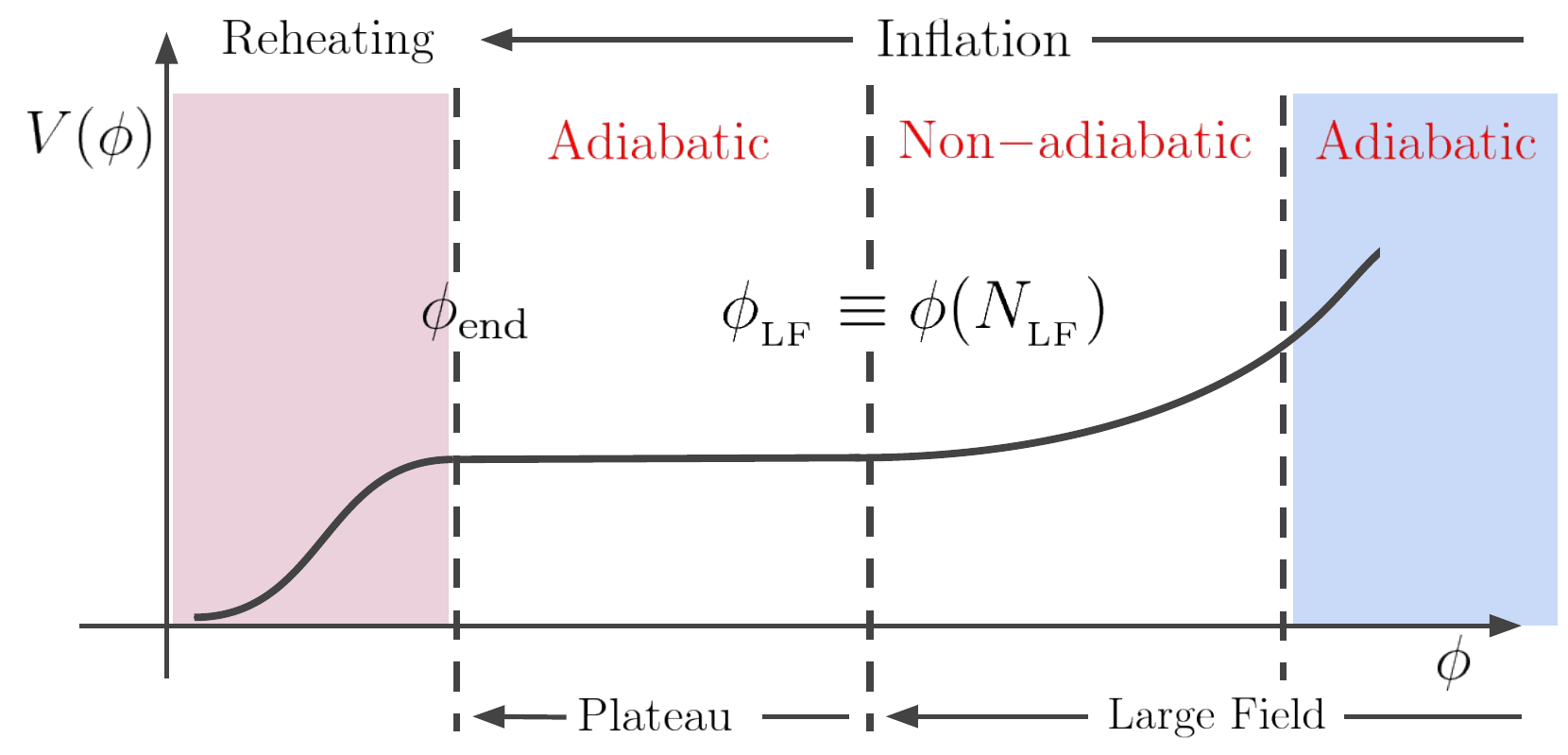}
\caption[Inflationary potential with large field corrections]{\label{fig:potential} Toy inflationary potential considered in this section, made of a plateau (\ie asymptotically constant) part between $\phi_\uend$ and $\phi_{\mathrm{LF}}$, and a monomial large-field part (where $V \propto \phi^p$) at $\phi>\phi_{\mathrm{LF}}$. CMB observations constrain the number of \efolds spent on the plateau to be $N_{\mathrm{plateau}}>60$, while the dynamics of spectator fields is sensitive on a much wider part of the inflationary potential. This is also the inflationary plateau potential with large-field corrections studied in \Sec{sec:vary-inflation}.}
\end{center}
\end{figure}

Another, less direct but more sensitive, cosmological probe sensitive to the early stages of inflation through quantum diffusion is the field displacement acquired by spectator fields~\cite{Enqvist:2012xn, Sanchez:2016lfw, Hardwick:2017fjo}. Let us consider the toy model depicted in \Fig{fig:potential} where the inflaton potential $V(\phi)$ is made of a plateau (\ie asymptotically constant) part between $\phi_\uend$ and $\phi_{\mathrm{LF}}$ and a monomial large-field (\ie $V\propto \phi^p$) part at $\phi>\phi_{\mathrm{LF}}$. The equation for such a potential would be
\begin{equation} \label{eq:potential}
V \left(\phi \right) = M^4 \left[ \left( 1- \ee^{-\sqrt{\frac{2}{3}} \frac{\phi}{\Mp}} \right)^2 + \left( \frac{\phi}{\phi_{{}_{\rm LF}}}\right)^p \right] \,,
\end{equation}
where, in this expression, $\phi_{{}_{\rm LF}} \gg  \Mp$ such that the potential is of the plateau type when observable scales leave the Hubble radius. Observations of the CMB constrain the potential to be of the plateau type in the last few \efolds of inflation~\cite{Martin:2013nzq} so in the standard setup, the only constraint one has is that $\phi_{\mathrm{LF}}$ should be located at least $\sim 60$ \efolds before the end of inflation.

\begin{figure}
\begin{center}
\includegraphics[width=7cm]{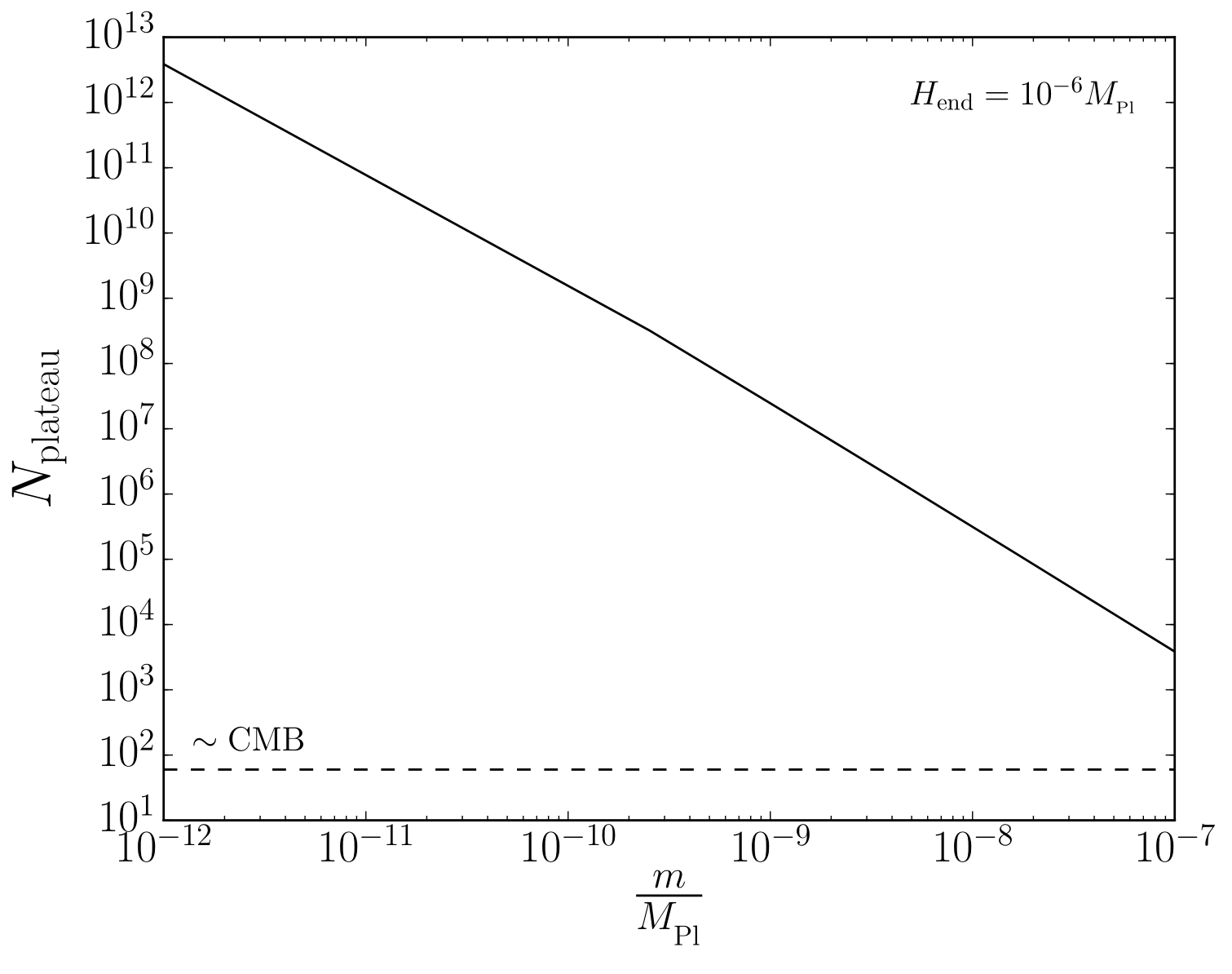}
\includegraphics[width=7cm]{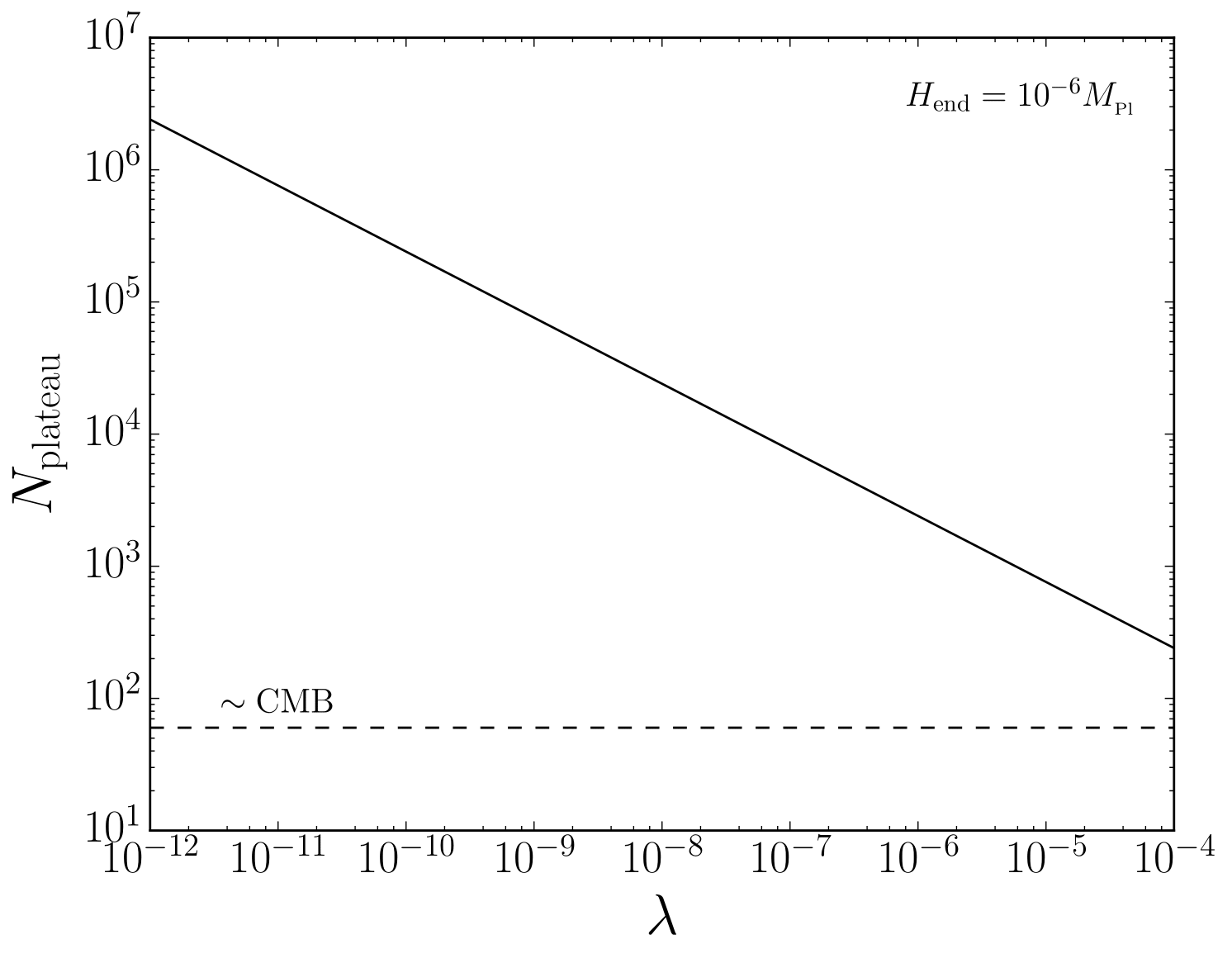}
\caption[Number of plateau \efold{s} in order to erase information]{\label{fig:deltaN} Minimum number of \efolds spent on the plateau part of the inflationary potential so that the spectator field displacement at the end of inflation is independent of the large-field correction to the inflaton potential. The left panel corresponds to a quadratic spectator, $V(\sigma) = m^2\sigma^2/2$ and the right panel corresponds to a quartic spectator, $V(\sigma)=\lambda \sigma^4$. Through CMB observations interpreted in the standard way, one gets the constraint $N_{\mathrm{plateau}}>60$ (denoted with the dashed line), while spectator fields are sensitive to a much wider part of the inflationary dynamics.}
\end{center}
\end{figure}
A spectator field $\sigma$ on top of this inflationary background evolves under its potential $V(\sigma)$ and according to \Eq{eq:Langevin} (where $\phi$ is to be replaced by $\sigma$). If $H$ is constant, the probability distribution $P (\sigma ,N)$ relaxes towards the de Sitter equilibrium solution of \Eq{eq:Pstat}, where any initial condition is erased. However, as we demonstrated in \Chap{sec:infra-red-divergences}, this does not always happen on the large-field part of the inflationary potential, since the relaxation time towards \Eq{eq:Pstat} can be larger than the variation time scale of $H$ there. For example, if the spectator potential is quadratic, $V(\sigma) = m^2 \sigma^2 /2 $, \Eq{eq:Pstat} can never be attained in the early phase of large-field evolution where the typical field displacement remains strongly dependent on initial conditions. By setting $\sigma=0$ at the exit point of eternal inflation (where the dynamics of $\phi$ is itself dominated by stochastic corrections), one can derive a lower bound on the number of \efolds $N_{\mathrm{plateau}}$ spent on the plateau part of the inflaton potential using Eqs.~\eqref{eq:plateau-quadratic-evolve} and~\eqref{eq:quadratic:sigmaend:pgt2} so that the details of the large-field phase are erased from the distribution of $\sigma$ at the end of inflation~\cite{Hardwick:2017fjo},
\bea
N_{\mathrm{plateau}} \geq \frac{3H_{\mathrm{plateau}}^2}{2m^2} \ln \left[ \frac{8\pi p \, m^2 \Mp^2 }{3H_{\mathrm{plateau}}^4 (p+2)} \right]
\eea
for $p\geq 2$. It is displayed in the left panel of \Fig{fig:deltaN} for $p=2$ (but the result depends only mildly on $p$). Compared to the standard constraint $N_{\mathrm{plateau}} \geq 60$, one can see that the observational window on the inflaton potential extends by orders of magnitude. In \Chap{sec:infra-red-divergences} we also found that for a quartic spectator $V(\sigma) = \lambda \sigma^4$, \Eq{eq:Pstat} is adiabatically tracked at early time in the large-field phase. In this case, initial conditions on the spectator field displacement can be erased during this adiabatic epoch, and, using Eqs.~\eqref{eq:quartic:plateau:Appr} and~\eqref{eq:quartic:sigmaend:adiab}, the minimal number of \efolds spent on the plateau such that no imprint is left from the large-field epoch on the distribution of $\sigma$ at the end of inflation is given by~\cite{Hardwick:2017fjo}
\bea
\label{eq:quartN}
N_{\mathrm{plateau}} \geq \frac{4\pi \Gamma \left( \frac{3}{4}\right)}{\Gamma \left( \frac{1}{4}\right)} \sqrt{\frac{2}{3\lambda}} \ln (2) ,
\eea
for $p \geq 2$. It is displayed in the right panel of \Fig{fig:deltaN} where one can see again that the observational window on the inflaton potential extends by orders of magnitude.

Thus the quantum dynamics of cosmological fields in the early Universe gives access to a vast range of scales that extend the classical window by orders of magnitude and allow us to explore high-energy gravity beyond the observable horizon.

\section{\textsf{Freeze-in dark matter}}
\label{sec:basic-argument}
Amongst the parameters that are relevant to inflationary perturbations, two have been measured: the amplitude of the curvature power spectrum, $\AmpS$, and the corresponding spectral tilt, $\nS$ (given for classical single field slow-roll evolution by \Eq{eq:Pzeta:class-intro} and \Eq{eq:spectral-ind}, respectively), which the {\it Planck} collaboration have recently measured to an accuracy of $\Delta \AmpS /\AmpS = \mathcal{O}(10^{-2})$ and $\Delta \nS/\nS = \mathcal{O}(10^{-3})$ \cite{Ade:2015xua}. However, the energy scale at which inflation --- or more accurately, the last $\sim 60$ \efolds of inflation --- happened is still unknown. The energy scale of inflation can be characterised by the value of the Hubble parameter during inflation, $H_*$. In single-field slow-roll models of inflation, this can be expressed by the primordial tensor-to-scalar ratio $r$ by rewriting \Eq{eq:tens-scalar-ratio} as 
\begin{equation}
\ H_* = 8\times 10^{13}\sqrt{\frac{r}{0.1}} {\rm GeV}\,.
\end{equation}
The current upper bound provided by the joint analysis of BICEP2/Keck Array and {\it Planck} data is $r <0.07$ ($95\%$ c.l.) \cite{Array:2015xqh, Ade:2015xua}, whereas no strict lower bound exists other than the requirement for realising successful BBN at $T\sim 1$ MeV \cite{Kawasaki:2000en,Hannestad:2004px,Ichikawa:2005vw,DeBernardis:2008zz}. Hence, there is a huge gap between the scales at which the dynamics of the Universe is understood. It is elementary then, and of great importance to understanding the physics between these scales, to quantify how large the gap is.

The next-generation experiments may be able to push the upper bound for the tensor-to-scalar ratio down to $r<0.03$ from BICEP3 \cite{Wu:2016hul} and $r<0.001$ from LiteBIRD \cite{Matsumura:2013aja} or COrE \cite{Finelli:2016cyd, DiValentino:2016foa}, or any of these may detect it above these limits. However, these numbers illustrate that if no detection is made, even in the best possible case the planned experiments cannot determine the inflationary scale by primordial tensor modes if it was smaller than $H_*\simeq 8\times 10^{12}$ GeV. It would therefore be interesting if one could find scenarios in which the inflationary scale could be determined by other means. This will be our aim in the next few sections.

Based on \Refs{Nurmi:2015ema,Kainulainen:2016vzv,Heikinheimo:2016yds}, we present a scenario where the scale of inflation $H_*$ is determined by three observables: the dark matter (DM) isocurvature perturbation amplitude, its mass and self-coupling constant. This determination is made completely independently of the tensor-to-scalar ratio $r$, increasing the range in $H_*$ that one can infer to values for the inflationary scale well below the current lower bound, or below the sensitivity of the next-generation experiments. Furthermore, we find that in this scenario the inflationary scale can be determined almost solely from the spectator field dynamics discussed in \Chap{sec:infra-red-divergences}.

As a representative example of this kind of scenario, we study a generic real singlet scalar extension to the SM. The new singlet scalar particle is a Feebly-Interacting Massive Particle (FIMP) \cite{McDonald:2001vt,Hall:2009bx,Bernal:2017kxu}, which we assume to constitute the DM abundance. Due to a feeble coupling between the singlet scalar and the SM sector, the singlet never thermalises with the SM and the DM abundance is produced by the ``freeze-in'' mechanism instead of the standard freeze-out. We discuss this in detail throughout the following sections.

We begin by presenting a simple version of the scenario where the energy scale of inflation can be determined without measuring the tensor-to-scalar ratio. The model we consider is a minimal extension to the SM Lagrangian, where in addition to the SM particle content there is a $\mathbb{Z}_2$-symmetric real singlet scalar, $s$, coupled to the SM via the Higgs portal \cite{Silveira:1985rk,McDonald:1993ex}
\begin{equation} \label{eq:model-potential}
\ \mathcal{L} = \mathcal{L}_{\rm SM} 
-\frac12 \partial^\mu s \partial_\mu s + \frac{m_s^2}{2}s^2 + \frac{\lambda_s}{4}s^4 + \frac{\lambda_{hs}}{2}\Phi^\dagger \Phi s^2\, . 
\end{equation}
In this expression, $\mathcal{L}_{\rm SM}$ is the SM Lagrangian\footnote{Radiative corrections in a curved background generate an extra term to the scalar potential, $\ V_{\rm G} = \xi_h h^2 R  + \xi_s s^2 R$, constituting of the non-minimal couplings to gravity $\xi_h$, $\xi_s$ of both the Higgs and singlet, respectively \cite{Callan:1970ze,Freedman:1974gs}. For this scenario, we shall consider the case where the singlet has negligible $\xi_s$. The value of the SM Higgs non-minimal coupling to gravity is not relevant for our purposes.} and the SM Higgs doublet in the unitary gauge is written as $\sqrt{2}\Phi^{\rm T} = (0,v + h)$, where $v$ is the vacuum expectation value of the Higgs field. We assume that the portal coupling takes a small value\footnote{Note that this does not impose a fine-tuning issue, as the running of the portal coupling is always very small in this model \cite{Alanne:2014bra,Heikinheimo:2017ofk}.}, $\lambda_{hs}<10^{-7}$, so that the singlet $s$ does not thermalise with the SM in the early Universe, but remains a FIMP DM candidate \cite{McDonald:2001vt, Hall:2009bx}. 
The $s$ particles can constitute all the DM if the Higgs field can produce sufficient number of $s$ particles from Higgs decay after electroweak symmetry breaking or, if the decay is not kinematically allowed, if Higgs-mediated gauge boson annihilations into $s$ particles are frequent enough \cite{McDonald:2001vt, Yaguna:2011qn,Bernal:2017kxu}. 
For the basic scenario, the exact production mechanism is not relevant, and we will discuss the low-energy dynamics in more detail in \Sec{sec:low-energy-dynamics}.

Let us see how to determine the scale of inflation with the known behaviour of spectator fields during inflation. During inflation, if $s$ is a spectator, it approaches the de Sitter equilibrium distribution \cite{Starobinsky:1986fx} characterised by $\langle V \rangle \sim H^4$ for a sufficiently slowly-varying Hubble rate~\cite{Enqvist:2012xn, Enqvist:2014zqa, Hardwick:2017fjo}, with a typical value obtained from \Eq{eq:quartic:deSitter}
\begin{equation} 
\label{eq:stationary_s}
\left. s \right\vert_{\rm typical} \simeq \sqrt{\langle s^2\rangle} =  \left[ \frac{3}{2\pi^2\lambda_s}\dfrac{\Gamma^2\left(\frac{3}{4}\right)}{\Gamma^2\left(\frac{1}{4}\right)} \right]^{\frac{1}{4}} H\, .
\end{equation}
In deriving \Eq{eq:stationary_s} we require that the quartic terms in the scalar potential dominate over the quadratic ones, $\lambda_s \langle s^2\rangle \gg 2m_s^2+\lambda_{hs}\langle h^2\rangle$; we will verify that this is always the case in \Sec{sec:vary-inflation}.

\begin{figure}
\begin{center}
\includegraphics[width=14cm]{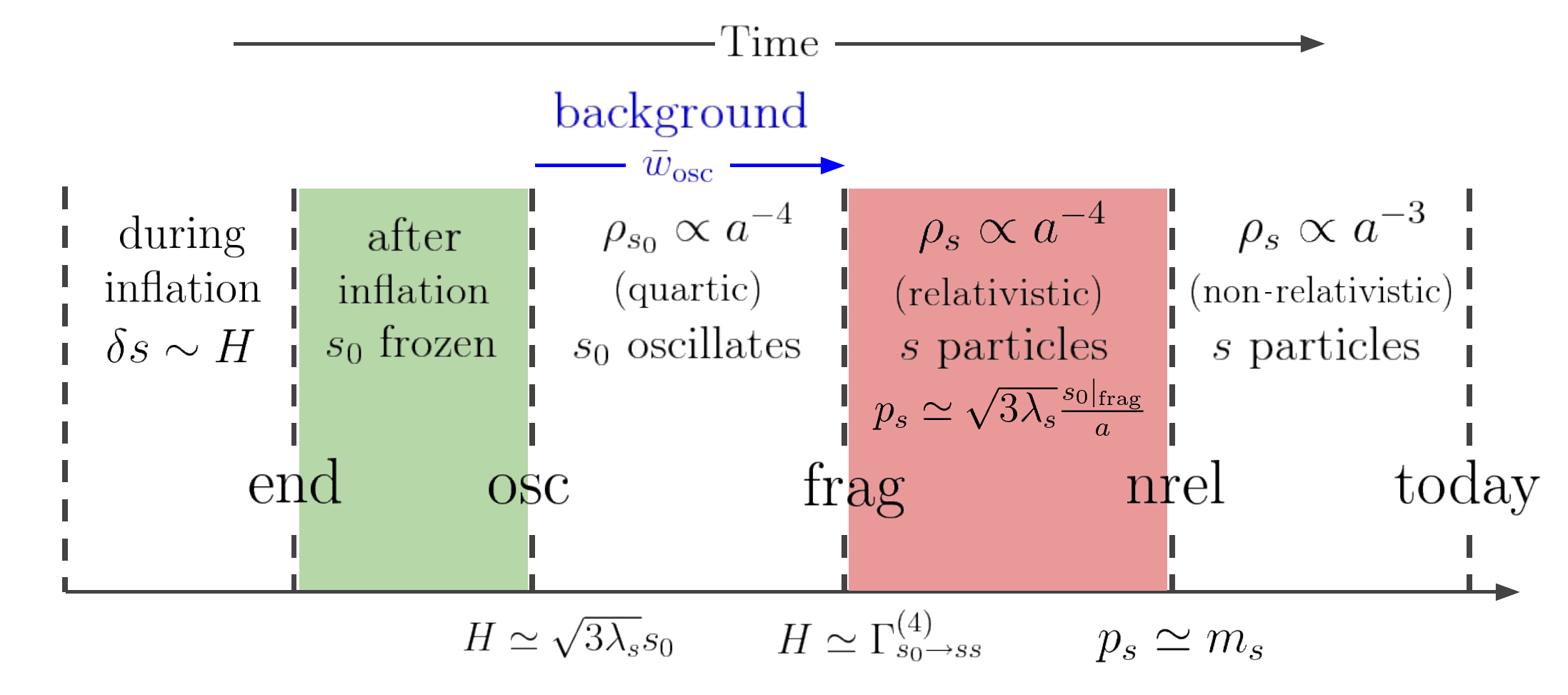}
\caption[Freeze-in dark matter reheating schematic]{\label{fig:reheating} Timeline for the dynamics of the singlet scalar field studied in this chapter. During inflation, $\delta s\sim H$ refers to the typical size of fluctuations during inflation. After inflation, when the Hubble rate $H$ drops below the effective mass of the scalar field $\sqrt{3\lambda_s}s_0$, it starts to oscillate at the bottom of its quartic potential, and $\rho_{s_0}$, the energy density of the singlet condensate, decays as $1/a^4$. The mean equation-of-state parameter of the background energy density during this oscillation period is denoted $\bar{w}_{\rm osc}$. When $H$ drops below the fragmentation rate $\Gamma_{s_0\rightarrow ss}^{(4)}$, the singlet condensate fragments into singlet $s$ particles with typical momentum $p\simeq \sqrt{3\lambda_s} s_0|_{\rm frag}$ that redshifts as $1/a$. When this momentum reaches the mass $m_s$, these particles become non-relativistic, and $\rho_s$, the energy density contained in the singlet particles, decays as $1/a^3$. Its final value in this multi-stage process, which determines the DM abundance, is derived in \App{app:general-scaling-energy-density}.}
\end{center}
\end{figure}

Both the Higgs and $s$ field fluctuations represent isocurvature perturbations relative to the adiabatic inflaton perturbations during inflation\footnote{Unless one of them is the inflaton, but in this chapter we do not consider this possibility.}. Soon after inflation the Universe becomes radiation-dominated; once the Hubble rate drops below their effective mass the fields start to oscillate about their minima. The Higgs field then decays into radiation quickly, typically within a few \efold{s}~\cite{Figueroa:2015rqa}, reaching thermal equilibrium and thus leaving only adiabatic perturbations in the SM radiation. However, due to the feeble coupling between the singlet scalar and the SM, the $s$ condensate (denoted by $s_0$ from now on) does not thermalise and therefore its fluctuations remain isocurvature perturbations relative to the adiabatic perturbations of the SM radiation. Even though the $s_0$ condensate is assumed not to decay into SM radiation, the condensate may fragment into $s$ particles which eventually become cold (non-relativistic) DM particles and inherit the primordial isocurvature perturbations from the condensate. This happens if $\lambda_s$ is large enough, so that the $s_0$ condensate fragments while still in an effectively quartic potential \cite{Nurmi:2015ema,Kainulainen:2016vzv}, and this condition will be carefully checked in \Sec{sec:vary-reheating}. We sketch the main sequence of events for this scenario in \Fig{fig:reheating}.

The CMB constraint on DM isocurvature matter perturbations (over $\beta \equiv \mathcal{P}_{\rm S}/(\mathcal{P}_\zeta + \mathcal{P}_{\rm S})$) can thus be expressed as an upper bound on the DM energy density sourced by the $s_0$ condensate as \cite{Kainulainen:2016vzv}
\begin{equation}
\label{isocurvature}
\frac{\rho_{\rm S}(T_{\rm CMB})}{\rho_{\rm A}(T_{\rm CMB})} \simeq \sqrt{\frac{\beta}{1-\beta}}\sqrt{\frac{\mathcal{P}_\zeta}{\mathcal{P}_{\rm S}}} \,,
\end{equation}
where $\rho_{\rm S}$ and $\rho_{\rm A}$ are the isocurvature and adiabatic contributions to the DM density evaluated at last scattering of the CMB at $T_{\rm CMB} \simeq 0.3$ eV ---  that is, in our case, the $s$ particle DM sourced by the primordial $s_0$ condensate and Higgs decays, respectively. In this expression, $\mathcal{P}_\zeta \simeq 2.2\times 10^{-9}$ is the primordial curvature power spectrum \cite{Ade:2015xua}, $\mathcal{P}_{\rm S}$ is the primordial isocurvature power spectrum and $\beta\leq 0.05$ is the isocurvature parameter constrained by the \emph{Planck} data \cite{Ade:2015lrj}. We require that the Higgs decays into $s$ particles dominate over the DM yield from the primordial $s_0$ condensate. 

By assuming that the comoving number densities of the singlet scalars produced by the decay of the primordial $s_0$ condensate and Higgs decays are separately conserved, and together constitute all of the observed DM, $\rho_{\rm A,\mathrm{today}}/(3\Mp^2H^2_\mathrm{today})\simeq 0.12$, where $\Mp$ is the reduced Planck mass, one finds that \cite{Kainulainen:2016vzv}
\begin{equation}
\label{isocurvature_no_thermalisation}
\frac{\Omega_{\rm DM}^{(s_0)}h_{100}^2}{0.12} \simeq 
0.642\, \Omega_\gamma^{\frac34}h_{100}^\frac32 \lambda_s^{-\frac14} \frac{m_s}{\rm GeV}\left(\frac{s_*}{10^{11}{\rm GeV}}\right)^{\frac{3}{2}} \simeq \sqrt{\frac{\beta}{1-\beta}}\sqrt{\frac{\mathcal{P}_\zeta}{\mathcal{P}_{\rm S}}}
 \,.
\end{equation}
In this expression, $h_{100} = H_{\rm today}/(100\,\mathrm{km}\, \mathrm{s}^{-1}\, \mathrm{Mpc}^{-1})$ parametrises the Hubble parameter today, $\Omega_\gamma$ is the dimensionless photon density parameter today and $s_*$ is the spectator field value during the last 60 \efolds of inflation (where it remains effectively constant).

By then using the typical value for $s_*$ given by \Eq{eq:stationary_s}, one obtains
\begin{equation}
\label{Pdeltas}
\mathcal{P}_{\rm S} = \frac{9}{4}\frac{H_*^2}{(2\pi)^2s_*^2} \simeq  \left[ \frac{27\lambda_s}{128 \pi^2}\dfrac{\Gamma^2\left(\frac{1}{4}\right)}{\Gamma^2\left(\frac{3}{4}\right)} \right]^{\frac{1}{2}}  \,,
\end{equation}
and we can determine the Hubble scale to be
\begin{align}
\frac{H_*}{10^{11}{\rm GeV}} &\simeq 
4.89 \frac{\calP_\zeta^{\frac13}}{h_{100}\Omega_\gamma^\frac12} \left(\frac{\beta}{1-\beta}\right)^{\frac{1}{3}} \lambda_s^{\frac{1}{4}}\left(\frac{m_s}{{\rm GeV}}\right)^{-\frac{2}{3}} \nonumber \\
& \simeq 0.97 \left(\frac{\beta}{1-\beta}\right)^{\frac{1}{3}} \lambda_s^{\frac{1}{4}}\left(\frac{m_s}{{\rm GeV}}\right)^{-\frac{2}{3}} \label{H*_no_thermalisation}\,,
\end{align}
given our previously outlined assumptions, where for the second line we have used ${\cal P}_\zeta = 2.2 \times 10^{-9}$, $h_{100} =  0.673$ and $\Omega_\gamma = 9.3 \times 10^{-5}$~\cite{Ade:2015xua}. This result for $H_*$ then allows one to determine the energy scale of inflation independent of the inflationary tensor perturbations.

\begin{figure}
\begin{center}
\includegraphics[width=14cm]{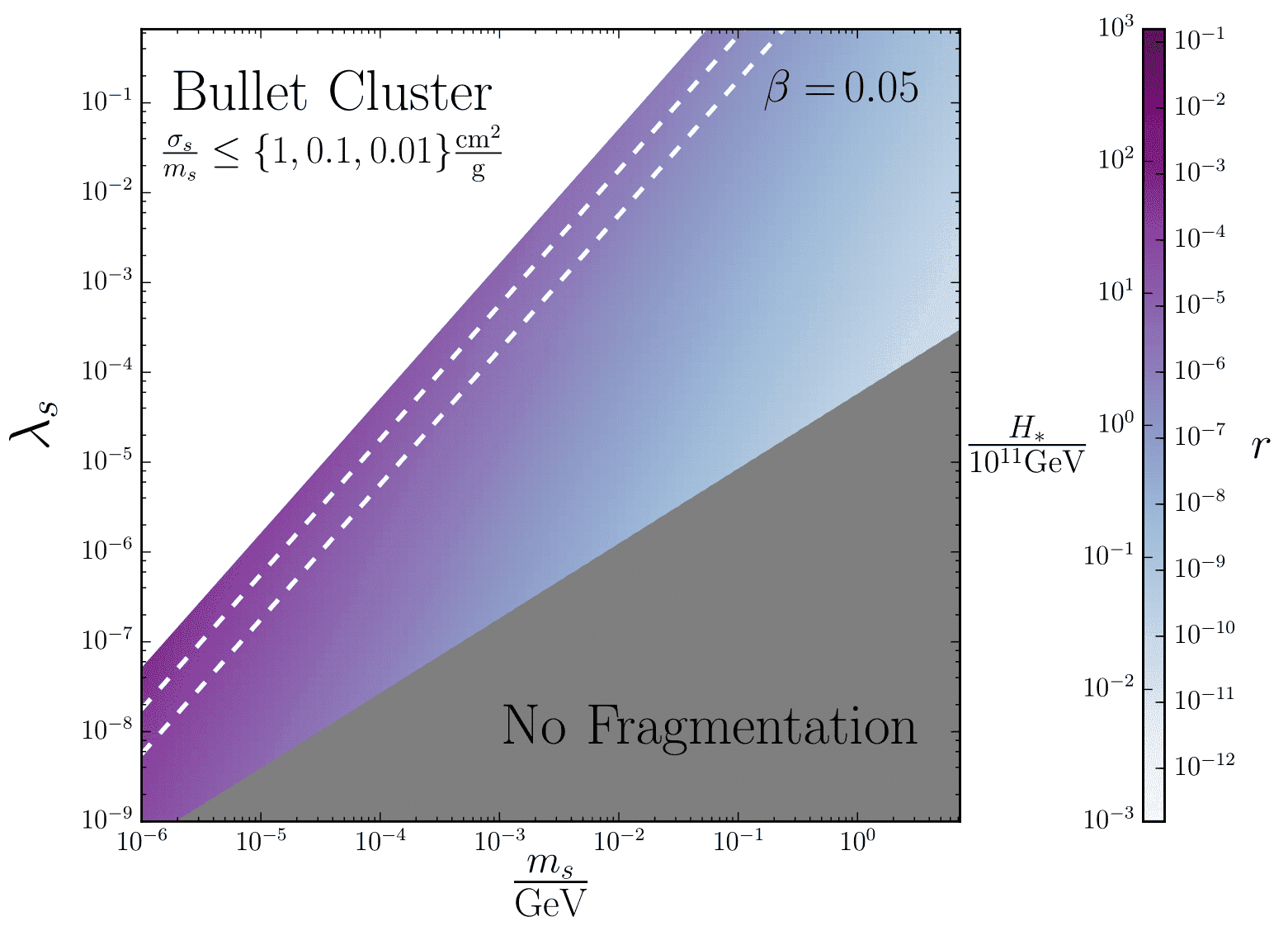}
\caption[Energy scale of inflation]{\label{fig:H_*results} The value of the inflationary energy scale $H_*$ as a function of the singlet scalar mass $m_s$ and self-interaction strength $\lambda_s$, where values have been fixed using \Eq{H*_no_thermalisation} and setting the DM isocurvature relative amplitude to the \emph{Planck}~\cite{Ade:2015lrj} upper limit $\beta = 0.05$ for demonstration. The white region in the top left-hand corner represents the constraint on the self-interaction cross-section provided by the Bullet Cluster in \Eq{scrosssection}, with the dotted white lines indicating how this constraint strengthens with decreasing upper limits on $\sigma_s/m_s$. The grey region in the bottom right-hand corner is a consistency bound related to the requirement of fragmentation occurring in the quartic potential (the requirement that fragmentation occurs before the non-relativistic transition of the field oscillations), \Eq{eq:consistency-frag-before-nrel}, which is computed in \App{app:general-scaling-energy-density}. In this figure, $\bar{w}_{\rm osc}=1/3$. 
}
\end{center}
\end{figure}

The value obtained for $H_*$, and the corresponding value for the tensor-to-scalar ratio $r$, are shown in \Fig{fig:H_*results}. The constraints on the DM self-interaction cross-section from observations of small-scale structure, namely the Bullet Cluster, have been superimposed. Indeed, in the limit where the singlet mass is much smaller than the Higgs mass, $m_s\ll m_h$, the singlet scalar self-interaction cross-section divided by its mass is given by~\cite{Kaplinghat:2015aga}
\begin{equation}
\label{scrosssection}
\frac{\sigma_s}{m_s} = \frac{9\lambda_s^2}{32 \pi m_s^3} \leq 1
 \frac{{\rm cm}^2}{{\rm g}}\,,
\end{equation}
where the upper bound applies when the $s$ particles constitute all DM. The exclusion zone that would be obtained from more stringent constraints on $\sigma_s/m_s$ is also displayed, in order to assess how parameter space could be even more reduced by improving the constraints on, or by measuring, the DM self-interaction cross-section. The result is not displayed in the grey region either, since it corresponds to values of the parameters for which fragmentation does not occur in the part of the potential dominated by the quartic term, and our calculation does not apply.

As shown in \App{app:general-scaling-energy-density} the result is valid for $\lambda_s\ll 1$, and for $m_s\gtrsim \mathcal{O}(1)$ keV because, otherwise, the $s$ particle DM is too hot and suppresses structure formation \cite{Murgia:2017lwo}. As discussed above, we also require $\lambda_{hs} < 10^{-7}$, as otherwise the singlet sector would thermalise with the SM sector and the primordial isocurvature perturbations would be washed away, $\beta=0$. Furthermore, despite the fact that the primordial singlet condensate yields only a subdominant contribution to the total DM abundance, the SM particle decays and annihilations can produce the rest of the DM abundance. This amounts to choosing a sufficiently large value for $\lambda_{hs}$, which for $m_s \in [10^{-6},1]$ GeV is roughly $\lambda_{hs} \in [10^{-12},10^{-9}]$ \cite{Yaguna:2011qn, Nurmi:2015ema,Kainulainen:2016vzv, Heikinheimo:2016yds}. The exact value, however, is not relevant for the minimal scenario (but will be in the extended one).

As discussed in \Refs{Nurmi:2015ema,Kainulainen:2016vzv,Heikinheimo:2016yds}, the above result for $H_*$ in \Eq{H*_no_thermalisation} is a generic consequence of a model where the additional scalar field is light and energetically subdominant during inflation and does not thermalise with the SM radiation after it. The result, however, is subject to a number of uncertainties related to dynamics in the inflaton sector, reheating history, and low-energy dynamics. We carefully consider these in the next section.

\section{\textsf{Freeze-in: extended scenario}}
\label{extended_scenario}

Due to uncertainties in the inflationary dynamics, reheating history, and low-energy dynamics, relaxing one or several assumptions we made above introduces modifications to our result \eqref{H*_no_thermalisation}. For example, even in slow-roll inflation, the Hubble rate may have a finite time dependence during inflation and the typical field displacement acquired by the spectator field at the end of inflation may change, or reheating might have taken a finite time, which introduces an arbitrary expansion history during which the primordial $s_0$ condensate grows its energy density with respect to the background, leading to different DM abundance today. 

These modifications can be effectively parameterised in \Eq{H*_no_thermalisation} as 
\begin{equation}
\label{H*_with_corrections}
\frac{H_*}{10^{11}{\rm GeV}} \simeq 0.97 \left(\frac{\beta}{1-\beta}\right)^{\frac{1}{3}} \lambda_s^{\frac{1}{4}}\left(\frac{m_s}{{\rm GeV}}\right)^{-\frac{2}{3}} \times \mu_{\rm inf}  \times \mu_{\rm reh} \times \mu_{\rm low} \,,
\end{equation}
where $\mu_{\rm inf}$, $\mu_{\rm reh}$, $\mu_{\rm low}$ are effective correction coefficients induced by inflationary dynamics, reheating history, and low-energy dynamics, respectively. Their detailed effect will be discussed one by one in the following subsections.

\subsection{\textsf{Varying the inflationary dynamics}} \label{sec:vary-inflation}

In this section we quantify the degree to which a finite time dependence of the Hubble rate during the early stages of inflation, \eg due to large-field corrections, $V\propto \phi^p$, to the inflaton potential, may affect our result. As was shown in \Ref{Hardwick:2017fjo}, the variance of $s$ at the end of inflation can be significantly larger than that given by \Eq{eq:stationary_s} depending on whether the distribution for $s$ has sufficient time to relax to the equilibrium distribution for a fixed value of $H$ --- the ``adiabatic'' regime in our terminology --- or whether, instead, the Hubble rate varies too fast (while still being in the slow-roll regime) for the system to relax to the equilibrium distribution. This can lead to a larger value for the variance than would be expected for a given, constant value of $H$. 

In order to illustrate this effect, we shall consider a potential for the inflaton field $\phi$ which interpolates between a plateau potential, consistent with \emph{Planck} constraints on the inflaton potential when observable scales leave the Hubble radius \cite{Martin:2013nzq, Ade:2015xua}, and a large-field model at early times when $\phi > \phi_{{}_{\rm LF}}$. This is the same potential as \Eq{eq:potential} where we have also sketched it in \Fig{fig:potential}. 

In both regimes we can identify the number of \efold{s} associated with two characteristic timescales: the relaxation timescale for the $s$ field to relax to the equilibrium distribution for a quartic potential from \Eq{eq:quartic:Nrelax} is $N_{\rm relax} = 1/\sqrt{\lambda_s}$~\cite{Enqvist:2012xn, Hardwick:2017fjo}, and the timescale associated with a variation in the Hubble parameter is, as usual, $N_H = 1/\epsilon_1$. Using these timescales, the effect of the inflationary background evolution (i.e., the inflaton field rolling down in the potential~(\ref{eq:potential})) on the variance of $s$ can be divided into three phases: 
\begin{enumerate}
\item{At early times in the large-field regime we know from \Eq{eq:slow-roll-param1-V} that $\epsilon_1 \sim (\Mp/\phi)^2$, and the $s$ field evolves adiabatically (hence the far-right label in \Fig{fig:potential}) because its relaxation timescale is shorter than the timescale associated with the variation of the Hubble parameter of the background, $N_{\rm relax}\ll N_H$.}
\item{Still within the large field regime, the value of $\epsilon_1$ gradually increases and $N_H$ decreases over time until $N_H < N_{\rm relax}$, at which point the evolution of $s$ ceases to be adiabatic and its variance effectively freezes in until the end of this phase~\cite{Hardwick:2017fjo} with a value we label $\langle s^2_{{}_{\rm LF}}\rangle$. }
\item{After the large-field regime ends, $N_H$ increases such that the condition $N_H> N_{\rm relax}$ is quickly fulfilled again. The $s$ field then begins to relax to its new equilibrium distribution on the plateau, but starting with an initial variance $\langle s^2_{{}_{\rm LF}}\rangle$ determined by the preceding large-field regime.}
\end{enumerate}

At the end of the large-field regime, the spectator field $s$ acquires a typical field displacement given by~\cite{Hardwick:2017fjo}
\begin{equation} \label{eq:eq-dist}
\langle s^2_{{}_{\rm LF}}\rangle = 12\left( 1-\frac{2}{p}\right)\frac{\Gamma^2 \left( \frac{3}{4} \right)}{\Gamma^2 \left( \frac{1}{4} \right)} \frac{H_\uend^2}{\lambda_s} \,,
\end{equation}
where we have used \Eq{eq:quartic:sigmaend:adiab}, $p$ has been defined in \Eq{eq:potential} and $H_\uend$ denotes the value of the Hubble parameter at the end of inflation (which is of the same order as the one along the plateau). The number of \efolds that must be realised to reach the stationary distribution~(\ref{eq:stationary_s}) is given by $N_{\rm relax}=1/\sqrt{\lambda_s}$. Therefore for the equilibrium distribution \Eq{eq:stationary_s} to be valid, we require a large number of \efolds on the plateau, $N_{\rm plateau} \gg N_{\rm relax} = 1/\sqrt{\lambda_s}$.

The variance of the spectator field at the end of the plateau phase, subject to the initial condition set by \Eq{eq:eq-dist}, can be written as
\begin{equation}
\left\langle s^2 \right\rangle  =
\dfrac{\Gamma\left(\frac{3}{4}\right)}{\Gamma\left(\frac{1}{4}\right)}\sqrt{\frac{3H_\uend^4}{2\pi^2\lambda_s}} \times {\mu_{\rm inf}^{-6} \left( N_{\rm plateau} \right) }\,,
\label{eq:corrected-variance}
\end{equation}
where $\mu_{\mathrm{inf}}$ defines the correction to \Eq{eq:stationary_s} and is given by~\cite{Hardwick:2017qcw}
\begin{align}
\mu_{\rm inf} &\left( N_{\rm plateau} \right) = \nonumber \\
& \left( \tanh \left\lbrace \sqrt{\dfrac{3\lambda_s}{8}}\dfrac{\Gamma\left(\frac{1}{4}\right)}{8\pi\Gamma\left(\frac{3}{4}\right)}N_{\rm plateau} +{\rm atanh} \left[  \frac{p}{3p-6} \frac{\Gamma\left(\frac{1}{4}\right)}{\Gamma\left(\frac{3}{4}\right)} \sqrt{\frac{3\lambda_s}{32\pi^2}} \right] \right\rbrace \right)^{\frac{1}{6}} \label{eq:muinf} \,,
\end{align}
where we have substituted \Eq{eq:eq-dist} into \Eq{eq:quartic:plateau:Appr}, obtaining the result by comparison with \Eq{eq:corrected-variance}. Note that since $s_*$ (hence $\mu_{\rm inf}$) appears in both sides of the last equality in \Eq{isocurvature_no_thermalisation}, through $s_{\rm *}$ directly and through $\mathcal{P}_{\rm S}$ indirectly, see \Eq{Pdeltas}, the power of $\mu_{\rm inf}$ in \Eq{eq:corrected-variance} indeed yields a factor $\mu_{\rm inf}$ in \Eq{H*_with_corrections}.

\begin{figure}
\begin{center}
\includegraphics[width=14cm]{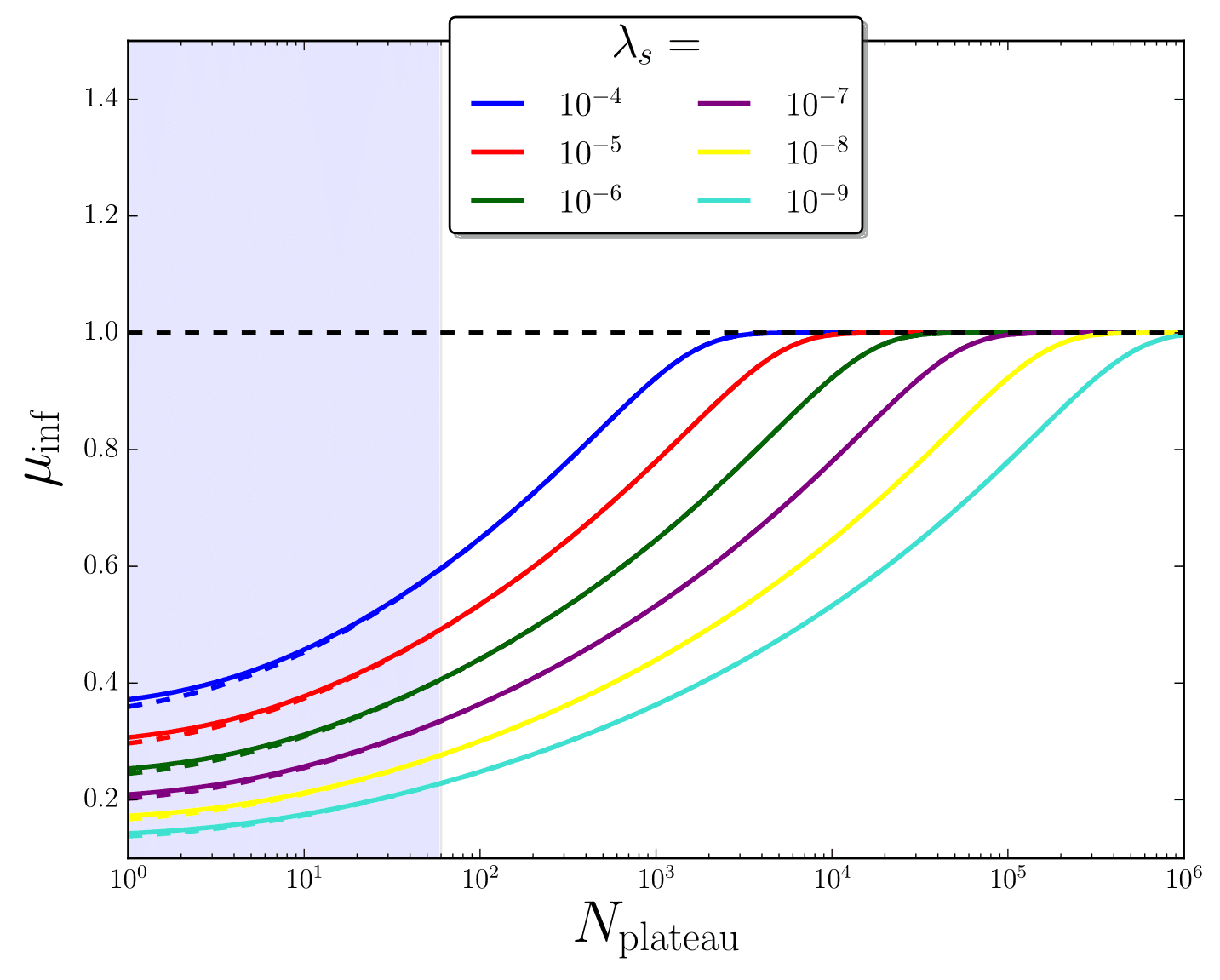}
\caption[Corrections factor from inflationary dynamics]{\label{fig:Hend_over_Hadiab} Correction factor to the value of $H_*$ from the dynamics of the spectator field during the early stages of inflation, $\mu_{\mathrm{inf}}$, as a function of the number of \efolds $N_{\mathrm{plateau}}$ spent on the plateau in the potential depicted in \Fig{fig:potential}. The solid lines stand for $p=4$ in \Eq{eq:potential} and the dashed lines for $p=6$, which shows that the result is almost independent of $p$. The light blue shaded region corresponds to values of $N_{\rm plateau}$ that are too small to let the observable scales leave the Hubble radius in the plateau phase, as required from observations.
}
\end{center}
\end{figure}
\begin{figure}
\begin{center}
\includegraphics[width=14cm]{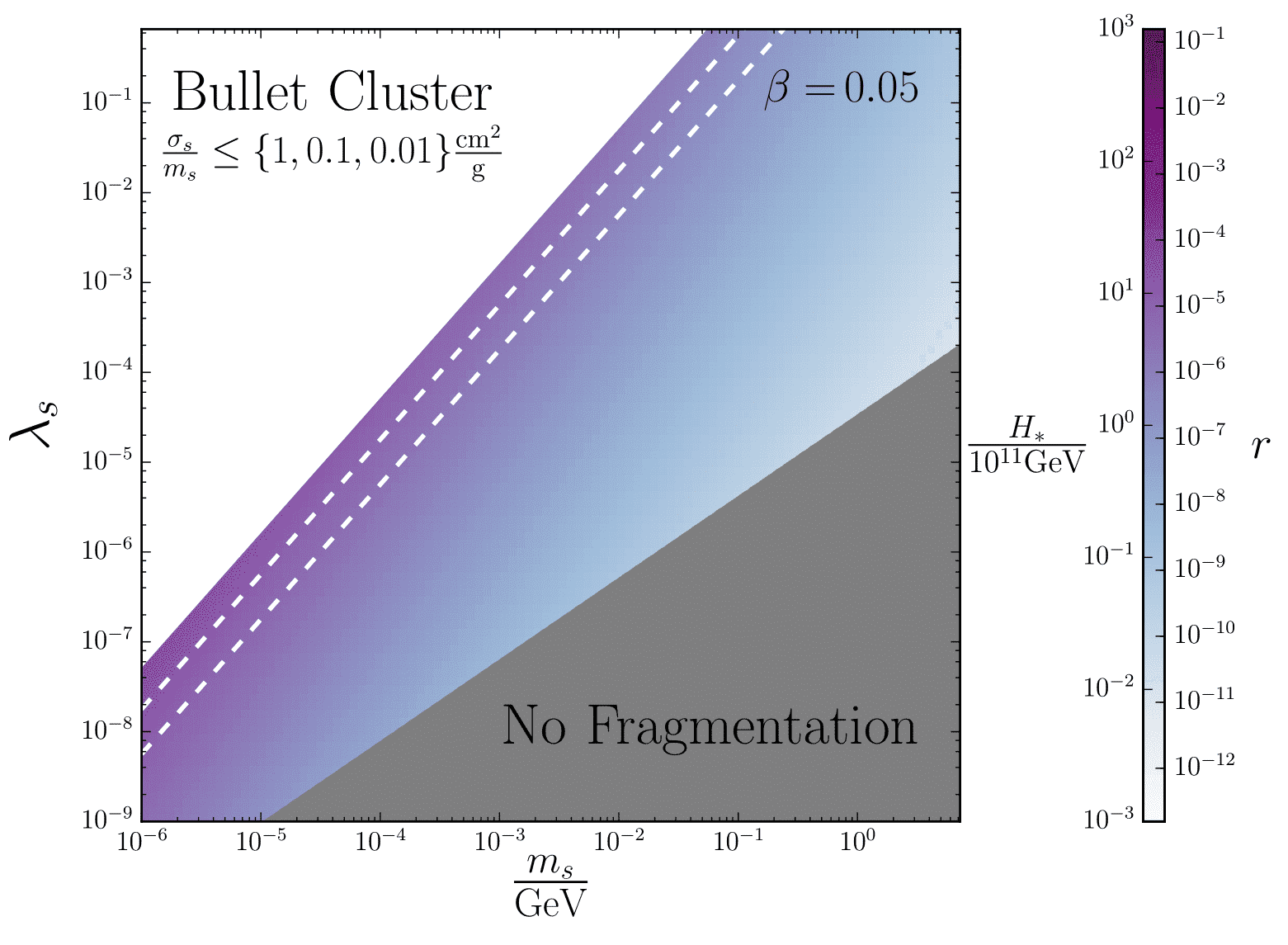}
\caption[Energy scale of inflation corrected by the inflationary dynamics]{\label{fig:H_*results_mu_inf} Same as \Fig{fig:H_*results}, but with the correction $\mu_{\rm inf}$ appearing in \Eq{H*_with_corrections} and defined in \Eq{eq:muinf} included, with $N_{\mathrm{plateau}}=100$ and $p=4$, yielding only small differences with \Fig{fig:H_*results}.
}
\end{center}
\end{figure}

The correction factor $\mu_{\rm inf}$ is displayed in \Fig{fig:Hend_over_Hadiab} as a function of the number of \efolds spent on the plateau, $N_{\mathrm{plateau}}$, for several values of $p$ and $\lambda_s$. One can check that when $N_{\mathrm{plateau}}$ is sufficiently large, $\mu_{\rm inf} \simeq 1$, and that the number of \efolds that need to be spent on the plateau in order to erase the imprint of the large-field early stage decreases with $\lambda_s$, in agreement with the formula $N_{\mathrm{relax}}=1/\sqrt{\lambda_s}$ given above. The result is almost independent of $p$. Since at least $\sim 60$ \efolds must be realised on the plateau, one can check that $\mu_{\rm inf}$ is always of order one, so that the value of $H_*$ computed from \Eq{H*_with_corrections} is impacted by the large-field corrections to the spectator field dynamics by at most an ${\cal O}(1)$ constant. This is also illustrated in \Fig{fig:H_*results_mu_inf}, where $H_*$ is displayed as a function of $m_s$ and $\lambda_s$ taking $N_{\mathrm{plateau}}=100$, and where the differences with \Fig{fig:H_*results} are very mild.

Finally, let us check that, as assumed in the above calculation, during inflation the quartic term in the scalar potential dominates over the quadratic one, $\lambda_s \langle s^2\rangle \gg 2m_s^2+\lambda_{hs}\langle h^2\rangle$. Using \Eq{eq:corrected-variance} to estimate $\langle s^2 \rangle$ in the first condition $\lambda_s \langle s^2\rangle \gg 2m_s^2$, one obtains $H_* \gg m_s \lambda_s^{-1/4} \mu_{\mathrm{inf}}^3$. In all following figures, we make sure that this condition is always satisfied. Using a relation similar to \Eq{eq:corrected-variance} to estimate $\langle h^2 \rangle$ in the second condition $\lambda_s \langle s^2\rangle \gg \lambda_{hs}\langle h^2\rangle$, one obtains  $\sqrt{\lambda_h \lambda_s} \gg \lambda_{hs}$, where $\lambda_h$ is the self-interaction strength of the Higgs. As noted above, to prevent the singlet $s$ from thermalising with the SM in the early Universe,  one must have $\lambda_{hs}<10^{-7}$, and since we assume $\lambda_h \gtrsim 10^{-5}$~\cite{Figueroa:2015rqa}, the lower bound on $\lambda_s$ used in all figures is such that this condition is always satisfied too.

\subsection{\textsf{Varying the reheating history}} \label{sec:vary-reheating}

We now turn our attention to the second possible modification in \Eq{H*_with_corrections}, namely the reheating expansion history. So far we have assumed that after inflation, the energy density of the background decays as radiation. If this is not the case, the abundance of DM obtained from the particles into which the condensate fragments is different, hence the inferred value of $H_*$ changes. 

\begin{figure}
\begin{center}
\includegraphics[width=14cm]{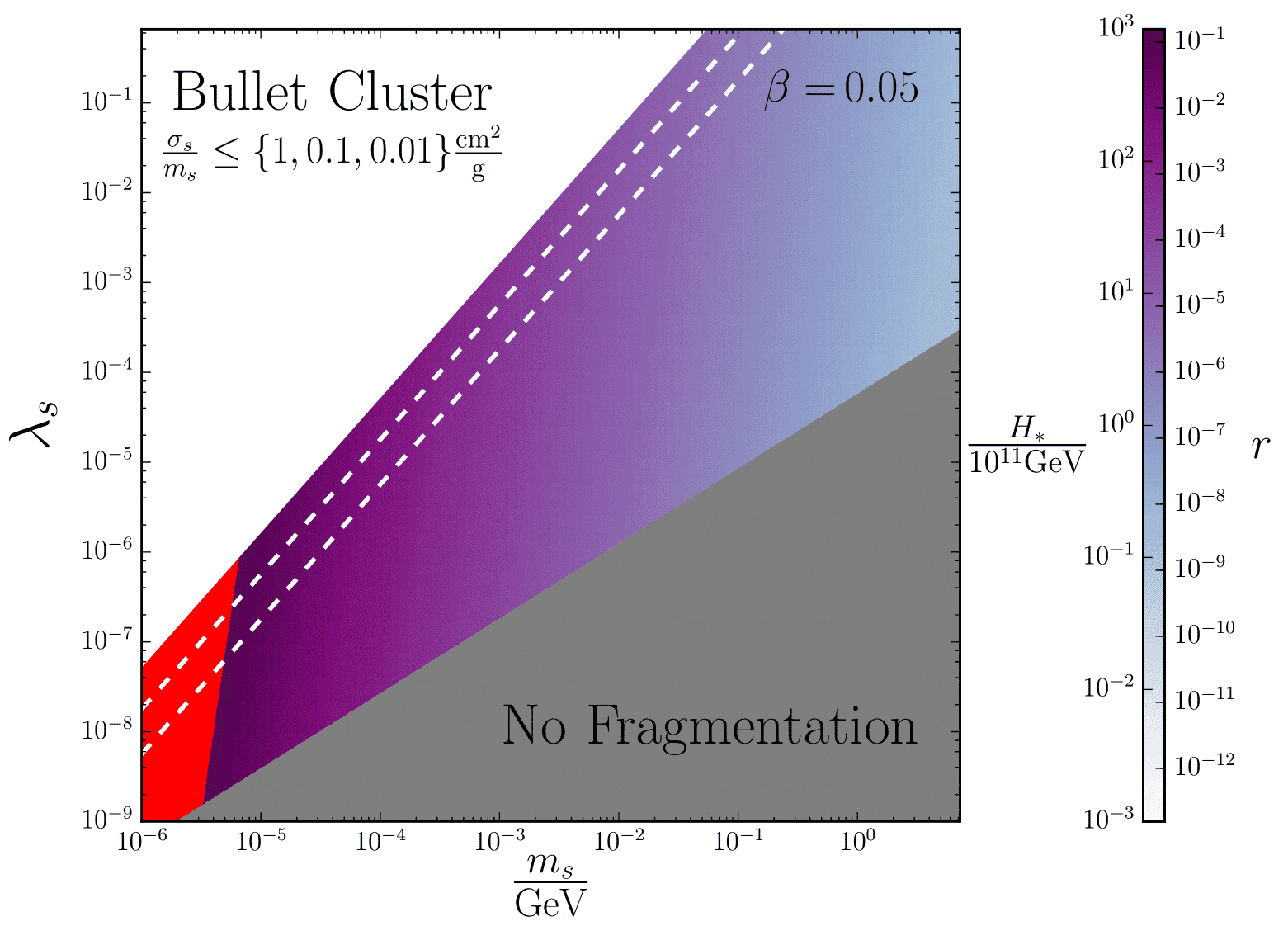}
\caption[Energy scale of inflation corrected by the reheating history]{\label{fig:H_*results_mu_reh} 
Same as \Fig{fig:H_*results}, but with the correction $\mu_{\rm reh}$ appearing in \Eq{H*_with_corrections} and defined in \Eq{eq:mu_reh}, included, with $\bar{w}_{\rm osc}=0.23$. The region shaded in red  is ruled out since it yields values for the tensor-to-scalar ratio that are larger than the observational upper bound $r< 0.12$~\cite{Ade:2015tva}.}
\end{center}
\end{figure}

In \App{app:general-scaling-energy-density} we provide a detailed calculation of the energy density contained in the singlet particles at the end of the multi-stage process depicted in \Fig{fig:reheating}, for an arbitrary background expansion history between the end of inflation and the fragmentation time (for the result we use about fragmentation rate to apply, the Universe needs to be in a radiation era at the fragmentation time). We find that the result only depends on the average equation-of-state parameter during the oscillation phase of the condensate, $\bar{w}_{\mathrm{osc}}$, and on the quartic coupling constant $\lambda_s$. More precisely, an analogous expression to \Eq{isocurvature_no_thermalisation} is obtained,
\begin{align} 
\frac{\Omega^{(s_0)}_{\rm DM}h^2_{100}}{0.12} &=  0.642\, \Omega_\gamma^{\frac34}h_{100}^\frac32 \lambda_s^{-\frac14} \frac{m_s}{\rm GeV}\left(\frac{s_*}{10^{11}{\rm GeV}}\right)^{\frac{3}{2}} \times \mu_{\rm reh}^{-\frac{3}{2}}(\lambda_s,\bar{w}_{\rm osc})  \label{eq:mu_reh_obtain_1}\,,
\end{align}
where we have defined
\begin{equation} \label{eq:mu_reh}
\mu_{\rm reh}(\lambda_s,\bar{w}_{\rm osc}) \equiv \left( \frac{\alpha \lambda_s}{\sqrt{3}}\right)^{\frac{3\bar{w}_{\rm osc}-1}{3\bar{w}_{\rm osc}+1}} \,.
\end{equation}
One can check that the power to which $\mu_{\mathrm{reh}}$ appears in \Eq{eq:mu_reh_obtain_1} is such that it appears with power one in \Eq{H*_with_corrections}. In this expression, $\alpha \simeq 0.023$ is a numerical constant that comes from the calculation of the fragmentation rate. When $\bar{w}_{\rm osc}=1/3$, $\mu_{\mathrm{reh}}=1$ and \Eq{isocurvature_no_thermalisation} is recovered. In \App{app:general-scaling-energy-density}, we also derive and carefully study the conditions under which the assumptions made in the timeline of \Fig{fig:reheating} are satisfied. In particular, this results in the ``no fragmentation'' exclusion zone in \Figs{fig:H_*results}, \ref{fig:H_*results_mu_inf}, \ref{fig:H_*results_mu_reh} and \ref{fig:H_*results_mu_low}.

\begin{figure}
\begin{center}
\includegraphics[width=14cm]{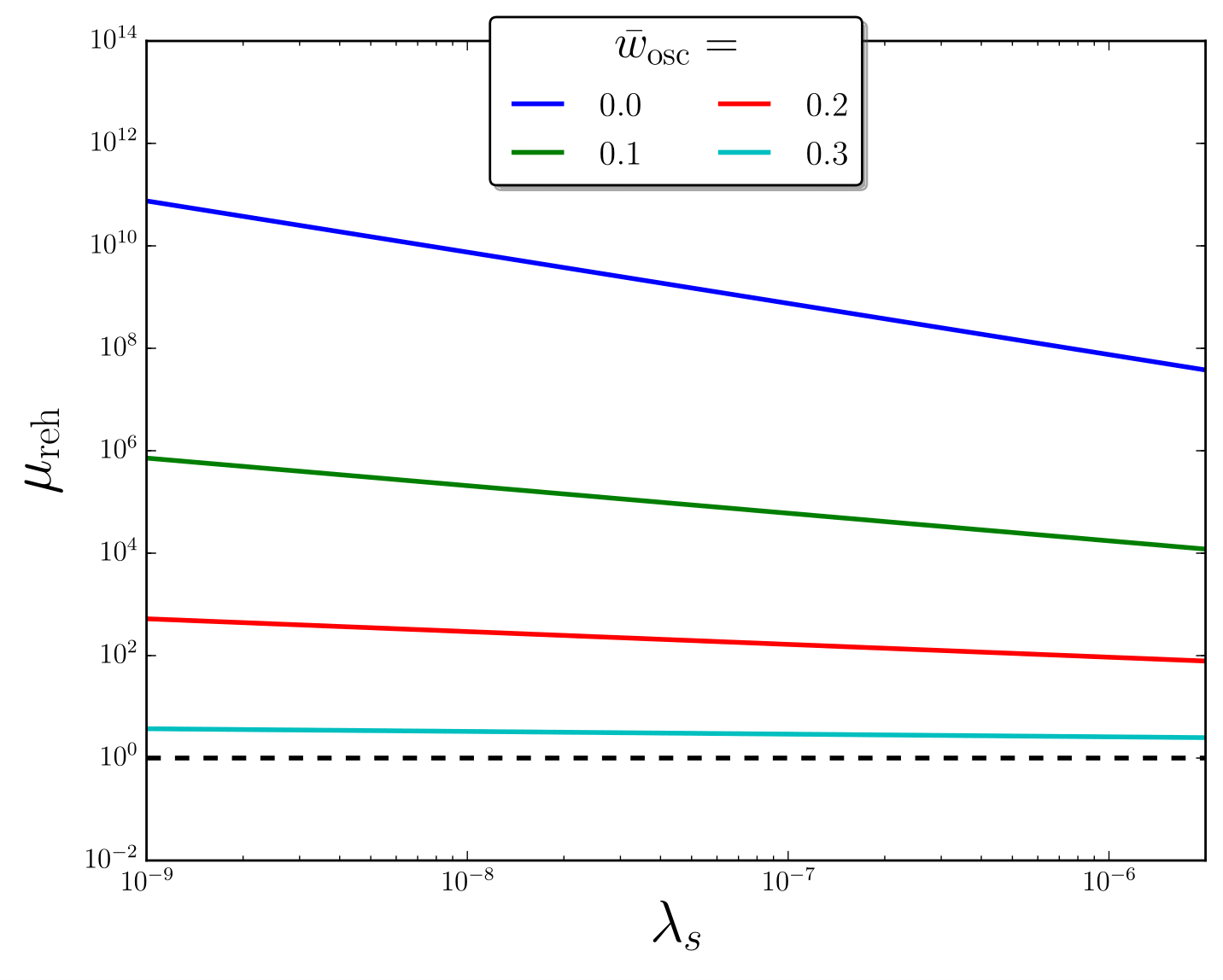}
\caption[Correction from reheating history]{\label{fig:mu_reh} Correction factor $\mu_{\mathrm{reh}}$ appearing in \Eq{H*_with_corrections} and accounting for an arbitrary expansion history during reheating, plotted as a function of the self-interaction strength of the singlet scalar $\lambda_s$ for a few values of the background average equation-of-state parameter $\bar{w}_{\mathrm{osc}}$ during the oscillation phase of the condensate after inflation.}
\end{center}
\end{figure}

The correction factor $\mu_{\mathrm{reh}}$ is plotted as a function of $\lambda_s$ for a few values of $\bar{w}_{\rm osc}$ in \Fig{fig:mu_reh}. Unlike the correction factor $\mu_{\rm inf}$ in the preceding subsection, we see that $\mu_{\rm reh}$ can vary by many orders of magnitude when $\bar{w}_{\rm osc}$ departs from $1/3$. This is also illustrated in \Fig{fig:H_*results_mu_reh}, where $H_*$ is displayed as a function of $m_s$ and $\lambda_s$ taking $\bar{w}_{\mathrm{osc}}=0.23$, and where the difference with \Fig{fig:H_*results} is quite large. There even are regions (in red) for which the predicted value of the tensor-to-scalar ratio is too large to satisfy observational bounds~\cite{Ade:2015tva}.

One notices that if $\bar{w}_{\mathrm{osc}}<1/3$, $\mu_{\mathrm{reh}}>1$ and the inferred value of $H_*$ in the minimal setup is smaller than the actual one, while if $\bar{w}_{\mathrm{osc}}>1/3$, $\mu_{\mathrm{reh}}<1$ and the inferred value of $H_*$ is larger than the actual one. The large effect from the reheating expansion history on our estimate of $H_*$ should be taken with a grain of salt since in practice, $\bar{w}_{\mathrm{osc}}$ may not depart too much from $1/3$. At the end of the oscillating phase indeed, one must have a background equation of state $w=1/3$ (for our expression for the fragmentation rate in \Eq{gamma} to apply), so $\bar{w}_{\mathrm{osc}}$ receives a contribution from values close to $1/3$. Let us also note that linear instabilities on small scales have been shown to yield $w=1/3$ very quickly after the end of inflation, in fact well before the inflaton field has effectively decayed~\cite{Lozanov:2016hid}. Such a mechanism would yield $\bar{w}_{\mathrm{osc}}=1/3$, leaving no imprint from the reheating expansion history on our result. 

\subsection{\textsf{Low-energy dynamics}} \label{sec:low-energy-dynamics}

In addition to corrections arising from dynamics during and immediately after inflation, there are corrections arising from particle dynamics at low energies, namely below the electroweak scale after the SM particle decays and annihilations have yielded the initial $s$ particle abundance. Contrary to the variations in the inflationary and the reheating dynamics studied in \Secs{sec:vary-inflation} and~\ref{sec:vary-reheating}, respectively, the low-energy dynamics effect is not a variation to the minimal setup but rather an inevitable correction that is inherent to it. It should therefore be understood as part of the minimal scenario.

\begin{figure}
\begin{center}
\scalebox{-1}[1]{\includegraphics[width=14cm]{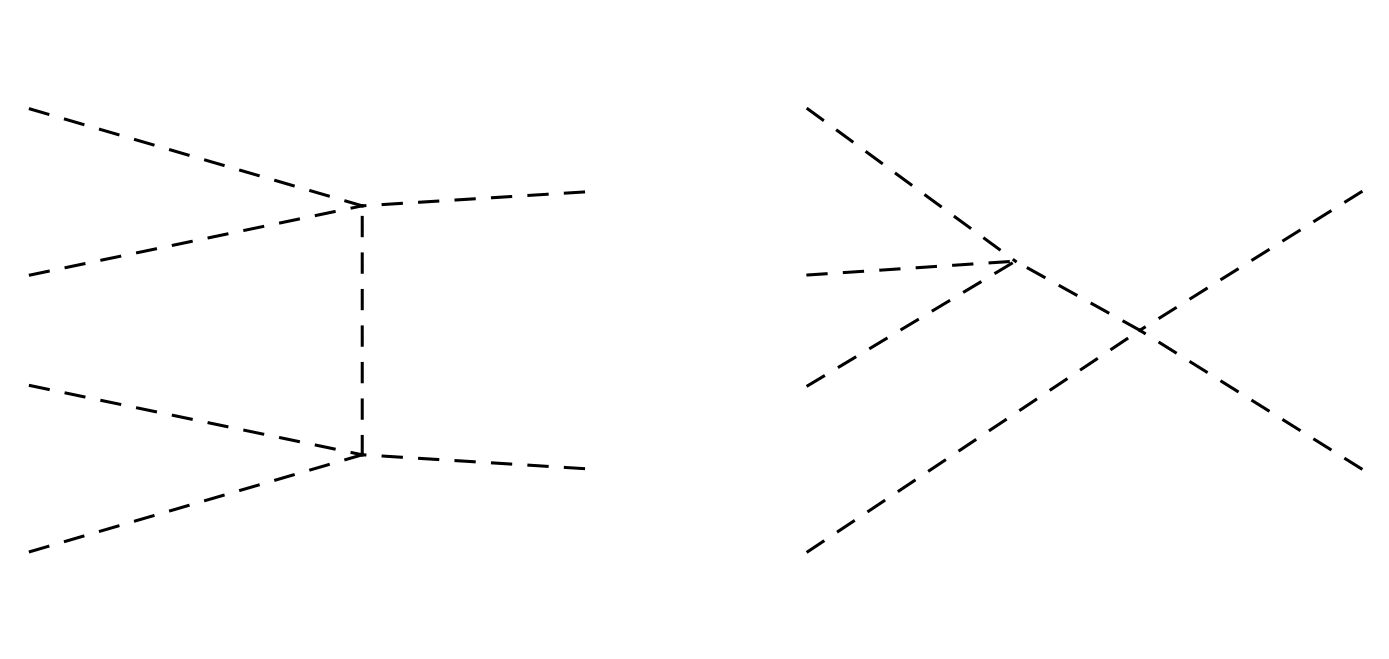}}
\caption[Feynman diagrams of dark matter cannibalism]{Examples of relevant Feynman diagrams for the $4\to 2$ scalar self-annihilation process at the limit $m_s \ll m_h$.}
\label{diagram}
\end{center}
\end{figure}

Even though the portal coupling between DM and the SM sector is assumed to be so small that the $s$ particles never enter thermal equilibrium with the SM particles, it may happen that the $s$ particles reach chemical equilibrium within the singlet sector if the singlet sector has sufficient self-interactions. This leads to a characteristic hidden sector temperature $T_s$ different from the SM photon temperature $T$. If the singlet self-interactions are sufficiently strong, they can maintain the equilibrium for some time also after the singlet particles have become non-relativistic, leading to so-called \textit{DM cannibalism} \cite{Carlson:1992fn}, where number-changing interactions, such as $4\rightarrow 2$ annihilations\footnote{The $2\rightarrow 3$ annihilations are in our case forbidden due to the assumed $\mathbb{Z}_2$ symmetry of the scalar field.} (see \Fig{diagram}), reduce the singlet particle number density and heat the singlet sector with respect to the SM sector. Depending on the strength of singlet self-interactions, the $s$ number density can be significantly depleted before its final freeze-out from the equilibrium in the singlet sector. Thus the final DM abundance depends not only on the portal coupling $\lambda_{hs}$ and the mass $m_s$, but on a combination of the parameters $\lambda_{hs}$, $\lambda_s$ and $m_s$. This production mechanism is called \textit{dark} or \textit{hidden freeze-out}~\cite{Carlson:1992fn, Bernal:2015ova, Bernal:2015xba, Heikinheimo:2016yds, Bernal:2017mqb, Heikinheimo:2017ofk, Bernal:2017kxu}.

The main result \eqref{H*_no_thermalisation} applies only if there are no number-changing interactions in the singlet sector, \ie if the quartic scalar self-interaction strength $\lambda_s$ is small enough. The critical value above which the number-changing interactions play a significant role in determining the final DM abundance is \cite{Heikinheimo:2016yds}
\begin{equation}
\label{lambdas_fi}
\lambda_s^{(\rm fi)} \simeq \sqrt{\frac{19.4\left[g_*\left(m_h\right)g_*\left(m_s\right)\right]^\frac14 \sqrt{m_h m_s}}{\lambda_{hs}^{(\rm fi)} (m_s) \Mp }} 
\simeq 2.3 \times 10^{-8}\left[ \lambda_{hs}^{(\rm fi)}(m_s)\right]^{-\frac{1}{2}}\left( \frac{m_s}{{\rm GeV}}\right)^{\frac{1}{4}}\,,
\end{equation}
where $g_*(T)$ is the effective number of relativistic degrees of freedom in the SM plasma at temperature $T$ and $\lambda_{hs}^{(\rm fi)}(m_s)$ is the value of the portal coupling that yields the observed DM abundance for a given mass $m_s$ in the usual freeze-in case. For $\lambda_s<\lambda_s^{(\rm fi)}$ the usual freeze-in picture and the result \eqref{H*_no_thermalisation} are sufficient. Recalling that we assume $m_s \ll m_h$ in order for \Eq{scrosssection} to hold, the value of $\lambda_{hs}^{(\rm fi)}$ is determined by the usual freeze-in relation \cite{McDonald:2001vt,Hall:2009bx,Bernal:2017kxu}
\begin{equation}
\lambda_{hs}^{(\rm fi)} \simeq 10^{-12}\sqrt{\frac{m_h}{m_s}}\,.
\end{equation}

If the number-changing interactions in the singlet sector become active, the $s$ particles equilibrate among themselves before the formation of the CMB. After the equilibration, the singlet scalar particles from both origins --- Higgs decays and primordial $s_0$ condensate fragmentation --- contribute to the thermal bath of DM, so that the relative abundance of the isocurvature component with respect to the total DM abundance remains constant from there on, as discussed in \Ref{Heikinheimo:2016yds}. We assume that the thermalisation of the $s$ particles takes place at $T_{\rm therm}\simeq m_s$, which is the latest moment when the $s$ particles can reach chemical equilibrium with themselves.

The abundances from the primordial isocurvature condensate source $\rho^{(s_0)}_{\rm DM}(T_{\rm therm})$ and the adiabatic Higgs freeze-in source $\rho^{({\rm fi})}_{\rm DM}(T_{\rm therm})$ at the time of the thermalisation can be found by scaling the result in \Eq{isocurvature_no_thermalisation} by $a^3$ from the CMB temperature today, $T_{\rm today}\simeq 2.725$ K, up to $T=m_{ s}$, and scaling the usual freeze-in abundance of scalars by $a^{-4}$ down to the same temperature. The isocurvature abundance is \cite{Kainulainen:2016vzv}
\begin{equation}
\label{rhoS}
\frac{\rho^{(s_0)}_{\rm DM}(T_{\rm therm})}{\GeV^4} \simeq 2.97 \times 10^{-52}\lambda_{s}^{-\frac{1}{4}}
\frac{m_{ s}}{{\rm GeV}}
\left(\frac{s_*}{10^{11}{\rm GeV}} \right)^{\frac{3}{2}}
\frac{g_{*{\cal S}}(m_{ s})}{g_{*{\cal S}}(T_{\rm today})} 
\left(\frac{m_{ s}}{T_{\rm today}}\right)^3 \,,
\end{equation}
where $g_{*{\cal S}}=g_{*{\cal S}}(T)$ is the effective number of entropy degrees of freedom in the radiation heat bath, and the usual freeze-in abundance of scalars is~\cite{Heikinheimo:2016yds}
\begin{eqnarray}
\label{rhoA}
\rho^{(\rm fi)}_{\rm DM}(T_{\rm therm})  &\simeq& m_{ s}n_{ s}(m_{ h})
\left[\frac{a(m_{h})}{a(m_{ s})} \right]^3 \\ \nonumber
&\simeq& \frac{3m_{ s}n_{h}^{\rm eq}(m_h)\Gamma_{h\to ss}}{H(m_{ h})}\frac{g_{*{\cal S}}(m_{ s})}{g_{*{\cal S}}(m_{ h})}\left(\frac{m_{ s}}{m_{ h}}\right)^3 \\ \nonumber
&=& \frac{3e^{-1}}{(2\pi)^{\frac{3}{2}}}\sqrt{\frac{90}{\pi^2}}\frac{g_{*{\cal S}}(m_{ s})}{g_{*{\cal S}}(m_{ h})\sqrt{g_{*}(m_{ h})}}\frac{\Gamma_{h \to ss}\Mp m_{ s}^4}{m_{ h}^2}\,,
\end{eqnarray}
where $n_s$ and $n_h^{\rm eq}$ are the singlet scalar and Higgs number densities, respectively, $a(m_i)$ is the scale factor at the time the photon temperature is $T=m_i, i=h,s$, and where in the limit $m_{ s}\ll m_{ h}$,
\begin{equation}
\label{gamma-higgs}
\Gamma_{h\rightarrow ss} = \frac{\lambda_{ hs}^2 v^2}{32\pi m_{ h}} \, .
\end{equation}
To derive this expression, we have assumed that the $h$ particles obey Maxwell-Boltzmann statistics after the electroweak symmetry breaking, that the singlet scalars are produced by $h\to ss$ at $T=m_{ h}$, and that the thermalisation of scalars takes place no earlier than $T=m_{ s}$.

\begin{figure}
\begin{center}
\includegraphics[width=14cm]{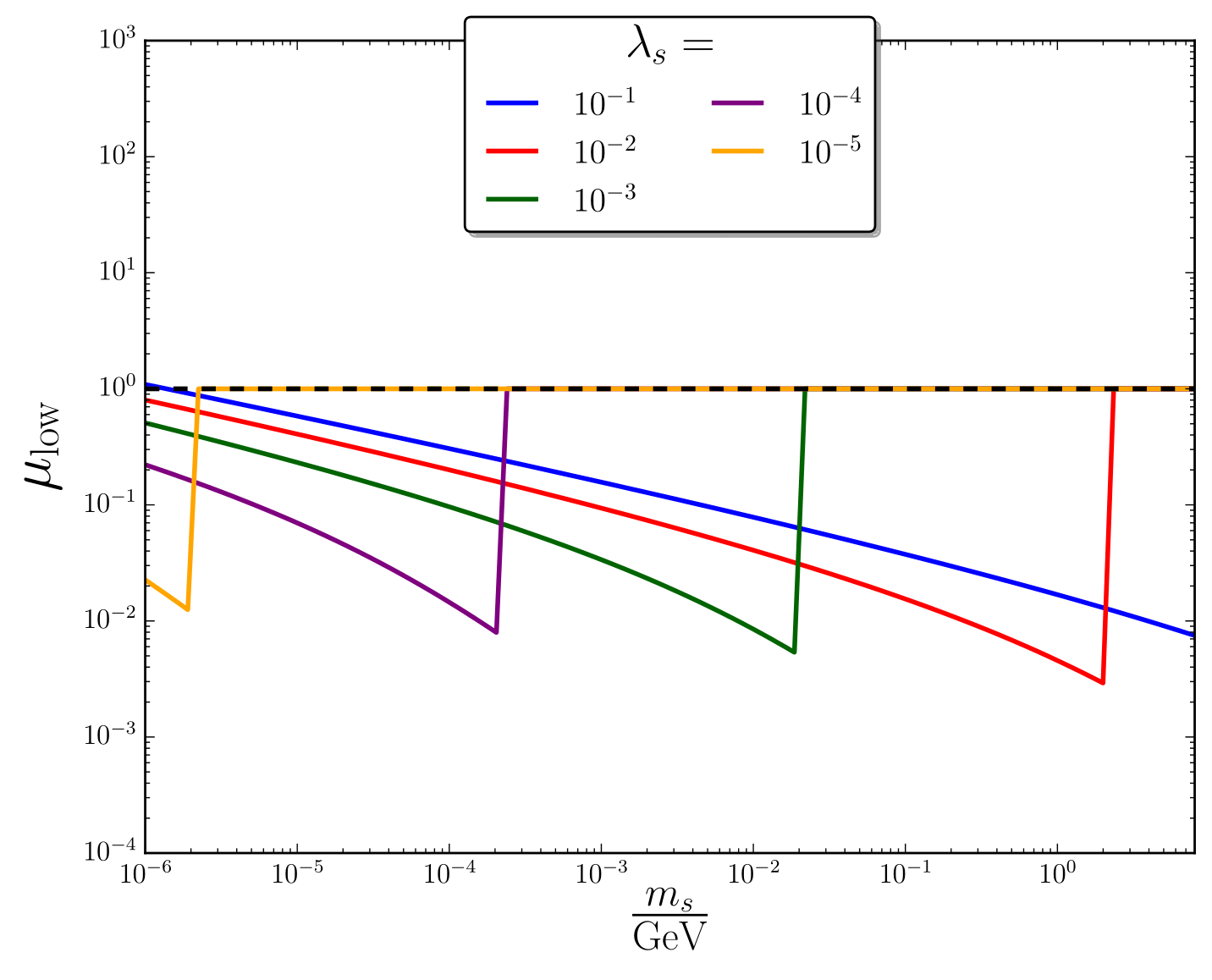}
\caption[Correction from low energy dynamics]{\label{fig:mu_low} Correction factor $\mu_{\mathrm{low}}$ appearing in \Eq{H*_with_corrections} and defined in \Eq{mulow:2}, accounting for variations in the low-energy dynamics, plotted as a function of the singlet mass $m_s$ for a few values of the self-interaction strength $\lambda_s$. The vertical jumps correspond to critical values of $m_s$ above which $\lambda_s^{(\mathrm{fi})}>\lambda_s$, where $\lambda_s^{(\mathrm{fi})}$ is given in \Eq{lambdas_fi}. In such a case the usual freeze-in picture applies and  $\mu_{\rm low}= 1$.}
\end{center}
\end{figure}

Plugging \Eqs{rhoS} and~(\ref{rhoA}) into \Eq{isocurvature}, and using \Eq{Pdeltas} as before, we then obtain
\begin{equation}
\label{H*_thermalisation}
\frac{H_*}{10^{11}{\rm GeV}} \simeq 0.97\left(\frac{\beta}{1-\beta}\right)^{\frac13}\lambda_s^{\frac14}\left(\frac{m_s}{{\rm GeV}}\right)^{-\frac23}\times \mu_{\rm low}(\lambda_{hs},m_s) ,
\end{equation}
where we have defined
\begin{equation}
\label{mulow:1}
\mu_{\rm low}(\lambda_{hs},m_s) \equiv 8\times 10^{13}\lambda_{hs}^{\frac43}\left(\frac{m_s}{{\rm GeV}}\right)^{\frac23} .
\end{equation}
Because the result now depends explicitly on $\lambda_{hs}$, its exact value becomes important. As discussed above, we require that the singlet particles constitute all DM, which allows us to fix $\lambda_{hs}$ in terms of $\lambda_{s}$ and $m_{s}$, as shown in \App{relating_lambdahs_to_lm}, see \Eq{eq:lambdahs:ms:lambdas}. This gives rise to
\begin{equation}
\label{mulow:2}
\mu_{\rm low}(\lambda_{s},m_s) = 4\times 10^{-3}
W_0^{\frac{8}{9}} \left[7.1\times 10^{4} \lambda_s^{\frac{24}{11}}\left( \frac{m_s}{{\rm GeV}}\right)^{-\frac{10}{11}} \right]
\left(\frac{m_s}{{\rm GeV}}\right)^{-\frac29} .
\end{equation}
The correction factor $\mu_{\rm low}$ is plotted in \Fig{fig:mu_low}, where one can see that the reduction in $H_*$ caused by variations in the low-energy dynamics is at most of order $\order{10^{-3}}$.

\begin{figure}
\begin{center}
\includegraphics[width=14cm]{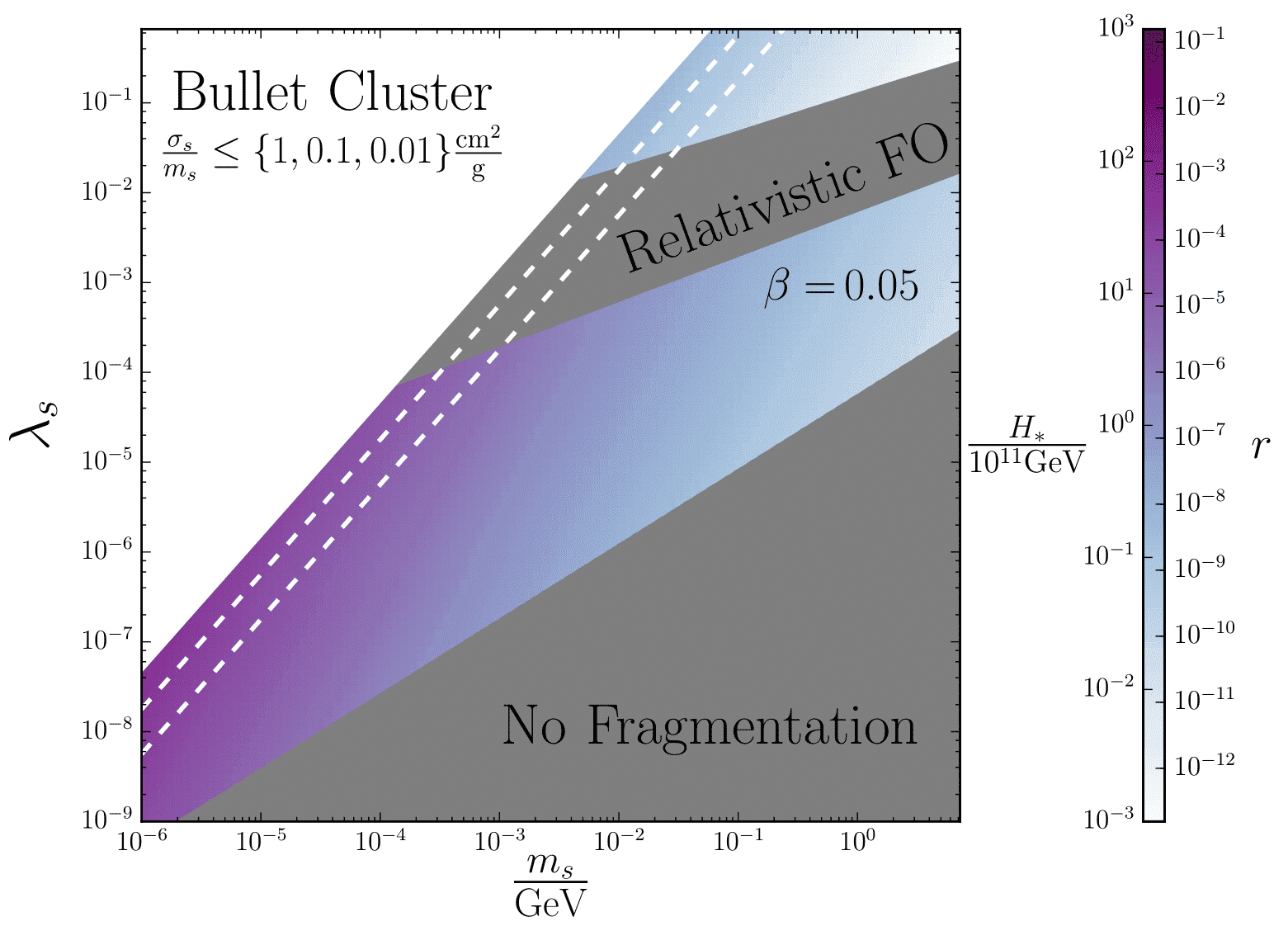}
\caption[Energy scale of inflation corrected by the low energy dynamics]{\label{fig:H_*results_mu_low} Same as \Fig{fig:H_*results}, but with the correction $\mu_{\rm low}$ appearing in \Eq{H*_with_corrections} and defined in \Eq{mulow:2} included. In the grey region labeled ``relativistic FO'' for ``relativistic freeze-out'' the dark freeze-out occurs while the singlet particles are still relativistic and the present calculation does not apply. Above this grey region, the singlet particles thermalise within themselves and the DM abundance is determined by dark freeze-out at $T\lesssim m_s/3$ instead of the usual freeze-in mechanism at $T\simeq m_h$. Below this grey region, the DM abundance is determined by the usual freeze-in mechanism, $\mu_{\rm low} = 1$, and one retains the results of \Fig{fig:H_*results}.}
\end{center}
\end{figure}

The corresponding effect on $H_*$ is shown in \Fig{fig:H_*results_mu_low}. In the grey region over the top right hand side of the plot shown in \Fig{fig:H_*results_mu_low}, the dark freeze-out occurs while the singlet particles are still (semi-)relativistic, $m_s/T_s^{({\rm fo})}\leq 3$, and finding a solution that yields the correct DM abundance in that region requires a detailed numerical analysis, as discussed in \Ref{Heikinheimo:2016yds}. We will postpone that for future work. Above the grey region the freeze-out occurs at temperatures where the DM is non-relativistic, $m_s/T_s^{({\rm fo})}> 3$, and below this region the singlet particles do not thermalise within the singlet sector and the usual freeze-in picture is sufficient. In general, thermalisation of the singlet sector increases the number density of the $s$ particles, resulting in a larger final DM abundance than in the standard freeze-in scenario, and in order to produce the observed DM abundance, a smaller initial abundance sourced by the SM particles is needed. Thus, an initial population of scalars produced from the decay of the primordial $s_0$ condensate contributes a larger fraction of the total DM energy density than it would in the standard freeze-in scenario, and hence the isocurvature contribution is larger. Thus, to keep the ratio \eqref{isocurvature} constant for fixed $\beta$, a smaller value for $\rho_{\rm S}$, \ie a smaller value for $H_*$, is needed. This explains why the correction factor $\mu_{\rm low}$ is always less than $1$.
\begin{figure}
\begin{center}
\includegraphics[width=13cm]{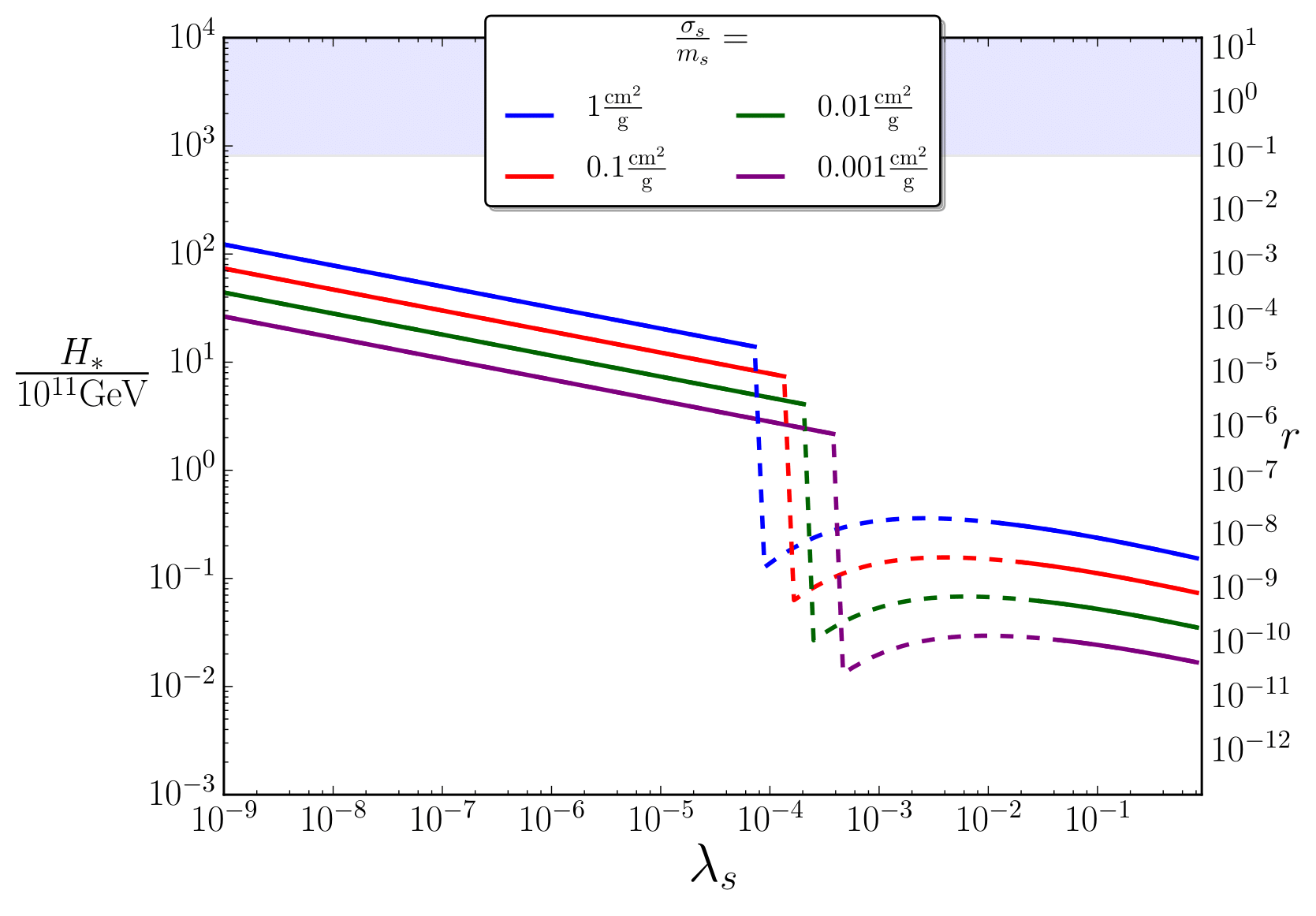}\\
\hspace{-0.7cm}\includegraphics[width=12cm]{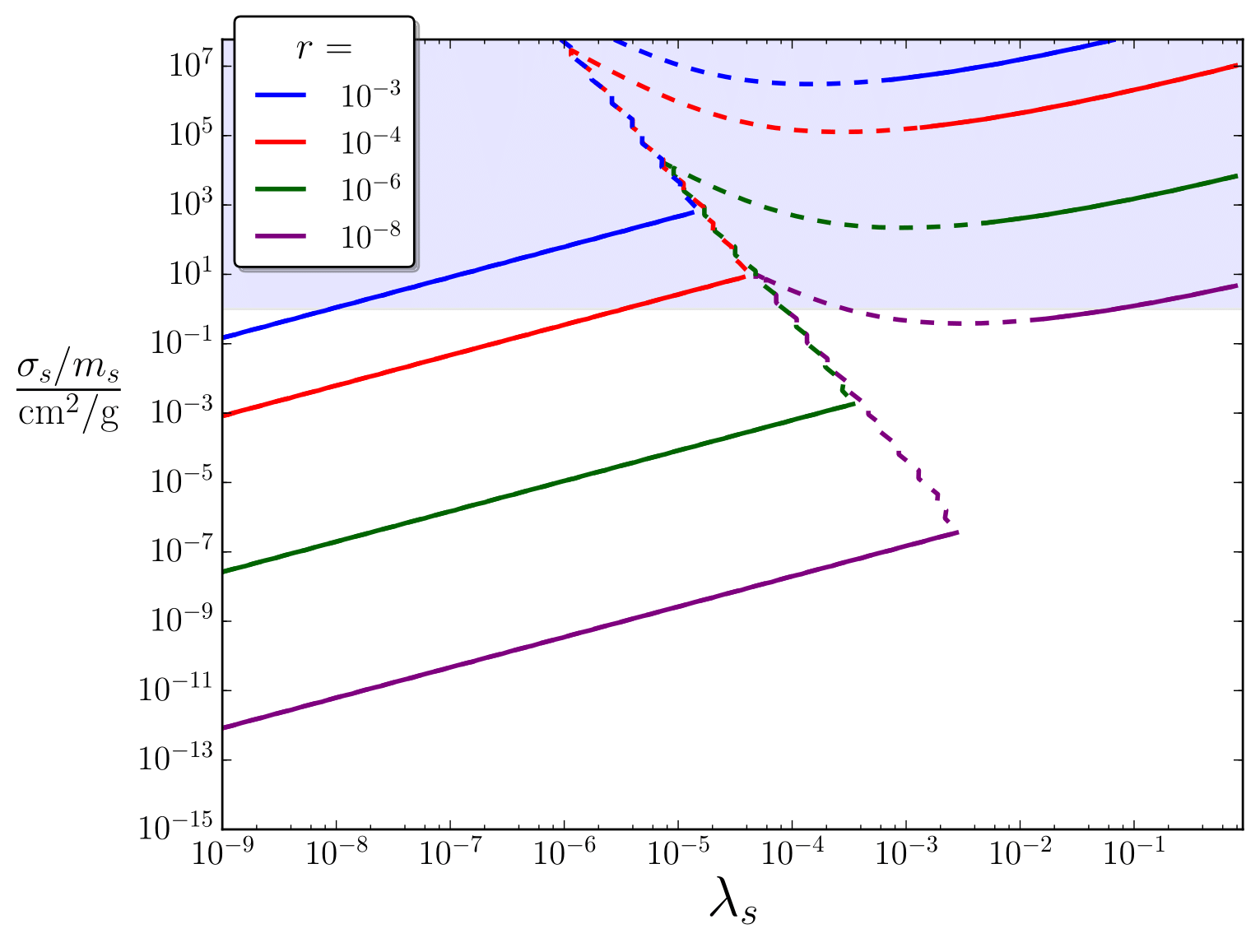}
\caption[Constraint on the tensor-to-scalar ratio from dark matter]{}
\end{center}
\end{figure}
\begin{figure}
\contcaption{\label{fig:constraint} 
Constraints on the tensor-to-scalar ratio $r$ for a few fixed values of the dark matter self-interaction cross-section divided by its mass, $\sigma_s/m_s$ (upper panel), and conversely, constraints on $\sigma_s/m_s$ for a few fixed values of $r$ (lower panel), as a function of the DM self-coupling constant $\lambda_s$. The dashed parts of the curves stand for the relativistic freeze-out regime (labeled ``relativistic FO'' in \Fig{fig:H_*results_mu_low}) where the calculation presented in this section does not apply. The shaded regions are observationally excluded, and correspond to the upper bound on $r$ obtained from CMB temperature and polarisation measurements~\cite{Ade:2015tva} on the upper panel and to the upper bound on $\sigma_s/m_s$ obtained from the ``Bullet Cluster'' constraint~\cite{Kaplinghat:2015aga} on the lower panel. Both plots assume that the background dynamics during inflation and reheating is standard, namely $\mu_{\rm inf}=\mu_{\rm reh}=1$, and set the DM isocurvature relative amplitude to the \emph{Planck}~\cite{Ade:2015lrj} upper limit $\beta = 0.05$ for demonstration.}
\end{figure}

\section{\textsf{The duration of inflation with a curvaton}}

The overall duration of inflation is not generally known, however, we have seen already in this chapter how one way to circumvent this cosmic amnesia is through spectator fields~\cite{Hardwick:2017qcw,Torrado:2017qtr}, whose field displacements are sensitive to a much longer phase of the inflationary epoch and which can be observationally accessible \cite{Linde:2005yw}. 

Since current CMB measurements are compatible with single-field models of inflation (if the potential is of the plateau type)~\cite{Ade:2015lrj, Martin:2013tda, Martin:2013nzq}, such extra fields may not be directly required by the data. It is of course always possible to fit the data with complex multi-field inflationary models, but the amount of fine tuning required in these models may be large, which is why models should be compared in a Bayesian framework (building from \Chap{sec:statistical-intro}) that correctly accounts for the quality of the fit and the waste of parameter space.

The questions we will seek to answer in the remaining sections of this chapter are therefore: Are there multiple-field models of inflation that are as favoured by the data as single-field plateau inflation from a Bayesian perspective? What insight can be gained on the inflationary history in these models?

We will investigate these questions with the curvaton model, whose potential is given by \Eq{eq:pot:gen:curvaton}. We outlined the general model in \Sec{sec:sourcing-cos-pert-curvaton}, however we shall briefly review it here for context.

After inflation, the inflaton field decays into radiation and the energy density contained in the curvaton field, $\rho_\sigma$, may grow relative to the background energy density, until it also decays into radiation. Assuming that no isocurvature perturbations persist~\cite{Lyth:2002my, Weinberg:2004kf, Smith:2015bln}, the total adiabatic power spectrum is given by combining \Eq{ps} and \Eq{eq:powerspecs}. Observations are also often discussed in terms of the spectral index $\nS$ and the tensor-to-scalar ratio $r$ in \Eq{eq:nsr:slowroll}. When the primordial density perturbation is entirely due to curvaton field fluctuations then the original curvaton model~\cite{Lyth:2001nq, Enqvist:2001zp, Moroi:2001ct} is realised. Hence, in this section we term situations where $\lambda > 0.9$ as the ``curvaton scenario''. 

At the pivot scale, the latest 2015 BICEP2/Keck Array and Planck \cite{Array:2015xqh, Ade:2015xua} combined observations give ${\mathcal{P}}^{\mathrm{total}}_{\zeta} \sim 2.2\times 10^{-9}$, $\nS =0.9667\pm0.008$ and $r <0.07$ ($95\%$ c.l.). If the inflaton potential is of the large-field type $U(\phi)\propto\phi^p$, in the curvaton limit $\lambda\simeq 1$, \Eq{eq:nsr:slowroll} implies that $\nS\simeq 1-p/120$, and the observed value of the spectral index means that the inflaton field potential must be close to quartic, $p=4$. The ``simplest'' curvaton scenario with a quadratic inflaton and 
curvaton field is now disfavoured by the data~\cite{Byrnes:2014xua,Hardwick:2015tma, Vennin:2015egh, Byrnes:2016xlk}. 

The observational constraints on $\nS$ and $r$ imply that when any inflaton potential is included in the analysis, only two classes of models with an additional spectator field are found to be favoured~\cite{Vennin:2015egh}: plateau inflation, which \emph{cannot} fit the data in the curvaton scenario (thereby requiring $\lambda\ll1$), and quartic inflation, which \emph{can only} fit the data in the curvaton scenario ($\lambda\sim1$). An advantage of a quartic potential is that the inflaton field energy decreases like radiation when it oscillates, making the model more predictive by removing the dependence of post-inflationary dynamics on the inflaton decay rate into radiation.

Another way to detect the curvaton is through primordial non-linearity of the density perturbations, of which the key observable is the local non-Gaussianity of the bispectrum, parametrised by $\fnl$. Its value in the sudden-decay approximation is given by~\Ref{Ichikawa:2008iq} and \Eq{eq:fnl}, where the observational non-Gaussianity constraint of $\vert \fnl \vert\lesssim10$ implies that either we predominantly observe inflaton perturbations, $\lambda\simeq0$, or the the spectator must have a non-negligible energy density at its decay, $\rdec\gtrsim0.1$.

The contribution from the curvaton to the primordial power spectrum crucially depends on its field value, $\sigma_*$, when observable modes exit the Hubble radius. Combining \Eqs{eq:powerspecs} and~(\ref{eq:lambda:def}), one can see that the curvaton dominates the perturbations, $\lambda>1/2$, if $\sigma_*/\Mp < \sqrt{\epsilon_{1*}}\rdec$. Therefore $\sigma_*$ must be sub-Planckian (if it is super-Planckian, it may drive a second phase of inflation and the above formulas do not apply, but below we show that this case is excluded). In practice, the value of $\sigma_*$ is determined by the details of the inflaton's potential $U(\phi)$ over the entire inflating domain, as we demonstrated in \Chap{sec:infra-red-divergences}. This makes the model more predictive since the typical value of $\sigma_*$ is not a free parameter anymore but depends on $U(\phi)$. This will play an important role in the Bayesian analysis below. In particular, the value of $\sigma_*$ also depends on the total duration of inflation, which will allow us to constrain it.

Most previous analyses of curvaton models assumed no knowledge a priori about spectator field values. Instead, we adopt a physical prior for the typical field displacement $\sqrt{\langle \sigma_*^2\rangle}$ of the curvaton. This prior depends on the inflaton potential $U(\phi)$ and the total duration of inflation, as one can see immediately from the results of \Sec{sec:quad_spec}. Using these results, in the presence of a plateau inflationary potential, if inflation lasts more than the relaxation timescale $N_{\rm relax}=H^2/m_\sigma^2$ \efold{s}~\cite{Starobinsky:1986fx, Enqvist:2012xn}, the vev of $\sigma$ reaches a Gaussian equilibrium distribution with a variance given by
\bea
\label{eq:hisigmaend}
\left\langle \sigma_*^2 \right\rangle = \frac{3 H_*^4}{8\pi^2 m_\sigma^2}
\,.
\eea
In the presence of a quartic large-field inflationary potential ($U(\phi ) \propto \phi^4$), we find that \Eq{eq:meansigma2:quadratic:generic:pGT2} can be rewritten to give
\bea
\label{eq:lfi4sigmaend}
\left\langle \sigma_*^2 \right\rangle =
\left\langle \sigma^2_{\mathrm{in}} \right\rangle +
\frac{H_*^2}{12\pi^2}N_{\mathrm{tot}}^3
\,,
\eea
with a strong dependence upon initial conditions. The distributions (\ref{eq:hisigmaend}) and (\ref{eq:lfi4sigmaend}) define the prior we take on $\sigma_*$ for plateau and quartic inflation, respectively. In \Eq{eq:lfi4sigmaend}, $N_{\mathrm{tot}}$ is the total number of \efolds~elapsed during quartic inflation and $\langle \sigma^2_{\mathrm{in}} \rangle$ denotes the variance of the curvaton vev distribution at the onset of inflation. 
In the following we will take $\langle \sigma^2_{\mathrm{in}} \rangle=0$ for the sake of simplicity. This is important as it means $N_{\rm tot}$ is now the \emph{maximum} number of \efold{s}. A more specific model for the curvaton could readily specify $\langle \sigma^2_{\mathrm{in}} \rangle$ from, e.g, a symmetry breaking mechanism, however we shall leave $\langle \sigma^2_{\mathrm{in}} \rangle=0$ such that our argument in this section represents a proof-of-principle.

The expansion history of reheating depends on the mass of the curvaton and the decay rates of the inflaton and the curvaton. We impose that the onset of the radiation-dominated period occurs after the end of inflation and before the electro-weak symmetry breaking. We also assume that the inflaton and the curvaton decay at least as fast as they would through their minimal coupling to the gravitational sector, given in \Eq{eq:Mstar-before}.\footnote{Here $\Gamma_\phi$ (or $\Gamma_\sigma$) denotes the value of $H$ below which the energy density contained in $\phi$ (or $\sigma$, respectively), or its decay products, redshift like radiation.} Using non-informative priors, as discussed in \Sec{sec:priors-intro}, this leads to
\begin{align}
\label{eq:newbaseline}
\Gamma_\sigma &\sim \log \uniform\left[\max\left(H_\mathrm{EW}, \frac{m_\sigma^3}{\Mp^2}\right), \min\left(H_\uend, m_\sigma\right)\right] \\
\Gamma_\phi   &\sim \log \uniform\left[\max\left(H_\mathrm{EW}, \frac{H_\mathrm{end}^3}{\Mp^2}\right), H_\uend\right] \\
m_\sigma &\sim \log \uniform\left[ H_{\mathrm{EW}}, H_\uend \right]\, ,
\end{align}
where $H_\uend$ is the Hubble scale at the end of inflation, $H_\mathrm{EW}=(150\,\mathrm{GeV})^2/\Mp$ is the Hubble scale at electro-weak symmetry breaking, and we remind the reader that $x\sim\log \mathcal{U}[a,b]$ means that $\log x$ is uniformly distributed between $\log a$ and $\log b$. 

As discussed in \Sec{sec:pert-reheat-intro}, if the inflaton has a quartic potential, its coherent oscillations around the minimum of its potential give rise to a radiation-like era of expansion immediately after inflation~\cite{Turner:1983he}. In this case we set $\Gamma_\phi=H_\uend$ and reheating can be described by two parameters only, the mass and decay rate of the curvaton.

\section{\textsf{Duration of inflation: results}}
\begin{figure}
\begin{center}
\includegraphics[width=14cm]{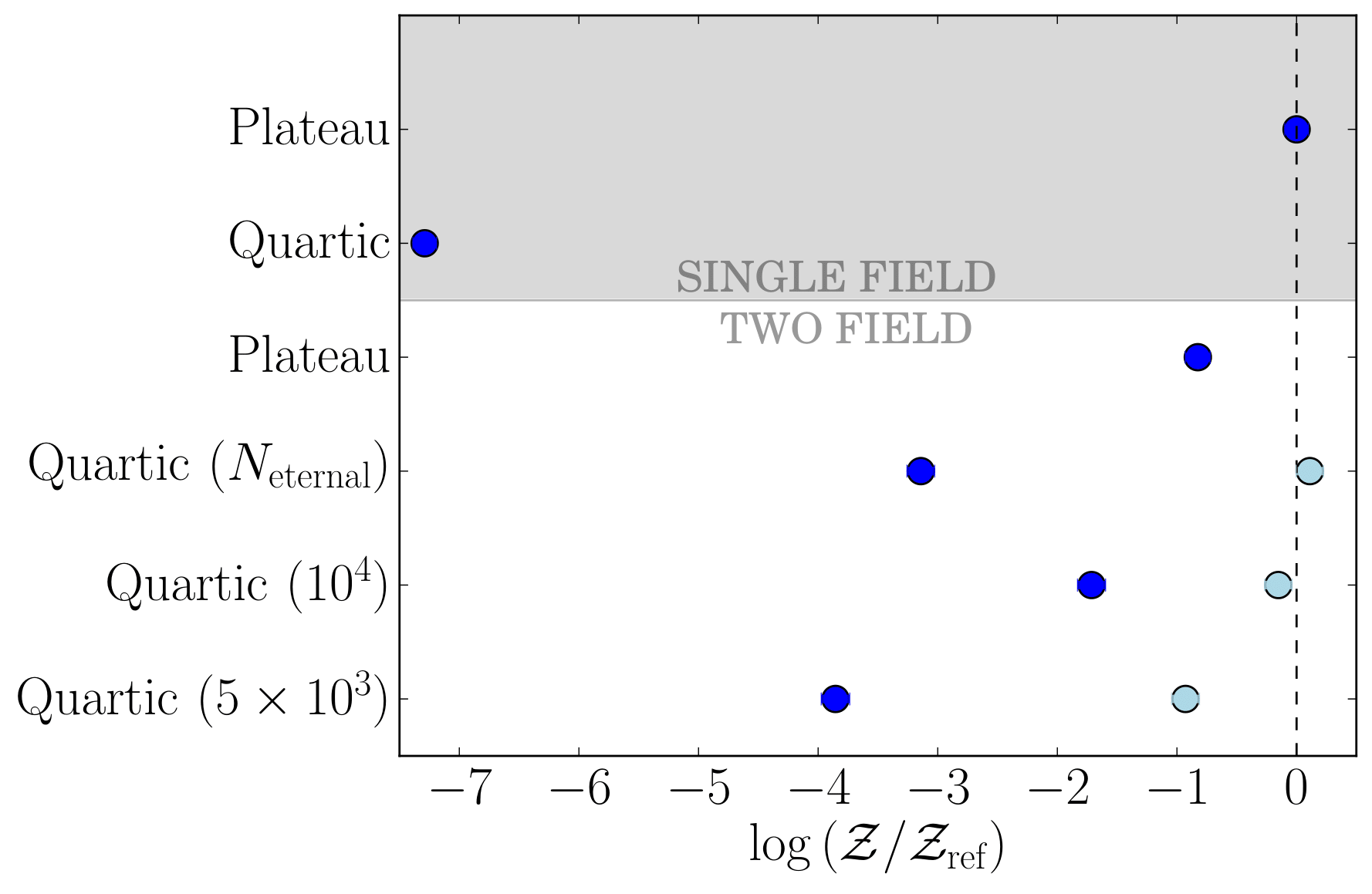}
\end{center}
\caption[Bayesian evidence of single field versus curvaton models]{Bayesian evidences $\mathcal{Z}$ of the single-field (inflaton) and two-field (inflaton plus spectator) models (inside and below the shaded region, respectively) considered in this section. The plateau model (taken as the reference here) is robust with respect to the introduction of an additional field. Quartic inflation with a spectator field (where the total number of \efolds~is written in parenthesis) has a higher evidence than its single-field version, but lower than the plateau model. Imposing the curvaton scenario (here defined as $\lambda >0.9$, see the main text) at the level of the prior (lighter blue points), the evidence becomes comparable with the one of plateau models.}
\label{fig:evidences} 
\end{figure} 

The Bayesian analysis is performed on the January 2015 BICEP2/Keck-Array/Planck data combination \cite{Ade:2015tva}, using the machine-learned effective inflationary likelihood described in \Ref{Ringeval:2013lea}, which has been marginalised over late-time background cosmology, reionisation, and astrophysical foregrounds. The predictions of the models are computed with the curvaton extension of the \texttt{ASPIC} library~\cite{aspic}, making use of the method presented in \Refs{Vennin:2015egh, Hardwick:2016whe}. The Bayesian evidences are integrated using the \texttt{MultiNest} algorithm \cite{Feroz:2007kg,Feroz:2008xx}; further technical details on the numerical integration can be found in the appendix. The Bayesian evidences are displayed in \Fig{fig:evidences} and the corresponding posterior distributions in \Fig{fig:observables}. 

\subsection{\textsf{Single-field versus spectator model}}

One can check in \Fig{fig:evidences} that for single-field models, plateau potentials are favoured while a quartic potential is strongly disfavoured (and even ruled out at the level of its maximum likelihood). When a light spectator field is included, the evidence of plateau potentials remains stable, and the two-field model cannot be distinguished from its single-field counterpart in terms of its Bayesian evidence~\cite{Jeffreys:1961}. This is because, in spite of the significant enlargement in prior parameter space caused by the introduction of the spectator field, most of the prior mass in the distribution~(\ref{eq:hisigmaend}) reproduces single-field phenomenology, which gives a very good fit to the data irrespective of the value of the reheating parameters. This result is consistent with what was found in \Refs{Vennin:2015egh, Hardwick:2016whe}. 

\begin{figure*}
  \centering
  \includegraphics[width=7cm]{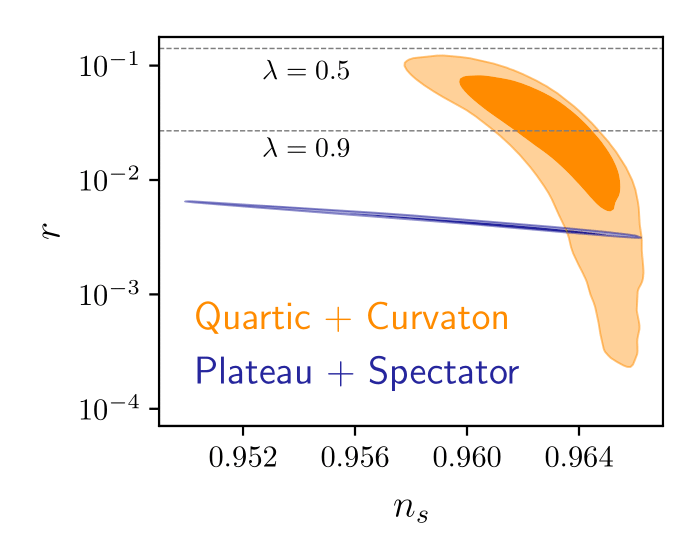}
  \includegraphics[width=7cm]{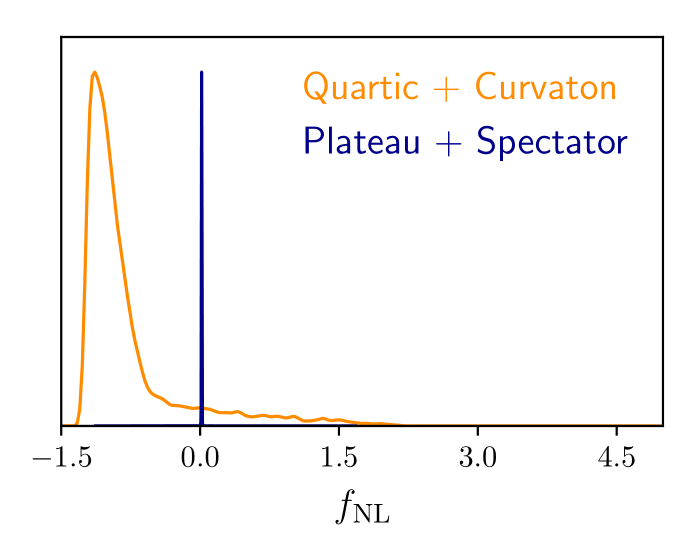} 
  \caption[Marginal posterior distributions for the curvaton models]{\label{fig:observables}
Marginal posterior distributions over the key observables from inflation for plateau-like inflation (blue, darker) and quartic inflation (orange, clearer) with a spectator field. In the quartic case, the posterior fraction below the lower (upper) dotted line has more than $90\%$ ($50\%$) of primordial density perturbations generated by the curvaton field. Post-2020 CMB experiments would likely distinguish between or rule out both scenarios in terms of $\nS$ and $r$. In combination with LSS data, the typical value of $\fnl = -5/4$ associated with the curvaton scenario could also be distinguished in the future from $\fnl\sim\mathcal{O}(10^{-2})$ in the inflaton scenario.}
\end{figure*}

For the quartic potential, the evidence obtained once a spectator field is included depends on the total duration of inflation, $N_{\mathrm{tot}}$, through the prior distribution~(\ref{eq:lfi4sigmaend}) for the curvaton vev. We give the Bayesian evidence for a few values of $N_{\mathrm{tot}}$ in \Fig{fig:evidences}. We take $N_{\mathrm{tot}}\sim 6\times 10^4$ as an upper bound, since for larger values the inflaton would initially be in the ``self-reproducing'' regime~\cite{Linde:2005ht, Winitzki:2008zz} where stochastic corrections to its dynamics become important and the calculation of \Ref{Hardwick:2017fjo} does not apply.

In all cases, one can check that quartic models with a spectator field are favoured with respect to their single-field counterpart, but are still moderately or strongly disfavoured with respect to the plateau potential. If one restricts the parameter space to the curvaton model, \ie if one imposes $\lambda>0.9$ at the level of the prior, one obtains an evidence similar to that of single-field plateau models irrespective of the duration of inflation (see the lighter blue points in \Fig{fig:evidences}), indicating that the dependence of the evidence on the number of \efolds~of inflation actually reflects the proportion of $\sigma_*$ values that correspond to the curvaton scenario in each case.

In terms of the observables shown in \Fig{fig:observables}, plateau inflation (the Higgs inflation or Starobinsky model in the present case) with a spectator field gives very similar predictions to its single-field counterpart, namely a small tensor-to-scalar ratio, a value for the spectral index that is in good agreement with observations, and a slow-roll suppressed value for $\fnl$ that is currently (and in the foreseeable future) undetectable. For quartic inflation, independently of the duration of inflation, the tensor-to-scalar ratio and the spectral index are correlated, with bluer spectra corresponding to reduced gravitational waves, and non-Gaussianity has the typical amplitude $\fnl\simeq-5/4$, which, from \Eq{eq:fnl}, corresponds to a preference for values $\lambda\simeq \rdec \simeq 1$, \ie to situations where the curvaton dominates the energy budget of the Universe when it decays and provides the dominant contribution to primordial density perturbations.

Post-2020 CMB experiments \cite{Matsumura:2013aja, Finelli:2016cyd, DiValentino:2016foa} will shrink the $1-\sigma$ constraints on the inflationary observables to $\Delta \nS \sim 2 \times10^{-3}$ and $\Delta r\sim 10^{-4}$, while cross-correlation with future LSS experiments should drive the constraint on local non-Gaussianity down to $\Delta \fnl\sim 0.4$ \cite{Schmittfull:2017ffw}. This would be enough to distinguish between plateau inflation (with or without a spectator field) and quartic inflation with a curvaton, or even to rule out both models.

\subsection{\textsf{Measuring the duration of inflation}}

For quartic potentials with a spectator field, the data shows strong preference for curvatonic phenomenology (see the difference between the dark and light points in \Fig{fig:evidences}), which corresponds to sub-Planckian spectator field values of a few $10^{-2}\,M_\mathrm{Pl}$. This yields an ``optimal'' value for the total number of \efolds~of quartic inflation such that it maximises the parameter volume that falls within this range of values.

A smaller variance for the prior distribution~(\ref{eq:lfi4sigmaend}) of $\sigma_*$ (requiring a shorter duration of inflation) limits the spectator field vev so that single-field quartic inflation is recovered, which is ruled out observationally. A larger variance (due to a longer duration of inflation or larger initial variance) locates most of the prior mass in spectator vevs so large that they drive a second phase of quadratic inflation, which is also ruled out.\footnote{If the light spectator field is displaced by $\sigma_*^2\gtrsim 2\Mp^2$ during inflation, then it may drive a second period of inflation, which lasts for $N_2\simeq \sigma_*^2/(4 \Mp^2)$ \efolds. The amplitude of the curvaton perturbations generated during the first period of inflation is~\cite{Vernizzi:2006ve}
\bea
{\mathcal{P}}^{\sigma}_{\zeta} = N_2\left(\frac{H_*}{2\pi\Mp}\right)^2 \nonumber \, .
\eea
Independently of the inflaton potential, the tensor-to-scalar ratio is given by
\bea 
r=\frac{{\mathcal{P}}_h}{{\mathcal P}^{\sigma}_{\zeta}+{\mathcal P}^{\phi}_{\zeta}} = \lambda \frac{{\mathcal P}_h}{{\mathcal P}^{\sigma}_{\zeta}} = \lambda \frac{8}{N_2} \nonumber \, , 
\eea
where ${\mathcal P}_h=8 [H_*/(2\pi \Mp)]^2$. The observational bound on $r$ then imposes
\bea 
\lambda \lesssim \frac12 \left(\frac{N_2}{60}\right) \nonumber \, .
\eea
Since we require $N_2<60$, because otherwise the first period of inflation would end before the observable modes exit the horizon, this implies that a quadratic spectator field that then inflates the Universe cannot generate the majority of the observed perturbations.\label{footnote}}

Adapting \Eq{eq:bayes-theorem-ppiL}, the posterior ${\cal P}$ on the total duration of inflation can be computed according to
\bea
\label{eq:nmaxpost}
\ {\cal P}\left(N_\mathrm{tot}\right\vert\left.\mathcal{D}\right) \propto
{\cal P}\left(\mathcal{D}\right\vert\left.N_\mathrm{tot}\right)\,\pi\left(N_\mathrm{tot}\right)
\, ,
\eea
where ${\cal P}(\mathcal{D}|N_\mathrm{tot}) = \mathcal{Z}(N_\mathrm{tot})$ is the evidence of the quartic plus spectator field model with prior~(\ref{eq:lfi4sigmaend}) on $\sigma_*$ corresponding to $N_\mathrm{tot}$, and $\pi(N_\mathrm{tot})$ is the prior we set on the duration of inflation. 

\begin{figure}
  \centering
  \includegraphics[width=10cm]{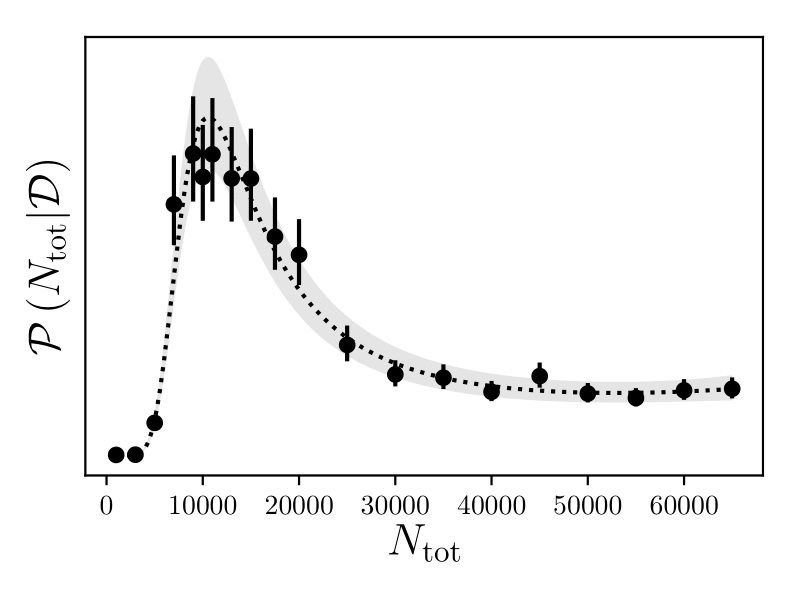}
  \caption[Marginal posterior over the number of \efold{s} of inflation]{\label{fig:Nefolds}
    Marginal posterior over the total number of \efolds~of inflation $N_{\mathrm{tot}}$ with a uniform prior, for quartic inflation with a spectator field. The upper limit corresponds approximately to the ``self-reproducing'' regime, $N_\mathrm{tot}\sim 6\times 10^4$. The dotted line and the grey band are respectively the mean and $1-\sigma$ confidence-level limit of a logarithmic Gaussian process interpolation with maximum-a-posteriori noise level, scale and correlation length~\cite{Rasmussen:2005:GPM:1162254}. The black dots and bars are the evidences and their error computed with \texttt{MultiNest}.}
\end{figure} 

We reconstruct this posterior in \Fig{fig:Nefolds}, where one can see that inflation is constrained to last less than a few tens of thousands of \efolds. In particular, cases where inflation starts close to the ``self-reproducing'' regime ($N_{\mathrm{tot}}\sim 6\times 10^4$), are strongly disfavoured~\cite{Winitzki:2010yz}. This is because in such cases, \Eq{eq:lfi4sigmaend} yields $\langle \sigma_*^2\rangle^{1/2}> \Mp$ (which is true in any large-field inflationary potential~\cite{Hardwick:2017fjo}) and the spectator field drives a second phase of inflation. Note that if the initial variance $\langle\sigma^2_{\rm in}\rangle$ does not vanish then the constraint that we have obtained is only an upper bound on the duration of inflation, but the conclusion that it should not start in the self-reproducing regime remains true.

\section{\textsf{Conclusions}} \label{sec:conclusions-isocurvaturefields}
In this chapter, we presented a novel way to determine the energy scale of inflation in the case where the DM component is a feebly-interacting singlet scalar. Assuming it is light and energetically subdominant during inflation, we have shown that the inflationary energy scale $H_*$ can be expressed as a function of the DM isocurvature perturbation amplitude $\beta$ and the DM self-interaction cross-section divided by its mass $\sigma_s/m_s$, with only a very weak dependence on the DM four-point self-coupling $\lambda_s$,
\bea
\frac{H_*}{10^{11}\mathrm{GeV}}\simeq 10.0 \left(\frac{\beta}{1-\beta}\right)^{\frac{1}{3}}
\lambda_s^{-\frac{7}{36}}
\left(\frac{\sigma_s/m_s}{\mathrm{cm}^2/\mathrm{g}}\right)^{\frac{2}{9}}\, .
\eea
This relation is obtained combining \Eqs{H*_no_thermalisation} and~(\ref{scrosssection}), and is valid for the case of freeze-in only. It connects observables that constrain two seemingly unrelated topics, namely the one of inflation and the one of DM. By doing so, it opens up the possibility to access the energy scale of inflation by studying the properties of DM, and vice versa.

To illustrate this, in the upper panel of \Fig{fig:constraint} we have displayed the value of $H_*$ (and the corresponding value of the tensor-to-scalar ratio $r$) one would infer from measuring $\sigma_s/m_s$ to certain fixed values, as a function of $\lambda_s$. One can see that because of the weak dependence on $\lambda_s$, if $\sigma_s/m_s$ were measured, the energy scale of inflation would be given up to a few orders of magnitude at most, a huge improvement compared to the $15$ orders of magnitude that are a priori allowed. One should also note that a detection of $\sigma_s/m_s$ close to the current threshold~(\ref{scrosssection}) would allow one to probe values of $r$ between $10^{-9}$ and $10^{-4}$, which cannot be reached by present day CMB technology. On the lower panel conversely, we have displayed the value of the $\sigma_s/m_s$ one would infer from measuring $r$ to certain fixed values. One can see that current constraints on $\sigma_s/m_s$ already almost rule out the target of the next generation of CMB experiments $r\sim 10^{-3}$~\cite{Matsumura:2013aja, Finelli:2016cyd, DiValentino:2016foa}. In fact, if such a value were detected, then in this model $\sigma_s/m_s$ would be predicted to be close to $0.1\, \mathrm{cm}^2/\mathrm{g}$. Since this value is within the reach of forthcoming observations~\cite{Tulin:2017ara}, that would open up the possibility to either confirm or rule out the scenario presented in this section. 

In addition to presenting the basic scenario, we have also discussed the robustness of this result and quantified how it changes under various effects related to inflation, reheating, and DM dynamics at low energies. We have characterised these effects by correction factors introduced in \Eq{H*_with_corrections}. We found that the change in the background evolution during inflation and the possible thermalisation of scalar particles within the singlet sector and the following ``DM cannibalism'' phase introduce only at most $\mathcal{O}(0.1)$ and $\mathcal{O}(10^{-3})$ corrections, respectively, to the result for $H_*$, whereas variations in the reheating history can in principle have a larger effect, depending on the duration of reheating.

Although the result obtained in this section is model dependent, it is generic to a large class of scenarios and allows one to measure or constrain the energy scale of inflation even in models where the associated predicted value for the tensor-to-scalar ratio is well below the current lower bound or sensitivity of the next-generation of CMB experiments. Conversely, a detection of the tensor-to-scalar ratio would allow one to infer a measurement for the DM self-interaction cross-section. This could represent a new promising chapter in constraining DM.

In this chapter we have also studied the observational consequences of \Chap{sec:infra-red-divergences} when applied to the curvaton. We found that if the inflationary potential is of the plateau type, the single-field limit is the preferred one (the predictions of the model are robust under the introduction of a spectator field), while quartic potentials are favoured only in the curvaton limit. Both options, plateau inflation in the single-field limit and quartic inflation in the curvaton limit, are equally favoured by current data, but we have shown that future CMB and LSS measurements may allow us to distinguish between them.

The contribution from spectator fields to cosmological perturbations strongly depends on their field values at the end of inflation \cite{Hardwick:2017fjo}. The accumulation of long-wavelength quantum fluctuations during the entire inflationary period gives rise to a distribution for the local field displacement that depends on the total duration of inflation. As a consequence, we found that the number of \efold{s} elapsed during inflation, $N_{\mathrm{tot}}$, enters as a parameter of the model due to the lack of early adiabatic regime. Hence, $N_{\mathrm{tot}}$ itself can be constrained by the data. 

In the curvaton limit, the inflationary potential is constrained to be close to the quartic type. In that case, $N_{\mathrm{tot}}$ cannot be too small otherwise the spectator field does not acquire a large enough field value to source cosmological perturbations, and cannot be too large otherwise the spectator field acquires too large a field value that drives a second phase of inflation. The posterior distribution on $N_{\mathrm{tot}}$ is displayed in \Fig{fig:Nefolds}, where we find that according to the data, inflation cannot last more than a few tens of thousands of \efold{s}. In particular, it is very unlikely that one starts quartic inflation in the so-called ``self-reproducing'' regime.

For the first time, we have thus quantified how much cosmological data can constrain the pre-inflationary history, much beyond the $N \gtrsim 60$ epoch probed by potential large scale CMB anomalies. One should note that the mechanism we presented is not only sensitive to the duration of inflation but also on the shape of the inflationary potential over its entire inflating domain, and on the spectator field displacement prior to inflation. This opens up a new observational window that extends the conventional scales by orders of magnitude and allows us to explore the physics of the very early Universe beyond our currently observable horizon.

\newpage

\begin{subappendices}

\section{\textsf{Calculation of the dark matter abundance}}
\label{app:general-scaling-energy-density}

In this appendix we track the energy density contained in the $s$ field from the end of inflation up until the measured abundance of dark matter today, in the sequence of events depicted in \Fig{fig:reheating}. This figure also a reference guide to the various subscripts used throughout this section.

At the end of inflation, we assume that $s_\uend$ takes a specific realisation resulting from its stochastic dynamics during inflation, $s_\uend \sim \sqrt{\langle s_\uend^2 \rangle}$, where $\langle s_\uend^2 \rangle$ is given in \Eq{eq:corrected-variance}. After the end of inflation, $s$ continues to be slowly-rolling (while quantum diffusion is shut off) until it becomes effectively massive, at the time $N_{\mathrm{osc}}$, when it starts to oscillate. One can check that the value of $s$ barely changes during this phase and to the approximation level at which the calculation is performed, it can be taken as effectively frozen, $\left.s_0\right\vert_{\mathrm{osc}}\simeq s_\uend$. The oscillations start when the effective mass of the condensate, $m_\ueff \sim \sqrt{3 \lambda_s} s_0$, becomes of order $H$. The time at which this happens can be calculated by introducing the mean equation-of-state parameter between the end of inflation and the beginning of the $s$ oscillations
\bea
\bar{w}_{\mathrm{frozen}} \equiv \frac{1}{N_{\mathrm{osc}}-N_\uend}\int^{N_{\mathrm{osc}}}_{N_\uend}w(N)\dd N\, .
\eea
The relation $\dd H/ H =-3/2(1+w) \dd N$ can then be integrated as
\bea
H_{\mathrm{osc}} = H_{\rm end} \exp\left[-\frac{3}{2}\left(1+ \bar{w}_{\mathrm{frozen}}\right) \left(N_{\mathrm{osc}}-N_\uend\right)\right]\, .
\label{eq:H:s_frozen}
\eea
By equating $H_{\rm osc} = \sqrt{3\lambda_s} \left.s_0\right\vert_{\mathrm{osc}} = \sqrt{3\lambda_s} s_\uend$, one obtains
\bea
N_{\mathrm{osc}}-N_\uend = \frac{2}{3\left(1+\bar{w}_{\mathrm{frozen}}\right)}\ln\left(\frac{H_\uend}{\sqrt{3\lambda_s}  s_\uend}\right)\, .
\label{eq:Nosc:s_frozen}
\eea
Let us note that for this number to be positive, the condition $H_\uend^2 > 3\lambda_s  s_\uend^2$ must be satisfied, which is always the case for the typical value of $s_\uend$ given by \Eq{eq:corrected-variance} if $\lambda_s\ll 1$.

After the condensate $s_0$ becomes effectively massive, it oscillates about the minimum of its quartic potential, so its energy density decays as the one of radiation, $\rho_{s_0}\propto a^{-4}$, until it fragments into $s$ particles. Fragmentation occurs when the fragmentation rate $\Gamma^{(4)}_{s_0 \rightarrow ss}$ is of order $H$. In \Ref{Tenkanen:2016idg}, it was found that
\bea
\label{gamma}
\Gamma_{s_0\rightarrow ss}^{(4)}(t)=\alpha \lambda_s^{\frac32}\sigma_0(t)\, ,
\eea
where $\alpha = 0.023$ is a numerical constant, and $\sigma_0$ is the envelope of the background $s_0$ time evolution, i.e. $s_0(t)=\sigma_0(t)\times F(t)$, where $F(t)$ is an oscillatory function. Notice that this expression is valid if the background is radiation-dominated, so reheating must have occurred at this stage for consistency. Since $\rho_{s_0}\propto s_0^4\propto a^{-4}$, during this epoch $\sigma_0 \propto 1/a$ and one has
\bea
\Gamma_{s_0\rightarrow ss}^{(4)}(N)=\alpha \lambda_s^{\frac32}   s_\uend \, \ee^{-\left(N-N_{\mathrm{osc}}\right)}\, .
\eea
On the other hand, similarly to \Eq{eq:H:s_frozen}, one has
\bea
H_{\mathrm{frag}} = H_{\mathrm{osc}} \exp\left[-\frac{3}{2}\left(1+ \bar{w}_{\mathrm{osc}}\right) \left(N_{\mathrm{frag}}-N_{\mathrm{osc}}\right)\right]\, ,
\label{eq:H:s_osc}
\eea
where $\bar{w}_{\mathrm{osc}}$ is the mean equation-of-state parameter in the oscillation phase. By equating the two previous formulas, one finds that
\begin{align}
N_{\mathrm{frag}}-N_{\mathrm{osc}} &= \frac{2}{1+3\bar{w}_{\mathrm{osc}}}\ln\left(\frac{H_{\mathrm{osc}}}{\alpha \lambda_s^{\frac32} s_\uend}\right) = -\frac{2}{1+3\bar{w}_{\mathrm{osc}}}\ln\left( \frac{{\alpha \lambda_s}}{\sqrt{3}}\right)\, ,
\label{eq:osc:Nfrag}
\end{align}
where in the second equality we have used that $H_{\rm osc} =  \sqrt{3\lambda_s} s_\uend$. One can see that in order for $N_{\mathrm{frag}}-N_{\mathrm{osc}}$ to be positive, one must have $\alpha \lambda_s < \sqrt{3}$, which is again always satisfied if $\lambda_s\ll 1$. Combining $H_{\rm osc} = \sqrt{3\lambda_s} s_\uend$, \Eqs{eq:H:s_osc} and~(\ref{eq:osc:Nfrag}), one then obtains
\bea
H_{\mathrm{frag}} = \sqrt{3\lambda_s} s_\uend \left( \frac{\alpha\lambda_s}{\sqrt{3}}\right)^{3\frac{1+\bar{w}_{\mathrm{osc}}}{1+3\bar{w}_{\mathrm{osc}}}}\, .
\eea
On the other hand, combining \Eq{eq:osc:Nfrag} with the formula $\left.\rho_{s_0}\right\vert_{\mathrm{frag}} = \rho_{s_0}\vert_{\mathrm{osc}} \ee^{-4(N_{\mathrm{frag}}-N_{\mathrm{osc}})}\simeq \left.\rho_{s_0}\right\vert_\uend \ee^{-4(N_{\mathrm{frag}}-N_{\mathrm{osc}})}$, one can further obtain
\bea
\rho_{s_0}\vert_{\mathrm{frag}} = \frac{\lambda_s}{4}  s_\uend^4 \left(\frac{\alpha \lambda_s}{\sqrt{3}}\right)^{\frac{8}{1+3\bar{w}_{\mathrm{osc}}}}\, .
\label{eq:rho_s0_frag}
\eea
Finally, let us note that at the time of fragmentation, we have assumed the singlet scalar potential to be still approximated as quartic. This means that $3\lambda_s \left.s_{0}^2\right\vert_{\mathrm{frag}} \gg m_s^2$, \ie $\rho_{s_0}\vert_{\mathrm{frag}}\gg (m_s^2/6) \sqrt{\left( \rho_{s_0}\vert_{\mathrm{frag}}\right) /\lambda_s}$, which implies the following consistency relation 
\bea \label{eq:quartic_frag_consist}
m_s \ll \sqrt{3\lambda_s}  s_\uend \left(\frac{\alpha\lambda_s}{\sqrt{3}}\right)^{\frac{2}{1+3\bar{w}_{\mathrm{osc}}}} = \left[ \frac{27\lambda_s}{2\pi^2}\frac{\Gamma^2\left( \frac{3}{4}\right)}{\Gamma^2\left( \frac{1}{4}\right)} \right]^{\frac{1}{4}}\left( \frac{\alpha \lambda_s}{\sqrt{3}}\right)^{\frac{2}{1+3\bar{w}_{\rm osc}}}H_\uend \, \mu_{\rm inf}^{-3} \, .
\eea
In the second equality, we have used \Eq{eq:corrected-variance}. As we will see below, this condition is in fact always satisfied if another condition, derived in \Eq{eq:consistency-frag-before-nrel}, is verified.

Moving on to the fragmentation products, the $s$ particles are created with a typical 3-momentum $p_s\simeq\sqrt{3\lambda_s} \left. s_0\right\vert_{\mathrm{frag}}$ \cite{Kainulainen:2016vzv}, which redshifts as the inverse of the scale factor, so that
\begin{align}
p_s &= \sqrt{3\lambda_s} \left(\frac{4\rho_{s_0}\vert_{\mathrm{frag}}}{\lambda_s}\right)^{\frac{1}{4}} \exp\left[-\left(N-N_{\mathrm{frag}}\right)\right] \nonumber \\
& = \sqrt{3\lambda_s} s_\uend \left( \frac{\alpha \lambda_s}{\sqrt{3}}\right)^{\frac{2}{1+3\bar{w}_{\mathrm{osc}}}} \exp\left[-\left(N-N_{\mathrm{frag}}\right)\right]\, ,
\end{align}
where in the second equality \Eq{eq:rho_s0_frag} has been used. When the energy becomes of order the mass $m_s$ of the particles, they stop being relativistic. This happens at the time $N_{\mathrm{nrel}}$ at which $E_s = \sqrt{m_s^2+p_s^2}\simeq m_s$ (or, roughly equivalently, when $p_s \simeq m_s$), which yields
\bea
N_{\rm nrel}- N_{\mathrm{frag}} \simeq \ln\left[  \frac{\sqrt{3\lambda_s} s_\uend}{m_s} \left( \frac{\alpha \lambda_s}{\sqrt{3}}\right)^{\frac{2}{1+3\bar{w}_{\mathrm{osc}}}}\right]\, .
\label{eq:Nms}
\eea
Requiring that $N_{\rm nrel}- N_{\mathrm{frag}}$ is positive, one finds another consistency relation, namely
\bea \label{eq:consistency-frag-before-nrel}
\frac{s_\uend }{m_s}> 3^{\frac{1-3\bar{w}_{\rm osc}}{2(1+3\bar{w}_{\rm osc})}}\alpha^{-\frac{2}{1+3\bar{w}_{\mathrm{osc}}}} \lambda_s^{-\frac{5+3\bar{w}_{\mathrm{osc}}}{2(1+3\bar{w}_{\mathrm{osc}})}}\, .
\eea
In practice, one can show that if this condition is satisfied, \Eq{eq:quartic_frag_consist} is always satisfied too. Hence, \Eq{eq:consistency-frag-before-nrel} guarantees that both consistency relations are verified, and corresponds to the grey region labeled ``no fragmentation'' in \Figs{fig:H_*results}, \ref{fig:H_*results_mu_inf}, \ref{fig:H_*results_mu_reh} and \ref{fig:H_*results_mu_low}.

During this epoch, the energy density of the $s$ particles decays as the one of radiation, so one has
\begin{align}
\rho_{s}\vert_{\rm nrel}  =  \rho_{s_0}\vert_{\mathrm{frag}} \exp\left[-4\left(N_{\rm nrel}-N_{\mathrm{frag}}\right)\right] = \frac{m_s^4}{36\lambda_s}\, ,
\end{align}
where in the second equality, we have combined \Eqs{eq:rho_s0_frag} and~(\ref{eq:Nms}). Let us also notice that since the Universe must have reheated before fragmentation in order for the result~(\ref{gamma}) to apply, at the fragmentation time it is radiation-dominated so one has $H_{\rm nrel} = H_{\mathrm{frag}} \exp[-2(N_{\rm nrel}-N_{\mathrm{frag}})]$, which gives rise to
\bea
H_{\rm nrel} = \frac{m_s^2}{\sqrt{3\lambda_s} s_\uend }\left( \frac{\alpha \lambda_s}{\sqrt{3}}\right)^{\frac{3\bar{w}_{\mathrm{osc}}-1}{3\bar{w}_{\mathrm{osc}}+1}} \,.
\label{eq:Hms}
\eea

Finally, when the particles are non-relativistic and their energy density decays as matter we can scale this up to the value it would take today, given by
\begin{align}
\rho_{s}\vert_{\rm today} = \rho_{s}\vert_{\rm nrel}\exp\left[-3\left(N_{\rm today}-N_{\rm nrel}\right)\right] &= \frac{m_s^4}{36\lambda_s} \left(\frac{a_{\rm nrel}}{a_{\rm today}}\right)^3 \nonumber \\
&= \frac{m_s^4}{36\lambda_s} \left(\frac{\left.\tilde{\rho}_\gamma\right\vert_{\mathrm{today}}}{\rho_{\rm nrel}}\right)^\frac34 \,.
\end{align}
In this expression, $\left.\tilde{\rho}_\gamma\right\vert_{\mathrm{today}}$ stands for the energy density of radiation today rescaled by the number of relativistic degrees of freedom, and $\rho_{\rm nrel}$ is the energy density of the Universe at the time when the $s$ particles became non-relativistic. This is because, as stated above, reheating must have occurred before fragmentation for consistency. Using the Friedmann equation, this gives rise to
\begin{equation}
\rho_{s}\vert_{\rm today} = \frac{m_s^4}{36\lambda_s} \left(\frac{\sqrt{\Omega_\gamma} H_{\rm today}}{H_{\rm nrel}}\right)^\frac32\, ,
\end{equation}
from which one obtains
\bea
\Omega^{(s_0)}_{\rm DM} = \frac{\rho_{s}\vert_{\rm today}}{\rho_{\rm today}} = \frac{\rho_{s}\vert_{\rm today}}{3\Mp^2 H_{\rm today}^2} = \frac{m_s^4 \Omega_\gamma^{\frac34}}{108\lambda_s\Mp^2 H_{\rm today}^2} \left(\frac{ H_{\rm today}}{H_{\rm nrel}}\right)^\frac32 \,.
\eea
By using \Eq{eq:Hms}, one finally has
\bea
\frac{\Omega^{(s_0)}_{\rm DM}h^2_{100}}{0.12} &=  0.642\, \Omega_\gamma^{\frac34}h_{100}^\frac32 \lambda_s^{-\frac14} \frac{m_s}{\rm GeV}\left(\frac{s_*}{10^{11}{\rm GeV}}\right)^{\frac{3}{2}}\left(\frac{\alpha \lambda_s}{\sqrt{3}}\right)^{\frac32 \frac{1-3\bar{w}_{\mathrm{osc}}}{1+3\bar{w}_{\mathrm{osc}}}}\, .
\eea
By comparing this expression with \Eq{eq:mu_reh_obtain_1}, one obtains the value for $\mu_{\mathrm{reh}}$ given in \Eq{eq:mu_reh}.

\section{\textsf{Calculation of the portal coupling}}
\label{relating_lambdahs_to_lm}

In all of the scenarios presented in this chapter, we require that the $s$ particles fully constitute the DM. Through this constraint, we demonstrate here that the value of the portal coupling $\lambda_{hs}$ can be determined directly from the value of the self-interaction strength $\lambda_s$ and mass $m_s$ of the scalar field.

The time at which the dark freeze-out happens in the usual units of $x\equiv m_{s}/T_s$ is \cite{Carlson:1992fn}
\begin{equation}
x_{\rm DM}^{\rm (fo)} = \frac{m_{s}}{3.6 {\rm eV}\,\Omega^{\rm total}_{\rm DM}h_{100}^2}\frac{{\cal S}^{\rm hid}}{{\cal S}} = 2.3\times 10^9\left(\frac{m_{s}}{{\rm GeV}}\right)\frac{{\cal S}^{\rm hid}}{{\cal S}} \,,
\end{equation}
where one can compute the ratio between the entropy density of the hidden sector and that of the SM degrees of freedom, ${\cal S}^{\rm hid}/{\cal S}$, once the scalars have reached chemical equilibrium within the singlet sector, as
\begin{align}
\frac{{\cal S}^{\rm hid}}{{\rm S}} &= \frac{g^{\rm hid}_{*{\cal S}}}{g_{*{\cal S}}} \left( \frac{T_s}{T}\right)^3 = \frac{g^{\rm hid}_{*{\cal S}}(m_s)}{g_{*{\cal S}}(m_s)} \left[ \frac{g_*(m_s)}{g_*^{\rm hid}(m_s)}\frac{\rho_s(m_s)}{\rho (m_s)}\right]^{\frac{3}{4}} 
\simeq 5.3 \times 10^{8}\lambda_{hs}^{\frac{3}{2}} \, .
\end{align}
To derive this expression, we have used that
\begin{align} 
\frac{\rho_s(m_s)}{\rho_{\rm tot} (m_s)} = \frac{\rho_s(m_s)}{3H^2(m_s)\Mp^2} 
&\simeq \frac{m_sn_s(m_h)}{3H^{2}(m_h)\Mp^2 } \frac{a(m_s)}{a(m_h)}  \nonumber \\
&\simeq \frac{m_sn_h^{\rm eq}(m_h)}{H^{2}(m_h)\Mp^2}  \frac{\lambda_{hs}^2v^2}{32\pi m_h} \frac{g_{*S}^{\frac{1}{3}}(m_h)}{g_{*S}^{\frac{1}{3}}(m_s)}\frac{m_h}{m_s} \nonumber \\
&\simeq \frac{\ee^{-1}\Mp}{m_h^{3}\pi^3g_*^{\frac{1}{2}}(m_h)} \left( \frac{45}{ \pi }\right)^{\frac{3}{2}}\left( \frac{\lambda_{hs}^2v^2}{32\pi }\right) \frac{g_{*S}^{\frac{1}{3}}(m_h)}{g_{*S}^{\frac{1}{3}}(m_s)} \nonumber \\
&\simeq 9.4\times 10^{11}\lambda_{hs}^2 \,,
\end{align}
where $v=246$ GeV is the vacuum expectation value of the Higgs field and where we take $g_{*}^{\rm hid}=g_{*{\cal S}}^{\rm hid}=1$. Thus, the time of the dark freeze-out is
\begin{equation}
\label{xDFO}
x_{\rm DM}^{\rm (fo)} \simeq 1.2\times 10^{18}\lambda_{hs}^{\frac{3}{2}}\left(\frac{m_{s}}{{\rm GeV}}\right)\,.
\end{equation}

On the other hand, the dark freeze-out temperature can be estimated as the temperature at which the $4\to 2$ interaction rate drops below the Hubble rate \cite{Heikinheimo:2016yds}
\begin{equation}
\label{xDFO2}
x_{\rm DM}^{\rm (fo)} = \frac{1}{3}{\rm ln}\left[\frac{\xi^2\lambda_{s}^4\Mp}{6.5\times 10^3\sqrt{g_*}m_{s}\left(x_{\rm DM}^{\rm (fo)}\right)^{\frac{5}{2}}} \right] \,,
\end{equation}
where $\xi\equiv [g_*(m_{s})\rho_s(m_{s})/\rho(m_{s})]^{1/4}$. Equating \Eq{xDFO2} with \Eq{xDFO} and requiring $\Omega^{\rm total}_{\rm DM}h_{100}^2=0.12$ then yields a relation between the model parameters $\lambda_{hs}$, $\lambda_{s}$, $m_{s}$ and allows one to fix $\lambda_{hs}$ in terms of the other two parameters. The value we find is
\begin{equation}
\label{eq:lambdahs:ms:lambdas}
\lambda_{hs} \simeq 6.3 \times 10^{-13}\left( \frac{m_s}{{\rm GeV}}\right)^{-\frac{2}{3}} W_0^{\frac{2}{3}} \left[7.1\times 10^{4} \lambda_s^{\frac{24}{11}}\left( \frac{m_s}{{\rm GeV}}\right)^{-\frac{10}{11}} \right]  \,,
\end{equation}
where $W_0$ is the 0-branch of the Lambert W function. When plugging this expression into \Eq{mulow:1}, one obtains \Eq{mulow:2}.

\section{\textsf{Statistical computation}} \label{sec:statistical-comp-importancesamp}

In the models presented in this chapter, the total power of the primordial density perturbations constitutes an additional free parameter, which we have omitted because it affects both models equally. For numerical purposes, we use a log-uniform prior which comfortably contains the posterior observed by Planck for this parameter. Thus, the total parameter space sampled is $(\Gamma_\phi, \Gamma_\sigma, m_\sigma, \sigma_\mathrm{end}, A_\mathrm{s})$, and our posteriors and evidences are conditioned to the model producing \emph{close to} the right amount of power.

In the quartic inflaton case, the radiation-like reheating of the inflaton, described as $\Gamma_\phi=H_\uend$, is imposed via a half log-normal $\log_{10}(\Gamma_\phi) \sim \normal_{1/2}\left[ \log_{10}(H_\mathrm{end}),(1/2)^2\right]$. This needs to be done for numerical purposes, since $H_\mathrm{end}$ is a derived quantity that depends of the full parameter combination and can only be computed a posteriori.

We ensure the correct normalisation of the evidences by dividing the marginal likelihood by the total prior mass in the same parameter domain, obtained with a quick \texttt{MultiNest} integration of a mock unit likelihood. All results are obtained with 1000 \emph{live points} and a very low \emph{sampling efficiency} of 0.01 (i.e.\ inverse of ellipsoid enlargement factor). A significant enlargement of the ellipsoids is needed to properly account for the hard edges of the prior and the fact that in the quartic case the mode of the spectator field value is located at the edge of the prior (otherwise if a mode at the edge of the prior is partially or totally missed by the initial sample of live points, the final evidence will be undervalued). This low efficiency produces a lot of rejected points that spoil the computation of the weights used by the Importance Nested Sampling estimator \cite{2013arXiv1306.2144F}, what makes it numerically unstable, most often severely undervalued. Thus, we use the standard nested sampled estimator in this chapter.

\end{subappendices}
\chapter{\textsf{The probable future}}
\label{sec:future-prospects} \niceline {\vskip+1ex} 

\begin{center}
\fbox{\parbox[c]{13cm}{\vspace{1mm}{\textsf{\textbf{Abstract.}}} In this chapter we shall consider a probabilistic perspective on the future of inflationary model building and selection. Building from \Chap{sec:statistical-intro}, we develop a new formalism to forecast the performance of an astronomical survey~\cite{Hardwick:2018zry} with respect to its expected information gain, capability to measure parameters and decisiveness in a space of pre-determined models. We also introduce a new computational forecasting code, \href{https://sites.google.com/view/foxicode}{\texttt{foxi}}, which is based on our formalism and can be readily applied to other experimental design problems.\vspace{1mm}}}
\end{center}

\section{\textsf{Introduction}}
\label{sec:intro-future-prospects}

We begin this chapter with a brief review of inflationary model selection using data from the CMB. The recent \emph{Planck} collaboration results~\cite{Ade:2013sjv,Adam:2015rua,Ade:2015lrj} marked a significant milestone. In the case of single-field models, the decreased upper bound on the tensor-to-scalar ratio combined with a red-tilted spectral index lead the analysis to mostly favour inflationary potentials with a plateau~\cite{Easther:2011yq,Planck:2013jfk,Martin:2013nzq,Ade:2015lrj,Vennin:2015eaa}. Additionally, multi-field inflation has also recently begun to be rigorously statistically analysed, e.g. in the context of curvaton models~\cite{Hardwick:2015tma,Vennin:2015egh,dePutter:2016trg}.

Despite the significant reduction in the number of observationally viable models, it has become abundantly clear that there are still quite a number of models that satisfy the \emph{Planck} constraints, especially those classed in the plateau category of potential. This dissatisfying state of affairs is only mitigated by the potential for other future surveys to augment the current constraints such as CMB Stage-4~\cite{Abazajian:2016yjj}, LiteBIRD~\cite{Matsumura:2013aja} and COrE~\cite{DiValentino:2016foa,Finelli:2016cyd}. Despite the promise of further observations, the future of inflationary model selection is still tremendously unclear. In the face of an uncertain future, we seek to answer the following question: To what extent can one be certain of a future survey being capable of deciding between models, or within the space of many models? The answer is probabilistic and clearly dependent not only on the particular model choice, but also on the current constraints made by the \emph{Planck} collaboration. Since a decision must be made, the natural framework to answer this question uses Bayesian probability.

\begin{figure}
\begin{center}
\includegraphics[width=12cm]{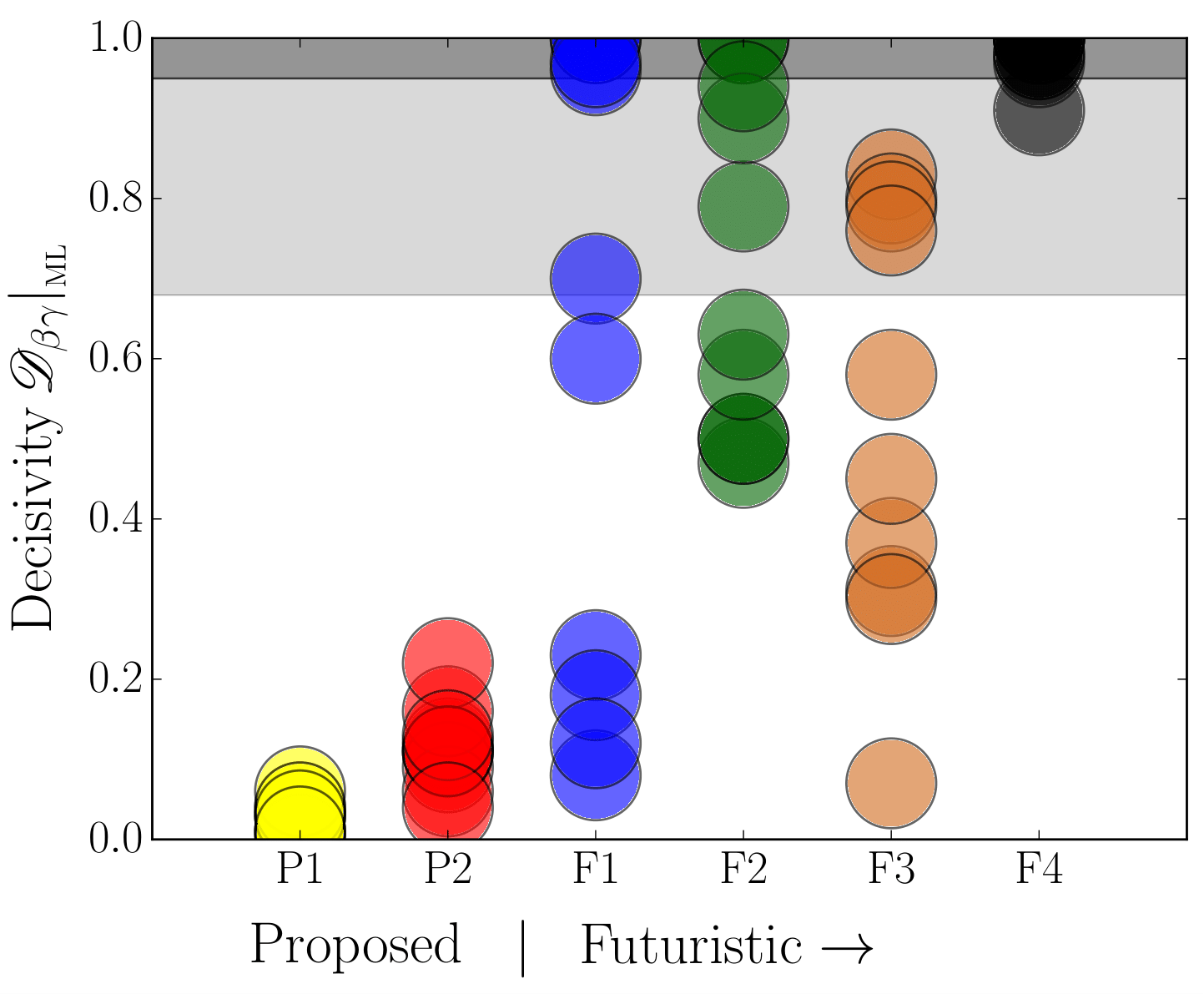}
\caption[Decisivity score for a selection of CMB experiments]{~\label{fig:allmodels_deci} A scatter plot of each model pair score in the decisivity utility $\deci_{\beta \gamma}\vert_{{}_{\rm ML}}$ (computed using the maximum-likelihood average, see \Eq{eq:ml-average}) using the Bayes factors of each of the possible pairs of models for each futuristic survey, and the $5$ representative single-field models used in this section. We have assumed a logarithmic prior over $\epsilon_1$ (\Eq{eq:predictive-priors-logeps1}) and a flat prior over $(\epsilon_2,\epsilon_3)$ --- see also the discussion in \Sec{sec:formalism}. The light and dark grey rectangles correspond to $\deci_{\beta \gamma}\vert_{{}_{\rm ML}}=0.68$ and $0.95$ i.e. to situations where the probability to rule out one model against the other is $68\%$ and $95\%$, respectively. The colours and labels on the horizontal axis correspond to the measurement configurations of Table \ref{tab:future-widths}. }
\end{center}
\end{figure}

It seems clear that there are many interesting unanswered questions one can pose relating to the predictive probabilities of future survey performance. In this section, we will restrict ourselves to focus on using a futuristic set of measurement 1-$\sigma$ error bars to compute our defined expected utilities for model distinguishability. Therefore, the specific question we pose for this chapter is as follows: \emph{How much more do we stand to learn about single-field inflationary models given a forecast set of future measurement 1-$\sigma$ error bars over the slow-roll parameters?} To this end, we set up six classes of survey over the space of slow-roll parameters $(\epsilon_1 , \epsilon_2, \epsilon_3 )$, defined in \Eq{eq:slowrollp}, where the corresponding choices of measurement 1-$\sigma$ error bars (for $i=1,2,3$) of each fictitious experiment are defined in Table \ref{tab:future-widths}. Our expectation will be a clear trend between decreasing measurement error bars and an improvement in the score from our utility functions, e.g. as can be seen from \Fig{fig:allmodels_deci}, where we have plotted the quantity $\deci_{\beta \gamma} \vert_{{}_{\rm ML}}$ --- defined as a score of decisive merit between models in later chapters --- against our mock surveys.

We acknowledge that the broad question we seek to answer in this chapter has been approached, to some degree, at various angles by Refs.~\cite{Martin:2014rqa,Hardwick:2015tma,Finelli:2016cyd} (though no work yet appears to apply this to CMB experiments and models of inflation). In each case, the authors target a slightly different problem with specific surveys in mind. Further to this, we note that some of the quantities we will later define (such as $\deci$) have already been introduced in similar works for Dark Energy models~\cite{Trotta:2010ug}, likelihood parameter inference for \emph{Planck}~\cite{Trotta:2007hy} and to classify the cosmic web in~\cite{Leclercq:2016fyj} --- yet the formalism will be extended and improved in this section to properly quantify the ability of future surveys to distinguish between models of inflation.
\begin{table}
\centering
\resizebox{12cm}{!}{
\begin{tabular}{|l*{7}{|c}|}\hline
\multicolumn{2}{|c}{Reference} & \multicolumn{3}{|c|}{Measurements} & \multicolumn{2}{c|}{$\langle \dkl \rangle$ }   \\\hline
Name & Colour &$\sigma^1 (\epsilon_1 )$ & $\sigma^2 (\epsilon_2 )$ & $\sigma^3 (\epsilon_3 )$  & $\pi (\bmuf \vert \, \epsilon_1)$ & $\pi (\bmuf \vert \, \log \epsilon_1)$
\\\hline\hline
\cellcolor{yellow!55} & Proposed 1 (P1) & $10^{-3}$ & $10^{-2}$ & $10^{-1}$ & 5.6 $\pm$ 0.3 & 0.6 $\pm$ 1.1 \\\hline
\cellcolor{red!55} & Proposed 2 (P2) & $10^{-4}$ & $10^{-2}$ & $10^{-2}$ & 9.9 $\pm$ 0.5 & 2.1 $\pm$ 2.1 \\\hline
\cellcolor{blue!35} & Futuristic 1 (F1) & $10^{-5}$ & $10^{-2}$ & $10^{-2}$ & $>11.4$ & 2.5 $\pm$ 2.5 \\\hline
\cellcolor{green!55} & Futuristic 2 (F2) & $10^{-4}$ & $10^{-3}$ & $10^{-2}$ &  $>11.4$ & 3.5 $\pm$ 2.7 \\\hline
\cellcolor{brown!90} & Futuristic 3 (F3) & $10^{-4}$ & $10^{-2}$ & $10^{-3}$ &  $>11.4$ & 4.1 $\pm$ 2.2 \\\hline
\cellcolor{black!65} & Futuristic 4 (F4) & $10^{-5}$ & $10^{-3}$ & $10^{-3}$ &  $>11.4$ & 5.7 $\pm$ 3.0 \\\hline
\end{tabular}}
\caption[Measurement configurations for each toy CMB experiment]{~\label{tab:future-widths} Measurement accuracy (in terms of the $1$-$\sigma$ error bars on the first three slow-roll parameters) and expected Kullback-Leibler divergence (information gain) between the prior and posterior distributions over the slow-roll parameters for the future toy surveys studied in this section. The first two are set with similar characteristics to potential surveys in the near future and are denoted P1 and P2 (CMB Stage-4 and COrE/LiteBIRD, respectively, where `P' stands for `Proposed'). In addition, we have exceeded these forecasts with our Futuristic categories 1-4 (F1-4) to indicate various (possibly absolute) limits. We direct the reader to \Sec{sec:formalism} for the discussion that motivates the $\epsilon_1$ flat ($\pi (\bmuf \vert \, \epsilon_1 )$) and the $\epsilon_1$ logarithmic ($\pi (\bmuf \vert \log \epsilon_1)$) priors. The $\langle \dkl\rangle > 11.4$ values using a flat prior over $\epsilon_1$ exceed a numerical threshold associated to the integral computation of \Eq{eq:calcDKL}.}
\end{table}

In this chapter we will outline a simple method to compute any expected utility for a future survey given a previous set of measurements on the same variables from an independent survey (which, in our case, shall always be the \emph{Planck} 2015 constraints). In \Sec{sec:formalism} we outline in detail our definition of the utility functions to be used throughout this section, as well as introducing some new methods of computation --- including our outline of the new \href{https://sites.google.com/view/foxicode}{\texttt{foxi}} algorithm.

The \href{https://sites.google.com/view/foxicode}{\texttt{foxi}} (Futuristic Observations and their eXpected Information) package is a general-purpose, publicly available, python class for use on any forecasting problem. It outputs \LaTeX \, compile-able tables and has a variety of plotting options. One can fork the code and other details through the website: \href{https://sites.google.com/view/foxicode}{https://sites.google.com/view/foxicode}. We have also included some robustness checks and a brief summary of the computational methods used by the algorithm in Appendix~\ref{sec:foxi-computation}.

\begin{figure}
\begin{center}
\includegraphics[width=12cm]{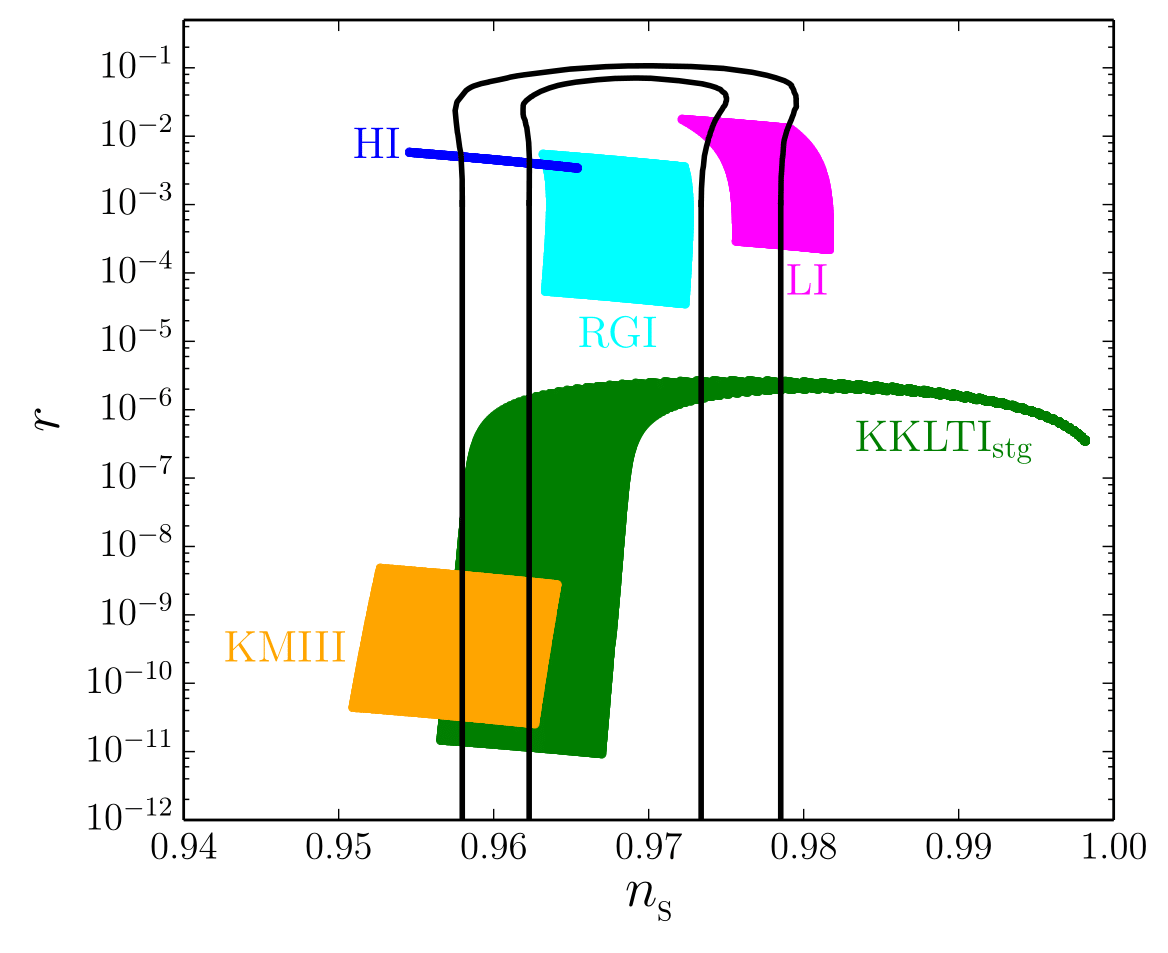}
\caption[Spectral index against tensor-to-scalar ratio]{~\label{fig:nsr-plot} An $(\nS , r)$-plot of the available parameter space to each of the models used in this section, where the solid black contours are the 68\% and 95\% limits currently imposed by the \emph{Planck} 2015 data~\cite{Ade:2015lrj}. $\nS$ on the horizontal axis is the scalar spectral index and $r$ on the vertical axis is the tensor-to-scalar ratio. }
\end{center}
\end{figure}

Since literally hundreds of single-field models have been proposed in the literature~\cite{Martin:2013tda}, including all of them in our analysis would be numerically too expensive. In order to infer results that are representative of the full model set one must therefore choose a variety of models that fill e.g. the $(\nS ,r)$ diagram using their calculated $\nS$ and $r$ values from \Eq{eq:spectral-ind} and \Eq{eq:tens-scalar-ratio}. In Appendix~\ref{sec:models} we list the 5 representative single-field models --- employed in the \href{http://cp3.irmp.ucl.ac.be/~ringeval/aspic.html}{\texttt{ASPIC}} library~\cite{Martin:2013tda,aspic}: Higgs Inflation (HI), K\"{a}hler Moduli Inflation II (KMIII), Kachru-Kallosh-Linde-Trivedi Inflation (${\rm KKLTI}_{\rm stg}$), Loop Inflation (${\rm LI}_{\alpha >0}$) and Radion Gauge Inflation (RGI) --- that we have chosen, neglecting many reasonable alternatives for the sake of brevity and capturing the essential information about the competition between models. Though no favouritism for these 5 is intended in this chapter,\footnote{\href{https://sites.google.com/view/foxicode}{\texttt{foxi}} copes relatively well with the inclusion of many models, though the number of model pairs to analyse scales with the Binomial coefficient $\frac{N!}{(N-2)!2!}$, where $N$ is the number of models. Already with $N=5$, we note that $10$ model pairs must be considered.} as they are merely representative of the explored parameter space shown by our representation of each prior volume over the $(\nS , r)$-plane in \Fig{fig:nsr-plot}, we nonetheless have provided very brief introduction for each (which includes both their potentials and priors on their parameters) in Appendix~\ref{sec:models}.

Our results can be found in \Sec{sec:results}, where we employ a comprehensive suite of expected utilities to analyse the future of model selection for inflation.
We have additionally included a small section (\Sec{sec:reheating}) on the interesting possibility of using our framework to examine the future prospects of inferring the reheating temperature in the example of the HI model as well as a computation of the probability in the future that each of the various survey configurations will be able to exceed a $2$-$\sigma$ detection of the running of the scalar spectral index $\alphaS$ in \Sec{sec:measuring-alphaS} (with a preliminary calculation in Appendix~\ref{sec:alphaS-calculation}). Both of these short examples are intended to give an impression of the possible scope of usage for our code \href{https://sites.google.com/view/foxicode}{\texttt{foxi}} with a model-focused question in mind. Finally, in \Sec{sec:concl} we present our conclusions.

\section{\textsf{Formalism}}
\label{sec:formalism}
\subsection{\textsf{Probability measures primer}}
Due to the fact that all of the models of inflation considered here are slow-roll models, there exists a general parameterisation of the power spectrum (which we observe) that includes $n$ slow-roll parameters ${\cal P}_{\zeta}={\cal P}_{\zeta} (\epsilon_1,\epsilon_2,\epsilon_3,\dots , \epsilon_n )$ that is sufficient to constrain their observational characteristics once the amplitude has been measured and fixed. The precise relationship between ${\cal P}_\zeta$ and single-field models of inflation is discussed in more detail in \Sec{sec:sourcing-cos-pert}. The current data, using \emph{Planck} CMB measurements~\cite{Ade:2013sjv,Adam:2015rua,Ade:2015lrj}, limits our capabilities to constrain up to essentially $n=3$ slow-roll parameters~\cite{Ringeval:2013lea,Ade:2013sjv,Adam:2015rua,Ade:2015lrj}. Even though future surveys may in principle be able to constrain parameters further up the slow-roll hierarchy, e.g. $\epsilon_4$, they will first need to constrain $\epsilon_3$ at the level that is consistent with slow roll, which we find to be difficult even for the most futuristic of our toy surveys considered here (see \Sec{sec:measuring-alphaS}). Hence, though all of the formalism in this section can be applied to any $n$-dimensional parameter spaces, we shall consider here only the space of slow-roll parameters $(\epsilon_1,\epsilon_2,\epsilon_3)$ as a first example. This space will subsequently be equipped with three distinct probability measures.

\subsection*{\textsf{The posterior given the current data}}
Hereafter, the fiducial point vector $\bmuf$ spans the real $n$-dimensional parameter space of central points for future measurements. This, naturally, has a probability measure associated to it which is derived from the current observations over each separate direction in the space. We can therefore define the integral measure over the domain of $\bmuf$ (such as will be used in \Eq{eq:exu}) as the posterior distribution of current data $p \, (\bmuf \vert {\cal D}_{\rm cur} ) \, \dd \bmuf$. There is a subtlety in obtaining $p \,(\bmuf |{\cal D}_{\rm cur} )$, that is revealed through Bayes' rule
\begin{equation} \label{eq:post-pred}
\ p \,(\bmuf |{\cal D}_{\rm cur} ) \propto   \pi_{{}_{\cal I}}  (\bmuf ) \, \like\,({\cal D}_{\rm cur} | \bmuf ) \,,
\end{equation}
which includes the prior information $\pi_{{}_{\cal I}}  (\bmuf )$ over the space of $\bmuf$, the former containing some initial information ${\cal I}$ about the sampling space.

Based on the principles outlined in \Sec{sec:priors-intro} and within the specific choice of parameterisation $(\epsilon_1,\epsilon_2,\epsilon_3)$, throughout this chapter we will make two choices of prior where $\muf^2 \in [0,0.09] \,,   \,\, \muf^3 \in [-0.2,0.2]$ and
\begin{align} \label{eq:predictive-priors-flateps1}
\pi_{\epsilon_1} \, (\bmuf ) \propto {\rm const.} \,, \quad &{\rm where} \quad \muf^1 \in [10^{-4},10^{-2}] \,, \\
\pi_{\log \epsilon_1} \, (\bmuf ) \propto \frac{1}{\muf^1} \,, \,\,\, \quad &{\rm where} \quad \log (\muf^1) \in [-13,-1] \,, \label{eq:predictive-priors-logeps1}
\end{align}
corresponding to either flat, or, flat in all dimensions except a log prior over the component $\muf^1$ i.e. the first slow-roll parameter $\epsilon_1$, respectively. By setting the hard prior limits in \Eq{eq:predictive-priors-logeps1}, we have artificially chosen the lower bound on $\epsilon_1 = 10^{-13}$, which seems reasonable when none of the models we study here are capable of lower values than this and, in the absence of an absolute lower fundamental limit\footnote{
We restrict $\epsilon_1\geq 10^{-13}$, otherwise we would need to include second-order effects in perturbation theory~\cite{Martineau:2007dj}. In addition, this lower bound encompasses the predictions from all of our chosen model priors.} on $r$, that limit is also placed so as to not overweight too much of the prior volume on very low values which will likely never be detectable. The upper limit on $\epsilon_1$ and the bounds on both $\epsilon_2$ and $\epsilon_3$ are set by slow-roll consistency.

To give an indication of the volume of permitted $\bmuf$ points used in this section, the $\pi_{\log \epsilon_1} \, (\bmuf )$ prior has been used in \Fig{fig:nsr-plot} to display the 68\% and 95\% contour limits (in solid black) for the current \emph{Planck} 2015 posterior marginalised over the $(\nS , r)$-plane. 

\subsection*{\textsf{The prior from each model}}

We define $\boldx$ as a real $n$-dimensional vector over the same observables represented by $\bmuf$ (hence, for this section, over $(\epsilon_1,\epsilon_2,\epsilon_3)$). To generate a model prior $\bar{\pi}$ over $\boldx$ one simply varies the parameters that are specific to the model (e.g. parameters in the inflationary potential --- see Appendix~\ref{sec:models}) over their priors and computes the distribution over the $\boldx$ domain that this generates.

Distributions denoted with a bar --- such as $\barpi$, $\barp$ and $\barlike$ --- are defined over each individual model observable value $\boldx$, with measure $\barpi (\boldx \vert {\cal M}_\alpha ) \, \dd \boldx$ and are typically twice integrated in order to compute the expected utility: once over the $\boldx$ space and the second time over the space of $\bmuf$ so as to take into account the uncertainty in the values that a future measurement may be centred on.

\subsection*{\textsf{The posterior given the future data}}
Finally, we shall also consider the likelihood (defined with $\bmuf$ and $\boldsigma$) and posterior probability from a future survey, with measure $\hatp  \left[ \, \boldy \, \vert \, {\cal D}_{\rm fut}(\bmuf ,\boldsigma )  \, \right] \dd \boldy$, which is specified over the $\boldy$ (another real $n$-dimensional parameter vector sharing the same space of observables represented by $\bmuf$) domain. The futuristic dataset ${\cal D}_{\rm fut}={\cal D}_{\rm fut}(\bmuf , \boldsigma )$ is centred on $\bmuf$ with a vector of mutually independent forecast error bars $\boldsigma$ which we can specify either `by hand' or through e.g. a Fisher forecasting method, given a specific survey.

All distributions denoted with a hat, such as $\hatpi$, $\hatp$ and $\hatlike$ are defined over $\boldy$. Through Bayes' rule, we can connect the posterior probability distribution given the current data (the same distribution as the one defined over $\bmuf$) to the probability distribution over the future data, once a future likelihood function has been specified
\begin{equation} \label{eq:bayes-rule-future}
\hatp  \left[ \, \boldy \, \vert \, {\cal D}_{\rm fut}(\bmuf ,\boldsigma )  \, \right] \propto  p \, (  \boldy \vert {\cal D}_{\rm cur} ) \, \hatlike \, [\, {\cal D}_{\rm fut}(\bmuf , \boldsigma ) \, \vert \, \boldy \, ] \,.
\end{equation}
Note that this distribution, and hence the points $\boldy$, are independent of the space of models $\boldsymbol{\calM}$ (although, of course, still dependent on an overall underlying cosmological model such as $\Lambda$CDM). Hence, this will be useful for defining model-independent utilities later e.g. the forecast information gain. In this chapter, we shall assume
\begin{equation} \label{eq:future-gaussian-assumption}
\hatlike \, [\, {\cal D}_{\rm fut}(\bmuf , \boldsigma )  \, \vert \, \boldy \, ] = {\cal N} (\boldy \vert \bmuf , \boldsigma ) \,,
\end{equation}
where the multivariate Gaussian distribution here can be defined generally as
\begin{equation}
\ {\cal N} (\boldsymbol{a} \vert \bmuf , \boldsigma ) \equiv (2\pi )^{-\frac{n}{2}} \left( \prod^n_{i=1} \sigma^i \right)^{-1} \exp \left[ - \sum^n_{i=1}\frac{(a^i-\muf^i)^2}{2(\sigma^i)^2} \right] \,,
\end{equation}
and where, crucially, we will be ignoring possible covariances and when a parameter restricted to a positive-only range is used (such as $\epsilon_1$) a half-Gaussian is used. Both this and \Eq{eq:future-gaussian-assumption} will prove to be a \emph{key} assumption of this section. It is clear that forecasting for proposed missions for which the configuration of the detectors and physics of the measurement is well-understood, realistic future likelihoods may be inferred and are probably extremely complex, rendering the Gaussian assumption possibly a poor fit (we check this assumption explicitly in Appendix~\ref{sec:gaussian-assumption-limitations}).

We consider this section to be a new step in developing a set of numerical forecasting tools, in which, the natural first step is to assume a Gaussian ansatz. Furthermore, we have two main reasons to focus initially on \Eq{eq:future-gaussian-assumption}:
\begin{enumerate}
\item{Our Gaussian mock forecasts represent the simplest first approximation to the full calculation where detector noises are carefully translated into error bars over the slow-roll parameters.}
\item{The narrow-variance limit of all possible $\hatlike$ distributions is well-modeled by a Dirac delta measure in $\bmuf$-space, hence the shape of our ansatz for $\hatlike$ becomes irrelevant when this limit is met (we will show that this shape-independence appears for our more futuristic surveys in \Sec{sec:results}). This is an important feature that can also be exploited for more rapid computation (see Appendix~\ref{sec:foxi-computation} for further details).}
\end{enumerate}
Hence, we shall implement \Eq{eq:future-gaussian-assumption} throughout this section. A more detailed discussion of the limitations of the Gaussian assumption is provided in Appendix~\ref{sec:gaussian-assumption-limitations}.

We have now clarified the important distinctions between the probability measures used within this section, so we are ready to introduce our formalism fully.

\subsection{\textsf{Defining the expected utility}}
We discuss the introductory principles of Bayesian experimental design in \Sec{sec:experimental-design-principles}. Building from this section, to correctly manipulate our probability spaces, it is natural to define a utility function $U$ which has a dependence on the target parameters $\boldsigma$ (e.g. parameterisations of the survey geometry, as discussed in~\Ref{Bassett:2004st}). One typically seeks to maximise the expected value of $U$ in achieving a goal e.g. optimising the expected information gain from a survey with a certain configuration. Using the posterior given the current data, we can define the expected utility $\langle U\rangle$ (which can be dependent on the set of indexed models $\boldsymbol{\calM} = \{ \calM_\alpha \}$, for example) as
\begin{equation} \label{eq:exu}
\langle U \rangle = \langle U (\boldsigma ) \rangle \equiv \int_{\bmuf \in \mathbb{R}^n} U\,[\, \boldsymbol{\calM}, {\cal D}_{\rm fut} (\bmuf , \boldsigma ) \, ] \, p \,(\bmuf |{\cal D}_{\rm cur} )\,{\rm d}\bmuf \,,
\end{equation}
and, given an appropriate $U$, its corresponding centred second-moment equivalent
\begin{equation} \label{eq:secondmomentu}
\left\langle \left( U-\langle U\rangle \right)^2\right\rangle \equiv \int_{\bmuf \in \mathbb{R}^n} \bigg \{ \,  U\,[ \, \boldsymbol{\calM}, {\cal D}_{\rm fut} (\bmuf , \boldsigma ) \, \, ] -\langle U\rangle \, \bigg \}^2 \, p \,(\bmuf |{\cal D}_{\rm cur} )\,{\rm d}\bmuf \,,
\end{equation}
where $p \,(\bmuf |{\cal D}_{\rm cur} )$ is defined as the measure of uncertainty in the value that the future measurement is centred on, $\bmuf$, which is conditioned on the current data ${\cal D}_{\rm cur}$ --- which in the present case is the \emph{Planck} data. Computing both \Eq{eq:exu} and \Eq{eq:secondmomentu} above is sufficient to answer all of the questions in this section through appropriate choice of utility $U$.

\begin{figure}
\begin{center}
\includegraphics[width=12cm]{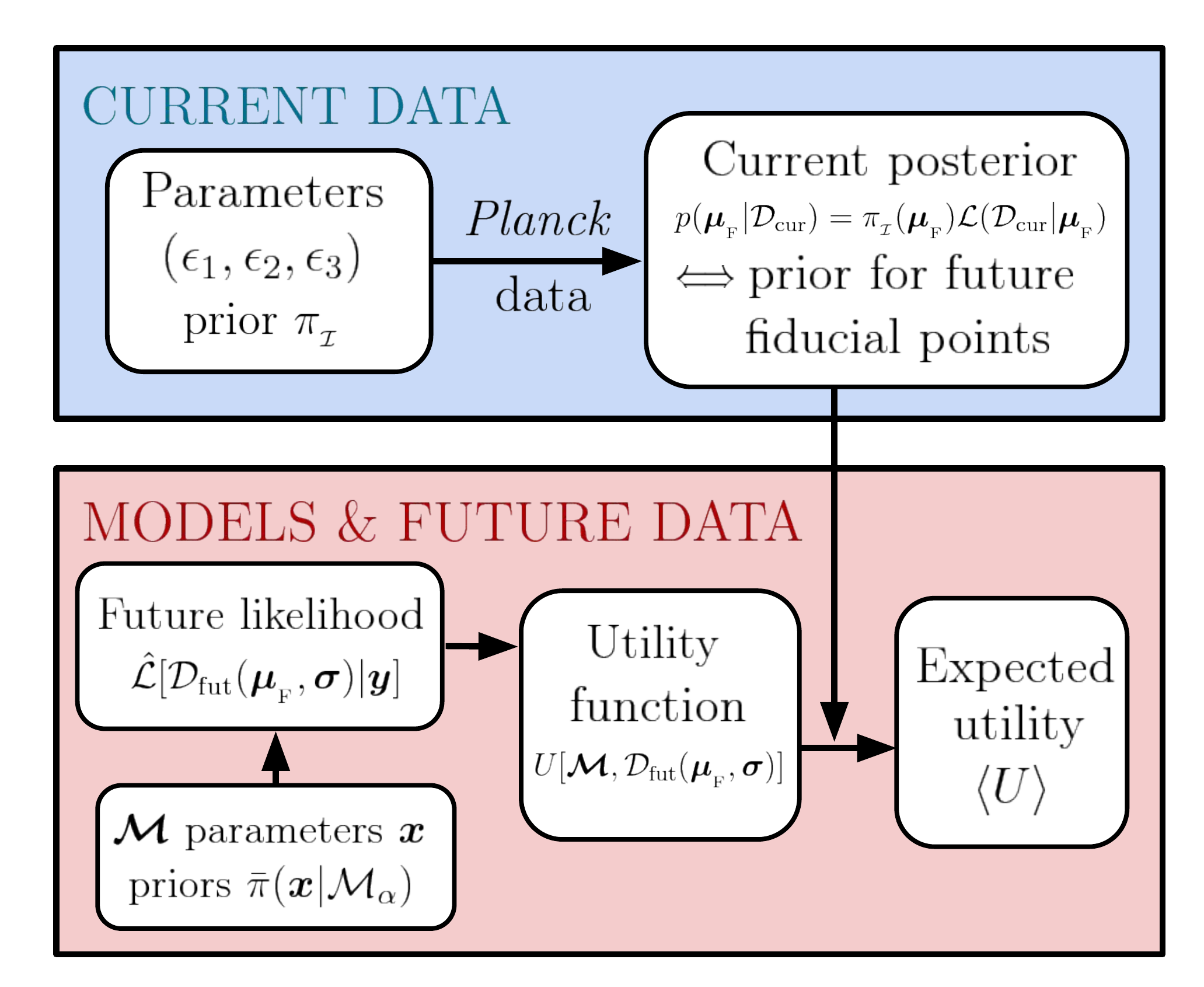}
\caption[Diagram of the Bayesian experimental design setup]{~\label{fig:utility-diag-dec} A schematic diagram of the dependencies implied by the experimental design formalism described in \Sec{sec:formalism}. The top left hand rectangle (within the blue region) is specific to inflation --- with single-field inflationary slow roll parameters $(\epsilon_1,\epsilon_2,\epsilon_3)$ and \emph{Planck} data --- but may be replaced by any current measurement for a given survey design problem.  }
\end{center}
\end{figure}

To clarify the formalism, we have illustrated the procedure defined in this section with \Fig{fig:utility-diag-dec}. We note that the top left hand rectangle (inside the blue region), which represents the input from the {\rm Planck} data~\cite{Ade:2013sjv,Adam:2015rua,Ade:2015lrj}, may in principle be replaced with data from any measurement design problem.

\subsection{\textsf{The utility functions}} \label{sec:utility}
We begin by defining ${\cal E}_{\beta}$ and ${\cal E}_{\gamma}$ which denote the Bayesian evidences for two models ${\cal M}_{\beta}$ and ${\cal M}_{\gamma}$ respectively, given a future survey (and a fiducial cosmology such as $\Lambda$CDM), whose form for $\alpha = \{ \beta , \gamma \}$ is given by adapting \Eq{eq:evidence:def:before} into
\begin{align} \label{eq:evidences}
\ {\cal E}_\alpha (\bmuf , \boldsigma ) & \equiv \int_{\boldx \in \mathbb{R}^n} \hatlike \, [\, {\cal D}_{\rm fut}(\bmuf , \boldsigma )  \, | \, \boldx \, ] \, \barpi \,(\boldx | \calM_\alpha ) \, {\rm d} \boldx \,,
\end{align}
which uses the likelihood function $\hatlike$ from some future dataset ${\cal D}_{\rm fut}$ (assumed to be \Eq{eq:future-gaussian-assumption} in this section) defined over the model point space $\boldx$, centred at $\bmuf$ and multiplied by the prior probability measure $\barpi$ for each model.

The key quantity for model comparison is the Bayes factor ${\rm B}_{\beta \gamma}$ between two models, defined, as in \Sec{sec:bayes-model-selection}, as the ratio of their evidences 
\begin{equation} \label{eq:BayesFactor}
\ {\rm B}_{\beta \gamma} (\bmuf , \boldsigma )  = \frac{\cal E_{\beta} (\bmuf , \boldsigma )}{\cal E_{\gamma} (\bmuf , \boldsigma ) }\,,
\end{equation}
which favours models that realise a good compromise between quality of fit and a lack of fine tuning.\footnote{In this context, the degree of `fine-tuning' corresponds to the degree to which only a narrow region of a given models' possible observable characteristics actually fit the data well.} Thus, one favours $\calM_\beta$ within the set $\boldsymbol{\calM}$ that extremizes ${\rm B}_{\beta \gamma}$ with respect to the others. In \Sec{sec:bayes-model-selection} we also introduced a threshold to rule ${\cal M}_{\beta}$ out with respect to ${\cal M}_{\gamma}$ --- the Jeffreys threshold~\cite{jeffprob,2008arXiv0804.3173R}, where one needs to satisfy ${\cal E}_{\beta}<{\rm e}^{-5}{\cal E}_{\gamma}$. Therefore, in logarithmic terms $\ln {\rm B}_{\beta \gamma} = -5$ marks the point at which $\cal M_{\beta}$ may be considered `strongly disfavoured' versus $\cal M_{\gamma}$.

Consider now the choices of utility
\begin{align} \label{eq:BayesU}
\ U &= \left\vert \ln {\rm B}_{\beta \gamma} \right\vert \,, \\
\ U &= \Theta \left( \left\vert \ln {\rm B}_{\beta \gamma}  \right\vert -5\right) \,,\label{eq:DeciU}
\end{align}
which --- though utilities in \Eq{eq:exu} may be defined generally over the indexed model space $\boldsymbol{\calM} = \left\{ \calM_\alpha\right\}$ --- we have defined individually for each pair of models $\calM_\beta$ and $\calM_\gamma$. Depending on how observationally separable the two models are, computing the expectation value through \Eq{eq:exu} of \Eq{eq:BayesU} may provide a strong indication of the most probable absolute value of the Bayes factor, where the typical spread away from this mean value can be estimated through the centred second-moment in \Eq{eq:secondmomentu}.

Turning our attention to the other utility defined by \Eq{eq:DeciU}, the decisiveness $\deci_{\beta \gamma}$ between $\calM_\beta$ and $\calM_\gamma$, is defined as
\begin{equation}
\deci_{\beta \gamma} \equiv \left\langle \Theta \left( \vert \ln {\rm B}_{\beta \gamma}\vert-5\right) \right\rangle \,, \label{eq:decisiveness}
\end{equation}
and $\deci_{\beta \gamma}=\deci_{\gamma \beta}$, where we note that this quantity has been previously defined in \Ref{Trotta:2010ug}. $\deci_{\beta \gamma}$ incorporates the Jeffreys threshold into the decision between models, where its value is that of a real number selected from the closed interval $[0,1]$ (or the odds of a clear decision). In this way, model pairings with a large decisiveness value will be imminently distinguishable in the future, with the opposite holding true for a low decisiveness value.

Our last, model-independent,\footnote{At least dependent only upon the background cosmology.} utility function is the information gained (in the same space of observables as $\bmuf$ and $\boldx$, e.g. $(\epsilon_1,\epsilon_2,\epsilon_3)$ for our single-field inflation problem) by improving the measurement with error bars $\boldsigma$ at each possible $\bmuf$
\begin{equation} \label{eq:DKLU}
\ U = \dkl \left\{ \, \hatp \left[ \, \boldy \, \vert \, {\cal D}_{\rm fut}(\bmuf ,\boldsigma )  \, \right] \, \right\vert \! \left\vert \, p \, (  \boldy \vert {\cal D}_{\rm cur} ) \, \right\} \,,
\end{equation}
also referred to as the Kullback-Leibler divergence~\cite{kullback1951} between the two distributions, which we define here as
\begin{align}  
& \dkl \left\{ \, \hatp \left[ \, \boldy \, \vert \, {\cal D}_{\rm fut}(\bmuf ,\boldsigma )  \, \right] \, \right\vert \! \left\vert \, p \, (  \boldy \vert {\cal D}_{\rm cur} ) \, \right\} = \nonumber \\
 & \qquad \quad \int_{\boldy \in \mathbb{R}^n} \hatp \left[ \, \boldy \, \vert \, {\cal D}_{\rm fut}(\bmuf ,\boldsigma )  \, \right] \ln \left\{ \frac{\hatp \left[ \, \boldy \, \vert \, {\cal D}_{\rm fut}(\bmuf ,\boldsigma )  \, \right] }{p \, (  \boldy \vert {\cal D}_{\rm cur} )  }\right\} \dd \boldy  \label{eq:calcDKL_general}\,.
\end{align}
By defining the normalisation
\begin{equation}
\ E \equiv \int_{\boldy \in \mathbb{R}^n} p \, (  \boldy \vert {\cal D}_{\rm cur} ) \, \hatlike \, [\, {\cal D}_{\rm fut}(\bmuf , \boldsigma )  \, \vert \, \boldy \, ] \, \dd \boldy \,,
\end{equation}
we can rewrite \Eq{eq:calcDKL_general}, using \Eq{eq:bayes-rule-future} and $E$, as
\begin{align}
\ & \dkl \left\{ \, \hatp \left[ \, \boldy \, \vert \, {\cal D}_{\rm fut}(\bmuf ,\boldsigma )  \, \right] \, \right\vert \! \left\vert \, p \, (  \boldy \vert {\cal D}_{\rm cur} ) \, \right\} =\nonumber \\
 & \,\, \frac{1}{E} \int_{\boldy \in \mathbb{R}^n} p \, (  \boldy \vert {\cal D}_{\rm cur} )  \, \hatlike \, [\, {\cal D}_{\rm fut}(\bmuf , \boldsigma )  \, \vert \, \boldy \, ] \ln \left\{ \frac{ \hatlike \, [\, {\cal D}_{\rm fut}(\bmuf , \boldsigma )  \, \vert \, \boldy \, ] }{E}\right\} \dd \boldy \label{eq:calcDKL}\,.
\end{align}

\subsection{\textsf{The maximum-likelihood average}} \label{sec:utility-expectation-values}
Throughout this section, we will use the notation $\langle \cdot \rangle$ to denote the current-data posterior averaging as in \Eq{eq:exu}. While this is perfectly adequate to obtain expected utilities, in the case of both model-dependent utility functions (defined by \Eq{eq:BayesU} and \Eq{eq:DeciU}), one should also consider averaging over only those $\bmuf$ points that generate future likelihood distributions which do not immediately rule both models out. Indeed, in cases where both models are ruled out, the fact that one model is even more ruled out than the other does not provide valuable information and one may wish to simply discard such situations from forecasts. The removal of such situations restricts the space of future scenarios to those for which a Bayesian model selection is even necessary to conduct.

An averaging scheme that can solve this problem removes the $\bmuf$ points for which the maximum likelihood of both models is too low in comparison to the global maximum likelihood. We will refer to this method hereafter as the `maximum-likelihood averaging' scheme, defined as
\begin{align}
\ \langle \cdot \rangle_{{}_{\rm ML}} \equiv  \frac{1}{1-r_{{}_{\rm ML}}} & \int_{\bmuf \in \mathbb{R}^n} \,\, \cdot \,\, \Theta \left[ \, \max_{i = \beta , \gamma} \left \{ \ln \hatlike \, ({\cal D}_{\rm fut}  |\boldy_* , {\cal M}_i) \right\} + t_{{}_{\rm ML}} - \ln \hatlike ({\cal D}_{\rm fut} |\bmuf ) \right]  \nonumber \\
& \qquad \qquad \qquad \qquad \qquad \times p \, (\bmuf \vert \, {\cal D}_{\rm cur})\, \dd \bmuf \label{eq:ml-average} \,,
\end{align}
where for this section we set $t_{{}_{\rm ML}}=5$ but this threshold value can be arbitrarily defined,\footnote{Hence, we are quite restrictive, permitting only those models for which the maximum likelihood is $\hatlike \, ({\cal D}_{\rm fut}  |\boldy_* , {\cal M}_i) \geq \ee^{-5}\hatlike \, ({\cal D}_{\rm fut}  | \bmuf )$, e.g. within roughly $\sqrt{5}\simeq 2.2$-$\sigma$ of the global maximum likelihood.} we have suppressed the dependence ${\cal D}_{\rm fut} = {\cal D}_{\rm fut} (\bmuf , \boldsigma )$ for brevity and $\boldy_*$ is the maximum likelihood point for a given distribution. Thus, expected utilities generated using $\langle \cdot \rangle_{{}_{\rm ML}}$ will effectively subsample all of those possible `futures' that still require a model selection procedure to provide new information. We have also defined a normalisation factor $1-r_{{}_{\rm ML}}$ in \Eq{eq:ml-average}, where $r_{{}_{\rm ML}}$ is defined as
\begin{align} 
\ r_{{}_{\rm ML}} \equiv & \int_{\bmuf \in \mathbb{R}^n} \Theta \left[ \, \ln \hatlike ({\cal D}_{\rm fut} |\bmuf ) - t_{{}_{\rm ML}} - \max_{i = \beta , \gamma} \left \{ \ln \hatlike \, ({\cal D}_{\rm fut}  |\boldy_* , {\cal M}_i) \right\}\, \right] \nonumber \\ 
& \qquad \qquad \qquad \qquad \qquad \times p \, (\bmuf \vert \, {\cal D}_{\rm cur})\, \dd \bmuf \label{eq:rML} \,,
\end{align}
hence in the limit of low accuracy $r_{{}_{\rm ML}}=0$, $\langle \cdot \rangle_{{}_{\rm ML}}=\langle \cdot \rangle$ and, in the limit of infinite accuracy, $1-r_{{}_{\rm ML}}$ measures the volume (weighted by the posterior of the current measurement) of the union of the priors between the two models. With \Eq{eq:rML} we may also keep track of the proportion of the $\bmuf$ space that has already ruled both models ${\cal M}_\beta$ and ${\cal M}_\gamma$ out with respect to the maximum likelihood point.\footnote{This choice is justified since the maximum likelihood point can be viewed as the optimal `benchmark' model to compare all other models in the space to.}

In \Eq{eq:decisiveness} we defined $\deci_{\beta \gamma}$ as the decisiveness between models $\calM_\beta$ and $\calM_\gamma$. Hence, using our newly developed maximum-likelihood averaging scheme in \Eq{eq:ml-average}, we define a new expected utility $\deci_{\beta \gamma}\vert_{{}_{\rm ML}}$ which we dub the `decisivity' between $\calM_\beta$ and $\calM_\gamma$. We shall make extensive use of this new quantity for the analysis \Sec{sec:results}.

\subsection{\textsf{A novel computational forecasting method}}
The utility functions we study here contain either of the two integrals \Eq{eq:calcDKL_general} and \Eq{eq:evidences}, which must be nested inside the integral over the $\bmuf$ point domain defined by \Eq{eq:exu} in order to compute the expected utility. The canonical approach would be to perform Nested-Nested sampling with a modification to the \texttt{MultiNest} algorithm~\cite{Feroz:2008xx}, but this would make this problem too computationally expensive due to the length of time required for (even efficient) Nested sampling to converge. Furthermore, in the particular case of the Bayes factor, we cannot always rely on the models being nested within one another, as in the implementation with the SDDR\footnote{The Savage-Dickey Density Ratio is a way to compute the Bayes factor --- valid only when the models involved are nested --- which reduces the often-intractable problem of computing the Bayesian evidence to a conditional prior volume ratio.}~\cite{dickey1971, Heavens:2007ka,Trotta:2010ug}, therefore we must still perform the integrals for the evidences of each model from \Eq{eq:evidences} explicitly.

This issue can, in fact, be resolved by with a relatively simple computational programme. By relaxing the infinitessimal element in \Eq{eq:calcDKL} to be finite, we may rewrite the integral as a discrete summation
\begin{align}
& \dkl \left\{ \, \hatp \left[ \, \boldy \, \vert \, {\cal D}_{\rm fut}(\bmuf ,\boldsigma )  \, \right] \, \right\vert \! \left\vert \, p \, (  \boldy \vert {\cal D}_{\rm cur} ) \, \right\} \simeq \nonumber \\
 & \,\, \sum_{\boldy_i \in \left\{ {\cal D}_{\rm cur} \, {\rm chains}\right\}} \hatlike_{{}_{\rm N}} \left[ \,{\cal D}_{\rm fut}(\bmuf , \boldsigma ) \, \vert \, \boldy_i \, \right] \ln \left\{ \hatlike_{{}_{\rm N}} \left[ \,{\cal D}_{\rm fut}(\bmuf , \boldsigma ) \, \vert \, \boldy_i \, \right] \right\} \label{eq:calcDKL_discrete} \,,
\end{align}
where we assume the $\boldy_i$ to be drawn from Markov chains that sample directly from $p \, (  \boldy \vert {\cal D}_{\rm cur} )$ and we have normalised the future likelihood $\hat{\like}$ in a particular way, such that
\begin{equation} \label{eq:norm_calcDKL_discrete}
\hatlike_{{}_{\rm N}} \left[ \,{\cal D}_{\rm fut}(\bmuf , \boldsigma ) \, \vert \, \boldy_i \, \right] \equiv \frac{\hatlike  \left[ \,{\cal D}_{\rm fut}(\bmuf , \boldsigma ) \, \vert \, \boldy_i \, \right]}{ \vphantom{\bigintss} \sum_{\boldy_j \in \left\{ {\cal D}_{\rm cur} \, {\rm chains}\right\}} \hatlike \left[ \, {\cal D}_{\rm fut}(\bmuf , \boldsigma ) \, \vert \, \boldy_j \, \right] }\,.
\end{equation}
Using \Eq{eq:norm_calcDKL_discrete}, \Eq{eq:calcDKL_discrete} and a sufficiently large number of points, one can efficiently compute \Eq{eq:DKLU} such that the expected utility integral in \Eq{eq:exu} --- which also must be approximated by a discrete summation --- is tractable over reasonable timescales.\footnote{$2$-$3$ days on the \href{http://www.sciama.icg.port.ac.uk/}{Sciama High Performance Compute} cluster, with $\sim 83000$ likelihood samples and $5$-$10$ models with $\sim 6000$ prior samples each.}

\Eq{eq:BayesFactor} may also be computed as a discrete summation with an appropriate weighting scheme implied by the priors of each model, where we find the following formula
\begin{equation} \label{eq:lnB-approx-discrete}
\ {\rm B}_{\beta \gamma} (\bmuf , \boldsigma ) \simeq K\frac{ \vphantom{\bigintss} \sum_{\boldx_i \in \left\{ \calM_\beta \, {\rm chains}\right\}}  \hatlike \left[ \, {\cal D}_{\rm fut}(\bmuf , \boldsigma ) \, \vert \, \boldx_i \, \right] }{ \vphantom{\bigintss} \sum_{\boldx_i \in \left\{ \calM_\gamma \, {\rm chains}\right\}}  \hatlike \left[ \, {\cal D}_{\rm fut}(\bmuf , \boldsigma ) \, \vert \, \boldx_i \, \right] } \,,
\end{equation}
in which the summations are over the Markov chains that sample directly from $\pi \, ( \boldx \vert \calM_\beta )$ (numerator) and $\pi \, ( \boldx \vert \calM_\gamma )$ (denominator) --- modulo a normalisation $K$ that exists due to varying the number of points within each chain, respectively. We note here that a related method to compute the Bayesian evidence for the Markov chains themselves was recently introduced by \Ref{Heavens:2017hkr}, whereas the goal for this chapter is forecasting with futuristic distributions which instead simplifies the integration procedure to multiple evaluations of a distribution function.

Our method can effectively construct the Bayesian evidence for any model defined by its prior over $\boldx$ and has been incorporated in our public code, \href{https://github.com/umbralcalc/foxi}{\texttt{foxi}}. The algorithm to compute whichever $\langle U\rangle$ is straightforward and robust (see appendices~\ref{sec:foxi-computation} and \ref{sec:gaussian-assumption-limitations}), requiring only a minimal number of samples. The main procedure of this computation is:
\begin{enumerate}
\item{\textbf{Draw a value} from the Markov chain representing the distribution $p\, (\bmuf \vert {\cal D}_{\rm cur})$. }
\item{\textbf{Compute} the utilities $U$ using either \Eq{eq:BayesFactor} or by integrating over the whole set of future posterior samples to compute the integral in \Eq{eq:calcDKL}, given the corresponding $\bmuf$ in $p\, (\bmuf |{\cal D}_{\rm cur})$. }
\item{\textbf{Store} the contribution to the integral \Eq{eq:exu}.}
\item{\textbf{If} the integral has not yet converged, go to 1.}
\item{\textbf{Compute} \Eq{eq:exu} and \Eq{eq:secondmomentu} using the contributions stored in 3.}
\end{enumerate}
In order to calculate expected utilities with the $\langle \cdot \rangle_{{}_{\rm ML}}$ average, one simply discards points at steps 1. and 4. which do not satisfy the condition within \Eq{eq:ml-average}. We also note that higher-order statistics such as \Eq{eq:secondmomentu} can be computed trivially from the samples generated by this algorithm.

We shall now progress to analyse the results obtained for the surveys introduced in Table~\ref{tab:future-widths}. We refer the interested reader to Appendix~\ref{sec:foxi-computation} for further details on the computational strategies and robustness checks we have implemented in the code.
\section{\textsf{Results and analysis}}
\label{sec:results}
In all of the analysis below we will consider probability distributions over the various utilities defined in the previous section given a set of futuristic measurement 1-$\sigma$ error bars. In Table \ref{tab:future-widths} we listed the different settings used for each futuristic scenario, where in each case we represented the characteristic measurement errors that might be forecast for a particular configuration of experiment. The specifications of the first two experiments are relatively close to being realised by either CMB Stage-4~\cite{Abazajian:2016yjj}, LiteBIRD~\cite{Matsumura:2013aja} or  COrE~\cite{DiValentino:2016foa,Finelli:2016cyd} and are therefore optimistically labeled `Proposed' with P1 (CMB Stage-4) and P2 (LiteBIRD/COrE). The other four configurations represent a futuristic order of magnitude improvement in the constraint on each of the three slow roll parameters (F1-3), where the final one represents the simultaneous improvement in all three previous configurations (F4).

\begin{figure}
\begin{center}
\includegraphics[width=7cm]{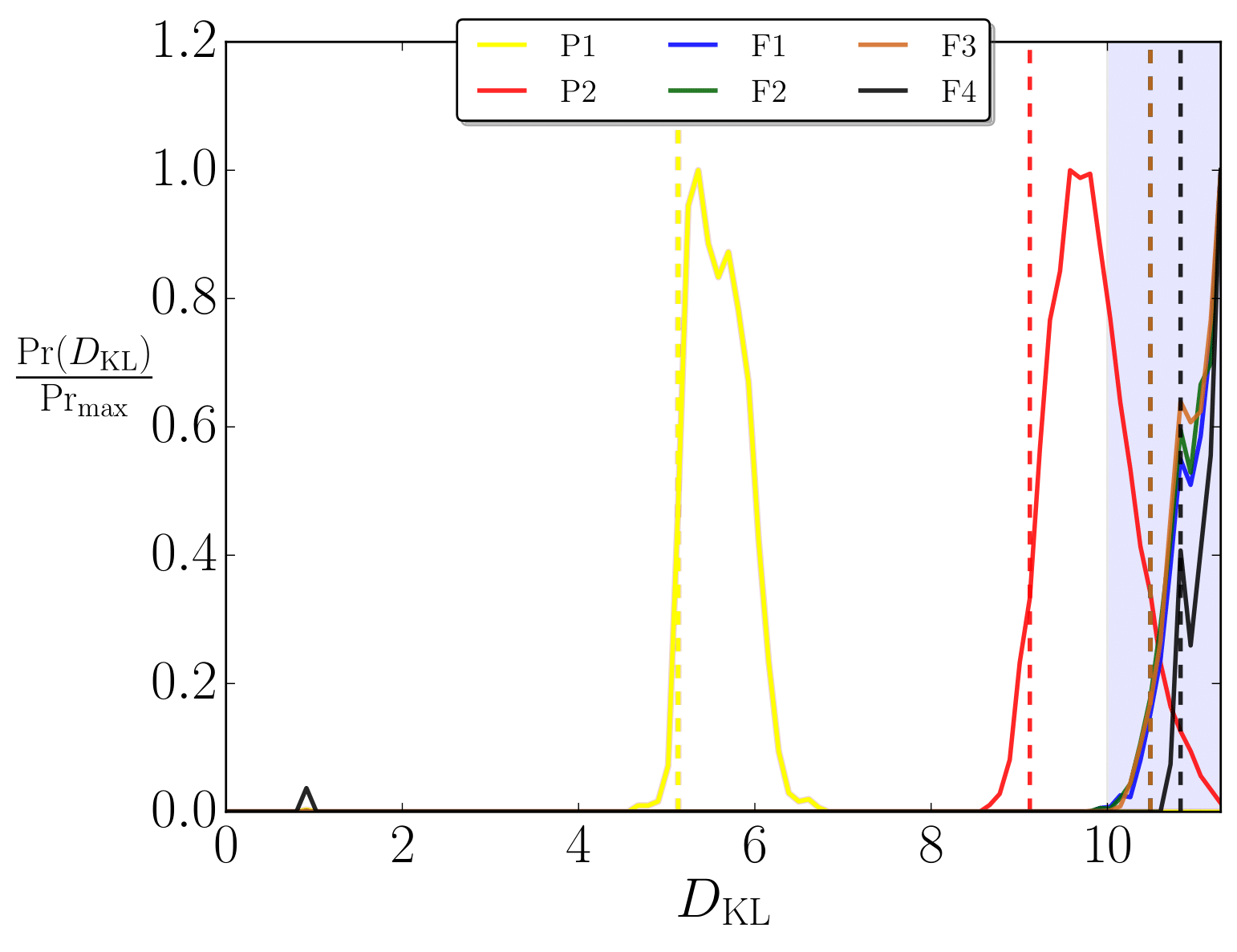}
\includegraphics[width=7cm]{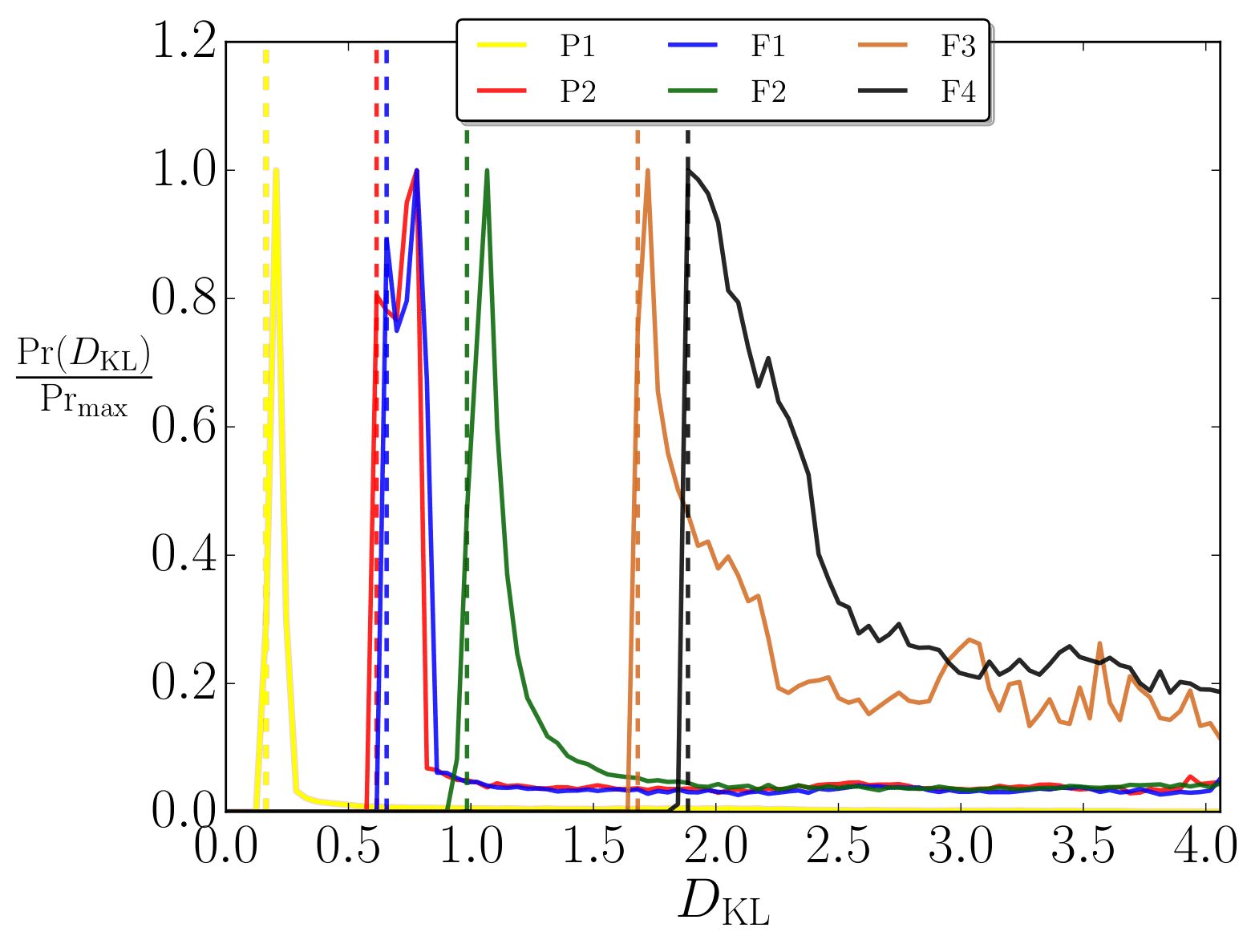}
\caption[Probability density of the Kullback-Leibler divergence in each case]{~\label{fig:info-gain-future} Binned probability density plots showing the distribution of values of the Kullback-Leibler divergence $\dkl$ corresponding to each set of futuristic 1-$\sigma$ error bars in Table \ref{tab:future-widths}. The vertical line associated to each colour is the 95\% lower bound for each experiment. The posterior samples are derived from the Planck data marginalised using the Machine Learning methods defined in~\Ref{Ringeval:2013lea} over $( \epsilon_1 , \epsilon_2, \epsilon_3 )$. The plot on the left uses the $\pi_{\epsilon_1} (\bmuf )$ prior (see \Eq{eq:predictive-priors-flateps1}) where one can see that $\dkl$ in this case is predominately $>11.4$ for F1-4. The plot on the right assumes the $\pi_{\log \epsilon_1} (\bmuf )$ prior (see \Eq{eq:predictive-priors-logeps1}). The grey region in the plot on the left side represents the region $\dkl > 11.4$ beyond the precision of our numerical procedure (see main text). }
\end{center}
\end{figure}

In Table \ref{tab:future-widths} we have also displayed the expectation value on the $\dkl$ (information gain) between the current Planck data and each future dataset in turn. The 95\% bound in each case is also depicted with the dashed lines in \Fig{fig:info-gain-future} where the solid lines represent the predicted probability density in the future of the $\dkl$ value. The distinction between a choice of prior is striking (left and right plots correspond to \Eq{eq:predictive-priors-flateps1} and \Eq{eq:predictive-priors-logeps1} respectively) where e.g. all of the F1-4 datasets saturate an effective numerical upper bound on the expected information gain achievable $\langle \dkl \rangle > 11.4$. Notice indeed that \Eq{eq:norm_calcDKL_discrete} is limited by the number of samples in the Markov chains representing ${\cal D}_{\rm cur}$, such that the typical number of samples used for computations over this space in this chapter ($\sim 85000$) yields this upper bound directly $\ln (85000) \simeq 11.4$.\footnote{This arises from equal-weight, normalised independent samples.}

The value of $\langle \dkl \rangle$ appears to rise far more quickly towards the numerical bound in the case of the flat prior over $\epsilon_1$ as opposed to logarithmic $\epsilon_1$, which can be attributed to the improvements in measurement errors that squeeze up to the hard prior lower bound in the former case, which is $\epsilon_1 \geq 10^{-4}$ from \Eq{eq:predictive-priors-flateps1}. Due to this strong hard prior bound dependence there is a large information gain, which is to be expected when the measurement precision over $\epsilon_1$ becomes of the same order as this bound. From \Fig{fig:nsr-plot} one can also see that two of the models are already ruled out by such a measurement (KMIII and ${\rm KKLTI}_{\rm stg}$) due to their tensor-to-scalar ratios (given by $16\epsilon_1$, see \Eq{eq:tens-scalar-ratio}) being both orders of magnitude below this bound. For this reason we will only consider the logarithmic prior over $\epsilon_1$ defined by \Eq{eq:predictive-priors-logeps1} when considering our model selection utilities, since it is a far more conservative choice.

Turning our attention now to the values of $\dkl$ sampled by the $\bmuf$ points using a logarithmic prior over $\epsilon_1$ in \Fig{fig:info-gain-future}, we see a clear trend and increase in information gain by each survey configuration, which is matched by the values of $\langle \dkl \rangle$ in Table \ref{tab:future-widths}. Notably, the optimal expected information gain (measured by $\langle \dkl \rangle$) between surveys F1-3 is achieved through improvements to the measurement over $\epsilon_3$ in F3. This is clearly due to the fact that the current constraints are the least constraining over $\epsilon_3$ when compared with the other two parameters in the slow-roll hierarchy. We shall return to this interesting point for further discussion in \Sec{sec:concl}.
\subsection{\textsf{General statements}}
\label{sec:gen-state}
The combined results of this chapter span Tables \ref{tab:p1-utilities}, \ref{tab:p2-utilities} and \ref{tab:f14-utilities}. We have performed the analysis computing $\langle \vert \ln {\rm B}_{\beta \gamma} \vert \rangle$, $\deci_{\beta \gamma}$, $\langle \vert \ln {\rm B}_{\beta \gamma} \vert \rangle_{{}_{\rm ML}}$ and $\deci_{\beta \gamma}\vert_{{}_{\rm ML}}$ as expected utilities using all possible pairs of the models defined in Appendix~\ref{sec:models}, where the latter two expected utilities make use of the maximum-likelihood average $\langle \cdot \rangle_{{}_{\rm ML}}$ from \Eq{eq:ml-average}. In addition, we have also provided the ratio of rejected points $r_{{}_{\rm ML}}$
according to this alternative averaging scheme defined by \Eq{eq:rML} in each table.

The increasing decisivity between models is best summarised in \Fig{fig:allmodels_deci}, where the general trend begins with survey P1, where no value of $\deci_{\beta \gamma}\vert_{{}_{\rm ML}}$ is above a probability of 0.1,  towards complete certainty of a decision between all model pairs ($\deci_{\beta \gamma}\vert_{{}_{\rm ML}}=1.0$) in survey F4. An important detail to note at this point is that between F1-3 the best decisive outcome between all model pairs is achieved by survey F2, which corresponds to an order of magnitude decrease in the measurement errors over the second slow-roll parameter $\epsilon_2$. This already gives a strong indication that the possible future directions for selection between inflationary models may rely more on increased precision over the spectral index $\nS$ and less on the tensor-to-scalar ratio $r$. We shall, once again, return to this discussion point later in \Sec{sec:concl}.

\subsection{\textsf{Forecasts using} {\rm P1} \textsf{and} {\rm P2}}
\label{sec:p1p2}

We first examine Tables \ref{tab:p1-utilities} and \ref{tab:p2-utilities} (P1 and P2 surveys, respectively corresponding to CMB Stage-4 and COrE/LiteBIRD-like surveys) which use the measurement error bars that are expected to be achievable in the relatively near future, whence, the label `P' for `Proposed'. For P1 the $r_{{}_{\rm ML}}$ values suggest that already $\sim 2-4\%$ of the possible future realisations will rule both models of each pair out at the level of either model's maximum likelihood given our threshold of $\ee^{-t_{{}_{\rm ML}}}\like_{\rm max}$ or above (see \Sec{sec:utility-expectation-values}), where $t_{{}_{\rm ML}}=5$. Note that this is not the same as \emph{all} of the model pairs being ruled out at once but instead reflects the specific decision question for each model in-turn. P2 has a far more striking result --- in $\geq 94\%$ of the possible future measurements, both models in each pair (in all 10 possible combinations) will have been eliminated at the maximum likelihood level. We can infer from these results alone that the upcoming future surveys of the P2-type will have strong decision-making capabilities even before any further analysis or detailed model selection program is initiated. This indicates that an important first threshold in the space of possible CMB missions exists, somewhere between the capabilities of P1 and P2, where most single-field model pairs will already be ruled out at the level of their maximum likelihoods. This threshold can be crossed in the future by a COrE/LiteBIRD-like mission.

Let us move on to the expected model selection utilities by improving measurement bounds by an order of magnitude on both $\epsilon_1$ and $\epsilon_3$. In doing so we advance from P1 to P2, where most model pairs receive a very large amplitude increase in $\langle \vert \ln {\rm B}_{\beta \gamma} \vert \rangle_{{}_{\rm ML}}$ e.g. all of the pairs that include the RGI model increase by an order of magnitude in $\ln$-scale. The uncertainties associated to this expected utility also become significantly larger in most cases. Though it is instructive to consider the expected Bayes factor utilities, the variance in their value for each model pair (especially in the case of survey P2) leads to significant uncertainty in assertions about the future that rely on these utilities alone. Therefore, we can support our claims by considering the decisivity $\deci_{\beta \gamma} \vert_{{}_{\rm ML}}$ for the same pairs of models, where most receive a greater-than factor of 4 increase in the odds of a decisive model selection with survey P2 when compared to P1.

\begin{table}
\centering
\resizebox{12cm}{!}{
\begin{tabular}{|c|c|c|c|c|c|}
\cline{2-6}
\multicolumn{1}{c}{\cellcolor{yellow!55}} & \multicolumn{5}{|c|}{P1 with $\pi_{\log \epsilon_1} (\bmuf )$}  \\      \hline
\backslashbox{${\cal M}_\beta$ - ${\cal M}_\gamma$}{$\langle  U\rangle$}  & $\langle \vert \ln {\rm B}_{\beta \gamma}\vert \rangle$ & $\langle \vert \ln {\rm B}_{\beta \gamma} \vert \rangle_{{}_{\rm ML}}$ & $\mathscr{D}_{\beta \gamma}$ & $\mathscr{D}_{\beta \gamma}\vert_{{}_{\rm ML}}$ & $r_{{}_\mathrm{ML}}$ \\     \hline\hline
${\rm KMIII}$ - ${\rm HI}$ & $2.42 \,(< 91.72)$ & $2.41 \,(< 92.88)$ & 0.01 & $0.0+\varepsilon$ & 0.04  \\      \hline
${\rm KKLTI}_{\rm stg}$ - ${\rm HI}$ & $3.20 \,(< 52.88)$ & $3.22 \,(< 53.36)$ & 0.03 & 0.03 & 0.03  \\      \hline
${\rm LI}_{\alpha >0}$ - ${\rm HI}$ & $3.21 \,(< 17.24)$ & $2.99 \,(\pm 1.15)$ & 0.06 & 0.04 & 0.03  \\      \hline
${\rm RGI}$ - ${\rm HI}$ & $3.09 \,(< 61.96)$ & $1.41 \,(< 4.74)$ & 0.01 & 0.01 & 0.03  \\      \hline
${\rm KKLTI}_{\rm stg}$ - ${\rm KMIII}$ & $5.33 \,(< 104.30)$ & $5.42 \,(< 105.64)$ & 0.03 & 0.03 & 0.03  \\      \hline
${\rm LI}_{\alpha >0}$ - ${\rm KMIII}$ & $5.59 \,(< 93.06)$ & $5.39 \,(< 92.26)$ & 0.08 & 0.06 & 0.03  \\      \hline
${\rm RGI}$ - ${\rm KMIII}$ & $5.48 \,(< 110.64)$ & $3.79 \,(< 92.22)$ & 0.03 & 0.01 & 0.02  \\      \hline
${\rm LI}_{\alpha >0}$ - ${\rm KKLTI}_{\rm stg}$ & $5.03 \,(< 55.04)$ & $4.85 \,(< 52.82)$ & 0.07 & 0.04 & 0.04  \\      \hline
${\rm RGI}$ - ${\rm KKLTI}_{\rm stg}$ & $5.04 \,(< 81.50)$ & $3.40 \,(< 53.26)$ & 0.04 & 0.03 & 0.03  \\      \hline
${\rm RGI}$ - ${\rm LI}_{\alpha >0}$ & $3.46 \,(< 59.12)$ & $1.83 \,(< 4.34)$ & 0.03 & 0.01 & 0.04  \\      \hline
\end{tabular}}
\caption[Experiment P1 expected utilities]{~\label{tab:p1-utilities} Computed expected utilities for a P1 experiment. All results correspond to a choice of the $\pi_{\log \epsilon_1} (\bmuf )$ prior in \Eq{eq:predictive-priors-logeps1}. Note that $\varepsilon$ reminds the reader that the value is subject to rounding errors of up to $0.005$. Values in brackets $\pm$ around each computed expected utility correspond to the $1$-$\sigma$ uncertainties, which are evaluated using \Eq{eq:secondmomentu}. This symmetric error about our different expected values for $\vert \ln {\rm B}_{\beta \gamma} \vert$ is replaced with a $2$-$\sigma$ upper bound (because it is positive by definition) if the lower error is greater than the expected value itself.}
\end{table}

\begin{table}
\centering
\resizebox{12cm}{!}{
\begin{tabular}{|c|c|c|c|c|c|}
\cline{2-6}
\multicolumn{1}{c}{\cellcolor{red!55}} & \multicolumn{5}{|c|}{P2 with $\pi_{\log \epsilon_1} (\bmuf )$}  \\      \hline
\backslashbox{$\calM_\beta$ - $\calM_\gamma$}{$\langle  U\rangle$}  & $\langle \vert \ln {\rm B}_{\beta \gamma}\vert \rangle$ & $\langle \vert \ln {\rm B}_{\beta \gamma} \vert \rangle_{{}_{\rm ML}}$ & $\deci_{\beta \gamma}$ & $\deci_{\beta \gamma}\vert_{{}_{\rm ML}}$ & $r_{{}_\mathrm{ML}}$ \\     \hline\hline
${\rm KMIII}$ - ${\rm HI}$ & $10.04 \,(< 105.64)$ & $43.76 \,(< 391.02)$ & 0.79 & 0.12 & 0.95  \\      \hline
${\rm KKLTI}_{\rm stg}$ - ${\rm HI}$ & $10.16 \,(< 55.06)$ & $3.06 \,(\pm 2.23)$ & 0.87 & 0.11 & 0.94  \\      \hline
${\rm LI}_{\alpha >0}$ - ${\rm HI}$ & $6.10 \,(< 77.44)$ & $2.42 \,(< 6.80)$ & 0.09 & 0.09 & 0.96  \\      \hline
${\rm RGI}$ - ${\rm HI}$ & $15.49 \,(< 185.90)$ & $18.73 \,(< 222.02)$ & 0.69 & 0.06 & 0.95  \\      \hline
${\rm KKLTI}_{\rm stg}$ - ${\rm KMIII}$ & $4.77 \,(< 91.26)$ & $39.57 \,(< 378.58)$ & 0.08 & 0.16 & 0.94  \\      \hline
${\rm LI}_{\alpha >0}$ - ${\rm KMIII}$ & $9.98 \,(< 133.10)$ & $41.38 \,(< 378.46)$ & 0.12 & 0.22 & 0.94  \\      \hline
${\rm RGI}$ - ${\rm KMIII}$ & $15.51 \,(< 214.72)$ & $55.34 \,(< 430.48)$ & 0.05 & 0.11 & 0.94  \\      \hline
${\rm LI}_{\alpha >0}$ - ${\rm KKLTI}_{\rm stg}$ & $9.98 \,(< 97.56)$ & $3.34 \,(< 12.84)$ & 0.65 & 0.11 & 0.94  \\      \hline
${\rm RGI}$ - ${\rm KKLTI}_{\rm stg}$ & $15.62 \,(< 194.58)$ & $19.28 \,(< 224.84)$ & 0.11 & 0.13 & 0.94  \\      \hline
${\rm RGI}$ - ${\rm LI}_{\alpha >0}$ & $10.68 \,(< 169.88)$ & $13.89 \,(< 198.60)$ & 0.04 & 0.04 & 0.95  \\      \hline
\end{tabular}}
\caption[Experiment P2 expected utilities]{~\label{tab:p2-utilities} Computed expected utilities for a P2 experiment. All results correspond to a choice of the $\pi_{\log \epsilon_1} (\bmuf )$ prior in \Eq{eq:predictive-priors-logeps1}. Note that $\varepsilon$ reminds the reader that the value is subject to rounding errors of up to $0.005$. Values in brackets $\pm$ around each computed expected utility correspond to the $1$-$\sigma$ uncertainties, which are evaluated using \Eq{eq:secondmomentu}. This symmetric error about our different expected values for $\vert \ln {\rm B}_{\beta \gamma} \vert$ is replaced with a $2$-$\sigma$ upper bound (because it is positive by definition) if the lower error is greater than the expected value itself.}
\end{table}

\subsection{\textsf{Forecasts using} {\rm F1-4}}
\label{sec:f14}

We begin our analysis of the results using surveys F1-4 in Table \ref{tab:f14-utilities} by noting that, from this point onward, because the measurement errors for each survey are so small it will no longer be informative to use $\langle \vert \ln {\rm B}_{\beta \gamma}\vert \rangle$ and  $\langle \vert \ln {\rm B}_{\beta \gamma}\vert \rangle_{{}_{\rm ML}}$ since their magnitudes are all above the Jeffrey's threshold $>5$ (and probably above the numerical precision). It is, however, far more illuminating to examine the values of $\deci_{\beta \gamma} \vert_{{}_{\rm ML}}$ and $r_{{}_{\rm ML}}$ together: firstly to assert whether or not the proportion of $\bmuf$ points remaining is already very small for which Bayesian model selection techniques are unnecessary (i.e. how large $r_{{}_{\rm ML}}$ is will dictate how likely it is in the future for a given model pair to be totally ruled out at the level of the maximum likelihood, and hence whether there are any likely futures for which Bayesian model selection will be required at all), and secondly in the event of model selection being required, whether or not $\deci_{\beta \gamma} \vert_{{}_{\rm ML}}$ gives good odds of successfully deciding between those models.

Survey F1 increases the measurement precision over $\epsilon_1$ from P2 by an order of magnitude. Using Table \ref{tab:f14-utilities}, for each pair of models this improvement is expected to leave a $\leq 0.06$ chance of avoiding a ruling-out with respect to the maximum likelihood of each model. Of the expected remaining $\bmuf$ points, there is varied performance by Bayesian model selection to be decisive --- one the one hand, KMIII - HI and ${\rm KKLTI}_{\rm stg}$ - HI are always decided between ($\deci_{\beta \gamma} \vert_{{}_{\rm ML}} = 1.0 - \varepsilon$ up to rounding errors $\varepsilon = 0.005$), whereas on the other hand, there are only chances of 0.12 and 0.18 to decide between RGI - ${\rm LI}_{\alpha >0}$ and RGI - ${\rm KKLTI}_{\rm stg}$, respectively.

In contrast, survey F2 increases the measurement precision over $\epsilon_2$ from P2 by an order of magnitude. For this improvement, one lowers slightly further the chance of avoiding a ruling-out with respect to the maximum likelihood of each model down to $\leq 0.05$.  Of the expected remaining $\bmuf$ points, there is a very impressive performance expected, yielding at worst chances of 0.47 and 0.5 to decide between the  pairs KMIII - HI and RGI - HI (also ${\rm KKLTI}_{\rm stg}$ - HI) respectively where, in fact, most other model pairs have high decisivity $\geq 0.68$. It is for this reason that we will conclude later that an F2 strategy for survey design is superior to F1 for single-field inflationary model selection.

Survey F3 increases the measurement precision over $\epsilon_3$ from P2 by an order of magnitude. Between F1-3 this survey configuration has the greatest chance of ruling out a given model pair at the level of the maximum likelihood, which is $\geq 0.96$. Of the remaining $\bmuf$ points, there is a wildly varied chance of a decisive conclusion between models e.g. 0.12 for RGI - ${\rm LI}_{\alpha >0}$, but conversely, a chance of $\geq 0.76$ for all model pairs including ${\rm KKLTI}_{\rm stg}$.

The decisiveness $\deci_{\beta \gamma}$ drops dramatically from F1 and F2 to F3 (and also F4 which inherits this feature from F3). This is as a feature that arises from situations where the Bayesian evidence of both models being too low to numerically evaluate, and hence the algorithm assigns $\vert \ln {\rm B}_{\beta \gamma} \vert = 0$, which results in a contribution of 0 to the decisiveness at that point. If this happens frequently enough then the value of $\deci_{\beta \gamma}$ drops accordingly, as is the case when the measurement precision over $\epsilon_3$ is improved enough for it to be a decisive observable. In principle this can be rectified by hand by assuming that $\vert \ln {\rm B}_{\beta \gamma} \vert > 5$ for all of these points, but this is not strictly correct, and hence we have not quoted $\deci_{\beta \gamma}$ for F3 and F4 accordingly. This numerical problem does not exist for the decisivity $\deci_{\beta \gamma} \vert_{{}_{\rm ML}}$, and hence provides another supporting argument for its use.

Finally, because using F4 always appears to give values of $r_{{}_{\rm ML}}\geq 0.97$, we can immediately conclude that the survey configuration F4 is close to the ultimate goal for, essentially, absolute certainty in deciding between the plateau models at the level of their maximum likelihood values alone. The fact that $r_{{}_{\rm ML}}$ saturates to a constant value for most model pairs in moving from F1-3 to F4 indicates that there is a second threshold in the space of CMB missions (the first being between P1 and P2). The value of $r_{{}_{\rm ML}}$ saturates to a constant when the measurement over $(\epsilon_1,\epsilon_2,\epsilon_3)$ is so precise that it is effectively a Dirac delta function when compared with the priors over a pair of models. Hence, the value of $1-r_{{}_{\rm ML}}$ in this limit (as discussed previously in \Sec{sec:utility-expectation-values}) corresponds to the total prior union volume of the two models relative to the total volume in the $(\epsilon_1,\epsilon_2,\epsilon_3)$ space that is weighted by the current likelihood $\like ({\cal D}_{\rm cur}\vert \bmuf )$.

Furthermore, in this limit, the Bayes factor between all model pairs reduces to a trivial prior point ratio
\begin{equation} \label{eq:trivial-bayes-factor}
\left. {\rm B}_{\beta \gamma} \right\vert_{\boldsigma \rightarrow 0} \rightarrow \frac{ \vphantom{\bigintss} \bigintsss_{\boldx \in \mathbb{R}^n} \delta (\boldx - \bmuf) \, \barpi \, ( \boldx | \calM_\beta )\, \dd \boldx  }{ \vphantom{\bigintss} \bigintsss_{\boldx \in \mathbb{R}^n} \delta (\boldx - \bmuf ) \, \barpi \, ( \boldx | \calM_\gamma )\, \dd \boldx  } = \frac{\barpi \, ( \bmuf | \calM_\beta )  }{ \barpi \, ( \bmuf | \calM_\gamma ) } \,,
\end{equation}
and note that this becomes independent of the future measurement error bars $\boldsigma$. Hence, to go any further than this measurement precision will require a reformulation of a new space of models $\boldsymbol{\calM}$ with priors that are coarse-grained to much finer detail so as to remain competitive.

\subsection{\textsf{Deciding between reheating scenarios}}
\label{sec:reheating}

Full statistical inference of the temperature of reheating for a given inflationary model is an exciting new research topic within early Universe cosmology~\cite{Martin:2014nya,Hardwick:2016whe,Finelli:2016cyd,Martin:2016oyk}. In principle, if one can infer a micro-physical parameter, such as temperature, from the thermal bath at high energies then the early Universe can become a laboratory for high-energy physics. In addition to this, one can potentially distinguish between inflationary models with the same potential, e.g. Higgs inflation~\cite{Bezrukov:2007ep} and Starobinsky inflation~\cite{Starobinsky:1992ts}, that are realised in different theoretical frameworks by using their possibly different reheating temperatures.

In this short section we use our formalism to study 3 nested models within the HI model: ${\rm HI}_{T-}$, ${\rm HI}_{T}$ and ${\rm HI}_{T+}$, which correspond to the HI potential at fixed reheating temperatures $T_{\rm reh}=10^{12}\, {\rm GeV}$, $10^{6}\, {\rm GeV}$ and $1\, {\rm GeV}$, respectively. Motivations for the reheating temperatures include the various relic species overproduction problems, e.g., the so-called `gravitino problem'~\cite{Dimastrogiovanni:2015wvk} for the lower temperature at $T_{\rm reh}= 1{\rm GeV}$, reheating temperatures of $T_{\rm reh}= 10^6{\rm GeV}$ are favoured by Supergravity channels for Starobinsky inflation~\cite{Terada:2014uia} and $T_{\rm reh}=10^{12}{\rm GeV}$ is typical for Higgs inflation~\cite{2009JCAP06029B}.

By performing the same analysis to compute the expected utilities for the comparison between these nested models, we will give a qualitative impression of how our formalism can be used to indicate the future performance of any survey with respect to carrying out inference on reheating.

Table \ref{tab:reheating-temps} lists our full results for this analysis. The chance of ruling out all of the reheating temperatures at the level of the maximum likelihood reaches 1.0 with surveys F1-4, and the reheating temperatures are essentially measured to extremely good precision, therefore we have not included these results in the table since they are essentially trivial.

Considering the results using the P1 configuration first, the chance of ruling out each pair of temperatures at the level of the maximum likelihood is low ($\leq 0.05$). In addition, we find that model selection offers no additional benefit of deciding between temperatures for the HI model since $\langle \vert \ln {\rm B}_{\beta \gamma} \vert \rangle_{{}_{\rm ML}}$ is well below $5$ (even with the typical standard deviation added) and $\deci_{\beta \gamma}\vert_{{}_{\rm ML}}$ supports this by indicating a 0.0 (up to rounding errors of 0.005) chance of decisive selection of temperature.

We now turn our attention to the P2 configuration. According to Table \ref{tab:reheating-temps}, the improvements to the measurement bounds in moving from P1 to P2 indicate that one can nearly be certain (chance of $\geq 0.97$) that they will be able to select away from each pair of reheating temperatures at the level of the maximum likelihood, boding well in this regard for the prospects of future surveys like COrE~\cite{Finelli:2016cyd}.\footnote{In addition, supporting the conclusions made by \Ref{Finelli:2016cyd}}

If one now considers the values of the $\langle \vert \ln {\rm B}_{\beta \gamma}\vert \rangle_{{}_{\rm ML}}$ utility for the P2 survey, these suggest that future values of $\vert\ln {\rm B}_{\beta \gamma}\vert \simeq 2$ occur more regularly at $2$-$\sigma$ for all three reheating temperatures, and hence they may be distinguished between, which is indeed consistent with \Ref{Finelli:2016cyd}. We note, however that this does not mean that such temperatures can be decisively ruled out with respect to one another --- a fully decisive future with $\vert\ln {\rm B}_{\beta \gamma}\vert = 5$ appears to occur only very infrequently at the beyond 5-$\sigma$ level.

We have demonstrated the versatility that our formalism has, as well as the range of applicable problems that the \href{https://sites.google.com/view/foxicode}{\texttt{foxi}} package can deal with. We continue to the next section with another example.

\begin{table}
\centering
\resizebox{13cm}{!}{
\begin{tabular}{|c|c|c|c|c|c|c|c|}
\cline{1-8}
\multicolumn{2}{|c|}{Survey} &\backslashbox{$\calM_\beta$ - $\calM_\gamma$}{$\langle  U\rangle$}  & $\langle \vert \ln {\rm B}_{\beta \gamma}\vert \rangle$ & $\langle \vert \ln {\rm B}_{\beta \gamma} \vert \rangle_{{}_{\rm ML}}$ & $\deci_{\beta \gamma}$ & $\deci_{\beta \gamma}\vert_{{}_{\rm ML}}$ & $r_{{}_\mathrm{ML}}$ \\     \hline\hline
\cellcolor{yellow!55} & P1  & ${\rm HI}_{T-}$ - ${\rm HI}_{T}$ & $0.39 \,(\pm 0.30)$ & $0.35 \,(\pm 0.20)$ & $0.0+\varepsilon$ & $0.0+\varepsilon$ & 0.05  \\      \hline
\cellcolor{yellow!55} & P1  & ${\rm HI}_{T-}$ - ${\rm HI}_{T+}$ & $0.79 \,(\pm 0.55)$ & $0.72 \,(\pm 0.38)$ & $0.0+\varepsilon$ & $0.0+\varepsilon$ & 0.04  \\      \hline
\cellcolor{yellow!55} & P1  & ${\rm HI}_{T}$ - ${\rm HI}_{T+}$ & $0.41 \,(\pm 0.25)$ & $0.37 \,(\pm 0.17)$ & $0.0+\varepsilon$ & $0.0+\varepsilon$ & 0.05  \\      \hline
\cellcolor{red!55} & P2 & ${\rm HI}_{T-}$ - ${\rm HI}_{T}$ & $2.09 \,(< 17.28)$ & $1.61 \,(\pm 0.52)$ & 0.04 & $0.0+\varepsilon$ & 0.98  \\      \hline
\cellcolor{red!55} & P2 & ${\rm HI}_{T-}$ - ${\rm HI}_{T+}$ & $4.17 \,(< 29.86)$ & $2.72 \,(\pm 0.93)$ & 0.12 & $0.0+\varepsilon$ & 0.97  \\      \hline
\cellcolor{red!55} & P2 & ${\rm HI}_{T}$ - ${\rm HI}_{T+}$ & $2.09 \,(< 24.12)$ & $1.0 \,(\pm 0.39)$ & 0.02 & $0.0+\varepsilon$ & 0.97  \\      \hline
\end{tabular}}
\caption[Expected utilities for post-Higgs inflation reheating]{\label{tab:reheating-temps} Computed expected utilities for the Higgs Inflation (HI) model (defined by the potential of \Eq{eq:higgs-pot}) fixed with 3 different reheating temperatures, where ${\rm HI}_{T-}$, ${\rm HI}_{T}$ and ${\rm HI}_{T+}$ each correspond to the model with reheating temperatures $T_{\rm reh}=1\, {\rm GeV}$, $10^{6}\, {\rm GeV}$ and $10^{12}\, {\rm GeV}$, respectively. The expected utilities have been computed with the first 2 survey configurations studied in this chapter (P1 and P2) and all results correspond to a choice of the $\pi_{\log \epsilon_1} (\bmuf )$ prior in \Eq{eq:predictive-priors-logeps1}. Note that $\varepsilon$ reminds the reader that the value is subject to rounding errors of up to $0.005$. Values in brackets $\pm$ around each computed expected utility correspond to the $1$-$\sigma$ uncertainties, which are evaluated using \Eq{eq:secondmomentu}. This symmetric error about our different expected values for $\vert \ln {\rm B}_{\beta \gamma} \vert$ is replaced with a $2$-$\sigma$ upper bound (because it is positive by definition) if the lower error is greater than the expected value itself. }
\end{table}

\subsection{\textsf{Measuring the scalar running}}
\label{sec:measuring-alphaS}

Another example of our formalism at work is in the forecasting of the probability that as-of-yet unobserved parameters will be measured in the future by a given survey with forecast error bars $\boldsigma$. Consider the running\footnote{This is also a good consistency check with our assumption that the $(\epsilon_1,\epsilon_2,\epsilon_3)$ is currently a sufficient space (and not including higher-order slow-roll parameters e.g. $\epsilon_4$) to characterise the single-field model selection capabilities of future CMB missions.} $\alphaS$ of the scalar spectral index in single-field inflation, defined in \Eq{eq:alphaS-intro}.

In Appendix~\ref{sec:alphaS-calculation} we derive a relation connecting the observed fiducial point and measurement 1-$\sigma$ error bar ($\muf^{\alphaS}$ and $\sigma^{\alphaS}$, respectively) over $\alphaS$ to the future error bars over the slow-roll parameters $\boldsigma$, which we compute for each given realisation over the measured $\bmuf$ points. We shall not quote the relation here, but by referring to the functional dependencies $\muf^{\alphaS} = \muf^{\alphaS} (\bmuf , \boldsigma )$ and $\sigma^{\alphaS} = \sigma^{\alphaS} (\bmuf , \boldsigma )$ we can show that the probability which we seek is implicitly
\begin{align}\label{eq:alphaS-probability}
\Pr{}_{\alphaS > 2\sigma} (\boldsigma ) &\equiv \int_{\bmuf \in \mathbb{R}^n} p \left( \vert \muf^{\alphaS} \vert - 2 \sigma^{\alphaS} > 0 \, \vert \, \bmuf , \boldsigma \right) \dd \bmuf \\
&= \int_{\bmuf \in \mathbb{R}^n} \Theta \left[ \vert \muf^{\alphaS} (\bmuf , \boldsigma ) \vert - 2  \sigma^{\alphaS} (\bmuf , \boldsigma ) \right] \, p \left( \bmuf \vert {\cal D}_{\rm cur}\right) \dd \bmuf \label{eq:alphaS-probability-2}\,,
\end{align}
where we have specified a $2\sigma$-measurement over $\alphaS$ to be identified as having `measured $\alphaS$'.

In Table \ref{tab:alphaS-probs} we quote the probabilities of measurement over $\alphaS$ for each of the survey configurations studied in this chapter. We find that for the survey P2 one obtains a substantial improvement over P1 in the probability of measuring $\alphaS$ --- moving from $\simeq 0.0$ to a probability of 0.93. When one reconsiders the posterior prediction, made this time when assuming that the Higgs Inflation model is `correct', we replace $p\left( \bmuf \vert {\cal D}_{\rm cur}\right)$ in \Eq{eq:alphaS-probability-2} with the posterior distribution $\bar{p}\left( \boldx \vert {\cal D}_{\rm cur},\calM_{{\rm HI}}\right)\propto \bar{\pi} \left( \boldx \vert \calM_{{\rm HI}} \right) \bar{\like}\left( {\cal D}_{\rm cur} \vert \boldx \right)$. From this change we see that there are significant probabilities for a detection of $\alphaS$ to be made by F2, F3 (and F4) surveys, hence improving the measurement over either $\epsilon_2$ or $\epsilon_3$ by an order of magnitude from the P2 survey. This can be seen explicitly through the relation in \Eq{eq:alphaS-width-formula}, where the otherwise relatively large term in the expression for $(\sigma^{\alphaS} )^2 \supset (\sigma^2)^2(\sigma^3)^2$ can only be reduced in size by decreasing either the measurement width over $\epsilon_2$ or $\epsilon_3$.

\begin{table}
\centering
\resizebox{12cm}{!}{
\begin{tabular}{|c|c|c|c|}
\cline{1-4}
\multicolumn{2}{|c|}{Survey ($\boldsigma$)}  & $\Pr{}_{\alphaS > 2\sigma} (\boldsigma )$ & $\Pr{}_{\alphaS > 2\sigma} (\boldsigma )$ (HI posterior prediction) \\     \hline\hline
\cellcolor{yellow!55} & P1 & $0.0 + \varepsilon$ & $0.0 + \varepsilon$ \\      \hline
\cellcolor{red!55} & P2 & 0.93 & 0.02 \\      \hline
\cellcolor{blue!35} & F1 & 0.93 & 0.02  \\      \hline
\cellcolor{green!55} & F2 & 0.96 & 0.85 \\     \hline
\cellcolor{brown!90} & F3 & 0.96 & $1.0 - \varepsilon$ \\
\hline
\cellcolor{black!65} & F4 & 0.99 & $1.0 - \varepsilon$ \\   \hline
\end{tabular}}
\caption[Probabilties to measure the running of the spectral index]{\label{tab:alphaS-probs} The probabilities of measurement over $\alphaS$ for each of the survey configurations studied in this chapter, where measurement is defined as the fiducial point $\muf^{\alphaS}$ exceeding the $2\sigma$-uncertainty bound for a given future realisation. Note that $\varepsilon$ reminds the reader that the value is subject to rounding errors of up to $0.005$. In the final column we assume that HI is the `correct' model (replacing $p \, (\muf \vert {\cal D}_{\rm cur})$ with $\barp \, (\muf \vert {\cal D}_{\rm cur},\calM_{\rm HI} )$ in \Eq{eq:alphaS-probability-2}) and forecast the probability of detection of $\alphaS$ for each survey. }
\end{table}

\section{\textsf{Concluding remarks}}
\label{sec:concl}
In this chapter we have outlined a simple method to compute any expected utility for a future survey given a previous set of measurements on the same variables from an independent survey. The tools that we have developed have all been included in \href{https://sites.google.com/view/foxicode}{\texttt{foxi}}, a publicly available python package that can be readily used in any survey forecasting problem. Crucially, our calculation relies on the assumption that the future likelihood can be modeled by an uncorrelated Gaussian distribution over the space of slow-roll parameters, hence, incorporating the level of detail required to tackle forecasting for proposed surveys like COrE/LiteBIRD must be an inevitable next step.

We have also modified the form of the expected utility in order to partition each possible future into either the rejection of models at the level of the maximum-likelihood or the decision between models using Bayesian model comparison. With the new expected utilities generated by this procedure, we have forecast the future of single-field inflationary model selection using 5 plateau potentials that are both indicative of the class and span the range of observables $(\epsilon_1,\epsilon_2,\epsilon_3)$ --- the slow-roll parameters --- that is typical for models of this type (see Appendix~\ref{sec:models} for their definitions). Our analysis finds two important thresholds in the space of missions:
\begin{enumerate}
\item{Increasing precision from a P1-type survey capabilities (like CMB Stage-4) to P2 (like LiteBIRD/COrE), we cross the first threshold where most of the possible future measurements that could be made will rule out both single-field models of each pair at the level of their maximum likelihoods. }
\item{Increasing precision from F1-3 to F4-type toy survey capabilities, we cross a second threshold where our utility functions saturate to constant values that do not depend on the precision of the measurement. In this limit, the error bars of the future likelihoods are much smaller than the prior volumes from the models that we consider. For both models of a given pair not to be rejected at the level of the maximum likelihood, the value of $\bmuf$ must fall within at least one of their prior volumes. If this is so then the Bayes factor becomes the ratio between their prior densities at that point (see \Eq{eq:trivial-bayes-factor}) which does not depend on the future measurement error bars.}
\end{enumerate}

The prior volume-dominated limit, arising from threshold 2 above, is analogous to the threshold reached within our computational procedure (outlined in Appendix~\ref{sec:foxi-computation}), where in the latter case we devise a method to calculate the Bayesian evidence that relies upon \Eq{eq:trivial-bayes-factor}. Once the threshold of this regime has been crossed it is \emph{essential} for more theoretical progress in the understanding of the remaining models to occur, which would result in more narrow priors on their parameters, before one builds a new survey to choose between them

Though the space of surveys that we explore in this section may be simplistic, the broad conclusions we draw are unlikely to change. Our results using only information theory considerations (the expected Kullback-Leibler divergence $\langle \dkl \rangle$) indicat1e that the greatest information to be gained is on $\epsilon_3$, since it is currently the least constrained of the three slow-roll parameters (and may also be used to detect a scalar running). However, our analysis also suggests that the most-likely decisive gains in selecting between single-field inflationary models are made by improving the second slow-roll parameter $\epsilon_2$ constraint (which can also potentially be used to detect a scalar running) --- which can be measured through more precision on the scalar spectral index $\nS$. Finally, as is suggested by many theoretical studies into the fundamental physics of quantum gravity, the tensor-to-scalar ratio $r$ might be the most important CMB observable and hence $\epsilon_1$ may be considered the most fundamentally attractive to theorists. Therefore, to order this trichotomy, we have compiled the following list:
\begin{enumerate}
\item{Improve the measurement over $\nS$, hence $\epsilon_2$ will be constrained to a greater degree and therefore one optimises the single-field slow-roll decisivity. Also we may potentially observe $\alphaS$.}
\item{Improve the measurement over $r$, hence $\epsilon_1$ will be constrained to a greater degree and we may learn more about fundamental physics.}
\item{Improve the measurement over $\alphaS$, hence $\epsilon_3$ will be constrained to a greater degree which is optimal from an information-theoretic standpoint.}
\end{enumerate}

We also considered the applications of our framework to forecasting the potential of surveys to infer the temperature of reheating, given the Higgs inflationary potential. This is an avenue which we only very briefly have explored in this section but a clear extension would be to conduct a more thorough analysis on reheating temperatures taking into account different choices of inflationary potentials that still match observations. This also serves to illustrate the next step in the challenges set to model-builders in the future: one must be more specific in predicting reheating temperatures that arise from a given inflationary potential as one approaches the second threshold.

In \Sec{sec:measuring-alphaS} we have promoted an additional application of our framework to obtaining probabilities of measuring a given parameter in the future. In this case, we considered the probability of measuring the scalar running $\alphaS$, initially when assuming no preferred model, and then subsequently when assuming that a slow-roll single-field model (the HI model in this case) is preferred and hence the current data is the posterior prediction of the model from \emph{Planck}. Our results broadly indicate that though a P2-like survey is generally expected to measure $\alphaS$, if the \emph{Planck} posterior is consistent with a slow-roll single-field model then the probability of such a measurement drops dramatically and it is only with more advanced mock surveys like F2 or F3 that the chances of measuring $\alphaS$ become significant once again. This can be traced to the fact that $\alphaS$ is typically small to be consistent with slow-roll single-field models, and hence a more advanced survey is required to measure its potential deviation away from 0.

\newpage

\begin{subappendices}

\begin{landscape}
\begin{table}
\centering
\resizebox{0.45\columnwidth}{!}{%
\begin{tabular}{|c|c|c|c|c|c|}
\cline{2-6}
\multicolumn{1}{c}{\cellcolor{blue!35}} & \multicolumn{5}{|c|}{F1 with $\pi_{\log \epsilon_1} (\bmuf )$}  \\      \hline
\backslashbox{$\calM_\beta$ - $\calM_\gamma$}{$\langle  U\rangle$}  & $\langle \vert \ln {\rm B}_{\beta \gamma}\vert \rangle$ & $\langle \vert \ln {\rm B}_{\beta \gamma} \vert \rangle_{{}_{\rm ML}}$ & $\deci_{\beta \gamma}$ & $\deci_{\beta \gamma}\vert_{{}_{\rm ML}}$ & $r_{{}_\mathrm{ML}}$ \\     \hline\hline
${\rm KMIII}$ - ${\rm HI}$ & $261.90 \,(\pm 110.24)$ & $295.11 \,(\pm 155.11)$ & 0.96 & $1.0-\varepsilon$ & 0.95  \\      \hline
${\rm KKLTI}_{\rm stg}$ - ${\rm HI}$ & $262.36 \,(\pm 104.38)$ & $262.26 \,(\pm 37.01)$ & 0.96 & $1.0-\varepsilon$ & 0.95  \\      \hline
${\rm LI}_{\alpha >0}$ - ${\rm HI}$ & $249.77 \,(\pm 99.15)$ & $242.08 \,(\pm 133.01)$ & 0.96 & 0.96 & 0.97  \\      \hline
${\rm RGI}$ - ${\rm HI}$ & $270.64 \,(\pm 133.67)$ & $277.18 \,(\pm 184.55)$ & 0.98 & 0.97 & 0.95  \\      \hline
${\rm KKLTI}_{\rm stg}$ - ${\rm KMIII}$ & $5.01 \,(< 97.20)$ & $41.17 \,(< 389.58)$ & 0.08 & 0.08 & 0.95  \\      \hline
${\rm LI}_{\alpha >0}$ - ${\rm KMIII}$ & $38.09 \,(< 323.48)$ & $89.15 \,(< 521.72)$ & 0.85 & 0.70 & 0.95  \\      \hline
${\rm RGI}$ - ${\rm KMIII}$ & $50.84 \,(< 408.62)$ & $123.43 \,(< 624.76)$ & 0.11 & 0.23 & 0.94  \\      \hline
${\rm LI}_{\alpha >0}$ - ${\rm KKLTI}_{\rm stg}$ & $37.81 \,(< 309.16)$ & $47.81 \,(< 364.24)$ & 0.91 & 0.60 & 0.95  \\      \hline
${\rm RGI}$ - ${\rm KKLTI}_{\rm stg}$ & $50.54 \,(< 397.36)$ & $84.83 \,(< 519.96)$ & 0.16 & 0.18 & 0.94  \\      \hline
${\rm RGI}$ - ${\rm LI}_{\alpha >0}$ & $21.77 \,(< 257.82)$ & $36.40 \,(< 353.30)$ & 0.40 & 0.12 & 0.95  \\      \hline
\end{tabular}}
\resizebox{0.45\columnwidth}{!}{%
\begin{tabular}{|c|c|c|c|c|c|}
\cline{2-6}
\multicolumn{1}{c}{\cellcolor{green!55}} & \multicolumn{5}{|c|}{F2 with $\pi_{\log \epsilon_1} (\bmuf )$}   \\      \hline
\backslashbox{$\calM_\beta$ - $\calM_\gamma$}{$\langle  U\rangle$}  & $\langle \vert \ln {\rm B}_{\beta \gamma}\vert \rangle$ & $\langle \vert \ln {\rm B}_{\beta \gamma} \vert \rangle_{{}_{\rm ML}}$ & $\deci_{\beta \gamma}$ & $\deci_{\beta \gamma}\vert_{{}_{\rm ML}}$ & $r_{{}_\mathrm{ML}}$ \\  \hline\hline
${\rm KMIII}$ - ${\rm HI}$ & $28.11 \,(< 172.58)$ & $71.59 \,(< 490.44)$ & 0.86 & 0.47 & 0.97  \\      \hline
${\rm KKLTI}_{\rm stg}$ - ${\rm HI}$ & $18.18 \,(< 149.76)$ & $51.05 \,(< 204.78)$ & 0.34 & 0.50 & 0.96  \\      \hline
${\rm LI}_{\alpha >0}$ - ${\rm HI}$ & $275.01 \,(\pm 172.90)$ & $96.64 \,(\pm 40.41)$ & 0.97 & $1.0-\varepsilon$ & 0.99  \\      \hline
${\rm RGI}$ - ${\rm HI}$ & $77.96 \,(< 254.88)$ & $26.79 \,(< 219.88)$ & 0.93 & 0.50 & 0.98  \\      \hline
${\rm KKLTI}_{\rm stg}$ - ${\rm KMIII}$ & $40.43 \,(< 222.38)$ & $86.58 \,(< 425.10)$ & 0.80 & 0.58 & 0.95  \\      \hline
${\rm LI}_{\alpha >0}$ - ${\rm KMIII}$ & $298.15 \,(\pm 185.73)$ & $210.27 \,(< 425.30)$ & 0.98 & $1.0-\varepsilon$ & 0.97  \\      \hline
${\rm RGI}$ - ${\rm KMIII}$ & $104.06 \,(< 303.76)$ & $86.86 \,(< 457.74)$ & 0.95 & 0.79 & 0.96  \\      \hline
${\rm LI}_{\alpha >0}$ - ${\rm KKLTI}_{\rm stg}$ & $266.92 \,(\pm 174.48)$ & $160.60 \,(< 385.16)$ & 0.96 & 0.90 & 0.96  \\      \hline
${\rm RGI}$ - ${\rm KKLTI}_{\rm stg}$ & $75.78 \,(< 246.76)$ & $63.35 \,(< 253.52)$ & 0.90 & 0.63 & 0.95  \\      \hline
${\rm RGI}$ - ${\rm LI}_{\alpha >0}$ & $217.27 \,(\pm 158.11)$ & $65.05 \,(< 189.44)$ & 0.97 & 0.94 & 0.98  \\      \hline
\end{tabular}}
\resizebox{0.45\columnwidth}{!}{%
\begin{tabular}{|c|c|c|c|c|c|}
\cline{2-6}
\multicolumn{1}{c}{\cellcolor{brown!90}} & \multicolumn{5}{|c|}{F3 with $\pi_{\log \epsilon_1} (\bmuf )$}   \\      \hline
\backslashbox{$\calM_\beta$ - $\calM_\gamma$}{$\langle  U\rangle$}  & $\langle \vert \ln {\rm B}_{\beta \gamma}\vert \rangle$ & $\langle \vert \ln {\rm B}_{\beta \gamma} \vert \rangle_{{}_{\rm ML}}$ & $\deci_{\beta \gamma}$ & $\deci_{\beta \gamma}\vert_{{}_{\rm ML}}$ & $r_{{}_\mathrm{ML}}$ \\     \hline\hline
${\rm KMIII}$ - ${\rm HI}$ & $5.64 \,(< 127.44)$ & $232.18 \,(< 834.90)$ & - & 0.58 & 0.99  \\      \hline
${\rm KKLTI}_{\rm stg}$ - ${\rm HI}$ & $25.68 \,(< 268.98)$ & $92.59 \,(< 216.98)$ & - & 0.83 & 0.97  \\      \hline
${\rm LI}_{\alpha >0}$ - ${\rm HI}$ & $4.41 \,(< 100.92)$ & $4.06 \,(\pm 3.81)$ & - & 0.31 & 0.99  \\      \hline
${\rm RGI}$ - ${\rm HI}$ & $6.28 \,(< 130.06)$ & $22.58 \,(< 226.72)$ & - & 0.37 & 0.99  \\      \hline
${\rm KKLTI}_{\rm stg}$ - ${\rm KMIII}$ & $24.75 \,(< 270.18)$ & $141.92 \,(< 474.28)$ & - & 0.80 & 0.97  \\      \hline
${\rm LI}_{\alpha >0}$ - ${\rm KMIII}$ & $3.81 \,(< 109.54)$ & $218.19 \,(< 817.54)$ & - & 0.45 & 0.99  \\      \hline
${\rm RGI}$ - ${\rm KMIII}$ & $5.12 \,(< 132.58)$ & $222.51 \,(< 812.02)$ & - & 0.30 & 0.99  \\      \hline
${\rm LI}_{\alpha >0}$ - ${\rm KKLTI}_{\rm stg}$ & $23.74 \,(< 261.16)$ & $88.34 \,(< 219.20)$ & - & 0.79 & 0.96  \\      \hline
${\rm RGI}$ - ${\rm KKLTI}_{\rm stg}$ & $24.88 \,(< 270.06)$ & $91.21 \,(< 243.78)$ & - & 0.76 & 0.96  \\      \hline
${\rm RGI}$ - ${\rm LI}_{\alpha >0}$ & $2.14 \,(< 83.62)$ & $15.23 \,(< 196.92)$ & - & 0.07 & 0.99  \\      \hline
\end{tabular}}
\resizebox{0.45\columnwidth}{!}{%
\begin{tabular}{|c|c|c|c|c|c|}
\cline{2-6}
\multicolumn{1}{c}{\cellcolor{black!65}} & \multicolumn{5}{|c|}{F4 with $\pi_{\log \epsilon_1} (\bmuf )$}  \\      \hline
\backslashbox{$\calM_\beta$ - $\calM_\gamma$}{$\langle  U\rangle$}  & $\langle \vert \ln {\rm B}_{\beta \gamma}\vert \rangle$ & $\langle \vert \ln {\rm B}_{\beta \gamma} \vert \rangle_{{}_{\rm ML}}$ & $\deci_{\beta \gamma}$ & $\deci_{\beta \gamma}\vert_{{}_{\rm ML}}$ & $r_{{}_\mathrm{ML}}$ \\     \hline\hline
${\rm KMIII}$ - ${\rm HI}$ & $36.00 \,(< 324.80)$ & $541.00 \,(\pm 334.40)$ & - & $1.0-\varepsilon$ & $1.0-\varepsilon$  \\      \hline
${\rm KKLTI}_{\rm stg}$ - ${\rm HI}$ & $43.24 \,(< 357.98)$ & $481.08 \,(\pm 184.16)$ & - & $1.0-\varepsilon$ & 0.97  \\      \hline
${\rm LI}_{\alpha >0}$ - ${\rm HI}$ & $22.82 \,(< 257.78)$ & $354.80 \,(\pm 140.62)$ & - & $1.0-\varepsilon$ & $1.0-\varepsilon$  \\      \hline
${\rm RGI}$ - ${\rm HI}$ & $34.43 \,(< 315.94)$ & $331.07 \,(\pm 224.09)$ & - & $1.0-\varepsilon$ & $1.0-\varepsilon$  \\      \hline
${\rm KKLTI}_{\rm stg}$ - ${\rm KMIII}$ & $15.58 \,(< 198.24)$ & $195.91 \,(< 449.30)$ & - & 0.98 & 0.97  \\      \hline
${\rm LI}_{\alpha >0}$ - ${\rm KMIII}$ & $29.18 \,(< 294.32)$ & $437.89 \,(\pm 386.77)$ & - & $1.0-\varepsilon$ & $1.0-\varepsilon$  \\      \hline
${\rm RGI}$ - ${\rm KMIII}$ & $18.94 \,(< 238.12)$ & $299.98 \,(< 847.90)$ & - & 0.91 & 0.99  \\      \hline
${\rm LI}_{\alpha >0}$ - ${\rm KKLTI}_{\rm stg}$ & $31.98 \,(< 300.82)$ & $307.15 \,(\pm 220.99)$ & - & $1.0-\varepsilon$ & 0.97  \\      \hline
${\rm RGI}$ - ${\rm KKLTI}_{\rm stg}$ & $20.93 \,(< 240.40)$ & $191.94 \,(\pm 183.46)$ & - & 0.97 & 0.97  \\      \hline
${\rm RGI}$ - ${\rm LI}_{\alpha >0}$ & $19.99 \,(< 231.98)$ & $131.58 \,(< 420.74)$ & - & 0.98 & $1.0-\varepsilon$  \\      \hline
\end{tabular}}
%
\caption[Expected utilities for F1-4 experiments]{~\label{tab:f14-utilities} Computed expected utilities for the F1-4 experiments in the case where the $\pi_{\log \epsilon_1} (\bmuf )$ prior is used (see \Eq{eq:predictive-priors-logeps1}). Note that $\varepsilon$ reminds the reader that the value is subject to rounding errors of up to $0.005$. Values in brackets $\pm$ around each computed expected utility correspond to the $1$-$\sigma$ uncertainties, which are evaluated using \Eq{eq:secondmomentu}. This symmetric error about our different expected values for $\vert \ln {\rm B}_{\beta \gamma} \vert$ is replaced with a $2$-$\sigma$ upper bound (because it is positive by definition) if the lower error is greater than the expected value itself.}
\end{table}
\end{landscape}
\newpage

\section{\textsf{The single-field models}} \label{sec:models}

The observational predictions from each of the models defined below have all been calculated using the publicly available \texttt{ASPIC} library: \href{http://cp3.irmp.ucl.ac.be/~ringeval/aspic.html}{http://cp3.irmp.ucl.ac.be/~ringeval/aspic.html}. The model priors were obtained from \Ref{Martin:2013nzq} and we have also provided arguments for the choice of each model as representatives of the full sample.

\textbf{Higgs Inflation (HI)}, as in \Eq{eq:starobinsky-pot-example}, has the following potential
\begin{equation} \label{eq:higgs-pot}
\ V = M^4 \left[ 1-\exp \left( -\sqrt{\frac{2}{3}}\frac{\phi}{\Mp}\right) \right] \,,
\end{equation}
and was chosen in our analysis of plateaus to represent models with a relatively large tensor-to-scalar ratio. In addition, the fact that it is effectively a 0-free-parameter model is attractive with respect to Bayesian inference.

\textbf{Loop Inflation (${\rm LI}_{\alpha >0}$)} with a particular prior choice for the $\alpha$ parameter
\begin{equation}
\ V = M^4 \left[ 1 + \alpha \ln \left( \frac{\phi}{\Mp} \right) \right] \,, \qquad \log (\alpha ) \in [\log (0.003), \log (0.3) ]\,,
\end{equation}
was considered here for its relatively large spectral index, thus ideally providing a decisive tension with the HI and KMIII models in particular.

\textbf{Radion Gauge Inflation (RGI)} was chosen with the following potential and prior
\begin{equation}
\ V = M^4 \frac{(\phi /\Mp )^2}{\alpha + (\phi /\Mp )^2}\,, \qquad \log (\alpha ) \in [-4,4] \,,
\end{equation}
and is a good all-round representative of a standard plateau model that is favoured by observations with a reasonably large tensor-to-scalar ratio. The model is also in a good position between HI and ${\rm LI}_{\alpha >0}$ in values of the spectral index.

\textbf{K\"{a}hler Moduli Inflation II (KMIII)} is a good example of a two-parameter plateau model with the following potential and choices of parameters
\begin{equation}
\ V = M^4 \left[ 1 - \alpha \frac{\phi}{\Mp}\exp \left( -\beta \frac{\phi}{\Mp}\right) \right] \,, \qquad \log ({\cal V}) \in [5,7]\,, \qquad \frac{\alpha}{\beta {\cal V}} \in [0.2,5]\,,
\end{equation}
where one calculates $\beta = {\cal V}^{2/3}$ and sets $\alpha$ through the ratio $\alpha /(\beta {\cal V})$. This model also has a much lower order of magnitude for the tensor-to-scalar ratio in comparison with the three above, mapping out a more complete region of the $(\nS ,r)$-diagram.

\textbf{Kachru-Kallosh-Linde-Trivedi Inflation (${\rm KKLTI}_{\rm stg}$)} phenomenologically interpolates between much of the currently available parameter space with the potential and the following potential and priors
\begin{equation}
\ V = \frac{M^4}{1+\left( \frac{\mu}{\phi} \right)^{4}} \,, \qquad \log \left( \frac{\mu}{\Mp}\right) \in [-6,\log (2)]\,,
\end{equation}
thus it is a good final addition to our small sample of models.

A summary plot of the available parameter space on the $(\nS , r)$-diagram for each of the models is shown in \Fig{fig:nsr-plot}, where it is immediately clear that we have selected a reasonable sample of single-field models to span the available parameter space.

\section{\textsf{Computational methods in} \texttt{foxi}}
\label{sec:foxi-computation}

\begin{figure}
\begin{center}
\includegraphics[width=14cm]{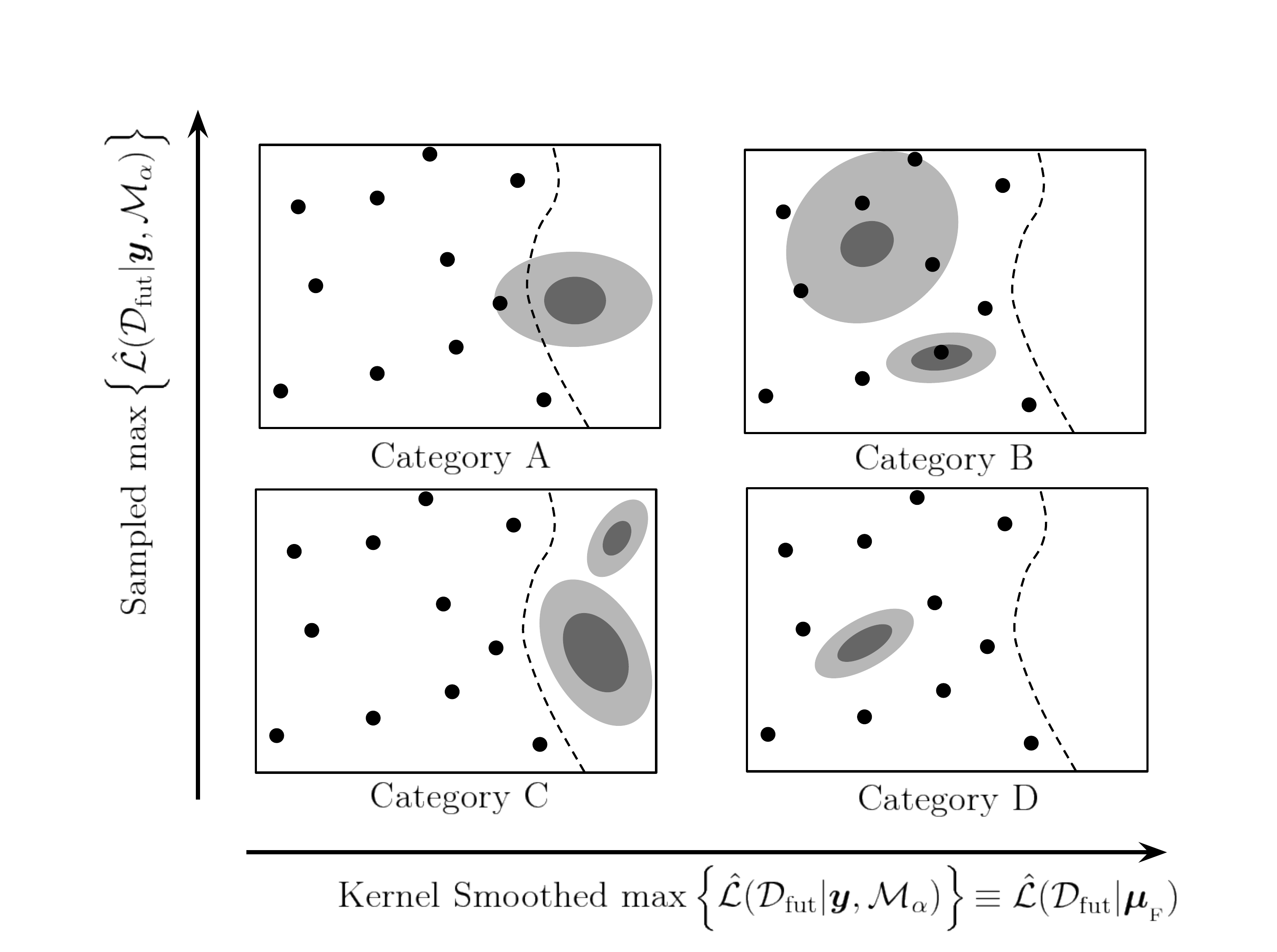}
\caption[Categories of situation in the numerical pipeline]{~\label{fig:foxi-diagram-categories} A diagram depicting 4 unique categories of scenario practically encountered in the computation of the Bayesian evidence using the approximation \Eq{eq:lnB-approx-discrete}. The black dots signify the prior chain samples, the shaded contours are the $1$-$\sigma$ and $2$-$\sigma$ limits of the future likelihood modeled with a Gaussian and the region to the left of the dotted curved line in all 4 boxes indicates the outer contour of the kernel-smoothed prior density using the samples and \Eq{eq:kde}. Boxes further to the right have larger maximum likelihood values contained within the prior obtained from kernel smoothing and boxes further upward have larger maximum likelihood values using the prior samples directly. Category A arises from only a mild overlap between the kernel-smoothed density and the future likelihood contour. Category B denotes either the future likelihood contour is quite large or is small but serendipitously centred directly over a $\bmuf$ point. Category C situations produce Bayesian evidences that are rightfully considered to be always ruled out beyond the Jeffrey's threshold. Category D situations have a very small future likelihood contour --- below the typical inter-point distance of the prior samples.  }
\end{center}
\end{figure}

In \Fig{fig:foxi-diagram-categories} we provide a reference diagram illustrating the various situations which arise during computation of the utility functions in the main body of the section. In particular, the Bayesian evidence approximation of \Eq{eq:lnB-approx-discrete}  practically requires the integration over the probability densities described by both a Gaussian function and prior samples. These distributions can be easily combined when the future likelihood described by the Gaussian function has relatively wide $1$-$\sigma$ contour limits compared to the typical inter-point distance of the prior chains --- such as is true for the category A situations depicted in \Fig{fig:foxi-diagram-categories} and some situations within category B.

Category D (and category B points with a relatively small error contour) represent situations where we must adopt a different computational approach. A convenient non-parametric method is to approximate the model prior probability density $\barpi (\boldx \vert \calM_\alpha )$ using Kernel Density Estimation
\begin{equation} \label{eq:kde}
\barpi (\boldx \vert \calM_\alpha ) \simeq \frac{1}{Z_\alpha} \sum_{\boldx_i \in \{ \calM_\alpha \, {\rm chains} \}} {\cal K}_{\boldsymbol{w}} (\boldx ,\boldx_i)\,,
\end{equation}
or `kernel smoothing', as illustrated in the right-hand column of boxes in \Fig{fig:foxi-diagram-categories}. $Z_\alpha$ in \Eq{eq:kde} is simply the number of samples within the Markov chains representing the prior of $\calM_\alpha$. In this section, the Kernel ${\cal K}_{\boldsymbol{w}}$ we select is simply a Gaussian function
\begin{equation}
\ {\cal K}_{\boldsymbol{w}}(\boldsymbol{a},\boldsymbol{b}) = (2\pi )^{-\frac{n}{2}} \left( \prod^n_{i=1} w^i \right)^{-1} \exp \left[ - \sum^n_{i=1}\frac{(a^i-b^i)^2}{2(w^i)^2} \right] \,,
\end{equation}
with bandwidth vector $\boldsymbol{w}$. Though Category D situations are easily identifiable because the maximum likelihood obtained from direct samples is much lower than the kernel-smoothed equivalent, in general, we have to use an optimal estimate\footnote{In our case we use the in-built Least-Squares Cross-Validation (LSCV) method implemented in the \href{http://www.statsmodels.org/dev/install.html}{\texttt{statsmodels}} package in python. LSCV is based on minimising the integrated square error between the estimated distribution $f_{\rm est}\propto \sum {\cal K}_{\boldsymbol{w}}$ and the underlying true distribution $f_{\rm true}$ i.e. minimising
\begin{equation}
\int_{\boldz \in \mathtt{R}^n} \left[ \frac{1}{M}\sum_{\boldz_i \in \left\{ {\rm Samples} \right\} } {\cal K}_{\boldsymbol{w}}(\boldz ,\boldz_i) - f_{\rm true}(\boldz )\right]^2 \dd \boldz \,,
\end{equation}
with $M$ samples, by minimising Silverman's~\cite{1986desd.book.....S} estimator
\begin{equation}
\ S = \int_{\boldz \in \mathtt{R}^n} \frac{1}{M^2} \left[ \sum_{\boldz_i \in \left\{ {\rm Samples} \right\} } {\cal K}_{\boldsymbol{w}}(\boldz ,\boldz_i) \right]^2 \dd \boldz - \frac{2}{M} \sum_{\boldz_j \in \{ {\rm Samples}\}} \sum_{ \forall \boldz_i\neq \boldz_j} {\cal K}_{\boldsymbol{w}}(\boldz_j ,\boldz_i) \,.
\end{equation}} of $\boldsymbol{w}$ to identify whether kernel smoothing is necessary in Category B i.e. if we are in regions where the local density of points is too sparse, we will find that one or more of the dimensions within $\boldsymbol{w}$ will fall outside the corresponding dimension of the $1$-$\sigma$ futuristic likelihood contour.

In the limit where the futuristic likelihood contour is very small compared with the typical $\boldsymbol{w}$ one finds for the smoothed prior chains, to good approximation we find that the local value $\barpi (\boldx \vert \calM_\alpha ) \propto {\rm const.}$ and therefore we need only compute the evidence (and the maximum likelihood point) using a single prior value centred at the $\bmuf$ point
\begin{equation} \label{eq:evidence-kde-category-b-d}
\left. {\cal E}_\alpha \, \right\vert_{\boldsigma \ll  \boldsymbol{w}} \simeq \barpi (\bmuf \vert \calM_\alpha ) \simeq \frac{1}{Z_\alpha} \sum_{\boldx_i \in \{ \calM_\alpha \, {\rm chains} \}} {\cal K}_{\boldsymbol{w}} (\bmuf ,\boldx_i) \,.
\end{equation}
Though this estimate can be shown to be very accurate, the \href{https://sites.google.com/view/foxicode}{\texttt{foxi}} algorithm itself actually computes the Bayesian evidence in the regime of some category B and all category D situations by implementing the combined approach of both \Eq{eq:evidence-kde-category-b-d} and drawing typically 1000 samples from the future likelihood (\Eq{eq:future-gaussian-assumption}) to sum over for the integral. This method is more computationally robust than \Eq{eq:evidence-kde-category-b-d} alone since it can accommodate for scenarios where the magnitudes of error in each dimension in $\boldsigma$ are very different, offering greater flexibility to the algorithm, at a cost of some additional computation time and efficiency.

\begin{table}
\centering
\resizebox{12cm}{!}{
\begin{tabular}{|c|c|c|c|c|c|}
\cline{1-6}
\multicolumn{2}{|c|}{Survey}  & {\rm Ave. Category A} & {\rm Ave. Category B} & {\rm Ave. Category C} & {\rm Ave. Category D} \\     \hline\hline
\cellcolor{yellow!55} & P1 & 84.1 \% & $0.0+\varepsilon$ \% & 21.9 \% & $0.0 +\varepsilon$ \% \\      \hline
\cellcolor{red!55} & P2 & 3.7 \% & 0.8 \% & 95.5 \% & $0.0 +\varepsilon$ \%  \\      \hline
\cellcolor{blue!35} & F1 & 2.8 \% & 0.8 \% & 96.4 \% & $0.0 +\varepsilon$ \%  \\      \hline
\cellcolor{green!55} & F2 & 1.3 \% & 0.6 \% & 97.9 \% & 0.2 \%  \\     \hline
\cellcolor{brown!90} & F3 & 0.5 \% & 0.4 \% & 98.8 \% & 0.3 \%  \\
\hline
\cellcolor{black!65} & F4 & $0.0 +\varepsilon$ \% & 0.1 \% & 99.3 \% & 0.6 \%  \\   \hline
\end{tabular}}
\caption[Percentages of realisation in each category of the numerical pipeline]{\label{tab:percentage-points-categories} The percentage number of $\bmuf$ points in the Markov chains representing the \emph{Planck} data that correspond to the computational situations defined in \Fig{fig:foxi-diagram-categories}. Note that $\varepsilon$ reminds the reader that the value is subject to rounding errors of up to $0.005$. }
\end{table}

\section{\textsf{Checking for numerical robustness}} \label{sec:gaussian-assumption-limitations}

This section aims to quantify empirically the accuracy of the Gaussian assumption used throughout this section with respect to the direct applicability of our mock forecasts to `real-world' surveys. Note that we are not suggesting that the assumption is `incorrect' in any sense, but that by definition, forecasting using the Gaussian assumption does not necessarily coincide with a true likelihood that would be obtained from a specific survey forecast.

We compared our results for each model pair using \Eq{eq:lnB-approx-discrete} with those obtained from the \href{http://www.mrao.cam.ac.uk/software/multinest/}{\texttt{MultiNest}}~\cite{Feroz:2008xx, Martin:2013nzq} algorithm in each case, where we obtained both $\bmuf$ and $\boldsigma$ for \Eq{eq:lnB-approx-discrete} through the prior samples and a Gaussian likelihood with mean and marginalised variances computed from the chains\footnote{The specifications used to forecast the likelihood for LiteCOrE are given in \Ref{Finelli:2016cyd} and correspond to what is referred to as `LiteCORE-120'.} used by \texttt{MultiNest}, respectively. A comparison is in Table \ref{tab:planck-foxi-comparison} for the \emph{Planck} 2015 data~\cite{Ade:2015oja}, where there is good general agreement up to the $\ln {\rm B}_{\beta \gamma} \pm 0.6$ level, and the forecast data for the LiteCOrE forecast dataset~\cite{DiValentino:2016foa,Finelli:2016cyd} using HI fixed with $T_{\rm reh}=10^6{\rm GeV}$ as the fiducial model, where there is less consistent agreement up to the $\ln {\rm B}_{\beta \gamma} \pm 5.0$ level, which is significantly smaller than the typical amplitude of the 2-$\sigma$ uncertainties over $\vert \ln {\rm B}_{\beta \gamma}\vert$ for a P2 experiment.

\begin{table}
\centering
\resizebox{12cm}{!}{
\begin{tabular}{|c|c|c|c|c|}
\cline{1-5}
 & \multicolumn{2}{c|}{\emph{Planck} 2015} & \multicolumn{2}{c|}{LiteCOrE (HI fiducial)}  \\      \hline
\backslashbox{$\calM_\beta$ - $\calM_\gamma$}{$\ln {\rm B}_{\beta \gamma}$}  & Gaussian & \href{ http://www.mrao.cam.ac.uk/software/multinest/}{\texttt{MultiNest}}~\cite{Feroz:2008xx, Martin:2013nzq} & Gaussian & \href{ http://www.mrao.cam.ac.uk/software/multinest/}{\texttt{MultiNest}} \\ \hline\hline
${\rm KMIII}$ - ${\rm HI}$ & 0.11 & 0.04 & -3.52 & -7.63   \\      \hline
${\rm KKLTI}_{\rm stg}$ - ${\rm HI}$ & -0.57 & -0.44 & -3.66 & -8.02    \\      \hline
${\rm LI}_{\alpha > 0}$ - ${\rm HI}$ & -2.33 & -2.48 & -18.11 & -17.89  \\      \hline
${\rm RGI}$ - ${\rm HI}$ & -0.92 & -0.68 & -4.32 & -4.63   \\      \hline
${\rm KKLTI}_{{\rm stg}}$ - ${\rm KMIII}$ & -0.68 & -0.48 & -0.14 & -0.39   \\      \hline
${\rm LI}_{\alpha > 0}$ - ${\rm KMIII}$ & -2.44 & -2.51 & -14.60 & -10.25  \\      \hline
${\rm RGI}$ - ${\rm KMIII}$ & -1.03 & -0.71 & -0.80 & 3.00  \\ \hline
${\rm LI}_{\alpha > 0}$ - ${\rm KKLTI}_{{\rm stg}}$ & -1.76 & -2.04 &  -14.45 & -9.86 \\      \hline
${\rm RGI}$ - ${\rm KKLTI}_{{\rm stg}}$ & -0.35 & -0.23 & -0.67 & 3.39  \\      \hline
${\rm RGI}$ - ${\rm L7I}_{\alpha > 0}$ & 1.41 & 1.81 & 13.79 & 13.25  \\      \hline
\end{tabular}}
\caption[Comparison of approximate method with known codes]{~\label{tab:planck-foxi-comparison} A comparison table showing the differences between Bayes factors approximated with a Gaussian assumption (denoted `Gaussian') to those obtained from the \href{ http://www.mrao.cam.ac.uk/software/multinest/}{\texttt{MultiNest}}~\cite{Feroz:2008xx, Martin:2013nzq, Ringeval:2013lea} algorithm in each case of model pair for the \emph{Planck} 2015 and forecast LiteCOrE~\cite{DiValentino:2016foa,Finelli:2016cyd} (with Higgs Inflation as a fiducial model) datasets. }
\end{table}

\begin{figure}
\begin{center}
\includegraphics[width=10cm]{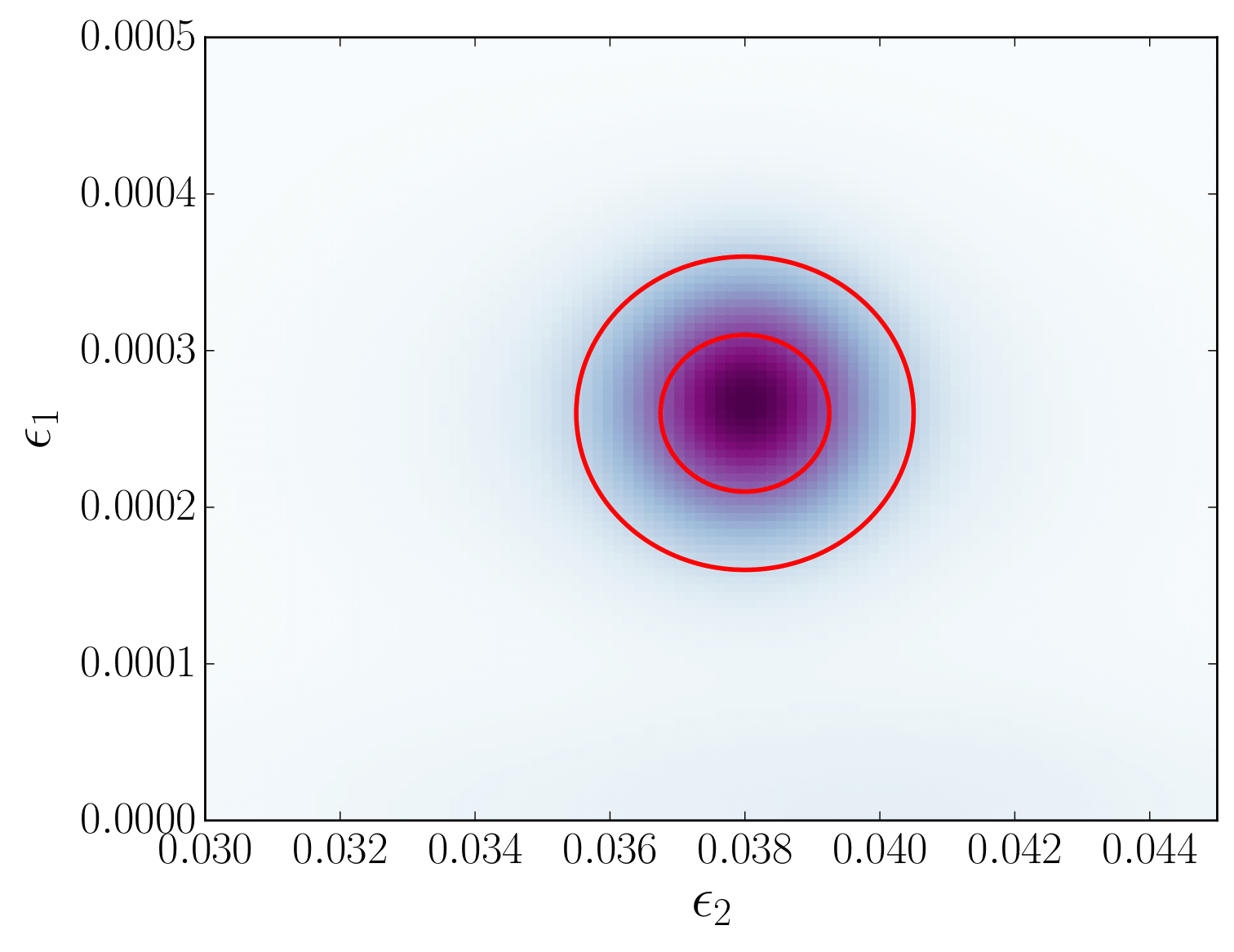}
\caption[Indicated forecast contours for the LiteCOrE experiment]{~\label{fig:plot_eps1_eps2} A probability density plot indicating the shape of the LiteCOrE forecast likelihood (in purple) over a $(\epsilon_1,\epsilon_2)$ surface marginalised from the full $(\epsilon_1,\epsilon_2,\epsilon_3)$ space, illustrating the comparison with our Gaussian likelihood (the red contours). }
\end{center}
\end{figure}
%



When comparing the values from \texttt{MultiNest} and our method, we note that the former method is permitted many more samples from the model (in order to converge the integral for the Bayesian evidence) than the latter (which must limit the number of samples because many more computations of the same integral are required). Hence, the disagreement in values between the two methods that is not limited by the Gaussian likelihood assumption itself is likely to originate from this limitation of our computational resources.

The differences between the uncorrelated Gaussian likelihood and the sampled likelihood forecast for LiteCOrE (using the $\log \epsilon_1$ prior) are minute in the slicing of $(\epsilon_1,\epsilon_2)$-space depicted by \Fig{fig:plot_eps1_eps2}. Therefore, inaccuracies that can appear in the Bayesian evidence that arise from an imprecise analogy between a more realistic likelihood forecast and our mock forecasts are clearly far smaller than the disagreement that comes from our limited computational resources. The points for which the methods are in most disagreement are Category B and D (see Appendix~\ref{sec:foxi-computation}), since they are characterised by a poor inter-point distance, but these points are sampled only very occasionally (see Table \ref{tab:percentage-points-categories}) and so we can expect minimal impact on our main conclusions in this section.

We shall leave the future application of our formalism to a proposed survey, such as COrE~\cite{Finelli:2016cyd}, for later work.

\section{\textsf{Identifying the constraint on $\alphaS$}}
\label{sec:alphaS-calculation}

To leading-order in the slow-roll expansion, the running of the scalar spectral index $\alphaS$ is given by \Eq{eq:alphaS-intro}. When no cross-correlations are observed --- as is the assumption in all of the forecast constraints in this section --- it can be shown that the generic cross-correlator from such a measurement reduces down to factors of correlators
\begin{align} \label{eq:alphaSection-cross-correlators}
\langle \epsilon_1^l \epsilon_2^m \epsilon_3^n \rangle &= \langle \epsilon_1^l\rangle \langle \epsilon_2^m\rangle \langle \epsilon_3^n\rangle\,.
\end{align}
For a Gaussian measurement on each of the slow-roll parameters, the fiducial point $\muf^{\alphaS} = \muf^{\alphaS}(\bmuf , \boldsigma ) \equiv \langle \alphaS \rangle$ can be derived from
\begin{align}
\muf^{\alphaS} &\simeq - 2\langle \epsilon_1 \epsilon_2 \rangle - \langle \epsilon_2 \epsilon_3 \rangle \,, \\
&\simeq  - 2\muf^1 \muf^2 - \muf^2 \muf^3 \label{eq:alphaS-fiducial-point-formula} \,.
\end{align}
The error bar of the measurement over $\alphaS$ can thus be unpacked into an expression containing only the fiducial points and error bars on the slow-roll parameters, i.e. $\sigma^{\alphaS} = \sigma^{\alphaS}(\bmuf , \boldsigma )$
\begin{align}
\ (\sigma^{\alphaS})^2 &\equiv \langle \alphaS^2\rangle - \langle \alphaS \rangle^2 \nonumber \\
&\simeq \left\langle (2\epsilon_1 \epsilon_2 + \epsilon_2 \epsilon_3)^2\right\rangle - \left(  2\langle \epsilon_1 \epsilon_2 \rangle + \langle \epsilon_2 \epsilon_3 \rangle \right)^2 \nonumber  \\
&\simeq 4\langle \epsilon_1^2\rangle \langle \epsilon_2^2\rangle + \langle \epsilon_2^2\rangle \langle \epsilon_3^2\rangle + 4 \langle \epsilon_1\rangle \langle \epsilon_2^2\rangle  \langle \epsilon_3\rangle - 4\langle \epsilon_1\rangle^2 \langle \epsilon_2\rangle^2 - \langle \epsilon_2\rangle^2 \langle \epsilon_3\rangle^2 - 4 \langle \epsilon_1\rangle \langle \epsilon_2\rangle^2  \langle \epsilon_3\rangle \nonumber \\
&\simeq 4 (\sigma^1 )^2(\sigma^2 )^2 + 4 (\muf^1)^2(\sigma^2)^2 + 4 (\muf^2)^2(\sigma^1)^2 + (\sigma^2 )^2(\sigma^3 )^2 \nonumber \\
& \qquad \qquad \qquad + (\muf^2)^2(\sigma^3)^2 + (\muf^3)^2(\sigma^2)^2 + 4 \muf^1 \muf^3 (\sigma^2)^2 \label{eq:alphaS-width-formula}\,.
\end{align}
Using \Eq{eq:alphaS-fiducial-point-formula} and \Eq{eq:alphaS-width-formula} for a specified collection of 1-$\sigma$ error bars on $\epsilon_1$, $\epsilon_2$ and $\epsilon_3$, we may identify all of the remaining fiducial points $\bmuf$ that satisfy a $2$-$\sigma$ measurement of $\alphaS$ and can therefore compute the probability defined in \Eq{eq:alphaS-probability}.

\end{subappendices}

\chapter{\textsf{Discussion and conclusions}}
\label{sec:conclusions-whole} \niceline {\vskip+1ex} 

\begin{center}
\fbox{\parbox[c]{13cm}{\vspace{1mm}{\textsf{\textbf{Abstract.}}} This thesis has demonstrated how, even when observations prove indecisive to learning about the inflationary paradigm, one can still extract valuable information about the physics of inflation by considering the precise predictions of well-motivated models. Adding to this careful study, we have explored the possible futures which observations might guide the theoretical developments toward. In this final chapter, we conclude with a summary of all of the results obtained in this thesis, an overview of their significance to the field of research and a discussion of future possible directions that the work could take.\vspace{1mm}}}
\end{center}

\section{\textsf{Outlining the results}} \label{sec:results-outline-main}

It is crucial to learning about the physics of inflation that the best inflationary models are studied carefully for their potentially unique observational characteristics and then compared to current observations in a statistically rigorous way. It has been the principle aim of this thesis to study the observational modifications to inflation that arise from the introduction of additional scalar degrees of freedom and, with those predictions, perform a statistical analysis in order to compare them to the available data.

In \Chap{sec:curvaton-reheating} we introduced the curvaton model as an alternative reheating model from which we obtained distinct observational predictions to the standard single-field setup. Using Bayesian inference, we demonstrated that the reheating temperatures one generally infers from CMB perturbations are lower in the case of curvaton models, where one also obtains more information on the exact value of the reheating temperature with the latter.

The initial conditions to the curvaton as well as all other scalar fields potentially sourced from an inflationary background were extensively reviewed and studied in detail to include new effects arising from couplings to multiple fields, alternative spectator potentials and generic slow roll inflationary backgrounds in \Chap{sec:infra-red-divergences}. The developments made in this chapter lead us to build detailed models for post-inflationary phenomenology (namely, the curvaton and freeze-in dark matter models) and to discover powerful new probes of inflation itself in \Chap{sec:isocurvature-fields}.

Lastly, in \Chap{sec:future-prospects} we discussed the future prospects for inflationary model selection. In the process we developed a new Bayesian experimental design formalism which incorporates toy survey configurations into a forecast for model selection and information gain performance. We found in particular that the most likely observable to optimise model selection between single-field inflationary models, through an order of magnitude precision improvement in the future, will be the scalar spectral index. 

\subsection{\textsf{Impact on the scientific community}}
\label{sec:impact-sci-comm}

The potential ramifications of the results here are broad with respect to building scalar field models of dark matter~\cite{Enqvist:2017kzh, Markkanen:2018gcw}, dark energy~\cite{Glavan:2017jye} and Higgs dynamics where the initial conditions must be specified from inflation. The effect of our work in \Sec{sec:axion_spec} on the QCD axion was recently taken into account in \Ref{Graham:2018jyp}, where low-scale inflation was found to permit axions with a lower mass range than previously thought ($\sim 10^{-12}{\rm GeV}$). 

In \Chap{sec:curvaton-reheating} we studied the effect on the reheating temperature inferred by CMB observations by including an additional field. A multi-field extension to our analysis was conducted in \Ref{Hotinli:2017vhx}, where it was found that post-inflationary curvaton behaviour obtained observables with the greatest distinguishability from standard single field reheating.

\section{\textsf{Future directions}} \label{sec:future-Directions-main}

We shall conclude here with a brief discussion of potential future areas of research based on the results of this thesis.

\subsection{\textsf{Dark matter initial conditions}}
\label{sec:dm-init-cond-fd}

The freeze-in real singlet scalar dark matter model of \Chap{sec:isocurvature-fields} is among the simplest possible cases of dark matter generation using inflation as the primary source for the field. Though its portal coupling to the Higgs is $\lambda_{hs} < 10^{-7}$ by construction, we have already demonstrated that this is well above the critical coupling value below which the two spectator fields can be treated as separable in the Fokker-Planck equation (this was calculated in \Sec{sec:dec-coup}). The initial conditions of each field should therefore be recalculated numerically to take this effect into account. In the same vein, one might consider the possibility of extending the model to many more fields and performing a Bayesian inference on its predictions with the same principles as in \Chap{sec:curvaton-reheating}. Fermionic extensions are also possible and interesting to consider~\cite{Heikinheimo:2016yds} as well as scenarios with a dominant non-minimal coupling term, as we discussed in \Sec{sec:nonmin_coup_spec}.

\subsection{\textsf{Higgs stability}}
\label{sec:higgs-stab-fd}

The SM Higgs vacuum is known to be unstable during inflation at a higher energy scale than $\sim {\cal O}(10^{10}){\rm GeV}$~\cite{Herranen:2014cua,Markkanen:2018bfx} unless there is an $\sim {\cal O}(1)$ non-minimal coupling. Due to their coupling to the Higgs, the gravitational generation of light top quarks has been shown to affect this instability criterion~\cite{Rodriguez-Roman:2018swn}. Similarly to~\cite{Gong:2017mwt}, it would be interesting to numerically explore the couplings of scalar dark matter required to do the same given the updates to the initial condition implied by \Chap{sec:infra-red-divergences}. We imagine this to be either a modification to the inflationary background that includes generic slow-roll in $H$ and thus a possibility to leave equilibrium, or in the equilibrium limit, the inclusion of additional scalars requiring full numerical evaluation due to the probability current issue described in \Sec{sec:non-vanish}. Trilinear couplings to the Higgs are also of interest to the question of stability~\cite{Ema:2017ckf} as well as a delay in the reheating decay efficiency of the inflaton through Higgs thermal blocking~\cite{Freese:2017ace}.

\subsection{\textsf{New gravitational wave signals}}
\label{sec:gw-new-sig-fd}

We also draw attention to a particular class of Axion-SU(2) model, originally proposed as `Natural Inflation'~\cite{Freese:1990rb}, which has evolved into what is known as `Chromo-Natural Inflation'~\cite{Adshead:2012kp} and has recently been studied as a spectator field during inflation~\cite{Dimastrogiovanni:2016fuu,Dimastrogiovanni:2018xnn}. Due to parity violation of the SU(2)-gauge field background that the axion is coupled to via a Chern-Simons term, this model is known to predict a chiral primordial gravitational wave spectrum. It remains an interesting project to further analyse the axion dynamics, and their effect on the spectrum for gravitational waves produced, in the context of our work in \Sec{sec:axion_spec}.

\subsection{\textsf{Survey design}}
\label{sec:survey-des-fd}

In \Chap{sec:conclusions-whole} we entered the new territory of Bayesian experimental design for model selection in the context of cosmological experiments. Our analysis could be performed for a specific survey by specifying more detail in the functional form of ${\cal D}_{\rm fut}$ in \Eq{eq:exu} that includes detector behaviour. Extensions in this regard might include analytic approximations such as those made by Refs.~\cite{Sellentin:2014zta,Sellentin:2015axa}. A more speculative, though interesting alternative may arise from the application of Information Geometry~\cite{amari2009information, Brehmer:2016nyr}.

Given a set of financial constraints and a fully characterised detector behaviour, it would also be straightforward to translate our formalism into searching for optimal specifications of a survey (e.g. number of detectors, frequency channels, noise sensitivity, angular resolution, telescope size, etc...). An optimisation problem of this kind would require some change in numerical methodology, however, due to the computational expense of efficiently scanning the search space of many survey designs.



\bibliographystyle{JHEP} 
\newpage
\phantomsection
\addcontentsline{toc}{chapter}{\textsf{Bibliography}}
\bibliography{ThesisRefs}

\end{document}